\documentclass[ALICE,manyauthors]{alice_public_notes}
\usepackage{geometry}
\usepackage[english,greek]{babel}
\usepackage[parfill]{parskip}
\usepackage{graphicx}
\usepackage{grffile}
\usepackage{float}
\usepackage{amssymb}
\usepackage{eurosym}
\usepackage{epstopdf}
\usepackage{multirow}
\usepackage{color}
\usepackage{footmisc}
\usepackage{cite}
\usepackage{booktabs}
\usepackage{url}
\usepackage{upgreek}
\usepackage{amsfonts}
\usepackage{amsmath}
\providecommand{\abs}[1]{\lvert#1\rvert} 
\usepackage{amssymb}
\usepackage{epsfig}
\usepackage{lineno}
\usepackage{hyperref}
\usepackage{lipsum}
\usepackage{subfigure}
\usepackage[pscoord]{eso-pic}
\usepackage[stamp=false]{draftwatermark}

\usepackage{float}
\usepackage{amssymb}
\usepackage{epstopdf}
\usepackage{color}
\usepackage[utf8]{inputenc}
\usepackage[normalem]{ulem} %
\usepackage[table]{xcolor}
\usepackage{pdflscape}
\usepackage{datetime}
\graphicspath{{./}}
\usepackage[binary-units,range-phrase=--,range-units=single,exponent-product=\times]{siunitx}[=v2]
\usepackage[noprefixcmds,sicmds]{hepunits}
\usepackage{siunitx}
\ifdefined\qty\else
  \ifdefined\NewCommandCopy
    \NewCommandCopy\qty\SI
  \else
    \NewDocumentCommand\qty{O{}mm}{\SI[#1]{#2}{#3}}
  \fi
\fi
\ifdefined\unit\else
  \ifdefined\NewCommandCopy
    \NewCommandCopy\unit\si
  \else
    \NewDocumentCommand\unit{O{}m}{\si[#1]{#2}}
  \fi
\fi
\DeclareSIUnit\clight{\text{\ensuremath{c}}}
\DeclareSIUnit\permille{\text{\textperthousand}}
\usepackage[margin=10pt,font=small,labelfont=bf,labelsep=colon]  {caption}
\newif\ifcomment
\newif\ifoldtext
\newif\ifcompat
\newif\ifextended
\newif\ifdraft
\newif\ifprefinal
\newif\ifirc
\newif\ifcost

\ifcomment
\newcommand{\comment}[1]{\textcolor{orange}{#1}}
\newcommand{\new}[1]{\textcolor{green!70!black}{#1}}
\DraftwatermarkOptions{text=Internal\\draft}
\newenvironment{revised}{\color{magenta}}{}
\else
\newcommand{\new}[1]{#1}
\newcommand{\comment}[1]{}
\DraftwatermarkOptions{text=Draft}
\newenvironment{revised}{}{}
\fi

\DraftwatermarkOptions{text=Internal\\draft}

\DraftwatermarkOptions{color=gray!20}

\usepackage{lineno}
\ifdraft
  \linenumbers
  \usepackage{fancyhdr}
\fi

\newcommand{\ALICETHR}    {{\ensuremath{{\rm ALICE\ 3}}}\xspace}
\newcommand{\Fig}[1]     {Fig.~\ref{#1}}

\newcommand{\detsigma}[1]  {\ensuremath{\sigma {\rm #1}}}
\newcommand{\richsigma}[1] {\ensuremath{\detsigma{_{RICH}}({#1})}}

\newcommand{\Nsigma}  {\ensuremath{n\sigma}}
\newcommand{\detNsigma}[1]  {\ensuremath{\Nsigma {\rm #1}}}
\newcommand{\richNsigma}[1] {\ensuremath{\detNsigma{_{RICH}}({#1})}}
\newcommand{\tofNsigma}[1]  {\ensuremath{\detNsigma{_{TOF}}({#1})}}
\newcommand{\InnerTOF}  {\ensuremath{{\rm bTOF1}}\xspace}
\newcommand{\OuterTOF}  {\ensuremath{{\rm bTOF2}}\xspace}
\newcommand{\RICH}  {\ensuremath{{\rm bRICH}}\xspace}

\begin{document}
\selectlanguage{english}

\newcommand{\RunsThreeFour}{Runs~3 \& 4\Xspace}
\newcommand{\ttbar}{\ensuremath{t\overline{t}}\Xspace}
\newcommand{\gaga}{\gamma\gamma}
\DeclareRobustCommand{\Pepem}{\HepParticle{\Pe}{}{+}\HepParticle{\Pe}{}{-}\Xspace} %
\DeclareRobustCommand{\PGmpGmm}{\HepParticle{\PGm}{}{+}\HepParticle{\PGm}{}{-}\Xspace} %
\newcommand{\sla}{\slash \hspace{-0.2cm}}
\newcommand{\slam}{\slash \hspace{-0.25cm}}
\newcommand{\no}{\nonumber}
\def\lsim{\mathrel{\rlap{\lower4pt\hbox{\hskip1pt$\sim$}}
    \raise1pt\hbox{$<$}}}         %
\def\gsim{\mathrel{\rlap{\lower4pt\hbox{\hskip1pt$\sim$}}
    \raise1pt\hbox{$>$}}}         %
\newcommand{\Dsc}{\ensuremath{D_{\rm s}}\xspace}
\newcommand{\twopiTDsc}{\ensuremath{2 \pi T D_{\rm s}}\Xspace}
\newcommand{\FIXME}{\textbf{FIXME}\xspace}
\newcommand{\JUSTADDED}{\textbf{JUSTADDED}\xspace}
\newcommand{\fixme}{\textbf{FIXME}\xspace}
\newcommand{\delphes}{\textsc{Delphes}\xspace}
\newcommand{\delphesotwo}{\textsc{Delphes}\otwo}
\newcommand{\otwo}{O\ensuremath{^{2}}\xspace}
\newcommand{\PYTHIA}{\textsc{PYTHIA8}\xspace}

\newcommand{\chisquared}{\ensuremath{\chi^{\rm 2}}\xspace}
\newcommand{\significance}{$\frac{\mathrm{S}}{\sqrt{\mathrm{S}+\mathrm{B}}}$\xspace}

\newcommand{\fm}{\ensuremath{\mathrm{fm}}\xspace}
\newcommand{\fmc}{\ensuremath{\mathrm{fm}/c}\xspace}
\renewcommand{\GeV}{\ensuremath{\mathrm{GeV}}\xspace}
\newcommand{\GeVc}{\ensuremath{\mathrm{GeV}/c}\xspace}
\newcommand{\gevc}{\GeVc\xspace}
\newcommand{\GeVcc}{\ensuremath{\mathrm{GeV}/c^{2}}\xspace}
\newcommand{\Tev}{\ensuremath{\mathrm{TeV}}\xspace}
\newcommand{\Mev}{\ensuremath{\mathrm{MeV}}\xspace}
\newcommand{\MeVc}{\ensuremath{\mathrm{MeV}/c}\xspace}
\newcommand{\mum}{\ensuremath{\mathrm{\mu}\rm m}\Xspace}

\newcommand{\enum}[2]{\ensuremath{#1\times10^{#2}}} %
\newcommand{\NQTY}[2]{\mbox{$[#1/{\rm #2}]$}}     %
\newcommand{\UQTY}[2]{\ensuremath{#1/\mathrm{#2}}}  %
\newcommand{\eqty}[3]{\qty{\enum{#1}{#2}}{#3}}  %

\newcommand{\elumi}[2]{\qty{\enum{#1}{#2}}{cm^{-2}s^{-1}}}
\newcommand{\murad}[1]{\qty{#1}{\mu rad}}
\newcommand{\intlumimub}[1]{\qty{#1}{\mu b^{-1}}}

\newcommand{\sqrts}{\ensuremath{\sqrt{s}}\xspace}
\newcommand{\sqrtsNN}{\ensuremath{\sqrt{\sNN}}\xspace}
\newcommand{\sNN}{\ensuremath{s_{\mbox{\tiny NN}}}}

\newcommand{\isotope}[3]{\ensuremath{^{#1}\mathrm{#2}^{#3}}}

\newcommand{\speciesheader}{ &
\isotope{16}{O}{8+}&
\isotope{40}{Ar}{18+}&
\isotope{40}{Ca}{20+}&
\isotope{78}{Kr}{36+}&
\isotope{129}{Xe}{54+}&
\isotope{208}{Pb}{82+}
}

\newcommand{\bfunc}{$\beta$-function}
\newcommand{\bstarval}[1]{$\bstar = #1\,\mbox{m}$}
\newcommand{\betarel}{\ensuremath{\beta_\text{rel}}}
\newcommand{\emittnx}{\ensuremath{\epsilon_{n,x}}}
\newcommand{\emittny}{\ensuremath{\epsilon_{n,y}}}
\newcommand{\emittnxy}{\ensuremath{\epsilon_{n,xy}}}
\newcommand{\emitts}{\ensuremath{\epsilon_s}}
\newcommand{\sigs}{\ensuremath{\sigma_s}}
\newcommand{\sigp}{\ensuremath{\sigma_p}}
\newcommand{\kb}{\ensuremath{k_b}}
\newcommand{\frev}{\ensuremath{f_0}}
\newcommand{\Nb}{\ensuremath{N_b}}
\newcommand{\Eb}{\ensuremath{E_b}}
\newcommand{\emittval}[1]{\ensuremath{\emittn=\qty{#1}{\mu m\,rad}}}
\newcommand{\Nbval}[2]{\ensuremath{\Nb=\enum{#1}{#2}}}
\newcommand{\taul}{\ensuremath{\tau_l}}
\newcommand{\taulval}[1]{\ensuremath{\taul=\qty{#1}{ns}}}
\newcommand{\sigzval}[1]{\ensuremath{\sigz=\qty{#1}{cm}}}
\newcommand{\etev}[1]{\ensuremath{\Eb=\qty{#1}{TeV}}}
\newcommand{\VRF}{\ensuremath{V_{\mathrm{RF}}}}
\newcommand{\lumival}[2]{\ensuremath{L=\qty{#1\times 10^{#2}}{cm^{-2} s^{-1}}}}

\newcommand{\aibsx}{\ensuremath{\alpha_{\mathrm{IBS},x}}}
\newcommand{\aibsy}{\ensuremath{\alpha_{\mathrm{IBS},y}}}
\newcommand{\aibsxy}{\ensuremath{\alpha_{\mathrm{IBS},x,y}}}
\newcommand{\aradd}{\ensuremath{\alpha_{\mathrm{rad}}}}
\newcommand{\aradds}{\ensuremath{\alpha_{\mathrm{rad},s}}}
\newcommand{\araddx}{\ensuremath{\alpha_{\mathrm{rad},x}}}
\newcommand{\araddy}{\ensuremath{\alpha_{\mathrm{rad},y}}}
\newcommand{\araddxy}{\ensuremath{\alpha_{\mathrm{rad},x,y}}}
\newcommand{\Z}{\ensuremath{Z_\text{ion}}}
\newcommand{\Circ}{\ensuremath{C_\text{ring}}}
\newcommand{\lumi}{\ensuremath{\mathcal{L}}}
\newcommand{\Lint}{\ensuremath{\mathcal{L}_{\text{int}}}}
\newcommand{\Lbint}{\ensuremath{L_{b,\text{int}}}}
\newcommand{\Lbpeak}{\ensuremath{\mathcal{L}_{b,\text{peak}}}}

\newcommand{\sigmaBFPP}   {\ensuremath{\sigma_\mathrm{BFPP}}\xspace}
\newcommand{\sigmaNN}   {\ensuremath{\sigma^{NN}_\mathrm{INEL}}\xspace}
\newcommand{\co}[1]       {\relax}
\newcommand{\nl}          {\newline}
\newcommand{\el}          {\\\hline\\[-0.4cm]}

\newcommand{\invnb}{\mathrm{nb}^{-1}}
\newcommand{\invpb}{\mathrm{pb}^{-1}}
\newcommand{\invfb}{\mathrm{fb}^{-1}}

\newcommand{\yNN}{\ensuremath{y_{\mbox{\tiny NN}}}}
\newcommand{\bstar}{\ensuremath{\beta^{*}}}
\newcommand{\emittn}{\ensuremath{\varepsilon_n}}
\newcommand{\LAA}{\ensuremath{L_\text{AA}}}
\newcommand{\LpA}{\ensuremath{L_{pA}}}
\newcommand{\Lpp}{\ensuremath{L_{pp}}}
\newcommand{\Lpeak}{\ensuremath{\hat{L}}}
\newcommand{\LNN}{\ensuremath{ L_{\text{NN}}}}

\newcommand{\ee}          {\ensuremath{\rm e^{+}e^{-}}\xspace}
\newcommand{\ep}          {\ensuremath{\rm e^{-}p}\xspace}
\newcommand{\pp}          {$\mathrm{pp}$\xspace}
\newcommand{\pPb}         {$\mathrm{p}$--$\mathrm{Pb}$\xspace}
\newcommand{\Pbp}         {$\mathrm{Pb}$--$\mathrm{p}$\xspace}
\newcommand{\pO}          {$\mathrm{p}$--$\mathrm{O}$\xspace}
\newcommand{\Op}          {$\mathrm{O}$--$\mathrm{p}$\xspace}
\newcommand{\OO}          {$\mathrm{OO}$\xspace}
\renewcommand{\pA}          {$\mathrm{pA}$\xspace}
\newcommand{\AOnA}        {$\mathrm{AA}$\xspace}
\newcommand{\PbPb}        {$\mathrm{Pb}$--$\mathrm{Pb}$\xspace}
\newcommand{\ArAr}        {$\mathrm{Ar}$--$\mathrm{Ar}$\xspace}
\newcommand{\XeXe}        {$\mathrm{Xe}$--$\mathrm{Xe}$\xspace}
\newcommand{\KrKr}        {$\mathrm{Kr}$--$\mathrm{Kr}$\xspace}
\newcommand{\AuAu}        {$\mathrm{Au}$--$\mathrm{Au}$\xspace}
\newcommand{\CuCu}        {$\mathrm{Cu}$--$\mathrm{Cu}$\xspace}
\newcommand{\pAu}         {$\mathrm{p}$--$\mathrm{Au}$\xspace}
\newcommand{\dAu}         {$\mathrm{d}$--$\mathrm{Au}$\xspace}

\newcommand{\Npart}{\ensuremath{N_{\rm part}}\xspace}
\newcommand{\aveNpart}{\ensuremath{<N_{\rm part}}>\xspace}
\newcommand{\Ncoll}{\ensuremath{N_{\rm coll}}\xspace}
\newcommand{\aveNcoll}{\ensuremath{<N_{\rm coll}}>\xspace}

\newcommand{\ToverTc}{\ensuremath{T/T_{\rm c}}\Xspace}
\newcommand{\Tc}{\ensuremath{T_{\rm c}}\Xspace}

\newcommand{\RpPb}{\ensuremath{R_{\rm pPb}}\xspace}
\newcommand{\RAA}{\ensuremath{R_{\rm AA}}\xspace}
\newcommand{\RpA}{\ensuremath{R_{\rm pA}}\xspace}
\newcommand{\TAA}{\ensuremath{T_{\rm AA}}\xspace}
\newcommand{\RCP}{\ensuremath{R_{\rm CP}}\xspace}
\newcommand{\vtwo}{\ensuremath{v_{\rm 2}}\xspace}
\newcommand{\vone}{\ensuremath{v_{\rm 1}}\xspace}
\newcommand{\vthree}{\ensuremath{v_{\rm 3}}\xspace}
\newcommand{\vfour}{\ensuremath{v_{\rm 4}}\xspace}
\newcommand{\vfive}{\ensuremath{v_{\rm 5}}\xspace}
\newcommand{\vsix}{\ensuremath{v_{\rm 6}}\xspace}
\newcommand{\vseven}{\ensuremath{v_{\rm 7}}\xspace}
\newcommand{\vn}{\ensuremath{v_{\rm n}}\xspace}

\newcommand{\pt}{\ensuremath{p\sb{\scriptstyle\mathrm{T}}}\xspace}
\newcommand{\pT}{\pt}
\newcommand{\ptee}        {\ensuremath{p_\matrhm{T,ee}}\xspace}
\newcommand{\nch}         {\ensuremath{N_{\mathrm {ch}}\xspace}}
\newcommand{\meannch}     {\ensuremath{\langle \nch \rangle\xspace}}
\newcommand{\meanpT}      {\ensuremath{\langle \pT \rangle}\xspace}
\newcommand{\mee}         {\ensuremath{m_\matrhm{ee}}\xspace}
\newcommand{\dNdeta}      {\ensuremath{\mathrm{d}N/\mathrm{d}\eta}\xspace}
\newcommand{\dNdy}        {\ensuremath{\mathrm{d}N/\mathrm{d}y}\xspace}
\newcommand{\dNchdeta}    {\ensuremath{\mathrm{d}N_\mathrm{ch}/\mathrm{d}\eta}\xspace}
\newcommand{\dNchdy}      {\ensuremath{\mathrm{d}N_\mathrm{ch}/\mathrm{d}y}\xspace}
\newcommand{\kT}          {\ensuremath{k\sb{\scriptstyle\mathrm{T}}}\xspace}
\newcommand{\kt}          {\ensuremath{k\sb{\scriptstyle\mathrm{T}}}\xspace}
\newcommand{\ptt}         {\ensuremath{p_{\mathrm{T, trig}}}}
\newcommand{\pta}         {\ensuremath{p_{\mathrm{T, assoc}}}}
\newcommand{\avedNdeta}   {\ensuremath{\langle \dNdeta \rangle\xspace}}
\newcommand{\avedNchdeta}   {\ensuremath{\langle \dNchdeta \rangle\xspace}}
\newcommand{\avedNdetaatzero}{<\mathrm{d}N_\mathrm{ch}/\mathrm{d}\eta>|_{\eta=0}}
\newcommand{\avedNdetaatzeronorm}{<\mathrm{d}N_\mathrm{ch}/\mathrm{d}\eta>|_{\eta=0} \times 2/<\Npart>}
\newcommand{\avedNdy}     {\ensuremath{\langle \dNdy \rangle\xspace}}
\newcommand{\avedNchdy}   {\ensuremath{\langle \dNchdy \rangle\xspace}}

\newcommand{\nequiv}{\ensuremath{1~\si{\mega\eV}~\mathrm{n}_\mathrm{eq} / \si{\cm^2}}}

\newcommand{\upsilonones} {\ensuremath{\rm \Upsilon(1S)}\xspace}
\newcommand{\upsilontwos} {\ensuremath{\rm \Upsilon(2S)}\xspace}
\newcommand{\upsilonthrees} {\ensuremath{\rm \Upsilon(3S)}\xspace}

\newcommand{\dzero}        {\ensuremath{\rm D^{0}}\xspace}
\newcommand{\dpl}        {\ensuremath{\rm D^{+}}\xspace}
\newcommand{\dpm}        {\ensuremath{\rm D^{\pm}}\xspace}
\renewcommand{\dm}        {\ensuremath{\rm D^{-}}\xspace}
\newcommand{\dstar}        {\ensuremath{\rm D^{*+}}\xspace}
\newcommand{\dspm}           {\ensuremath{\rm D^{\pm}_{\rm s}}\xspace}
\newcommand{\dsp}           {\ensuremath{\rm D^{+}_{\rm s}}\xspace}
\newcommand{\dsm}           {\ensuremath{\rm D^{-}_{\rm s}}\xspace}
\newcommand{\bp}        {\ensuremath{\rm B^{+}}\xspace}
\newcommand{\bm}        {\ensuremath{\rm B^{-}}\xspace}
\newcommand{\bspm}           {\ensuremath{\rm B^{\pm}_{\rm s}}\xspace}
\newcommand{\bsp}           {\ensuremath{\rm B^{+}_{\rm s}}\xspace}
\newcommand{\bsm}           {\ensuremath{\rm B^{-}_{\rm s}}\xspace}
\newcommand{\lc}      {\ensuremath{\rm \Lambda_{\rm c}}\xspace}
\providecommand{\Lc}{\lc}
\newcommand{\lcp}      {\ensuremath{\rm \Lambda_{\rm c}^{+}}\xspace}
\newcommand{\lcm}      {\ensuremath{\rm \Lambda_{\rm c}^{-}}\xspace}
\newcommand{\lcpm}      {\ensuremath{\rm \Lambda_{\rm c}^{\pm}}\xspace}
\newcommand{\lb}      {\ensuremath{\rm \Lambda_{\rm b}\xspace}}
\newcommand{\lbz}      {\ensuremath{\rm \Lambda_{\rm b}^{0}}\xspace}
\newcommand{\xicp}          {\ensuremath{\rm \Xi_{\rm c}^{+}}\xspace}
\newcommand{\xicm}          {\ensuremath{\rm \Xi_{\rm c}^{-}}\xspace}
\newcommand{\xicpm}          {\ensuremath{\rm \Xi_{\rm c}^{\pm}}\xspace}

\newcommand{\Dzero}        {\ensuremath{\rm D^{0}}\xspace}
\newcommand{\Dzerobar}        {\ensuremath{\overline{\rm D^{0}}}\xspace}

\newcommand{\dzerotokpi}        {\ensuremath{\rm D^{0} \rightarrow K^{-}\pi^{+}}\xspace}
\newcommand{\bptodbarpi}        {\ensuremath{\rm B^{+} \rightarrow \bar{D^{0}} \pi^{+}}\xspace}

\newcommand{\xiccpp}{\ensuremath{\Xi^{++}_{cc}}\xspace}
\newcommand{\xiccp}{\ensuremath{\Xi^{+}_{cc}}\xspace}
\newcommand{\xicc}{\ensuremath{\Xi_{cc}}\xspace}
\newcommand{\Xicc}{\xicc}
\newcommand{\bc}{\ensuremath{\rm B_{c}}\xspace}
\newcommand{\omegacc}{\ensuremath{\Omega^{+}_{cc}}\xspace}
\newcommand{\omegaccc}{\ensuremath{\Omega^{++}_{ccc}}\xspace}

\newcommand{\gmom}{\ensuremath{\mathrm{GeV}\kern-0.05em/\kern-0.02em c}\xspace}
\newcommand{\antip}{\ensuremath{\overline{\mathrm{p}}}\xspace}
\newcommand{\antid}{\ensuremath{\overline{\mathrm{d}}}\xspace}
\newcommand{\deuterium}{\ensuremath{{\mathrm{d}}}\xspace}
\newcommand{\tritium}{\ensuremath{{}^{3}\mathrm{H}}\xspace}
\newcommand{\antitritium}{\ensuremath{{}^{3}\overline{\mathrm{\mathrm{He}}}}\xspace}
\newcommand{\hethree}{\ensuremath{{}^{3}\mathrm{He}}\xspace}
\newcommand{\hefour}{\ensuremath{{}^{4}\mathrm{He}}\xspace}
\newcommand{\antihethree}{\ensuremath{{}^{3}\overline{\mathrm{He}}}\xspace}
\newcommand{\antihefour}{\ensuremath{{}^{4}\overline{\mathrm{He}}}\xspace}
\newcommand{\hesix}{\ensuremath{{}^{6}{\mathrm{He}}}\xspace}
\newcommand{\antihesix}{\ensuremath{{}^{6}\overline{\mathrm{He}}}\xspace}
\newcommand{\lisix}{\ensuremath{{}^{6}{\mathrm{Li}}}\xspace}
\newcommand{\antilisix}{\ensuremath{{}^{6}\overline{\mathrm{Li}}}\xspace}

\newcommand{\hyp}        {\ensuremath{^{3}_{\Lambda}\mathrm{H}}\xspace}
\newcommand{\antihyp}{\ensuremath{^{3}_{\overline{\Lambda}}\overline{\mathrm{H}}}\xspace}
\newcommand{\hypfour}    {\ensuremath{^{4}_{\Lambda}\mathrm{H}}\xspace}
\newcommand{\antihypfour}{\ensuremath{^{4}_{\overline{\Lambda}}\overline{\mathrm{H}}}\xspace}
\newcommand{\hyphefour}    {\ensuremath{^{4}_{\Lambda}\mathrm{He}}\xspace}
\newcommand{\antihehypfour}{\ensuremath{^{4}_{\overline{\Lambda}}\overline{\mathrm{He}}}\xspace}
\newcommand{\sigmahyp}     {\ensuremath{^{3}_{\Sigma^{0}}\mathrm{H}}\xspace}
\newcommand{\antisigmahyp} {\ensuremath{^{3}_{\bar{\Sigma}^{0}}\overline{\mathrm{H}}}\xspace}

\newcommand{\Anucl}{$\mathrm{A}$\Xspace}
\newcommand{\isospin}{$I$\Xspace}
\newcommand{\spinJ}{$J$\Xspace}
\newcommand{\BA}{$B_{\mathrm{A}}$\Xspace}
\newcommand{\Tchem}{\ensuremath{T_{\mathrm{chem}}}\Xspace}

\newcommand{\exoticx}{\ensuremath{X(3872)}\xspace}

\def\Bs{{\overline{B}}_s}
\def\R{\mathcal{R}}
\newcommand{\e}{\epsilon}
\newcommand{\tce}{\frac{t_{\rm cool}(\e)}{t_{\rm esc}(\e)}}
\newcommand{\tcer}{\frac{t_{\rm c}(\R)}{t_{\rm esc}(\R)}}
\def\Xe{X_{\rm esc}}
\def\X{X_{\rm esc}}
\def\te{t_{\rm esc}}
\def\tc{t_{\rm cool}}
\def\nb{n_{\rm B}}
\def\nc{n_{\rm C}}
\def\ni{n_{i}}
\def\rism{\rho_{\rm ISM}}
\def\nism{n_{\rm ISM}}
\def\x{(\R,\vec r,t)}
\def\xo{(\R,\vec r_\odot,t_\odot)}
\def\ap{\overline{\rm p}}
\def\ad{\overline{\rm d}}
\def\Qep{Q_{e^+}}
\def\epm{$e^\pm$\ }
\def\ah{\overline{\rm ^3He}}
\def\at{\overline{\rm t}}
\def\s{$(*)$}
\newcommand{\dd}{\text{d}}
\newcommand{\Rp}{\mathcal{R}^\prime}
\newcommand{\Lp}{L^{\prime}}

\newcommand{\qqbar}{\ensuremath{\rm q\bar{q}\xspace}}
\newcommand{\bbbar}{\ensuremath{\rm b\bar{b}\xspace}}
\newcommand{\ccbar}{\ensuremath{\rm c\bar{c}\xspace}}
\newcommand{\DDbar}{\ensuremath{\rm D\overline{D}\xspace}}
\newcommand{\Dbar}{\ensuremath{\rm \overline{D}\xspace}}
\newcommand{\BBbar}{\ensuremath{\rm B\overline{B}\xspace}}
\newcommand{\dgamma}{\ensuremath{\rm D\gamma\xspace}}
\newcommand{\dzerodzerobar}{\ensuremath{\rm D^{0} \overline{\rm D^{0}}\xspace}}
\newcommand{\LctopKpi}{\ensuremath{\rm \Lambda_{\rm c}^{+} \to pK^{-}\pi^{+}}\xspace}

\newcommand{\dPhi}{\ensuremath{\Delta\varphi}}
\newcommand{\dEta}{\ensuremath{\Delta\eta}}

\newcommand{\jpsi}{\ensuremath{{\rm J/\psi}\xspace}}
\newcommand{\chiX}{\ensuremath{\chi_{c1}(3872)\xspace}}
\newcommand{\chicone}{\ensuremath{\chi_{c1}\xspace}}
\newcommand{\chictwo}{\ensuremath{\chi_{c2}\xspace}}
\newcommand{\chibone}{\ensuremath{\chi_{b1}\xspace}}
\newcommand{\chibtwo}{\ensuremath{\chi_{b2}\xspace}}

\newcommand{\mlna}{\langle \ln\!A \rangle}
\newcommand{\nmu}{N_\mu}
\newcommand{\lnnmu}{\ln\!\nmu}
\newcommand{\xmax}{X_\text{max}}
\newcommand{\nmult}{N_\text{mult}}
\newcommand{\tocite}{{\bf REF}}

\begin{titlepage}
  \PHyear{2022}
  \ifdraft
    \PHnumber{\color{red}DRAFT \dvers, SVN \sref\color{black}}
  \else
    \PHnumber{009}         %
  \fi
  \PHdate{\today}              %
  \ifprefinal
    \title{ALICE 3 LOI}
  \else
    \title{Letter of intent for ALICE 3:\\[.25cm]     A next-generation heavy-ion experiment at the LHC\\[.25cm]    Version 2}
  \fi


  \setcounter{tocdepth}{4}
  \setcounter{secnumdepth}{5}

  \ShortTitle{Letter of intent for ALICE 3 (CERN-LHCC-2022-009)}   %
  \Collaboration{ALICE  Collaboration
    }
  \ShortAuthor{ALICE Collaboration}      %
\end{titlepage}
\setcounter{page}{2}

\section*{Executive summary}
\label{sec:exec}

The goal of the ALICE physics programme is to determine the properties of strongly interacting matter and to discern how they arise from the underlying interactions as described by quantum chromodynamics, i.e. to understand the condensed matter of QCD.
Collisions of heavy nuclei at the LHC provide unique experimental access to the hottest and longest-lived quark--gluon plasma available in the laboratory with abundant production of heavy flavour probes.

A rich programme of theoretical and experimental studies targets key questions in this area:
\begin{itemize}
\item What is the nature of interactions between high-energy quarks and gluons and the quark--gluon plasma? How do transport properties arise from first-principle quantum chromodynamics?

\item 
Which mechanisms drive strongly-interacting matter towards equilibrium? To what extent do quarks of different mass reach thermal equilibrium within the plasma? How do they behave close to the diffusion regime?

\item 
How do partons transition to hadrons as the quark--gluon plasma cools down?
How does this process differ from hadron formation in elementary collisions?

\item What are the mechanisms for the restoration of chiral symmetry in the quark--gluon plasma?
\end{itemize}

Despite the progress expected in the current decade,
it will not be possible to fully answer these questions with the present or currently planned detectors. 
A number of key measurements will still be missing after Runs 3 and 4, among which:
\begin{itemize}

\item accurate measurements of charm and beauty hadrons, including their correlation over a wide rapidity range, to determine the interactions of heavy quarks of different mass in the quark--gluon plasma down to the thermal scale.

\item systematic measurements of multiply heavy-flavoured hadrons for which the production from the quark--gluon plasma is expected to be enhanced by orders of magnitude, providing sensitivity to how quarks combine into hadrons depending on their degree of thermalisation.

\item comprehensive data on the production and behaviour of the charmed exotic states in the quark--gluon plasma and their structure, for example by determining the strong interaction potential between hadrons from  measurements of their momentum correlations.

\item high-precision, multi-differential measurements of electromagnetic radiation from the quark--gluon plasma to probe its early evolution and the restoration of chiral symmetry through the coupling of vector and axial-vector mesons.

\item measurements of net-quantum number fluctuations over a wide rapidity range to constrain the susceptibilities of the quark--gluon plasma and to test the realisation of a cross-over phase transition as predicted by lattice QCD.

\end{itemize}

To pursue this physics programme, we propose a novel detector with high readout rate, superb pointing resolution and excellent tracking and particle identification over a large acceptance, using 
advanced silicon detectors. 
To optimise the pointing resolution, the first tracking layer must be placed as close as possible to the interaction point. To provide the larger aperture required for the beams at injection energy, the vertex detector must be retractable.
In the proposed apparatus, the detection layers are constructed from wafer-scale CMOS Active Pixel Sensors thinned to \SI{\sim 30}{\um} and bent into cylinders to minimise the material. 
An outer tracker with barrel and endcap layers provides a relative momentum resolution of 1-2\% over a large acceptance by measuring about 10 space points.
The large active area of the outer tracker requires the exploitation of commercially available high-volume production processes, ranging from CMOS technology for the sensors to highly automated bonding techniques for module integration.
For particle identification, a time-of-flight detector and a ring-imaging Cherenkov detector cover a broad momentum range, both relying on novel silicon timing and photon sensors. R\&D programmes are being set up to push current technological limits of silicon sensors for tracking, timing, and photon detection.
Photon detection and lepton identification at higher momentum are provided by an electromagnetic calorimeter and a muon identifier, both of which exploit established detector technologies. 
A forward conversion tracker 
measures photons at very low transverse momentum, through their conversion into electron-positron pairs at forward rapidity.

In order to maximise the impact of the physics programme, further improvement of the luminosities available with ion beams at the LHC should be explored.
Preliminary studies show that the current limits stem from effects in the injector chain, which could be alleviated with ion species lighter than Pb. 
Additional simulations and machine development studies are needed to refine the luminosity projections and to determine the optimal ion species for the AA programme.
Smaller collision systems, including high statistics proton-proton collisions, also provide the opportunity to study collective effects and the approach to thermal equilibrium.
The requirements for operation with additional systems, such as pA and OO, will be evaluated based on the results from LHC Run~3 and 4.
In this document, the physics performance for ALICE~3 is demonstrated for \PbPb{} collisions, which represent the most challenging environment in terms of per-event multiplicity while being the most conservative choice in terms of luminosity.

With ALICE 3 we propose an extensive programme to fully exploit the LHC for the study of the properties of the quark--gluon plasma. In addition, the novel and unique detector will also
open up important new physics opportunities in other areas.

\cleardoublepage
\tableofcontents
\clearpage
\listoffigures
\clearpage
\section{Introduction}
\label{sec:introduction}

The Standard Model of particle physics describes the fundamental constituents of matter and the laws governing their interactions. It also accounts for how collective phenomena and equilibrium properties of matter arise  from elementary interactions. Theory makes quantitative statements about the equation of state of Standard Model matter, about the nature of the electroweak and strong phase transitions, and about fundamental properties such as transport coefficients and relaxation times. In addition, there has been considerable theoretical progress in describing how out-of-equilibrium evolution drives non-abelian matter towards equilibrium. Collisions of nuclei at ultra-relativistic energies offer a unique possibility for testing some key facets of the rich high-temperature thermodynamics of the Standard Model in laboratory-based experiments. They test the strong interaction sector of the Standard Model at energy densities at which partonic degrees of freedom dominate equilibration processes. They thus give access to the partonic dynamics that drives fundamental non-abelian matter towards equilibrium and that determines the properties of the QCD high temperature phase, the quark--gluon plasma (QGP).

The field of ultra-relativistic nuclear collisions has seen enormous progress since its inception in the mid-eighties, from the first signals of colour deconfinement at the SPS~\cite{Heinz:2000bk} to the evidence, at RHIC, for a strongly-coupled QCD medium that quenches hard partons~\cite{Arsene:2004fa, Adcox:2004mh, Back:2004je, Adams:2005dq}.
Nuclear collisions at the LHC offer an ideal environment for a broad programme of characterization of the properties of this unique state of matter. Besides providing access to the highest-temperature, longest-lived experimentally accessible QCD medium, they also offer an abundant supply of self-calibrating heavy-flavour probes. In addition, the very low net baryon density eases the quantitative connection between experimental measurements and lattice QCD calculations significantly.

\subsection{Present status of heavy-ion physics at the LHC}

All four main LHC experiments now take part in the ultrarelativistic heavy-ion physics programme, and the Run 1 and Run 2 campaigns have already led to crucial further advances in our understanding of the properties of the QCD phase diagram and of the QGP. 
Measurements of the total yields of different particle species at an unprecedented level of precision over eight orders of magnitude appear to be consistent with originating from the statistical partition function.
Measurements in the heavy flavour sector have allowed the observation of a
quark-mass dependence of energy loss in the QGP, opening the way to a systematic programme of investigation of the medium properties using charm and beauty quarks as calibrated probes.
Fundamental discoveries were made in the study of J/$\psi$ production, with evidence for a new mechanism of production  by combination of deconfined charm and anti-charm quarks. The notion of deconfined, largely thermalised c-quarks in the QGP is strongly supported by the five-sigma observation of azimuthal asymmetry in J/$\psi$ production, a direct signal of asymmetry in the distribution of c-quarks themselves. 
Striking signals of collective behaviour were observed in small collision systems (proton-lead, proton-proton), prompting the search for the development of possible QGP precursor phenomena as a function of the collision-system size. Direct access to fully-reconstructed jets has yielded valuable data on the modification of parton showers in the medium and on the medium-induced energy asymmetries in di-jets.
High-precision measurements of azimuthal asymmetries in bulk particle production have established a precise connection between the harmonics of the azimuthal distribution and fluctuations in the initial collision geometry.
(These are only a few major examples of the important findings so far of the LHC heavy-ion campaign. For a detailed assessment of the main findings of the ALICE experiment in Run 1 and Run 2 see~\cite{alicerevpap}.)

In spite of the huge progress made so far, several fundamental questions still remain open, and need to be addressed experimentally.
For instance, low-\pT{} collective flow measurements are modeled successfully in terms of QCD hydrodynamics operating at sub-fermi time scales, and the generic suppression of high-\pT{} hadron yields can be accounted for by jet quenching models in terms of medium-modified parton showers. However, the microscopic dynamics that gives rise to the fast hydrodynamisation is untested experimentally, and so is the question of how efficiently partons lose momentum, hydrodynamise, and thermalise in dense QCD matter. Also, the dynamical mechanisms that lead to a characteristic, event-multiplicity dependent hadrochemical composition remain unknown, and our understanding of heavy-flavour hadronisation, in particular in the baryon sector, is still incomplete. 

\subsection{Heavy-ion physics at the LHC beyond Run 4}

Building on the very successful Run 1 and 2 campaigns, the LHC experiments are gearing up in preparation for the upcoming Runs 3 and 4, in which significant further progress is expected. The ALICE collaboration has just completed a major upgrade programme specifically targeted at the physics of ultrarelativistic nuclear collisions, focussing on the improvement of its event-rate capabilities and on a significant enhancement of the tracking and vertexing performance at low momentum.
The upgrades underway or planned by the ATLAS, CMS and LHCb experiments, while not specifically aimed at QGP physics, will nevertheless also lead to important improvements in their heavy-ion capabilities. As a consequence, substantial further progress is expected in the course of the present decade. Precision measurements of  heavy flavour hadrons will allow us to achieve a more precise characterisation of the mass dependence of the energy loss of quarks traversing the medium.  High-statistics studies of the relative abundance of heavy-flavour particles will bring into view a large fraction of the charm hadronisation table. $\gamma$/Z$^0$-jet correlations will allow us to access the initial transverse momentum balance of the propagating parton. Fully reconstructed heavy-flavour jets will enable the study of modifications of the heavy-flavour fragmentation functions. Precision measurements of the azimuthal asymmetries of quarkonium states, the first measurement of the temperature of the thermal dilepton spectrum and other novel studies will also become possible (see~\cite{Citron:2018lsq} for a review).

ALICE in particular, aims, in Runs 3 and 4, to improve substantially the precision of heavy flavour production measurements and to measure for the first time the thermal emission of dileptons in heavy--ion collisions at the LHC. Existing measurements at the LHC have already shown that charm quarks lose energy as they propagate through the QGP and that charm baryon production is enhanced in central Pb--Pb collisions, suggesting that new hadronisation mechanisms, for example via combination of independently produced quarks, are at work, and that 
charm quarks may largely equilibrate in the plasma.
The improved pointing resolution and readout rate of 
the upgraded apparatus 
will allow to measure baryon to meson ratios in the charm sector, as well as dual flavour mesons, like the $D_s$, with sufficient precision to disentangle charm transport and hadronisation effects and to constrain the transport coefficients of charm quarks in the QGP. With the ITS 3 upgrade, ALICE will also perform first measurements with beauty hadrons. Beauty quarks are not expected to reach full thermal equilibirum, thus providing not only an important qualitative test of heavy quark transport, but also of hadronisation away from equilibrium. However, the precision of the expected results from Runs 3 and 4 in the beauty sector is limited, and a new step in pointing resolution, acceptance and event rates is needed to determine the transport and hadronisation properties of beauty quarks. 

The study of QGP hadronisation via the combination of uncorrelated quarks is key to understand the dynamics connecting hadronisation and collective flow. In spite of the significant jump in the number of heavy-flavour particle species expected to be accessible during Runs 3 and 4, such studies will still be severely limited by the reliance on hadrons for which the confounding admixture of other competing hadronisation mechanisms (string fragmentation, initial production, etc.) is still significant. Unambiguous access to the physics of the hadronising QGP requires the capability to measure the production of particles for which the production by other mechanisms is severely suppressed, by as many as two orders of magnitude for hadrons with two charm quarks, such as the 
$\Xi_{cc}$, and up to three order of magnitude for three charm quarks, for the $\Omega_{ccc}$.

Another important goal for Runs 3 and 4 is to obtain the first measurement of thermal dilepton emission at the LHC, both to determine the temperature in the early phase of the collisions and to study the mechanisms for chiral symmetry restoration in the regime of vanishing baryon density. 
At intermediate dilepton mass, modifications of the $\rho$ spectral function will provide insight in chiral symmetry restoration at high temperature.
Thermal dilepton production at high mass provides a unique window on the temperature reached in the early stages of the collision before hadronisation takes place. With the present ALICE apparatus, however, these measurements will still suffer from a sizeable background from correlated dileptons from heavy-flavour decays, which limit the precision in the mass range above the $\rho$ peak, where $\rho-\mathrm{a}_{1}$ mixing is expected to leave a specific signature. Improved precision at high mass as well as larger statistics are also required to perform differential measurements of dielectron production and elliptic flow to trace the evolution of the emission of electromagnetic radiation with the temperature and to disentangle emission from the QGP and from the hadronic phase.

In spite of the ambitious scientific programme for the upcoming Runs 3 and 4, therefore, crucial questions will still remain unanswered with the present detector concepts.
The overarching goal of our field is to achieve textbook understanding of the rich phenomenology of QCD matter, connecting parton energy loss, collective flow, hadronisation and electromagnetic radiation in a unified description.  
Achieving this goal will require a novel experimental approach, most notably in the areas most relevant to the capabilities for heavy flavour and electromagnetic radiation studies. A few key examples are given below.
\begin{itemize}
\item 
{\bf High-precision beauty measurements.}
In order to establish a firm connection between parton transport, collective flow and hadronisation, the study of parton energy loss needs to be extended down to the momenta typical of diffusion phenomena. To this end precision measurements of the spectra and flow coefficients of beauty hadrons are indispensable. 
In particular, due to their significantly larger masses, the description of 
processes involving beauty quarks using pQCD may be more appropriate. 
Furthermore, while a significant fraction of charm hadrons originate from 
beauty decays, the beauty sector poses a much cleaner field of study with much 
less contributions from feeddown. 
This calls for very high readout-rate capabilities combined with a secondary vertexing performance far exceeding those of the present inner trackers.
\item
{\bf $D \overline{D}$ correlations.}
In order to discriminate between the different regimes of in-medium energy loss and reveal the onset of charm isotropisation,
more differential measurements, such as that of
correlations between fully reconstructed charm hadron pairs over a wide rapidity range, are needed. As will be seen in Chapter~\ref{sec:performance}, this calls for very high efficiency down to very low transverse momentum, pushing the impact parameter resolution to the technological limit, and for the ability to operate at rates that significantly exceed the capabilities of the present ALICE detector.

\item
{\bf Multi-charm baryons, P-wave quarkonia, exotic hadrons.}
A key goal of the high-energy heavy-ion programme is the study of the formation of hadrons from the deconfined QGP. In this regard,
the measurement of the production of multi-heavy-flavour hadrons, P-wave quarkonia and exotic states will offer unprecedented sensitivity.
To achieve high statistics
for multi-heavy-flavour particles a novel experimental approach is needed to track all their decay products, typically including hyperons, before they decay. This calls very high tracking/vertexing precision very close to the interaction point, particle identification capabilities over a wide transverse momentum range, and high readout rates. Moreover, large acceptance is required, not only for reasons of statistics, but also in order to investigate the dependence of the production of multi-heavy-flavour hadrons on the variation of the heavy quark density with rapidity. Access to large statistics for P-wave quarkonia and exotic states calls for quarkonium reconstruction capabilities down to zero \pt,
combined with very high data acquisition rates, and, for electromagnetic decays, excellent soft photon capabilities.

\item
{\bf Azimuthal asymmetry of electromagnetic radiation.}
The study of the azimuthal asymmetries in the production of different particle species is a very powerful tool, %
which provides access to fundamental physical parameters of the QGP such as its shear and bulk viscosities
by correlating final-state anisotropies to the geometry of the strongly-interacting medium produced in the collision.
Measurements planned for Runs 3 and 4 will strongly constrain the initial conditions and geometry of the collision, allowing us, for instance, to ascertain whether all hadron species truly emerge from a single, common flow field and to make precision measurements of the QGP viscosities, but only insofar as their values integrated over the history of particle emission are concerned. In order to gain access to the full time evolution of these fundamental transport parameters (and hence to their temperature dependence), electromagnetic probes are required.
This calls for a comprehensive study of the azimuthal asymmetry of the production of dileptons as a function of their transverse momentum and their mass providing a direct connection to the emission time.
Such a fully-differential measurement of the dilepton azimuthal asymmetries requires very high statistics and very low background, calling for a tracker with extremely low mass
to minimise photon conversions, and with very high pointing resolution for the rejection of electrons from the decays of heavy flavour particles.
\item
{\bf Chiral symmetry restoration.}
The deconfinement phase transition is expected to be accompanied by a partial restoration of chiral symmetry, leading to a modification of the dilepton spectrum in the light vector meson mass range.
The high-statistics, low-background capabilities of the detector concept discussed in the present document will be essential for a high-precision measurement of the medium modification of the dilepton spectrum at the LHC, particularly in the region around the mass of the chiral partner of the $\rho$: the a$_1$ meson. Since the a$_1$ is an axial meson, it does not couple to the dilepton channel in the vacuum, but, as chiral symmetry is restored in the plasma, it is expected to start mixing with the $\rho$, leading to a signature modification of the mass spectrum that would provide a direct signal of the admixture of the axial state. The detection of this signal would represent a crucial confirmation of chiral symmetry restoration in the medium. The expected magnitude of this effect, however, of the order 15\%, will still be out of reach of the ALICE capabilities in Run 3 and 4. 
As will be discussed in Section~\ref{sec:performance}, the improvements offered by the design proposed in this document, in particular the ultra-low mass and the very high resolution vertexing capabilities, will provide the required precision.

\item
{\bf Collectivity in small systems.}
The discovery of collective effects in small collision systems has been one of the most remarkable LHC results so far. The investigation of the origin of these effects will be one of the major goals of the Run 3/4 campaign, with 
a dedicated high-multiplicity pp programme in ALICE, a new p--Pb run and a short oxygen-oxygen run in the plans. 
However, high-multiplicity pp studies in Runs 3 and 4 will still be affected by the biases introduced by the limited phase-space acceptance of the present detectors, and any future measurement campaigns with lighter ions that may be warranted by the results of Runs 3 and 4 (for instance by the upcoming oxygen-oxygen studies) will require the capability of operating at data acquisition rates significantly beyond those of the existing experiments.

\item
{\bf High-statistics hadronic physics.}
The approach proposed here is also ideal to address key topics at the interface of QGP, hadronic, and nuclear physics. High rate capabilities will allow to access rare phenomena, such as the production of nuclei, hypernuclei and, combined with excellent heavy-flavour capabilities, possibly the as-yet undiscovered supernuclei (nuclei in which a nucleon is replaced by a charmed baryon).
Wide acceptance, again combined with outstanding heavy-flavour performance at low transverse momenta, will allow to extend the studies of hadron-hadron potentials via two-particle correlations to the charm sector, providing a powerful tool to investigate the structure of the newly-discovered charmed exotic states. The combination of wide acceptance and strong particle identification capabilities will also open up the possibility to explore more complex vector meson decays than previously possible in ultra-peripheral collisions.
In addition, a special detector arrangement to reconstruct photon conversion at ultra-low transverse momenta will allow testing the infrared limit of QED as a gauge theory.

\end{itemize}

\textit{To address the physics topics discussed above, we propose an innovative detector concept that will be described in detail in the following.   It features a high-speed silicon tracker with very high resolution and minimum mass, positioned very close to the interaction point.  Its first layers are inside the LHC beam pipe and provide a very wide rapidity acceptance, complemented with hadron, electron, muon and photon identification capabilities over a very broad range of transverse momenta}

Such a unique detector at the LHC would also open other physics opportunities, in addition to the topics discussed so far.
Some further examples are discussed in the document,  such  as  the  possibility of searching for critical behaviour in the study of event-by-event fluctuations of conserved charges, the prospect of measuring  the  production  of  antinuclei  in  the  decay  of beauty  baryons  (with important  implications  for  dark  matter  searches)  and  the  possibility of tightening significantly the limits on the production of axion-Like particles in a region of the parameter space that has recently received particular attention.

In the remainder of this chapter, after some considerations on the expected machine conditions at the LHC after LS 4 with various nuclear beams, a possible design is outlined and its uniqueness and competitiveness are discussed.
In Chapter 2 the physics motivation, only broadly outlined in this introduction, is discussed in detail.
The expected performance of the proposed detector concept is then discussed in more detail in Chapter 3. The status of the development for the detector systems under consideration to meet the specifications discussed in this chapter is then discussed in Chapter 4 and, finally, some preliminary considerations on planning and resources are presented in Chapter 5.%

\subsection{Ion collisions at the LHC}
\label{sec:introduction:machine_considerations}

The measurements outlined in the previous section rely on the collisions of ions to create fireballs of strongly-interacting matter. 
So far, the LHC has operated with Pb ions for about one month per operational year, a pattern expected to be continued for the ALICE~3 programme.
While Pb ions provide the largest possible collision system with the strongest QGP effects, the luminosity is limited. 
Lighter ions, because of their lower charge, allow the injection of bunches with higher intensities and, thus, higher luminosities. 
So far, the operation of the LHC with lighter ions was studied with a successful Xe pilot run in 2017.
Projections for a variety of ion species can be derived from the bunch intensities that can be expected to become available from the injector complex\footnote{These are the most recent projections from~\cite{Bruce2021}. Improved estimates will require further studies, both through simulations and machine developments with different ion species.}, see Tab.~\ref{tab:ion_lumi}.
Lighter ions than Pb are expected to provide higher nucleon--nucleon luminosities, which determine the rate of hard processes.
However, the magnitude of QGP effects scales differently for different observables and are generally more significant in heavier collision systems. Based on these projections, Runs~5 and 6 will allow us to study more than five times the nucleon--nucleon luminosity available with \PbPb collisions in Runs~3 and 4 ($\sim \SI{0.5}{\femto\invbarn}$).
This increase in luminosity, in combination with vastly improved detector capabilities, enables the ALICE~3 physics programme.
Several measurements would benefit or become possible through a further increase in luminosity, which motivates efforts to identify suitable measures, e.g. for the operation with lighter ions in the injector complex.

To meet the needs of the \pp programme of ALICE 3, the experiment shall be operated in \pp collisions at an instantaneous luminosity of $L = \SI{3.e32}{cm^{-2}s^{-1}}$ corresponding to an interaction rate of \SI{24}{\mega\hertz}. 
This leads to the accumulation of about \SI{3}{fb^{-1}} per operational year and results in particle fluxes and data rates comparable to those in the operation with ion collisions.
Higher \pp luminosities would lead to significant limitations on the detector, which should provide the best possible performance for the heavy-ion programme.
With this rate, the radiation load at a given distance from the interaction point in ALICE~3 will be two orders of magnitude lower than for ATLAS and CMS.
This will allow us to operate the first tracking layer at a distance of only \SI{5}{\mm} from the interaction point.

For the heavy-ion programme, the optimal running scenario will depend on the luminosities achievable with the various species.
In this letter, we show projections for the most challenging environment, i.e. \PbPb collisions.
Based on the current luminosity estimates, Xe--Xe could be the better option providing twice the nucleon--nucleon luminosity with a relatively large collision system.
Runs with smaller collision systems, as well as p--A runs, are also being considered.
The optimisation in terms of possible intermediate species and luminosity goals will also depend on new information on collective phenomena in small systems from the Run 3 and 4 campaign.
A dedicated \pp reference sample recorded at the same centre-of-mass energy as the A--A runs is planned.
In addition, the programme foresees the accumulation of \pp statistics throughout the year.
Table~\ref{tab:running_scenario} shows the resulting integrated luminosities.
It should be noted that one year of Run~5 and 6 could be used for a run at reduced magnetic field to allow analyses requiring particularly low \pt reach, e.g. the measurement of low-mass dielectrons.

\begin{landscape}
\begin{table}
    \centering
    \small
    \renewcommand{\arraystretch}{1.5}
    \begin{tabular}{@{}p{3.5cm} *{8}{S[table-format=6.2]}@{}}
        \toprule
         Quantity & {pp} & {O--O} & {Ar--Ar} & {Ca--Ca} & {Kr--Kr} & {In--In} & {Xe--Xe} & {Pb--Pb} \\
         \midrule
         $\sqrt{s_\mathrm{NN}}$ (\si{\tera\eV}) & 14.00 & 7.00 & 6.30 & 7.00 & 6.46 & 5.97 & 5.86 & 5.52 \\
         $L_\mathrm{AA}$ $(\unit{cm^{-2}s^{-1}})$ & 
         \num{3.0e32} & \num{1.5e30} & \num{3.2e29} & \num{2.8e29} &
         \num{8.5e28} & \num{5.0e28} & \num{3.3e28} & \num{1.2e28}\\
         $\langle L_\mathrm{AA} \rangle$ $(\unit{cm^{-2}s^{-1}})$ & 
         \num{3.0e32} & \num{9.5e29} & \num{2.0e29} & \num{1.9e29} & 
         \num{5.0e28} & \num{2.3e28} & \num{1.6e28} & \num{3.3e27} \\
         $\mathcal{L}_\mathrm{AA}^\mathrm{month}$ $(\unit{nb^{-1}})$ & 
         \num{5.1e5} & \num{1.6e3} & \num{3.4e2} & \num{3.1e2} & 
         \num{8.4e1} & \num{3.9e1} & \num{2.6e1} & \num{5.6e0} \\
         $\mathcal{L}_\mathrm{NN}^\mathrm{month}$ $(\unit{pb^{-1}})$ & 
         505 & 409 & 550 & 500 & 510 & 512 & 434 & 242\\
         $R_\mathrm{max} (\unit{kHz})$ 
         & 24000 & 2169 & 821 & 734 & 344 & 260 & 187 & 93 \\
         $\mu$ 
         & 1.2 & 0.21 & 0.08 & 0.07 & 0.03 & 0.03 & 0.02 & 0.01\\
         $\mathrm{d}N_\mathrm{ch} / \mathrm{d}\eta$ (MB) 
         & 7 & 70 & 151 & 152 & 275 & 400 & 434 & 682 \\
        \cmidrule{2-9}
        & \multicolumn{8}{c}{at $R = \SI{0.5}{\cm}$}\\
        \cmidrule{2-9}
         $R_\mathrm{hit}$ $(\si{\mega\hertz\per\cm^{2}})$ 
         & 94 & 85 & 69 & 62 & 53 & 58 & 46 & 35\\

         NIEL (\si{\nequiv})
         & \num{1.8e14} & \num{1.0e14} & \num{8.6e13} & \num{7.9e13} 
         & \num{6.0e13} & \num{3.3e13} & \num{4.1e13} & \num{1.9e13} \\

         TID (\unit{Rad})
         & \num{5.8e6} & \num{3.2e6} & \num{2.8e6} & \num{2.5e6} 
         & \num{1.9e6} & \num{1.1e6} & \num{1.3e6} & \num{6.1e5} \\
         \cmidrule{2-9}
        & \multicolumn{8}{c}{at $R = \SI{100}{\cm}$}\\
        \cmidrule{2-9}
         $R_\mathrm{hit}$ $(\si{\kilo\hertz\per\cm^{2}})$ 
         & 2.4 & 2.1 & 1.7 & 1.6 & 1.3 & 1.0 & 1.1 & 0.9 \\

         NIEL (\si{\nequiv})
         & \num{4.9e9} & \num{2.5e9} & \num{2.1e9} & \num{2.0e9} 
         & \num{1.5e9} & \num{8.3e8} & \num{1.0e9} & \num{4.7e8}\\

         TID (\unit{Rad})
         & \num{1.4e2} & \num{8.0e1} & \num{6.9e1} & \num{6.3e1}
         & \num{4.8e1} & \num{2.7e1} & \num{3.3e1} & \num{1.5e1}\\
         \bottomrule
    \end{tabular}
    \caption[Projected LHC performance with ions]{Projected LHC performance: For various collision systems, we list the peak luminosity $L_\mathrm{AA}$, the average luminosity $\langle L_\mathrm{AA} \rangle$, the luminosity integrated per month of operation $\mathcal{L}_\mathrm{AA}^\mathrm{month}$, also rescaled to the nucleon--nucleon luminosity $\mathcal{L}_\mathrm{NN}^\mathrm{month}$ (multiplying by $A^2$). Furthermore, we list the maximum interaction rate $R_\mathrm{max}$, the minimum bias (MB) charged particle pseudorapidity density $\mathrm{d}N/\mathrm{d}\eta$, and the interaction probability $\mu$ per bunch crossing. 
    For the radii \SI{0.5}{\cm} and \SI{1}{m}, we also list the particle fluence, the non-ionising energy loss, and the total ionising dose per operational month (assuming a running efficiency of 65\%).}
    \label{tab:ion_lumi}
\end{table}
\end{landscape}

\begin{table}
\centering
    \renewcommand{\arraystretch}{1.5}
    \begin{tabular}{@{}llrr@{}}
        \toprule
         \multicolumn{2}{l}{System} & $\mathcal{L}^\mathrm{month}$ & $\mathcal{L}^\mathrm{Run 5 + 6}$ \\
         \midrule
         \multicolumn{2}{l}{\pp} & \SI{0.5}{\femto\invbarn} & \SI{18}{\femto\invbarn}\\
         \multicolumn{2}{l}{\pp reference} & \SI{100}{\pico\invbarn} & \SI{200}{\pico\invbarn} \\
         \multicolumn{2}{l}{A--A}\\
         & Xe--Xe & \SI{26}{\nano\invbarn} & \SI{156}{\nano\invbarn}\\
         & \PbPb & \SI{5.6}{\nano\invbarn} & \SI{33.6}{\nano\invbarn}\\
         \bottomrule
    \end{tabular}
    \caption[Int. luminosities for collision systems]{Integrated luminosities for different collision systems}
    \label{tab:running_scenario}
\end{table}

\subsection{Detector concept}
\label{sec:introduction:experimental_layout}

\begin{table}
    \centering
    \rowcolors{2}{gray!10}{}
    \sisetup{range-phrase={\ \text{to}\ }}
    \renewcommand{\arraystretch}{1.5}
    \begin{tabular}{p{6cm} p{5cm}}
        \toprule
        \textbf{Observables} & \textbf{Kinematic range} \\
        \midrule
        Heavy-flavour hadrons & $\pt \to 0$, \newline $\abs{\eta} < 4$\\
        Dielectrons & $\pt \approx \SIrange{0.05}{3}{\giga\eVc}$, \newline
        $M_\mathrm{ee} \approx \SIrange{0.05}{4}{\giga\eVcsq}$ \\
        Photons & $\pt \approx \SIrange{0.1}{50}{\giga\eVc}$, 
        \newline $-2 < \eta < 4$ \\
        Quarkonia and exotica & $\pt \to 0$, \newline $\abs{\eta} < 1.75$ \\
        Ultrasoft photons & $\pt \approx \SIrange{1}{50}{\mega\eVc}$, \newline $3 < \eta < 5$ \\
        Nuclei & $\pt \to 0$, \newline $\abs{\eta} < 4$\\
        \bottomrule
    \end{tabular}
    \caption[Overview of key observables]{Overview of key physics objects and the respective kinematic ranges of interest for ALICE~3.}
    \label{tab:intro:observables}
\end{table}

\begin{figure}
    \centering
    \includegraphics[width=.8\textwidth]{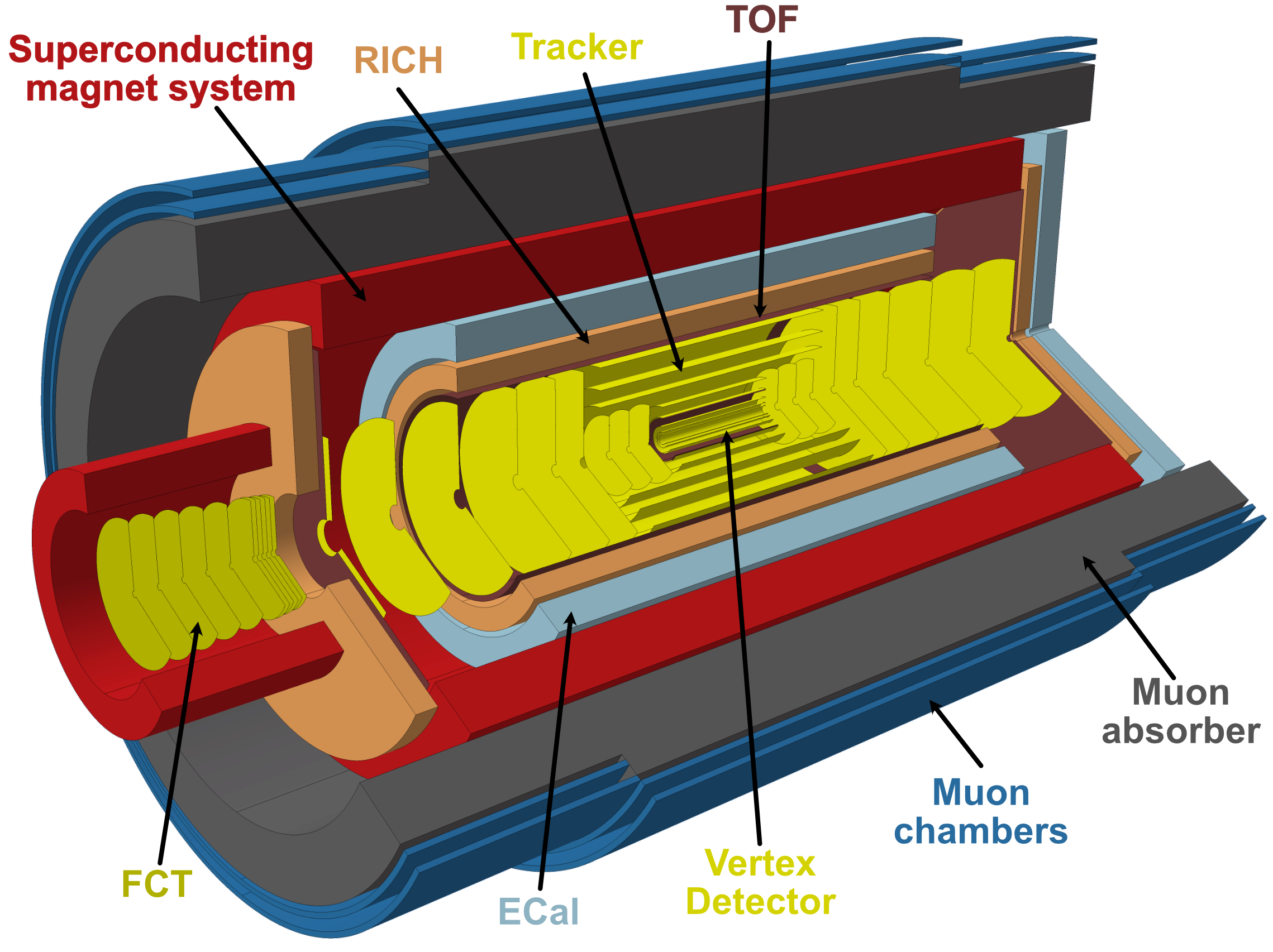}
    \caption[Detector concept]{ALICE 3 detector concept: A silicon tracker composed of cylinders and disks serves for track reconstruction in the magnetic field provided by a super-conducting magnet system. 
    The vertex tracker is contained within the beam pipe. 
    For particle identification a time-of-flight detector, RICH detector, photon detector, and a muon system are employed.
    The forward conversion tracker is housed in a dedicated dipole magnet.}
    \label{fig:alice3_overview}
\end{figure}

The measurements planned to address the physics goals outlined in Sec.~\ref{sec:introduction} rely on a series of observables, which are summarised in Tab.~\ref{tab:intro:observables}.
These observables determine the requirements on the detector design.
Foremost, they rely on charged particle tracking providing a transverse momentum resolution in the order of \SIrange{1}{2}{\percent} with excellent efficiency over a large pseudorapidity range.
The studies in this letter have been carried out up to $\abs{\eta} = 4$.
The precision of the \pt measurement enters in the mass resolution, which is particularly important for the extraction of signals for decays with a large background.
Besides the gain in statistics, the large rapidity coverage is crucial for the low \pt reach as well as studies of long-range correlations and for studies of the rapidity dependence of the yields.
In addition, ultimate pointing resolution is required ($\sim \SI{2}{\micro\meter} / p [\si{\giga\eVc}]$).
This forms the basis of the reconstruction of secondary vertices and decay chains, e.g. of heavy-flavoured hadrons.
It is also a key ingredient for the rejection of heavy-flavour background for the dielectron measurements.
Particle identification is essential for the ALICE~3 programme.
Electrons must be identified in a \pt range from $\sim \SI{50}{\mega\eVc}$ up to about $\sim \SI{2}{\giga\eVc}$.
For muons the challenge is to achieve an efficient identification starting from \pt $\sim \SI{1.5}{\giga\eVc}$ at $\eta = 0$.
The combination of electrons and muons at low and high transverse momenta, respectively, ensures access to leptonic channels over a wide \pt range.
Hadron identification must span the range up to a few \si{\giga\eVc} to identify the products of heavy-flavour decays, while retaining high efficiency. 
For the detection of photons, the requirements are different for different measurements.
While the measurement of photon-jet correlations calls for a large acceptance, the reconstruction of states such as the $\chi_c$ relies in addition on the measurement of photons at low energies.
The detection and energy determination of very low-\pt (\SI{\geq 1}{\mega\eVc}) photons calls for a specially optimised detector in the forward direction to exploit the longitudinal boost.
In addition, the apparatus must enable us to fully exploit the luminosities discussed in the previous section.
These requirements are summarized in Tab.~\ref{tab:detector_requirements} and have led to the detector concept shown in Fig.~\ref{fig:alice3_overview}.
 
\begin{figure}
    \centering
    \includegraphics[width=.95\textwidth]{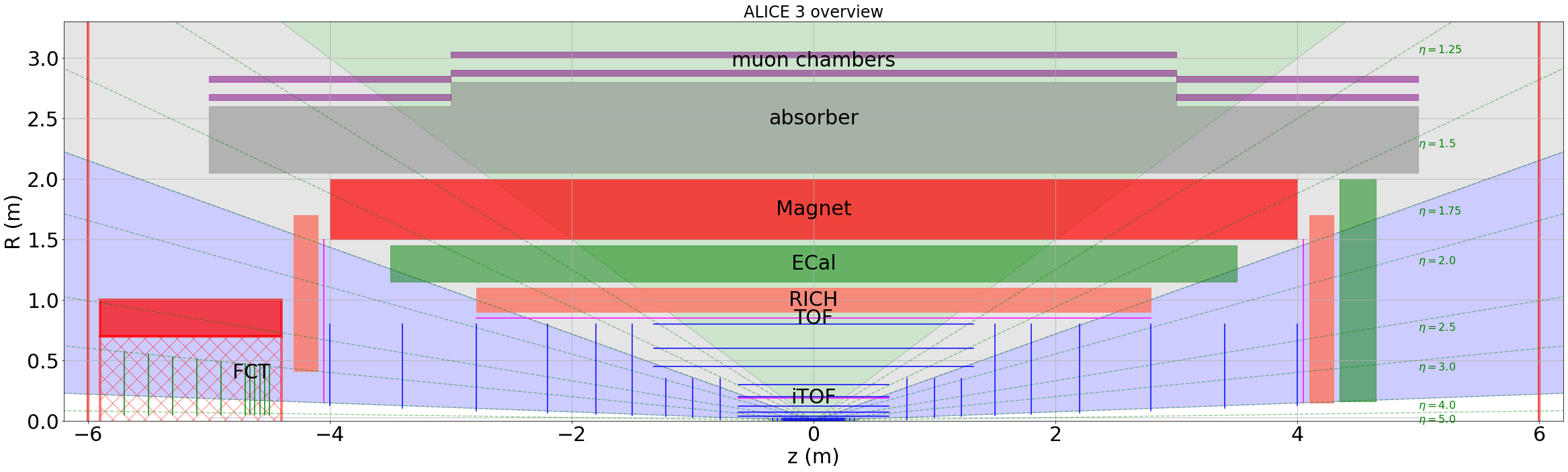}
    \caption[Longitudinal cross section of ALICE~3 detector]{Longitudinal cross section of the ALICE 3 detector: The MAPS-based tracker is complemented by PID detectors (inner and outer TOF, RICH), all of which are housed in the field from a superconducting magnet system. In addition, the electromagnetic calorimeter (ECal), the muon identifier, and the Forward Conversion Tracker (FCT) are shown.}
    \label{fig:alice3_v2}
\end{figure}

\paragraph*{Tracking.}
\begin{revised}
The heart of ALICE~3 is charged particle reconstruction based on a silicon pixel tracker, with sensors arranged in barrel layers and forward disks, see Fig.~\ref{fig:alice3_overview} and~\ref{fig:alice3_v2}.
The tracker is installed in the volume of \SI{80}{\cm} radius and \SI{\pm 4}{\metre} length around the interaction point.
The outer tracker consists of 8~cylindrical layers and 9~forward discs on either side of the interaction point, with each of the layers contributing about \SI{1}{\percent} of a radiation length in material.
The momentum is reconstructed from the curvature in the magnetic field provided by a superconducting magnet system, for which different field configurations are under study.
A superconducting solenoidal magnet providing a field of $B = \SI{2}{\tesla}$ leads to a relative \pt resolution of \SI{\sim 0.6}{\percent} at midrapidity and of \SI{\sim 2}{\percent} at $\abs{\eta} = 3$.
Additional forward dipoles can be considered to maintain a relative \pt resolution of \SI{\sim 1}{\percent} up to $\abs{\eta} = 4$. 
To assess the full physics potential of the experiment, the performance studies presented in this document have been performed for the latter configuration.
We expect the impact of the reduced performance with just an optimised solenoid on the physics performance to be small and, thus, consider this the baseline configuration.
A detailed evaluation of the optimal magnet configuration will be carried out for the Technical Design Reports.
\end{revised}

In order to achieve the required pointing resolution, the first hit must be measured as close as possible to the interaction point and with as little material as possible in front of the first layer to reduce multiple scatterings.
The minimal radial distance from the interaction point is determined by the aperture required for the LHC beam.
While this amounts to $\sim \SI{5}{\milli\metre}$ at top energy, $\sim \SI{15}{\milli\metre}$ are necessary at injection energy.
Therefore, having the first detection layer at a radial distance of $\sim \SI{5}{\milli\metre}$ for data taking is possible only with a detector that can be retracted for injection and inserted for data taking at collision energy.
This implies that the vertex detector, consisting of 3~cylindrical layers and 3~disks on either side, must be installed in a secondary vacuum within the beampipe.
The requirements on the pointing resolution are met by a vertex detector with an inner radius of \SI{5}{\milli\metre}, about \SI{1}{\permille} of a radiation length for the first layer, and a position resolution of $\sim \SI{2.5}{\micro\metre}$.

\paragraph*{Particle identification.}
The tracker is complemented by systems for particle identification.
The exact specifications are still being studied.
With a time resolution of \SI{20}{\pico\second}, a time-of-flight layer outside of the tracker at a radius of \SI{85}{\cm} allows us to identify electrons and hadrons up to transverse momenta of \SI{500}{\mega\eVc} and \SI{2}{\giga\eVc} ($\pi$/K separation), respectively. 
For particles below \SI{300}{\mega\eVc}, which do not reach this TOF layer, an inner TOF layer is foreseen at a radius of \SI{20}{\centi\meter}.
In the forward direction, TOF disks are installed following the last tracking disk.

To further extend the PID capabilities beyond the momentum reach of the TOF, a Ring Imaging Cherenkov detector (RICH) is foreseen behind it. In the central barrel, an aerogel-based detector with a refractive index of $n = 1.03$ and an angular resolution of \SI{\sim 1.5}{mrad} enables the separation of electrons and pions up to \SI{2}{\giga\eVc} and of protons from e, $\pi$, K up to \SI{14}{\giga\eVc}. 
In the forward region, a smaller refractive index will be used to account for the Lorentz boost.

For the identification of muons, a steel-absorber of about \SI{70}{\centi\metre} thickness would be installed outside of the magnet. 
Two layers of muon detectors are used to detect and match the muon tracklets to tracks in the silicon pixel tracker, which will provide the information on the transverse momentum.
The absorber is optimised to be efficient for the reconstruction of $\jpsi$ at rest (muons down to pt \SI{\sim 1.5}{\giga\eVc}) at $\eta = 0$. 

\begin{revised}
In addition, an electromagnetic calorimeter covering the barrel acceptance and one forward direction enables a variety of measurements, including those of photon-jet correlations and radiative decays of $\chi_c$ and ${\rm D}^{*0}$. 
The large acceptance of the calorimeter in particular enables energy balance measurements of dijets and photon-jet pairs, in particular those involving heavy-flavour jets. In addition, it allows to detect large samples of $\chi_c$, ${\rm D}^{*0}$ and provides access to a unique region of parameter space in searches for axion-like particles. %
While complete performance studies are still pending, such measurements require a large acceptance with moderate energy resolution as achievable with Pb-scintillator sampling technology.
In addition the calorimeter system extends the electron identification capabilities of the experiment.
The detection of direct photons in the thermal emission regime, and the separation of decays of different $\chi_c$ states into a J/$\psi$ and a photon require higher energy and spatial resolutions, which can be provided by instrumenting the region around midrapidity with PbWO$_4$ crystals.
These states could also be separated by reconstructing the photons through their conversion in the material, benefiting from the good momentum resolution for charged particles. Simulation studies indicate very good performance in the measurement of the $\chi_{c}$ states utilising photon conversions.
\end{revised}

The reconstruction of photons with very low transverse momenta, down to \SI{1}{\mega\eVc}, poses a particular challenge. 
Their measurement is only possible by exploiting the Lorentz boost at large pseudorapidities around $\eta \sim 4$.
The tracking of $e^\pm$ pairs from photon conversion could be achieved with a Forward Conversion Tracker (FCT), an array of silicon pixel disks installed in the forward direction ($3 < \eta < 5$).
In this way, both the photon direction and energy can be reconstructed precisely.

The apparatus is planned to be installed at Point~2 during the Long Shutdown 4 of the LHC and will be housed around the nominal interaction point within the existing L3 magnet (not used to provide the magnetic field). 
Further details of the implementation are discussed in Chapter~\ref{sec:systems}.

\begin{landscape}
\begin{table}
    \centering
    \small
    \rowcolors{2}{gray!10}{}
    \renewcommand{\arraystretch}{1.5}
    \begin{tabular}{@{}p{1.8cm} p{3cm} p{4.8cm} p{4.8cm} p{4.8cm}@{}}
        \toprule
        \textbf{Component} 
        & \textbf{Observables} 
        & {\textbf{Barrel} ($\abs{\eta} < 1.75$)} 
        & {\textbf{Forward} ($1.75 < \abs{\eta} < 4$)} 
        & \textbf{Detectors} 
        \\
        \midrule
        Vertexing
        & (Multi-)charm baryons, \newline dielectrons
        & Best possible DCA resolution, \newline
        $\sigma_\mathrm{DCA} \approx \SI{10}{\micro\metre}$ at \newline
        $\pt = \SI{200}{\mega\eVc}$, $\eta = 0$
        & Best possible DCA resolution, \newline
        $\sigma_\mathrm{DCA} \approx \SI{30}{\micro\metre}$ at \newline
        $\pt = \SI{200}{\mega\eVc}$, $\eta = 3$
        & retractable Si-pixel tracker: \newline
        $\sigma_\mathrm{pos} \approx \SI{2.5}{\micro\metre}$, \newline
        $R_\mathrm{in} \approx \SI{5}{\mm}$, \newline
        $X / X_0 \approx \SI{0.1}{\percent}$ for first layer
        \\
        Tracking
        & (Multi-)charm baryons, \newline 
        dielectrons, \newline
        photons \dots
        & \multicolumn{2}{c}{
            $\sigma_{\pt} / \pt \approx \SIrange{1}{2}{\percent}$
        }
        & Silicon pixel tracker: \newline
        $\sigma_\mathrm{pos} \approx \SI{10}{\micro\metre}$, \newline
        $R_\mathrm{out} \approx \SI{80}{\cm}$, \newline
        $L \approx \pm \SI{4}{\metre}$ \newline
        $X / X_0 \approx \SI{1}{\percent}$ per layer
        \\
        Hadron ID 
        & (Multi-)charm baryons 
        & \multicolumn{2}{c}{
            $\pi/K/p$ separation up to a few \si{\giga\electronvolt/\clight}
        }
        & Time of flight: $\sigma_\mathrm{tof} \approx \SI{20}{ps}$ \newline
        RICH: $n \approx \numrange[range-phrase={-}]{1.006}{1.03}$, \newline 
        $\sigma_\theta \approx \SI{1.5}{mrad}$
        \\
        Electron ID 
        & Dielectrons, \newline quarkonia, \newline $\chi_{c1}(3872)$ 
        & pion rejection by 1000x \newline 
        up to \SIrange{2}{3}{\giga\electronvolt/\clight}
        & 
        & Time of flight: $\sigma_\mathrm{tof} \approx \SI{20}{ps}$ \newline
        RICH: $n \approx \numrange[range-phrase={-}]{1.006}{1.03}$, \newline 
        $\sigma_\theta \approx \SI{1.5}{mrad}$
        \\
        Muon ID 
        & Quarkonia, \newline $\chi_{c1}(3872)$
        & reconstruction of \jpsi{} at rest, \newline
        i.e.~muons from \pt $\sim$ \SI{1.5}{\giga\eVc} at $\eta = 0$
        & 
        & steel absorber: $L \approx \SI{70}{\cm}$ \newline
        muon detectors
        \\
        ECal
        & Photons, \newline jets
        & \multicolumn{2}{c}{large acceptance}
        & Pb-Sci sampling calorimeter
        \\
        ECal
        & $\chi_c$
        & high-resolution segment
        &
        & PbWO$_4$ calorimeter
        \\
        Soft photon \newline detection
        & Ultra-soft photons 
        & 
        & measurement of photons \newline 
        in \pt range \SIrange{1}{50}{\mega\electronvolt/\clight}
        & Forward conversion tracker \newline
        based on silicon pixel tracker
        \\
        \bottomrule
    \end{tabular}
    \caption[Detector requirements]{Detector requirements}
    \label{tab:detector_requirements}
\end{table}
\end{landscape}

\subsection{Uniqueness and competitiveness}
\label{sec:introduction:uniqueness_competitivene}

The proposed ALICE 3 experiment combines excellent particle identification capabilities with a tracking system that has very low mass and unique pointing resolution, covering a much larger rapidity range than the current ALICE setup.  This combination provides unique access to thermal dielectron production and heavy flavour probes of the quark--gluon plasma.

ALICE 3 will be able to cleanly identify dielectrons over a broad range in mass and \pt{}, thus providing unique access to the temperature evolution of the early stage of the collision, as well as signatures of chiral symmetry restoration. These measurements will be unique at LHC, since the other LHC experiments mainly rely on muon identification for lepton probes. 
In this channel, the momentum threshold for muon identification in ATLAS and CMS of $\pt \gtrsim 3$ \GeVc{} does not allow to access the mass and \pt scales for thermal production. 
The forward geometry of the LHCb detector provides access to lower momenta, down to about $\pt = 1$ \GeVc, which is still above typical thermal momentum scales. 
The misidentification background in the muon channel is also significantly larger than in ALICE 3, which makes precision measurements of the thermal emission spectrum very challenging. A more detailed discussion of these points and the comparison of the dielectron performance in ALICE 3 with the current ALICE setup is given in Section~\ref{sec:performance:physics:dileptons:comparisonITS3}.

The retractable inner tracker that is envisaged for ALICE 3 provides unique pointing capabilities, enabling heavy flavour hadron detection with very high purity and efficiency down to low \pt (see Section~\ref{sec:benchmarkHF}) enabling high-precision measurements of the charm and beauty mesons and baryons (see Section~\ref{sec:performance:physics:heavy_flav:baryonflow}), including azimuthal \DDbar{} correlations measurements that provide unique direct access to heavy quark interactions with the QGP. 
The latter are not accessible to ATLAS and CMS due to the larger material budget of their tracking systems and the distance of the first layers to the interaction point, see Section~\ref{sec:performance:ddbar} for a more detailed discussion and performance study. 
In addition, the close proximity of the first layer of the ALICE 3 vertex tracker provides very high detection efficiency for direct tracking of weakly decaying strange baryons (``strangeness tracking'') which are key to the detection of multi-charm baryons in heavy-ion collisions, see Section~\ref{sec:performance:detector:strangeness_tracking}.

The proposed muon identifier system for ALICE 3 is able to identify muons down to low transverse momentum to provide unique \pT{} reach not only for quarkonia, including P-wave states with photon detection in the calorimeter (see Section~\ref{sec:performance:physics:quarkonia:chis}), but also for exotic hadrons like the \chiX{}, to explore formation and dissociation of these states in heavy-ion collisions at thermal momentum scales (see Section~\ref{sec:performance:physics:quarkonia:exo}). Measurements of the \chiX{} by CMS, so far are limited to the range $\pt \gtrsim 10$~\GeVc.

In addition, ALICE 3 has unique capabilities for the identification of nuclei, and for searches for super-nuclei (see Section~\ref{sec:performance:physics:nuclear_states:charm}). Particle identification with the TOF and RICH systems over a larger rapidity range provide unique opportunities for the study of event-by-event fluctuations of the net-baryon number which probe the dynamical effects of baryon number conservation and are sensitive to critical behaviour of QCD matter around the phase transition (Section~\ref{sec:perf:fluctuations}). ALICE 3 will also provide unique coverage for for physics beyond the Standard Model, such as the search for axion-Like particles using ultra-peripheral collisions (Section~\ref{sec:performance:physics:bsm:upc:alp}). The photon conversion tracker that is proposed as a part of ALICE 3 provides unique access to very soft photons, to test fundamental aspects of field theory in this regime, see Sections~\ref{sec:physics:ultra-soft-photons:low} and~\ref{sec:performance:detector:photon_id}.

\clearpage
\section{Physics motivation and goals}
\label{sec:physics}

In this chapter we present the main physics goals that motivate the present proposal.
The performance studies that have been carried out to define the specifications presented in Section~\ref{sec:introduction:experimental_layout} will be presented in Chapter~\ref{sec:performance}.

\subsection{Heavy quark propagation and hadronisation}
\label{sec:qgp_physics:partonprop}

In studies of heavy-ion collisions at high energy, partons produced in hard scatterings serve as powerful probes of the deconfined medium (QGP)~\cite{Bjorken:1982tu}. High-\pt partons and heavy quarks are produced in processes with large momentum transfer that can be described by perturbative QCD calculations and that take place on short time scales compared to QGP formation. Once created, they interact with the QGP, leading to energy loss and momentum broadening perpendicular to the direction of propagation.  By measuring the effect of the QCD medium on the propagation of such partons, one can constrain fundamental properties of the QGP and characterise its inner structure~\cite{Wiedemann_2010}. \\ \\
The study of heavy-quark interactions with the QGP plays a unique role in this respect. The masses of the heavy quarks  ($m_{c} \approx $1.3 \GeV, $m_{b} \approx$ 4.2 \GeV~\cite{ParticleDataGroup:2020ssz}) set well-defined perturbative scales, which are significantly larger than both the non-perturbative scale of QCD ($\Lambda_{\rm QCD}$) and the typical temperature reached in the medium~\cite{Busza_2018}. Therefore, heavy quark production is essentially limited to initial hard scatterings or gluon splittings on short time scales, and the number of heavy quarks is conserved throughout the QGP evolution~\cite{Andronic_2016}. In addition, heavy quarks can be traced from heavy-flavour hadrons in the final state, serving as probes of the properties of the medium at different scales.
\\ \\
The interaction of a heavy quark with a constituent of the medium is typically classified according to whether it proceeds via elastic processes, without particle production, or via inelastic processes~\cite{Wiedemann_2010}. %
Due to the large mass of heavy quarks, low-momentum elastic scatterings with the QGP give rise to a Brownian motion process that determines the long-wavelength diffusion properties of the QGP~\cite{Rapp:2008qc,He:2013zua}. Inelastic interactions with large momentum transfer probe the QGP at short wavelengths, accessing the regime where deconfined quasi-particles of the medium are expected to act as scattering centers for the heavy quarks.
\\ \\
The interactions of high-energy partons with the medium are characterised by the rate of transverse momentum broadening $\hat{q}$ , i.e. the mean square transverse momentum kick per collision $\langle q_\perp^2 \rangle$ divided by the mean free path $\lambda$.
\begin{equation}
    \hat{q} = \frac{\langle q_\perp^2 \rangle}{\lambda}.
\end{equation}
One of the prominent effects of the interactions of energetic partons with the QGP constituents is medium-induced energy loss, which for high-momentum partons is expected to be dominated by inelastic processes, i.e.\ gluon radiation~\cite{Wiedemann_2010}.
Within the BMDPS-Z formalism~\cite{BDMPSZ1998}, in the large momentum transfer regime and for a short path length $L$ through the medium, the average amount of radiative energy loss of a parton is proportional to $\hat{q}$: $\langle \Delta E \rangle \propto \alpha_s \hat{q} L^2$.

In the low-momentum regime, the interactions of heavy quarks with the QGP are described in transport models in terms of drag and momentum diffusion coefficients, which depend on the medium temperature and the parton momentum.
The pertinent transport parameter is the spatial diffusion coefficient $D_s$, which is defined as the limit for $\rm p \rightarrow$~0 of the drag coefficient~\cite{Moore:2004tg,Rapp:2008qc,Rapp:2018qla} and it relates the average of the square displacement of the heavy quark $\langle \vec{x}^2 \rangle$ to the diffusion time $t$ through the expression:
\begin{equation}
    \langle \vec{x}^2 \rangle = 6 D_s t.
\end{equation}

The $D_{\rm s}$ coefficient provides direct access to the in-medium QCD force, and thus to the coupling strength of the QGP, and can be calculated from first-principle Lattice QCD calculations~\cite{Moore:2004tg,Francis:2015daa,Altenkort:2020fgs,Ding:2021ise}. Constraining its values, and more generally the temperature and momentum dependence of the drag coefficient, allows one to discriminate among different descriptions of the QGP medium~\cite{Rapp:2008qc}.

The diffusion coefficient is also related to the relaxation time $\tau_Q$ for a heavy quark of mass $m_Q$ in a medium at temperature $T$ by the expression:
\begin{equation}
\label{eq:tau_Q}
    \tau_Q = (m_Q/T) D_s.
\end{equation}
Eq.~\ref{eq:tau_Q} implies that beauty quarks approach equilibration in the QGP about three times slower ($m_c/m_b$)~than charm quarks. \\ \\
Two widely used observables for the study of the interactions of heavy quarks with the QGP are the nuclear modification factor \RAA{} and the elliptic flow $v_2$.
The nuclear modification factor is defined as the ratio between the measured yields (usually $\pt$ differential) in A--A and pp collisions, normalised to the average number of nucleon--nucleon collisions, $N_{\rm coll}$, in the considered A--A centrality interval:
\begin{equation}
\label{eq:Raa}
R_{\rm AA}(\pt)=
{\frac{1}{\langle {N_{\rm coll}} \rangle }} \times 
{\frac{{\rm d} N_{\rm AA}/{\rm d}\pt}{{\rm d} N_{\rm pp}/{\rm d}\pt}}\
\end{equation}
The anisotropic flow coefficients $v_n$ are the coefficients of the Fourier expansion of the particle azimuthal distribution with respect to the corresponding symmetry planes of the participant distribution $\Psi_n$:
\begin{equation}
\label{eq:vn}
v_{\rm n} = \langle {\rm cos}[n(\varphi-\Psi_{n})]\rangle
\end{equation}
The second harmonic coefficient $v_2$ is the elliptic flow and the corresponding symmetry plane is the reaction plane, i.e.\ the plane defined by the impact parameter of the colliding nuclei and the beam direction.\\ \\
Measurements of D-meson $R_{\rm AA}$ and $v_n$ from RHIC and LHC Run 2 samples allowed for a first estimate of the $D_{\rm s}$ coefficient via a data-to-model comparison with calculations that provided a good description of the experimental data~\cite{PHENIX:2010xji,STAR:2017kkh,ALICE:2017pbx,ALICE:2020iug,ALICE:2021rxa}. This allowed to constrain the $D_{\rm s}$ value to the interval $1.5 < 2\pi D_\mathrm{s} T_\mathrm{c} < 4.5$ at the pseudocritical temperature $T_\mathrm{pc} = 155$~MeV, in good agreement with Lattice QCD predictions~\cite{Ding:2021ise}.
This estimate is currently limited both by experimental uncertainty, in particular at low $p_{\rm T}$, and by the different possible choices for some of the model ingredients.
Among them, the mechanisms governing the interaction of heavy quarks with the medium and the hadronisation of the heavy quarks at the QGP phase boundary (in-vacuum fragmentation vs.~recombination with quarks from the bulk) were shown to play a decisive role in the determination of the transport coefficient~\cite{Rapp:2018qla}.
A more precise estimation of the spatial diffusion coefficient will be obtained from the charm-hadron measurements that will be performed during Run 3 and 4 of the LHC~\cite{Citron:2018lsq}. However, in order to fully exploit the potential of nuclear collisions at the LHC to constrain the values of the QGP transport coefficients, the inclusion of high-precision data on the transverse momentum spectra and elliptic flow of beauty mesons and baryons down to $\pt{}$ close to 0~is required (see Section~\ref{sec:physics:beauty}).
Measurements of the beauty-hadron $R_{\rm AA}$ and $v_n$ coefficients together with the relative abundances of different beauty-hadron species, and in particular the baryon-to-meson ratios, down to low $p_{\rm T}$ will therefore play a crucial role to simultaneously constrain the heavy-quark diffusion coefficient and the hadronisation mechanisms in the beauty sector (more detail in Section~\ref{sec:physics:qgp_hadronisation}).

In order to pin down the relative importance of elastic and inelastic interactions in different \pt{} intervals and to access the nature of the QGP at different momentum scales, a precision measurement of the transverse momentum broadening of $D\overline{D}$ pairs is a key input, as discussed in Section~\ref{sec:physics:hfcorrelations}.
Another essential ingredient in this programme is a comprehensive study of the QGP hadronisation mechanisms, by measuring the relative abundances of the different heavy-flavour species, including hadrons containing multiple heavy-flavour quarks, as discussed in Section~\ref{sec:physics:qgp_hadronisation}.

\subsubsection{Beauty hadron production and flow}
\label{sec:physics:beauty}

ALICE~3 would allow to perform high-accuracy measurements of both the production and collective modes of expansion (quantified e.g. by the $v_{n}$ coefficients) of beauty mesons and baryons down to \pt=~0 with extreme accuracy. Such measurements would provide a key set of new and independent constraints on heavy-flavour transport and hadronisation, in particular on the diffusion coefficient $\Dsc$. Thanks to their larger mass compared to charm quarks, beauty quarks also provide better theoretical control on the modelling of the quark transport in the expanding QGP~\cite{Das:2013kea,Rapp:2018qla}. Beauty measurements also provide increased experimental constraints on the temperature and momentum dependence of the drag coefficient, which is of critical importance to achieve a microscopic description of heavy-quark interactions as a probe of the QGP medium~\cite{Rapp_2018}. With beauty measurements, one has better access to the properties of the drag coefficient for vanishing momenta, since the velocity of beauty quarks at the same momentum is lower by a factor $\rm m_{b}/m_{c} \approx~$3 with respect to charm quarks. Since their relaxation time is predicted to be somewhat larger than the lifetime of the QGP (which is approximately 10 fm/$c$)~\cite{Busza_2018}, beauty quarks provide better constraints on the heavy-quark propagation at the late stages of the medium evolution, where the coupling strength is expected to be larger.
\\ \\
In Run~3 and 4, beauty measurements in heavy-ion collisions will still be affected by sizeable statistical uncertainties~\cite{Citron:2018lsq}. In particular, the precision on the $\lb$ production yields and $v_n$ coefficients, critical to establish the mechanisms of hadronisation of beauty sector in the QGP, will be limited~\cite{ALICE:ITS2:2014,ALICE:ITS3:2019,Andreou:2020oir}.
Preliminary studies of the expected ALICE~3 performance for the measurement of the production of B mesons and of the azimuthal asymmetry of the $\Lambda_b$ baryon are shown in Section~\ref{sec:performance:physics:heavy_flav}. ALICE~3 will provide, in particular, a large improvement in the reconstruction accuracy and the selection purity of both B-mesons and $\Lambda_{\rm c}^{+}$-baryons, especially at very low \pt.

A large acceptance allows for a larger separation between the phase-space region used for measurements and the one considered for event characterization and classification. In particular, the large rapidity coverage of the ALICE~3 apparatus will open the possibility of fully exploiting the Event Shape Engineering (ESE) technique to measure the $p_{\rm T}$-differential yield and $v_2$ of heavy-flavour hadrons in events with the same centrality and different magnitude of the average bulk elliptic flow, which is related to the initial state geometry~\cite{Schukraft:2012ah}.
This extended coverage will allow for the use of widely separated rapidity intervals to reduce the bias from non-flow correlations between the region in which heavy-flavour hadrons are reconstructed and the one used to estimate the elliptic flow of the bulk from charged-particle tracks. This will significantly improve the power of ESE-based measurements to discriminate among  models~\cite{ALICE:2020iug,Citron:2018lsq}.

\subsubsection{\revised Azimuthal decorrelation and energy balance measurements} 
\label{sec:physics:hfcorrelations}

As discussed above, the measurements of single-particle \pt{} spectra and harmonic coefficients of azimuthal anisotropy of charm and beauty hadrons to very low transverse momenta will provide critical constraints for a precision determination of the transport coefficients of the QGP.
By averaging over many events, such measurements still integrate over many different initial configurations as well as different amounts of energy loss, providing only partial sensitivity to the propagation of heavy-flavour quarks in the QGP. More differential measurements are therefore needed to quantitatively characterise such phenomena. {\revised Measurements of azimuthal and \pt{} correlations between the heavy-flavour quark and the recoil parton produced in hard-scattering processes } 
provide additional sensitivity to the propagation of heavy quarks since they constrain the modification of both their momentum and direction and provide insight into the in-medium path-length dependence of energy loss. Moreover, they offer unique access to the modifications of the heavy-quark transverse momentum, which cannot be constrained with {\revised single-inclusive hadron and jet} measurements and are sensitive to the microscopic description of the QGP~\cite{PhysRevC.92.054909,Nahrgang:2013saa}. By performing such a ``Rutherford-like experiment”, one can therefore constrain the inner structure of the QGP by observing how it affects the relative direction and the momenta of pairs of heavy quarks that are traversing it.
{\revised Recent theoretical and experimental studies~\cite{Renk:2013xaa,ALICE:2021kpy} demonstrate that the energy loss of heavy quarks in the hot QCD medium differs from that of light partons. In this context, one should also outline the importance of performing photon-jet measurements, where the photon and the jet provide information on the heavy quark kinematics before and after interaction with the medium, respectively, allowing for a quantitative characterization of the energy loss in the medium and the fragmentation to mesons.}

At leading order, charm-anti-charm quark ($\ccbar$) pairs are produced in a back-to-back configuration in azimuth. Higher-order processes, including initial and final state radiation or fragmentation effects, modify such configurations and result, in particular, in a moderate broadening around $\Delta\varphi = \pi$. These effects can be precisely measured in pp collisions, and further broadening in heavy-ion collisions can then be attributed to medium interactions. 
Radiative energy loss mostly involves small-angle radiation and therefore leads to limited broadening, while collisional processes, in particular at small transverse momentum, are expected to lead to  significant broadening of the initial correlation~\cite{PhysRevC.92.054909,Nahrgang:2013saa}.
In the limit of full equilibration, the flight directions of the charm quarks would be fully randomised, and no remnant of the initial correlation would be visible. A measurement of the correlation of $\DDbar$ pairs in $\Delta \varphi$ measured differentially in \pt, $\Delta \pt$ and $\Delta \eta$ would provide, therefore, novel constraints on the mechanisms of heavy-quark parton propagation and, in particular, on the process of charm equilibration in the QGP.

\begin{figure}
    \centering
    \includegraphics[width=0.6\textwidth]{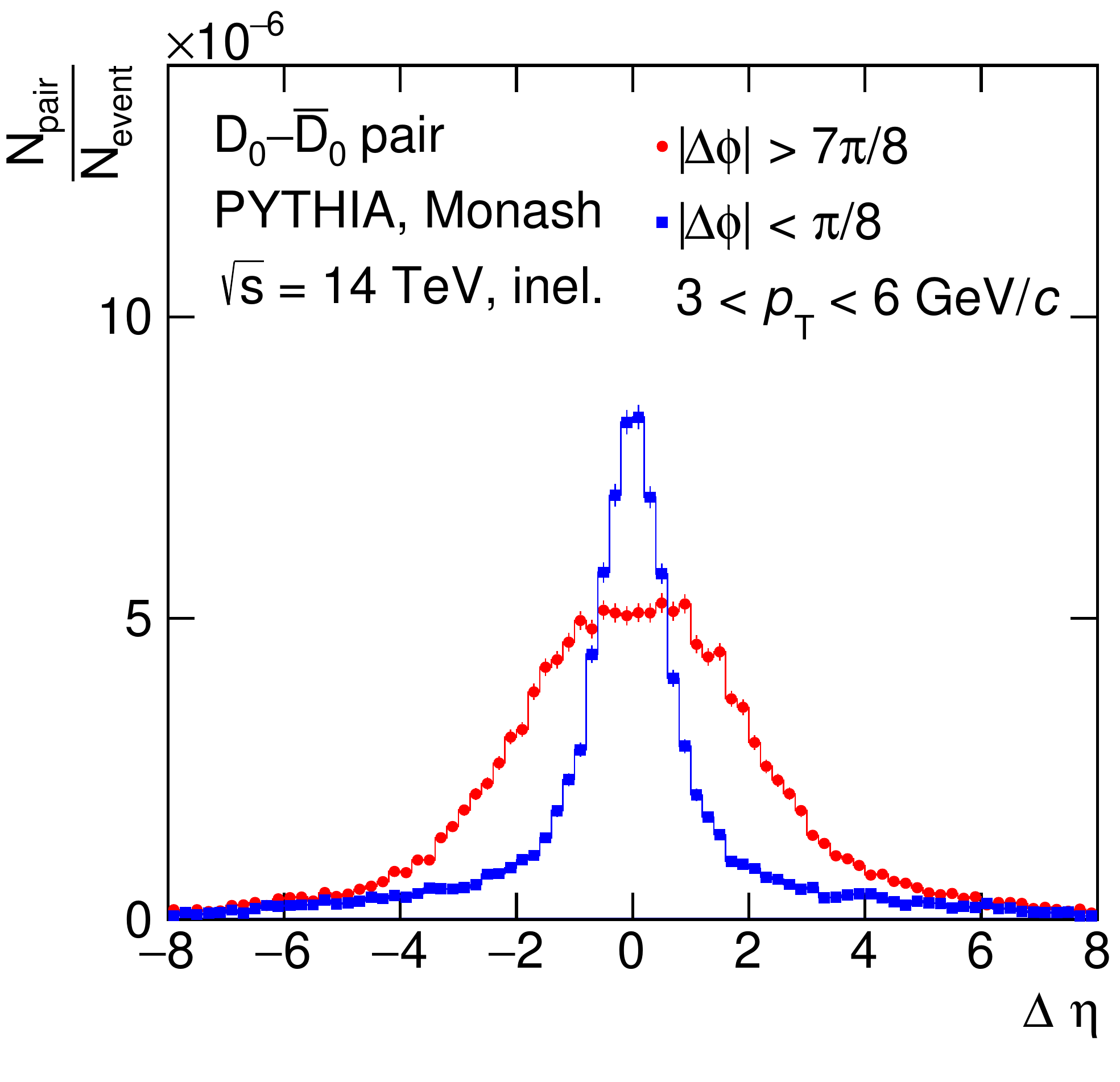}
    \caption[\DDbar{} pair yield vs $\Delta\eta$]{Distribution of pseudorapidity difference between $\rm D^0$ and $\rm \overline{D^0}$ mesons for pairs which are close in azimuth (near side, blue points) and pairs around the back-to-back configuration (away-side, red points).}
    \label{fig:physics:DDbaryieldvsdeta}
\end{figure}

A study of the expected ALICE~3 performance for the measurement of $\Delta \varphi$ correlations of $\DDbar$ pairs is presented in Section~\ref{sec:performance:ddbar}. Prospects for measurements of $\BBbar$ correlations are also being studied. 
Measurements of heavy-flavour correlations in heavy-ion collisions pose stringent requirements on the detector design. In particular, measurements at low $\pt$ have to rely on very large samples of untriggered events. In addition, a wide pseudorapidity coverage is needed to be efficient for the detection of back-to-back pairs, which have an intrinsic separation of up to a few units of rapidity, further broadened by decay kinematics, see Figure~\ref{fig:physics:DDbaryieldvsdeta}.
Such measurements will still be severely limited in Run~3 and 4 by the pseudorapidity coverage of the current ALICE apparatus ($|\eta|< 0.9$), and only come into reach with ALICE~3.

High signal purity at low and intermediate transverse momentum is needed to reduce the large combinatorial background, which enters quadratically in such analyses. 
To perform these measurements, the ALICE~3 detector needs to provide high readout rates, high-precision tracking, excellent pointing resolution, and particle identification capabilities for pions, kaons and protons across the widest possible pseudorapidity range. 

{\revised Measurements of jet pairs and photon-jet pairs allow to map the energy loss in addition to transverse momentum broadening, via the energy-imbalance between the jets or the photon and the jet~\cite{Stavreva_2013}. An electromagnetic calorimeter is foreseen to detect photons from hard scatterings, as well as the neutral energy from decay photons in jets, to perform these measurements. Reconstruction and identification of charm jets is based of neutral jet component in a calorimeter, charge component in the tracking system and jet tagging with charm hadrons~\cite{Vertesi:2020gqr}. These measurements make use of standard techniques in the field. More detailed physics performance studies will be pursued in the context of the Technical Design Report.}

Precision measurements of correlations in the heavy-flavour sector will also provide a powerful tool for the investigation of the strong interaction potentials between heavy-flavour hadrons. This topic is discussed in Section~\ref{sec:physics:HFstrong}.

\subsubsection{QGP hadronisation and multi-charm hadrons}
\label{sec:physics:qgp_hadronisation}
The hadronisation of a final state parton cannot be described adopting a perturbative approach, and is usually modelled through a phase of string breaking and/or cluster formation, which is considered to be independent of the surrounding parton density~\cite{Andersson:1983ia}. 
However, as also confirmed by measurement of e.g.~baryon yields in pp collisions at the LHC~\cite{ALICE:2016fzo,ALICE:2017thy,ALICE:2020jsh,ALICE:2020wfu}, additional hadronisation mechanisms may exist, whereby quarks that are close in phase space can combine into colourless hadrons. 

In heavy-ion collisions, where partons may travel freely over distances much larger than the typical hadron sizes and a dense system of partons close to thermal equilibrium is formed, such mechanisms become dominant,
making the production of baryons and other heavy hadrons more favourable than in pp collisions. 
Measurements in ultra-relativistic nuclear collisions 
of light and multi-strange baryon yields normalised to pions indeed show a significant increase with respect to the corresponding ones in \pp collisions~\cite{ALICE:2016fzo,ALICE:2017thy,ALICE:2020jsh,ALICE:2020wfu}.
Most of the measured yields are well described by the Statistical Hadronisation Model (SHM), in which the abundances of light and strange hadrons follow the equilibrium populations of a hadron-resonance gas at the freeze-out temperature $T_{\mathrm{ch}}$ of about \SI{155}{\mega\eV}~\cite{Cleymans:1992zc,AGSThermalJpsiSHM,BraunMunzinger:1999qy,Becattini:2000jw,BraunMunzinger:2001ip,Andronic:2017pug}.

At the LHC, the large cross section for heavy-flavour production allows to extend these studies with high precision to the charm sector. One of the key discoveries of the LHC heavy-ion programme is that a novel hadronisation process for J/$\psi$ production, beyond the single-scattering mechanisms that describe its production in \pp collisions, is at play in the presence of large partonic densities. Indeed, \pt, centrality and rapidity dependence of the J/$\psi$ production cross section and the J/$\psi$ azimuthal asymmetry in Pb--Pb collisions indicate that at low \pt, charmonia can be produced by the combination of $\rm c$ and $\bar{c}$ quarks that originated from two independent hard scattering processes~\cite{ALICE:2012jsl,ALICE:2019lga,Andronic:2007bi,Du:2015wha} (see also the discussion in Section~\ref{sec:physics_chis}).
\\ \\
First measurements of baryon/meson ratios in the charm sector also indicate a low-\pt enhancement consistent with such a picture.
A comprehensive campaign of precision measurements of charm baryon production yields across collision systems is planned for Run 3 and Run 4. Such measurements, however, will be limited to baryons containing only one charm quark.
In this context, the measurement of the production yields of multi-charm baryons, which can only be produced by combination of uncorrelated charm quarks, would provide a qualitatively new handle on the production of heavy-flavour hadrons.
\begin{figure}
    \centering
    \includegraphics[width=0.8\textwidth]{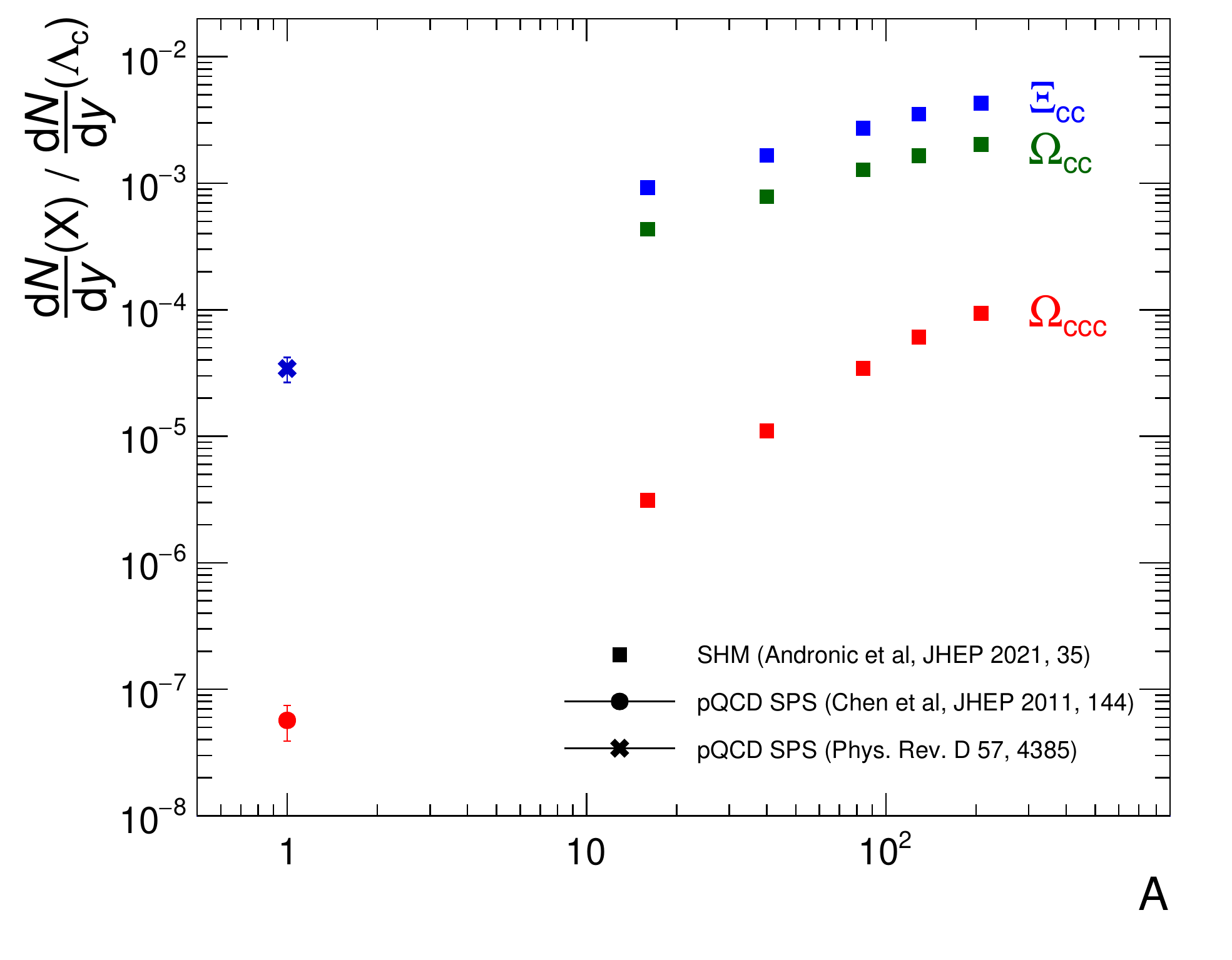}
    \caption[Ratio of multi-charm baryon yields]{Ratio of multi-charm baryon to single-charm baryon ($\Lambda_{c}$) yield as a
function of mass number, using single parton scattering (SPS)
calculations~\cite{Berezhnoy:1998aa,He:2014tga} at $\sqrts=14$~TeV, shown at $A=1$, and a statistical hadronisation model
calculation~\cite{Andronic:2021erx} for $AA$ collisions at $\sqrtsNN = 5.02$~TeV with $A \neq 1$. The $\Lambda_{c}$ yield for the SPS calculation is obtained from a measurement performed by ALICE in pp collisions at $\sqrt{s}$~=~5.02~TeV~\cite{ALICE:2017thy}, scaled by 1.7 to account for the
increase in beam energy.}
    \label{fig:physics:multicharm}
\end{figure}

Measurements of multi-charm hadrons, such as the $\Xi_{\textrm{cc}}^{+}$ (ccd), $\Xi_{\textrm{cc}}^{++}$ (ccu), $\Omega_{\textrm{cc}}^{+}$ (ccs), $\Omega_{\textrm{ccc}}^{++}$ (ccc), and exotic states such as the newly-discovered $\rm T_{cc}^{+}$ (ccud)~\cite{LHCb:2021vvq,LHCb:2021auc}, would provide a direct window on hadron formation from a deconfined quark--gluon plasma. 
In fact, the yields of multi-charm baryons relative to the number of produced charm quarks are predicted to be significantly enhanced in AA relative to pp collisions. First predictions in this area were made in~\cite{Becattini:2005hb} and recent calculations are discussed in
~\cite{He:2014tga,Zhao:2016ccp,Yao:2018zze,Cho:2019syk,Andronic:2021erx}.
Enhancements are expected by as much as a factor 10$^2$ for the recently observed $\Xi_{cc}$ baryon~\cite{LHCb:2017iph} and even by as much as a factor 10$^3$ for the as yet undiscovered $\Omega_{ccc}$ baryon. The observation and precise quantification of such effects would represent a quantum jump for the study of the properties of deconfined matter. 

Figure~\ref{fig:physics:multicharm} shows 
the wide span between the single parton scattering expectation and the statistical hadronisation model prediction for the ratio between the yields of multi-charm baryons and that of the $\Lambda_c$ as a function of the colliding system size, indicating the large sensitivity of these yields
to the underlying production mechanisms. For example, the SHM predictions shown in the figure assume full equilibration of the charm quarks, and the comparison of the measured yields to these predictions would provide a very sensitive measure of the degree of equilibration of charm quarks in the medium, including its system-size dependence. 

Similarly, beauty thermalisation can also be studied with ALICE~3. More specifically, measuring the yields of beauty hadrons with different masses, such as B mesons and $\Lambda_{b}$, $\Xi_{b}$ and $\Omega_{b}$ baryons, 
will offer a new insight into hadronisation. 
Beauty hadrons are not expected to be completely thermalised, i.e.\ their yields may not follow mass exponentials displaced with a beauty quark fugacity factor. As such, their yield measurements represent the ultimate stress test of statistical hadronisation models and may be a unique opportunity for obtaining insight on the hadron yield equilibration process itself. 
Beauty baryon measurements will be feasible down to low momentum in central nucleus-nucleus collisions at midrapidity also 
via the application of the innovative strangeness tracking technique. 
Studies of the expected performance of ALICE~3 for $\Xi_{b}$ and $\Omega_{b}$ baryons are presented in Section~\ref{sec:performance:physics:heavy_flav:beautybaryons}.
For both charm and beauty, it must ultimately be possible to relate the equilibration level to the transport properties of the QGP via kinetic transport modeling in order to provide a consistent description of the entire phenomenology.

Such studies are currently far beyond experimental reach. A comprehensive programme of measurements in the multi-charm sector requires a new detector, featuring unprecedented pointing resolution, ultra-low material thickness, high-rate capabilities, large rapidity coverage, high mass resolution and particle identification over a large \pt interval and the full rapidity range of the apparatus.
Moreover, it also requires a special layout close to the interaction region, with the 
tracking layers spaced very closely, in order to enable a novel experimental technique (``strangeness tracking''), that relies on measuring the trajectories of hyperons before their decay.
The large pseudorapidity coverage of ALICE~3, about four times larger than that of the current ALICE experiment,
will also allow for rapidity-dependent measurements of multi-charm production to explore the dependence of the production yields on the charm density.
Studies of the expected performance of ALICE~3 for multi-charm baryons are presented in Section~\ref{sec:performance:physics:heavy_flav:multicharm}.

\subsection{Bound states}
\label{sec:physics:qgp_properties:potential}

A complete microscopic description of the dynamics of heavy quarks in the QGP must include a detailed and comprehensive study of the bound state formation and dissociation mechanisms.

\subsubsection{Quarkonium states}
\label{sec:physics_chis}

Quarkonia represent one of the most valuable probes of the QGP since the very first experimental studies with ultrarelativistic heavy-ion collisions~\cite{NA38:1989okg, NA50:2000brc}. Quarkonia were traditionally considered to be privileged ``thermometers'' of the deconfined phase of QCD because of a direct relation between the screening of the colour force in a QGP and the temperature attained by the medium~\cite{Matsui:1986dk,Digal:2001ue}. According to this picture, the attractive force between a quark and an antiquark immersed in the QGP medium weakens via a Debye-like screening, possibly leading to the dissociation of the bound state. The observation of a suppression of the production yields of charmonium and bottomonium states in heavy-ion collisions with respect to proton-proton or proton-nucleus collisions was therefore proposed as an experimental signature for the existence of deconfined quarks and gluons~\cite{Matsui:1986dk}. 
As discussed in Section~\ref{sec:physics:qgp_hadronisation}, in a deconfined medium coloured partons may travel freely over distances much larger than the typical hadron sizes. Quarkonium states could then be formed via the recombination (or regeneration) of heavy quark/anti-quark pairs produced in independent hard scatterings. The contribution of this mechanism to the total yield is expected to increase rapidly with the heavy quark multiplicity.

In the charmonium sector, the role of suppression in the QGP was first studied in detail in the 1990s by the NA38 and NA50 experiments in O--U and Pb--Pb collisions at the SPS ($\sqrtsNN = 17.8$ and 21.0~GeV)~\cite{NA38:1989okg,NA50:2000brc}. Measurements at RHIC showed a suppression similar to the one measured at the SPS, despite the larger medium density and temperature at the higher collision energy ($\sqrtsNN=200$~GeV), in line with predictions by theoretical calculations that include recombination mechanisms~\cite{Andronic:2007bi,Du:2015wha}. At the LHC, the suppression of J/$\psi$ production is significantly smaller than the one observed at the SPS and RHIC and shows a characteristic \pt-dependence, which represents a significant piece of evidence for the importance of regeneration for \jpsi{} production at low \pt~\cite{ALICE:2012jsl,ALICE:2019lga}. 
In the bottomonium sector, studies of the $\Upsilon(1S)$ and the 2S, and 3S radial excitations at the LHC have shown a clear hierarchy in the suppression, that becomes stronger for the less bound states, a behaviour expected in presence of a high-temperature deconfined medium~\cite{Sirunyan:2018nsz,ALICE:2020wwx}.

From a theory perspective, recent progress has been made in the description of the modifications of the various quarkonium states in the QGP, largely driven by advancements in Lattice QCD studies~\cite{Asakawa:2000tr,Burnier:2013nla,Aarts:2014cda}. However, such ab-initio calculations still have difficulties in relating their output, via unfolding techniques, to the quarkonium spectral functions and their evolution as a function of temperature~\cite{Kim:2018yhk,Kelly:2018hsi}.  Alternatively, lattice-supported potential models provide robust calculations of the in-medium spectral functions but potentially miss relevant field theory corrections that cannot be captured by a simple instantaneous potential.
Spectral studies on pseudoscalar and P-wave ($L=1$) states like the $\chi_c$ and $\chi_b$ are very important in this context, thanks to their binding energy and spatial extent lying in-between those of ground and excited vector states, and the non-zero orbital angular momentum altering the structure of their wave function. These characteristics are expected to influence the stability of P-wave states with respect to other bound states, so that a comparison of their suppression patterns represents a important source of information on formation and propagation of bound states in the QGP. Current theoretical results for P-wave states are still ambiguous, with potential models indicating a survival of the resonance peaks, up to relatively high $T$~\cite{Burnier:2016kqm} ($\approx 1.2 T_{\rm c}$ for the $\chi_{\rm c}$ and $\approx 1.5 T_{\rm c}$ for $\chi_{\rm b}$ 1P states), while relativistic lattice QCD approaches~\cite{Kelly:2018hsi} indicate significantly lower temperatures for the disappearance of the ground P-wave states.

The theory framework of open quantum systems~\cite{Brambilla:2016wgg,Brambilla:2020qwo,Yao:2021lus} has recently gained much interest as a way to provide a dynamical description of a compact quarkonium system immersed in a hot QCD environment. This formalism lays out a clear and systematic chain of approximations, leading from QCD to the individual models of quarkonium evolution, where the different underlying assumptions are identified each with their own range of validity. Once a complete open-quantum-systems framework has been derived from first principles, one can try to find the most sensitive observable of each quarkonium state with respect to medium properties such as viscosity, opacity to charges etc., by interrogating the reduced density matrix of the quarkonium probe system. In this way, one should be able to describe the full quantum dynamics of strongly interacting systems and therefore the time evolution of quarkonia in the medium, including the recombination phase.
Establishing such a programme for quarkonium has just started and is work in progress.

Access to P-wave states would provide important additional information to discriminate between different approaches that have proven successful in describing results on vector S-wave states~\cite{Brambilla:2020qwo,Blaizot:2018oev}. Precise measurements of the \pt and rapidity dependence of $L=1$ states ($\chi_{\rm c}$, $\chi_{\rm b}$) are expected to provide stronger constraints on the spectral properties of bound states in the QGP and allow for a more accurate description of the dynamics of quarkonium interactions with the medium, from dissociation to regeneration. The interest of the specific case of bottomonium has been recently discussed in~\cite{Petreczky:2021zmz}, where estimates of the melting temperatures of various states including also pseudo-scalar, scalar and axial-vector states are given, showing that the fate of P-wave bottomonia is not yet completely settled. The ongoing development of the theory calls for a new experiment designed to access quarkonia with a variety of quantum numbers. Comparisons between S-wave and P-wave states both for charmonium and bottomonium would also allow for a more precise characterization of dissipative effects, which could significantly affect the early time evolution of relativistic heavy-ion collisions.

On top of the above arguments, assuming full charm thermalization occurring in the most energetic collisions at the LHC, with the population of the various quarkonium states described by Boltzmann factors, the measurement of the yields of multiple charmonium states would allow to test whether the freeze-out temperature of $c\bar{c}$ systems is the same as that of light quarks. 

As of today, pseudoscalar ($\eta_{\rm c}$,$\eta_{\rm b}$) and $L=1$ states remain largely unexplored in heavy-ion collisions, mainly because of difficulties in their detection in the complex experimental environment, and the $\eta_{\rm b}(1S)$ has not even been measured in pp collisions at the LHC. 
In LHC Run~3 and Run~4, the precision of measurements of $L=1$ states will be still marginal. ALICE~3 will have unique capabilities for the reconstruction of quarkonium states down to $\pt = 0$ and excellent performance for low energy photons ($0.5$~GeV and below), enabling accurate measurements of $\chi_{\rm c} \to \mathrm{J}/\psi~\gamma$ and $\chi_{\rm b} \to \Upsilon~\gamma$ in \pp and heavy-ion collisions over the widest kinematic range at the LHC. The expected performance for $\chi_c$ analysis is presented in Section~\ref{sec:performance:physics:quarkonia:chis}.
As for the pseudoscalars, the performance of ALICE~3 has not yet been studied in detail, but hadronic decays such as $\eta_{\rm c}\rightarrow p{\bar p}$ and $\eta_{\rm c}\rightarrow \Lambda{\bar \Lambda}$ may represent a viable channel, in spite of the relatively small branching ratios ($1.45\times10^{-3}$ and $1.1\times 10^{-3}$, respectively).

\subsubsection{Exotic hadrons}

The study of the existence and the nature of ``exotic'' states is a very active and challenging field in hadron physics, which aims at answering fundamental questions about how quarks arrange themselves in bound states. 
The study of exotic hadronic states containing a charm and an anti-charm quark, originally dubbed ``X, Y, Z'' states, has become a very exciting field, as many new hadrons have been discovered since the first observation of \chiX{} in 2003~\cite{Belle:2003nnu}. Several candidates for exotic states have been found in $\ee$ data in the last two decades, and the existence of some of them has been confirmed in hadronic collisions at the LHC~\cite{Aaij:2019evc}. New states have also been found at the LHC, see e.g.~\cite{Aaij:2019vzc,LHCb:2021vvq}. At the moment, the number of observed states is limited with respect to the potential richness of the spectroscopy of states with more than 3 valence quarks, and a clear explanation of the underlying quark structure of these particles in terms of tetraquarks, pentaquarks, or charm-hadron molecules is still missing~\cite{Lebed:2016hpi,Guo:2017jvc,Liu:2019zoy,Albuquerque:2018jkn,Brambilla:2019esw,Pepin:1996id,Vijande:2007rf,Hudspith:2020tdf}. Further experimental input is needed to achieve a coherent theoretical picture, from which clues to novel properties in the strong interaction may emerge.

Despite the results cited above, the experimental study of exotic hadronic states is still in its early stages and needs inputs on multiple fronts: discovery of new states, investigation of the nature and the quantum numbers of known states, characterization of the hadronisation mechanisms in a deconfined QCD medium. In this respect, $\ee$ and high-energy pp and p$\bar{\mathrm{p}}$ collisions offer the best opportunities for the discovery of new states and the determination of their quantum numbers, with $\ee$ collisions also allowing for precise measurements of the mass widths, and high-energy hadronic collisions enhancing the probability of observing states containing multiple heavy quarks. On the other hand, $\ee$ collisions are subject to intrinsic constraints on the quantum numbers of the investigated states, both in prompt production and measurements in decay chains, while in high-energy pp collisions an important constraint comes from the limited probability of producing heavy quarks lying sufficiently close in phase space to create bound states, due to the small size of the collision system.

Complementing these opportunities, relativistic heavy-ion collisions provide the only available tool to study the formation and interaction of exotic hadronic states inside a medium filled with deconfined colour charges, putting new constraints on the properties of the states, the characterization of their binding potential, and the details of their hadronisation mechanisms. Nuclear collisions at the LHC energies, in particular, offer a unique opportunity to test these mechanisms in the presence of large charm-quark densities~\cite{ExHIC:2010gcb,Fontoura:2019opw,Zhang:2020dwn,Wu:2020zbx,Chen:2021akx,Hu:2021gdg}. In this environment, while the screening of colour charges may reduce the yield of certain bound states, medium-induced or medium-enhanced mechanisms may enhance their production in the low-$\pt$ region ($\pt \lesssim 5-6$~GeV/$c$), as predicted by statistical hadronisation models~\cite{Andronic:2019wva}.
The interest of these observations is two-fold. On one hand, the rates of formation and dissociation of bound states depend on their binding energy and size~\cite{Esposito:2020ywk}, and the measurement of the production rates and anisotropic flow will allow one to distinguish compact multi-quark configurations from molecular states. On the other hand, states whose nature can be determined by other means, for example by measuring interaction potentials through final state interactions, as discussed in Section~\ref{sec:physics:HFstrong}, could be exploited as a calibrated probe in nuclear collisions, shedding light on the underlying mechanisms and timescales driving the hadronisation in a deconfined medium, and testing the properties of the deconfined medium itself. 

In addition to that, relativistic heavy-ion collisions also have a discovery potential for exotic hadronic states which still has to be fully explored, thanks to the long-lived deconfined medium and the large cross sections for heavy quarks available at the LHC energies, providing an ideal playground for the production of exotic states with e.g.~multiple b quarks. The most notable case is the $T^{-}_{bb}$, whose experimental detection would profit from the long predicted lifetime ($c\tau \approx 2.3$~mm) due to the stability of the state with respect to strong decays~\cite{Hernandez:2019eox} (contrary to the recently discovered, shortly-lived $T^{+}_{cc}$ state). Discovery opportunities also exist for compact bound hidden-charm hexaquarks also predicted to be stable with respect to strong interaction decays~\cite{Liu:2021gva}, and molecular states composed of three D mesons.

Studying exotic QCD states in nuclear collisions is therefore of central importance for QCD physics. 
The feasibility of such studies has been demonstrated by a recent measurement of \chiX{} production in \PbPb collisions by the CMS collaboration~\cite{CMS:2021znk} in the range $\pt > 10$~\GeVc. 
An immediate goal for ALICE~3 is to measure the production of \chiX{} down to the \pt region $\pt \lesssim 5-6$~GeV/$c$, where a significant enhancement of the yield was predicted~\cite{Andronic:2019wva} and which is not accessible by other LHC experiments. This can be achieved by providing muon identification down to $\pt\sim 1.5$~\GeVc{} at $\eta = 0$ and high efficiency for the detection of hadronic decay products in a large pseudorapidity acceptance. %

\subsection{Electromagnetic radiation}
\label{sec:physics:qgp_properties:qgp_radiation}

Photons and dileptons (virtual photons) are emitted during all phases of the collision and do not interact strongly with the medium. Since thermal emission rates increase strongly with temperature, real and virtual photon measurements are particularly sensitive to the temperature and flow field of the early stages of the collision. This makes them crucial probes for the earliest phases of the collision,  opening up the unique possibility to probe the pre-hydrodynamic phase~\cite{Coquet:2021lca}.

Real and virtual photons provide important complementary information on the temperature and the radial expansion velocity of the QGP. For real photons, the exponential slope of the transverse momentum spectrum in the range $1 \lesssim \pt \lesssim 3$~\GeVc reflects the apparent temperature of the QGP in the lab frame. As the plasma cools, the expansion velocity increases, resulting in a blue shift of the temperature. As a result, the measured slope depends both on the temperature and the expansion velocity~\cite{Shen:2013vja}. The mass distribution of dileptons (virtual photons) on the other hand, provides a Lorentz-invariant measure of the temperature that is not affected by collective radial expansion. By combining the information from both measurements, the temperature and radial flow velocity can both be determined.

Previous measurements of dilepton and photon emission at SPS and RHIC energies resulted in apparent temperatures in the range 210--240~MeV
~\cite{Arnaldi:2008er,Specht:2010xu,WA98:2000vxl,Rapp:2014hha,PHENIX:2014nkk}, which is already above the phase transition temperature to the QGP ($T_c = 156.5 \pm 1.5$~MeV~\cite{HotQCD:2018pds}). While the results on the photon \pt{} spectra are clearly in line with early emission from the hot phase of the system evolution, measurements of the elliptic flow of photons at RHIC~\cite{PHENIX:2011oxq} show a large azimuthal asymmetry, indicating significant photon emission at later times. These two observations do not fit naturally into a single hydrodynamical description of the pre-hadronic evolution of the system~\cite{Shen:2013cca} (the 'photon puzzle'). Measurements at the LHC, where the initial temperature is larger and the life time of the system is longer, will provide significant additional input. Existing measurements of direct photons in the thermal emission regime at LHC show a significantly larger apparent temperature of \SIrange{300}{400}{\mega\eV}~\cite{ALICE:2015xmh}, which is in line with theoretical expectations~\cite{Shen:2013vja}, albeit with large uncertainties. In Run~3 and 4, ALICE will collect a sufficiently large data sample to perform a first measurement of thermal dilepton emission to determine the temperature of the early stage. With larger sampled luminosity, larger acceptance and excellent  tracking of photon conversion products, ALICE~3 will further improve on these measurements of the temperature from the emission of photons and their elliptic flow. 

For dileptons, the double-differential analysis of the production in both transverse momentum and mass is a unique and so far largely uncharted terrain, which provides more differential access to the time-evolution of the temperature.
In the low invariant mass region $M < \SI{1}{\giga\eVcsq}$ (LMR), 
dilepton production is largely mediated by the light vector mesons $\rho$, $\omega$ and $\phi$. In the intermediate mass region $1 < M < \SI{3}{\giga\eVcsq}$ (IMR), after the subtraction of the open heavy flavour contribution, the spectrum is expected to be dominated by thermal radiation from
both hadronic and QGP sources, with the latter becoming dominant for  $M > \SI{1.5}{\giga\eVcsq}$~\cite{Rapp:2014hha}. Dilepton measurements in the intermediate mass range  provide a handle to disentangle the contributions of hadronic and QGP emission, unlike real photons. Precision measurements of virtual photons were performed at the CERN SPS by the CERES\,\cite{CERESNA45:1997tgc,CERESNA45:2002gnc,CERES:2006wcq} and NA60\,\cite{Arnaldi:2006jq,Arnaldi:2008er,Specht:2010xu,na60imr} experiments in the dielectron and dimuon channels, respectively. At those energies, dileptons from the Drell-Yan process contribute significantly to the measured spectra down to small invariant masses ($M$ $\approx$ 1.5\,\GeVcc{}). This background is experimentally irreducible and difficult to calculate in the non-perturbative regime, limiting the accuracy of the determination of the thermal contribution in the IMR. At RHIC energies, where the Drell-Yan contribution becomes already less relevant, the STAR and PHENIX collaborations are facing a large correlated heavy-flavour background, which requires high-precision vertex detectors to isolate the thermal contributions. Both experiments did not manage up to now to disentangle the prompt and non-prompt dilepton production and suffer from significant uncertainties in their results on thermal radiation\,\cite{STARdielectrona,Phenixdielectronb}. At LHC energies, the situation is even more challenging but at the same time more exciting since the in-medium modifications of vector mesons can be probed under conditions that are most closely related to the regime accessible by lattice QCD (vanishing net baryon-density) and the thermal radiation yield is expected to be the largest. The Drell-Yan contribution is completely negligible in this region. Dedicated experiments are required to reach precision dilepton measurements at the LHC. Significant improvements are expected with the ALICE upgrades\,\cite{ALICE:ITS3:2019,ALICEupgradeits2,ALICE:ITS2:2014} compared to the first ALICE dielectron results\,\cite{ALICE:2018ael}. Nevertheless, as shown in Section~\ref{sec:performance:physics:dileptons:comparisonITS3}, qualitative steps in detector performance are still needed to reduce the expected systematic uncertainties from the background subtraction to a negligible level. The expected improvements with ALICE 3 are presented in Section~\ref{sec:performance:physics:dileptons:chiral} with performance studies for double-differential dielectron analyses via temperature measurements as a function of transverse momentum.

Measurements of elliptic and higher harmonic flow coefficients of hadrons reflect the flow field at freeze-out. The study of the elliptic flow of dileptons provides independent constraints on the transport coefficients of the QGP fluid: the shear viscosity $\eta$ and the bulk viscosity $\zeta$
as a function of time and therefore of temperature\,\cite{Vujanovic:2019yih,Kasmaei_2019}, and is expected to shed further light on the photon puzzle.

The measurement of photons and dileptons from thermal radiation is experimentally challenging due to the low rates compared to hadronic processes, which produce a large background of decay photons and electrons. 
ALICE~3 aims to provide qualitative improvements in dilepton and photon measurements, by combining a very thin and light tracker to minimize the background from photon conversions, with excellent low-\pt tracking performance to reconstruct and reject the remaining electrons from conversions in the beam pipe and the first layer of the tracker. Moreover, the rejection of electron-positron pairs from $\pi^0$ Dalitz decays based on the invariant mass and opening angle will benefit from this. 
The excellent pointing resolution allows us to remove the correlated and uncorrelated background from heavy-flavour decays. Performance studies for the measurement of elliptic flow of dileptons are presented in Section~\ref{sec:performance:physics:dileptons:azimuth}.

{\revised In addition, thermal radiation of real photons will be measured via conversions in the tracking system and with the electromagnetic calorimeter.}
The insertion of a layer with calibrated conversion probability could be foreseen to improve the precision of the conversion photon measurement. {\revised The low-\pt{} capabilities of both the tracker and the ECAL of ALICE~3  enable a precise determination of} the decay photon background leading to a reduction of the systematic uncertainties. In particular at forward rapidity, tracks can be reconstructed down to very low \pt with high efficiency, %
which allows us to perform measurements of the $\eta$ meson yield at low \pt, the dominant source of systematic uncertainties in this region, with unprecedented precision.

\subsection{Chiral symmetry restoration}
\label{sec:physics:qgp_properties:chiral_symm}

In the massless limit, the QCD Lagrangian is symmetric under chiral transformations between left- and right-handed states. The mass spectrum of hadrons, however, indicates that this symmetry is strongly broken. In fact, chiral symmetry breaking, with the associated formation of quark and gluon condensates, generates most of the mass of ordinary matter, while the contribution of the mass generated by the Higgs field is almost negligible. Chiral symmetry breaking is a fundamental aspect of QCD, which is closely related to confinement: lattice QCD calculations show that the deconfinement phase transition takes place at temperatures close to chiral phase transition. In fact, it is the chiral transition that is more precisely established in lattice QCD calculations~\cite{Aoki:2009sc,Borsanyi:2010bp,Bazavov:2011nk,HotQCD:2018pds,Borsanyi:2020fev}. 

Dilepton production from the medium in the vicinity of the transition temperature is sensitive to effects of chiral symmetry restoration via the spectral functions (the imaginary part of the propagators) of the vector mesons, which mediate the interactions in the hot hadron gas~\cite{Pisarski:1995xu}. The contribution of the $\rho(770)$ is by far the most important, making it the prime probe for in-medium modifications of hadron properties. Its strong coupling to the
$\pi^+\pi^-$ channel and its life time of only \SI{1.3}{fm/\clight}, make it subject to regeneration in the longer-lived fireball at the LHC.

Experimental investigations of these effects started with fixed target experiments, which led to measurements of $\rho$  properties at the CERN SPS by the CERES experiment~\cite{CERESNA45:1997tgc,CERESNA45:2002gnc,CERES:2006wcq} and NA60~\cite{Arnaldi:2006jq,Arnaldi:2008er,Specht:2010xu}.
The CERES and NA60 data are dominated by the medium-modified $\rho$ contribution for $M < 1$ GeV and
agree well with a microscopic many-body model predicting a very strong broadening with essentially no mass shift~\cite{Rapp:1999us,vanHees:2007th}. %

\begin{figure}
    \centering
    \includegraphics[width=\textwidth]{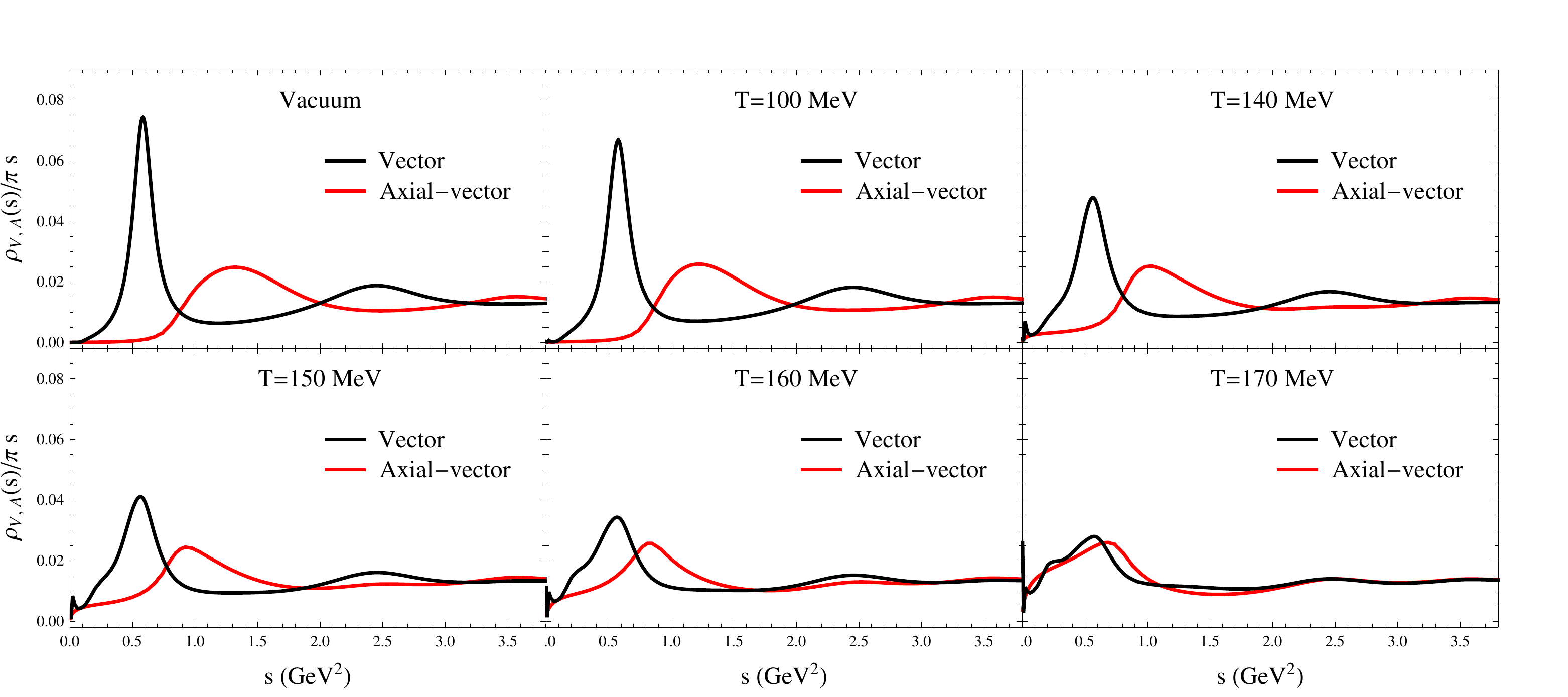}
    \caption[Temperature evolution of spectral functions]{Temperature evolution of vector and axial-vector spectral functions (non-linear realization)~\cite{Hohler:2013eba}.}
    \label{fig:physics:spectralfunctions}
\end{figure}
In recent years, robust theoretical evaluations have been performed to understand if this melting is compatible with chiral symmetry restoration for the realistic case of a hadron resonances gas. These calculations assume the decrease of the chiral condensate calculated in lattice QCD and use first-principle Weinberg sum rules which relate (moments of) the difference between the vector and axial-vector spectral functions to chiral order parameters.
Figure~\ref{fig:physics:spectralfunctions} shows the finite-temperature vector (black) and axial vector (red curve) spectral functions from the calculation~\cite{Hohler:2013eba}. They display strong broadening, and the vector and axial vector mesons become degenerate close to the critical temperature. The broadening of the $\rho$-meson spectral function observed at SPS\,\cite{CERESNA45:1997tgc,CERESNA45:2002gnc,CERES:2006wcq,Arnaldi:2006jq,Arnaldi:2008er,Specht:2010xu} and later at RHIC energies\,\cite{STARdielectrona,Phenixdielectronb} is therefore consistent with chiral symmetry restoration. However an unambiguous way to observe chiral symmetry restoration would be to measure the $\rho$ meson and its chiral partner $a_{\rm 1}$. The latter is very challenging in heavy-ion collisions.
\begin{figure}
    \centering
    \includegraphics[width=0.6\textwidth]{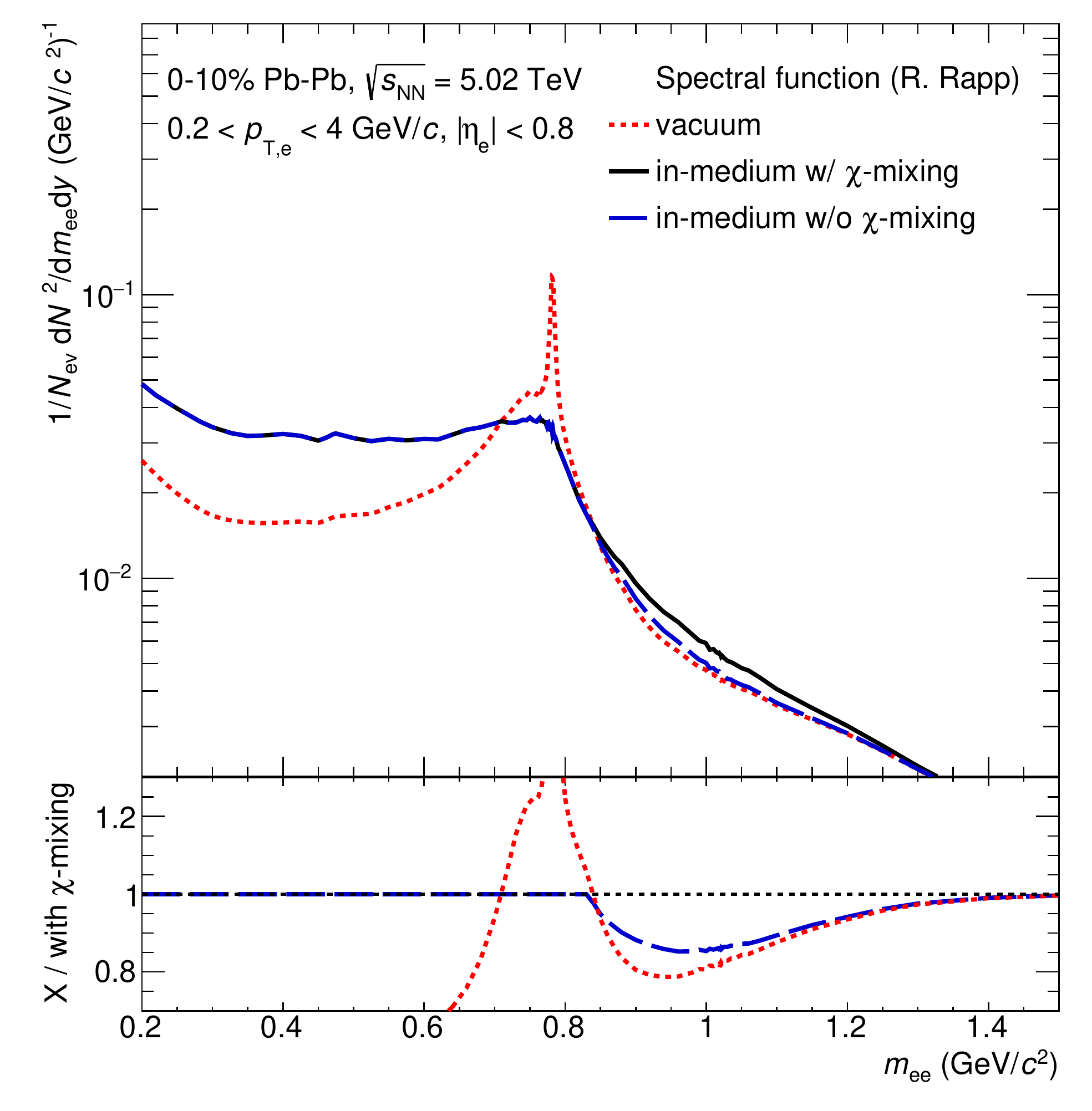}
    \caption[Thermal dilepton mass spectra]{Thermal dilepton mass spectra for three different scenarios: no medium effects, only $\rho$ broadening and $\rho$ broadening with chiral mixing of the $\rho$ and a$_1$.}
    \label{fig:physics:dileptons}
\end{figure}

The $\rho - {\rm a_{1}}$ chiral mixing mechanism provides access to the properties of the $a_{\rm 1}$, albeit indirectly. Figure~\ref{fig:physics:dileptons} shows the expected impact of $\rho-{\rm a}_1$ mixing on the thermal dilepton spectrum at LHC energies. The effect of the mixing can be identified by a precise study of the mass region between \SI{0.85}{\giga\eVcsq} and \SI{1.2}{\giga\eVcsq}.
There is no direct coupling of axial states to the dilepton channel in vacuum and, thus, the vector spectral function has a pronounced dip in the ${\rm a}_1$ mass range (See thermal dilepton mass spectrum calculated with vacuum spectral functions in Fig.~\ref{fig:physics:dileptons}).
Close to the critical temperature, the vector spectral function flattens in the mass range around 1~\GeVcc as a consequence of the $\rho\text{--}{\rm a}_1$ chiral mixing\,\cite{DEY1990620} and the approach to chiral symmetry restoration. In the medium, the presence of pions (both real and virtual) will “admix” the axial channel into the dilepton channel, most notably through 4-pion states (e.g. via $\pi {\rm a}_1$ annihilation) in the “dip region”. Albeit a small effect, the change in this region is sensitive to the %
mixing of the chiral partners $\rho$ and $a_1$ and therefore of chiral symmetry restoration.
This will lead to an enhancement of the yield of 20-25\% %
with respect to calculations performed with the vacuum %
spectral functions, and about 15\% compared to calculations with an in-medium spectral functions without mixing. In order to discern it, an experimental accuracy of 5$\%$ for the yield measurement will be needed in this mass range and is not expected to be reached with the ALICE upgrades. An overall understanding of the $\rho$ spectral function is at the same time mandatory. Precision measurements at lower mass ($m_{\rm ee}$ $\leq$ $m_{\omega}$) in combination with the Weinberg (and QCD) sum rules constrain the theoretical predictions at larger mass and are sensitive to additional mechanisms of the chiral symmetry restoration. %
Performance studies in central Pb--Pb collisions are presented in Section~\ref{sec:performance:physics:dileptons:chiral}. Measurements in elementary collision  systems (pp) are nevertheless crucial to establish a vacuum reference. Furthermore, multi-differential measurements as a function of system size, pair transverse momentum, azi\-muthal angle ($v_{\rm 2}$) and polarization variables\,\cite{SPERANZA2017282} will help to establish a clear connection between $\rho$-$a_{\rm 1}$ chiral mixing and a dielectron excess above the $\omega$ mass by mapping radial and elliptic flow effects. In addition, studies of dielectron polarization may help to to determine the origin of the enhancement (partonic or hadronic).

At the LHC, the in-medium modifications of vector mesons can be probed under conditions that are most closely related to what can be computed with lattice QCD (vanishing net baryon-density). In this sense, dilepton measurements at the LHC are ideally suited to determine how hadronic degrees of freedom transit into a QGP.
Full chiral mixing entails the observation of a purely exponential thermal spectrum starting from $M \sim \SI{1}{\giga\eVcsq}$. The onset of the  exponential part of the mass spectrum can be studied experimentally, and deviations from that limit are sensitive to the strength of the mixing effect.

\subsection{Electrical conductivity}
\label{sec:physics:qgp_properties:el_conductivity}

The electrical conductivity $\sigma$ is one of the key bulk properties of the QGP. It relates the electric current density $j_\mu$ with the electric field:
\begin{equation}
\langle j_i \rangle = \sigma \langle E_i \rangle.    
\end{equation}

Various theoretical approaches have been used to calculate the electrical conductivity of the QGP including lattice QCD~\cite{Aarts:2020dda}, perturbative QCD matrix elements in kinetic theory~\cite{Arnold:2000dr,Arnold:2003zc,Greif:2014oia}, effective models~\cite{Cassing:2013iz,Marty:2013ita}, and the AdS/CFT correspondence~\cite{Policastro:2002se,Teaney:2006nc}. 

The electrical conductivity is expected to be approximately proportional to the temperature $T$~\cite{Arnold:2000dr}.
Most theoretical calculations predict a moderate temperature-dependence of the ratio $\sigma/T$ for the quark--gluon plasma in the range of up to a few times the critical temperature $T_\mathrm{c}$. However, there is a large spread in the predictions from different theoretical approaches which can be approximately characterised by $0.001 \lesssim \sigma/T \lesssim 0.1$~\cite{Greif:2014oia}. 

Given the large theoretical uncertainty, an experimental constraint is highly desirable.  Moreover, the electrical conductivity of the QGP determines the time evolution of electromagnetic fields generated by spectator protons in non-central heavy-ion collisions~\cite{Huang:2015oca,Tuchin:2015oka}. A precise knowledge of the electrical conductivity is therefore important for the understanding of phenomena like the Chiral Magnetic Effect related to the presence of these strong magnetic fields. 

In the theoretical modeling, the electrical conductivity is associated with the so-called conductivity peak of the spectral function~\cite{Ding:2016hua}. The conductivity peak gives rise to an enhanced production of photons and dielectron pairs with very low masses and momenta. Different electrical conductivities correspond to different heights and widths of the conductivity peak~\cite{Moore:2006qn,Floerchinger:2021xhb}. 

{\revised So far the electrical conductivity is largely unconstrained by experimental data.
A major experimental challenge in measuring the electrical conductivity are background photons and dielectrons from the decay of neutral pions, $\eta$ mesons, and other short-lived hadrons. Preliminary studies of the background distributions show that by careful selections in transverse momentum and mass, the background from neutral pion Dalitz decays can be suppressed. With such selections and the identification of electrons at low momentum with the inner TOF layer, the thermal conductivity peak may be accessible in the dielectron channel with ALICE~3. In addition, the use of photon Hanbury Brown-Twiss correlations~\cite{WA98:2003ukc} using pairs of photons detected using the electromagnetic calorimeter and the conversion method in the tracker, as well as pairs reconstructed with the Forward Conversion tracker, can be explored to access the signal in the photon channel. More detailed physics performance studies are being performed.}

\subsection{Fluctuations of conserved charges}
\label{sec:phys:fluctuations}
Phase transitions in strongly interacting matter can be addressed by investigating the response of the system to external perturbations via measurements of fluctuations of conserved charges in heavy-ion collisions, see, e.g., refs. ~\cite{Ejiri:2005wq,HotQCD:2012fhj}. Such measurements can provide information on critical behaviour near the phase boundary between quark--gluon plasma and hadronic matter. The fluctuations can be directly related to generalized susceptibilities computed in lattice QCD (lQCD). Specifically, the susceptibilities are obtained from the derivatives of the pressure with respect to the chemical potentials corresponding to the conserved charges. The relevant charges are conserved quantum numbers such as electric charge Q, baryon number B, strangeness S, charm C and so on. At vanishing chemical potential, i.e., precisely under the conditions probed at the LHC, these susceptibilities are defined as  
\begin{equation}
    \label{eq:mixed_cumulant}
\chi_{klmn}^{\mathrm{B,S,Q,C}}=\frac{\partial^{(k+l+m+n)} (P(\hat{\mu}_{B},\hat{\mu}_{S},\hat{\mu}_{Q},\hat{\mu}_{C})/T^{4})}{\partial \hat{\mu}_{B}^{k}\partial \hat{\mu}_{S}^{l}\partial \hat{\mu}_{Q}^{m}\partial \hat{\mu}_{C}^{n}    }\Big|_{\vec{\mu}=0},
\end{equation}
and can be directly related to the experimental measurements of the cumulants\footnote{The cumulants, $\kappa_{n}$, of net-baryon number, $\Delta \rm N_{\rm B}=N_{\rm B}-N_{\rm \Bar{B}}$, are defined as the coefficients in the Maclaurin series of the logarithm of the characteristic function of $\Delta \rm N_{\rm B}$~\cite{Braun-Munzinger:2016yjz}.} of net-charge distributions. 

Recent lQCD calculations exhibit a rather strong signal for the existence of a pseudo-critical temperature at about 156 MeV~\cite{HotQCD:2018pds,Borsanyi:2020fev}.
Already the magnitude of the fourth order cumulant of net-baryon number fluctuations obtained from lQCD calculations is significantly below the expectation from non-critical Poissonian fluctuations of baryons and antibaryons. Critical fluctuations due to the vicinity of the cross-over line to a 2nd order phase transition of O(4) universality at vanishing u, d quark masses are expected to strongly modify the 6th and higher order cumulants of the net-baryon distribution~\cite{Friman:2011pf,Almasi:2017bhq}. Therefore, measuring 6th and higher order cumulants is one of the primary goals in ALICE~3. 

Even though the lQCD calculations suggest that the critical fluctuations are visible for 6th and higher order susceptibilities, one can address different dynamical signals by studying lower-order susceptibilities. For instance, a prominent effect that one has to consider is the global/local baryon number conservation, which plays a significant role in the interpretation of the net-charge cumulants. Generally, such fluctuation measurements are also influenced by volume fluctuations. They can, however, be controlled as discussed in ~\cite{Braun-Munzinger:2016yjz}. 

In experiments over the full acceptance, baryon number is conserved in each event. Therefore, fluctuations in net baryon number can only be  seen in measurements over a limited acceptance. Relevant results on conservation laws and the higher order cumulants of net-protons are discussed in~\cite{Braun-Munzinger:2016yjz,ALICE:2019nbs,Braun-Munzinger:2019yxj,Pruneau:2019baa,Arslandok:2020mda} with particular emphasis on the global/local baryon number conservation. See also the earlier work in~\cite{Capella:1978rg} where new data from the CERN ISR led to a first theoretical investigation on long-range rapidity correlations. Long-range rapidity correlations were also investigated more recently in~\cite{Andronov:2018xom} and in references given there. In this work, particular emphasis was placed on looking for correlations obtained by measuring charged particle multiplicities in different and widely separated rapidity windows. Such measurements could be obtained in ALICE~3 over an unprecedented window of 8 units in rapidity.

Analysis of net-proton data from ALICE ~\cite{ALICE:2019nbs} indeed indicate the presence of long-range rapidity correlations ($\Delta y_{corr}>5$) between protons and antiprotons arising from the early phase of the collision~\cite{Dumitru:2008wn}. In contrast, HIJING model calculations result in a significantly smaller correlation length ($\Delta y_{corr}=2$) presumably due to string breaking, as is implemented using the Lund string model. If this result can be substantiated by measurements over a much larger rapidity acceptance as is planned in ALICE 3 this would call into question, at least for baryon production, the physical basis of the Lund string model that is at the heart of all LHC event generators.
\begin{figure}[htbp]
	\centering
	\includegraphics[width=9cm]{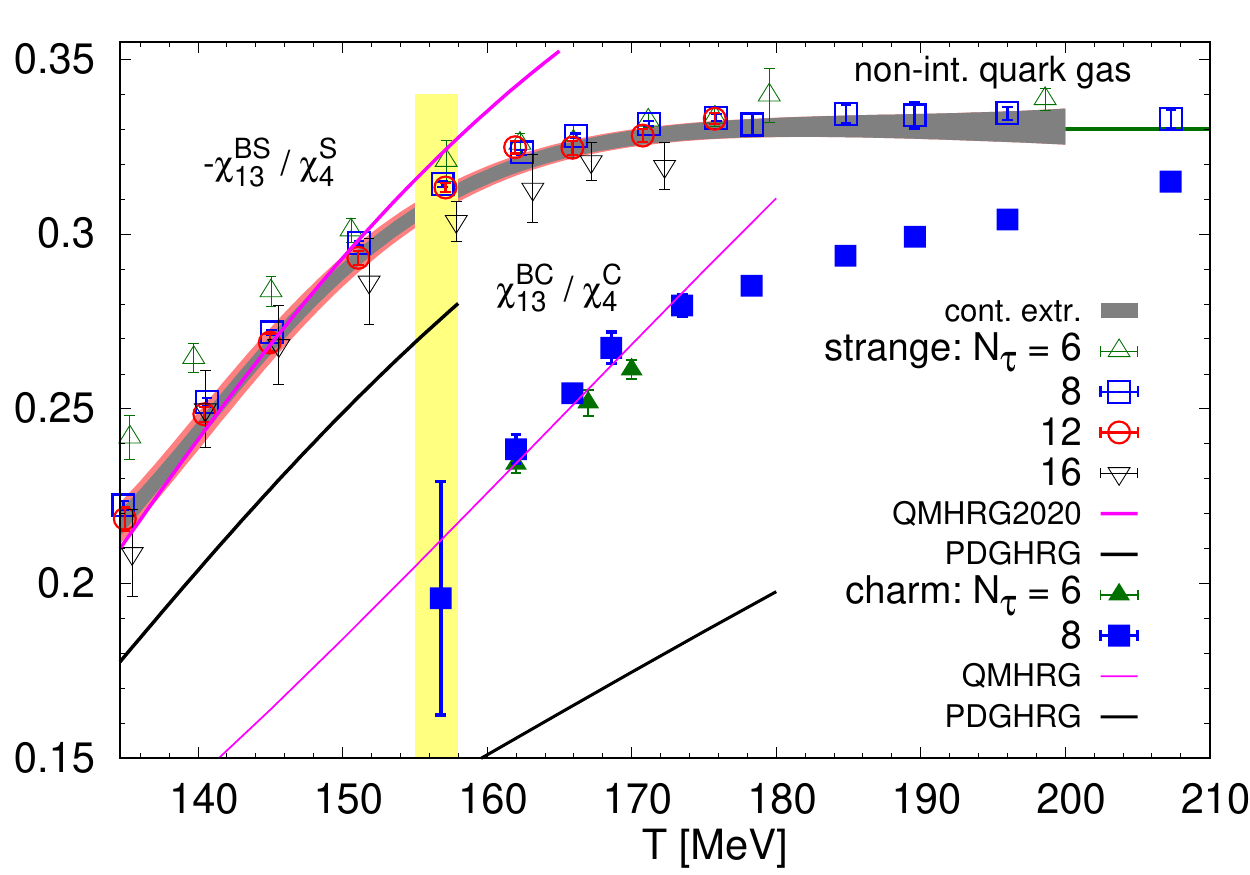}
	\caption[Ratios of 4th order cumulants of strangeness and charm fluctuations. ]{(Colour online) Ratios of 4th order cumulants of strangeness and charm fluctuations. The ratio $\chi_{13}^{\rm BS}/\chi_{\rm S}^{4}$ is dominated by the ratio of strange baryon number fluctuations over net kaon number fluctuations, and similarly $\chi_{13}^{\rm BC}/\chi_{\rm C}^{4}$. Lattice QCD results are compared to HRG model calculations using baryons and mesons listed in the particle data tables (PDGHRG) and augmented by hadrons calculated in a relativistic quark model (QMHRG), respectively.}  
	\label{fig:BS-BC}
\end{figure}

In ALICE~3, these studies can be extended to the strangeness and charm sector~\cite{Goswami:2020yez,Bazavov:2020bjn,Bazavov:2014yba}. In Figure~\ref{fig:BS-BC} we show 4th order cumulant ratios in the strangeness and charm sector that can be studied experimentally by measuring the cumulant ratios of strange (charmed) baryons and mesons. The comparison to HRG model calculations clearly shows the large discrepancy between lQCD calculations and HRG model calculations in the charm sector when only the experimentally established charmed hadron resonances are used. This suggests that additional charm states exist in nature. Indeed, by adding additional states in the HRG model, the agreement can be significantly improved, see Figure~\ref{fig:BS-BC}. Experimental measurements of net-charm fluctuations require high efficiency and high purity for open charm hadron reconstruction. Such studies are out of reach for current generation detectors, and will be seriously explored with ALICE~3. 

The physics performance for studies of second and higher order cumulants with ALICE 3 will be presented in Sec.~\ref{sec:perf:fluctuations}.

\clearpage

\subsection{Collective effects in small collision systems}
\label{sec:physics:small_systems}

Striking signals of collective behaviour have been observed in small collision systems at the LHC. Flow-like correlations that are both long-range in $\eta$ and collective in nature, very reminiscent of the long-range azimuthal correlations present in heavy-ion collisions, have also been observed in \pPb and in high-multiplicity \pp collisions~\cite{CMS:2010ifv,ALICE:2012eyl,ATLAS:2015hzw,CMS:2016fnw,ATLAS:2017rtr,ALICE:2019zfl}. Enhancements in the production of multi-strange baryons similar to those observed in Pb--Pb collisions, signalling the approach to chemical equilibration, were found to also be present in p--Pb and pp collisions at the LHC, with a magnitude that increases with (and appears to be chiefly dependent on) the multiplicity of particles produced in the collision\cite{ALICE:2016fzo}.

At this time, the origin of these effects is being debated~\cite{Schenke:2015aqa,Dusling:2017dqg,Greif:2017bnr,Schenke:2019pmk}. One intriguing possibility is that the effects have the same origin as in \PbPb{} collisions, i.e.\ that also in high-multiplicity \pp{} and \pPb{} collisions 
a dense system, and that it thermalises 
to some extent~\cite{Kurkela:2019kip,Schenke:2020mbo}. 
On the other hand, no effects of the parton energy loss expected to be associated to the formation of a coloured medium have yet been observed in small collision systems.

The study of collectivity in small systems will be a central topic in the upcoming Run~3 and Run~4 campaigns. However, the correct interpretation of these measurements relies on the assessment of possible biases which are particularly important for multiplicity-dependent yield measurements. For example, the interpretation of collisions with low multiplicity requires understanding the interplay between soft and semi-hard processes which may produce an apparent threshold in the multiplicity dependence of rare particles, such as multistrange baryons~\cite{Loizides:2021ima}. At high multiplicities and high $\pT$, correlated particle production can lead to a stronger than linear increase of yields with multiplicity unrelated to collective effects~\cite{Weber:2018ddv}.
It has been shown that with the limited pseudorapidity coverage of the current ALICE detector, the multiplicity selection can be biased by the presence of a single high-multiplicity jet~\cite{Ortiz:2021peu,Bencedi:2021tst}. Quantifying these biases currently relies strongly on input from MC models. 
The large rapidity coverage of ALICE~3 will be key to study and control the correlations between central and forward rapidity particle production, differentiating high-multiplicity events with uniform high-particle density from those with only a few high-multiplicity parton-parton interactions (see also~\ref{sec:physics:hi_mult_pp} below).

\begin{revised}
Runs with intermediate collision systems, as well as pA runs, are also being considered, in view of increasing the nucleon--nucleon luminosities for the study of hard processes, as a tool for the investigation of the system-size dependence of QGP observables  and as an important input for the study of the strong interaction potentials via femtoscopy measurements discussed in Section~\ref{sec:physics:HFstrong}.
Progressing on the investigation of the origin and system-size dependence of collective effects in small collision systems will be an important goal of the upcoming Run 3 and Run 4 campaigns, and substantial progress on this front is expected in the coming years. 
An optimisation of the ALICE 3 strategy in terms of possible intermediate species and luminosity goals will depend critically on the results of those studies.

\end{revised}

\subsection{Characterization of high-multiplicity pp events}
\label{sec:physics:hi_mult_pp}

Event-structure and event-shape studies in pp collisions have been undertaken in ALICE since a long time~\cite{ALICE:2012cor}. In particular, the observation of rare pp events in which the final state particles are essentially uniformly distributed in $\eta-\varphi$ has been reported~\cite{ALICE:2019dfi}. Such studies, however, are limited by the restricted acceptance of the current ALICE detector~\cite{Bencedi:2020qpi,Bencedi:2021tst}.
Rare high-multiplicity $p\overline{p}$
events not involving high-transverse momentum jets were also observed at the Tevatron, and dubbed ``hedgehog'' events, but were not studied in detail~\cite{Quigg:2010nn}.
A detailed characterisation of such events would be of great interest for the study the evolution of jet production in pp collisions with increasing energy density and, more generally, of collective effects in small systems, as discussed in Section~\ref{sec:physics:small_systems} above.
A comprehensive study of the properties of such special high-multiplicity events with a uniformly high density of rather low transverse momentum particles over a broad rapidity range can only be achieved with ALICE~3, due to its unique capabilities for tracking at low momentum over a broad pseudorapidity range ($|\eta|<4$).

The wide phase-space coverage offered by ALICE~3 would also provide a unique opportunity for the study of long-range correlations as a tool to investigate the dynamics of the initial states of hadronic collisions as a function of the system size~\cite{ALICE:2015kal,Kovalenko:2016bcx}.

\subsection{Ultra-soft photons}
\label{sec:physics:ultra-soft-photons}

At relativistic energies photons are copiously produced in hadron--hadron  as well as in nuclear collisions. A large fraction of these photons results from decays of the many mesons (in particular $\pi^0$ and $\eta$) and baryons  produced in such collisions, which is considered for the rest of this section as background to the signal of direct photons. The sources for direct photons comprise photons from inelastic parton--parton collisions as well as from bremsstrahlung emitted in the process of production of prompt charged hadrons that are stable against strong decays. The former have generally large transverse momentum \kt, typically from a few hundred MeV/$c$ to the GeV/$c$ range and higher, while bremsstrahlung exhibits a characteristic  1/\kt shape with a divergence towards zero \kt.

In general, the yield of direct photons is strongly process-dependent: photons produced in QCD processes such as the QCD Compton effect can be calculated in a perturbative approach as long as the \kt is sufficiently large. This applies both to direct photons produced in initial hard collisions as well as to thermal production of photons from a QGP. In the low \kt region one has to resort to non-perturbative methods such as lattice QCD methods with their own inherent uncertainties.

\subsubsection{Low's theorem and the infrared limit of gauge theories}
\label{sec:physics:ultra-soft-photons:low}

It was recognized early on ~\cite{Low:1958sn,Weinberg:1965nx} that,  for vanishing photon transverse momentum or photon energy,  i.e. in the ultra-soft regime, the bremsstrahlung spectrum can be computed in a model- and process-independent way if one knows, for the collision under consideration,  the momenta of all incoming and outgoing charged particles. This results in the leading 1/\kt term for the photon spectrum and is a direct consequence of the conservation of electric charge. The resulting Low theorem~\cite{Low:1958sn} can thus be used to quantitatively test the infrared limits of quantum field theories such as QED and QCD.

Low's theorem arises from the fact that in order to go from the process without a photon in the final state to the same process including bremsstrahlung photons one has to modify the Feynman amplitude of the original process by multiplying it with the particle propagator and particle--$\gamma$ vertex, and sum over all incoming and outgoing charged particles. Denoting the particle and photon 4-momenta by $p$ and $k$, respectively, the propagator contains a factor $[(p \pm k)^2 - m^2]^{-1} = ( \pm 2 p \cdot k)^{-1}$, where the particle mass $m$  cancels, and the $+$ sign is for outgoing and the $-$ sign for incoming charged particles. The resulting amplitude of the process with a bremsstrahlung photon has a pole whenever $p \cdot k \rightarrow 0$. This means that, in the limit of vanishing photon energy (in any reference system), the amplitude diverges as 1 over the photon energy. The limit of photon energy going to zero is usually realized in experiment by $\kt \rightarrow 0$, and one can show that the cross section for photon bremsstrahlung for $\kt \rightarrow 0$ then diverges as 1/\kt. For a modern view of the role of Low's theorem in the context of quantum field theories see ~\cite{Strominger:2017zoo} and references therein.

Importantly, one should  recognize that, in the soft photon limit, all Feynman diagrams where the soft photon line is connected to an internal line, i.e. a virtual charged particle, yield a non-diverging and, hence, 
negligible contribution to the soft photon production cross section, 
see Section II of~\cite{Weinberg:1965nx}. Consequently, it is not necessary to evaluate the contribution of all possible internal loops to the cross section and therefore, Low's leading term is 'tree-level' correct~\cite{Strominger:2017zoo} . 
 One of the main aims of studying ultra-soft photon production with the ALICE~3 detector is to test Low's theorem for both exclusive collisions with two charged particles in the final state such as those studied in ~\cite{Lebiedowicz:2021byo}, as well as for collisions leading to the production of more than two hadrons. %

In addition to the standard Low term there is also a sub-leading term of order zero in \kt that was also computed in the original Low paper. Recently, it was shown ~\cite{Lebiedowicz:2021byo} that this term needs a correction due to the requirement to incorporate exact energy and momentum conservation at this order. Corrections to the leading-power expression for the soft photon yield were recently also studied in~\cite{Bonocore:2021cbv}. Measurements in the sub-leading \kt range are important if one wants to study the approach to the Low limit as we plan to do with ALICE~3.

The model independence of the leading term in the expansion of the cross section in powers of \kt, i.e.
\begin{equation}
  \frac{\mathrm{d}\sigma}{\mathrm{d}\kt} = \frac{\sigma_0 }{\kt} + \sigma_1 + \kt \sigma_2 + \ldots
\end{equation}
can be understood, since for \kt close to zero the photon wavelength far exceeds the dimensions of any hadronic or nuclear system. For \kt = 1~MeV/$c$, the reduced wavelength $\lambda/(2 \pi)$ is close to 200~fm, implying that only the charges and momenta matter, not the structure of the particles carrying them. This simple estimate also provides information on how low in \kt one has to go to reach Low's limit: for pp collisions where the system size is of order 1~fm, measurements down to \kt = 20~MeV/$c$ should be sufficient. For Pb--Pb collisions the system size can approach 10~fm, implying that the Low limit is reached only around \kt = 2~MeV/$c$. \footnote{The somewhat arbitrary safety factor of 10 between photon wavelength and system size provides a rough estimate. Note, however, that one anyway  needs to go down to \kt $\approx$ 5~MeV/$c$, to avoid contaminations from decay photons, see the discussion in the experimental section below.}

It will also be interesting to explore the infrared limit reachable with virtual photons, i.e. in the production of electron pairs. Due to the finite electron mass, there is no divergent term for the production of virtual photons with pair mass or $\kt \rightarrow 0$. Since the electron mass is, however,  very small on a typical QCD mass scale, it will nevertheless be important to explore the region around small pair mass and vanishing pair \kt. For a discussion of virtual photon production near the Low limit, see also ~\cite{Lebiedowicz:2021byo}. We consider it one of the strengths of our plans to explore with ALICE~3 this limit simultaneously with real and virtual photons since since some of the crucial systematic uncertainties in the two types of measurements are very different.

We further note that the infrared limit of gauge theories in general and of gravity specifically has received considerable theoretical attention recently, see the detailed discussion in~\cite{Strominger:2017zoo} and references therein, lending further support to our efforts to explore experimentally  the regime of soft (virtual and real) photon production with the state-of-the-art detector technology of ALICE~3. The interest in the infrared limit of field theories has to do with the recognition that, due to the Low divergence, each elementary particle is always 'dressed' by a cloud of photons and gravitons. In the Low limit these soft clouds should be independent of the properties of the particle. In this context a detailed experimental test of ultra-soft photon production is also  of fundamental importance.

\subsubsection{Experimental situation and proposed measurements}

There have  been a significant number of attempts to experimentally test the predictions of Low's theorem~\cite{ Goshaw:1979kq,CHLIAPNIKOV:1984276,EHSNA22:1991sdp, Antos:1993wv,BANERJEE1993182,Tincknell:1996ks,BELOGIANNI:1997487, BELOGIANNI:2002122,BELOGIANNI:2002129,DELPHI:2005yew,DELPHI:2008,DELPHI:2010cit, Ardashev:2016}. Experiments were performed with different collision systems and at different accelerator facilities including the BNL AGS, CERN SPS and LEP. 

A basic assumption underlying the derivation of Low's theorem~\cite{Low:1958sn} is that the momenta of the incoming and outgoing particles should far exceed the momentum of the soft photon. An increase in centre-of-mass energy by more than a factor of 50 as obtained at the LHC, will not only fulfil this condition better than at other existing or planned accelerators but importantly bring along many new opportunities. Among them are: (i) the wide range of charged particle multiplicities for studies of soft photon production in inclusive pp collisions, 
(ii) many available exclusive final states in pp induced central diffractive production with large rapidity gaps, where soft photon production can be measured,  and (iii)     vector meson production along with a soft photon in ultra-peripheral pPb and Pb--Pb collisions. ALICE 3 will open a new chapter in ultra-soft photon measurements by investigating such different reaction channels  with the availability of new technologies for photon detection at far forward rapidity and with a nearly massless tracking and vertexing detector which can operate also in a high multiplicity environment .

The large majority of the tests of Low's theorem mentioned above resulted in the observation of significant excesses over what was predicted, typically by factors between 4 to 8. A noteworthy exception is the purely leptonic collision $e^+e^- \rightarrow \mu^+\mu^-\gamma$ in which the DELPHI collaboration, working at the $Z$ pole, reported good agreement between experimental results and predictions based on Low's theorem ~\cite{DELPHI:2008}. In contrast, the DELPHI measurement using hadronic $Z$ decays (jets) yields an excess by a factor 4~\cite{DELPHI:2005yew}. Surprisingly, this excess increases much faster with the neutral-hadron multiplicity than with that of charged hadrons in jets~\cite{DELPHI:2010cit}.

Several theoretical investigations  were undertaken to explain the observed excess, see e.g.~\cite{Barshay:1989yd,SHURYAK:1989175, LICHARD:1990605, PhysRevD.50.6824, Czyz:1993ti, Botz:1994bg, Hatta:2010kt, Wong:PhysRevC.81.064903, Kharzeev:PhysRevD.89.074053}. Most of the proposed explanations introduce an additional source of soft photons such as: propagation of charged particles through a pion condensate~\cite{Barshay:1989yd }; soft-pion reflection from a boundary under random collisions~\cite{SHURYAK:1989175}; a blob of cold quark--gluon plasma of low temperature~\cite{LICHARD:1990605, PhysRevD.50.6824};
bremsstrahlung of partons from string fragmentation~\cite{Czyz:1993ti}; synchrotron radiation from quarks in the stochastic QCD vacuum~\cite{Botz:1994bg}; production in ADS/CFT supersymmetric Yang-Mills theory~\cite{Hatta:2010kt}; oscillation of electric charge densities in a flux tube~\cite{ Wong:PhysRevC.81.064903}; induced current in Dirac sea~\cite{Kharzeev:PhysRevD.89.074053}. The authors of the latter references clearly emphasise that the DELPHI results suggest an additional mechanism for soft photon production due to nonperturbative QCD evolution. The proposed processes might be relevant in the approach to the Low region, but they generally vanish as $\kt \rightarrow 0$. Currently, there is no agreement on the possible origin of the observed excess yields.   Clearly, it would be very important to make a new effort to measure ultra-soft photon production in a systematic way in hadronic collisions.

According to Low's theorem the bremsstrahlung photons follow the scale-free $1/k_T$ form. The natural approach to test Low's theorem therefore is to measure photons down to the lowest possible $k_T$. The photon conversion method is particularly suited for measuring low-momentum photons and with this method the lower limit for reconstructing photons in ALICE 3 is about $k_T \approx 1~\mathrm{MeV}/c$. Photons from the decay of neutral pions and $\eta$ mesons pose a significant background in inelastic hadronic collisions. For $k_T \lesssim 10~\mathrm{MeV}/c$, however, the decay photon yields drops quickly as this $k_T$ range is well below the Jacobian peak of the photons from neutral pion decays. This makes the range $1 \lesssim k_T \lesssim 10~\mathrm{MeV}/c$ ideal for testing Low's theorem.

Our plan would be to start with exclusive diffractive collisions of the type $p + p \rightarrow p + p + \pi^+ + \pi^- +\gamma$ by requiring that the final state  contains, in addition to the soft photon, only the charged pion pair without any other hadronic activity, while the scattered protons escape unmeasured in both directions. In ~\cite{Lebiedowicz:2021byo} the authors have given an exact expression for the case of soft photon production in diffractive $\pi\pi$ scattering. The exact expression for the $pp$ scattering case is currently being calculated. In ALICE~3, we plan to investigate the soft photon limit for a number of such diffractive reactions where an exact expression of the cross section exists.

The reaction $p + p \rightarrow p + p + \pi^+ + \pi^-$ has already been studied by the CMS experiment ~\cite{CMS:2020jbb}. The CMS collaboration measured central exclusive production of $\pi^+\pi^-$ pairs in pp collisions at 5.02 and 13 TeV. Central diffractive events were selected by identifying two oppositely charged pions in the CMS tracking detectors and by vetoing energy deposits in the CMS calorimeters which originated from the decay of additional neutral mesons (leading to 2 or more photons) produced in the collision. The CMS collaboration did not investigate events with an additional soft photon as is our plan here.

We have started a programme to simulate this reaction following the information given in the CMS paper for selecting events belonging to this class of exclusive diffractive collisions without and with inclusion of an additional soft photon due to bremsstrahlung from the two produced pions. 

The simulation
is still in an initial state but first results imply that the soft photon in the Forward Conversion Tracker is
measurable with signal/background ratio of order 1. This is achieved with appropriate cuts on the calorimeter signals due to  photon decays from additional neutral mesons
and with adding rapidity gaps to enhance the diffractive
$\pi^+ \pi^- \gamma$ signal while suppressing other diffractive and non-diffractive processes.

As a next step we plan to gradually increase the complexity of the final state under study by investigating inclusive collisions  such as $p + p \rightarrow n \text{ charged hadrons}+\gamma +X$ as function of multiplicity $n$. Inclusive relativistic nuclear collisions would then provide the test bed for soft photon production in multi-particle inclusive collisions with hundreds to thousands of charged particles in the final state.

An interesting addition might be to measure the exclusive reaction $p + p \rightarrow p + p + J/\psi +\gamma$, with $J/\psi$ decaying into a lepton pair. In this case, the contributions to soft-photon production involving incoming and outgoing protons in the Low limit mostly vanish (due to the interference), and we are left with an essentially purely leptonic final state. Such measurements can shed light on the extent to which the anomalous soft photon production is due to nonperturbative QCD evolution.

For LHC energies a very recent study was undertaken to compute the bremsstrahlung emitted in the initial part of a Pb--Pb collision leading to a strong deceleration and rapidity shift of the initial baryons (or quarks) ~\cite{Park:2021ljg}. Although still schematic, the results indicate a significant source of photons in the range $20 < \kt < 200$~MeV/$c$. While not exactly a test of the Low theorem, it would be important to measure these photons as a test of the deceleration mechanism which is a prerequisite for strong energy transfer into the central region of the collision.

{\it The ALICE~3 experiment is the ideal place for a new state-of-the-art measurement in the endeavour to test and establish experimentally the infrared limit of two of the most important quantum field theories of the standard model, namely QED and QCD. This requires an additional dedicated detector for ultra-soft photons in the forward direction along with charged particle coverage over 8 units of rapidity. In addition, this allows to select  exclusive diffractive and ultra-peripheral collisions for ultra-soft photons studies. The expected levels of backgrounds for ultra-soft photon measurements with the proposed detector are estimated in Section~\ref{sec:performance:physics:soft_photons}.}

\subsection{Hadronic physics}
\label{sec:physics:hadronic_physics}
The unique performance and versatility of the \ALICETHR detector make it the perfect tool for the study of key topics at the interface of QGP,  hadronic, and nuclear physics.
In particular, thanks to the LHC high sustainable interaction rate and to the large amount of data that can be collected, it will be possible to study rare processes such as the formation of light nuclei, hyper-nuclei, and super-nuclei (bound states of charm baryons such as \lc with ordinary nucleons). The large expected yields for rare hadrons and nuclear objects at low transverse momenta will also provide unprecedented access to strong-interaction potentials via the measurement of two-particle momentum correlations.
These measurements are crucial for an experimental investigation of the nature of exotic states such as the newly-discovered $T_{cc}^+$. Some of these measurements also have fundamental implications for the understanding and interpretation of astrophysical observations. 
For most of these studies it is necessary to perform the measurements in both small and large collision systems because theoretical predictions are either sensitive to the source size (coalescence models) or are more reliable in central collisions of large ions where a larger number of parton-parton re-scatterings leads to a complete thermalisation of the system (statistical-thermal models).
The large rapidity and transverse momentum acceptance of the proposed apparatus will also provide an ideal environment for the study of the hadronic physics in ultra-peripheral collisions.

\subsubsection{Study of the strong interaction between heavy flavour hadrons}
\label{sec:physics:HFstrong}
\newcommand{\ks}{\ensuremath{k^*}\xspace}
\newcommand{\rs}{\ensuremath{r^*}\xspace}

In the last decades, many new hadronic states containing charm and beauty quarks have been discovered at lepton and hadron colliders~\cite{Guo:2017jvc,Hosaka:2016pey}. The newly discovered states such as \chiX{}~\cite{Chen:2016qju}, $\mathrm{T}^+_\mathrm{cc}$~\cite{LHCb:2021vvq,LHCb:2021auc} or $\mathrm{P}_\mathrm{c}(4450)$~\cite{Aaij:2015tga} could be either multi-quark states (e.g., tetraquark and pentaquark) or hadronic molecules.
The properties of hadronic molecules can be predicted in a controlled way following different theoretical approaches, but experimental evidence of the molecular structure is not affirmative yet.
Many candidates for molecular states cannot be clearly distinguished from classical quark model states due to tunable ingredients in the theoretical models used to predict them  and possible large mixing of several coupled channels that contribute to the molecule formation.

The relationship between the scattering length governing the interaction of the two hadrons composing a molecule and the properties of the molecule can be obtained using Weinberg’s compositeness criterion~\cite{PhysRev.131.440}, which pinned down the nature of the deuteron as a proton-neutron bound state. There, the pole location in the corresponding hadron-hadron scattering S-matrix identifies unequivocally the presence of the bound state. Normally, the presence of such a pole translates into a sign inversion of the scattering length close to the mass threshold. 
A prominent example considering only light quarks is the $\Lambda(1405)$ molecule. Its pole mass is located slightly below the $\mathrm{K^-p}$ state and it is described as naturally emerging from the meson-baryon coupled channel system composed of $\mathrm{\overline{K}N}$ and $\pi\Sigma$~\cite{Hyodo:2011ur}. The nature of the $\Lambda(1405)$ could recently be pinned down~\cite{ParticleDataGroup:2016lqr} only by combining information from the decays into $\Sigma\pi$ with measurements of the scattering length of the $\mathrm{K^-p}$ channel above the threshold via scattering experiments and the study of kaonic hydrogen~\cite{SIDDHARTA:2011dsy}.
Here, we propose to extend the same approach to the study of hadronic molecules with heavy quarks, i.e. to measure the scattering parameters for the constituents of the molecule to shed light on the nature of the molecule candidates.
For hadrons containing charm and beauty quarks, scattering experiments are not feasible and no exotic atoms have yet been discovered, therefore the only way to access the strong interaction is the femtoscopy technique~\cite{Fabbietti:2020bfg,ALICE:2018ysd}. It consists of the measurement of  correlations in momentum space for any hadron-hadron pair and can be used to extract the corresponding scattering parameters. 
The ALICE3 experiment at the LHC will allow the measurement of several hadron combinations including $\mathrm{DD^*}$, $\Lambda_\mathrm{c}^+\Sigma_\mathrm{c}^{0,+,++}$ and $\mathrm{BB^*}$ in pp, p--Pb and Pb--Pb collisions and thereby shed light on the nature of many, which are so far not understood exotic hadrons.

The fundamental quantity to be measured in femtoscopy is the two-particle correlation function.
Experimentally, the correlation function is defined as $C(\ks)\,=\,\xi(\ks)\otimes\frac{N_{\mathrm{same}}(\ks)}{N_{\mathrm{mixed}}(\ks)}$, where $\xi(\ks)$ denotes the corrections for experimental effects,
$N_{\mathrm{same}}(\ks)$ is the number of pairs with a given \ks obtained by combining particles produced in the same collision (event), which constitute a sample of correlated pairs, and $N_{\mathrm{mixed}}(\ks)$ is the number of uncorrelated pairs with the same $\ks$, obtained by combining particles produced in different collisions. 
\begin{figure}
\centering
\includegraphics[width=0.8\textwidth]{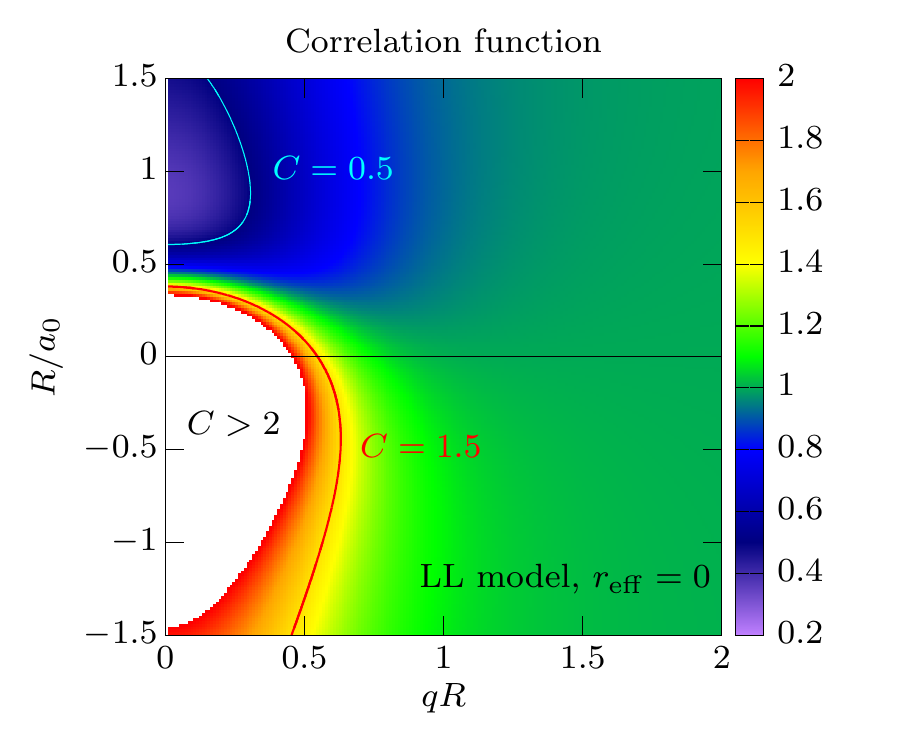}
\caption[Correlation function amplitudes]{Predicted correlation function as a function of the $R/\mathrm{a_0}$ ratio ($R$ being the source size and $a_0$ the s-wave scattering length) and the $q$R variable (q is the invariant relative momentum for a chosen pair)~\cite{Kamiya:2021haz}. The effective range of the interaction $d_{0}$ is set to $0$ in this calculation.}
\label{fig:ScatLen}
\end{figure}
Attractive interactions will be mapped into correlation amplitudes above unity, values between zero and one will indicate a repulsive interaction. The presence of bound states will be mapped on the correlation functions by values above unity for small $k^*$ values and below unity for larger $k^*$ values, with the crossing position depending on the interaction potential.
For all interactions the correlation function converges to unity for $k^*$ values above 200-300 MeV$/c$. %
The correlation functions can be computed 
using the Lednický and Lyuboshitz (LL) model~\cite{Lednicky:1981su}. The LL model relies on an effective field range approximation for the formulation of the wave function and provides an analytical formula for the correlation function depending on the scattering parameters (scattering length $a_0$ and effective range $d_0$) of the s-wave and the size of the particles emitting source.
If the scattering length $a_0$ is comparable to or larger than the emission source size $R$ the correlation function is sensitive to the interaction potential and to the presence of bound states~\cite{Morita:2019rph,Kamiya:2021haz}.
In ALICE 3, $R$ can be varied by measuring pp, p--nucleus and nucleus--nucleus collisions. 
Figure~\ref{fig:ScatLen} shows the expected correlation function calculated with the LL analytic model for an attractive potential and $d_0=\,0$ as a function of the $R/\mathrm{a_0}$ and $qR$ variables, with $q$ the invariant relative momentum.
The region of negative $R/\mathrm{a_0}$ corresponds to an attractive interaction without bound state formation. One can see that the correlation function is always larger than one, with a pronounced peak structure for small relative momenta, for any value of the radius. The region of positive  $R/\mathrm{a_0}$  corresponds to the bound region and there the correlation function will show the same features of the attractive-only interaction for small radii, but it will undershoot unity and form a dip for larger values of $R$. Given a fixed value of the scattering length $a_0$, the source size $R$ can be varied different colliding systems at the LHC. Indeed for pp collisions the radii are in the order of 1 fm, these increase to 1.5 fm for p--Pb collisions up to 4-5 fm for semi-central Pb--Pb collisions.

\begin{table}
\centering
\small
\renewcommand*{\arraystretch}{1.3}
  \sisetup{separate-uncertainty,table-space-text-post={test}}
\begin{tabular}[t]{l S[table-format=5.1(2)] S[table-format=2.1(2)] p{2.5cm} p{2.cm}}
  \toprule
  State & {Mass} & {Width} 
  & \multicolumn{1}{c}{S-wave threshold} 
  & \multicolumn{1}{c}{Coupled Channels} \\[-5pt]
  & {(\si{\mega\eVcsq})} & {(\si{\mega\eVcsq})} & \multicolumn{1}{c}{(\si{\mega\eVcsq})} & \\
  \midrule
  X(3872)~\cite{Zyla:2020zbs}  & 3872 \pm 0.2  &  1.19 \pm 0.21 & ${\rm D}^{*0}\bar{{\rm D}}^{0}(-0.04)$, \newline ${\rm D}^{*+}\bar{{\rm D}}^{-}(-8.11)$ 
    & $\pi^{+}\pi^{-}\mathrm{ J/\psi}$, \newline $\pi^{+}\pi^{-}\pi^{0}\mathrm{ J/\psi}$\\

  X(3940)~\cite{Zyla:2020zbs} & 3942\pm 9   & 37 & $\mathrm{D^{*}\overline{D}^*}$ (-75 $\pm$9) & ${\rm D}^{*}\bar{{\rm D}}$\\

  X(4140)~\cite{Zyla:2020zbs}& 4147 \pm 4.5   & 83 \pm 21 & $\mathrm{D_s\overline{D}^{*}_s}$ (-66$^{+4.9}_{-3.2}$) & $\phi\mathrm{ J/\psi}$  \\

  X(4274)~\cite{Zyla:2020zbs} & 4273 \pm 8.3   & 56 \pm 11  & $\mathrm{D_s\overline{D}^*_s}$ (-49.1$^{+19.1}_{-9.1}$) & $\phi\mathrm{ J/\psi}$\\

  $\mathrm{Z}_\mathrm{b}$(10610)~\cite{Zyla:2020zbs} & 10607 \pm 2.0   & 18.4 \pm 2.4  &  $\mathrm{B\overline{B}^*}$(4$\pm$3.2) & $\pi^{\pm} \Upsilon(\mathrm{nS})$ $\pi^{\pm} \mathrm{h_b}(\mathrm{nP})$  \\

  ${\rm Z}_{b}^{\pm}(10650)$~\cite{Zyla:2020zbs} 
  & 10652.2 \pm 1.5 & 11.5 \pm  2.2
  & ${\rm B}^{*}\bar{{\rm B}}^{*}(+2.9)$ 
  & $\pi^{\pm} \Upsilon(\mathrm{nS})$ $\pi^{\pm} \mathrm{h_b}(\mathrm{nP})$ \\

 $\mathrm{P_c}^{+}(4312)$~\cite{Aaij:2019vzc} 
  & 4311.9 \pm 0.7$^{+6.8}_{-0.6}$
  & %
  & $\Sigma_\mathrm{c}\bar{{\rm D}}(-9.7)$
  & $\mathrm{pJ/\psi}$ \\
  
$\mathrm{P_c}^{+}(4440)$~\cite{Aaij:2019vzc} 
  & 4440.3 \pm 1.3$^{+4.1}_{-4.7}$ & 20.6 \pm 4.9$^{+8.7}_{-10.1}$
  & $\Sigma_\mathrm{c}\bar{{\rm D}}^{*}(-21.8)$
  & $\mathrm{pJ/\psi},\Sigma_\mathrm{c}\bar{{\rm D}}$$\Sigma_\mathrm{c}^{*}\bar{{\rm D}}$ \\
  
$\mathrm{P_c}^{+}(4457)$~\cite{Aaij:2019vzc} 
  & 4457.3 \pm 0.6$^{+4.1}_{-1.7}$ & 6.4 \pm 2.0$^{+5.7}_{-1.9}$
  & $\Sigma_\mathrm{c}\bar{{\rm D}}^{*}(-4.8)$
  & $\mathrm{pJ/\psi},\Sigma_\mathrm{c}\bar{{\rm D}}$$\Sigma_\mathrm{c}^{*}\bar{{\rm D}}$ \\
 
$\mathrm{T_{cc}}^{+}$~\cite{LHCb:2021vvq} & 3874.827 & 0.410 &
${\rm D}^{*+}\bar{{\rm D}}^{0}(-0.273)$, ${\rm D}^{*0}\bar{{\rm D}}^{+}(-1.523)$ 
  & ${\rm D}^{0}\rm{D}^{0}\pi^{+}$ \\ 
\bottomrule
\end{tabular}
\caption{Selection of candidates for hadronic molecules with a mass close to a hadron-hadron mass threshold~\cite{Zyla:2020zbs,Aaij:2019vzc,LHCb:2021vvq}.}
\label{tab:channels}	
\end{table}

Table~\ref{tab:channels} provides a selected list of heavy-flavoured multiquark/molecule candidates with a mass close to a hadron-hadron mass threshold, which are good candidates for molecular states and are promising combinations for investigation by ALICE 3.\\
As an example of the projected physics capability of ALICE 3 in this area, preliminary studie of the expected performances for the study of $\mathrm{D^0D^{*+}}$ correlations close to the mass of the $\mathrm{T_{cc}^+}$ 
and of the $\rm D^{0(+)} \overline{D}{}^{*0(-)}$ correlations to unravel the nature of the \chiX{}
are presented in Sections ~\ref{sec:performance:physics:ddstar} and ~\ref{sec:performance:physics:ddbarstar} .

\subsubsection{Search for exotic anti-, hyper- and super-nuclei}
\label{sec:physics:exotica}

Little is known about the hyperon-nucleon interaction and even less about the charmed-baryon-nucleon interaction.
The \ALICETHR programme allows to shed light on this sector with a set of unique measurements and potential discoveries in the area of exotic anti-, hyper-, and super-nuclei.
It is worth noting that the underlying production cross sections are poorly understood on the theoretical side and there is a general lack of experimental data to cover this sector.

Collisions at LHC energies produce particles and anti-particles within experimental uncertainties in approximately equal abundances.
For this reason, we refer in the following typically to both particle and anti-particle even if only one is explicitly mentioned. %
For some studies, it is important to use anti-particles which do not suffer from  background from particle knock-out in the detector material. In addition, anti-nuclei and anti-hyper-nuclei with A~$\geq$~5 such as ${^{5}_{\overline{\Lambda}}\mathrm{\overline{He}}}$ or $^{6}\mathrm{\overline{Li}}$ have yet to be discovered and may well be in reach of \ALICETHR.

\newcommand{\hyperhefive}{\ensuremath{^{5}_{\Lambda}\mathrm{He}}\xspace}

\newcommand{\hyperhfour}{\ensuremath{^{4}_{\Lambda}\mathrm{H}}\xspace}

\begin{figure}[t]
  \centering
  \includegraphics[width=0.6\textwidth]{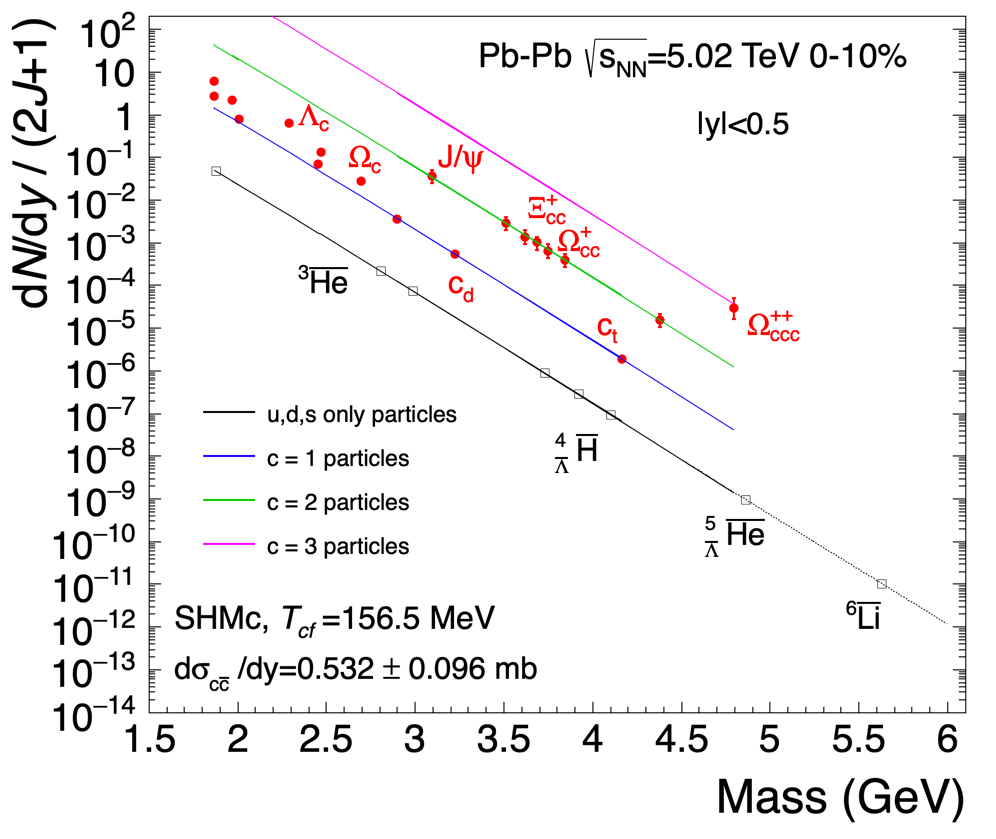}
  \caption[(Anti-)(hyper-)nuclei production in the SHM]{Statistical-thermal model predictions for (anti-)(hyper-)nuclei in black and \mbox{(multi-)}charm states in red.
  For each additional charm quark an enhancement in the yield by the charm fugacity factor $g_c$ appears at the same hadron mass.
  All states depicted here are potentially in reach of \ALICETHR.
  Figure taken from~\cite{Andronic:2021erx} with slight adaptations.
  }
  \label{fig:physics:nuclei:penalty_factor}
\end{figure}

The \ALICETHR apparatus is ideally suited for the observation of the $A = 4$ or $A = 5$ hyper-nuclei like \hyperhfour or \hyperhefive.
Their measurement is interesting in its own right for precision tests of particle production models~\cite{Bellini:2019zqc} and to constrain hyperon-nucleon potentials~\cite{Haidenbauer:2019thx}, but is also of interest as a baseline for the study of multi-charm baryon production in the QGP.
Hadron yields in high-energy heavy-ion collisions are well described by the statistical hadronisation model (SHM), where the particle mass enters decisively, i.e. the production yields follow a thermal distribution.
The \hyperhefive provides an interesting test case, with a mass ($\approx 4.867$~\GeVcc) close to the mass of the $\mathrm{\Omega_{ccc}}$. In the SHM, the latter are expected to be about a factor $g_{c}^3 \simeq 30000$ more abundant than the former (see Fig.~\ref{fig:physics:nuclei:penalty_factor} ), due to the large value of the charm fugacity factor $g_{c}$ at LHC energies~\cite{Andronic:2021erx}.

The measurement of $A = 6$ nuclei would provide precision tests for the formation of bound clusters thanks to the special nature of $^{6}\mathrm{He}$ and $^{6}\mathrm{Li}$.
$^{6}\mathrm{He}$ is the lightest known (anti-)halo-nucleus and its production is therefore expected to be suppressed in coalescence models with respect to thermal-statistical models due to its much larger size.
$^{6}\mathrm{Li}$ is a stable isotope with a spin of $J=1$.
With respect to the helium isotopes $^{4}\mathrm{He}$ and $^{6}\mathrm{He}$ with $J=0$, $^{6}\mathrm{Li}$ production is therefore expected to be enhanced by the degeneracy factor $g$ of its spin-substates: $g = 2J+1=3$.
Similarly, it should be possible to study the spin dependence and wave-function dependence of the production of nuclei by measuring the formation of the strongly decaying $^{4}\mathrm{Li}$ in the  $^{3}\mathrm{He}$--$\mathrm{p}$ channel via correlations techniques such as those discussed in Section~\ref{sec:physics:HFstrong}.

As can be seen from Fig.~\ref{fig:physics:nuclei:penalty_factor}, the expected production yields \dNdy for $A = 5$ hyper-nuclei and $A =6$ nuclei (\lisix and \hesix) are in the $10^{-9}$ to $10^{-11}$ range in Pb--Pb collisions. The corresponding requirements for their observation in terms of luminosity as well as purity of the particle identification are discussed in detail in Sections~\ref{sec:performance:detector:hadron_id} and~\ref{sec:performance:physics:nuclear_states:nuclei}.

The unique combination of high rates, wide acceptance (both in $\eta$ and \pt), strong particle identification capabilities, and excellent secondary vertex determination also provide an excellent toolbox for the search for light nuclei with charm, the so-called super-nuclei.
At LHC energies, the most promising candidates are the c-deuteron, c-triton and c-$^{3}$He as shown in Fig.~\ref{fig:physics:nuclei:penalty_factor}.
In these nuclei, an up quark is replaced by a charm quark in the proton that is bound to one or two nucleons.
The existence of such weakly decaying bound states with lifetimes similar to those of other unbound charmed baryons is being debated in the literature~\cite{Maeda:2015hxa,Haidenbauer:2017dua,Garcilazo:2019ryw,Guven:2021pkn}.
We propose to search for the c-deuteron in its $c_{d} \rightarrow d+K^{-}+\pi^{+}$ decay channel and for the c-triton in the $c_{t} \rightarrow\ ^{3}{\rm H} + K^{-}+\pi^{+}$ channel. In case these states are bound and under the assumption that their abundance in collisions of nuclei is described by the SHM, the yields are large enough to bring their experimental discovery within reach. A study of the expected ALICE~3 performance for the measurement of c-deuteron production is discussed in Section~\ref{sec:performance:physics:nuclear_states:nuclei}.

In addition to allowing for direct searches for the bound states, \ALICETHR should also make it possible to measure hadronic interaction potentials between charmed baryons and nucleons (and hadrons in general) by studying directly femtoscopic correlations in pp collisions, e.g. $\Lambda_c p$ correlations for which Lattice QCD predictions can be directly tested~\cite{Haidenbauer:2020kwo}, following the techniques discussed in Section~\ref{sec:physics:HFstrong}.

\subsubsection{Study of b-quark decays into $^3\mathrm{He}$}
\label{sec:physics:Lb_to_Antinuclei}

\newcommand{\Lambdab}{\ensuremath{\Lambda_\mathrm{b}^0}\xspace}
\newcommand{\antiLambdab}{\ensuremath{\overline{\Lambda}_\mathrm{b}^0}\xspace}
\newcommand{\He}{\ensuremath{{}^3\mathrm{He}}\xspace}
\newcommand{\hyperT}{\ensuremath{^3_{\Lambda}\mathrm{H}}\xspace}
\newcommand{\hyperTtoHe}{\ensuremath{\hyperT\to\He\uppi^-}\xspace}
\newcommand{\antihyperT}{\ensuremath{^3_{\overline{\Lambda}}\mathrm{\overline{H}}}\xspace}

\newcommand{\antiHe}{\ensuremath{^3\mathrm{\overline{He}}}\xspace}
\newcommand{\LambdabToHe}{\ensuremath{\Lambdab \to \He+ \mathrm{X}}\xspace}
\newcommand{\antiLambdabToHe}{\ensuremath{\antiLambdab \to \antiHe+\mathrm{X}}\xspace}
\newcommand{\mev}{\ensuremath{\mathrm{MeV}}\xspace}
\newcommand{\gev}{\ensuremath{\mathrm{GeV}}\xspace}
\newcommand{\tev}{\ensuremath{\mathrm{TeV}}\xspace}
\newcommand{\dedx}{\ensuremath{d{\rm E}/d{\rm x}}\xspace}
\newcommand{\antiLambdabToDominant}{\ensuremath{\antiLambdab \to \mathrm{\overline{d}u\overline{u}(\overline{ud})_0}}\xspace}

The excellent capabilities of the \ALICETHR apparatus for the measurement of nuclei also open new opportunities in interdisciplinary studies for one of the most interesting astrophysical questions.
The detection of cosmic-ray antinuclei such as antihelium is considered as one of the most promising signatures of the existence of weakly-interactive mass particles (WIMP), which represent an important candidate for dark matter~\cite{Donato:1999gy}. In fact, background sources given by antinuclei produced by the hadronic interactions of primary rays are expected to be negligible~\cite{Duperray:2005si}.
Recently, the Alpha Magnetic Spectrometer Collaboration (AMS-02) reported a preliminary evidence of $\mathcal{O}(10)$ \antiHe events. In~\cite{Winkler:2020ltd}, a previously neglected process, the production of \antiLambdab baryons in dark-matter annihilation, and their subsequent decay into \antiHe nuclei, was proposed as a possible explanation for the observation. However, the decay rates of \antiLambdab to antinuclei are not experimentally measured and serve as crucial inputs to understand the AMS data.
The large beauty-quark production cross section at LHC energies and the identification capabilities for nuclei over practically the full phase space, make ALICE 3 the ideal experiment for the measurement of these decays in a controlled environment. A study of the expected ALICE~3 performance for the measurement of the \LambdabToHe branching ratio is discussed in Section~\ref{sec:performance:physics:nuclear_states:beauty}.

\subsubsection{Photoproduction of vector mesons}
\label{sec:physics:vectormesons}

Ultra-Peripheral Collisions (UPC), for which the impact parameter exceeds the sum of two nuclear radii, are dominated by interactions involving quasi-real virtual photons from the strong electromagnetic fields that accompany ultrarelativistic heavy nuclei. Ultra-peripheral collisions are used to probe the target nuclei and can also be used as a source of vector mesons to study meson physics. One of the main UPC processes is vector-meson photoproduction, in which the incident photon fluctuates to a quark-antiquark pair that then scatters quasi-elastically from a target nucleon, emerging in the final state as a vector meson.

Most UPC studies require a complete reconstruction of the final state and the rejection of background from events with additional particles. Large acceptance is key to ensuring high efficiency and high purity.
With its wide acceptance in pseudorapidity and strong PID, ALICE 3 can explore more complex vector meson decays than were previously possible, allowing us to explore the spectroscopy of higher excitations of the $\rho$, $\omega$ and $\phi$. The decay $\rho'\rightarrow\pi^+\pi^-\pi^+\pi^-$, for instance, is of particular interest. The STAR collaboration has analysed about 500 $\pi^+\pi^-\pi^+\pi^-$ events, and found a shape consistent with the $\rho'$ being a single resonance~\cite{STAR:2009giy}. An analysis of multiple data sets, including  2$\pi$ and 4$\pi$ final states, however, found indications that the peak contains two states~\cite{rhoReview,ParticleDataGroup:2018ovx}. 
ALICE 3 will allow to study the $\rho'$ over a broad rapidity range (enabling a measurement of the energy dependence of the cross section) with vastly improved statistics.  This would also allow for a detailed study of the $4\pi$ substructure in different $M_{4\pi}$ ranges.  By comparing data from different targets (protons and lead), one could compare the mass spectra; a change in the $M_{4\pi}$ distribution would be a sign that there are two different resonances, with different nucleon interaction cross sections. A preliminary evaluation of the expected performance of the proposed ALICE~3 apparatus for the study of this channel is presented in Section~\ref{sec:performance:physics:bsm:upc:photoproduction}. The large statistics would also facilitate the study of rarer decays of the $\rho'$, such as that to $\pi^+\pi^-$~\cite{Klein:2016dtn,ALICE:2020ugp}. 

The $\phi$ meson is also of particular interest because it has a simple $s\overline s$ quark content and is intermediate in mass between `light' mesons and the $J/\psi$. 
Photoproduction of $\phi$ mesons at collider energies has not been observed yet, because $K^\pm$ from $\phi\rightarrow K^+K^-$ are too soft. With momenta of 135 MeV/$c$ in the $\phi$ rest frame, they lose energy rapidly, and do not leave detectable tracks.  The proposed apparatus would allow the study of $\phi$ production at large rapidity, where the kaons are longitudinally Lorentz boosted, and so lose energy less rapidly. 
More complicated final states, including 4-prong and 6-prong final states, and those involving kaon pairs and/or neutral mesons like $\pi^0$, could also come into reach.  These data will help in separating exotic and conventional mesons in the 1.5 - 2.5 GeV$/c^2$ mass  region.  Direct pair production is also of great interest; ALICE 3 may be able to probe $p\overline p$ pairs.  

Of even greater interest is the study of the production of pairs of vector mesons by exchanging more than one photon between the two ions~\cite{Klein:1999qj,Klusek-Gawenda:2013dka}.  Since the vector mesons are produced coherently, with small $\pt$, the phase space is very limited, and it should be possible to observe stimulated production of identical mesons into the same quantum states.  This should increase the production of pairs with similar rapidity and $\pt$.  Stimulated decays may even be visible.  These effects should be most visible with $\rho^0\rho^0$ pairs, although there is a large 4-pion background from $\rho'$ photoproduction.  ALICE 3 will have a unique opportunity to study this channel, by looking for $\rho$ pairs with large separation in rapidity.  These pairs will have a much larger $\rho$-$\rho$ invariant mass than is expected from a $\rho'$ or other single meson decay.
 
Other pairs may also be promising.  Measurements of $J/\psi - J/\psi$ would be the most striking, but the rates are likely to be low.  Pairs of $\phi$ mesons are more copiously produced, but are experimentally challenging because of the soft kaons.  In addition, $\omega-\omega$ pair production would be of particular interest because of the different observable final states, allowing the comparison of events with the same and with different final states, while $\rho'\rho'$ pairs are relatively common, but suffer from the 8-particle final states.  Non-identical pairs, like $\rho J/\psi$ are more easily accessible, and should provide a good baseline to probe for quantum enhancements.

Finally, photoproduction of exotic hadrons is a useful probe of their nature; meson molecules and tetraquarks, which are also discussed in Sections~\ref{sec:physics:exotica} and~\ref{sec:physics:HFstrong}, have different couplings to photons. Here, again, large acceptance is required.

\subsubsection{Photoproduction of dijets and open heavy flavour pairs}
\label{sec:physics:dijets}

The photoproduction of dijets and open heavy flavour hadron pairs is sensitive to gluon distributions at low Bjorken$-x$ at a range of $Q^2$, with relatively small theoretical uncertainties.  ATLAS has already performed some such studies~\cite{ATLAS:2017kwa}, and further analyses are expected during Runs 3 and 4.  Open charm is an attractive target for ALICE Runs 3 and 4, but the limited acceptance will likely preclude reconstruction of the full (charmed and anti-charmed hadron) final state. This is required to determine the Bjorken$-x$ and $Q^2$ on an event-by-event basis~\cite{Klein:2002wm,Goncalves:2017zdx}.  While studies of the performance of ALICE 3 are yet to be performed, we expect the large acceptance of the proposed apparatus to allow for the reconstruction of the full final state. Reconstruction of open bottom mesons should also be in reach.

Fully reconstructed exclusive dijets or open charm hadron will allow to go beyond one-dimensional parton distributions.  Measurements of the distributions of $\pt$ sum and difference of the two jets (or charmed and anticharmed hadrons) are related to the Wigner distribution of gluons (how they are distributed in nuclear targets, in impact parameter and $\pt$)~\cite{Hatta:2016dxp,Klein:2019qfb}.  Similar studies are planned at the proposed U.S. electron-ion collider, but the study of ultra-peripheral collisions at the LHC would reach down to significantly lower Bjorken$-x$ values.

To reach even lower Bjorken$-x$ values, we are considering to make use of the Forward Calorimeter~(FoCal)~\cite{ALICE:2020mso} that is planned to be installed for Run~4.
The FoCal is a highly granular Si+W electromagnetic calorimeter combined with a conventional metal-scintillator based hadron calorimeter.
FoCal could be placed at a similar location as in Run~4, at about $z=-7$~m, to provide access to direct photon, jet and heavy-quark processes at $3.4<\eta<5.8$.
If ALICE~3 is operated with a lighter ion species, this would also allow us to study the $A$ dependence of photo-production and the nuclear gluon density.
Reconstruction of electrons~(and charged particle in general) using the FoCal will be facilitated by the tracking layers in-front of the FoCal, which should significantly improve
the reconstruction capabilities for J/$\psi$ and $\Upsilon$, giving access to these probes at Bjorken-$x$ as low as $10^{-6}$.

\subsection{BSM studies}
\label{sec:physics:bsm_studies}

The search for physics beyond the Standard Model (BSM) is one of the main goals of the LHC. Compared to standard proton-proton collision studies, heavy-ion collisions provide unique and complementary means to search for new phenomena~\cite{Bruce:2018yzs}. In particular, ultra-peripheral collisions (UPCs) of heavy ions offer a natural environment for the studies of photon-mediated processes, such as light-by-light scattering~\cite{dEnterria:2013zqi}, axion-like particle searches~\cite{Bauer:2017ris} and $\tau$ $g-2$ measurements~\cite{Beresford:2019gww,Dyndal:2020yen}. ALICE 3 provides an opportunity to extend these studies down to low transverse momenta that are not accessible by other LHC experiments. The wide pseudorapidity coverage of the ALICE 3 detector provides an additional advantage for the selection of exclusive final states in UPCs.

\subsubsection{Light-by-light scattering measurements}

The elastic scattering of two photons proceeds at leading order via virtual one-loop box diagrams involving charged fermion (leptons
and quarks) and $W^\pm$ boson loops. The box diagram may also contain virtual contributions from new charged particles, therefore the light-by-light scattering process is particularly sensitive to various extensions of the Standard Model such as Born-Infeld Theory~\cite{Ellis:2017edi}, anomalous gauge couplings~\cite{Brodsky:1994nf}, supersymmetry~\cite{Ohnemus:1993qw}, monopoles~\cite{Ginzburg:1998vb}, unparticles~\cite{Cakir:2007xb}, low-scale gravity~\cite{Cheung:1999ja} and non-commutative interactions~\cite{Hewett:2000zp}. 

Ultra-peripheral collisions at the LHC provide a clean environment to study the light-by-light scattering~\cite{dEnterria:2013zqi}. The first evidence of this process has been reported by the ATLAS~\cite{ATLAS:2017fur,ATLAS:2020hii} and CMS~\cite{CMS:2018erd} Collaborations. The fiducial cross sections were found to be consistent within 2 standard deviations with the Standard Model predictions, however, the measurements were restricted to photon-photon invariant masses above 5 \GeVcc with precision limited by statistical uncertainties.

ALICE 3 provides a possibility to extend these measurements down to low photon-photon invariant masses where $s$, $t$ and $u$--channel meson exchanges may play an important role~\cite{Lebiedowicz:2017cuq,Klusek-Gawenda:2019ijn}. These measurements appear to be particularly interesting since hadronic contributions to the light-by-light scattering process introduce one of the largest theoretical uncertainties in the calculation of the muon anomalous magnetic moment (see~\cite{Muong-2:2021ojo} and references therein). Feasibility studies on light-by-light scattering measurements with the ALICE 3 apparatus are discussed in Section~\ref{sec:performance:physics:bsm:upc:lbl}.

\subsubsection{Axion-like particle searches}

There has been increasing interest in searches for axion-like particles (ALPs) emerging as pseudo Nambu–Goldstone bosons of a new spontaneously broken global symmetry in many BSM scenarios like supersymmetry, Higgs extensions and models based on composite dynamics~\cite{Bauer:2017ris,dEnterria:2021ljz}. 
Light pseudoscalars were also proposed as promising dark matter candidates or dark-sector mediators. In most of the considered scenarios, ALPs naturally couple to photons via the following effective Lagrangian:
\begin{equation}
{\cal L} = -\frac{1}{4}g_{a\gamma}\,a\,F^{\mu\nu}\tilde F_{\mu\nu},
\end{equation}
where $a$ is the ALP field, $F^{\mu\nu}$ is the photon field strength tensor, and $g_{a\gamma}=1/\Lambda_a$ is the dimensional ALP-$\gamma$ coupling constant related to the high-energy scale $\Lambda$ associated with the broken symmetry.  Thus the production and decay rates of axion-like particles are fully defined in the two-dimensional parameter space of the axion mass $m_a$ and the corresponding coupling $g_{a\gamma}$.

Existing constraints on the ALP mass and coupling are shown in Fig.~\ref{fig:bsm:alps}.
Cosmological, astrophysical, and low-energy accelerator studies have set stringent limits on the existence of ALPs at $m_a<50$\,MeV covering 13 orders of magnitude in $g_{a\gamma}$ from $10^{-10}$ to $10^3\;{\rm TeV}^{-1}$. The region of masses above 50 MeV$/c^2$ and couplings $g_{a\gamma}>1\;{\rm TeV}^{-1}$ has been investigated at $e^+e^-$ colliders via studies of diphoton and triphoton final states.
The best limits on ALPs in the $m_a$ range from $5$ to $100$\, \GeVcc{} have been set in light-by-light scattering studies performed in ultraperipheral collisions of lead ions by the CMS~\cite{CMS:2018erd} and ATLAS~\cite{ ATLAS:2020hii} collaborations. ATLAS and CMS have limited abilities to explore lighter masses due to the difficulties in the triggering and reconstruction of photons with transverse energy below 2 GeV.\footnote{Photon energies are equal to $m_a/2$ at midrapidity.} The ALICE and LHCb experiments, however, can potentially improve those limits in a mass region $m_a \approx 1 -5$\,\GeVcc{} which is otherwise very challenging to access experimentally.

The ALICE 3 experiment, in particular, has a unique opportunity to fill the gap in the intermediate ALP mass range from 50 MeV$/c^2$ to 5 \GeVcc{} and set limits on ALP couplings well below $1\;{\rm TeV}^{-1}$. This region is particularly interesting since it includes the unexplored parameter space in which ALPs can explain the anomalous magnetic moment of the muon ~\cite{Marciano:2016yhf,Bauer:2017ris}. 
The expected sensitivity of the proposed apparatus for ALP searches in ultra-peripheral collisions is discussed in Section~\ref{sec:performance:physics:bsm:upc:alp}.

\ifirc
\else
\cleardoublepage
\section{Performance}
\label{sec:performance}

In order to illustrate the physics reach of the detector, we present studies on the detector and physics performance in this chapter.
Based on the geometry and fundamental properties of the detector systems, key parameters, such as the resolutions and efficiencies, are evaluated.
Dedicated tools have been developed to simulate the detector response for physics events from Monte Carlo event generators.
The various strategies, implementing detector effects at different levels of detail, are discussed in the next sections.

\subsection{Simulation strategies}
\label{sec:performance:introduction}

The performance of the tracker can be characterised on an analytical level.
To this end, the detector is assumed to consist of cylindrical tracking layers (around midrapidity) and flat disks (in the forward region), which detect the passage of charged particles with a given efficiency and intrinsic position resolution.
This allows the calculation of tracking efficiency and momentum resolution from the likelihood to assign wrong hits and from a fit to the measured space points.
This evaluation takes the local hit density into account, which determines the occupancy in the detector.
We will report the detector performance for \pp collisions and (central) \PbPb collisions.

While analytical tools are useful to understand the performance on a conceptual level, our physics analyses rely on a set of reconstructed tracks, reflecting the estimated properties of particles.
To test their performance, Monte Carlo tools implement the generation of realistic events in different collision systems and the propagation of these particles through a concrete detector geometry, in which they deposit energy.
For the most realistic approach, a reconstruction step is needed to translate the measurements in the various detector parts to tracks.
This approach is called \textit{full simulation}.
An alternative procedure, referred to as \textit{fast simulation}, is to directly smear the parameters of the particles from an event generator, which avoids the computing-intense step of particle propagation through the detector and subsequent reconstruction. 

The studies presented in the following rely on both strategies, depending on the particular goal. 
For a few specific scenarios, a combination of the two approaches has been used in a technique called \textit{hybrid simulation}.
Hereafter, a brief overview of the techniques used in the estimation of the performance of the experimental apparatus is presented. 

\subsubsection{Full simulation}
\label{sec:performance:full-simulation}

The full simulation makes use of the ALICE framework for simulation, reconstruction, and analysis for Run~3, called~\otwo~\cite{ALICE:2014lor}.
The framework has been extended to support more flexible simulations and more complex detector layouts in order to implement the geometry of the ALICE~3 tracker.
Physics events are generated with the \textsc{Pythia8} Monte Carlo event generator~\cite{Sjostrand:2014zea}. 
The final-state particles are transported through the experimental apparatus by \textsc{Geant3}~\cite{Brun:1987ma} 
transport tools.%
These tools provide a detailed description of the passage of particles through matter, including their decays and interactions, which result in detectable energy loss in the active volumes of the detector. 
As such it accounts for all effects which determine the performance and is ideally suited for the evaluation of occupancies, including effects of secondary particles, for refined estimates of the impact of material, as well as for physics studies which depend on the most realistic description of the detector response.
The output of the detector transport is represented by hits in the sensitive volumes of the detector, i.e. space-points which correspond to the interaction points of particles and material in a sensitive detector volume.
The position resolution of the tracker is emulated by applying Gaussian smearing to the hit position.
The smearing reflects both the intrinsic pixel resolution and the resolution resulting from the clusterisation of multiple adjacent pixels.

The reconstruction step to match hits from the same particle and extract the track parameters was based on an extension of the tracking code used for the ITS2~\cite{Puccio:2016biw}. 
The tracking comprises three main phases:
\begin{description}
    \item\textit{Tracklet and cell finding} associates hits on adjacent layers to form pairs, which are called tracklets.
    Pairs of consecutive tracklets which share a cluster are then matched into cells.
    Overall, this phase considers hits from three layers at a time to form cells and therefore the execution time of the algorithm scales linearly with the number of layers. 
    \item\textit{Road finding} refers to the process of combining cells which share two hits (a tracklet) to identify track candidates, called roads.
    The tree of possible connections of cells is computed and roads are identified by means of a cellular automaton algorithm.
    The final roads represent the best candidates.
    Tracks with a number of hits smaller than the number of layers are good candidates to be reconstructed as secondary particles.
    \item\textit{Track fitting} is the final step, in which the track parameters are calculated from the road using a Kalman-filter-based algorithm.
    This fitting method takes multiple scattering and energy loss into account. By construction, it is a local process that can scale to a variable number of hits (layers). 
    Finally, track candidates undergo posterior selections based on the quality of fit. 
\end{description}
This sequence may be run multiple times with different parameters in order to assign as many hits as possible to tracks.

\subsubsection{Fast simulation}
\label{sec:performance:fast-simulation}

For the fast simulation a dedicated software package called \textsc{Delphes}\otwo has been developed based on the \textsc{Delphes}~\cite{deFavereau:2013fsa} fast-simulation package, with additional custom routines for ALICE~3 and an interface to produce analysis object files in the ALICE \otwo~\cite{o2tdr} data format. %
\delphesotwo provides an interface to events in the HepMC format, e.g. generated with the \textsc{Pythia8} Monte Carlo event generator, and delivers an intermediate data model for processing.
Final-state particles injected into the fast simulation framework are propagated by means of \textsc{Delphes} modules which take into account the curvature in the magnetic field and decays of unstable particles up to the desired radius.
This information is then stored for the next processing stage which applies detector effects via a smearing of kinematic variables.

The tracking efficiency and the elements of the covariance matrix have been calculated using a fast Monte Carlo tool. The code, which accounts for multiple scattering, detector occupancy and deterministic energy loss, provides accurate determination of the tracking resolution as a function of the detector configuration for both the spatial and the momentum components and a reliable estimate of the tracking efficiency.
The efficiency and full covariance matrix for the track parameters are stored in multi-dimensional look-up tables as a function of particle mass, transverse momentum, pseudorapidity and event charged-particle density. %
The use of the full covariance matrix (instead of diagonal element only) is an important extension of \textsc{Delphes} for the determination of the heavy-flavour performance.
Finally, the tracks are converted into the ALICE Analysis Data model and processed with standard analysis tools and routines, also developed for Run~3.

The fast simulation of the particle identification of the time-of-flight detector is based on the smearing of the track arrival time at the TOF layer after the \textsc{Delphes} propagation stage according to the desired time resolution.

The RICH performance is simulated using data from a detailed \textsc{Geant4}~\cite{GEANT4:2002zbu} simulation of the detector. 
The refractive index, length of the radiator, single-photon angular resolution and the single-photon detection efficiency are the input parameters of the fast-simulation model. 

The fast simulation of the electromagnetic calorimeter and of the muon identifier follows a similar approach, with their performance included into the framework from a parametrization of the results obtained from dedicated \textsc{Geant4} studies.

\subsubsection{Hybrid simulation}
\label{sec:performance:hybrid-simulation}

Some performance studies, e.g.~strangeness tracking explained in Sec.~\ref{sec:performance:detector:strangeness_tracking}, require both large statistics and an accurate description of weak decays.
With the full simulation it is not possible to generate a sufficiently large sample, while the fast simulation framework currently lacks an accurate description of weak decays, which is of special importance for this use case.
The two approaches have been combined into a technique called hybrid simulation.
Only the weakly decaying hadrons $\rm{K}^{0}_{S}$, $\Lambda$, $\Xi^{-}$ and $\Omega^{-}$ are simulated and transported through the detector material by the full simulation.
All other particles are only smeared within the fast simulation framework.
Since most particles do not require the compute-intensive propagation through the detector, such a hybrid approach allows the production of large samples while retaining the full description of weak decays and the subsequent evolution of the decay daughters.
Compared to the full simulation, a ten-fold increase in speed is achieved.

\subsection{Detector performance}
\label{sec:performance:detector}

\begin{revised}
In this section, the expected detector performance based on the detector specifications outlined in Section~\ref{sec:introduction:experimental_layout} is discussed.
The performance studies are based on a parametrisation of the momentum resolution with a solenoid and two dipole magnets, to assess the ultimate physics reach of the experiment. 
The impact of the reduced performance as achieved with a solenoidal magnet alone will be studied to finalise the design of the magnet system.
Some studies, e.g. on the reconstruction of low-mass dielectrons, assume a reduced magnetic field to increase the acceptance at lower transverse momenta. 
This could be accommodated in a dedicated run as done in Run~2.
\end{revised}

\subsubsection{Tracking and vertexing}
\label{sec:performance:detector:track_vtx}

\begin{figure}[p] %
  \centering
     \includegraphics[width=0.49\textwidth]{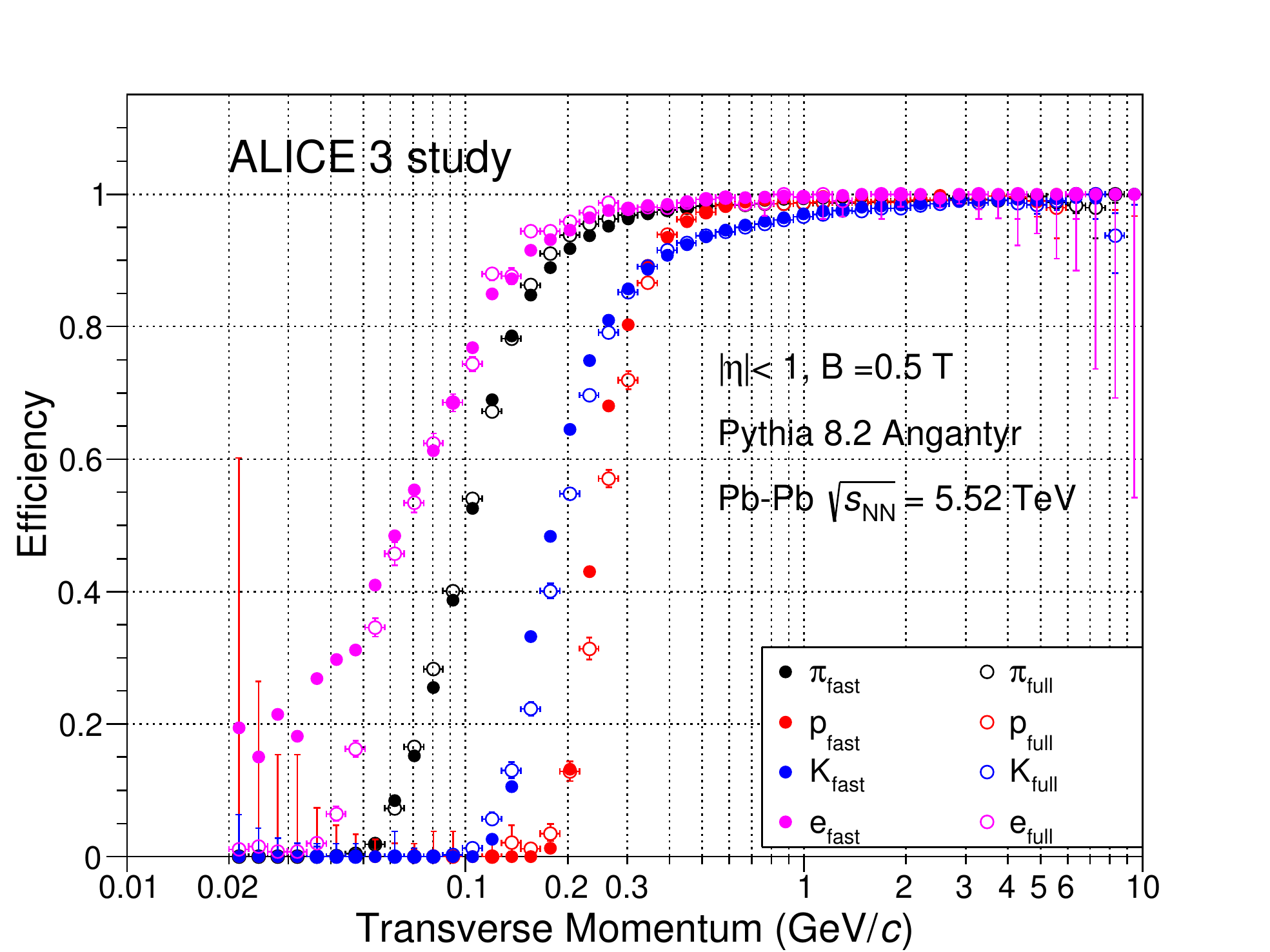}
     \includegraphics[width=0.49\textwidth]{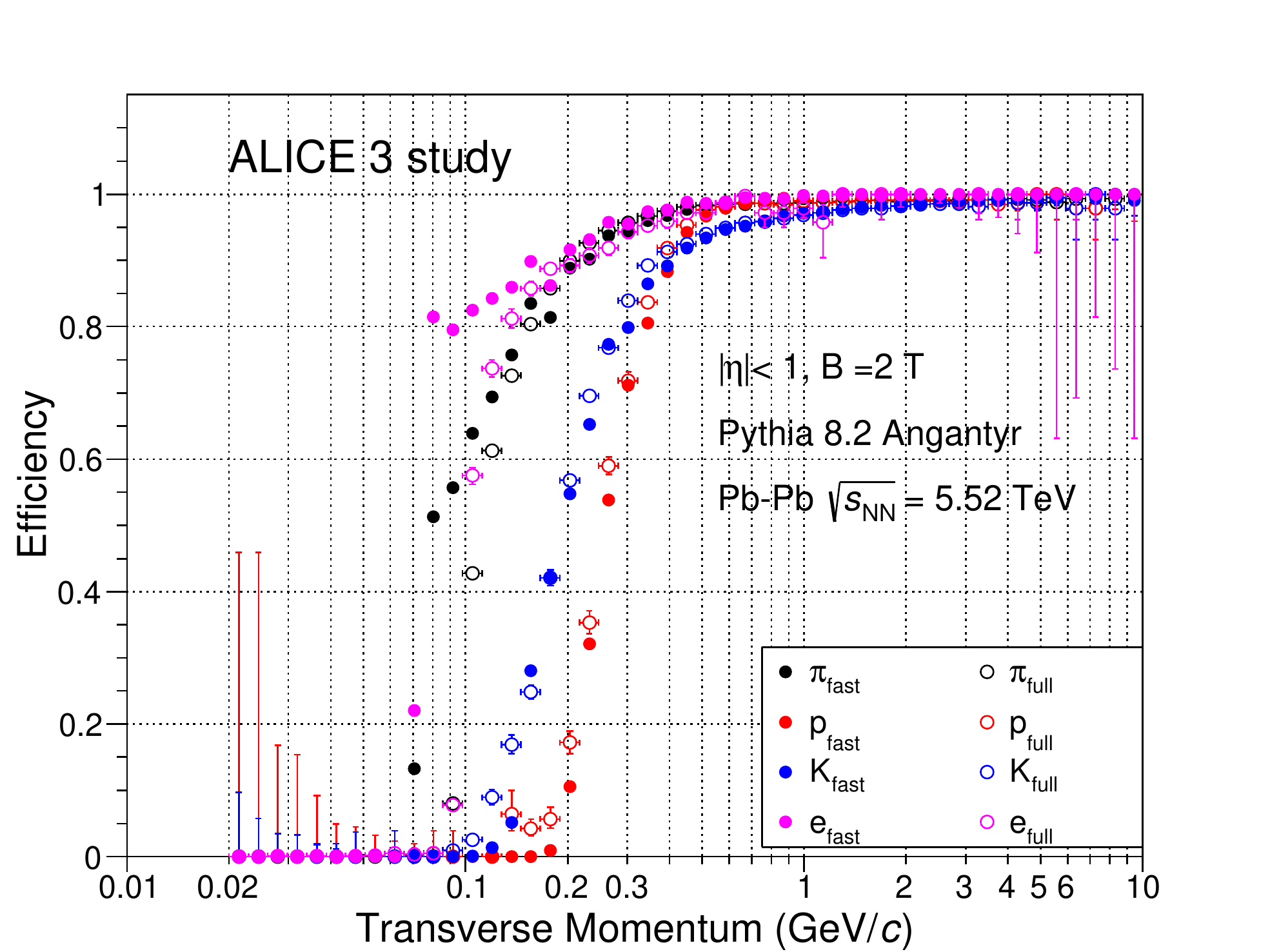}
   \caption[Tracking efficiencies from fast and full simulation]{Comparison of the tracking efficiency as a function of the transverse momentum obtained with fast and full simulations of minimum bias \PbPb{} collisions, with minimum radius of the tracks $R = 20$ cm for $B = 0.5$~T (left) and $B=2$~T (right).}
  \label{fig:performance:dectector:introduction:full_fast_efficiency_vs_pt_comparison}
\end{figure}

The expected tracking efficiencies for primary pions, kaons, protons and electrons for the proposed layout is shown in Fig.~\ref{fig:performance:dectector:introduction:full_fast_efficiency_vs_pt_comparison}, as a function of the transverse momentum. Particles are required to reach at least the seventh tracking layer, just outside the inner TOF layer, corresponding to a minimal radius $R = \SI{20}{cm}$. 
The performance is shown for both the full and the fast simulation. The agreement between the two is reasonable, except at lower transverse momenta, where the efficiency expected on the basis of the analytical approach is larger than that currently achieved using a preliminary version of the cellular automaton tracking described in Section~\ref{sec:performance:full-simulation} on full simulation events.
\new{Further optimisation of the track finding in the full simulation is ongoing.}

The reconstruction efficiency exhibits mass ordering below 800 MeV/$c$, above which the detector is able to reconstruct any particle species with an efficiency of nearly 100\%. It is worth mentioning that the B = 0.5 T field configuration has a larger acceptance for electrons and pions at low \pt than the B = 2 T field configuration. For the heavier kaons and protons the low momentum efficiency turn on is similar for the two field configurations, being dominated by the interactions (energy loss, multiple scattering) of the particles with the detector material.

\begin{figure} %
\begin{minipage}{0.49\textwidth}
  \centering
     \includegraphics[width=\textwidth]{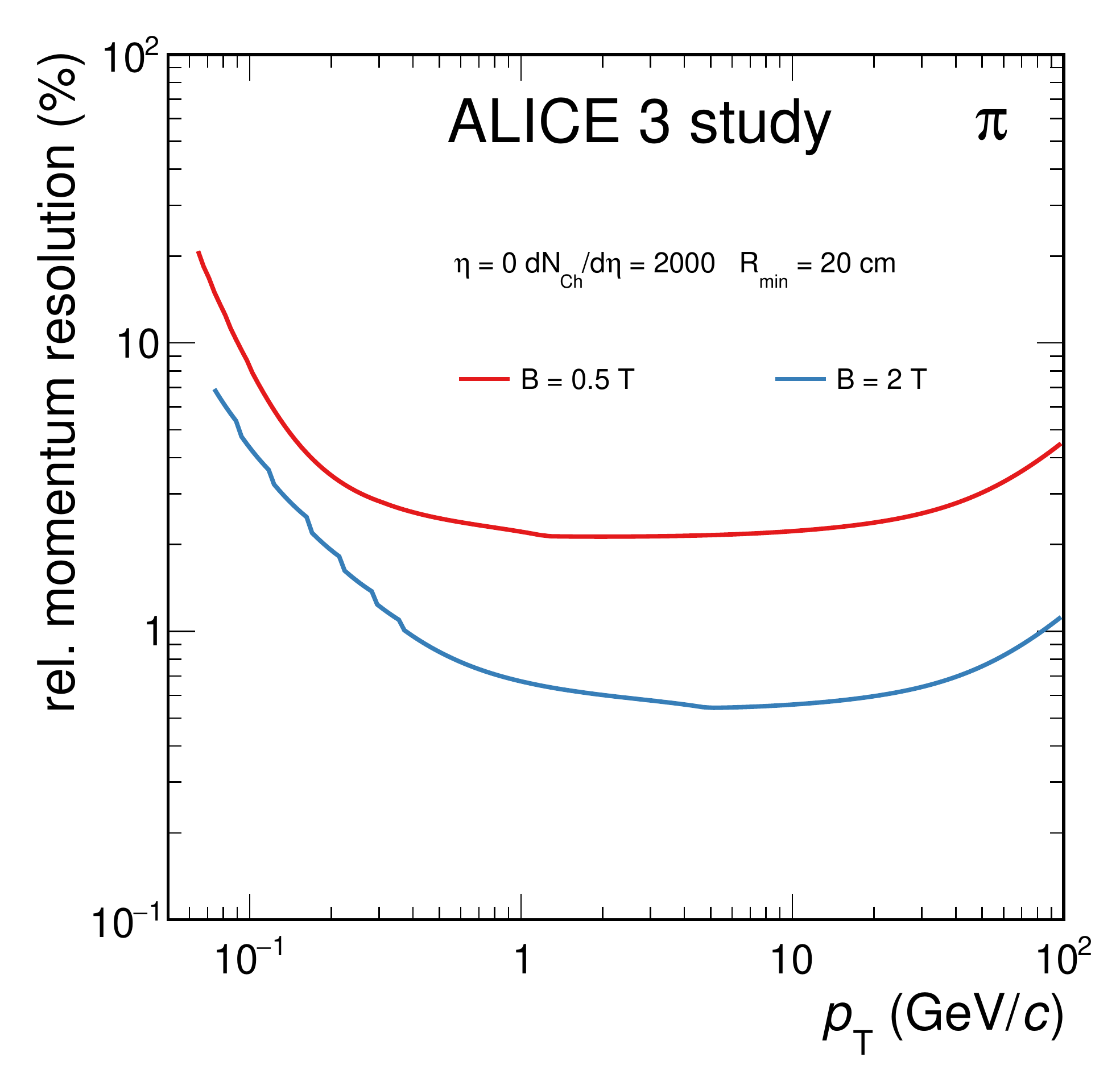}
     
   \caption[Transverse momentum resolution]{Comparison of the relative transverse momentum resolution for pions as a function of the \pt obtained with the fast analytical tool for $B = \SI{0.5}{\tesla}$ (red) and $B = \SI{2}{\tesla}$ (blue).
   }
  \label{fig:performance:dectector:introduction:full_fast_momentum_resolution_vs_pt_comparison}
\vfill
\end{minipage}\hfill%
\begin{minipage}{0.49\textwidth}
  \centering
    \includegraphics[width=\textwidth]{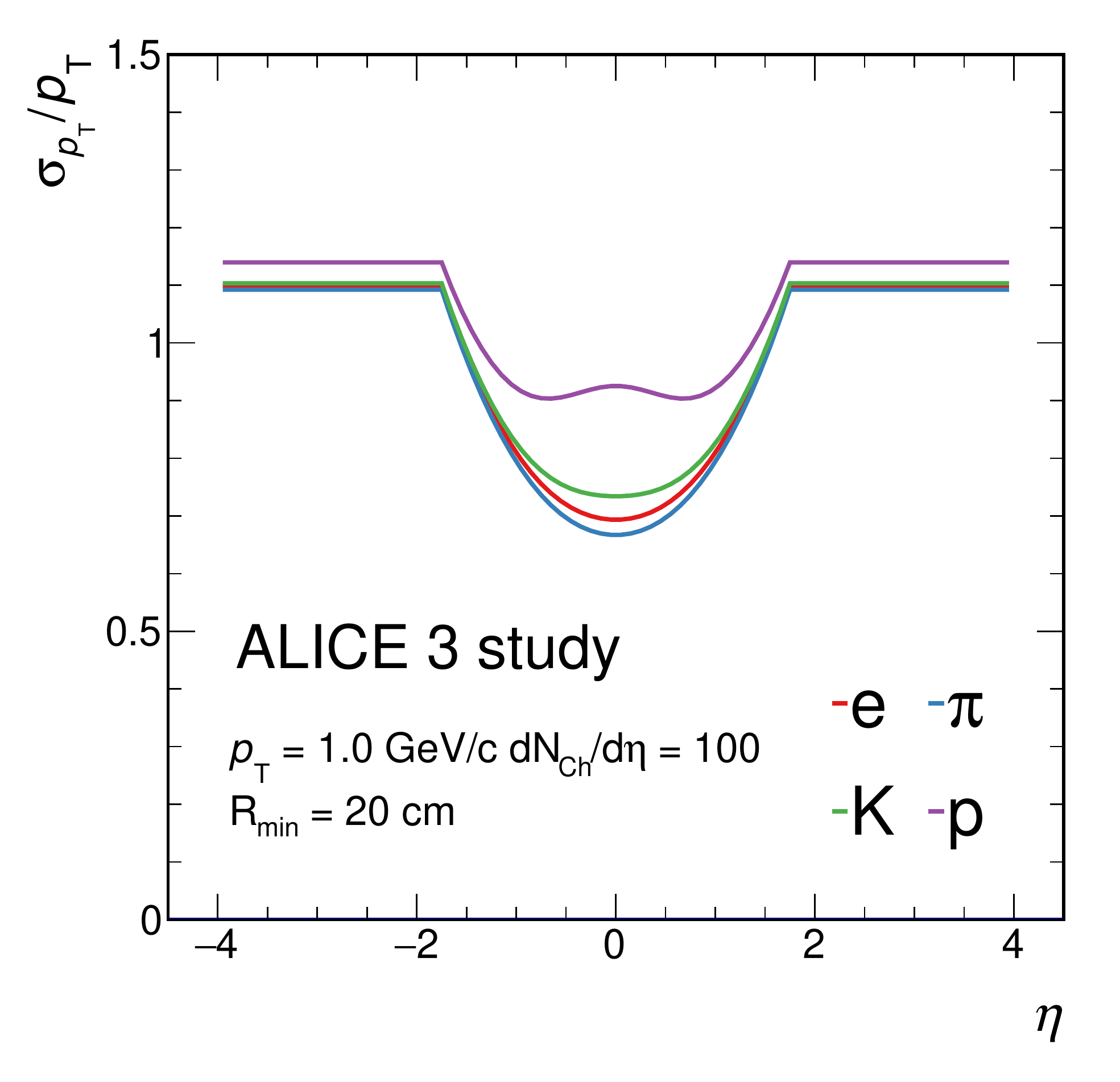}
   \caption[Transverse momentum resolution]{Relative transverse momentum resolution (in \si{\percent}) as a function of the pseudorapidity obtained with FAT for electrons, pions and protons with $\pt = 1$~\GeVc and minimum radius of the tracks R~=~20~cm in a magnetic field B~=~2~T. The continuation from $|\eta| = 2$ out to $|\eta| = 4$ is set to be constant as achievable  with forward dipoles. \comment{Add unit to vertical axis label; this information is now included in the figure before.}}
  \label{fig:performance:dectector:introduction:momentum_resolution_vs_eta_comparison}
\end{minipage}
\end{figure}
Fig.~\ref{fig:performance:dectector:introduction:full_fast_momentum_resolution_vs_pt_comparison} shows the expected transverse momentum resolution for pions as a function of \pt. We have considered a magnetic field of \SI{0.5}{\tesla} and \SI{2}{\tesla}, which results in a relative resolution of \SI{\sim 2.5}{\percent} and \SI{\sim 0.6}{\percent} at midrapidity, respectively. \new{The ideal field for the overall physics programme has yet to be established and the two configurations cover the considered range of performance.}

\begin{figure}
    \centering
    \includegraphics[width=.7\textwidth]{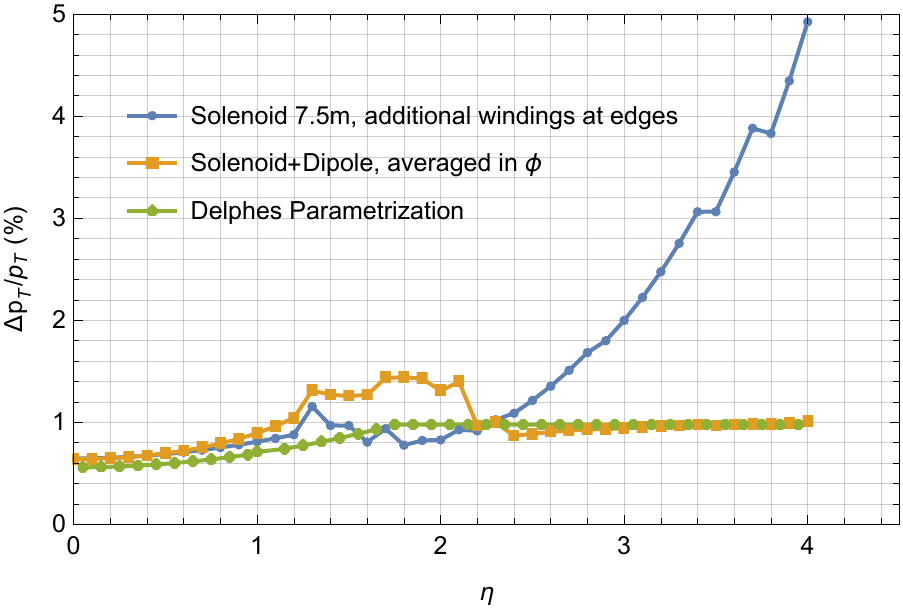}
    \caption[\pt resolution with different magnetic fields]{Momentum resolution for muons of $\pt = \SI{1}{\giga\eVc}$ with different field configurations together with the Delphes parametrisation used for the physics performance studies. The magnet designs are discussed in Sec.~\ref{sec:systems}.}
    \label{fig:perf:bfield}
\end{figure}

The pseudorapidity dependence of the transverse momentum resolution for  electrons, pions, kaons, and protons with $\pt$~=~1~$\GeVc$ in a magnetic field $B = 2.0$~T is shown in Fig.~\ref{fig:performance:dectector:introduction:momentum_resolution_vs_eta_comparison}. It rises from about 0.7\% (0.9\%) in the most central region for electrons and pions (protons) to about 1.1\% (1.15\%) at $|\eta|=2$. It is then assumed to be constant from there on out to $|\eta|=4$. This could be achieved with an appropriate magnet design allowing for a dipole component at large $z$ (see Fig.~\ref{fig:perf:bfield}).
\new{As the detailed plans for the magnet system are still evolving, the assumption of constant performance allows us best to study the impact of the \pt resolution on the physics performance also in the forward region.}

\paragraph{Determination of the primary vertex position} 

\begin{figure} %
  \centering
    \includegraphics[width=0.7\textwidth]{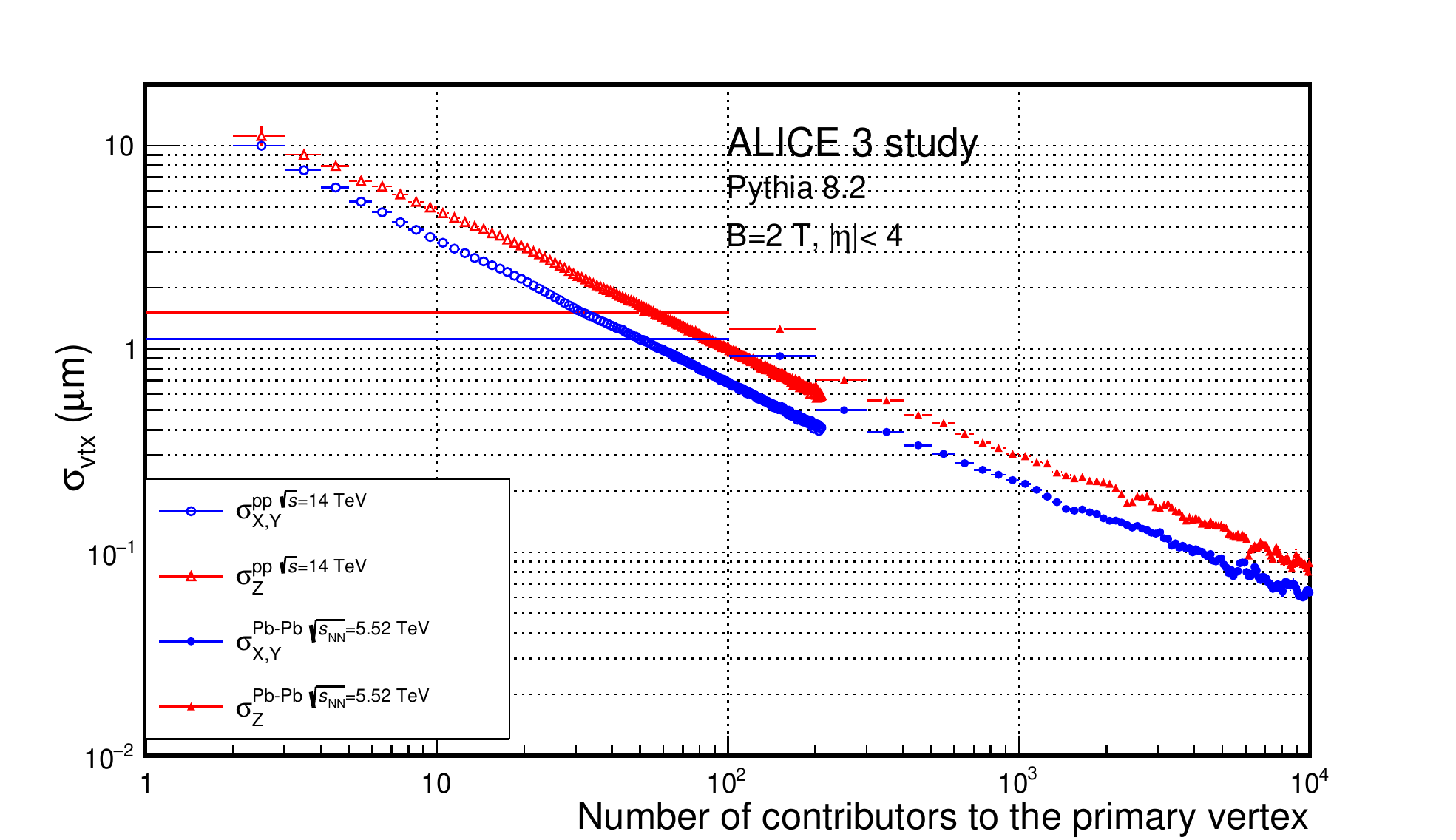}
   \caption[Primary vertex resolution]{Resolution of the primary vertex reconstruction in the transverse plane and along the beam axis as a function of the number of contributors to the primary vertex for $B = \SI{2}{\tesla}$.}
  \label{fig:performance:dectector:introduction:primary_vertexing}
\end{figure}

Figure~\ref{fig:performance:dectector:introduction:primary_vertexing} shows the primary vertex resolution of the detector. 
The close proximity of the silicon layers 
to the primary interaction vertex allows the reconstruction of the primary vertex to better than \SI{10}{\micro\metre}.
For this study, primary vertices are determined using 
a track distance minimization procedure
that follows the same algorithm that is currently employed 
in ALICE, based on the performance for primary track reconstruction discussed before. The difference between the vertex position resolution in pp and \PbPb{} collisions at equal multiplicity is due to a different momentum distribution of the particles in the generated events.

The impact parameter resolution along the longitudinal and the transverse directions, evaluated with the fast simulation, is shown in Fig.~\ref{fig:performance:dectector:tracking:dcaxyz} for charged tracks in pp collisions with $B = 2$~T. It is evaluated including the resolution for the reconstruction of the primary vertex position discussed above. Notably, it is observed
to be approximately a factor of five better than the ITS3+TPC 
combination for particles with \pt of \SI{\sim 500}{\mega\eVc}~\cite{ALICE:ITS3:2019}. 

\begin{figure} %
  \centering
  \includegraphics[width=.49\textwidth]{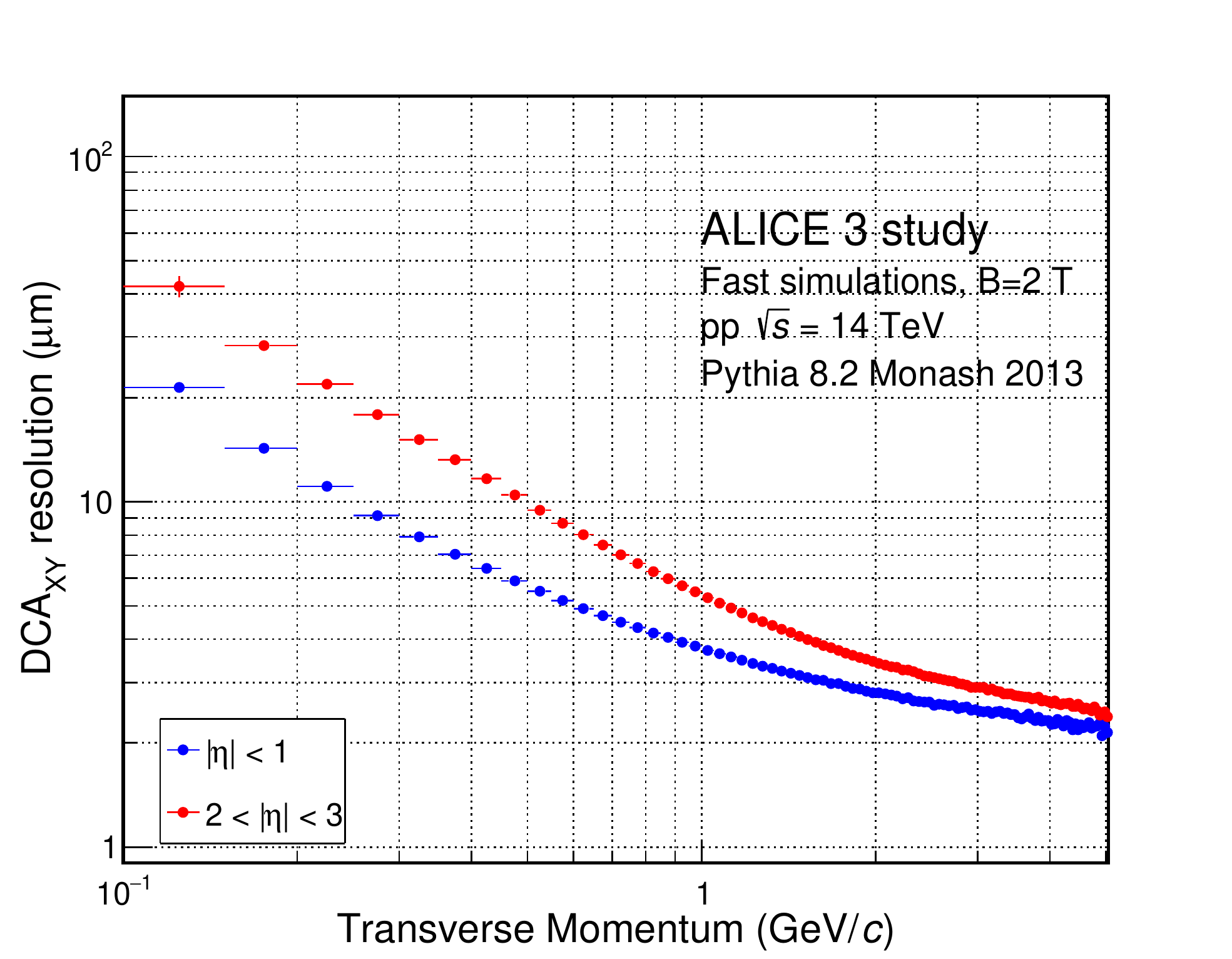}
   \includegraphics[width=.49\textwidth]{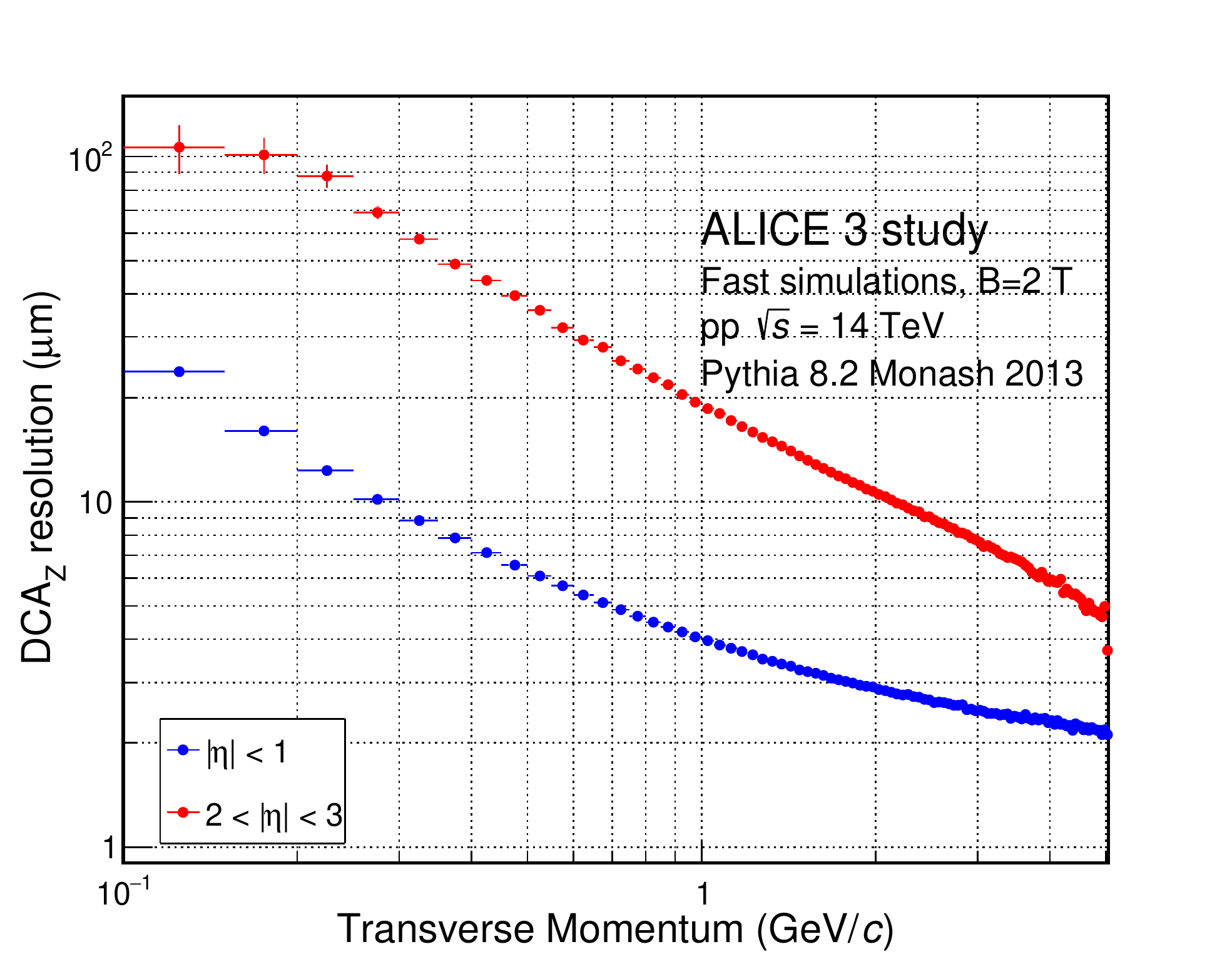} 
  \caption[Impact parameter resolution]{ 
    The impact parameter resolution for a magnetic field $B = \SI{2}{T}$ in the transverse (longitudinal) direction is shown in the left (right) panel. %
  }
  \label{fig:performance:dectector:tracking:dcaxyz}
\end{figure}

\paragraph{Secondary vertexing}

The secondary vertex position is determined employing track-distance minimisation methods. 
The residuals of the secondary vertex position distributions along the $x$, $y$ and $z$ coordinates are reported in Fig.~\ref{fig:performance:dectector:introduction:secondary_vertexing_resolution_D0_Lc} for two examples: $D^{0}~\to~K^{-}\pi^{+}$ and $\Lambda_{c}^{+}~\to~pK^{-}\pi^{+}$, reconstructed in the transverse momentum interval $1~<~\pt~<~2~\GeVc$. The widths of the distributions are below 5 $\mu$m in all three dimensions for both the $D^{0}$ and $\Lambda_{c}^{+}$ decays.

\begin{figure} %
  \centering
    \includegraphics[width=0.49\textwidth]{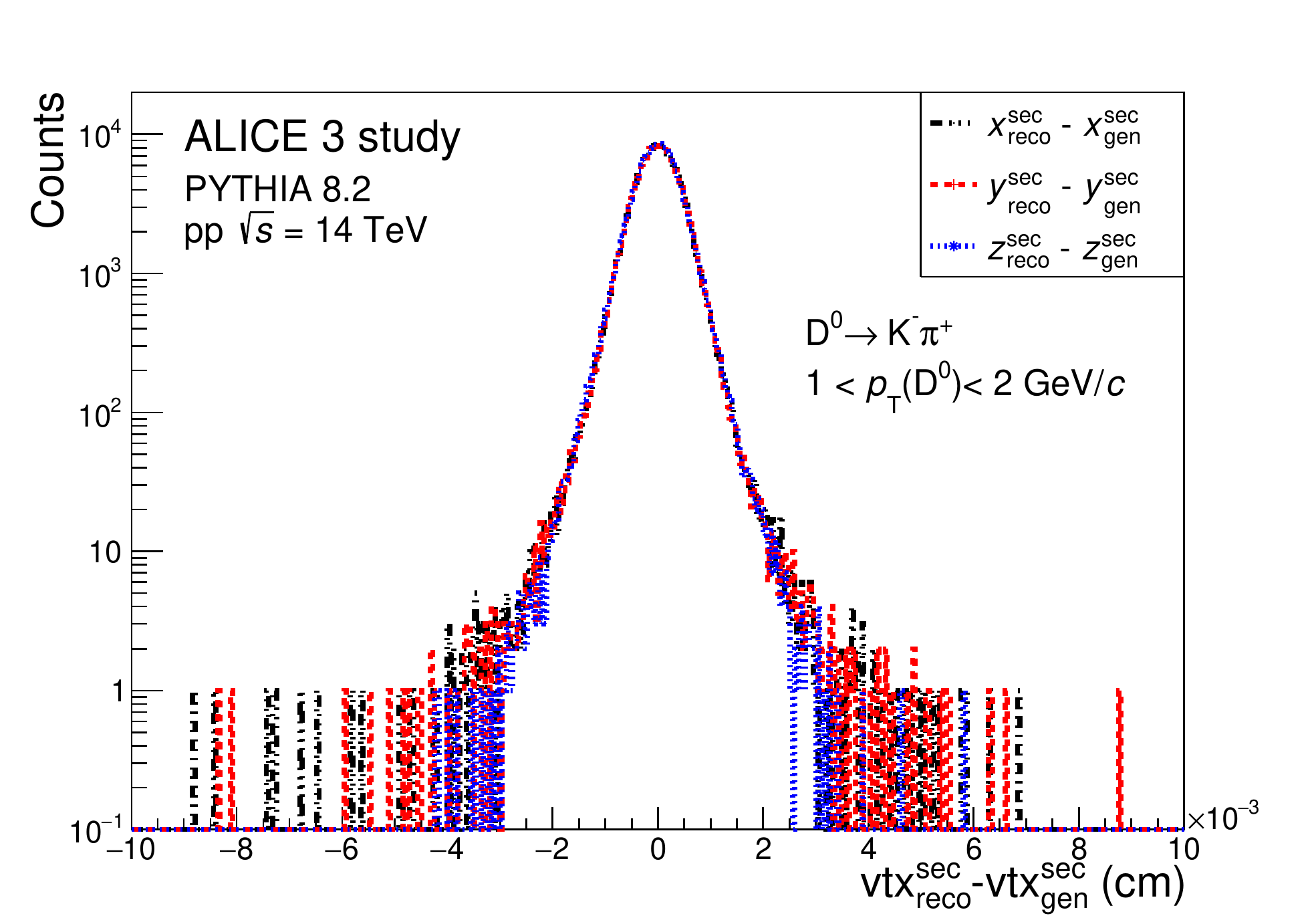}
    \includegraphics[width=0.49\textwidth]{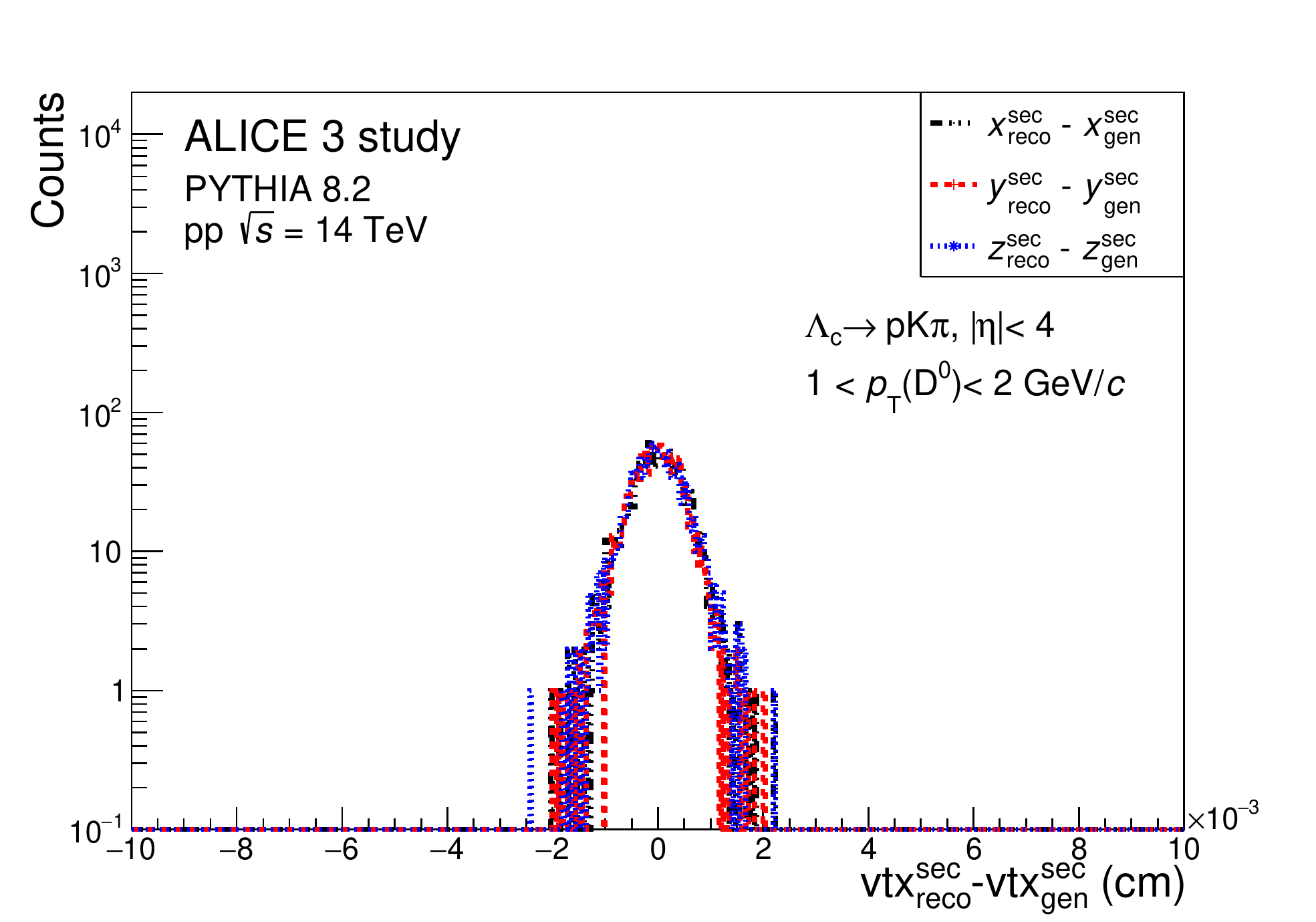}
   \caption[Secondary vertex residuals]{Distribution of the secondary vertex residuals for true  $D^{0}~\to~K^{-}\pi^{+}$ and $\Lambda_{c}^{+}~\to~pK^{-}\pi^{+}$ candidates along $x$, $y$ and $z$ direction in the transverse momentum range 1~$<~\pt~<~$2~\GeVc over the full pseudorapidity acceptance $|\eta|<4$ in pp collisions with B~=~2~T.}
  \label{fig:performance:dectector:introduction:secondary_vertexing_resolution_D0_Lc}
\end{figure}

\paragraph{Strangeness tracking}
\label{sec:performance:detector:strangeness_tracking}

\begin{figure} %
  \centering
  \includegraphics[width=.99\textwidth]{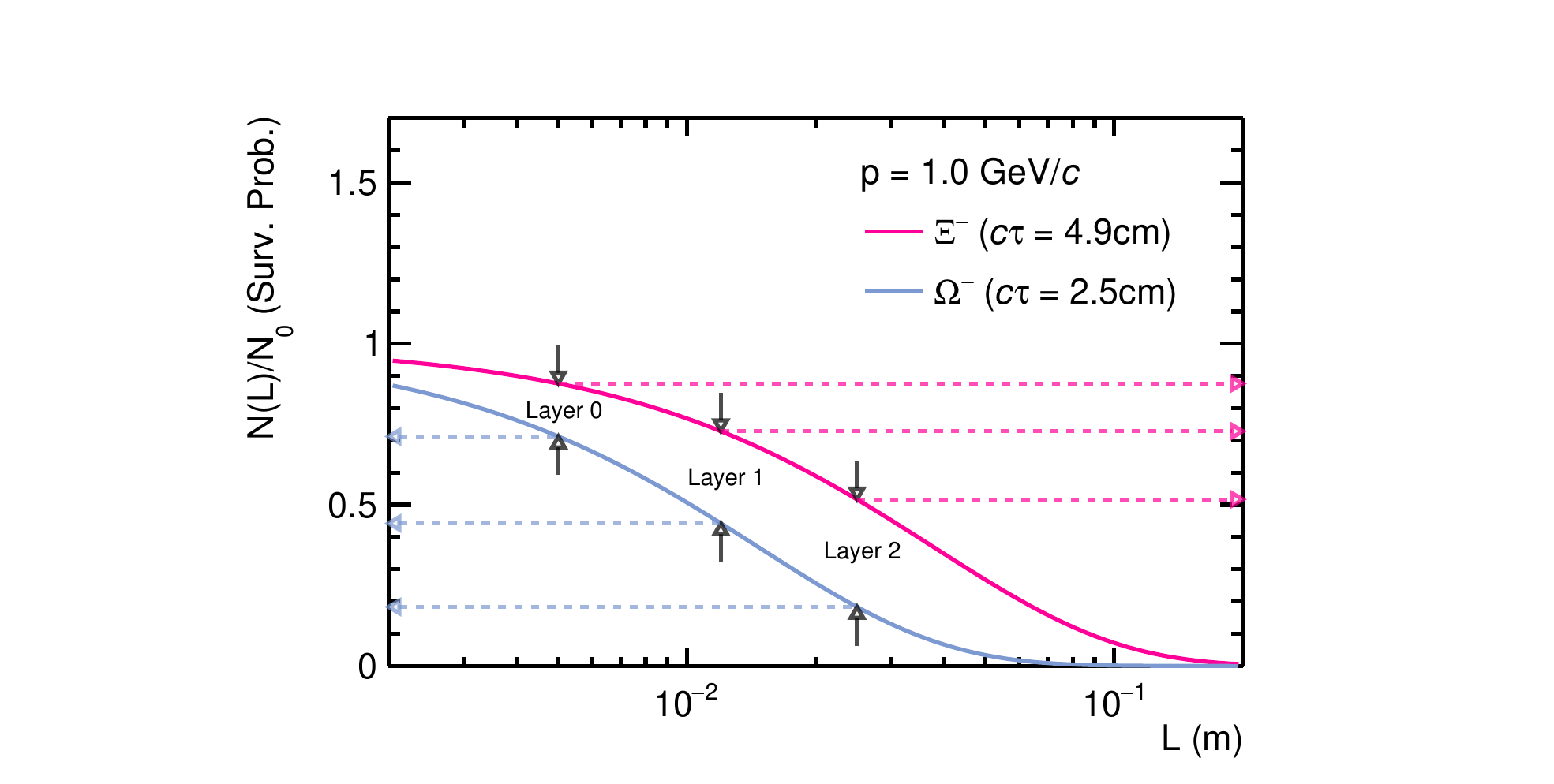}
  \caption[Survival probability of $\Xi^{-}$ and $\Omega^{-}$ for strangeness tracking]{
    Survival probability of a $\Xi^{-}$ (magenta line) and $\Omega^{-}$ (blue line) with a momentum of \SI{1}{\giga\eVc} as a function of the distance to the interaction point. The locations of the different layers of the \ALICETHR detector at $\eta=0$ are indicated by the arrows. 
  }
  \label{fig:performance:dectector:tracking:stra_decayLength}
\end{figure}

The ALICE 3 experiment has been designed so that weakly decaying strange particles can be tracked before they decay. 
Figure~\ref{fig:performance:dectector:tracking:stra_decayLength} shows the survival probability of $\Xi^{-}$ and $\Omega^{-}$ baryons with a momentum of 1~\GeVc as a function of the distance to primary vertex. These particles have a probability of at least 90\% and 70\%, respectively, to cross the first layer and leave a direct hit. This method, called `strangeness tracking', decisively improves the reconstruction of weakly decaying particles that stem from multi-charm baryon decays. 

\begin{figure} %
  \centering
  \includegraphics[width=.49\textwidth]{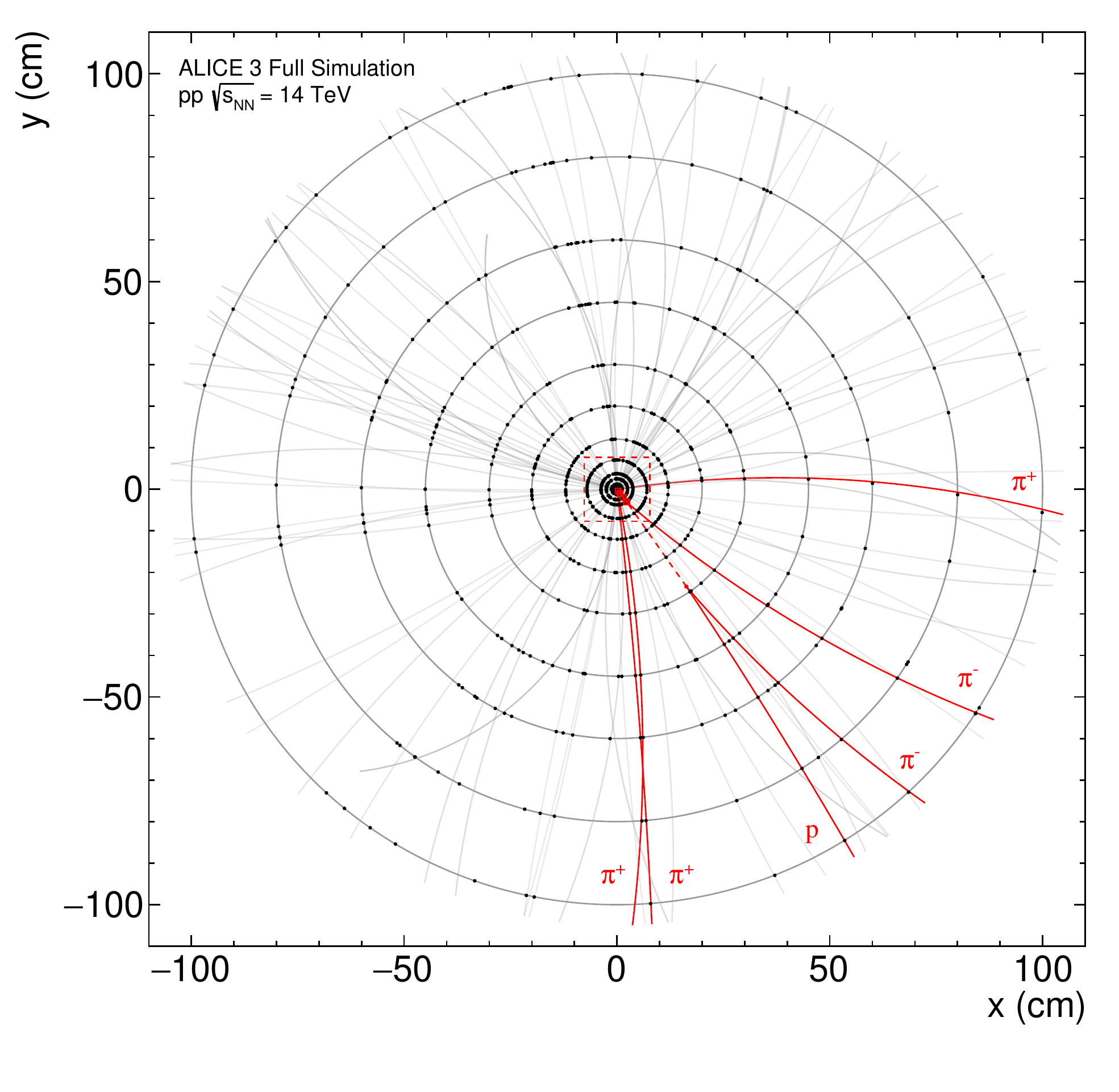}
  \includegraphics[width=.49\textwidth]{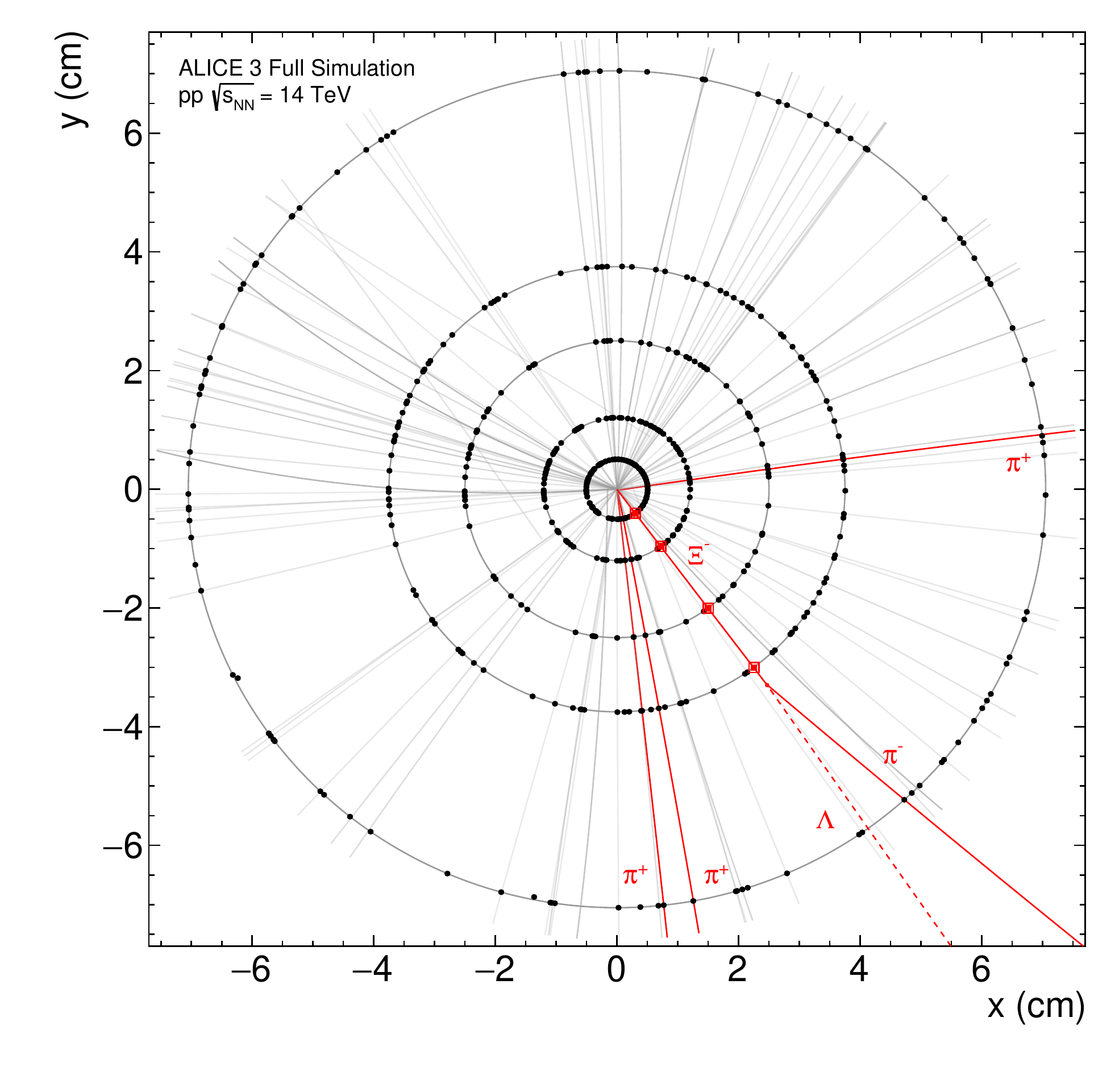}
  \caption[Illustration of strangeness tracking]{
    (left) Illustration of strangeness tracking from full detector simulation 
    of the $\Xi_{cc}^{++}$ decay into $\Xi_{c}^{+}+\pi^{+}$ with the successive decay $\Xi_{c}^{+}\rightarrow\Xi^{-}+2\pi^{+}$. (right) Close-up illustration of the region marked with a red dashed box in the left figure, containing the 
    five innermost layers of \ALICETHR and the hits that were added to the $\Xi^{-}$ trajectory (red squares).
  }
  \label{fig:performance:dectector:tracking:stratrack}
\end{figure}

Strangeness tracking is implemented as a second stage that follows after 
secondary track finding. The trajectory information of each weak decay candidate 
is calculated and utilized to extrapolate back in the direction of the primary interaction 
vertex. If hits are detected that are sufficiently close to this trajectory, as illustrated 
in Fig.~\ref{fig:performance:dectector:tracking:stratrack}, they 
are added to the track parametrisation, increasing the precision of the
weak decay candidate reconstruction. This increase is mostly related to the position 
of the candidate trajectory, since the momentum information is predominantly acquired 
from the longer daughter tracks. This method provides a much better 
distance-of-closest approach measurement of the weak decay trajectory to the primary 
vertex (see Fig.~\ref{fig:performance:physics:heavy_flav:mcPbPb:DCADemo}), greatly improving the ability to distinguish weak decays from primary or secondary sources.

An important aspect of strangeness tracking is that the resolution in determining the weak 
decay trajectory has to be sufficiently high to ensure a low fake hit association rate. 
In \ALICETHR, full simulations indicate an average expected occupancy of approximately 10-12 hits per square
millimeter in the innermost layer for central \PbPb{}, corresponding to one hit every
\SI{0.1}{\mm^2}. 
The expected precision of the inwards track propagation results in a search window size of approximately \SI{5e-3}{\mm^2} to be 
used in the innermost layers.
With this method, a impact parameter resolution for multi-strange baryons is achieved, that is similar to that of primary charged particles.

\new{Potential applications of strangeness tracking include the reconstruction of multi-charm and beauty baryons. Performance studies for 
multi-charm baryons are reproduced in Sec.~\ref{sec:performance:physics:heavy_flav:multicharm} and further studies using
this technique will explored in the future.}

\subsubsection{Hadron identification}
\label{sec:performance:detector:hadron_id}

\begin{figure}[tbp] %
  \centering
  \includegraphics[width=\textwidth]{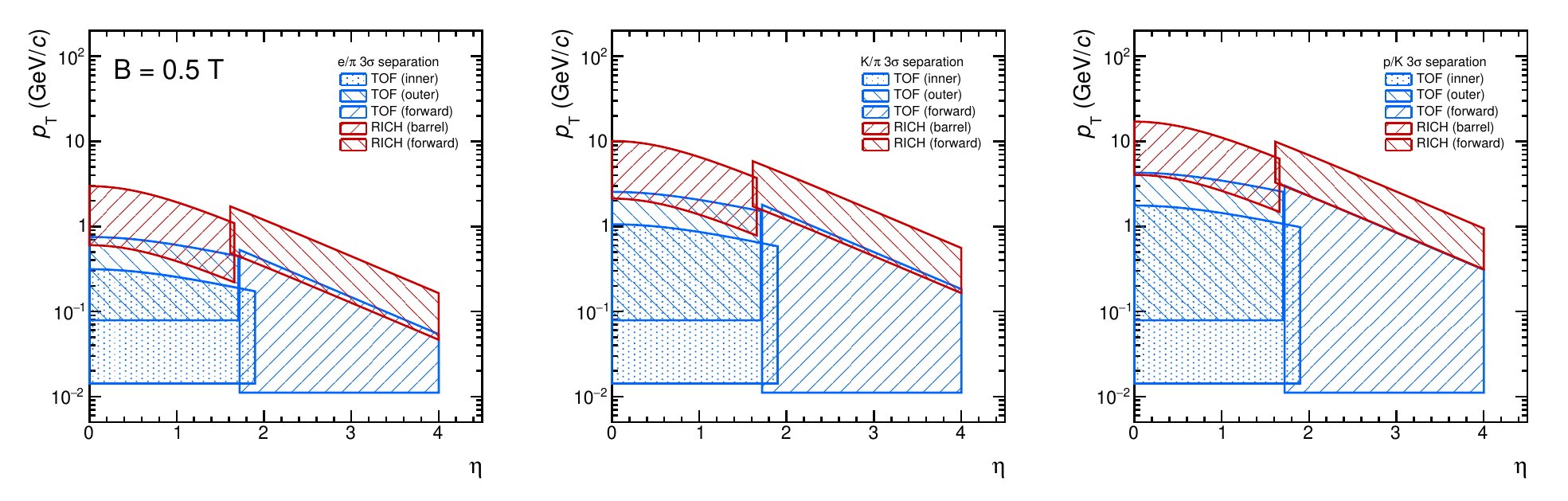}
  \caption[Particle identification (analytical, \SI{0.5}{\tesla})]{
    Analytical calculations of the $\eta-\pt$ regions in which particles can be separated by at least $3\sigma$ for the ALICE 3 particle-identification subsystems embedded in a 0.5~T magnetic field. Electron/pion, pion/kaon and kaon/proton separation plots are shown from left to right.
  }
    \label{fig:performance:detector:hadron_id:summary:5kgauss}
 \includegraphics[width=\textwidth]{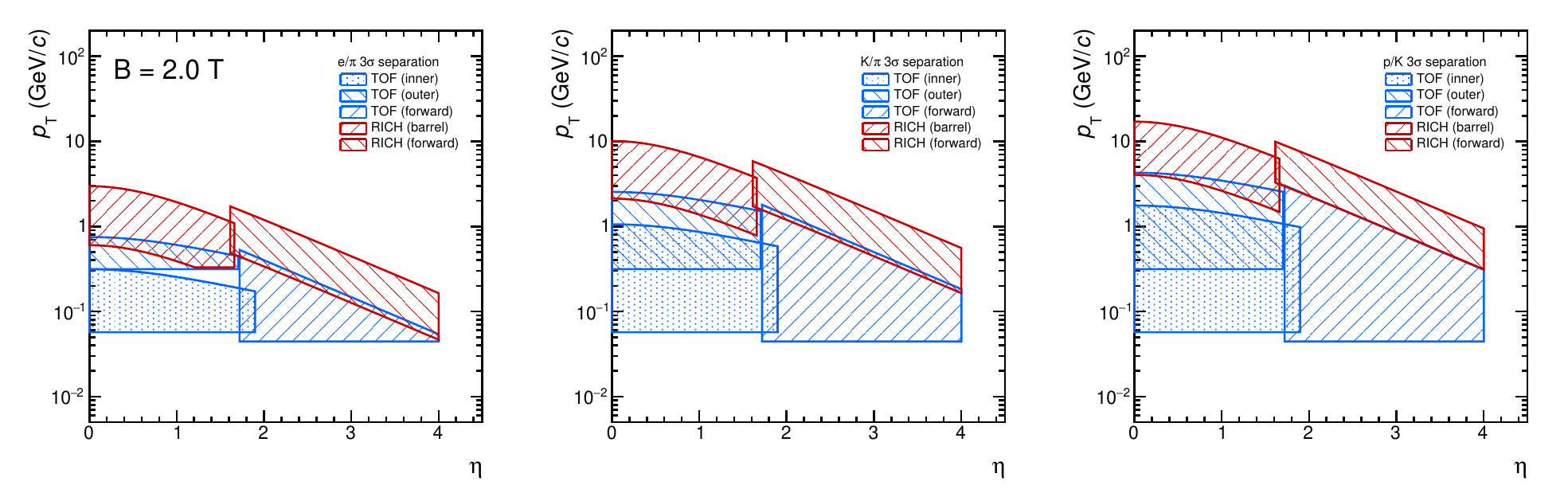}
  \caption[Particle identification (analytical, \SI{2}{\tesla})]{
    Analytical calculations of the $\eta-\pt$ regions in which particles can be separated by at least $3\sigma$ for the ALICE 3 particle-identification systems embedded in a 2.0~T magnetic field. Electron/pion, pion/kaon and kaon/proton separation plots are shown from left to right.
  }
  \label{fig:performance:detector:hadron_id:summary:2tesla}
\end{figure}

Hadron identification in the central barrel detector and in the forward region is performed via time-of-flight and Cherenkov imaging techniques. Figure~\ref{fig:performance:detector:hadron_id:summary:5kgauss} and~\ref{fig:performance:detector:hadron_id:summary:2tesla} summarise the PID capabilities of the proposed detector, highlighting the $\eta-\pt$ regions where particle identification with a better than 3$\sigma$ separation is possible. It is worth noticing the expected larger low-\pt acceptance in the B = 0.5 T field configuration when compared to the 2 T case, however it is important to note that effects due to interaction of particles with the detector material are not taken into account in this study.

Two layers of time-of-flight detectors, \InnerTOF and \OuterTOF, are located at 20~cm and 105~cm from the beam pipe and measure the arrival time of particles with a resolution of 20~ps. Forward TOF walls with the same time resolution are located at 405~cm on either side of the interaction point.
A description of the implementation and technology for the TOF systems is given in detail in Section~\ref{sec:systems:tof}.
The start time is self-determined by the system of time-of-flight detectors with a resolution better than 3~ps in high-multiplicity events, which adds a negligible contribution to the time-of-flight measurement.
\new{It should be noted that the combination of bTOF1 and bTOF2 also allows the true measurement of the time of flight and, thus, also the precise determination of the start time even with few tracks.}
The length of the track is measured by the tracking system and its resolution is assumed to be negligible.

Figure~\ref{fig:performance:detector:hadron_id:tof2:beta} shows the response of the barrel TOF systems for simulated \PbPb{} events in a magnetic field of $B = \SI{2}{T}$. 
It can be seen that due to the larger occupancy, the fake hit association is larger for the \InnerTOF layer, causing an increase in the background due to track-TOF mismatch. 

\begin{figure} %
  \centering
  \includegraphics[width=0.47\textwidth]{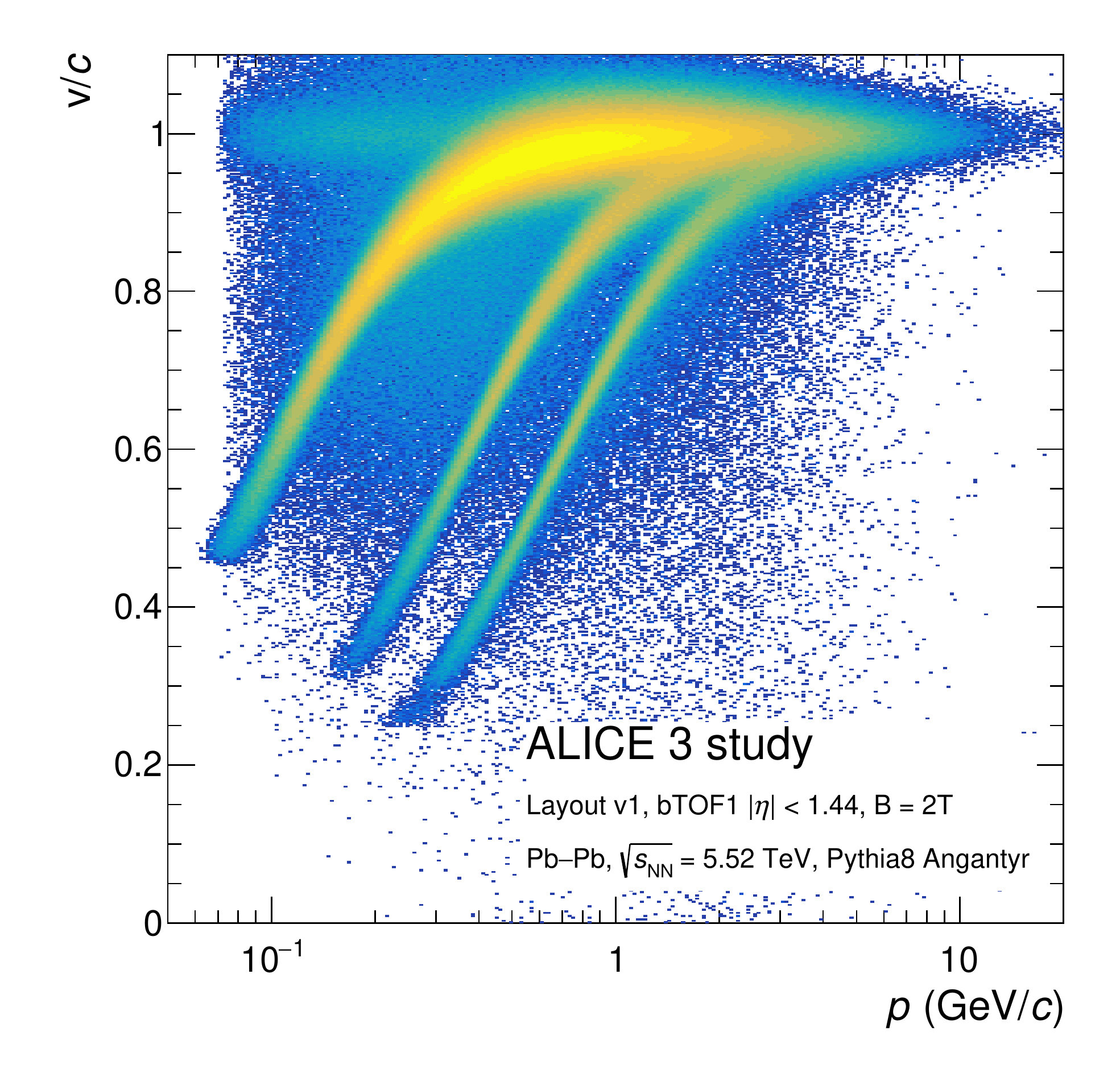}\hfill%
  \includegraphics[width=0.47\textwidth]{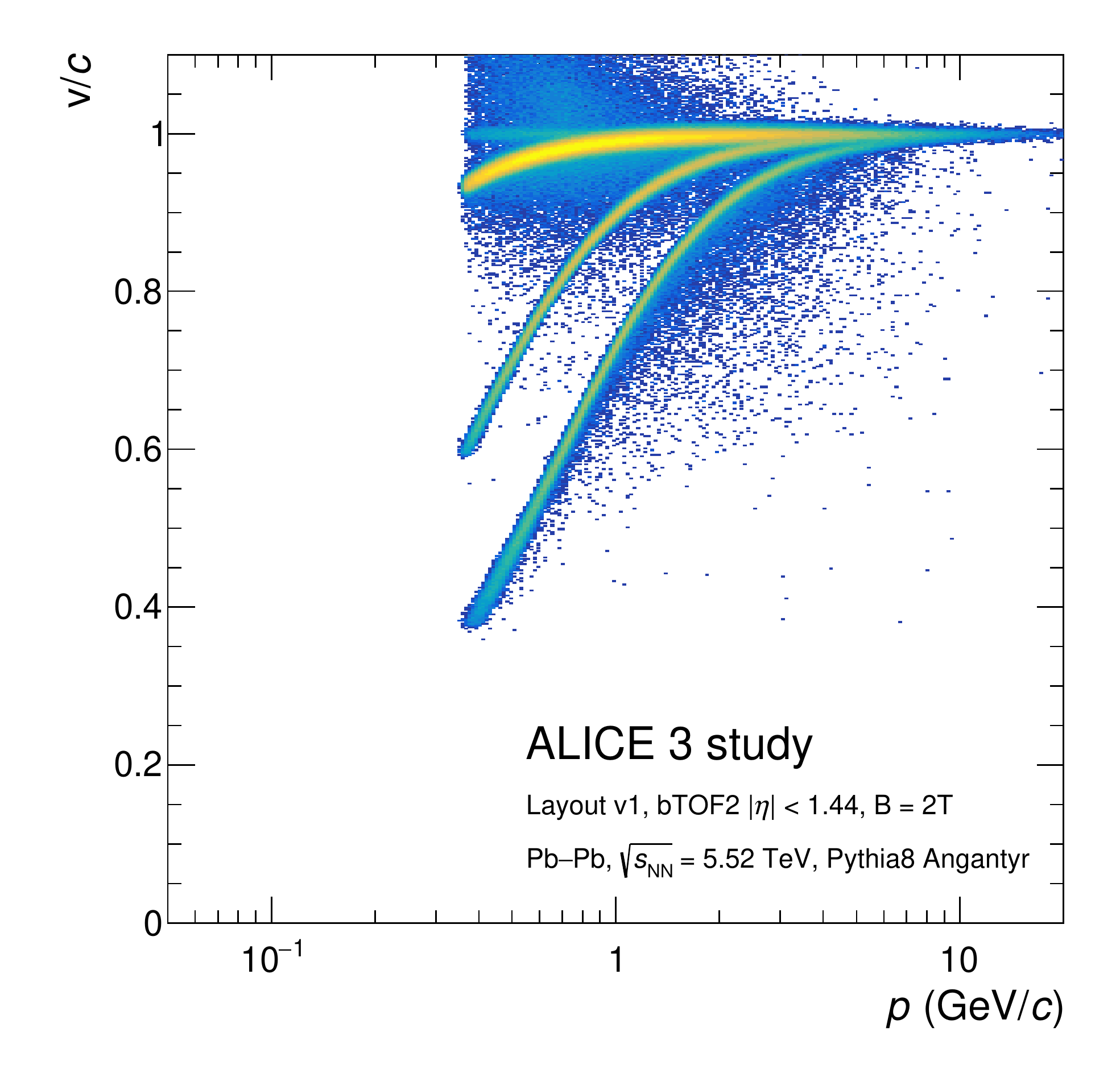}
  \caption[Particle velocity from inner and outer TOF] {
    Momentum dependent distribution of the particle velocity measured with the \InnerTOF (left) and \OuterTOF (right) in \PbPb collisions with the B~=~\SI{2}{T} configuration.
    The momentum thresholds for the different particle species are clearly visible for the two detector configurations as well as the effect of the improved momentum resolution.
  }
  \label{fig:performance:detector:hadron_id:tof2:beta}
\end{figure}

The barrel RICH detector, \RICH, is composed of a 2~cm thick aerogel radiator with refractive index $n = 1.03$ located at 110~cm from the beam line and followed by a 20 cm expansion volume with photo-detector readout.
Such a system is expected to provide a measurement of the Cherenkov angle with an average angular resolution of 1.5~mrad. 
For the forward region, RICH detectors with the matching performance but smaller refractive index are assumed at a longitudinal distance of 410~cm on either side of the interaction point.
This is still subject to optimisation.

\begin{figure}[tbp] %
  \centering
  \includegraphics[width=0.45\textwidth]{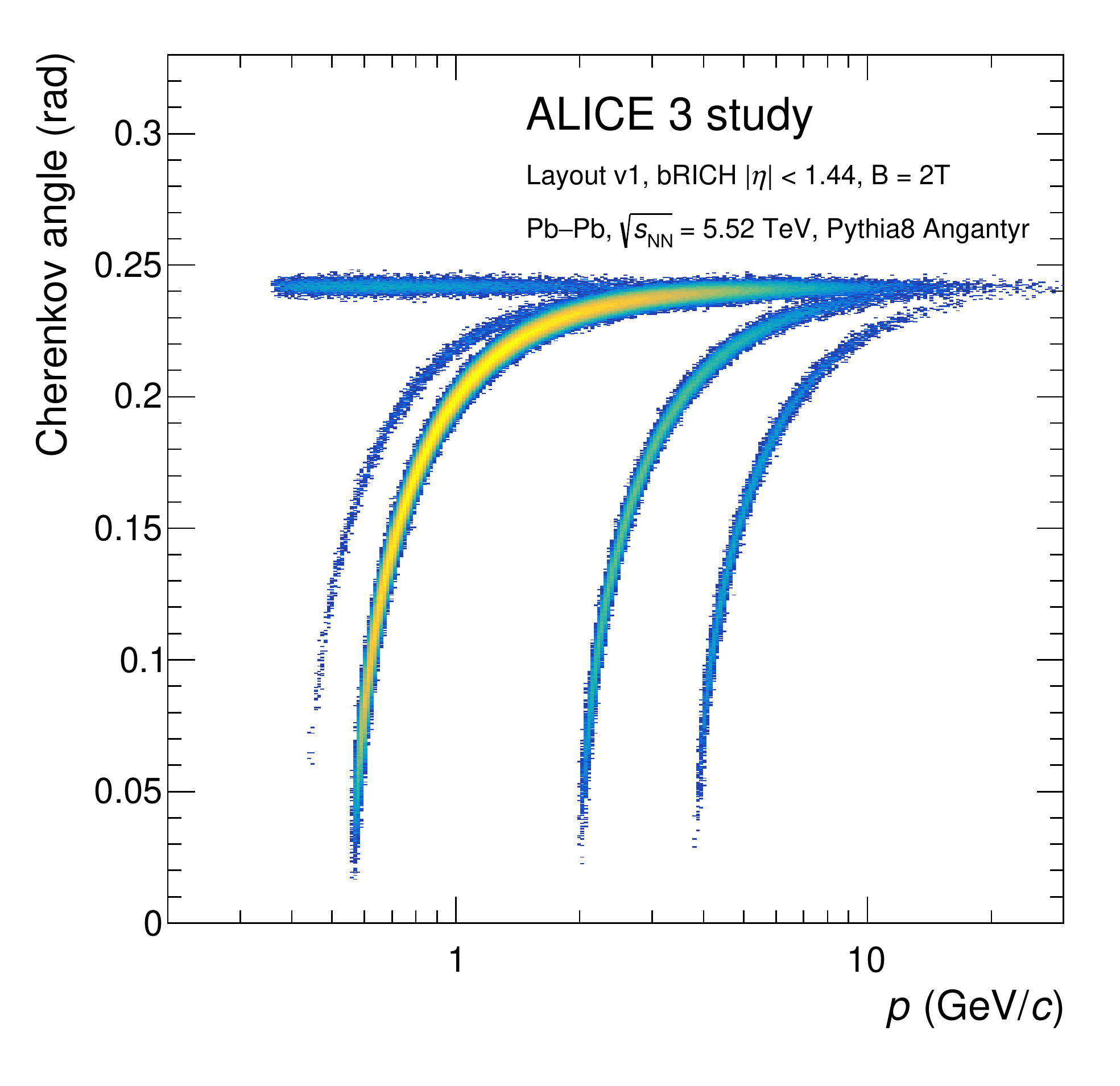}
  \caption[Cherenkov angle from the barrel RICH] {
    Momentum dependent distribution of the Cherenkov angle measured with the aerogel-based barrel RICH in \PbPb collisions.
    The momentum thresholds for the different particle species are clearly visible. 
    This is the combined result of the effect of the magnetic field and intrinsic threshold for the emission of Cherenkov light in the radiator.
  }
  \label{fig:performance:detector:hadron_id:rich:angle}
\end{figure}

The performance of the \RICH detector is reported in \Fig{fig:performance:detector:hadron_id:rich:angle}, where the measured Cherenkov angle is shown as a function of the reconstructed track momentum.
As for the TOF systems, the identification thresholds of the different particle species can be seen. In this case, the Cherenkov emission threshold determines the low-momentum cut-off for particle identification.

\begin{figure} %
  \centering
  \includegraphics[width=0.46\textwidth]{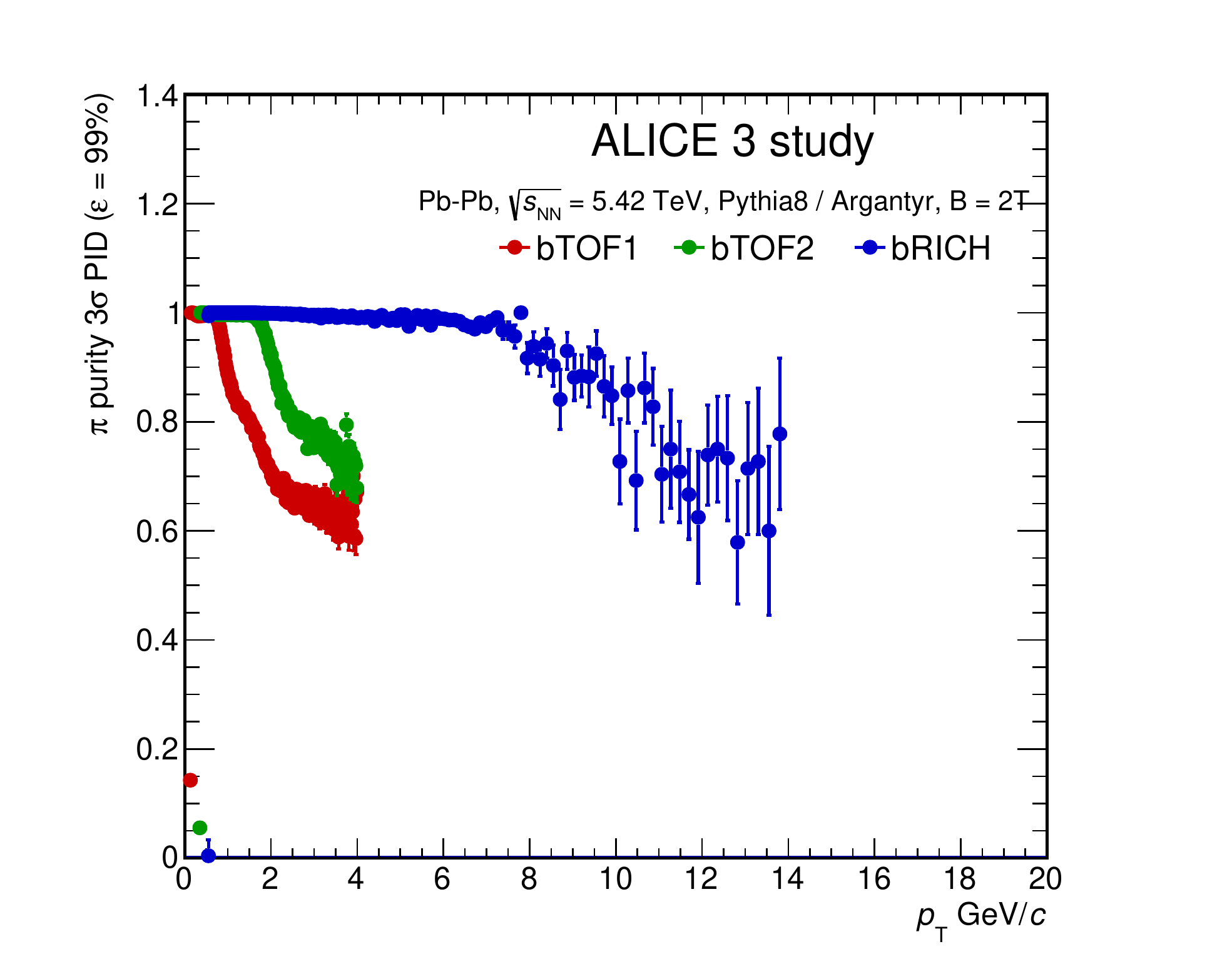}
  \includegraphics[width=0.46\textwidth]{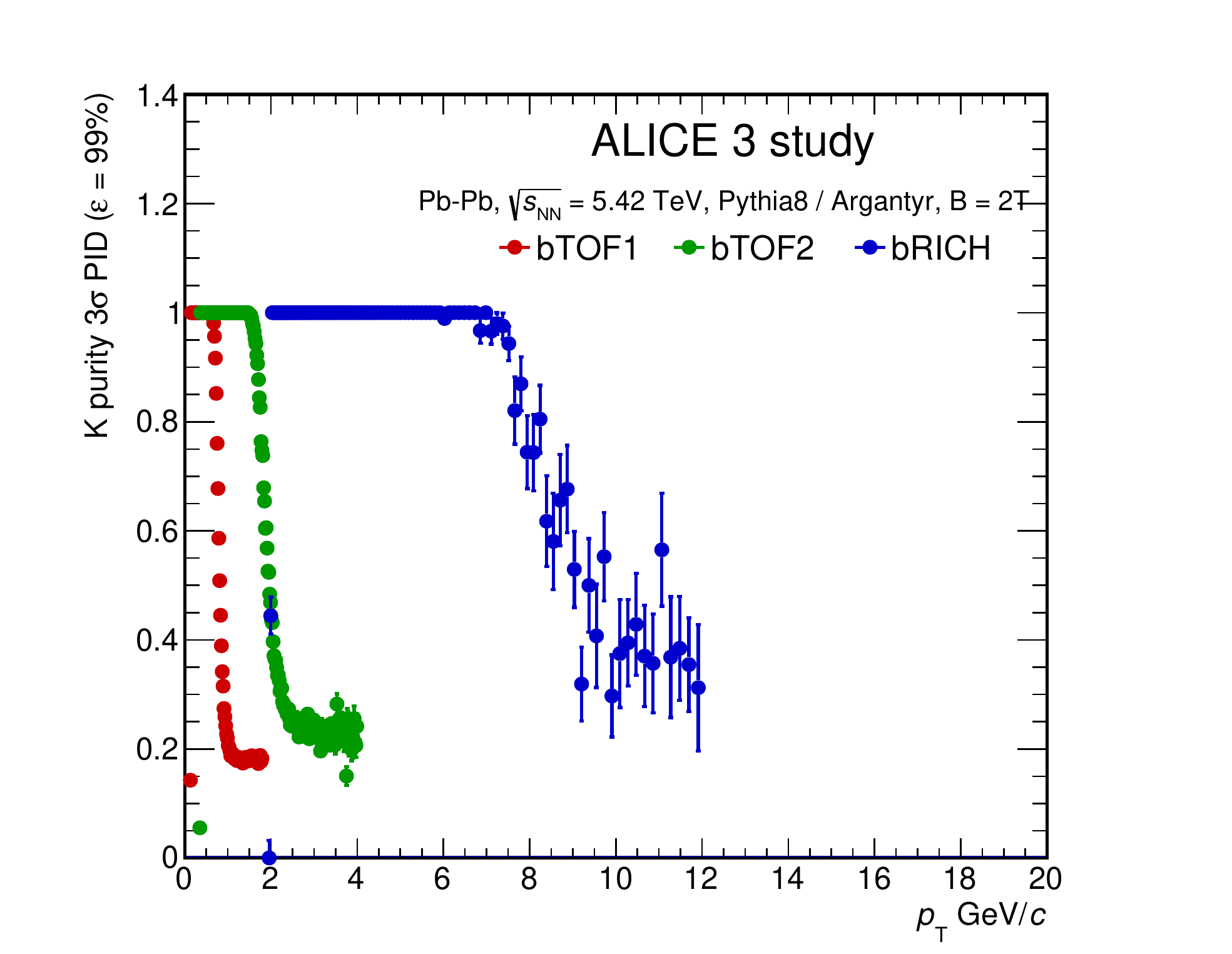}
  \includegraphics[width=0.46\textwidth]{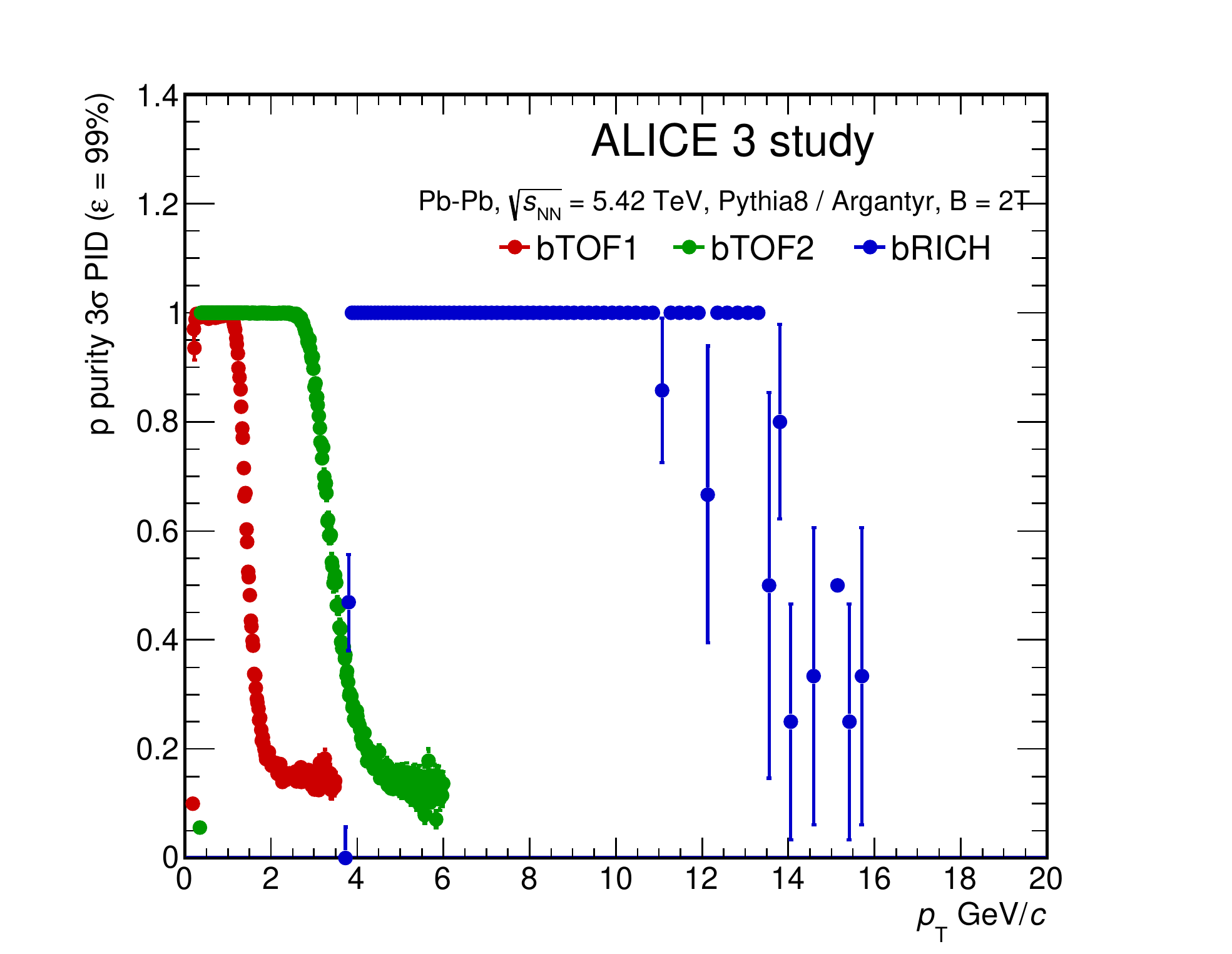}
  \caption[PID purity] {
  Purity of pions, kaons and protons obtained with the \InnerTOF, \OuterTOF and \RICH in \PbPb collisions with a $3\sigma$ cut with the $B = \SI{2}{\tesla}$ configuration.
  }
  \label{fig:performance:detector:hadron_id:tof:inner_separation_purity}
\end{figure}

Figure~\ref{fig:performance:detector:hadron_id:tof:inner_separation_purity} shows the selection purity with a $3\sigma$ cut on each particle hypothesis, i.e. keeping the PID efficiency above 99\%, for the three different identification systems.

\begin{figure}[t] %
  \centering
  \includegraphics[width=.45\textwidth]{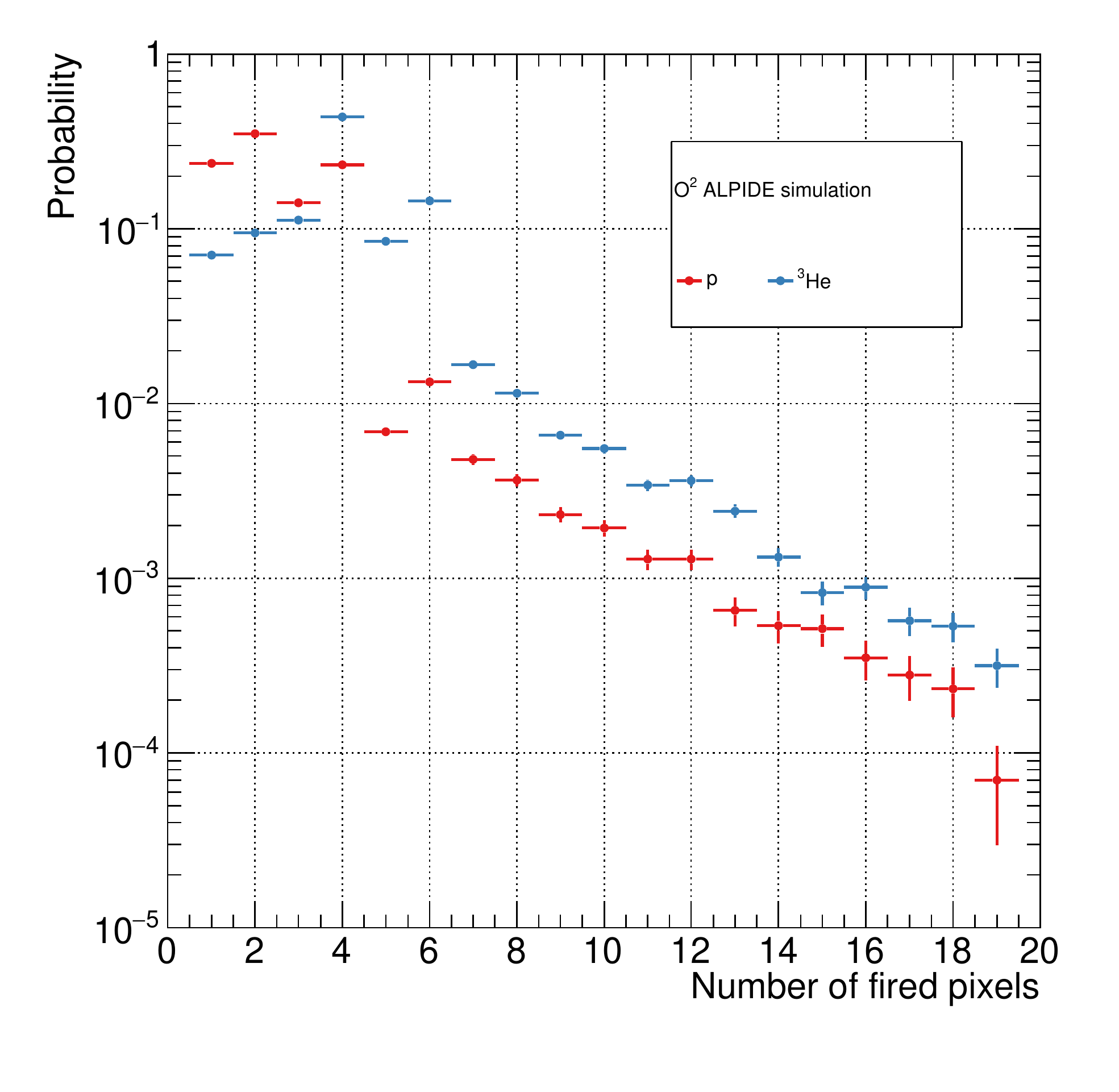}
  \caption[Cluster sizes for $z=1$ and $z=2$ particles]{
    Cluster size distribution for $z=1$ and $z=2$ charged nuclei with momenta corresponding to minimum ionizing particles obtained from Monte Carlo simulations of the ALPIDE chip.
  }
  \label{fig:performance:detector:nuclei:clustersize}
\end{figure}

The identification of light nuclei will be mostly based on the TOF measurement. The measurement
allows the distinction of light nuclei with different $m/z$, leaving a residual ambiguity for nuclei sharing the same $m/z$ (e.g. triton and \hesix).
Given the very large differences in the production yields of nuclei with same $m/z$, it is imperative to resolve this ambiguity. This calls for a charge-sensitive measurement.
The feasibility of such a separation has been studied based on the cluster-size of the hits in the tracking layers.
While the details of the response of the tracking layer chips for \ALICETHR are still subject to optimisation (see Section~\ref{sec:systems:tracking} for details), for this study, we employed the simulation of the ALPIDE chip implemented in the O2 framework, as a proof of concept.
The probability distributions of the cluster size corresponding to particles with $z=1$ (protons) and $z=2$ (\hethree) is shown in Fig.~\ref{fig:performance:detector:nuclei:clustersize} for momenta that correspond to minimum ionizing particles.
The distribution for \hethree is shifted towards larger cluster sizes as expected from the $z^{2}$ dependence of the energy loss in the material.
The average cluster size distribution expected when sampling from this distribution over all the tracking layers should therefore allow some separation between the two particles.
While the effectiveness of these considerations clearly depends on the final implementation of the detector, it can be taken as a proof of concept and further developed for the final detector, possibly with the usage of neural network algorithms.

\subsubsection{Electron identification}
\label{sec:performance:detector:electron_id}

In this section we consider a magnetic field in the solenoid of \SI{0.5}{\tesla}. Further studies for other magnetic field configurations are under way.
For measurements of thermal dilepton emission, high-purity electron identification at low transverse momentum is required. With the $B=0.5$ T field configuration, the \InnerTOF can provide electron identification starting from $\pt = 0.015$ \GeVc up to $p = 0.1 \GeVc$. At higher momenta the $\beta$ of pions and muons are close to that of electrons, and they cannot be clearly separated anymore. Electron identification at momenta larger than 0.1 \GeVc is provided by the \OuterTOF, which allows the separation of electrons from the other charged particles up to 0.5 \GeVc. In the same momentum range as the \OuterTOF, the \RICH signals of electrons are well separated from the pion and muon bands, extending the electron identification up to 1 \GeVc.

The combination of the information from different detector systems allows to obtain high electron identification efficiencies over a large momentum range while keeping the hadron contamination at a low level. An example of such a combination, this time including an estimate of the tracking efficiency, is shown in \Fig{fig:performance:dectector:electon_id:contamination_and_eff} for the \OuterTOF and the \RICH.

\begin{figure}
  \centering
  \includegraphics[width=.49\textwidth]{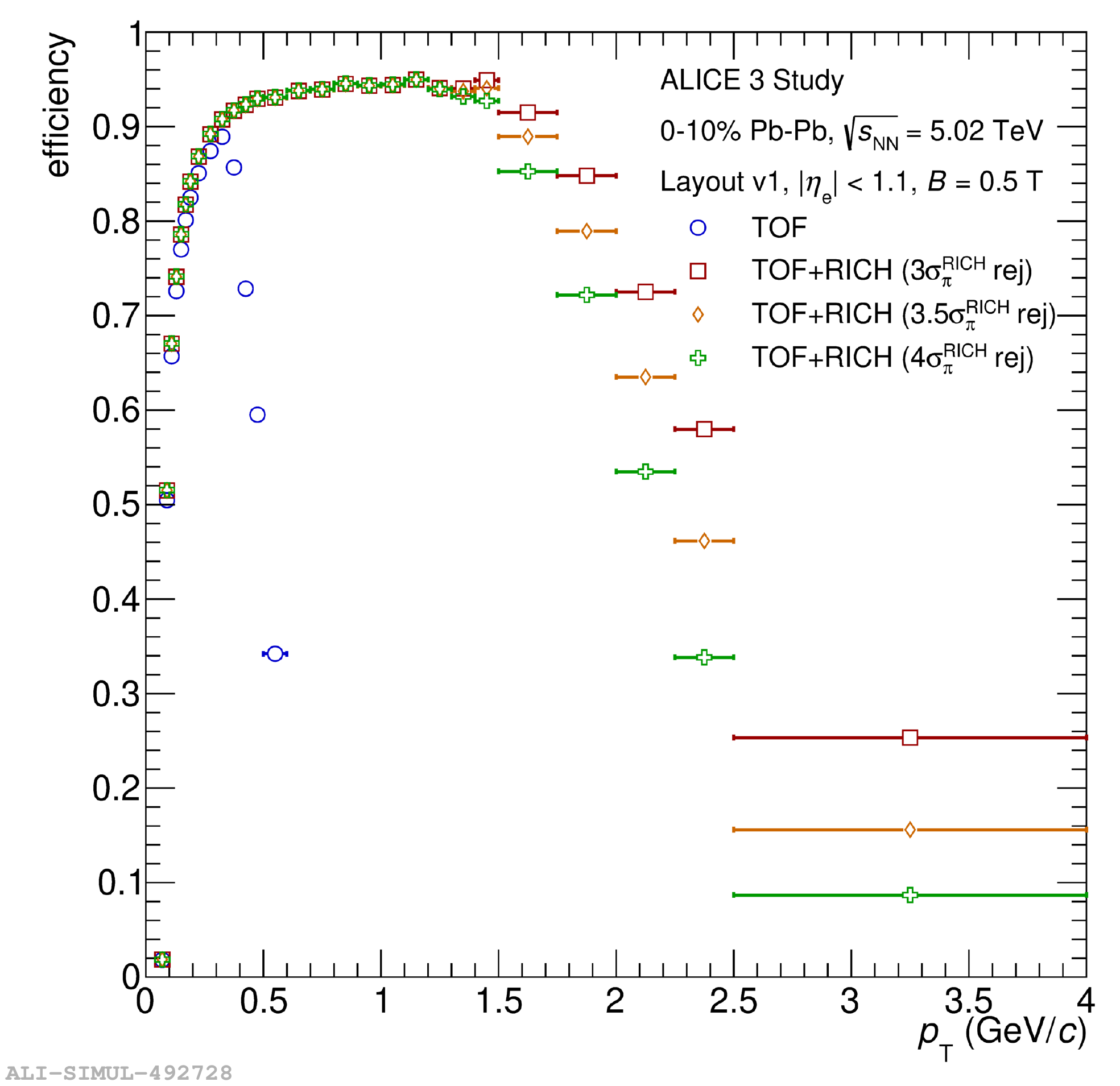}
  \includegraphics[width=.49\textwidth]{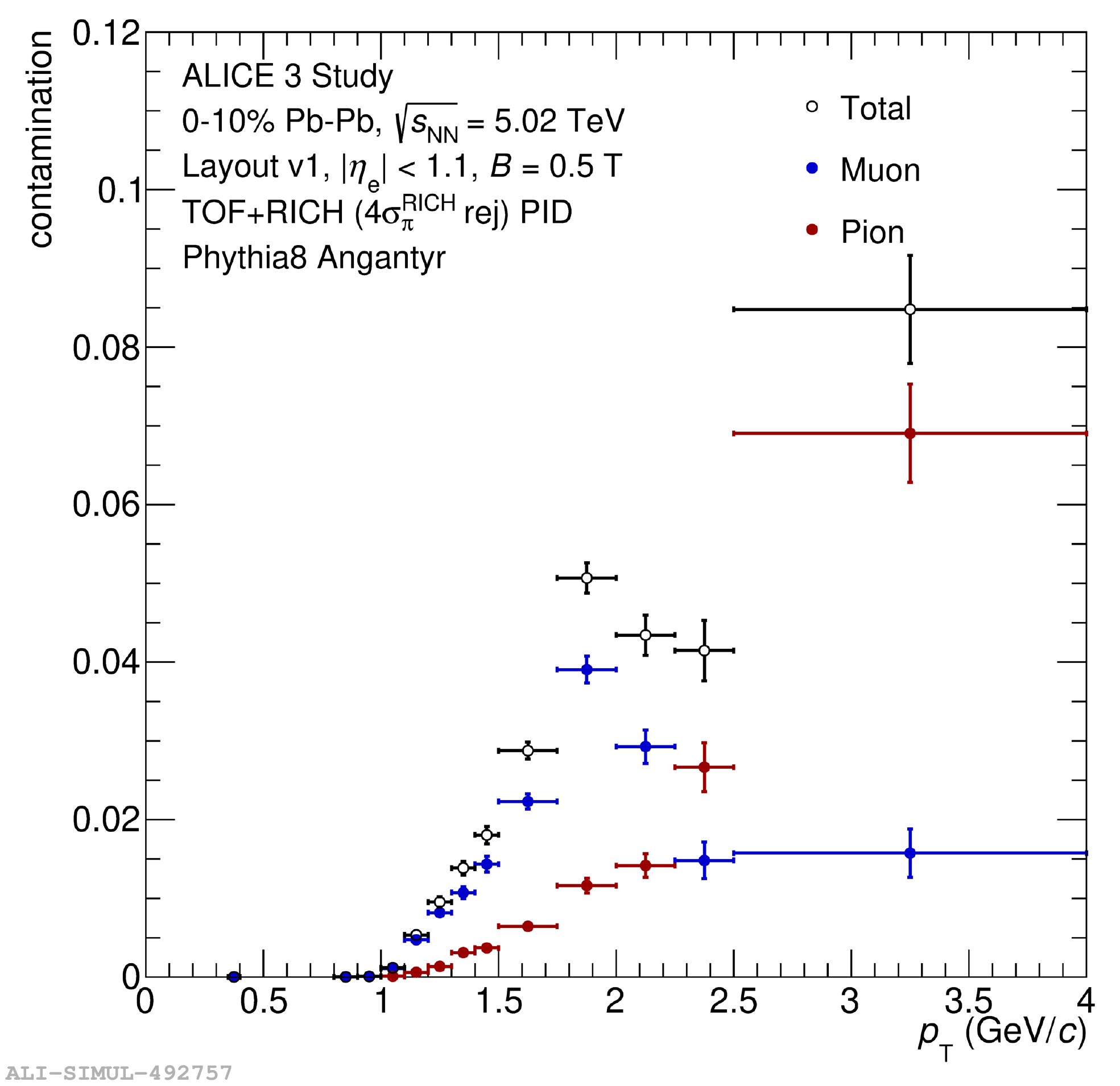}
  \caption[Electron efficiency and contamination] {
    Left: track efficiency of electrons and positrons as a function of \pt, considering different PID scenarios in a magnetic field of 0.5 T.
    In blue circles, using only a TOF detector and in red squares, yellow diamonds and green crosses, using a TOF and RICH detector with a 3.0, 3.5 and 4.0 \richsigma{\pi} rejection cut.
    Right: separation of contamination contributions as a function of \pt, using a TOF+RICH scenario with a 4.0 \richsigma{\pi} rejection cut in a magnetic field of B = 0.5 T.
    In black the total contamination, in blue the contamination from muons and in red the contamination from pions is shown.
    No mismatch was considered in these projections.
  }
  \label{fig:performance:dectector:electon_id:contamination_and_eff}
\end{figure}

The efficiency obtained by applying a selection of electrons with $\tofNsigma{e} < 3$ is shown as a function of \pt in \Fig{fig:performance:dectector:electon_id:contamination_and_eff} (left) for different rejection criteria on the pions.
The contamination, corresponding to a selection of $\richNsigma{\pi} < 4$ is shown in \Fig{fig:performance:dectector:electon_id:contamination_and_eff} (right).
In the region with sufficient efficiency for a measurement, below $\pt = 2$ \GeVc, the contamination is found to be below 5\%.
The $4\sigma$ rejection translates to a rejection factor of $10^5$ for the pions and $10^2$ for the muons.

At high \pt, the ECAL can be used to identify electrons with the $E/p$ method. This method generally provides sufficient purity to measure decays of quarkonia and open heavy-flavour hadrons, while the misidentification background may be too large for measurements of the thermal continuum.

\clearpage %
\clearpage %
\subsubsection{Muon identification}
\label{sec:performance:detector:muon_id}

Muon identification in ALICE~3 is based on the matching of the information from the tracker and the muon identifier. 
The design of the muon identifier is described in Section~\ref{sec:systems:muon}.
The thickness profile of the hadron absorber has been designed with the goal of achieving enough ($\gtrsim 20\,\%)$ acceptance down to $\pt = 0$ for dimuon invariant masses of $M_{\mu\mu} \approx 3$~GeV/$c^2$, in the whole rapidity region $\abs{\eta} < 1.5$ covered by the detector. %

The muon identification strategy starts with the reconstruction of tracklets in the muon identifier: a tracklet is defined as a straight segment connecting a pair of space points in the two detector layers. Some basic selection criteria on the orientation of the tracklets are imposed, under the hypothesis of particles coming from the interaction region, in order to reject a large fraction of the tracklets built out of combinations of uncorrelated space points. These cuts are optimized with the help of Monte Carlo simulations, with the goal of minimizing the loss of reconstruction efficiency for good tracklets.

The matching between the information of the internal tracker and the muon identifier is then performed by extrapolating each track to the innermost layer of the muon identifier system, taking into account the effect of the magnetic field and the energy loss in the traversed materials. The extrapolated track is then compared with the available tracklets. Only tracklets falling within a search window defined around the extrapolated tracks are considered as valid candidates for the definition of a ``global'' track. The size of the search region is optimized with the help of Monte Carlo simulations, with the goal of minimizing the loss of reconstruction efficiency for good global tracks, where the information from the inner tracker and the muon identifier corresponds to the same particle. Whenever more than one compatible tracklet is found within the search window, a global track is built out of each of them, and the one corresponding to the best global fitting $\chi^2$ is retained. The default strategy for the definition of the global fitting $\chi^2$ is based on a Kalman filter approach, combining the space points reconstructed in the inner tracking system and the muon identifier. 
Alternative criteria for the selection of the best global track candidate are under study.

By definition, a global track (a track from the inner tracker having found at least one valid extrapolation in the muon identifier) is identified as a muon track. The performance of the muon identification system was studied with GEANT4 simulations of pp events and the muon efficiency and hadron rejection factors were determined and parametrised for use in \delphesotwo. It was verified that the occupancy in \PbPb simulations is small enough that the expected impact on the performance is negligible. The efficiency and contamination from fast simulation is shown in Fig.~\ref{fig:performance:dectector:muonID:kineBeforeAfterMID} for muons, pions, and kaons, and the MID is found to be nearly $100\,\%$ efficient in identifying muons starting from $\pt \sim 1.5$~GeV/$c$ at $\eta = 0$, providing pion and kaon rejection factors of \numrange{50}{100}. Comparative studies on the physics performance in the muon and electron channels are ongoing.

\begin{figure}[htbp]
  \centering
  \includegraphics[width=0.55\textwidth]{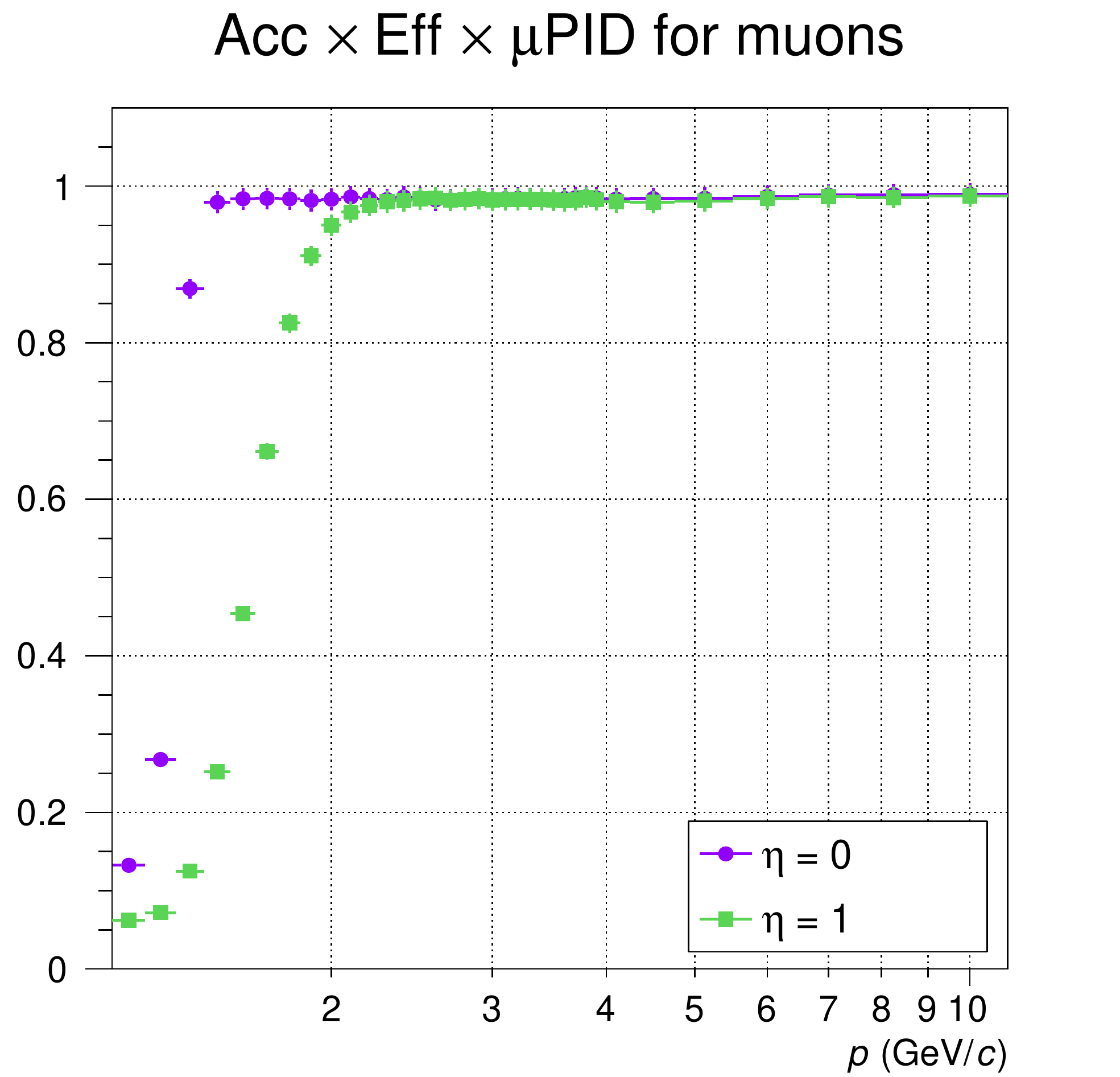}
  \includegraphics[width=0.65\textwidth]{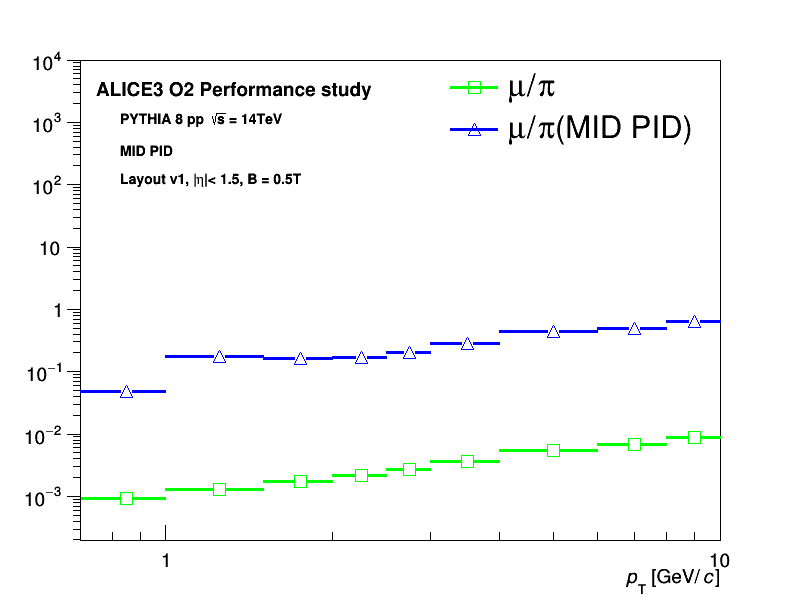}
  \includegraphics[width=0.65\textwidth]{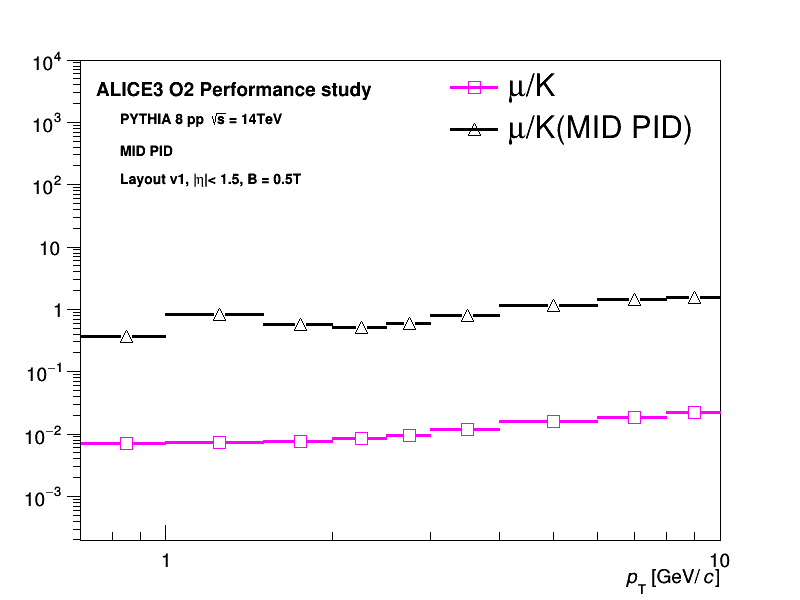}

  \caption[Muon efficiency and contamination] {\comment{Figure needs graphical improvement}MID rejection factor for muons as a function of $\pt$; $\mu/\pi$ ratio and $\mu/$K ratio, before and after MID identification, as a function of $\pt$.}
  \label{fig:performance:dectector:muonID:kineBeforeAfterMID}
\end{figure}

\subsubsection{Performance of the forward conversion tracker}
\label{sec:performance:detector:photon_id}

As discussed in Section~\ref{sec:physics:ultra-soft-photons:low}, access to the physics in the range relevant for testing Low's theorem poses the challenge of detecting photons down to transverse momenta in the MeV range.
At low photon energies, the photon conversion method generally provides a better energy resolution than a calorimeter measurement. However, also with the conversion method, limitations arise for low-energy photons due to interactions and difficulties in the tracking of the electrons and positrons produced at low momenta. 
These limitations become less problematic for photon energies $E_\gamma \gtrsim 50\text{--}100\,\mathrm{MeV}$. Experimentally, a photon measurement in the MeV/c transverse momentum range is therefore only feasible at larger rapidity. The ratio $E_\gamma / p_{T,\gamma} = \cosh \eta_\gamma$ is 10.1, 27.3, and 74.2 for $\eta_\gamma = 3,\,4,\,5$, respectively. As can be readily seen, experimental access to the MeV/c transverse momentum range becomes progressively easier as the photon rapidity is increased.

We have studied a setup in which a photon converts in front of a Forward Conversion Tracker (FCT) which subtends the pseudorapidity range $3 \lesssim \eta \lesssim 5$. The trajectories of the electron and positron are then measured by a sequence of silicon tracking layers whose spacing increases from below a centimeter to several centimeters. This allows us to measure tracks at low transverse momentum from the MeV range up to the GeV range. 
A transverse component of the magnetic field is necessary to provide sufficient momentum resolution.
A possible location of the FCT would be at a distance of about 3.5\,m away from the primary vertex. 

\begin{figure}
\centering
\includegraphics[width=\linewidth]{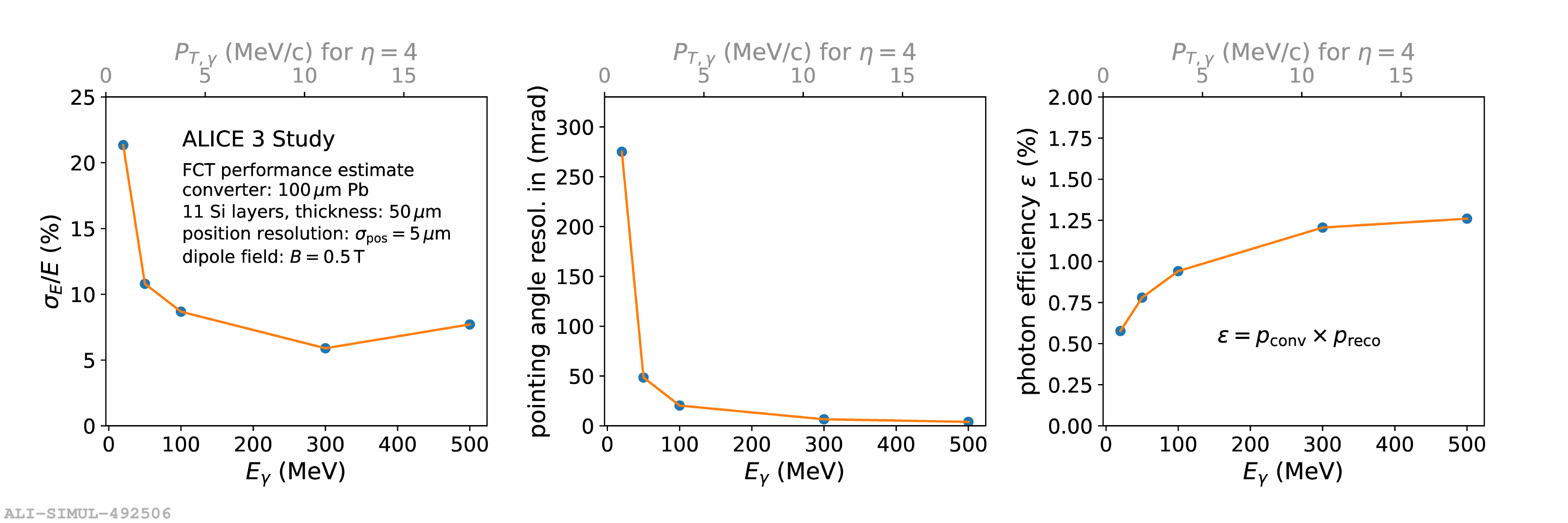}
\caption[Soft photon performance]{Photon energy resolution (left), pointing resolution (middle), and photon efficiency (right) of the FCT consisting of 11 silicon tracking layers. The GEANT4 simulation was performed for a dipole field of $B = 0.5\,\mathrm{T}$. A position resolution of $\sigma = 5\,\mathrm{\mu m}$ was assumed for the trackers. The tracker thickness was taken to be $50\,\mathrm{\mu m}$.}
\label{fig:fct_performance}
\end{figure}

The energy resolution, pointing resolution, and photon efficiency for such a setup are determined with a GEANT4 simulation with a simplified track momentum reconstruction and shown in Fig.~\ref{fig:fct_performance}. The photon efficiency $\varepsilon$ is dominated by the photon conversion probability in the $100\,\mathrm{\mu m}$ lead converter. The pointing resolution defines how well photon candidates can be constrained to originate from the primary vertex in order to reject background. Below a photon energy of $E_\gamma \approx 50\,\mathrm{MeV}$ both the energy resolution and the pointing resolution deteriorate significantly reflecting the effect of multiple scattering on the resolution on the momentum of the electron and positron tracks. 

In summary, the FCT as discussed in Section~\ref{sec:systems:fct} provides a photon measurement in the range $E_\gamma \gtrsim 50\,\mathrm{MeV}$. To reach transverse momenta below $p_T \lesssim 10~\mathrm{MeV}/c$, a measurement in the forward region up to $\eta \approx 5$ is necessary.

\subsection{Physics performance}
\label{sec:performance:physics}

\subsubsection{Open heavy flavours}
\label{sec:performance:physics:heavy_flav}
The ALICE~3 experiment is specifically optimized for measurements of rare heavy-flavour hadrons and heavy-flavour correlations in heavy-ion collisions, over a wide rapidity interval and for $\pt$ close to 0~(see Section~\ref{sec:introduction:experimental_layout}).
In this section, we will present performance studies for selected heavy-flavour signals with ALICE~3. For some of the key observables, the performance will be compared to the ones expected for other experiments at the LHC, such as CMS and LHCb. When available, a direct comparison to the expected results with ALICE in Run 3 and 4 (ALICE~2)~\cite{ALICEupgradeits2}, will be provided.

In Section~\ref{sec:benchmarkHF} studies performed with $\Dzero$ and $\Lambda_c$ in pp and Pb--Pb collisions are presented to illustrate the outstanding performance of ALICE~3 in the reconstruction and selection of basic heavy-flavour decays. These studies provide a benchmark for the excellent pointing resolution and PID capabilities of the proposed ALICE~3 detector.
In Section~\ref{sec:performance:physics:heavy_flav:multicharm}, the results expected for the measurements of multicharm hadrons with strangeness tracking and with traditional reconstruction techniques are discussed. In Section~\ref{sec:performance:physics:heavy_flav:Bmeson}, the performance expected for the reconstruction of beauty mesons in \pp and \PbPb collisions are presented. In Section~\ref{sec:performance:physics:heavy_flav:Bmeson}, the predictions for the measurements of the elliptic flow coefficient $\vtwo$ of \lc and \lb~baryons in semi-central \PbPb collisions are shown. The results for the study of fully reconstructed $\DDbar$ correlations in heavy-ion collisions are presented in Section~\ref{sec:performance:ddbar}. 
Finally, in Section~\ref{sec:performance:physics:ddstar} and Section~\ref{sec:performance:physics:ddbarstar}, the predictions for the measurements of exotic hadrons with charm-hadron correlations are discussed.

For most studies, background samples were generated using \PYTHIA inelastic pp events with SoftQCD processes and then propagated through the \textit{Fast Simulation} workflow to mimic the track reconstruction and PID detector response. A signal sample was generated using the same tune of \PYTHIA{}, selecting only events that have at least one candidate of interest. %
A sample of \texttt{PYTHIA8 Angantyr} minimum-bias \PbPb{} events with $\sqrt{\sNN} = 5.5$~TeV was used to estimate the combinatorial background expected in \PbPb{} collisions.

\paragraph{Benchmarks of tracking and PID performance with $\Dzero$ and $\Lambda_c$} ~\\ 
\label{sec:benchmarkHF}
\begin{figure}[htb!]
\centering
\includegraphics[width=0.8\textwidth]{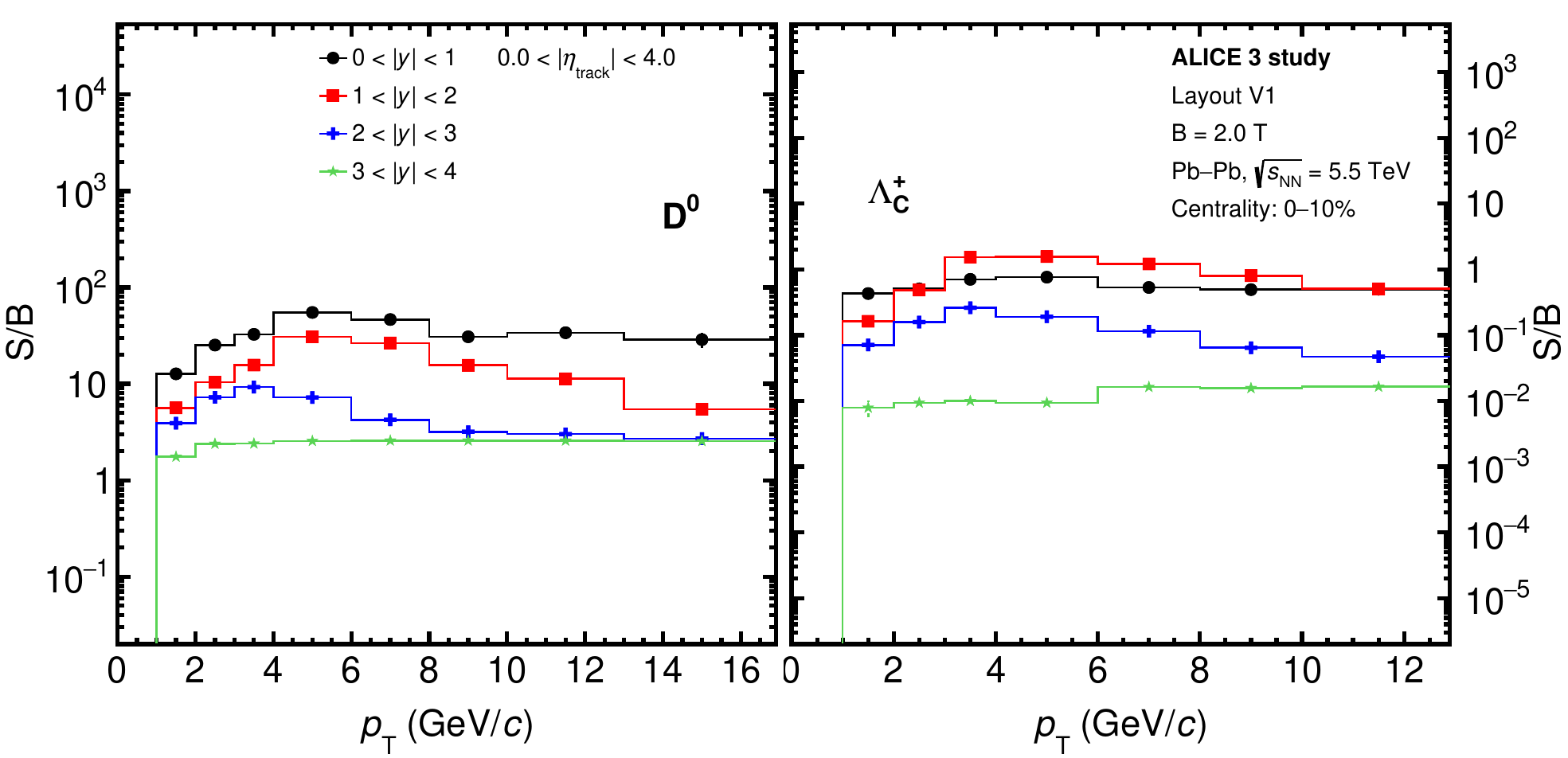}
\includegraphics[width=0.8\textwidth]{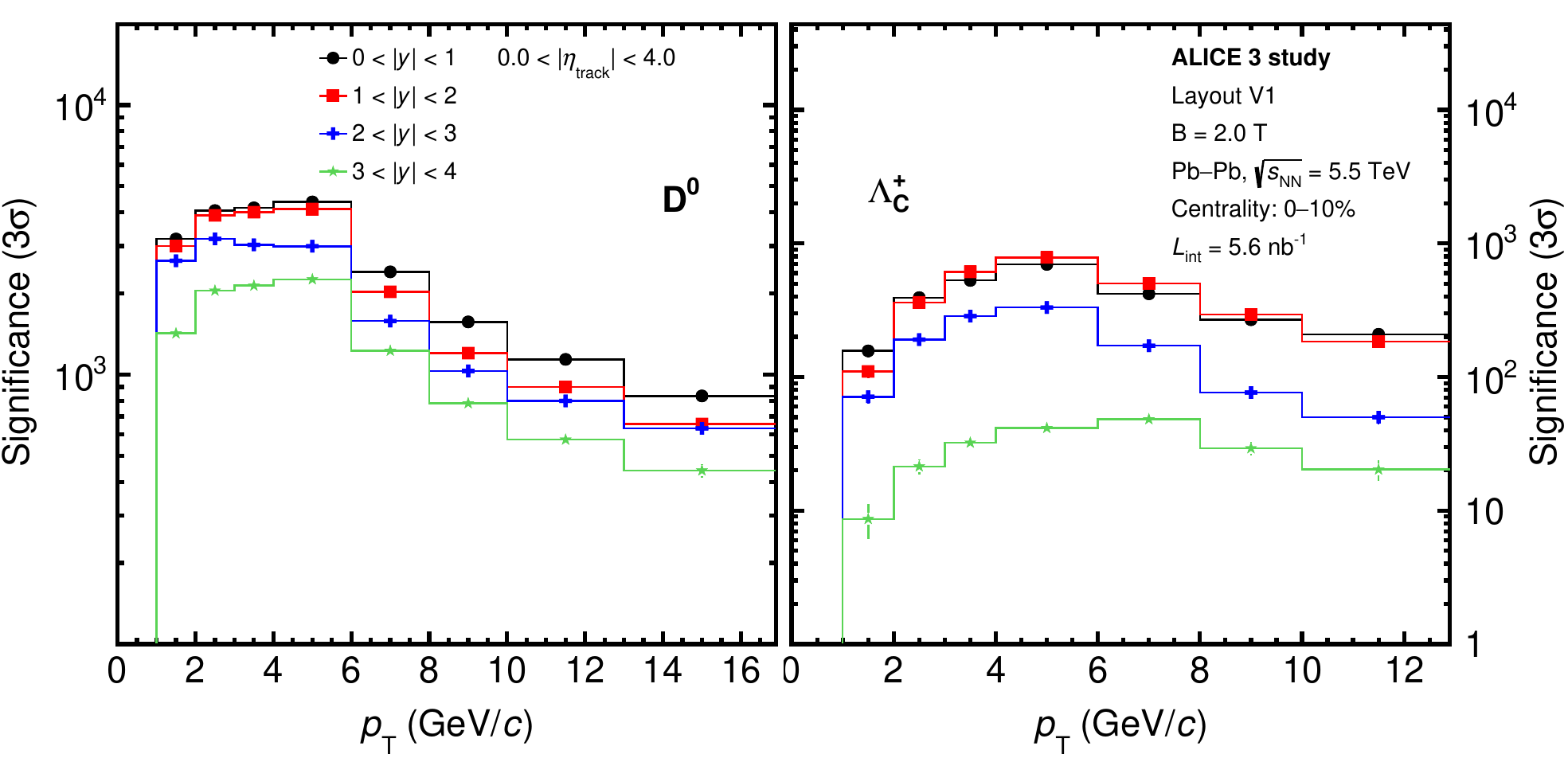}
\caption[Signal to background for $\Dzero$ and $\lc$]{
Signal to background ratio for $\Dzero$ (left) and $\Lambda_{c}^{+}$ (right) in intervals of rapidity as a function of \pt in different rapidity intervals in 0-10\% central \PbPb collisions. Bottom panels: The significance evaluated by integrating signal and background in a $3\sigma$ window in invariant mass for 1 running year ($\Lint = \SI{5.6}{\nano\barn^{-1}}$).}
\label{fig:performance:heavy_flav:D0}
\end{figure}
The hadronic decay channels $\rm \Lambda_c \to pK\pi$ and $\Dzero \to K\pi$ were used to benchmark the capabilities of the ALICE~3 detector in the reconstruction of simple heavy-flavour decays. For this study, the $\Dzero$ and $\Lambda_c$ candidates were selected according to topological criteria, which exploit the different kinematical and geometrical properties of signal and background candidates. An additional reduction of the background candidates is obtained by applying PID selections, which are based on the TOF and RICH detectors, on the daughter tracks.

Figure~\ref{fig:performance:heavy_flav:D0} shows the signal-to-background (S/B) ratio and significance $\rm S/\sqrt{(S+B)}$ for \Dzero and $\Lambda_{c}^{+}$ for central \PbPb{} collisions for different rapidity intervals. %
Thanks to its outstanding tracking capabilities, ALICE~3 will reconstruct and select \Dzero and $\Lambda_{c}^{+}$ hadrons with unprecendented accuracy over eight units of rapidity. The background reduction provided by the TOF detector and, at larger \pt, by the RICH detector is critical to obtain a very pure samples of both \Dzero and $\Lambda_{c}^{+}$ candidates from \pt close to 0 to $\pt>$~10 \GeVc. The excellent purity for the reconstruction of charm-hadron decays represents the starting point for more differential measurements, like the elliptic flow presented in Section~\ref{sec:performance:physics:heavy_flav:baryonflow}, as well as azimuthal correlation measurements with $\DDbar{}$ (Section~\ref{sec:performance:ddbar}) or DD$^*$ momentum correlations (Section~\ref{sec:performance:physics:ddstar}).
\\ \\
\textit{\textbf{Comparison to ALICE~2 performance in Run~3 and Run~4.}}
In Fig.~\ref{fig:competitivity:Dzeromeson_SoverB} (left), the expected S/B ratios for ALICE~3, estimated in intervals of $\Dzero$ \pt and rapidity, are compared to those expected for ALICE~2. The performance studies for Run~3 and Run~4 were performed only at midrapidity ($\rm |y|<$0.5) due to the limited pseudorapidity coverage of the detector. The tracking resolution for ITS~2~\cite{ALICEupgradeits2} was used for the results presented here. All studies were performed for central (0--10$\%$) \PbPb collisions at $\sqrtsNN=$5.5 TeV. The results obtained in a recent ALICE measurement with Run~2 data at 5.02 TeV~\cite{ALICE:2021rxa} are also superimposed. ALICE~3 significantly outperforms ALICE~2 over the entire \pt interval. With ALICE 3, the S/B 
ratio increases by $\mathcal O $($10^{2}$) at low \pt and by $\mathcal O$($10$) at high \pt.
For \pt $<$ 3--4 \GeVc, this improvement is largely due to the improved pointing resolution for single tracks, which allows for a more effective selection on the pointing angle of the heavy-flavour candidates. At intermediate $\pt$, the extended reach for daughter PID with the RICH detector provides a large reduction of the background with respect to the ALICE~2 case. This provides a substantial advantage to ALICE~3 in the study of mass-dependent effects for quenching with respect to both CMS and ATLAS. 
\\ \\
As shown in the right panel of Fig.~\ref{fig:competitivity:Dzeromeson_SoverB}, ALICE~3 will also provide a much larger signal efficiency. The excellent tracking and PID capabilities of ALICE~3 provide an excellent separation between background and signal candidates without the need of applying very tight selections, which results in an increase of the efficiency by about a factor 2--5 with respect to ALICE~2~(right panel of Fig.~\ref{fig:competitivity:Dzeromeson_SoverB}), in addition to a much reduced background contamination. The larger efficiency is especially critical for the measurements of higher-order cumulants of charmed-hadron multiplicity distributions (see Section~\ref{sec:perf:fluctuations} for more details). 
\\ \\
\textit{\textbf{Comparison to CMS and LHCb performance}}~The CMS detector in Run~4 will be equipped with a timing detector (MTD), which will provide hadron-PID capabilities at low-\pt~\cite{CMSMTDTDR} at midrapidity. The expected S/B for $\dzerotokpi$ candidates in the interval 5$<\pt<$6 \GeVc was estimated to be approximately 0.03--0.1. These estimations were based on results presented in the MTD Technical Design Report~\cite{CMSMTDTDR} and on an extrapolation that relies on a Run~2 CMS measurement~\cite{CMS:2017qjw}.
These values are significantly lower than those extracted in the same \pt interval with ALICE~3. %
A direct comparison with the performance expected for LHCb is currently unavailable, since the LHCb Collaboration is still assessing its capabilities in central \PbPb collisions. 
In p--Pb collisions, the estimated S/B ratio~\cite{LHCb:2017yua} is about 2.7 and 6.1 for $\Dzero$ with $2<\pt<3$ \GeVc in the rapidity interval -4$\rm <y<$-3.5 and 2.5$\rm <y<$3.0, respectively.
This value is similar to the ALICE~3 one obtained in very high-multiplicity \PbPb collisions in the same rapidity interval. In addition, this estimate is lower than the one obtained by ALICE~3 in central \PbPb collisions at central rapidity, where the ALICE~3 PID selection is much more effective. \\ \\

\begin{figure}
    \centering
    \includegraphics[width=0.99\textwidth]{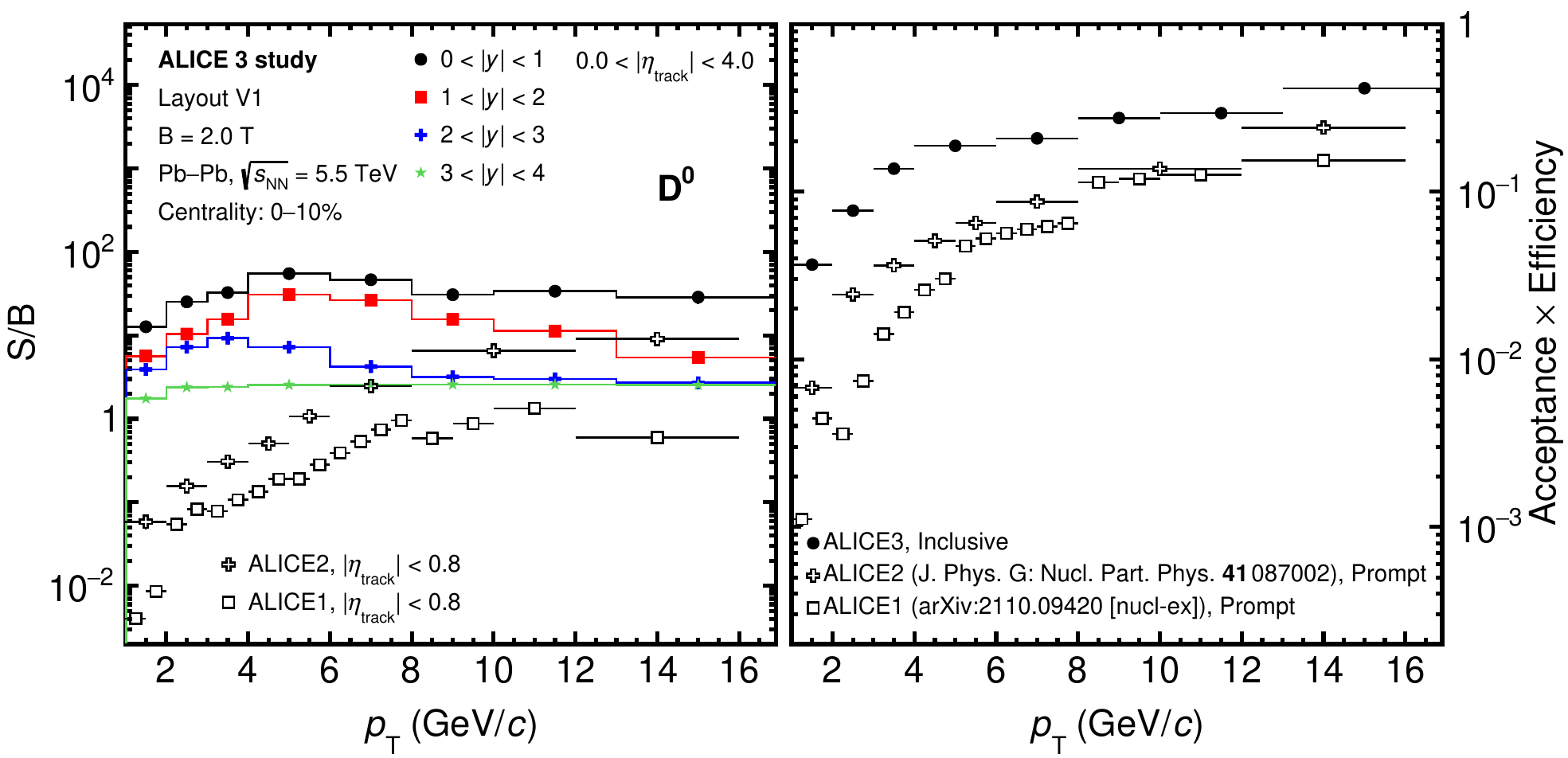}
    \caption[S/B and efficiency for D mesons in ALICE 1, 2, 3]{Left panel: signal to background ratio for $\Dzero$ in intervals of rapidity as a function of \pt in 0-10\% central \PbPb collisions. Results are compared to the measurements from ALICE1 and ALICE2 at midrapidity and shown by open marker. Right panel: $\Dzero$ meson reconstruction acceptance $\times$ efficiency at midrapidity for ALICE~1, ALICE~2 and ALICE~3. 
}
    \label{fig:competitivity:Dzeromeson_SoverB}
\end{figure}
\paragraph{Multi-charm baryons: $\Xi_{cc}^{++}$ and $\Omega_{cc}^{+}$} ~\\
\label{sec:performance:physics:heavy_flav:multicharm}
Measurements of the multi-charm baryons are a central part of the ALICE~3 physics programme (see Sec.~\ref{sec:physics:qgp_hadronisation}). In the following, we demonstrate the physics performance for the $\Xi_{cc}^{++}$ and $\Omega_{cc}^{+}$ baryons. ALICE~3, which is specifically designed for these studies, can effectively reconstruct these rare hadrons in proton-proton and nucleus-nucleus collisions using strangeness tracking, described in Section ~\ref{sec:performance:detector:strangeness_tracking}. The performance of the standard approach based on the reconstruction of the decay daughters in pp collisions is presented for comparison.

\paragraph*{$\Xi_{cc}^{++}$ and $\Omega_{cc}^{+}$ reconstruction using strangeness tracking} ~\\
The ALICE 3 apparatus is ideally suited to perform the direct tracking of multi-strange
hyperons, which are the decay products of multi-charm baryons in the channels: 
\begin{equation}
\label{lab:multicharmstratrack}
\Xi_{cc}^{++} \rightarrow \Xi_{c}^{+} + \pi^{+} \rightarrow \Xi^{-} + 3\pi^{+} \quad \rm{and} \quad \Omega_{cc}^{+} \rightarrow \Omega_{c}^{0} + \pi^{+} \rightarrow \Omega^{-} + 2\pi^{+}.
\end{equation} 
The properties of the charm and multi-charm baryons considered in these studies are reported in Table~\ref{tab:performance:physics:heavy_flav:multicharm}. %

\begin{table}
    \centering
    \renewcommand{\arraystretch}{1.3}
    \begin{tabular}{@{} l S[table-format=1.3] S[table-format=3, table-space-text-post={ (assumed)}] l l @{}}
        \toprule
         Particle & {Mass (\si{\GeVc})} & {$c\tau$ (\si{\um})} 
         &  Decay Channel & Branching Ratio (\%)  \\
         \midrule
         $\Omega_{cc}^{+}$ & 3.746 & 50 { (assumed)}  & $\Omega^{0}_{c} + \pi^{+}$ & 5.0 (assumed) \\
         $\Omega_{c}^{0}$ & 2.695 & 80  & $\Omega^- + \pi^{+}$ & 5.0 (assumed) \\
         $\Xi_{cc}^{++}$ & 3.621 & 76  & $\Xi^{+}_{c} + \pi^{+}$ & 5.0 (assumed) \\
         $\Xi_{c}^{+}$ & 2.468 & 137  & $\Xi^{-} + 2\pi^{+}$ &   $(2.86\pm1.27)$\\
         $\Xi_{c}^{+}$ & 2.468 & 137  & $\rm{p}+\rm{K}^{-}+\pi^{+}$ &   $(6.2 \pm 3.0) 10^{-3}$\\
        \bottomrule
    \end{tabular}
    \caption{Particles and decay channels used in the reconstruction of the $\Xi_{cc}^{++}$ and $\Omega_{cc}^{+}$ analyses using strangeness tracking. Values from~\cite{Zyla:2020zbs}. Where no measurement is available, a branching ratio of 5\% is assumed.}
    \label{tab:performance:physics:heavy_flav:multicharm}
\end{table}

Monte Carlo samples for the performance studies of $\Xi_{cc}^{++}$ and $\Omega_{cc}^{+}$ reconstruction in these channels were generated using the hybrid simulation technique described in Section~\ref{sec:performance:hybrid-simulation}. Separate background and signal samples 
were generated, with the background samples following a different strategy 
in pp and \PbPb{} collisions. In pp collisions, three types of background samples were generated using PYTHIA 8.3: two samples with events in which at least one a) $\Xi^{-}$ or b) $\Xi^{+}_{c}$ were generated and c) one sample of inelastic pp collisions. To estimate the background, the three samples were combined, with weights to account for the cross sections for the production of each species. 
The expected yield of $\Xi_{cc}^{++}$ in pp collisions was calculated based on the production cross section from~\cite{Berezhnoy:1998aa}, shown in Fig.~\ref{fig:physics:multicharm}. Based on Pythia calculations of charm baryons, it is assumed that 33\% of the total yield is seen the rapidity range $|\eta|<1.5$. The branching ratios are listed in Table~\ref{tab:performance:physics:heavy_flav:multicharm}.
The \xicp production cross section is assumed to be equal to the one of $\Xi_{c}^{0}$ reported in~\cite{ALICE:2021bli} and the $\Xi^{-}$ production cross is taken from~\cite{ALICE:2020jsh}.

In \PbPb{} collisions, large samples of inelastic PYTHIA Angantyr~\cite{Bierlich:2018xfw} Pb--Pb events were generated for background estimates. Because a dominant source of background in the strangeness tracking channels comes from primary $\Xi^{-}$ and $\Omega^{-}$ baryons and these have been found to be severely underpredicted by PYTHIA, additional multi-strange baryons are injected into PYTHIA in a multiplicity-dependent manner, to reproduce the $p_{\rm{T}}$-integrated $\Xi/\pi$ and $\Omega/\pi$ yield ratios measured by ALICE. In \PbPb{} collisions no further scaling is needed. %

Multi-charm baryon candidates are built by combining multi-strange baryon candidates with pions for each decay in the decay chain. Several selection criteria are available on the level of the prongs, strange- and charm-baryon candidates. Pion, kaon and proton identification is performed for all decay products using the inner time-of-flight detector. 
For strange decay products, the position of the decay vertex is explicitly taken into account in the calculation of the arrival time of $\pi$, K and p from either $\Xi^{-}$ or $\Omega^{-}$ baryons. Effectively, this allows the separation of primary and secondary 
pions (from the decay of a heavier object), as the latter arrive at the TOF detector with a delay.
Simulations indicate that even in central \PbPb{} collisions a purity of more than \SI{90}{\percent} can be reached for both $\Xi^{-}$ and $\Omega^{-}$, using this criterion alone. Decay products are then selected with a minimum DCA to the primary vertex, and a minimum \pt{} for positive pions produced in the decay of charmed baryons.

A characteristic feature of weakly decaying particles is the offset of the decay vertex from the primary interaction vertex. Selections are applied on the basis of the transverse distance between the primary vertex and the secondary decay vertices to reduce combinatorial background. 
Furthermore, strangeness tracking requires that the $\Xi^{-}$ or $\Omega^{-}$ produces a hit in at least one layer of the detector, i.e. the decay point is at a distance of more than \SI{0.5}{\cm} from the primary vertex. Direct hits of the charged hyperon not only improve its own tracking resolution, but also that of all parent particles. To illustrate this, Fig.~\ref{fig:performance:physics:heavy_flav:mcPbPb:DCADemo} shows the DCA distribution to the primary vertex of the $\Xi_{cc}^{++}$ for reconstruction with and without strangeness tracking. Further selections are applied on the DCA among the decay products at the decay vertex of the parent as well as on the invariant mass of the parent particles.

\begin{figure}
    \centering
    \includegraphics[width=.59\textwidth]{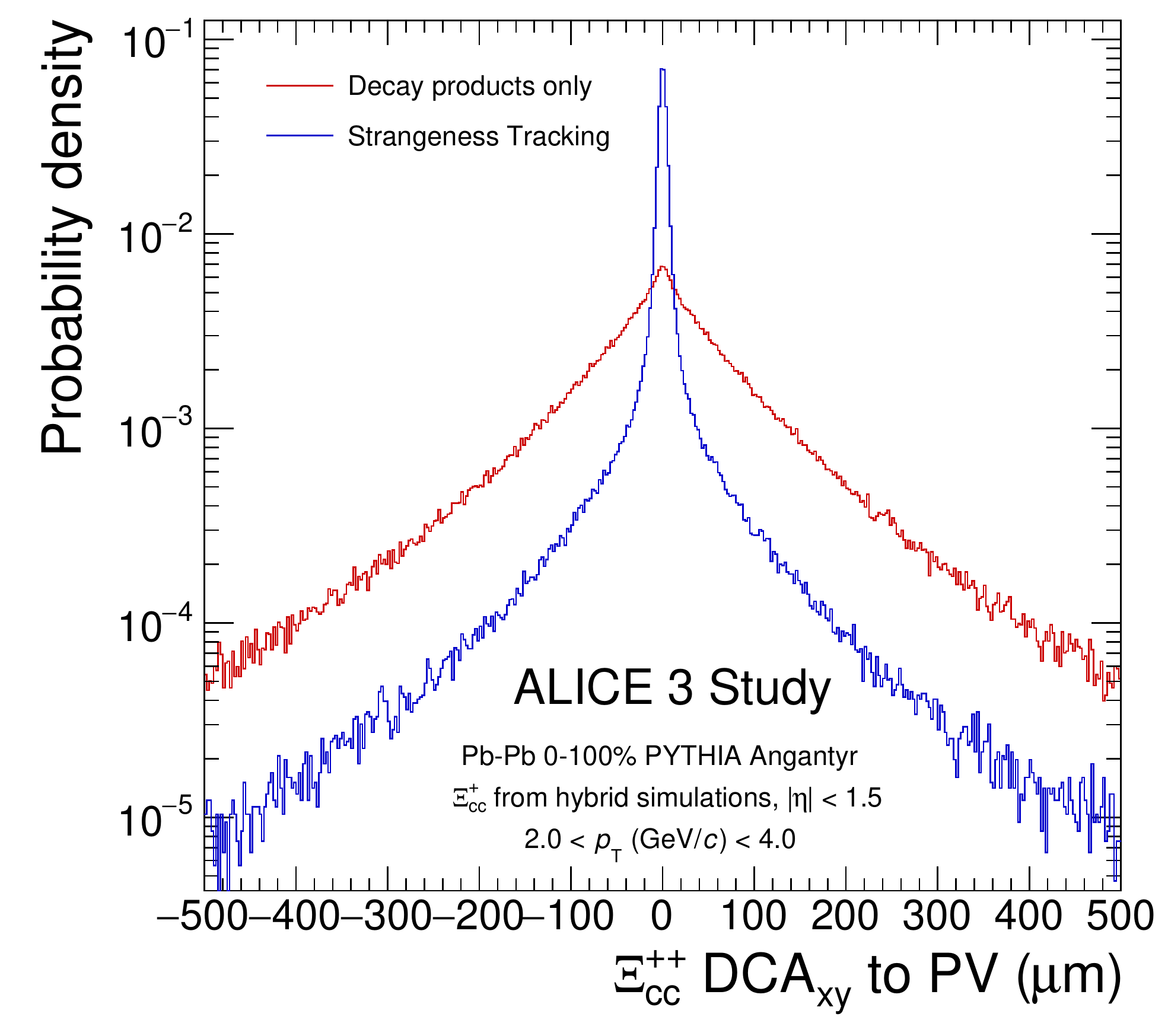}
    \caption[DCA distribution for $\Xi_{cc}^{++}$]{Distance of closest approach (DCA) distribution of reconstructed $\Xi_{cc}^{++}$ to 
    the primary interaction vertex in Pb--Pb collisions using decay product information 
    only or strangeness tracking for the $\Xi^{-}$ baryon in the decay chain of the $\Xi_{cc}^{++}$.}
    \label{fig:performance:physics:heavy_flav:mcPbPb:DCADemo}
\end{figure}

The expected $\Xi_{cc}^{++}$ and $\Omega_{cc}^{+}$ significance for central \PbPb collisions from a total sampled luminosity of (35 $\invnb$) is shown in Fig.~\ref{fig:performance:physics:heavy_flav:mcPbPb:significance}. The production yields are taken from the statistical hadronisation model shown in Fig.~\ref{fig:physics:multicharm}.

Starting from approximately 3-4~GeV/$c$, a significance of 10 or more can be seen for both multi-charm particle species, which is sufficient for a \pt-differential measurement with good statistical accuracy. 
With further tuning of the selection parameters by means of machine-learning techniques, the statistical significance is still expected to substantially improve. As an example, initial testing using machine learning algorithms based on Boosted Decision Tree for the $\Xi_{cc}^{++}$
has already achieved improvement in  
significance of up to a factor of $4-5\times$ for $\pt<2$~GeV/$c$, as shown in Fig.~\ref{fig:performance:physics:heavy_flav:mcPbPb:significanceML}, indicating that multi-charm baryons
will be measurable in central Pb--Pb collisions down to zero \pt with ALICE~3. 

More differential observables are likely to be within reach, as illustrated by Fig.~\ref{fig:performance:physics:heavy_flav:mcPbPb:v2Xicc}, which shows a first projection for the measurement of the elliptic flow of $\Xi_{cc}^{++}$ in semi-peripheral \PbPb collisions. The central values for the prediction were computed by fitting the mass dependence of the $v_{2}$ coefficients measured for lighter hadrons (p, K, $\pi$, $\Lambda$ and D mesons) and extrapolating the fit function to the $\Xi_{cc}^{++}$ mass. 
Also in this case, machine learning is expected to bring
significant improvements, allowing for an assessment of the collective properties of $\Xi_{cc}^{++}$ baryons.

\begin{figure}
    \centering
        \includegraphics[width=.75\textwidth]{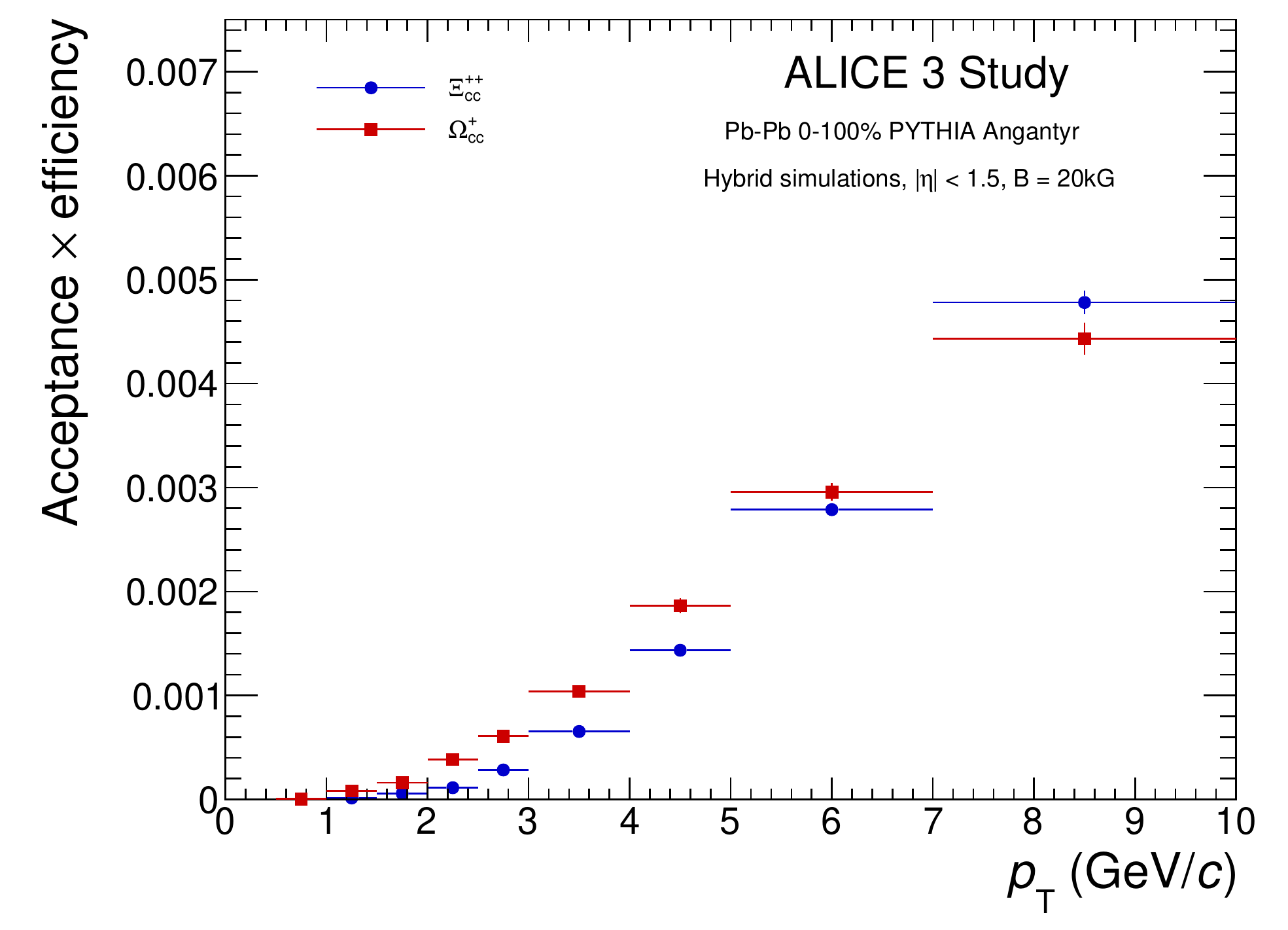}
    \caption[Efficiency of $\Xi^{++}_{cc}$ and $\Omega_{cc}^{+}$ reconstruction]{$\Xi^{++}_{cc}$ and $\Omega_{cc}^{+}$ efficiency as a function of \pt with a 2.0~T magnetic field, in the strangeness-tracking channel. Branching ratios of the various channels are given in Table~\ref{tab:channels} and are not taken into account here.
    }
    \label{fig:performance:physics:heavy_flav:mcPbPb:Efficiency}
\end{figure}

\begin{figure}
    \centering
    \includegraphics[width=.79\textwidth]{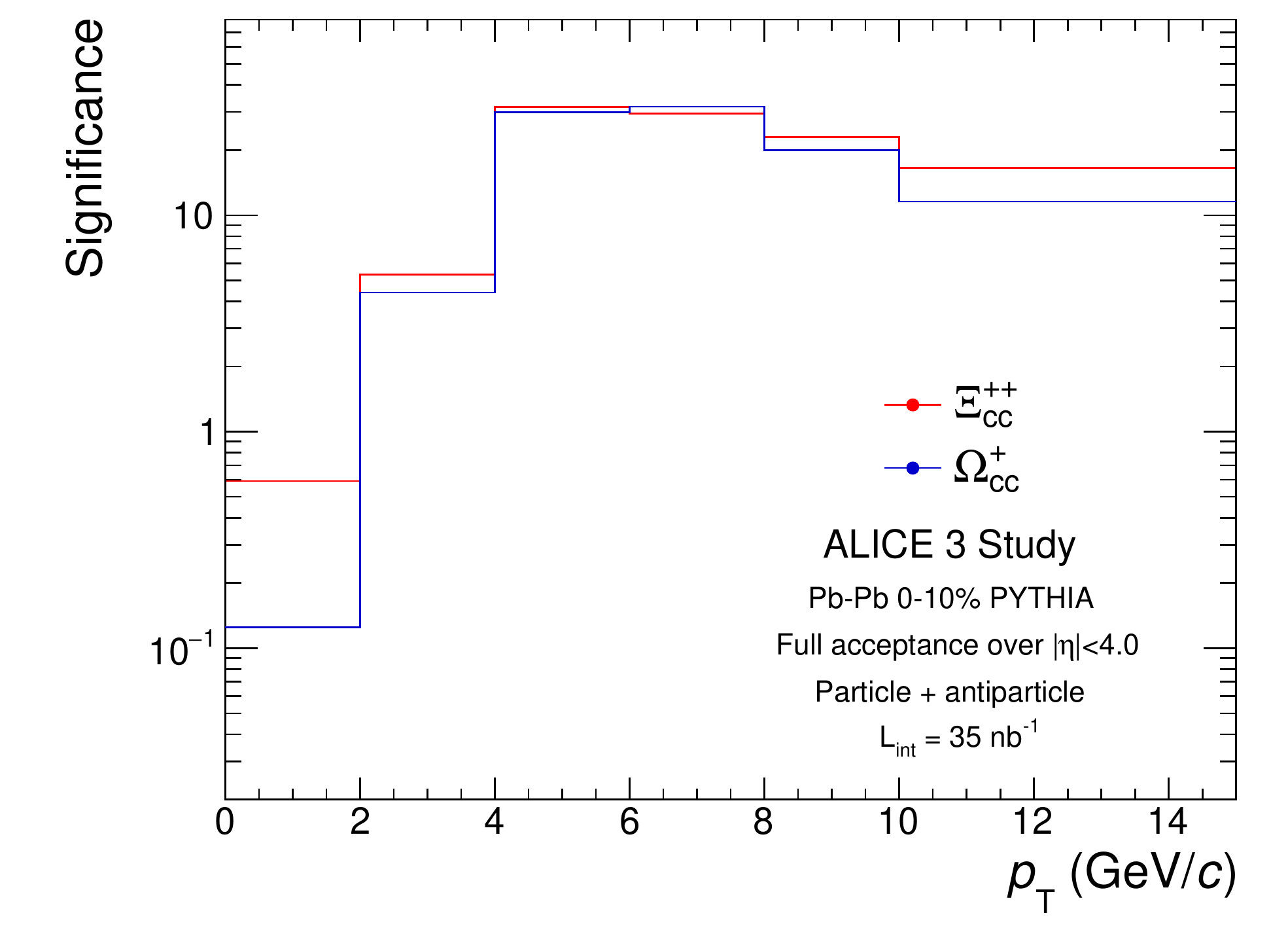}
    \caption[Significance for $\Xi^{++}_{cc}$ and $\Omega_{cc}^{+}$]{$\Xi^{++}_{cc}$ and $\Omega_{cc}^{+}$ significance in 0-10\% central Pb--Pb collisions at $\sqrt{s_{\rm{NN}}}$~=~5.52~TeV as a function of \pt with a 2.0~T magnetic field.}
    \label{fig:performance:physics:heavy_flav:mcPbPb:significance}
\end{figure}

\begin{figure}
    \centering
    \includegraphics[width=.79\textwidth]{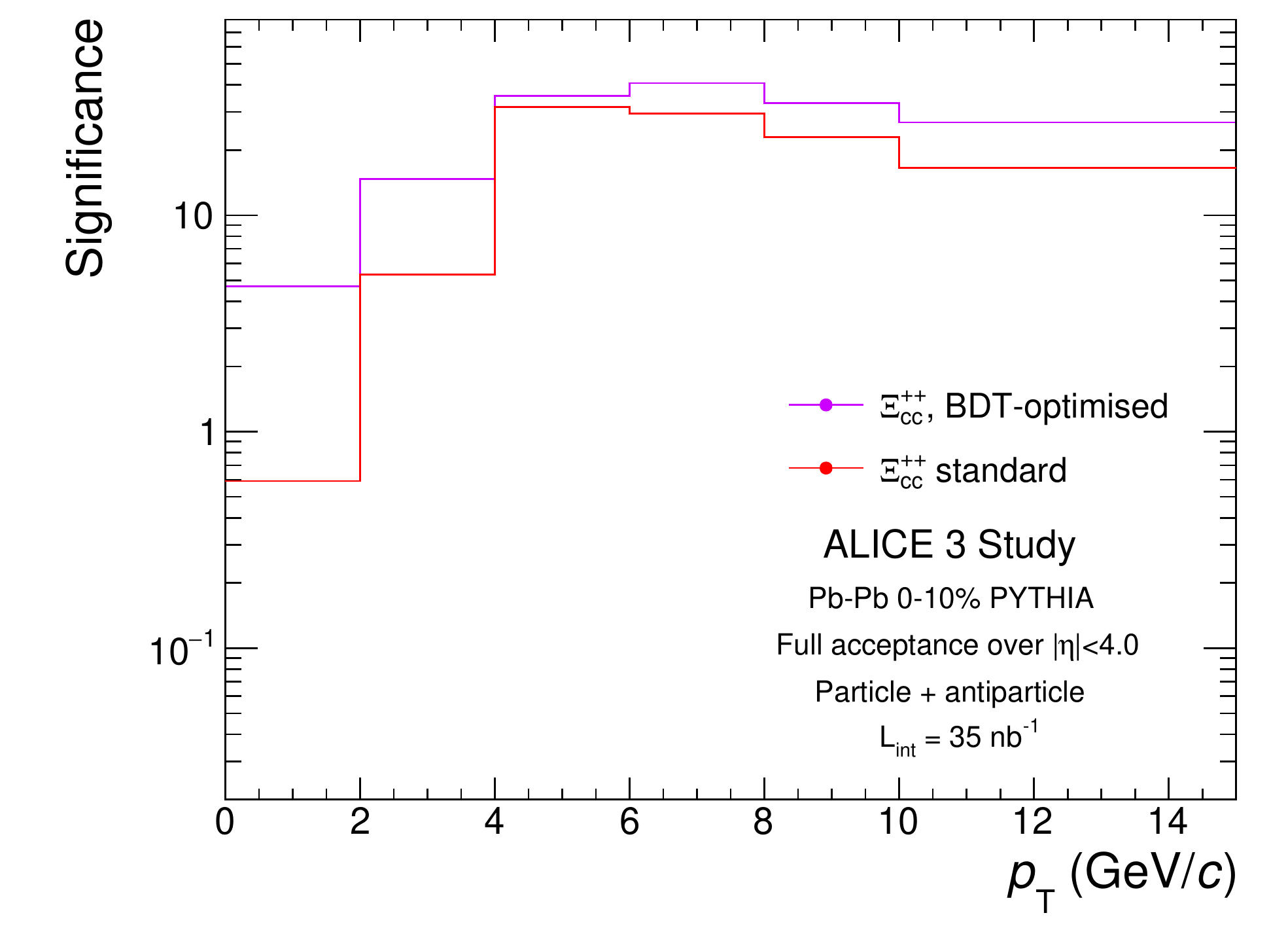}
    \caption[Significance for $\Xi^{++}_{cc}$ with machine learning]{$\Xi^{++}_{cc}$ significance in 0-10\% central Pb--Pb collisions at $\sqrt{s_{\rm{NN}}}$~=~5.52~TeV as a function of \pt with a 2.0~T magnetic field using standard selections and using machine learning.}
    \label{fig:performance:physics:heavy_flav:mcPbPb:significanceML}
\end{figure}

\begin{figure}
    \centering
    \includegraphics[width=.7\textwidth]{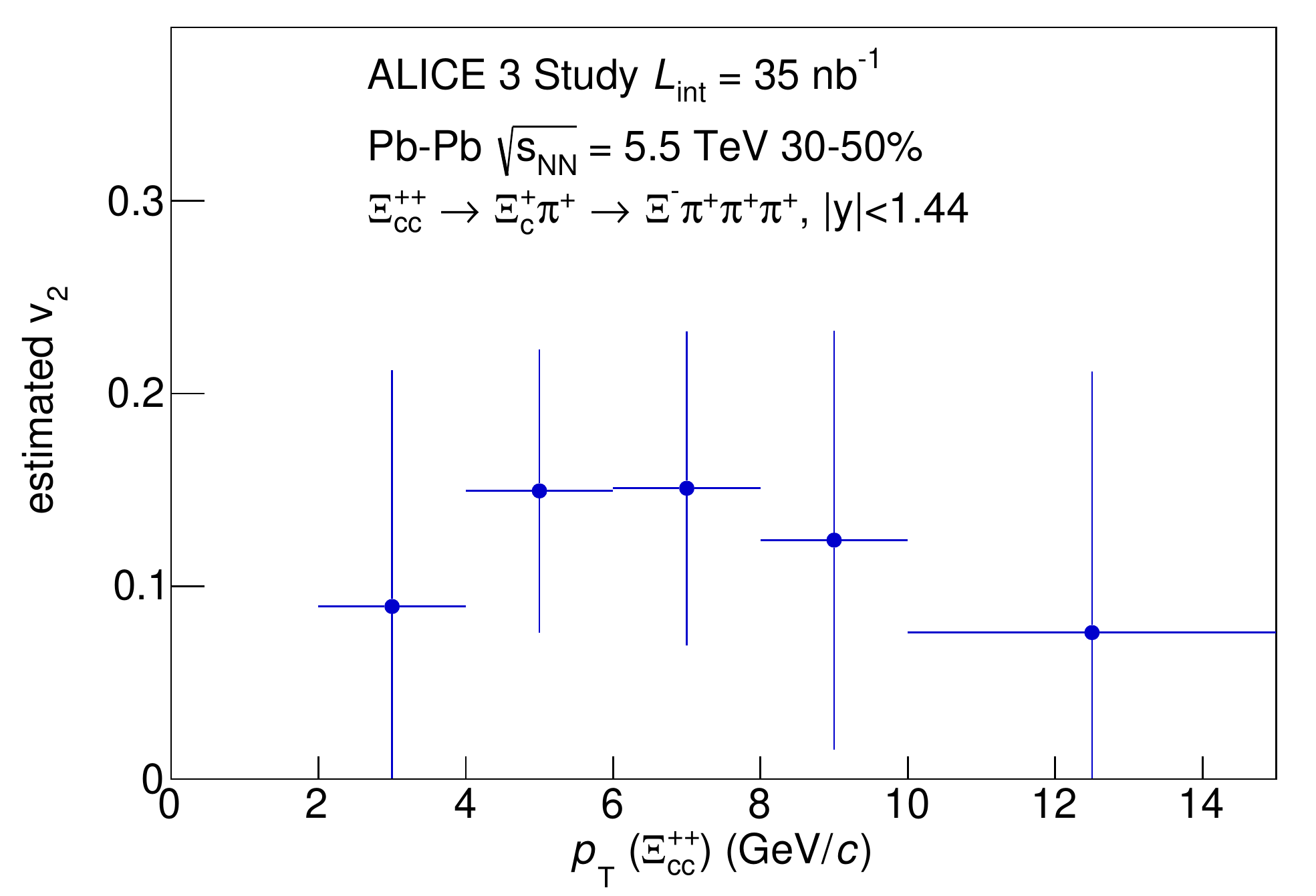}
    \caption[Elliptic flow $v_2$ of $\Xi_{cc}^{++}$]{Elliptic flow coefficient $v_{2}$ of $\Xi_{cc}^{++}$ in 30--50$\%$ Pb--Pb collisions at $\sqrt{s_{\rm{NN}}}$~=~5.52~TeV as a function of \pt.}
    \label{fig:performance:physics:heavy_flav:mcPbPb:v2Xicc}
\end{figure}

The expected reconstruction performance for $\Xi_{cc}^{++}$ in pp collisions can be seen in Fig.~\ref{fig:performance:physics:heavy_flav:xi_cc_pp_mass}, where the expected invariant mass spectrum of $\Xi_{cc}^{++}$ baryons for $0 < \pt < 20$ \GeVc is presented. With the current simulated background sample, only a few counts remain after selections, over the full analysed \pt range. A total significance of about 7 is expected for the full \pt range with an integrated luminosity of \SI{18}{\femto\barn}$^{-1}$.

\begin{figure}
    \centering   %
    \includegraphics[width=.86\textwidth]{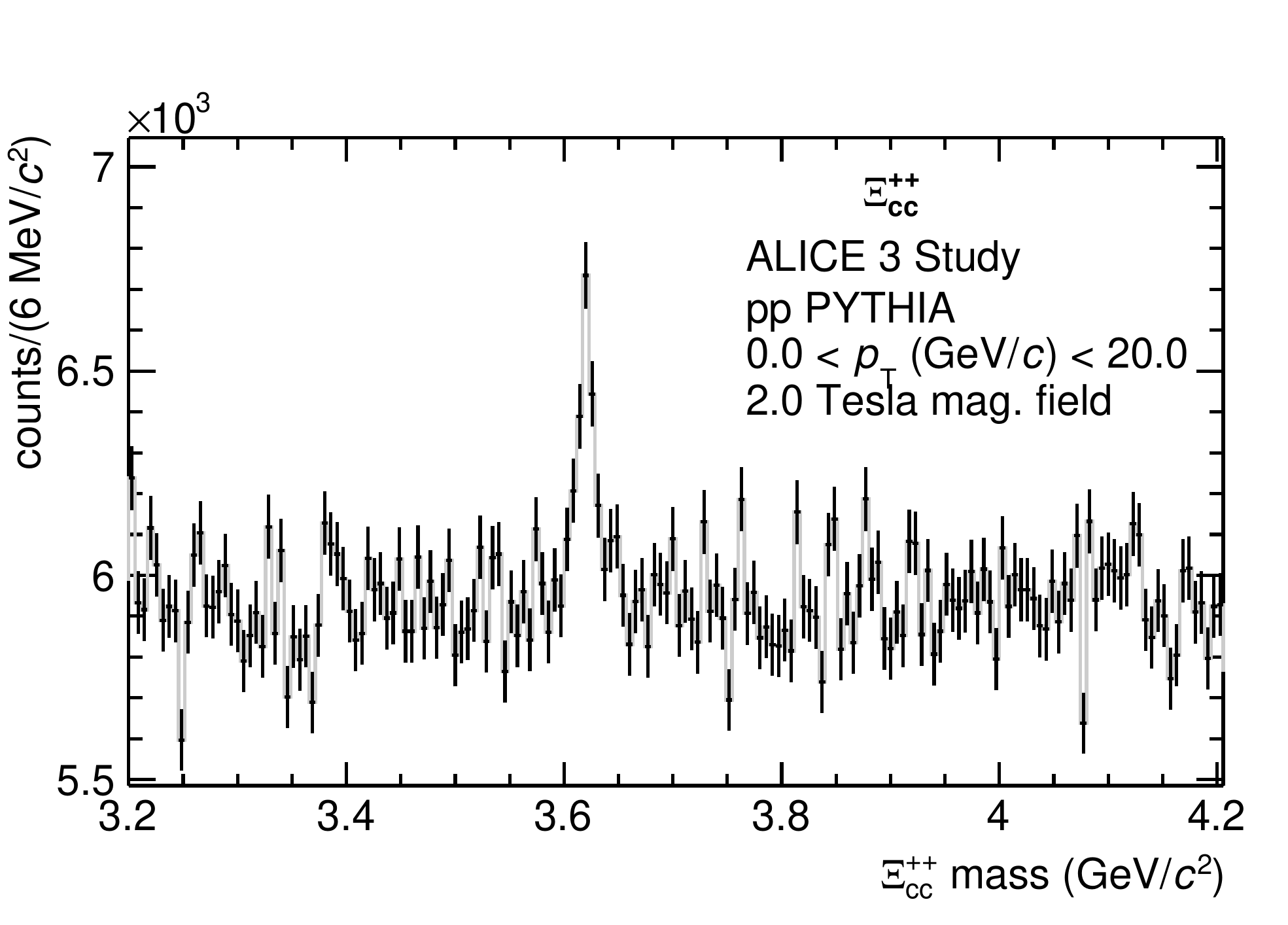}
    \caption[Expected $\Xi^{++}_{cc}$ mass peak and background]{Expected $\Xi^{++}_{cc}$ mass peak and background in \pp collisions with $\Lint=\SI{18}{\femto\barn^{-1}}$.}
     \label{fig:performance:physics:heavy_flav:xi_cc_pp_mass}
\end{figure}

\paragraph*{$\Xi_{cc}^{++} \to \Xi_{c}^{+} \pi^{+} \to p K^{-} \pi^{+} \pi^{+}$ reconstruction}~\\
In order to compare the performance of multicharm baryon reconstruction with strangeness tracking to the case of channels without multi-strange baryons, we have evaluated the performance using the decay channel $\Xi_{cc}^{++} \to \Xi_{c}^{+} \pi^{+}$, with $\Xi_{c}^{+} \to pK^{-}\pi^{+}$. The efficiency and expected significance of the signal in pp collisions are shown in Fig.~\ref{fig:performance:physics:heavy_flav:xi_cc_pp:all}. With the current candidate selection cuts, the reconstruction of $\Xi_{cc}^{++}$ in this channel is more efficient than the $\Xi_{cc}^{++}$ decay channel with strangeness tracking, while the background rejection is larger in the strangeness tracking analysis in pp collisions. In order to estimate the expected significance, the total cross section from~\cite{Berezhnoy:1998aa} and branching ratios from Table~\ref{tab:performance:physics:heavy_flav:multicharm} are used. A significance of around 3 for $\pt > \SI{4}{\giga\eVc}$ is currently projected with $\Lint=  \SI{18}{fb^{-1}}$ of \pp collisions. %

\begin{figure}
    \centering
    \includegraphics[width=.49\textwidth]{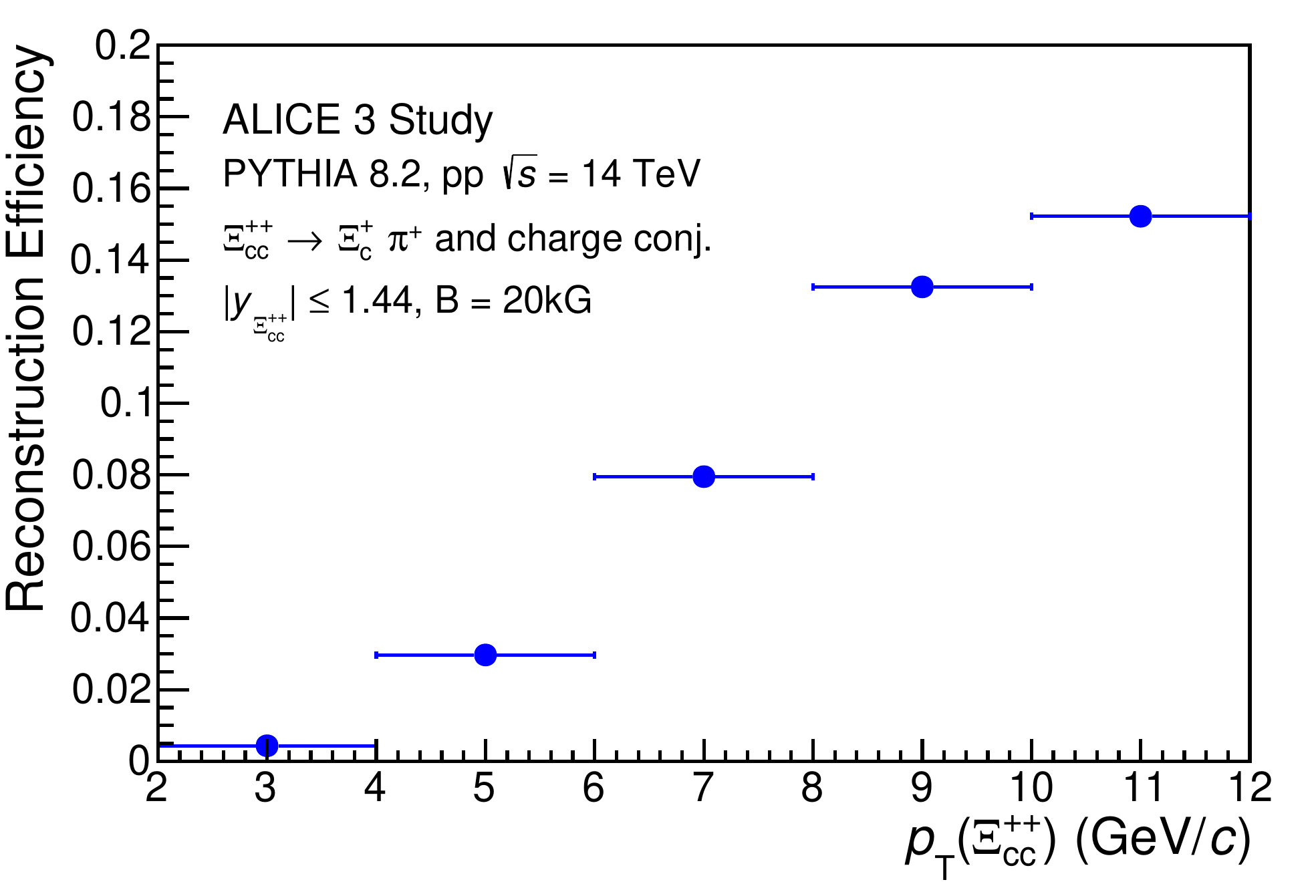}
    \includegraphics[width=.49\textwidth]{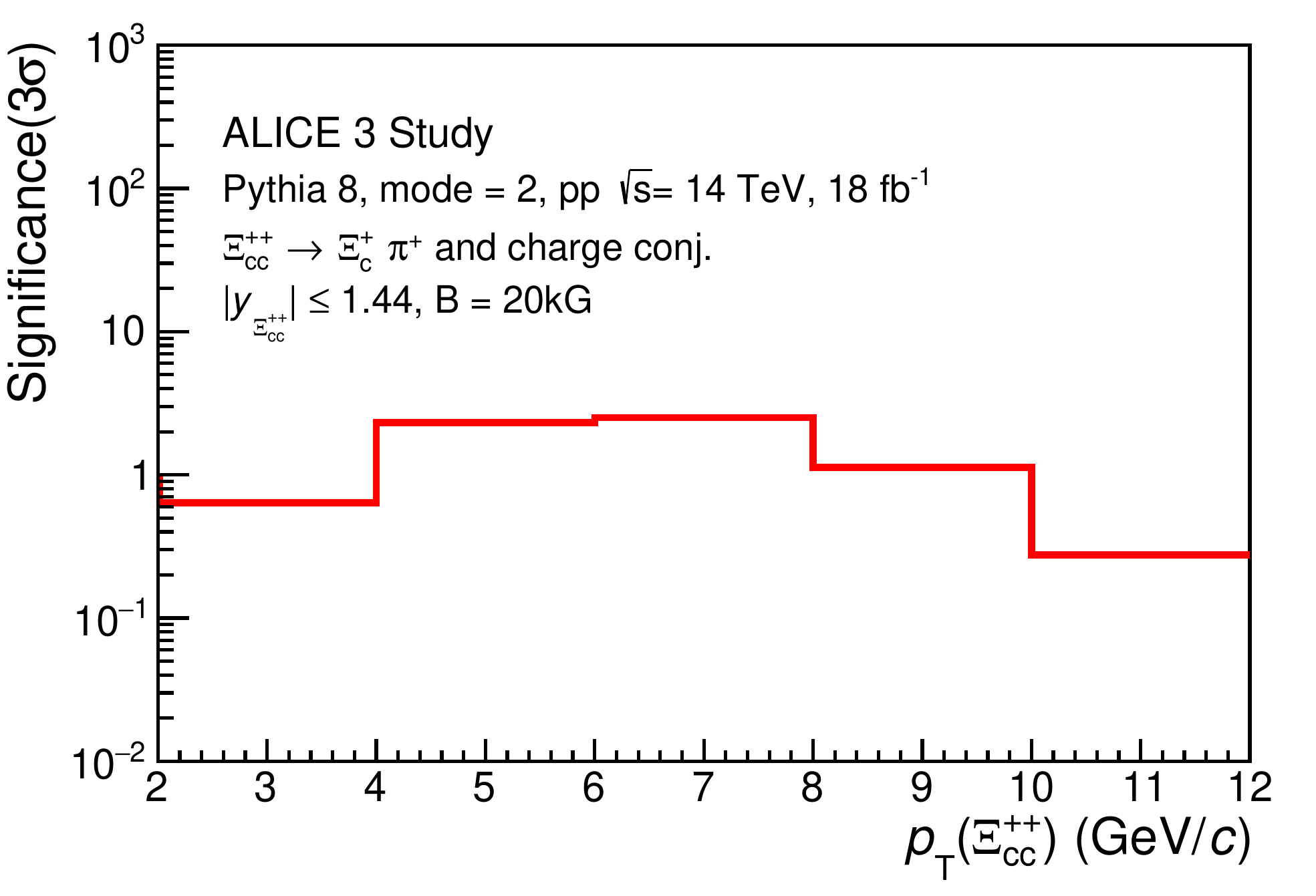}
    \caption[$\Xi^{++}_{cc}$ efficiency]{(Left) $\Xi^{++}_{cc}$ efficiency as a function of \pt after the reconstruction and selection (soft criteria) of candidates. Branching ratios of the decay $\Xi^{++}_{cc}\rightarrow\Xi^{+}_{c}+\pi^{+}$ and $\Xi^{+}_{c}\rightarrow$ p + K + $\pi^{+}$ are not taken into account. (Right) $\Xi^{++}_{cc}$ expected significance as a function of \pt in pp collisions at 14 TeV for a sample of $\Lint=\SI{18}{\femto\barn^{-1}}$ with the TOF information for decay prongs. }
    \label{fig:performance:physics:heavy_flav:xi_cc_pp:all}
\end{figure}

\textit{\textbf{ALICE~3 uniqueness and comparison with ALICE~2.}} The ability of the ALICE~3 apparatus for the detection of multi-charm baryons will be unique compared to other experimental setups at the LHC, especially in nucleus-nucleus collisions. In particular, the ability to reconstruct daughter tracks at low momenta will 
maximize efficiency. High-precision silicon 
pixel tracking layers that are very close to the interaction vertex significantly increase the efficiency to track weakly decaying hadrons prior to their decay 
(``strangeness tracking``) compared to ALICE, even with the ITS3 upgrade. 
For example, $\Omega^{-}$ baryons at 1~GeV/$c$ will 
be tracked with two hits with a 50\% probability, while the probability of that happening with ITS3 is less than 20\%, as shown in Fig.~\ref{fig:competitivity:heavy_flav:multistrangeaccept}. Those $\Xi^{-}$ and $\Omega^{-}$ candidates that are tracked will 
have a significantly improved $\rm{DCA}_{\rm{xy}}$ resolution with ALICE~3, with 
$\sigma(\rm{DCA}_{\rm{xy}})$ being at least 3-4 times better with ALICE~3 than 
with the combination of ITS3 and TPC. %
The combination of the increased strangeness tracking efficiency and the improved impact parameter result in a much improved capability of reconstructing heavy-flavour baryons that decay into multi-strange baryons in ALICE~3 compared to the current ALICE detector. 
\begin{figure}
    \centering
    \includegraphics[width=0.48\textwidth]{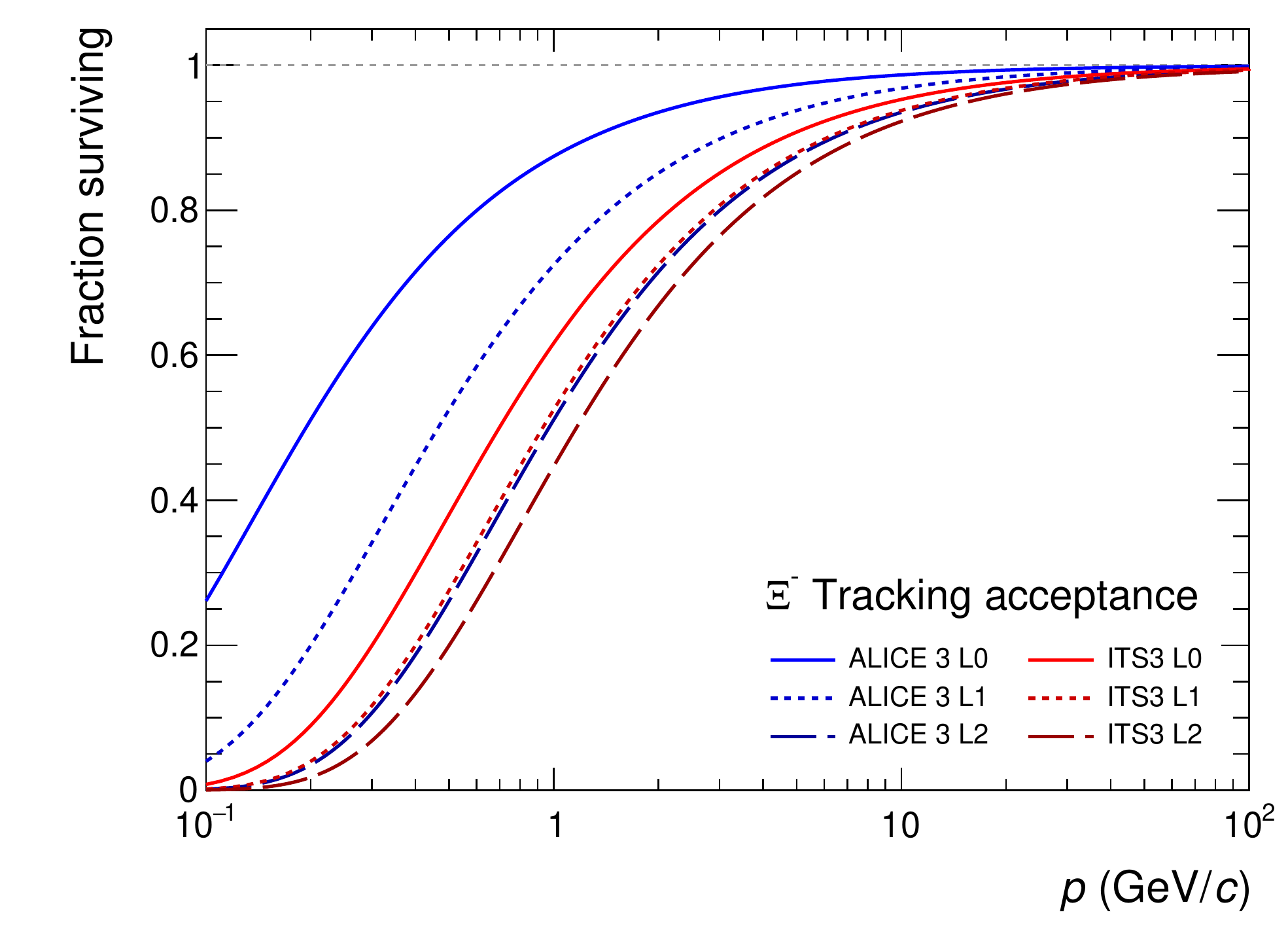}
    \includegraphics[width=0.48\textwidth]{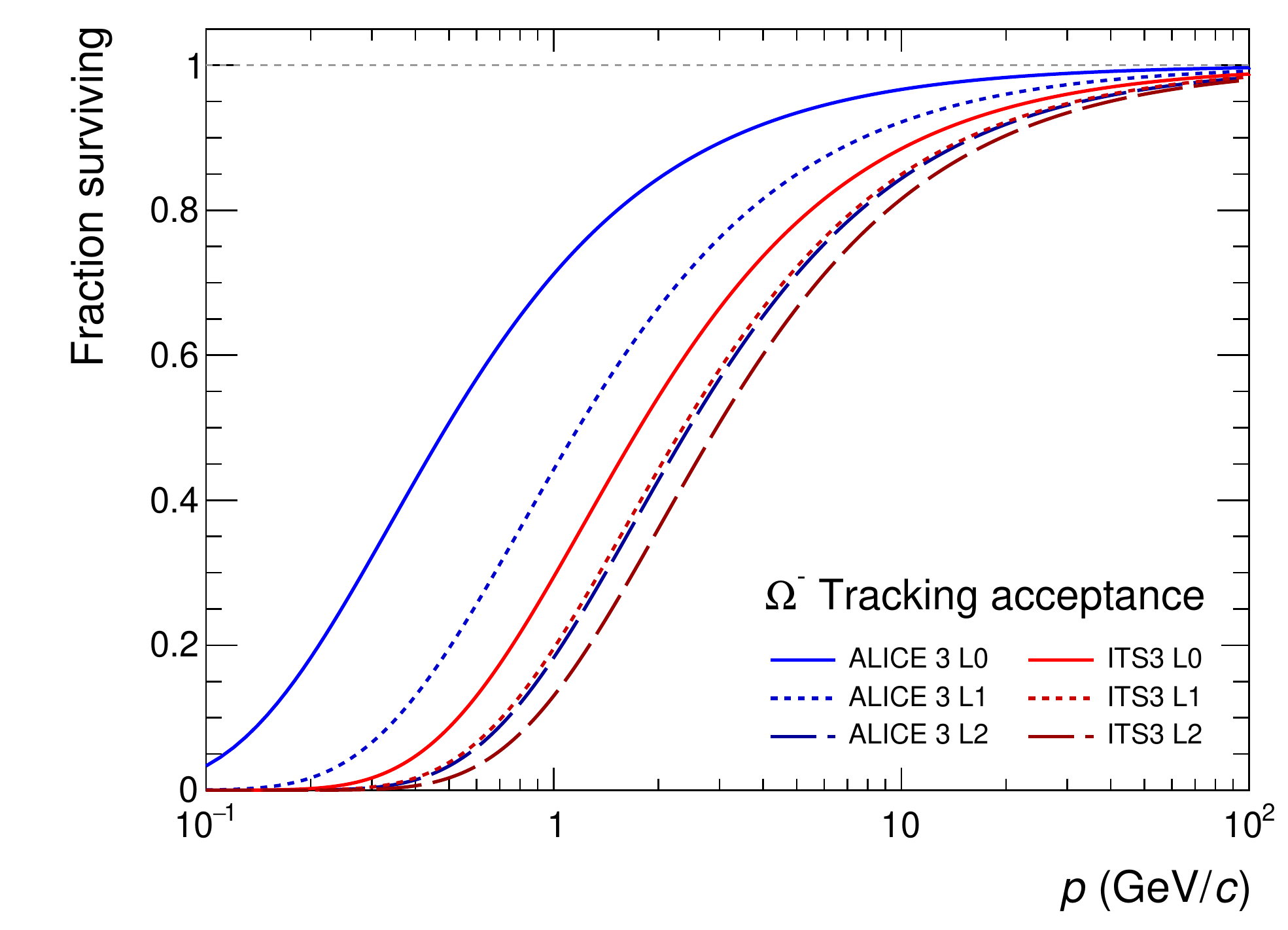}
    \caption[Multi-strange acceptance]{Fraction of $\Xi^{-}$ (left) and $\Omega^{-}$ (right) baryons reaching a specific layer of either ITS3 or the ALICE 3 tracker as a function of momentum.}
    \label{fig:competitivity:heavy_flav:multistrangeaccept}
\end{figure}
Similarly, ALICE~3 would have distinctive advantages with respect to the other LHC 
experiments, which have the first layer of the trackers further away from the interaction point than ALICE~2.
In addition, ALICE~3 is able to select and identify secondary particles with the TOF system, leading to %
a purity for multi-strange baryon 
detection which is close-to-ideal ($>95\%$) even at low transverse momenta. 
This is in contrast to the identification of the decay daughters based on TPC d$E$/d$x$ as used in previous ALICE setups, which achieved purity values of no more than 30-40\%. Since multi-strange baryons are relatively rare, requiring
the presence of non-prompt multi-strange baryons as part of the multi-charm baryon decay chain rejects a very 
large amount of background, with only around 0.5-1.0$\times10^{-4}$ candidates per event left at analysis level even in central \PbPb{} collisions.

The midrapidity coverage and high-multiplicity 
detection capabilities for multi-prong analyses provide sensitivity to charm recombination in the midrapidity region, where such effects 
are most pronounced, a region which is inaccessible to other experiments. 
The large acceptance would also allow for a rapidity scan of the 
multi-charm baryon yield, systematically probing how recombination 
dynamics changes in regions with different charm densities. The performance of ALICE~3 for multi-charm baryons was also be studied in decay channels without multi-strange baryons, for which a competing analysis exists in LHCb in proton-proton collisions. These show that ALICE~3 will be able to reconstruct $\Xi_{cc}^{++}$ into the $\Xi_{c}\pi$ channel with an efficiency that is at least an order of magnitude larger than the one from LHCb.

\paragraph{B-meson production in \pp and \PbPb collisions}~\\
\label{sec:performance:physics:heavy_flav:Bmeson}
Measurements of B-meson production with very high accuracy are key ingredients to constrain the diffusion properties of the QGP with beauty quarks and to provide a reference for studies of beauty hadronisation with $\lb$ baryons, as discussed in Section~\ref{sec:qgp_physics:partonprop}. In Fig.~\ref{fig:perf:heavy_flav:Bplus}, the expected significance as a function of $\pt$ for the hadronic decay channel $\bptodbarpi$, with $\dzerotokpi$, estimated in 0-10\% central \PbPb{} collisions is shown. For this study, only topological selections have been applied to reduce the contamination of background $\bp$ candidates. Even in the absence of PID selection, the $\bp$ production can be measured with very good uncertainties down to $\pt=0$. A substantial increase in the significance is expected at low \pt, when a TOF-based PID selection will be included.

\begin{figure}[htbp]
\centering
\includegraphics[width=0.65\textwidth]{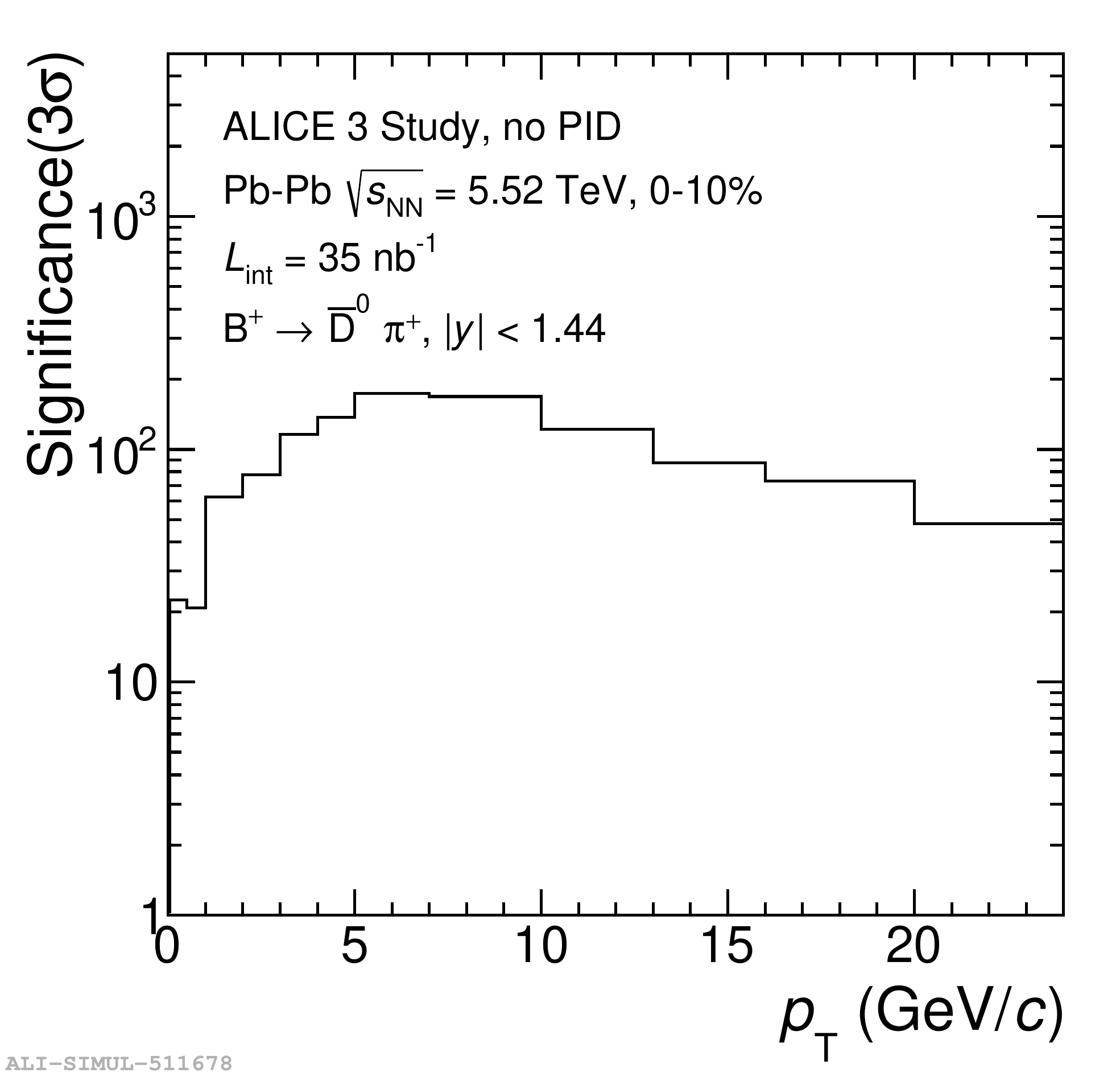}
\caption[Expected \bp significance]{Expected \bp significance as a function of \pt obtained in 0-10\% central \PbPb collisions with $\Lint = \SI{35}{\nano\barn^{-1}}$. The current result does not include PID selection.}
\label{fig:perf:heavy_flav:BplusPbPb}
\end{figure}

\begin{figure}[htbp]
\centering
 \includegraphics[width=0.65\textwidth]{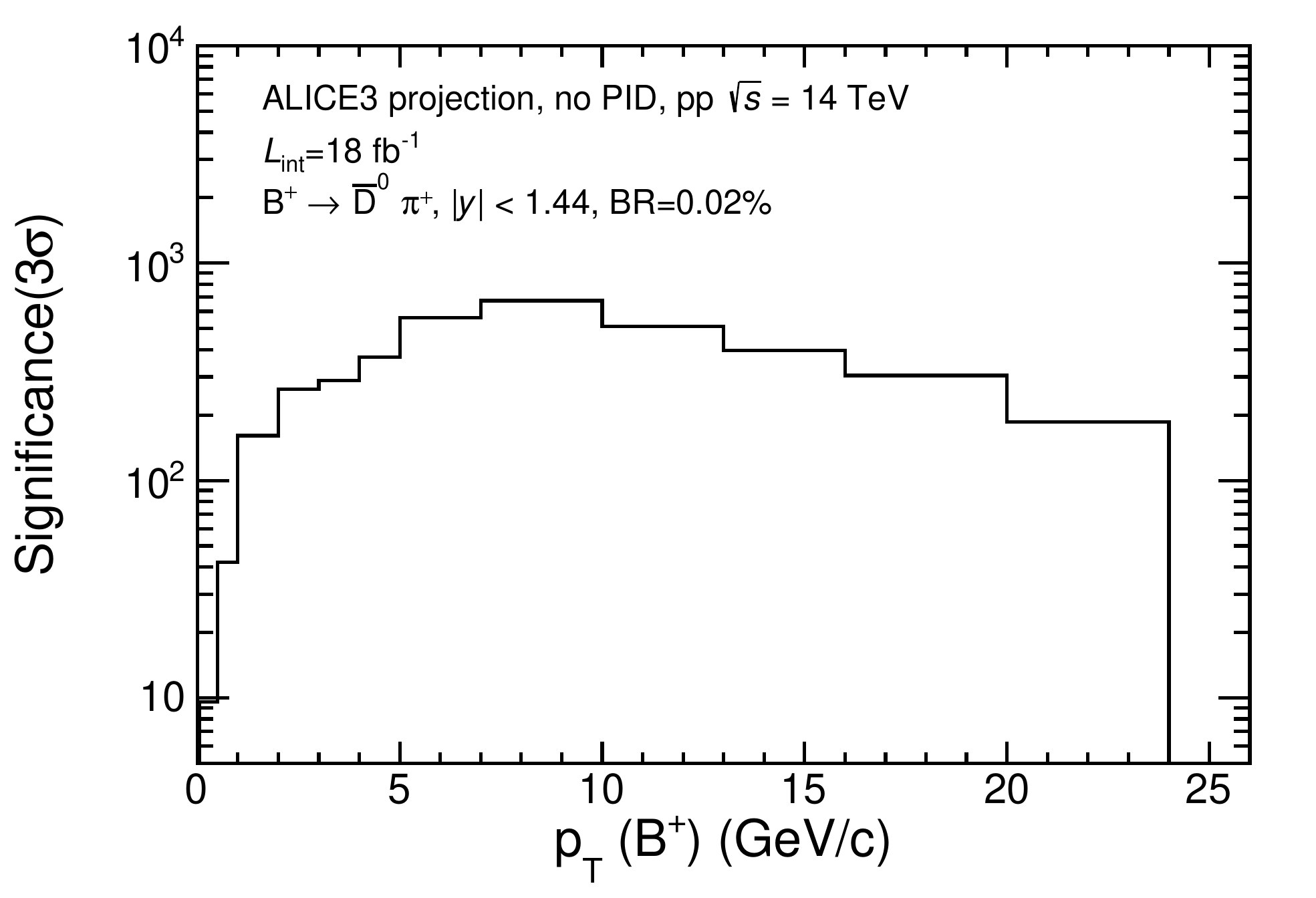}
\caption[Expected \bp significance]{Expected \bp significance a function of \pt obtained in pp collisions at 14 TeV for a sample of $\Lint=\SI{18}{\femto\barn^{-1}}$. The current result does not include PID selection.}
\label{fig:perf:heavy_flav:Bplus}
\end{figure}

Similar studies have also been conducted in pp collisions at 14 TeV, 
with the
corresponding significance as a function of transverse momentum being shown in 
Fig.~\ref{fig:perf:heavy_flav:BplusPbPb}. Results indicate very high precision, 
even without particle identification techniques, pointing to the uniqueness
of ALICE~3 with coping with very high-multiplicity nucleus-nucleus collisions. 

\paragraph{Elliptic flow of $\lc$ and $\Lambda_{\mathrm{b}}$ baryons in \PbPb collisions}~\\
\label{sec:performance:physics:heavy_flav:baryonflow}
\begin{figure}[htbp]
\centering
\includegraphics[width=0.65\textwidth]{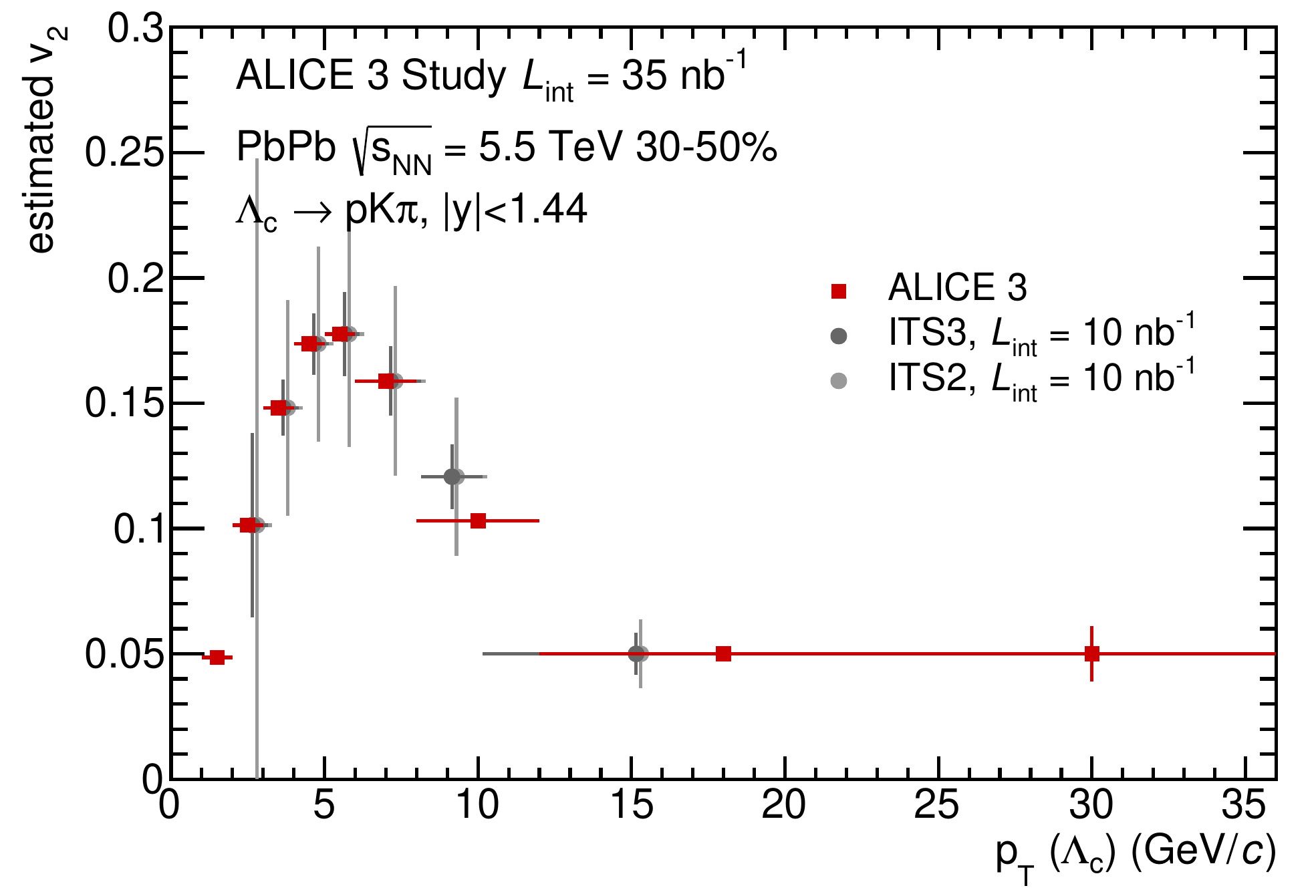}
\caption[Elliptic flow of $\lc$]{Elliptic flow $v_2$ of $\Lambda_c$ with the expected uncertainties for 30-50\% central \PbPb collisions with $\Lint = \SI{35}{\nano\barn^{-1}}$, compared to the performance expected in Run~3 and 4 with the ITS2 and ITS3, respectively.}
\label{fig:perf:heavy_flav:Lcflow}
\end{figure}
Measurements of elliptic flow of heavy-flavour baryons are critical to constrain with high accuracy the diffusion coefficient $\Dsc$ and the mechanisms of heavy-quark hadronisation in the QGP, as discussed in Section~\ref{sec:qgp_physics:partonprop}. In this section, we present the ALICE~3 performance for the $\vtwo$ measurement of the $\lc$ and $\lb$ baryons down to low \pt{}, where the sensitivity to medium interactions and hadronisation effects is largest. In Fig.~\ref{fig:perf:heavy_flav:Lcflow}, we present the expected ALICE~3 performance (red markers) for the measurement of the elliptic flow of the $\lc$ baryon in semi-central (30-50\% centrality) \PbPb{} events. The central values for the $\vtwo$ projection are based on the measured $v_2$ of p, K, $\pi$, $\Lambda$, and D mesons, corrected to account for the expected mass dependence of the $v_{2}$ parameter.  The statistical significance of the measurement was estimated following the method described in~\cite{ALICE:ITS2:2014}, starting from the accuracy expected for the measured $\lc$ yields as a function of \pt, for an integrated \PbPb luminosity $\Lint = \SI{35}{\nano\barn^{-1}}$. %
The uncertainty on the measured elliptic flow parameter $v_2$ shown in Fig.~\ref{fig:perf:heavy_flav:Lcflow} has been estimated as the inverse of the statistical significance, in each \pt interval. A similar study was performed also for the $\lb$ baryon, reconstructed in the decay channel $\Lambda_{\mathrm{b}}^{0} \rightarrow \Lambda_{\mathrm{c}}^{+} + \pi^{-} \rightarrow p + K^- + \pi^{+} + \pi^{-}$. Fig.~\ref{fig:perf:heavy_flav:Lbflow} presents the expected performance (red markers) for the measurement of $\lb$ $v_2$ as a function of \pt, for the 30-50\% centrality class of Pb--Pb collisions and $\Lint = \SI{35}{\nano\barn^{-1}}$. ALICE~3 will allow, for the first time, to measure the $\vtwo$ coefficients of both $\lc$ and $\lb$ baryons with high accuracy down to \pt close to 0.
\begin{figure}[htbp]
\centering
\includegraphics[width=0.65\textwidth]{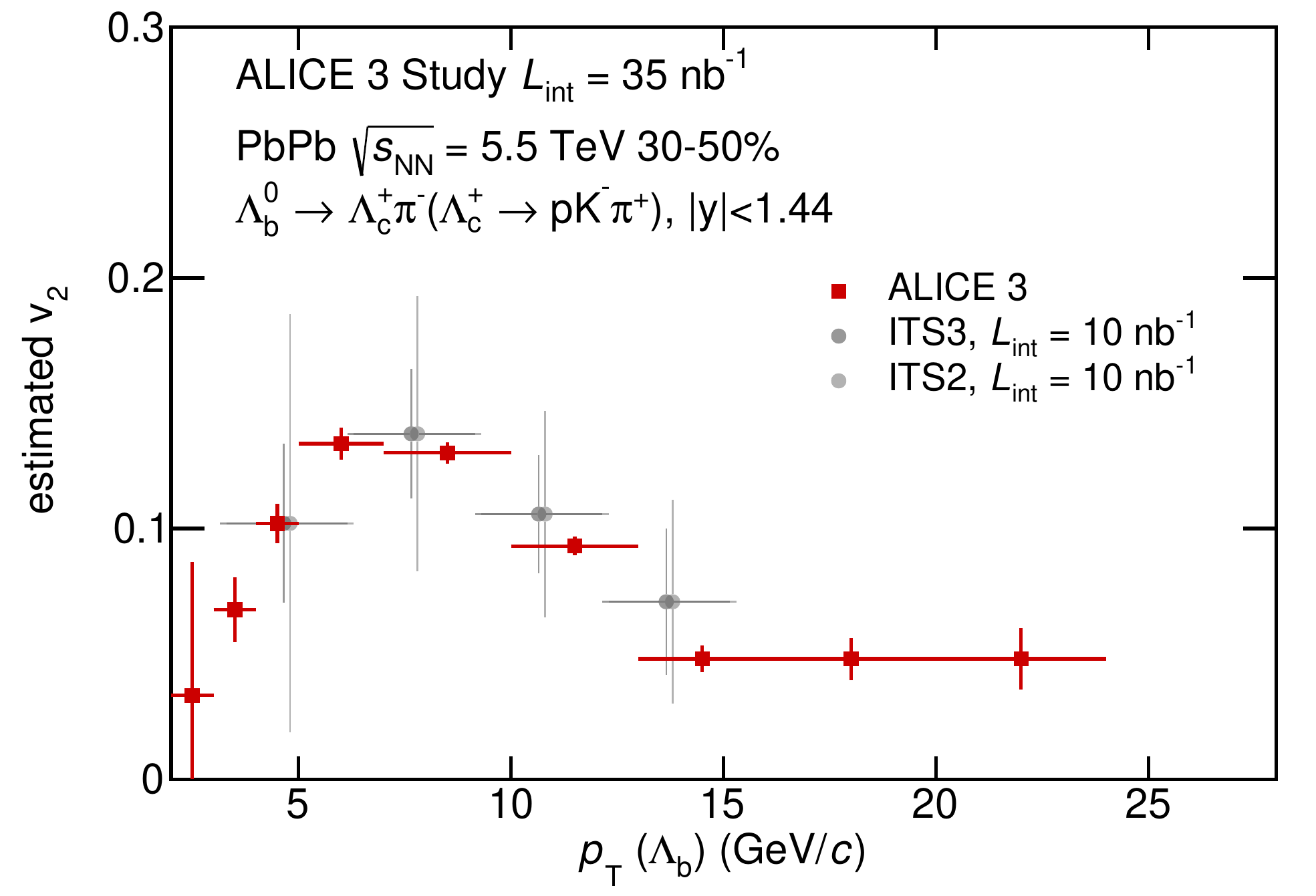}
\caption[Elliptic flow of $\Lambda_{\mathrm{b}}^{0}$]{
Estimated \pt~differential $v_2$ of $\Lambda_{\mathrm{b}}^{0}$ with the expected uncertainties for 30-50\% central \PbPb collisions with $\Lint = \SI{35}{\nano\barn^{-1}}$, compared to the performance expected in Run 3 and 4 with the ITS2 and ITS3.}
\label{fig:perf:heavy_flav:Lbflow}
\end{figure}
\\ \\
\textit{\textbf{Comparison with ALICE~2 and CMS.}}
The uniqueness of ALICE~3 in the measurement of decays at short distances from the primary vertex, such as the $\lc$ ($\rm c\tau \approx$ 60~$\rm \mu m$), is primarily thanks to its unprecedented tracking performance. ALICE~3 largely outperforms ALICE~2, even when equipped with the high-resolution ITS3 pixel detector. For tracks with $\pt=0.7$~\GeVc and $\eta=0$, the pointing resolution ($\rm \sigma_{DCA}$) of ALICE~3 is about 4 $\rm \mu m$, compared to about 30 and 50 $\rm \mu m$ of ALICE~2 with ITS3 and ITS2, respectively. The CMS detector will reach only $\rm \sigma_{DCA} \approx $~\SIrange{50}{60}{\micro\meter} for tracks in the same kinematic range, even with the new high-resolution tracker developed for Run~4~\cite{CMSTrackerTDRPhase2}. In Figs.~\ref{fig:perf:heavy_flav:Lcflow} and~\ref{fig:perf:heavy_flav:Lbflow}, the expected ALICE~3 performance for the measurement of the $\lc$ and $\lb$ elliptic flow coefficient ($\vtwo$) are compared to those predicted for ALICE~2 in Run~3 (ITS 2) and ALICE~2 in Run~4 (ITS 3). %
A significant improvement of the uncertainties is achieved with ALICE 3 for the charm and the beauty baryons. In particular, for the measurement of the $\lb$ $\vtwo$, the relative uncertainty with ALICE~2 will still be larger than 20\%. Such accuracy will not allow placing significant constraints on hadronisation models. Based on projections presented in~\cite{CMSMTDTDR}, the upgraded CMS detector will measure  the $\lc$ $\vtwo$ down to $\pt=3.5$~\GeVc{} with Run~3 and Run~4 data. In the lowest \pt interval accessible, 3.5$<\pt<$4.0~$\GeVc$, the statistical uncertainty is about 15$\%$. In the same kinematic region, the expected accuracy with ALICE~3 is expected to be well below 1$\%$.

\paragraph{Beauty baryons: $\Xi_{b}^{-}$ and $\Omega_{b}^{-}$}~\\
\label{sec:performance:physics:heavy_flav:beautybaryons}

The expected performance for yield measurements of the beauty baryons $\Xi_{b}^{-}$ and $\Omega_{b}^{-}$ in ALICE~3 have been studied in the following decay channels: 
\begin{equation}
\Xi_{b}^{-} \rightarrow \Xi_{c}^{0}+\pi^{-} \rightarrow \Xi^{-}+\pi^{+}+\pi^{-}
\end{equation}
\begin{equation}
\Omega_{b}^{-} \rightarrow \Omega_{c}^{0}+\pi^{-} \rightarrow \Omega^{-}+\pi^{+}+\pi^{-}    
\end{equation}

In what follows, the $\Xi_{b}^{-}$ and $\Omega_{b}^{-}$ baryons
were assumed to have masses of 5.797~GeV/$c^{2}$ and 6.046~GeV/$c^{2}$, respectively,
as measured by the LHCb collaboration~\cite{LHCb:2019sxa, LHCb:2016coe}, and all unknown branching ratios
are assumed to be 5\%. In simulations, production rates were assumed to follow the predictions of the statistical 
hadronisation model with the inclusion of beauty by A.~Andronic et al~\cite{AndronicQM22}.
Due to the specific layout of the ALICE~3 tracker, and in a similar manner as
for multi-charm baryon decay products (see Section~\ref{sec:performance:physics:heavy_flav:multicharm}), 
the multi-strange baryons from the decay chains above can be readily tracked in the 
innermost layers of the detector. Furthermore, the comparatively
large lifetime of beauty baryons ($c\tau \approx\SI{500}\mu m$) leads to 
background discrimination that is even more effective than for multi-charm baryons. 
Employing the same simulation and analysis strategy as for the case of multi-charm baryons, 
the expected significance in 0-10\% central Pb--Pb collisions in ALICE~3 is the one shown in Fig.~\ref{fig:performance:physics:heavy_flav:bbPbPb:significance}. 
Yields in the lowest transverse momentum interval considered, from 0-2~GeV/$c$, 
will be measurable with a significance of at least $\approx 10$ for the $\Xi_{b}^{-}$ 
and $\approx 20$ for the $\Omega_{b}^{-}$. It is important
to emphasise that the low-$p_{\rm{T}}$ reach is
is particularly relevant to get a complete picture of beauty production, which, 
in turn, is necessary to assess thermalisation in the beauty sector.
As in the multi-charm baryon 
case, further quantities of interest, such as the beauty 
baryon elliptic flow coefficients, will be accessible
as well and will also help to address the question of beauty equilibration in the medium. 

\begin{figure}
    \centering
    \includegraphics[width=.79\textwidth]{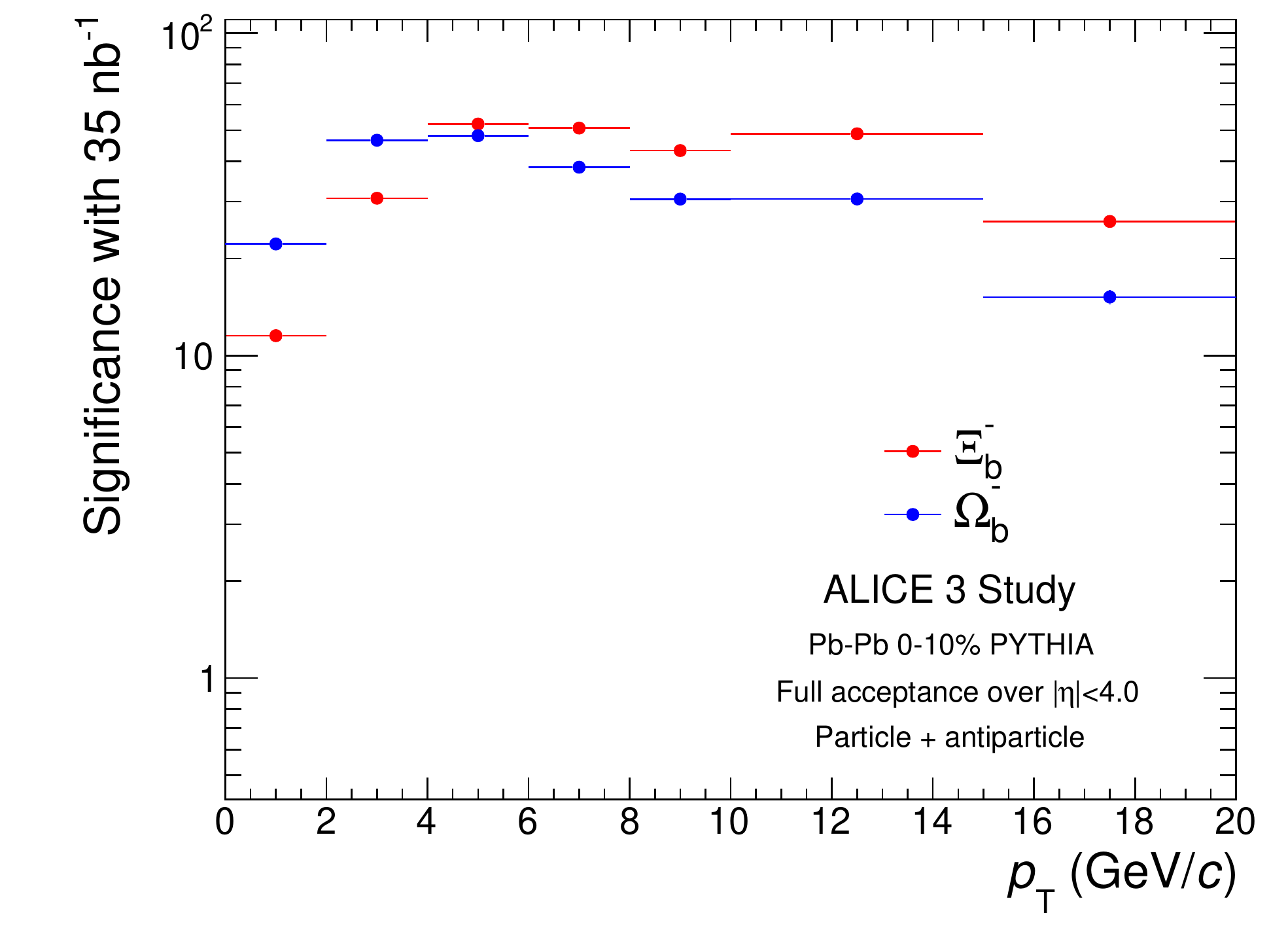}
    \caption[Significance for $\Xi^{-}_{b}$ and $\Omega_{b}^{-}$]{$\Xi^{-}_{b}$ and $\Omega_{b}^{-}$ significance in 0-10\% central Pb--Pb collisions at $\sqrt{s_{\rm{NN}}}$~=~5.52~TeV as a function of \pt with a 2.0~T magnetic field.}
    \label{fig:performance:physics:heavy_flav:bbPbPb:significance}
\end{figure}

\paragraph{$\rm D\bar{D}$ azimuthal correlations}~\\
\label{sec:performance:ddbar}

\begin{figure}
    \centering
    \includegraphics[width=0.48\textwidth]{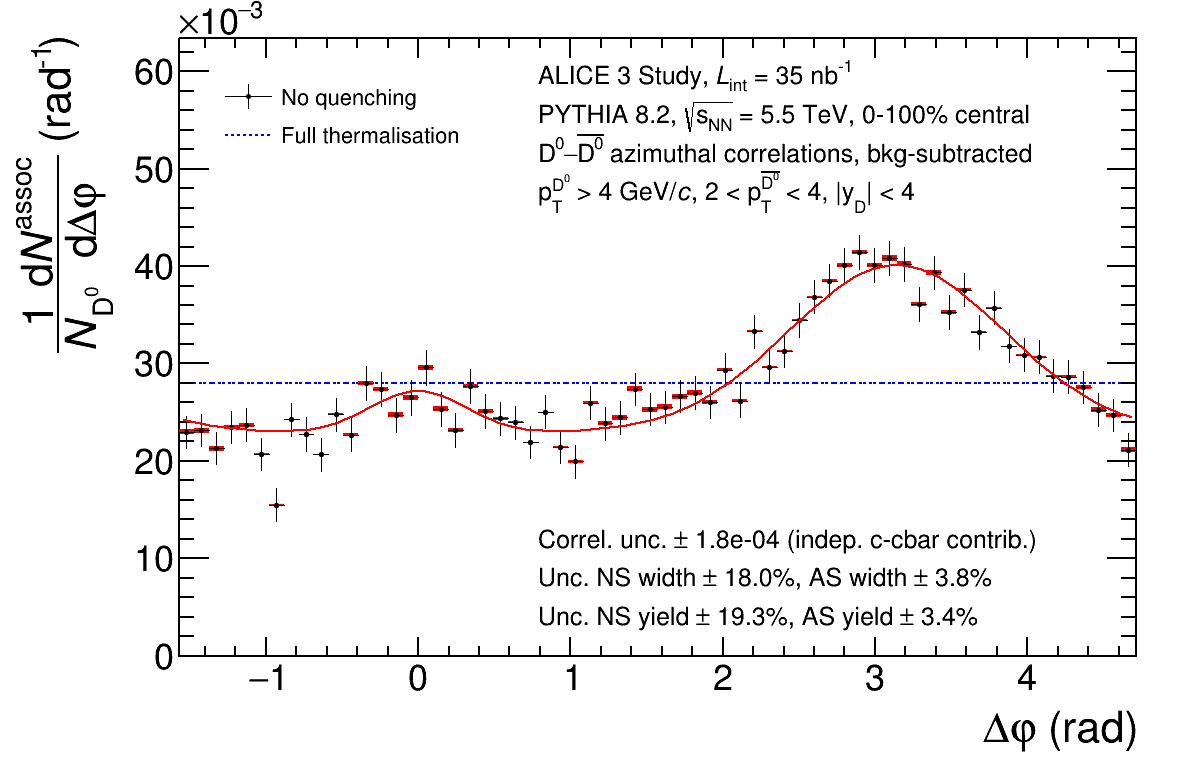}
    \includegraphics[width=0.48\textwidth]{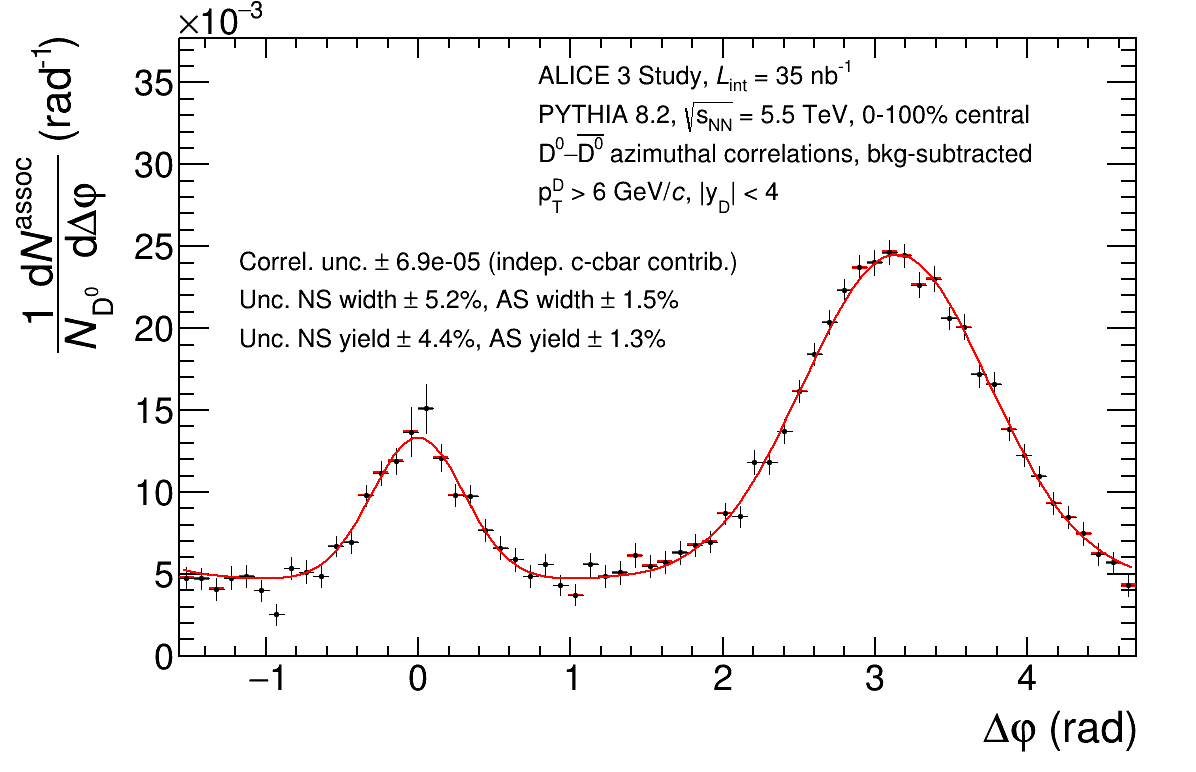}
     \caption[Azimuthal distribution of \DDbar{} pairs]{Azimuthal distribution of \DDbar{} pairs with ${\pt}_{1}>\SI{4}{\giga\eVc}$, $\SI{2}<{\pt}_{2}<\SI{4}{\giga\eVc}$ (left panel) and $\pt>\SI{6}{\giga\eVc}$ (right panel) and $|y| < 4$ in minimum bias \PbPb collisions. The combinatorial background of \DDbar~not coming from the same hard scattering has been subtracted. The uncertainties shown are for a total luminosity of \SI{35}{\nano\barn^{-1}}.}
    \label{fig:perf:heavy_flav:DDbar}
\end{figure}

Azimuthal correlations of \dzerodzerobar{} pairs in \PbPb collisions provide a direct measure of momentum broadening by the QGP, which is sensitive to the nature of the energy loss mechanisms and to the degree of charm thermalisation in the medium, as discussed in Section~\ref{sec:physics:hfcorrelations}. Projections for measurements of the azimuthal distributions of \dzerodzerobar{} pairs  in minimum bias \PbPb collisions are presented  Fig.~\ref{fig:perf:heavy_flav:DDbar}, for pairs with ${\pt}_{1}>\SI{4}{\giga\eVc}$, $\SI{2}<{\pt}_{2}<\SI{4}{\giga\eVc}$ (left) and ${\pt}_{1-2}>\SI{6}{\giga\eVc}$ (right). The statistical uncertainties are estimated for a \PbPb luminosity of \SI{35}{\nano\barn^{-1}}. Two types of combinatorial background were taken into account in the determination of the statistical precision, and subtracted to produce the distribution in Fig.~\ref{fig:perf:heavy_flav:DDbar}. D mesons coming from independent hard scatterings form a sizable background that was calculated by combining $N_{coll}=382$ Pythia pp events. The second source of background are  $K\pi$ pairs that are not coming from D meson decays. For this measurement, an excellent signal purity for heavy flavour signals as shown in Fig.~\ref{fig:performance:heavy_flav:D0} is a key feature of the ALICE~3 setup to keep this source of background under control. The relative statistical uncertainties on the azimuthal distribution range from 2\% to about 10\%. By comparing the amplitudes and widths of measured distribution in \PbPb{} collision with a reference measured in pp collisions, the effects of transport broadening and thermalisation can be quantitatively assessed.

\begin{figure}
    \centering
    \includegraphics[width=0.6\textwidth]{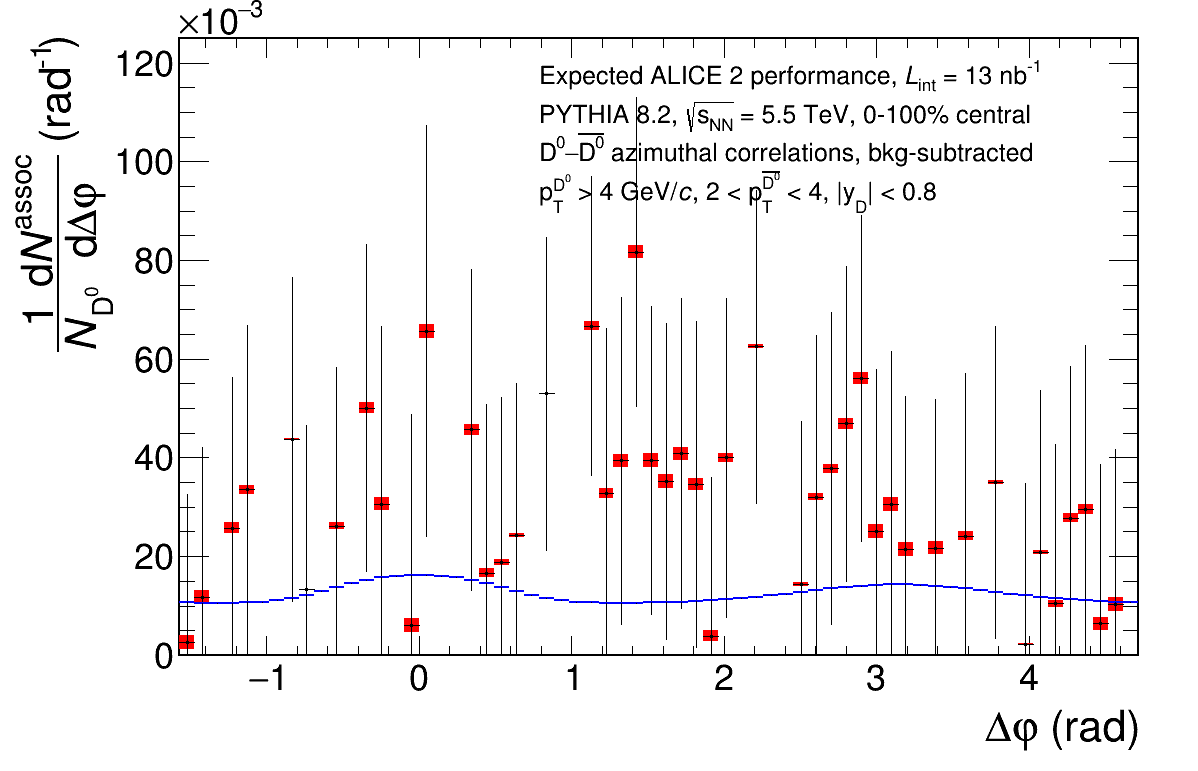}
    \caption[Azimuthal distribution of \dzerodzerobar{} pairs]{Azimuthal distribution of \dzerodzerobar{} pairs with ${\pt}_{1}>\SI{4}{\GeVc}$, $\SI{2}<{\pt}_{2}<\SI{4}{\GeVc}$ in minimum bias \PbPb collisions expected with ALICE~2 in $|y|<$0.8. The combinatorial background of \dzerodzerobar~not coming from the same hard scattering has been subtracted.}
    \label{fig:competitivity:heavy_flav:DDbar}
\end{figure}

\textit{\textbf{Uniqueness and comparison with ALICE 2 and CMS.}}~The large purity and efficiency of ALICE~3 in the reconstruction and selection of $\dzerotokpi$ decays are critical for HF-correlation measurements, and unique among current and future LHC experiments. In Fig.~\ref{fig:competitivity:heavy_flav:DDbar}, the projections for the azimuthal distribution of \dzerodzerobar{} pairs with ${\pt}_{1}>\SI{4}{\GeVc}$ and $\SI{2}<{\pt}_{2}<\SI{4}{\GeVc}$ in minimum bias \PbPb with ALICE~2 is presented and can be compared to Fig.~\ref{fig:perf:heavy_flav:DDbar} above. The measurement of $\dzerodzerobar$ correlations down to low \pt, which is critical to observe the effect of charm isotropization, will not be feasible with ALICE~2 as a consequence of the lower purity and signal efficiency and the narrower detector acceptance. 
\\
{\revised The expected S/B and efficiency for D meson reconstruction with the CMS detector can be estimated based on Run 2 performance~\cite{CMS:2017qjw} and the improvements expected with the MTD~\cite{CMSMTDTDR}. Based on this information, the overall performance for D meson reconstruction is expected to be slightly worse than for ALICE~2, with however, a larger rapidity coverage. %
According to our studies, the 10--50 times larger S/B of ALICE~3 is critical to obtain a precise measurement of \DDbar{} azimuthal correlations.}

\paragraph{$\rm D^0 D^{*+}$ momentum correlations} ~\\
\label{sec:performance:physics:ddstar}
\begin{figure}[hbt]
\centering
\includegraphics[width=0.50\textwidth]{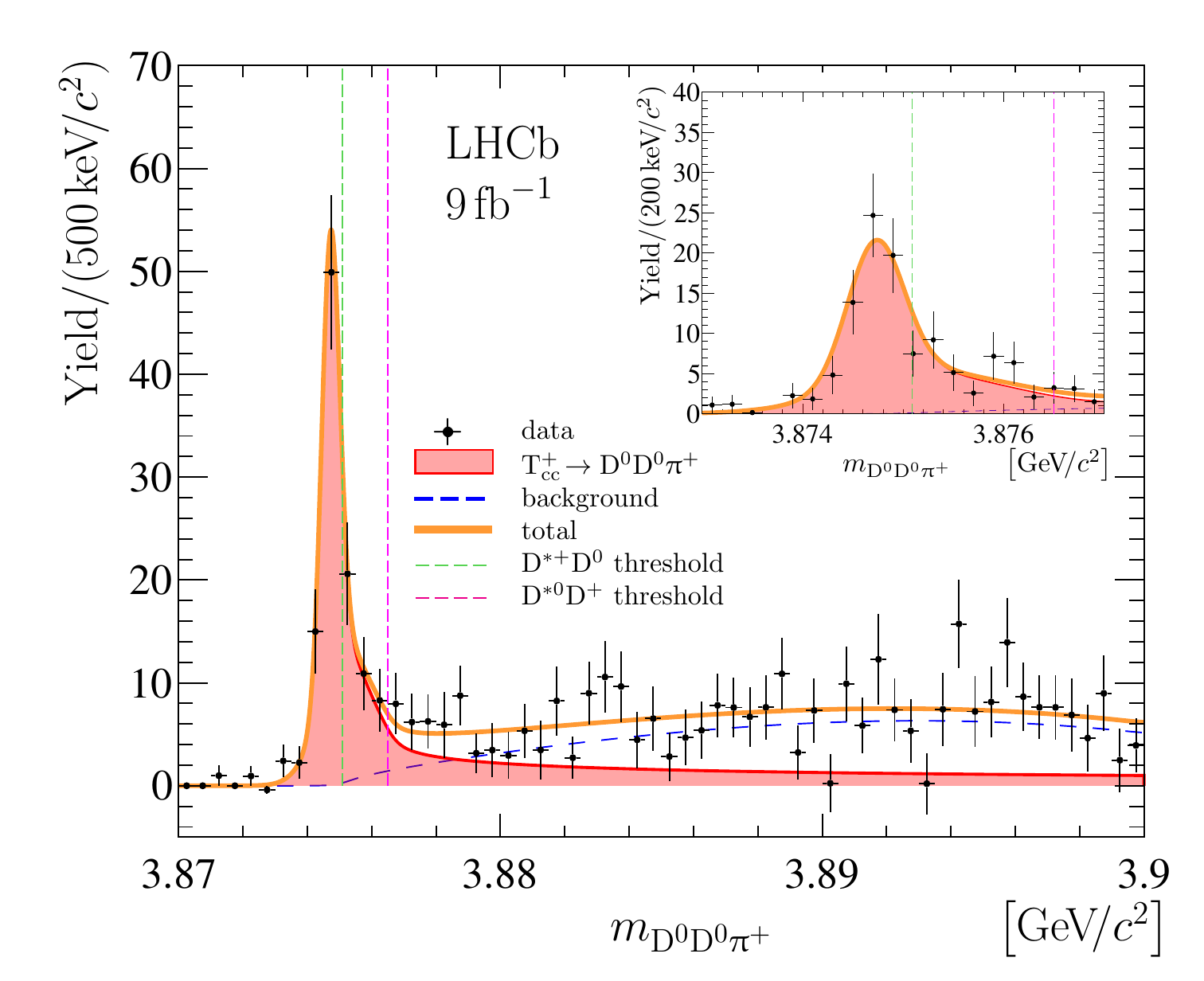}
\includegraphics[width=0.45\textwidth]{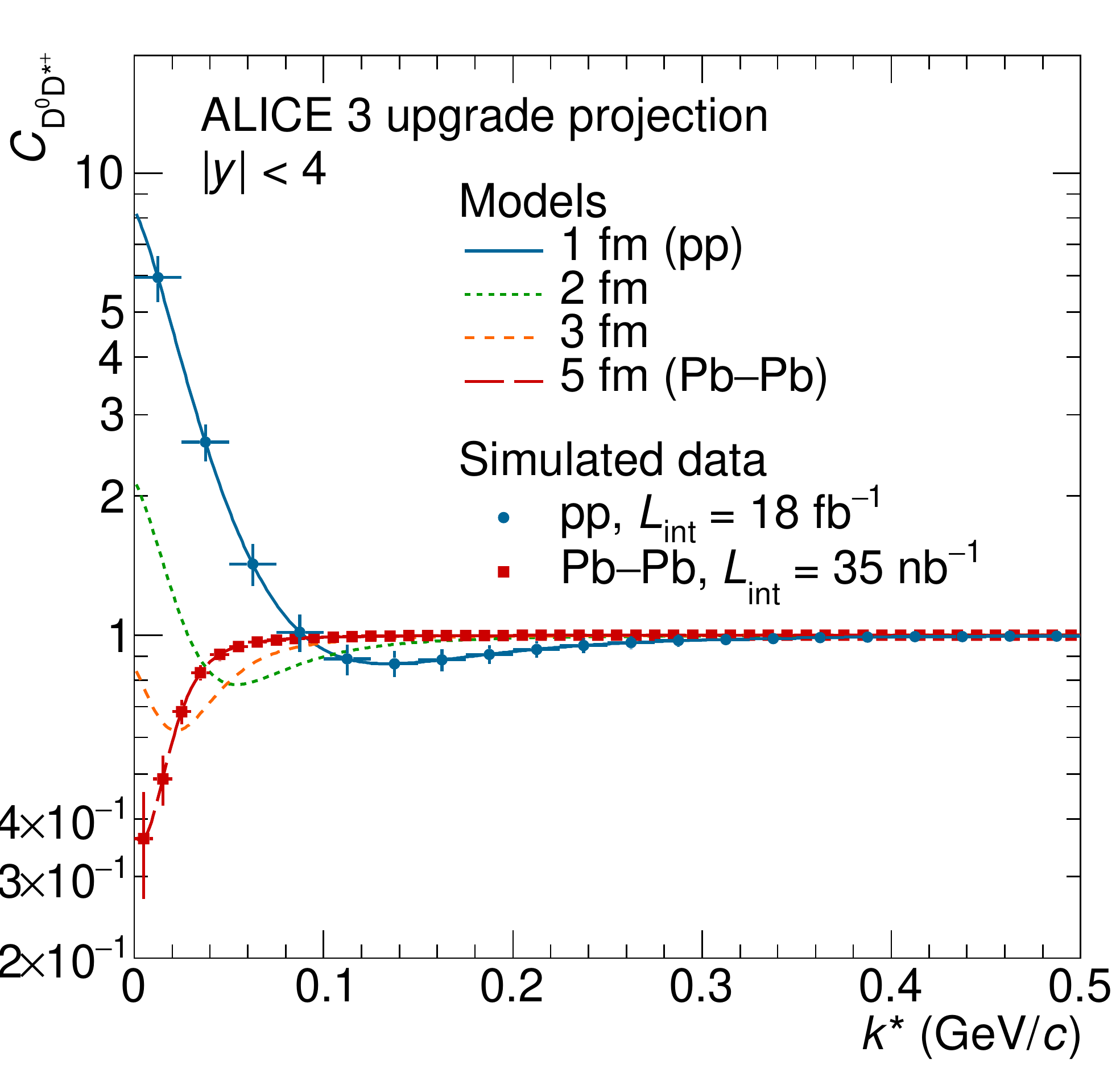}
\caption[$\mathrm{D^0}\mathrm{D^0}\pi^+$ reconstruction and $\mathrm{D^0}\mathrm{D^{*+}}$ correlation]{Left panel: $\mathrm{D^0}\mathrm{D^0}\pi^+$ invariant mass distribution measured by LHCb~\cite{LHCb:2021auc}. The $\mathrm{D^0}\mathrm{D^{*+}}$ and $\mathrm{D^+}\mathrm{D^{*0}}$ mass-threshold are indicated by the dashed green and magenta lines, respectively. Right panel: predicted $\mathrm{D^0}\mathrm{D^{*+}}$ correlation as a function of the relative momentum $k^*$ evaluated for four different source sizes. The expected statistical precision on the measurement of the $\mathrm{D^0D^{*+}}$ correlation function $C_\mathrm{D^0D^{*+}}$ in pp collisions at $\sqrt{s}=14~\mathrm{TeV}$ has been evaluated for an integrated luminosity of $\Lint=18~\mathrm{fb^{-1}}$ assuming an emitting source radius of about $r=1~\mathrm{fm}$. In the 10\% most central Pb--Pb collisions at $\sqrt{s_\mathrm{NN}}=5.5~\mathrm{TeV}$ an integrated luminosity of $\Lint=\SI{35}{\nano\barn^{-1}}$ assuming an emitting source radius of about $r=5~\mathrm{fm}$ has been considered. The predicted $C_\mathrm{D^0D^{*+}}$ for different emitting source radii are also shown for comparison.}
\label{fig:perf:DDstar}
\end{figure}

In this section we present an example analysis of femtoscopy measurements involving charmed hadrons with the scope of investigating the molecular nature of exotic states.
As a proof of principle, we discuss here the case of the $\mathrm{T^+_\mathrm{cc}}$ state newly observed by the LHCb collaboration in the $\mathrm{D^0}\mathrm{D^0}\pi^+$ spectrum with a mass of about \SI{3875}{\mega\eVcsq} and a width of \SI{400}{\kilo\eVcsq}~\cite{LHCb:2021vvq}. 
The left panel of Fig.~\ref{fig:perf:DDstar} shows the measured $\mathrm{D^0}\mathrm{D^0}\pi^+$ invariant mass distribution together with the $\mathrm{D^0}\mathrm{D^{*+}}$ and $\mathrm{D^+}\mathrm{D^{*0}}$ mass-thresholds.
Considering the  $\mathrm{D^0}\mathrm{D^{*+}}$ pair as possible constituents of a hadronic molecule with the same properties as $\mathrm{T_\mathrm{cc}}^+$, a binding energy of $360$ keV would be evaluated and this corresponds to a scattering length equal to $\mathrm{a_0} = -7.16 +i 1.85$ fm for the DD pairs~\cite{Kamiya:2022thy}. This was obtained via a study of the $\mathrm{T_\mathrm{cc}}^+$ lineshape, as described in Ref.~\cite{LHCb:2021auc}. Such a large value of the scattering length (the negative sign implies the presence of the bound state) would manifest itself clearly in the inversion of strength of the correlation function with respect to unity as a function of the system size $R$. Considering a single-channel Gaussian potential for the $\mathrm{D^0}\mathrm{D^{*+}}$ tuned to reproduce the mass and width of the sub-threshold molecular state, we have computed the $\mathrm{D^0}\mathrm{D^{*+}}$ correlation functions expected for four different values of the system size $R$~\cite{Kamiya:2022thy}. The results are shown in the right panel of Fig.~\ref{fig:perf:DDstar}.
The presence of the bound state manifests itself with an inversion of the correlation shape when modifying the radius $R$ from 1 fm to 5 fm. The interplay between the real part of the scattering length $a_0$ and the size of the source $R$, brings the correlation below unity for sufficiently large sources.

The expected performance for the measurement of $\mathrm{D^0D^{*+}}$ correlation function ($C_\mathrm{D^0D^{*+}}$) was computed by simulating pp collisions at $\sqrt{s}=14~\mathrm{TeV}$ with the PYTHIA~8 event generator. The $\mathrm{D}^{*+}$ and $\mathrm{D}^{0}$ mesons were selected in the rapidity interval $|y| < 4$ via the decay channels $\mathrm{D^{*+}}\rightarrow\mathrm{D^0}\pi^+$ and $\mathrm{D^0}\rightarrow\mathrm{K^-}\pi^+$, having branching ratios $(66.7\pm0.5)\%$ and $(3.951\pm0.031)\%$~\cite{Zyla:2020zbs}, respectively. $\mathrm{D}^{0}$ mesons coming from $\mathrm{D}^{*+}$ decays were rejected by offline selections on the decay topology. The reconstruction and selection efficiencies, as well as the signal-to-background ratios, were evaluated using the Fast Simulation tool described in Sec.~\ref{sec:performance:introduction}. For each selected pair of $\mathrm{D}^{*+}$ and $\mathrm{D}^{0}$ mesons, the relative momentum $k^*\,=\,|\rm \mathbf{p^*_2}-\rm \mathbf{p^*_1}|/2$ in the pair rest frame was computed. The total number of $\mathrm{D^0D^{*+}}$ pairs as a function of $k^*$ was calculated by scaling the number obtained from the PYTHIA~8 simulation in order to match the expected integrated luminosity of $\Lint=\SI{18}{\femto\barn^{-1}}$ and to reproduce the predicted $C_\mathrm{D^0D^{*+}}$ for an emitting-source radius of $1~\mathrm{fm}$. The number of $\mathrm{D^0D^{*+}}$ pairs in the 10\% most central Pb--Pb collisions at $\sqrt{s_\mathrm{NN}}=5.5~\mathrm{TeV}$ was obtained analogously for the expected integrated luminosity of $\Lint=\SI{35}{\nano\barn^{-1}}$, considering in addition that the D mesons produced in each Pb--Pb event scale with the number of binary nucleon--nucleon collisions ($N_\mathrm{coll}$) compared to the corresponding number in pp collisions. In this case, the expected $C_\mathrm{D^0D^{*+}}$ for an emitting-source radius of $5~\mathrm{fm}$ was considered.
The right panel of Fig.~\ref{fig:perf:DDstar} shows the expected statistical precision for the $C_\mathrm{D^0D^{*+}}$ measurement with the ALICE3 detector.  In particular, in case of bound state formation, the expected statistical uncertainties will allow for a significant measurement of a $C_\mathrm{D^0D^{*+}}$ lower than unity in Pb--Pb collisions and higher than unity in pp collisions. Hence, this would give the possibility to shed light on the molecular or tetraquark nature of the $\mathrm{T_{cc}^+}$ state.
In the same way, a systematic scan of light-to-heavy colliding systems will allow for a crucial test of the hadronic molecule hypothesis for the candidates listed in Table~\ref{tab:channels}.

\paragraph{$\rm D^{0(+)} \overline{D}{}^{*0(-)}$ momentum correlations} ~\\
\label{sec:performance:physics:ddbarstar}

\begin{figure}[hbt]
\centering
\includegraphics[width=0.45\textwidth]{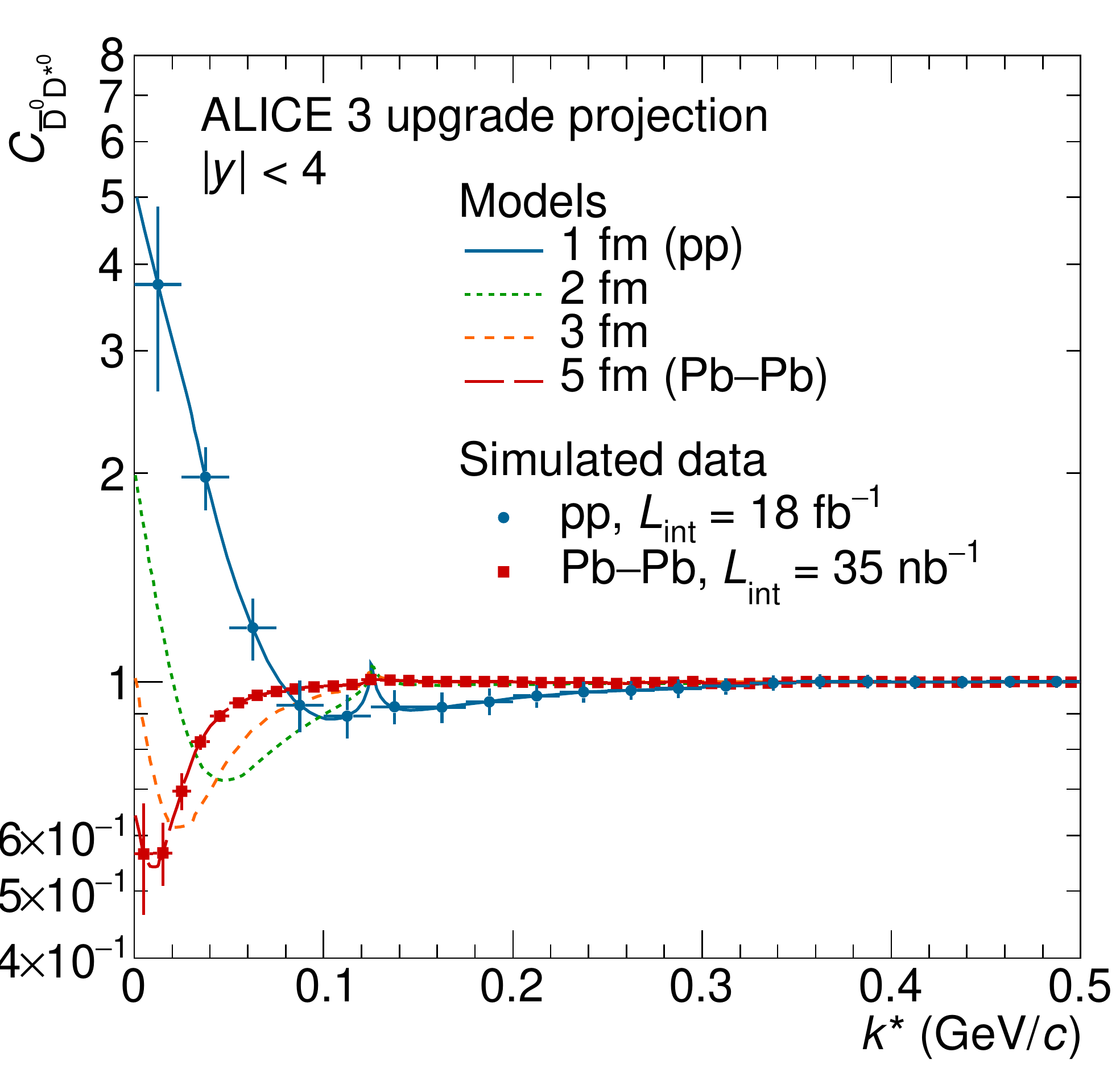}
\includegraphics[width=0.45\textwidth]{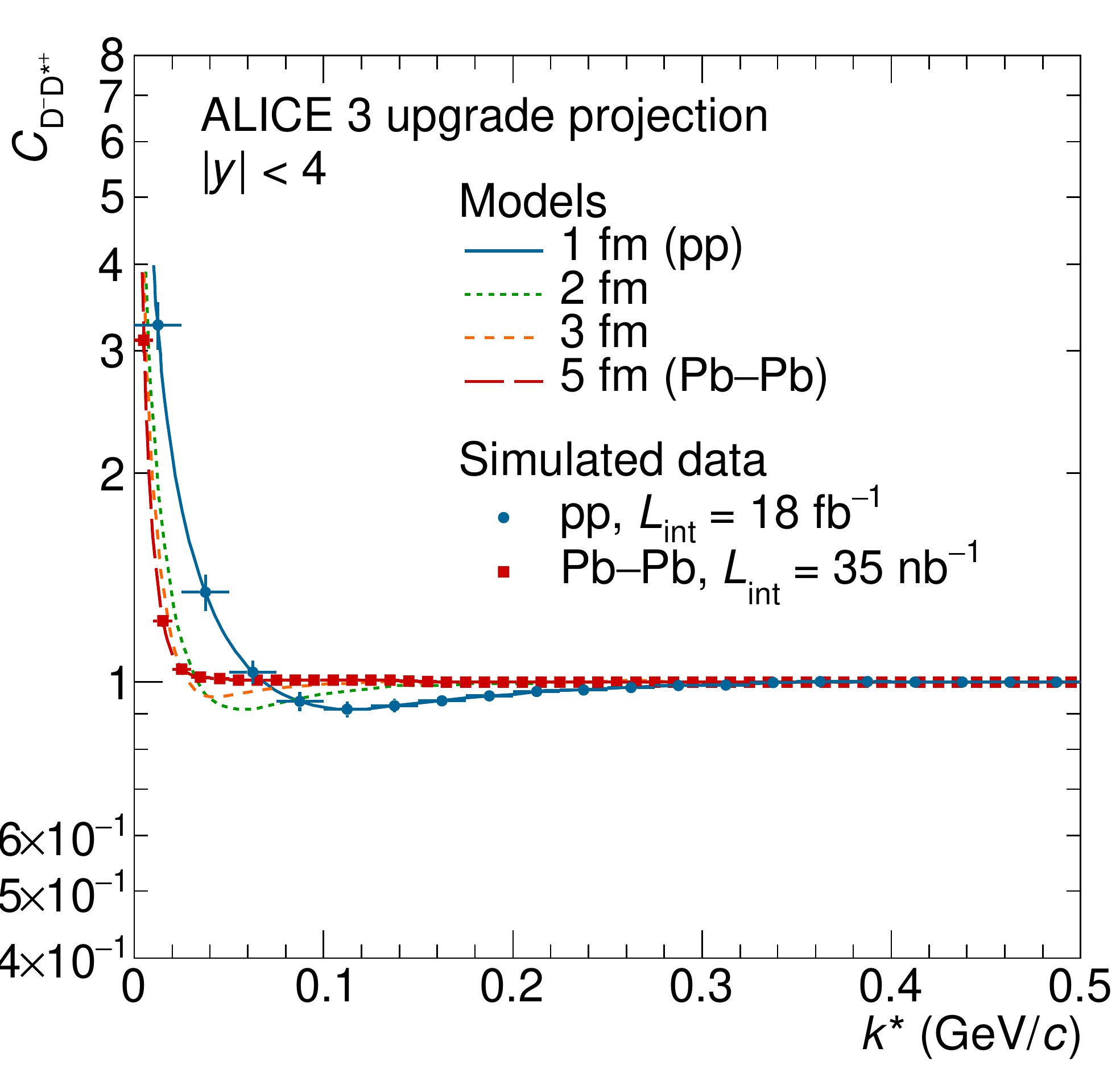}
\caption[]{$\mathrm{D^0\overline{D}^{*0}}$ and $\mathrm{D^+\overline{D}^{*-}}$ correlation function predictions and projections for the ALICE3 detector shown in the left and right panels. Different colours refer to different system radii. The total luminosity considered for pp and Pb--Pb collisions is indicated in the legend.}
\label{fig:perf:DDstarbar}
\end{figure}

Also the nature of the \chiX{} state is subject of a longstanding discussion as far as its molecular nature is concerned. The \chiX{} state ($J^{PC}= \, 1^{++}$ and $I=\,0$) couples to the $\mathrm{D\overline{D}^*}$ and $\mathrm{D^*\overline{D}^*}$, in particular its mass is located  below the $\mathrm{D^0\overline{D}^{*0}}$ pairs ($-40$ keV) and $\mathrm{D^+\overline{D}^{*-}}$ ($-8.27$ MeV) pairs. Since the \chiX is farther away from the $\mathrm{D^+\overline{D}^{*-}}$ threshold, its coupling to the state $\mathrm{D^0\overline{D}^{*0}}$ can be assumed to be dominant. In order to estimate the possible impact of a $\mathrm{D\overline{D}^*}$  molecular state on different $\mathrm{D\overline{D}^*}$  correlation functions a Gaussian potential has been considered for the $\mathrm{D\overline{D}^*}$ interaction constructed such to include  the coupling of the channels  $\mathrm{D^0\overline{D}^{*0}}$--$\mathrm{D^+\overline{D}^{*-}}$ and assuming that the \chiX{} is a molecular state with a binding energy of $40$ keV~\cite{Kamiya:2022thy}. The obtained value of the potential amplitude and the resulting scattering parameters for the $\mathrm{D^0\overline{D}^{*0}}$ and $\mathrm{D^+\overline{D}^{*-}}$ are summarised in Table~\ref{tab:channels}. One can see that the presence of the molecular state close to the $\mathrm{D^0\overline{D}^{*0}}$  threshold impacts the magnitude of the scattering length which is larger with respect to the $\mathrm{D^+\overline{D}^{*-}}$ case.
The correlation functions for different colliding systems and hence different source sizes have been computed for $\mathrm{D^0\overline{D}^{*0}}$ and $\mathrm{D^+\overline{D}^{*-}}$ pairs  and are shown in the left and right panels of Fig.~\ref{fig:perf:DDstarbar}, respectively. One can see on the left panel the presence of the Cusp due to the $\mathrm{D^0\overline{D}^{*0}}$--$\mathrm{D^+\overline{D}^{*-}}$ coupling visible for small radii values at a relative momentum of 120 MeV/c and   the modification of the correlation shape as a function of the increasing radius size. The presence of the molecular state modifies the shape of the correlation function that is mostly above unity for $r=\, 1$ fm and gets only below unity for the largest radius value of $r=\, 5$ fm.
This behaviour is similar to the $\mathrm{T_{cc}^+}$ prediction shown in Fig.~\ref{fig:perf:DDstar}.
Being the \chiX{} more far away from the $\mathrm{D^+\overline{D}^{*-}}$ threshold, despite of the coupled channel contributions, the correlation functions shown in the right panel of Fig.~\ref{fig:perf:DDstarbar} have similar shapes for all considered radii: they rise above unity for small relative momenta as it is the case for an attractive interaction without the presence of a bound state.
By measuring at the same time both correlations in different colliding systems, we will be able to establish the nature of the \chiX{}, since the modification of the correlation function shape across different system size must be observed for the $\mathrm{D\bar{D^{*}}}$ pairs that composes the molecular state.

\begin{figure}[hbt]
\centering
\includegraphics[width=1.\textwidth]{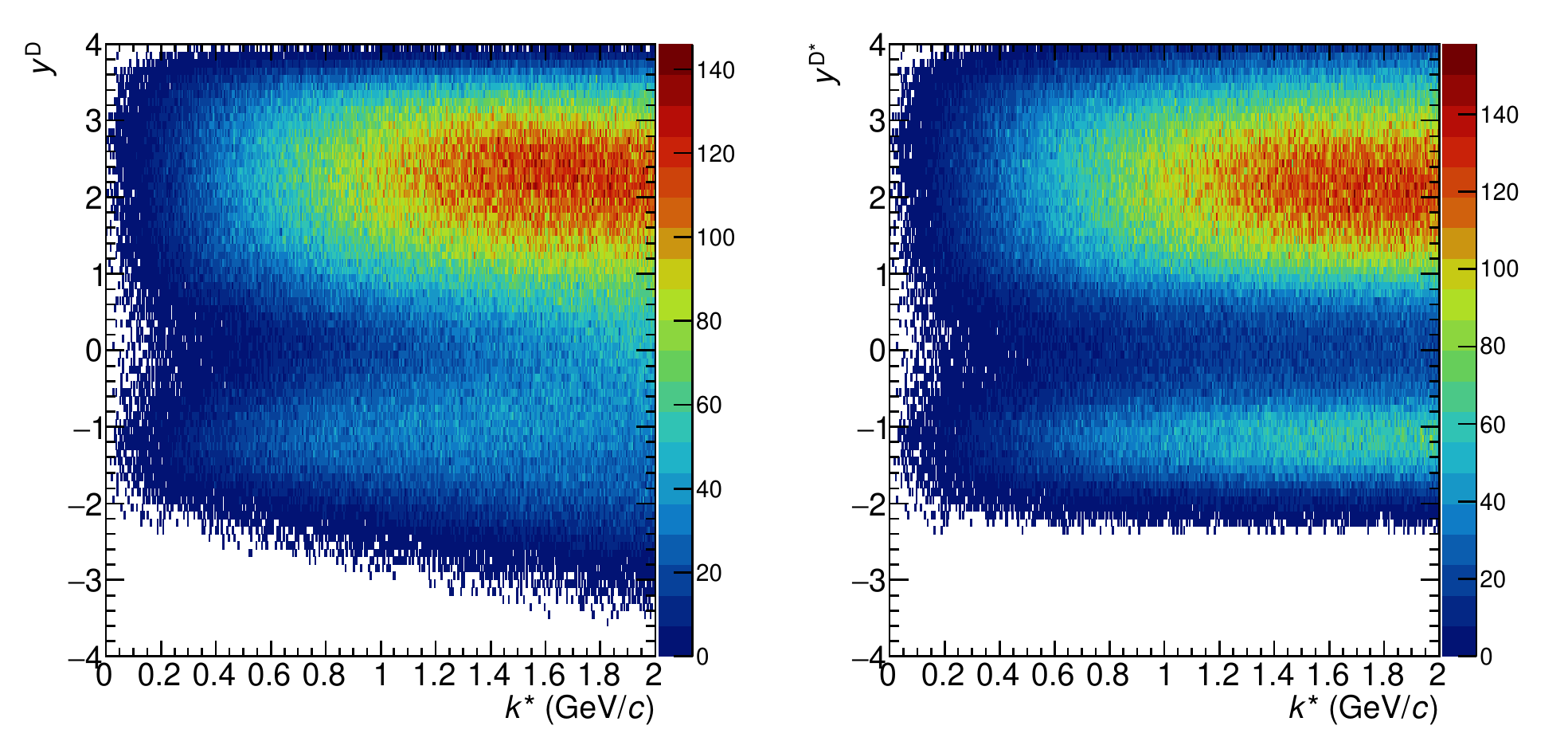}
\caption[]{$\mathrm{D}^0$ and $\mathrm{D}^{0*}$ rapidity vs. $k^*$ distributions for selected $\mathrm{D^0D^{*0}}$ pairs. The depletion at midrapidity is due to the requirement of $E_{\gamma}>400$~MeV imposed to the photon originating from the $\mathrm{D^{*0}}$ decays.}
\label{fig:perf:DDstarbarr_rapidity}
\end{figure}

In order to estimate the expected precision of the measurement of the $\mathrm{D^0\overline{D}{}^{*0}}$ and $\mathrm{D^+D^{*-}}$ correlation functions $C(k^*)$, a similar strategy as the one followed for $\mathrm{D^0D^{*+}}$ system described in Sec.~\ref{sec:performance:physics:ddstar} was pursued. In particular, pp collisions at $\sqrt{s}=14~\mathrm{TeV}$ were simulated using the PYTHIA~8 event generator. The same decay channels were considered for $\mathrm{D}^0$ and $\mathrm{D}^{*+}$ mesons, while $\mathrm{D}^{+}$ and $\mathrm{D}^{0*}$ mesons were selected via the decay channels $\mathrm{D^{+}}\rightarrow\mathrm{K^-}\pi^+\pi^+$ and $\mathrm{D^{*0}}\rightarrow\mathrm{D^0}\gamma$, with a branching ratio of $(9.38\pm0.16)\%$ and $(35.3\pm0.5)\%$~\cite{Zyla:2020zbs}, respectively. While the evaluation of the projected performance for the $\mathrm{D^+D^{*-}}$ correlation function is completely analogous to that of the $\mathrm{D^0D^{*+}}$ system, for $\mathrm{D^0\overline{D}{}^{*0}}$ pairs, the reconstruction of the photon originating from the $\mathrm{D}^{*0}$ decay has to be taken into account. For this purpose, the electromagnetic calorimeter (ECal) described in Sec.~\ref{sec:systems:ecal} was considered. Only photons within the acceptance of the ECal ($-1.6<\eta<4$) and with a minimum energy of $E_\gamma>400$~MeV were considered to guarantee sufficient efficiency. The rapidity distributions for the $\mathrm{\overline{D}{}^{*0}}$ and $\mathrm{D}^{0}$ mesons forming the pairs of particles to be considered for the computation of the correlation function are reported as a function of the relative momentum in the particle-pair rest frame $k^*$ in the right and left panels of Fig.~\ref{fig:perf:DDstarbarr_rapidity}, respectively. The depletion of pairs at midrapidity is caused by the selection of the minimum energy of the photons, provided that photons in the $\mathrm{D^{*0}}\rightarrow\mathrm{D^0}\gamma$ decay are very soft. Hence, for this measurement, the presence of an electromagnetic calorimeter at forward rapidity is crucial.
In Fig.~\ref{fig:perf:DDstarbar} the expected statistical precision for the $C_\mathrm{D^0\overline{D}{}^{*0}}$ (left panel) and $C_\mathrm{D^+D^{*-}}$ (right panel) measurement with the ALICE3 detector is shown. In the case of the $\mathrm{D^0\overline{D}{}^{*0}}$ system, the expected statistical uncertainties will allow for a significant measurement of a $C_\mathrm{D^0D^{*+}}$ lower than unity in Pb--Pb collisions and higher than unity in pp collisions, if a bound-state (\chiX{}) is formed. In the case of the charged mesons instead, the $C(k^*)$ is never expected to be lower than unity because of the attractive Coulomb interaction, which implies an increase at low $k^*$. Nevertheless, also in the $\mathrm{D^+D^{*-}}$ channel a difference in the $C(k^*)$ for different source sizes, which can be attributed to the formation of the \chiX{} bound state, can be resolved. Hence the measurement of the $C_\mathrm{D^+D^{*-}}$ represents an important and complementary study to the one of the neutral channel.The considered integrated luminosities for the different colliding systems are the same quoted in Section~\ref{sec:performance:physics:ddstar}.

\subsubsection{Quarkonia and Exotica}
\label{sec:performance:physics:quarkonia}

In this section, performance studies for some exemplary analyses in quarkonium and heavy-flavour exotica physics are presented. All studies in this section make use of the muon decay channels.

The performance studies for quarkonium and heavy-flavour exotic states are performed in pp collisions at $\sqrt{s} = 14$~TeV and \PbPb{} collisions at $\sqrtsNN = 5.5$~TeV. The signals for this analysis were generated with \texttt{PYTHIA8 SoftQCD} pp events at $\sqrt{s} = 14$~TeV, with a preselection of events that have a particle of interest, which is then forced to decay in the dimuon channel for $\jpsi$, and the $\jpsi\, \pi^+ \pi^-$ channel for the \chiX{}.
For background events, the same generators were used as for the charm hadron performance studies (\texttt{PYTHIA8 SoftQCD} and \texttt{PYTHIA8 Angantyr}).

The performance of the muon identifier presented in Section~\ref{sec:performance:detector:muon_id} is used in these studies.

\paragraph{Benchmark of quarkonium reconstruction with \jpsi}~\\
\label{sec:performance:physics:quarkonia:jpsi}
The signal to background ratio and significance for \jpsi{} reconstruction with ALICE~3 has been studied to demonstrate the detector capabilities in the quarkonia sector.

Decays of \jpsi{} in the muon channel are reconstructed by selecting tracks with positive muon ID in the MID, implying a minimum transverse momentum of $\sim$ \SI{1.5}{\giga\eVc} at $\eta = 0$.
Signal and background are then estimated from the invariant mass distribution of the candidates, within a $3\,\sigma$ mass window around the nominal $\jpsi$ mass, and normalised to the total luminosity for six running years.

\begin{figure}[htbp]
\centering
\includegraphics[width=0.49\textwidth]{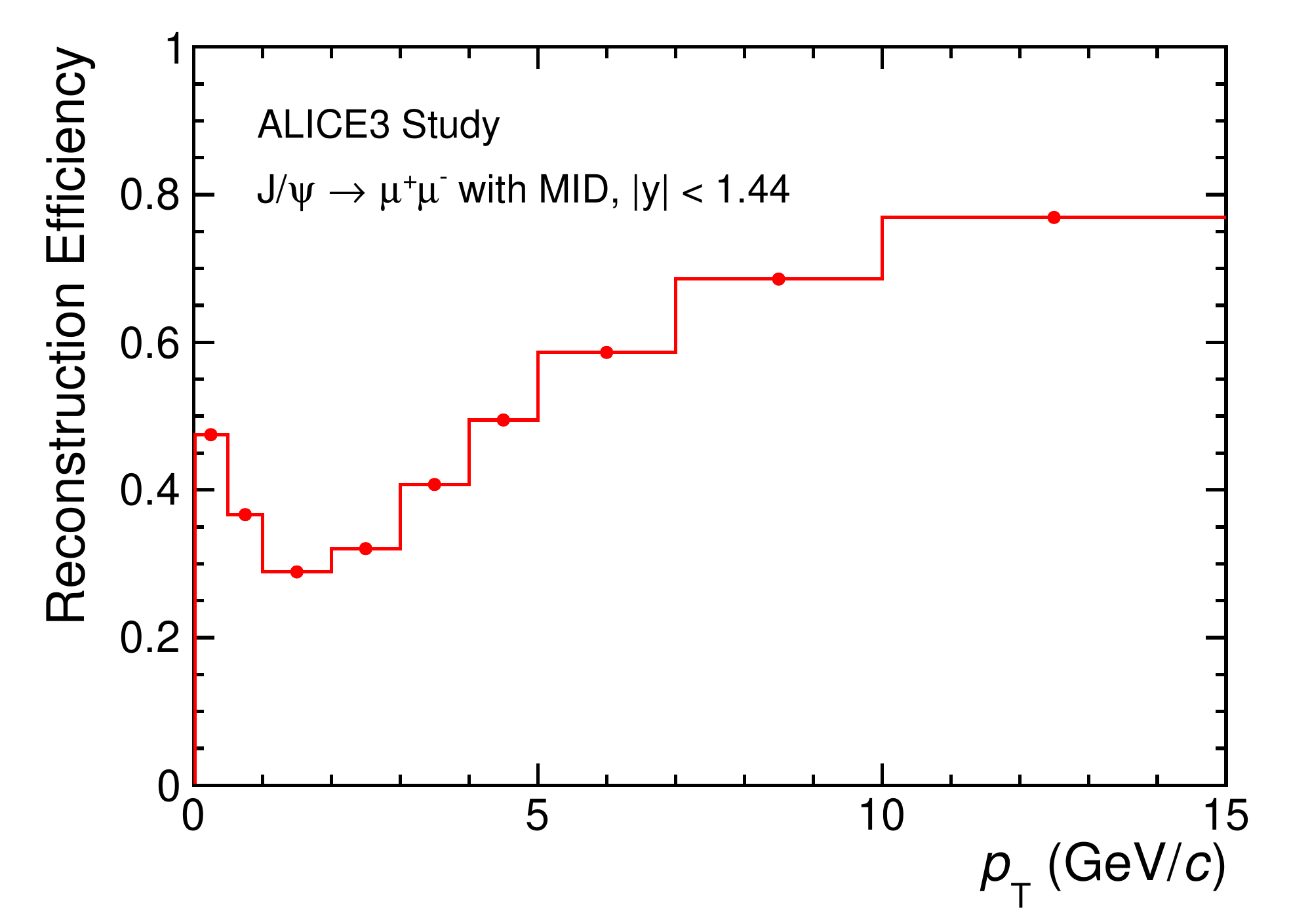}
\includegraphics[width=0.49\textwidth]{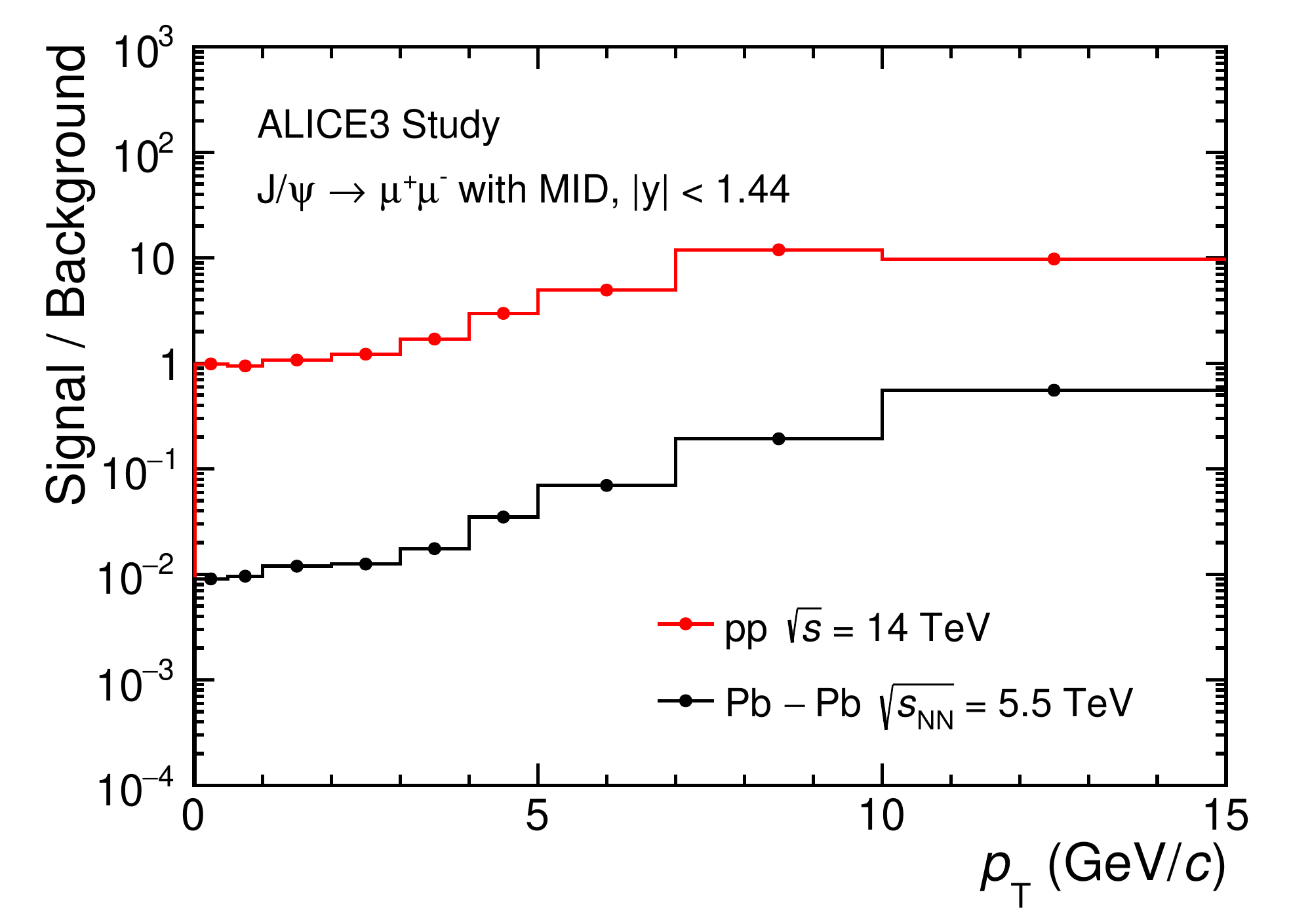}
\includegraphics[width=0.49\textwidth]{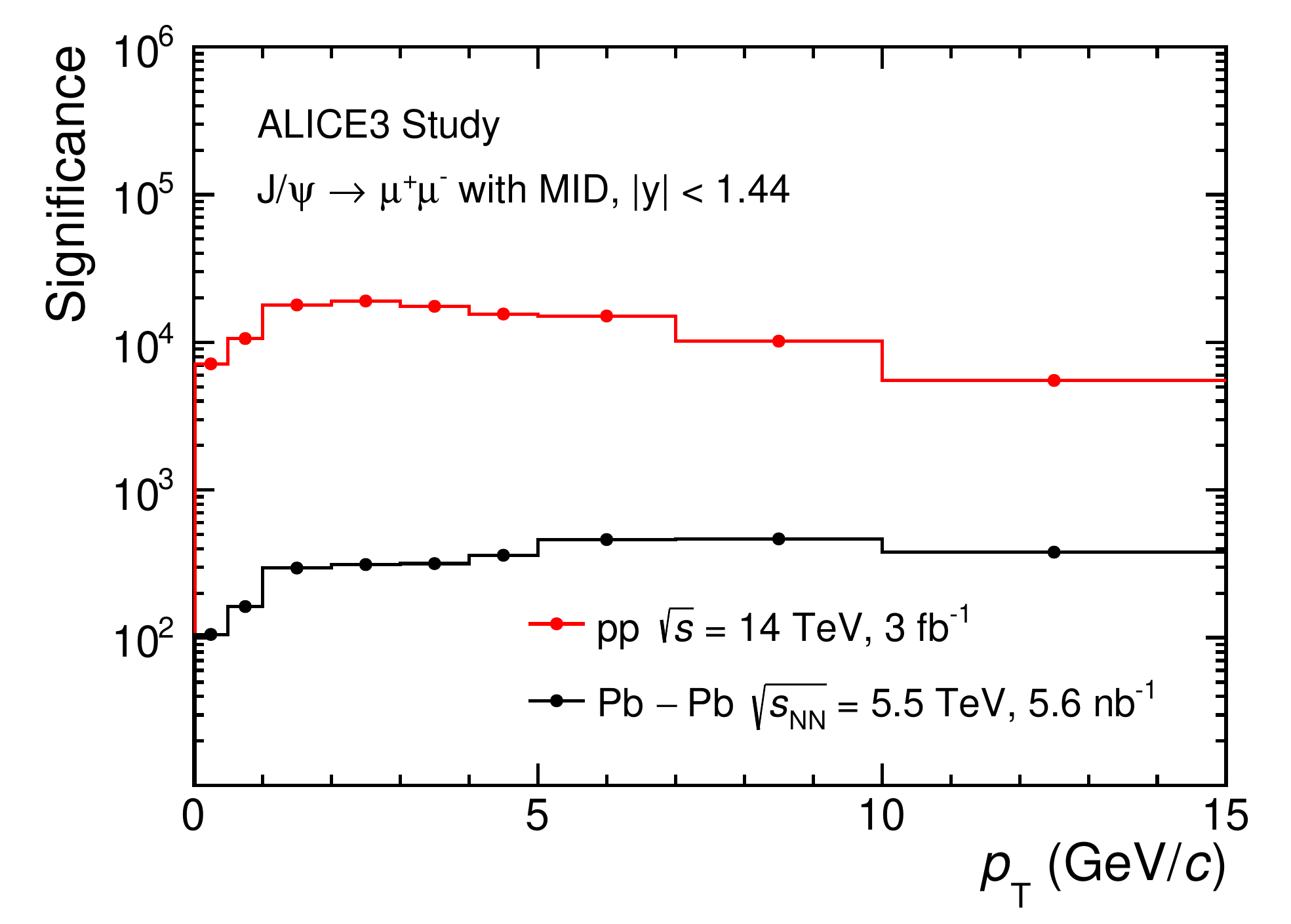}
\caption[Reconstruction efficiency of \jpsi]{$\jpsi$ reconstruction efficiency at mid rapidity as a function of $\pt$ (top-left panel), signal-to-background (top-right panel), and significance (bottom panel) for the signal extraction in pp collisions at $\sqrt{s} = 14$~TeV ($\Lint = \SI{3}{\femto\barn}^{-1}$) and in Pb--Pb collisions at $\sqrt{\sNN} = 5.5$~TeV ($\Lint = \SI{5.6}{\nano\barn}^{-1}$), corresponding to one-year data taking.}
\label{fig:jpsiEffAndSigOverBkg}
\end{figure}

The top-left panel of Fig.~\ref{fig:jpsiEffAndSigOverBkg} shows the $\jpsi$ reconstruction efficiency at mid rapidity as a function of $\pt$, and the signal to background ratio for \pp{} and \PbPb{} collisions is shown the top-right panel. In \pp{} collisions, the signal-to-background ratio is larger than $\sim 1$ over the entire studied range. For \PbPb{} collisions, even though the signal-to-background ratio is only around 0.05 at very low \pt, the significance of the signal is 200-300 per year, i.e. around 1000 for the full run period, see bottom panel of Fig.~\ref{fig:jpsiEffAndSigOverBkg}. %

\paragraph{$\chi_c$ and $\chi_b$ states}~\\
\label{sec:performance:physics:quarkonia:chis}
\begin{figure}
    \centering
    \includegraphics[width=0.48\textwidth]{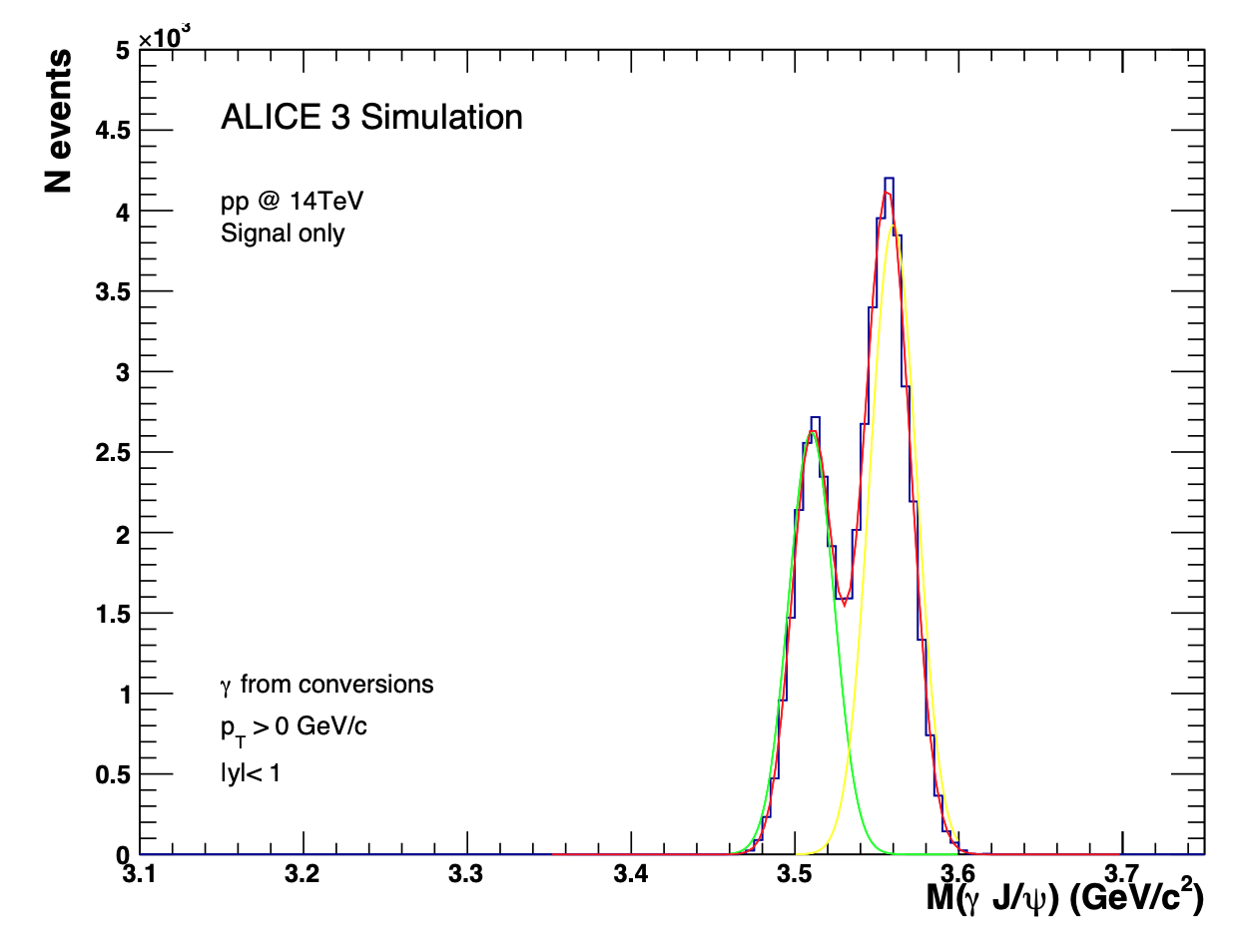}
    \caption[$\chi_c$ mass resolution]{Invariant mass distribution of $\chi_{c1}$ and $\chi_{c2}$ signals obtained reconstructing the $\chi_{c}$ photon daughter using the photon conversion method (PCM).}
    \label{fig:performance:chic_viapcm}
\end{figure}
\begin{figure}
    \centering
    \includegraphics[width=0.48\textwidth]{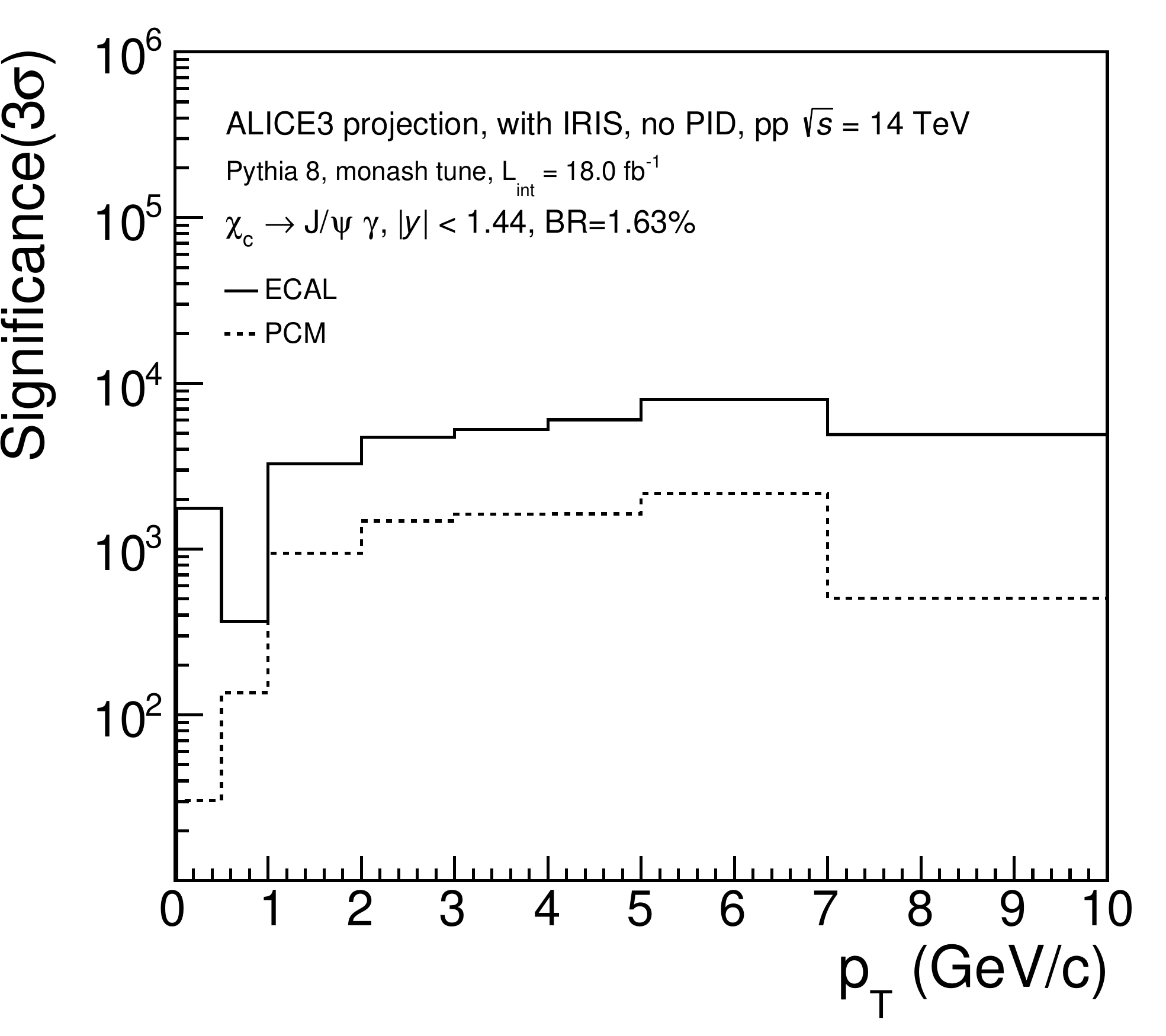}
    \includegraphics[width=0.48\textwidth]{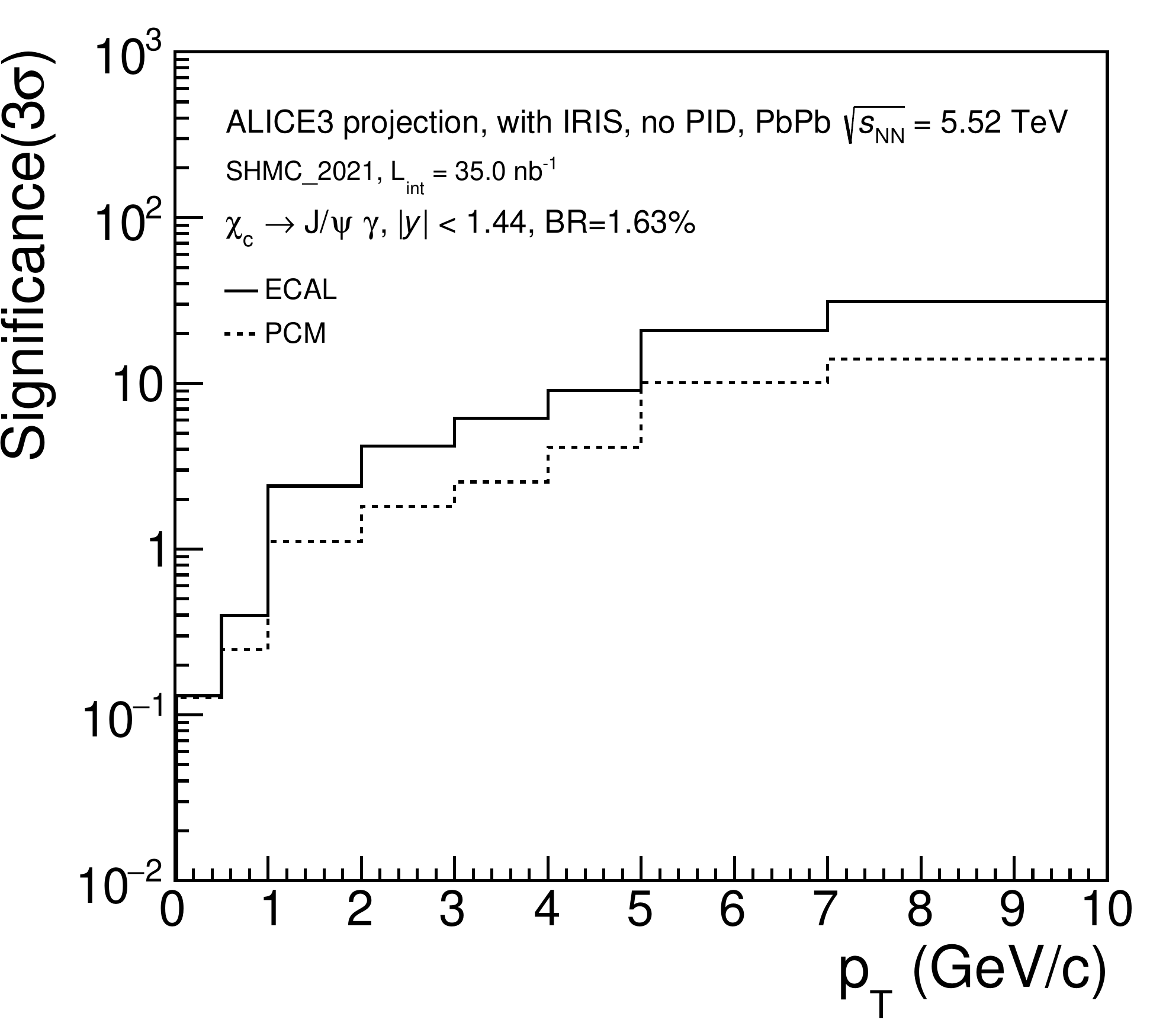}
    \caption[Significance of $\chi_c$ in pp and \PbPb{} collisions]{Significance of the $\chi_c$ signal (sum of the $\chi_{c1}$ and $\chi_{c2}$ signals) as a function of transverse momentum in pp collisions at $\sqrt{s} = 14$~TeV ($\Lint = \SI{18}{\femto\barn}^{-1}$ using the ECal and the PCM method 
    for photon detection and reconstruction. The fluctuation in the first bins is due to limited Monte Carlo statistics for the background) and in Pb--Pb collisions at $\sqrt{\sNN} = 5.5$~TeV ($\Lint = \SI{35}{\nano\barn}^{-1}$).}
    \label{fig:performance:chic_significance}
\end{figure}
As discussed in Section~\ref{sec:physics_chis}, a comprehensive study of charmonium states in heavy-ion collisions should include a characterization of P-wave states such as $\chi_c$ and $\chi_b$. With ALICE~3, this can be achieved by reconstructing decays in the $\chi_{cJ} \to J/\psi + \gamma$ decay channel using the muon identifier to reconstruct $\jpsi$ and the electromagnetic calorimeter to detect decay photons (see Section~\ref{sec:systems:ecal:specs}) or, alternatively, the
photon conversion method (PCM), as described in Sec.~\ref{sec:introduction:experimental_layout}. The invariant mass
distributions of the $\chi_{c1}$ and $\chi_{c2}$ states reconstructed with the latter method can be seen in Fig.~\ref{fig:performance:chic_viapcm}

The reconstruction of the $\chi_c$ candidates starts with the reconstruction of the $\jpsi$ candidates following the strategy discussed in Section~\ref{sec:performance:physics:quarkonia:jpsi}. Each selected $\jpsi$ candidate is then combined with the available photons detected in the same event. In order to select a clean sample of $\chi_c$ candidates and to reduce the combinatorial background, the cuts reported for the $J/\psi$ analysis are applied, combined with a $2\sigma$ cut on the invariant mass of the $\jpsi$ candidate and a lower cut on the photon energy $E_\gamma > 400$~MeV for the ECal-based analysis. Figure~\ref{fig:performance:chic_significance} shows the significance of the $\chi_c$ measurement (sum of the $\chi_{c1}$ and $\chi_{c2}$ signals) corresponding to the currently implemented kinematic cuts, as a function of transverse momentum in pp collisions at $\sqrt{s} = 14$~TeV ($\Lint = \SI{18}{\femto\barn}^{-1}$) and in Pb--Pb collisions at $\sqrt{\sNN} = 5.5$~TeV ($\Lint = \SI{35}{\nano\barn}^{-1}$). A~dedicated optimization of the cuts is currently undergoing, involving machine learning techniques based on boosted decision trees. It is important to note that extending the acceptance to the full pseudorapidity coverage of ALICE~3 will 
improve the significance reported in Fig.~\ref{fig:performance:chic_significance}. Moreover,
dedicated runs with a converter may be considered to increase the significance for the PCM-based measurements in heavy-ion collisions.

\paragraph{Exotic states: \chiX{}}~\\
\label{sec:performance:physics:quarkonia:exo}
The capability to reconstruct $\jpsi$ decays down to zero~$\pt$ provides the potential to measure \chiX{} decays to below $\pt \sim 5$~GeV/$c$, where the largest medium effects on the production yields are expected.

The reconstruction of \chiX{} candidates starts with the reconstruction of the $\jpsi$ candidates following the strategy discussed in Section~\ref{sec:performance:physics:quarkonia:jpsi}. Each selected $\jpsi$ candidate is then combined with the available $\pi^+ \pi^-$ pairs reconstructed in the same event.
In order to select a clean sample of \chiX{} candidates and to reduce the combinatorial background, a series of  selections are applied, namely a $3\sigma$ cut on the invariant mass of the $\jpsi$ candidates, a lower cut on its $\pt$ with the threshold ranging from 0 to 6~GeV/$c$ as a function of the $\pt$ of the \chiX{} candidate, and a $\pt > 0.5$~GeV/$c$ cut on the pions.

Since no \texttt{PYTHIA} prediction is available for the \chiX{}, the $\pt$ distribution is taken to be the same as that of the $\psi(\rm{2S})$.
The number of signal and background candidates are estimated from the invariant mass spectrum of the $\jpsi\pi^+\pi^-$ candidates within a $3~\sigma$ mass window around the \chiX{} nominal mass.

\begin{figure}[htbp]
\centering
\includegraphics[width=0.65\textwidth]{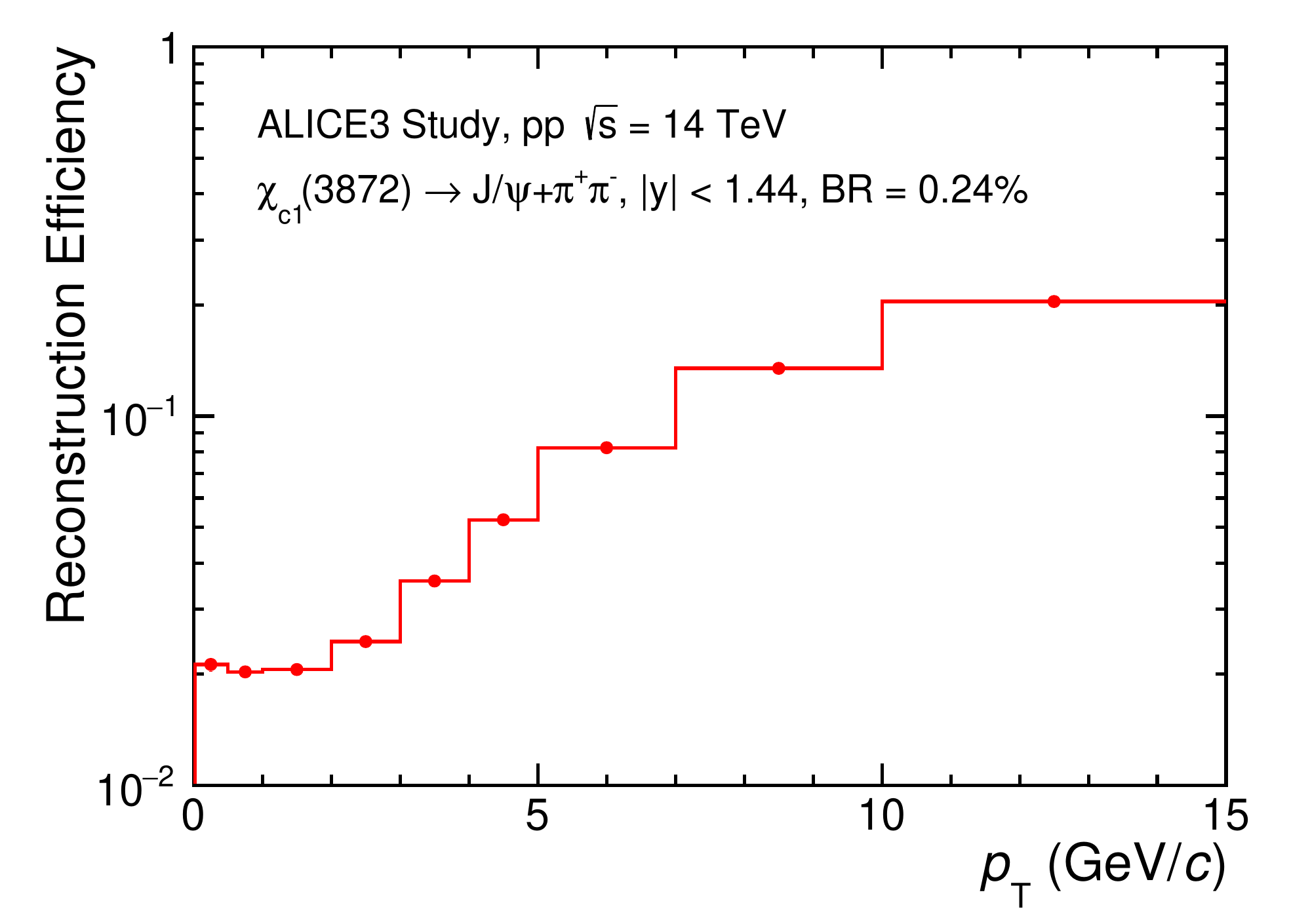}
\caption[Reconstruction efficiency of \chiX]{Reconstruction efficiency of \chiX, in the $\jpsi \pi^+ \pi^-$ decay channel.}
\label{fig:X3872Efficiency}
\end{figure}

\begin{figure}[htbp]
\centering
\includegraphics[width=0.65\textwidth]{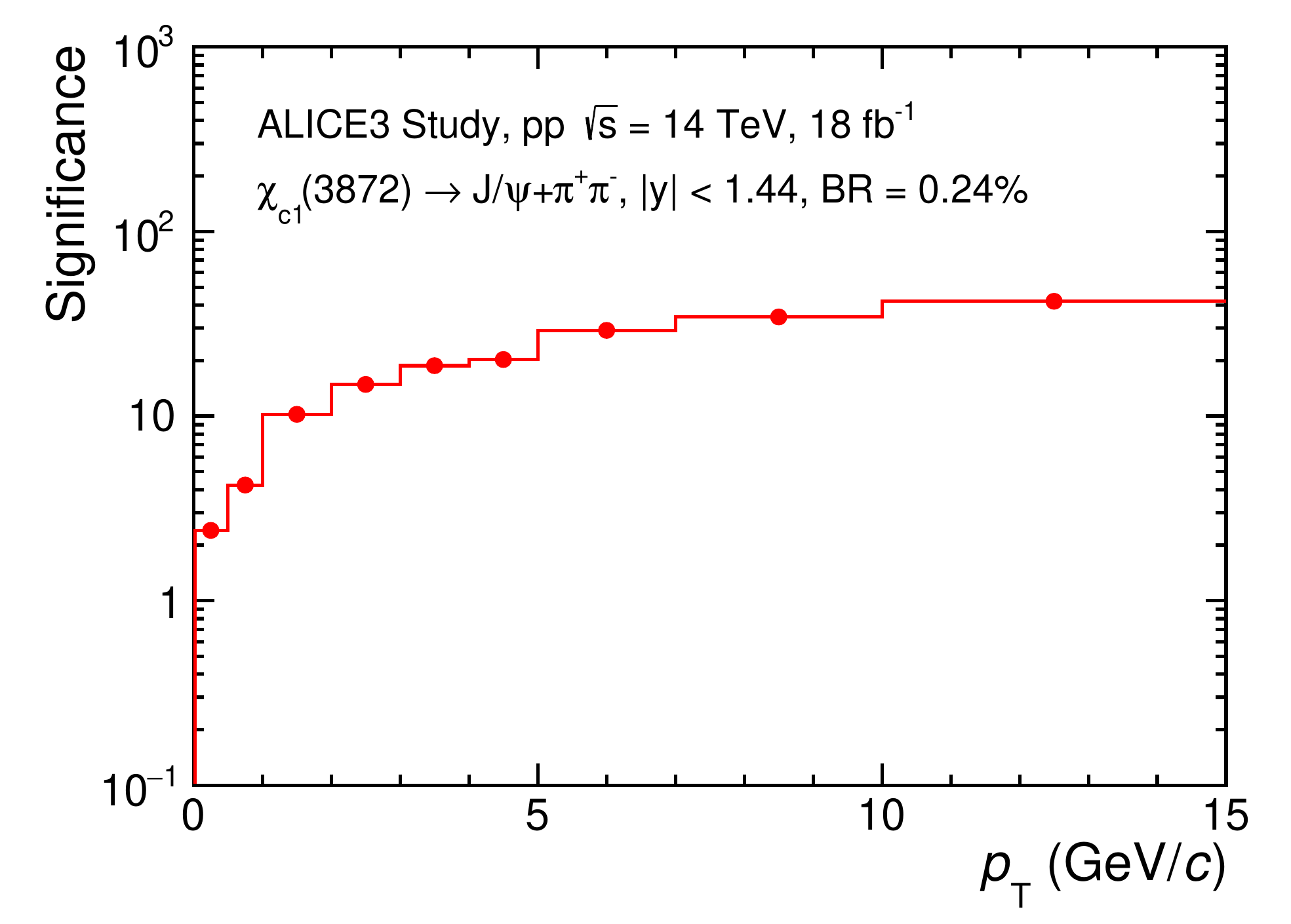}
\caption[Significance of the \chiX{} measurement]{\chiX{} significance, for $\rm{J}/\psi$ decaying to $\mu^{+}\mu^{-}$ in pp collisions at $\sqrt{s} = 14$~TeV with an integrated luminosity of 18~fb$^{-1}$.}
\label{fig:X3872Significance_pp}
\end{figure}

Figure~\ref{fig:X3872Efficiency} shows the \chiX{} reconstruction efficiency at midrapidity as a function of $\pt$. The corresponding significance of the \chiX{} signal is shown in Fig.~\ref{fig:X3872Significance_pp} for a total integrated luminosity of 18~fb$^{-1}$ of pp collisions at $\sqrt{s} = 14$~TeV. The significance of the signal is above 3 for $pt>2$~\GeVc. Further tuning of the selections is being pursued to increase the significance, involving machine learning techniques based on boosted decision trees.

\textit{\textbf{Uniqueness and comparison with CMS and ATLAS}} 
ALICE~3 will have the unique capability to measure $\jpsi$ decays down to zero~$\pt$ at midrapidity with high accuracy especially in central AA collisions, thanks to a muon identification system (MID) which is optimized for the reconstruction of muons down to $\pt \sim 1.5$~\GeVc at $\eta = 0$. No existing LHC experiment will be capable of measuring $\jpsi$ decays below 6--7 \GeVc at central rapidity with high purity. In CMS and ATLAS, muons can be reconstructed in the central barrel only down to $\pt \approx$ 3--4~\GeVc, as a consequence of the large energy loss of muons inside the calorimeters (with total material budget above 10 interaction lengths). LHCb also covers down to $\jpsi$ $\pt=0$, but only at forward rapidity. 

The aforementioned limitations of the CMS and ATLAS detectors on $\jpsi$ reconstruction at low-\pt also affect the reconstruction of decays of bound states as $\chicone$ or $\chiX$ at midrapidity. The acceptance for the reconstruction of $\chicone \rightarrow \jpsi \gamma \rightarrow \mu^+\mu^- \gamma$ decay channel in pp collisions, with the ATLAS and CMS detectors, is larger than zero only for $\pt > 10$ \GeVc~\cite{chic12ATLAS2014,chic12CMS}. The measurements of the $\chiX \rightarrow \jpsi \pi^+ \pi^-$ are also limited to the same kinematic region ($\pt > 10$~\GeVc)~\cite{CMS:2013fpt}.

Moreover, the measurement of $\chi_{c,b}$ at low $\pt$ largely relies on the capability of having a reliable detection and identification of soft photons with energy $E < 1$~GeV. The ALICE~3 setup provides the ideal conditions for such a measurement, thanks to the very limited material budget of the inner barrel detectors in front of ECAL, which is estimated to be $X/X_0 \approx 0.2$. For comparison, the material budget in front of the CMS ECAL detectors varies from $0.5~X_0$ to $1~X_0$ in the pseudorapidity range $0<\eta<1$~\cite{CMS:2013lxn}, and the ATLAS ECAL has up to $2~X_0$ in front of it~\cite{Cavallari:2011zza}.

\subsubsection{Dielectrons}
\label{sec:performance:physics:dileptons}

The large rapidity coverage of the ALICE 3 setup, 8 units of rapidity, will provide for the first time at LHC energies the means to study dielectron production as a function of rapidity. Such a measurement poses formidable challenges due to the large background of correlated electron pairs from heavy-flavour hadron decays. In addition, there is a large combinatorial background originating from other hadronic decays. Both contributions need to be either experimentally suppressed or subtracted. The measurement of dielectron production with the ALICE 3 setup at midrapidity will take advantage of the approximately five times better pointing resolution to reject heavy-flavour background, a factor 2 reduction in the amount of material in front of the second tracking layer, and high-precision tracking to remove conversion electrons by a prefilter technique compared to the ALICE Inner Tracking System for the LHC Run 4.

In the following we present the performance for key measurements with electron-positron pairs.

\paragraph{Thermal radiation and chiral symmetry restoration}
\label{sec:performance:physics:dileptons:chiral}

The strategy to estimate the expected statistical and systematic uncertainties on the measured yield of e$^{+}$e$^{-}$ pairs from thermal radiation, including thermal production of the $\rho$ meson is similar to that discussed in the ITS3 LoI\,\cite{ALICE:ITS3:2019}. The different contributions to the correlated dielectron yield are:
\begin{itemize}
    \item the decay of light-flavour hadrons ($\pi^{0}$,$\eta$,$\eta\prime$,$\omega$ and $\phi$) which have a long life time compared to that of the hot and dense medium. Their decays are simulated with the phenomenological event generator EXODUS\,\cite{PHENIX:2009gyd} using the measured $\pi^{\pm}$ spectra in Pb--Pb collisions at $\sqrt{s_{\rm NN}}$ $=$ 5.02\,TeV\,\cite{ALICE:2019hno} as input. This method is described in \,\cite{ALICE:2018fvj} and commonly called 'hadronic cocktail'.
    \item correlated semileptonic decays of pair-produced open-charm and open-beauty hadrons that are simulated with the PYTHIA 6 event generator (Perugia 2011 tune) and normalized using the number of binary collisions and the cross sections in pp collisions at $\sqrt{s_{\rm NN}}$ $=$ 5.02 TeV as measured in the dielectron channel\,\cite{ALICE:2020mfy} with the same event generator. Initial-state effects are incorporated by utilising the EPS09 nuclear parton distribution functions in the calculations.
    \item thermal radiation from the QGP ('Rapp QGP' in Fig.~\ref{expectedyields}) and the hadronic phase ('Rapp in-medium SF' in Fig.~\ref{expectedyields}) estimated with a hadronic many-body approach consistent with partial chiral restoration in the hot hadronic medium\,\cite{Rapp:1999us,vanHees:2007th,Rapp:2013nxa}.
\end{itemize}
The expected raw signal dielectron yield at midrapidity is shown on the left panel of Fig.~\ref{expectedyields} together with the different input distributions for central (0-10\%) \PbPb{} collisions at $\sqrt{s_{\rm NN}}$ = 5.02\,TeV. The tracking and electron identification efficiency using the outer TOF and RICH detectors are applied, as well as the momentum and opening angle resolution for a magnetic field of \SI{0.5}{\tesla}. Bremsstrahlung effects on the reconstructed $m_{\rm ee}$ and $p_{\rm T,ee}$ are not yet included. 

The contribution from correlated open heavy-flavour decays dominates the spectrum for invariant masses larger than 0.9\,GeV/$c^{\rm 2}$. The large lifetime of the open-charm and open-beauty hadrons can be used to separate the correlated heavy-flavour background from prompt dielectron sources via the pair distance of closest approach $DCA_{\rm ee}$ defined as:
\begin{equation}
    DCA_{\rm ee} = \sqrt{\frac{(DCA_{\rm xy,1}/\sigma_{\rm xy,1})^{2}+(DCA_{\rm xy,2}/\sigma_{\rm xy,2})^{2}}{2}},
\end{equation}
where $DCA_{\rm xy,i}$ and $\sigma_{\rm xy,i}$ are the measured $DCA$ and $DCA$-resolution of the reconstructed electron or positron track in the plane transverse to the beam direction. A maximum requirement on $DCA_{\rm ee}$ is chosen such that 95\% (98\%) of the correlated e$^{+}$e$^{-}$ pairs from open-charm (open-beauty) hadron decays are rejected, while 76\% of the prompt dielectrons are kept. For comparison, for the same rejection factor of $c\bar{c}\xrightarrow{}$e$^{+}$e$^{-}$ an efficiency for prompt pairs of about 30\% (17\%) is achieved with the ALICE ITS3 (ITS2) (See Section~\ref{sec:performance:physics:dileptons:comparisonITS3}). The expected raw  signal dielectron yield after the $DCA_{\rm ee}$ selection criterion is applied, is shown in the right-hand panel of Fig.~\ref{expectedyields}. Dielectrons from thermal radiation dominate the e$^{+}$e$^{-}$ spectrum for $m_{\rm ee}$ $>$ 0.4 GeV/$c^{\rm 2}$ up to $m_{\rm ee}$ $>$ 2.3 GeV/$c^{\rm 2}$ with the exception of the $\omega$ and $\phi$ peak regions. 

\begin{figure} [hb!]
\begin{center}
  \subfigure{\includegraphics[width=0.34\paperwidth]{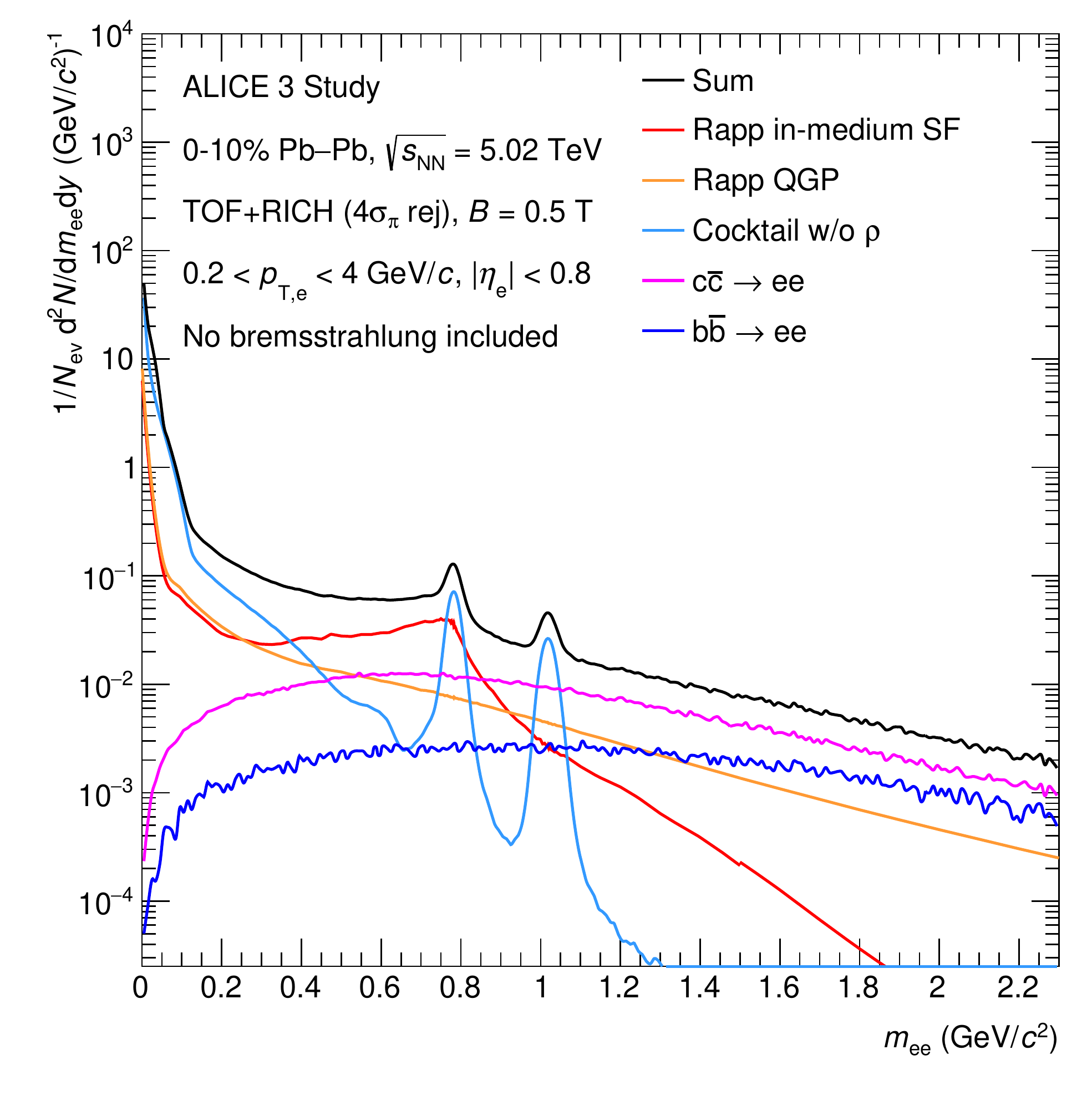}}
 \subfigure{\includegraphics[width=0.34\paperwidth]{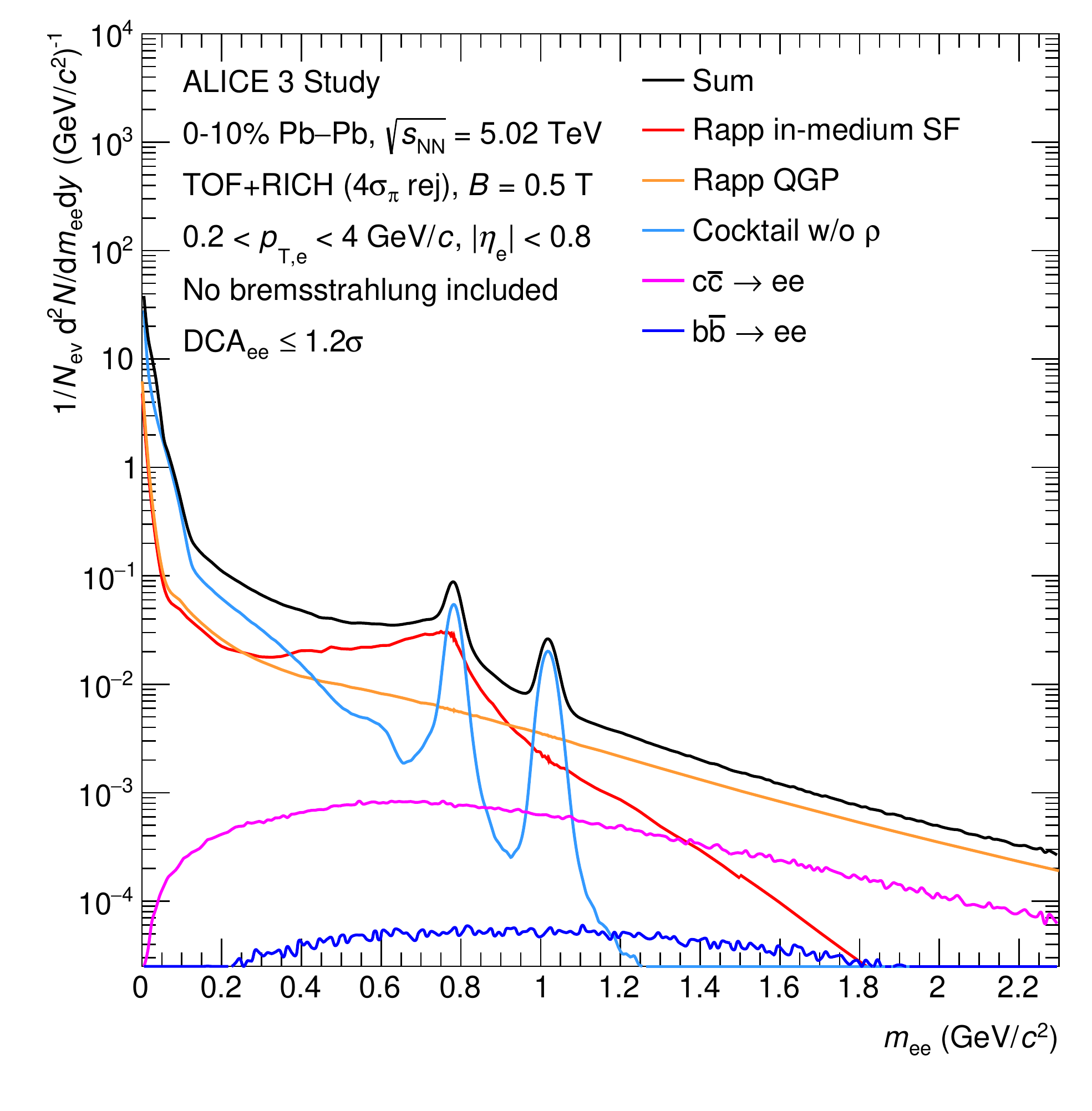}}
\caption[Expected raw signal dielectron yields]{Expected raw signal dielectron yield using the outer TOF and RICH particle identification at midrapidity in 0-10\% most central Pb--Pb collisions at $\sqrt{s_{\rm NN}}$ $=$ 5.02 TeV as a function of invariant mass $m_{\rm ee}$ before (left) and after (right) a maximum pair $DCA_{\rm ee}$ requirement.}
\label{expectedyields}
\end{center}
\end{figure}

The signal dielectron yield ($S$) is obtained in real data by forming all opposite-charge pairs of identified electrons positrons from the same event and then subtracting the contribution from combinatorial pairs ($B$). The combinatorial background is computed using like-sign pairs for each event and summing over all events\,\cite{ALICE:2018fvj}. The like-sign spectrum is self-normalized and contains all residual correlations arising from charge-symmetric processes that are present in the background. The statistical significance of the measured spectra is computed as $S$/$\sqrt{S+2B}$.

In order to estimate $B$, central Pb--Pb collisions are simulated with the PYTHIA 8 Angantyr event generator and propagated through the detector using the fast simulation. Electrons are weighed such that the calculations reproduce the measured pion ($\pi^{\pm}$ $\approx$ $\pi^{0}$) \,\cite{ALICE:2019hno} and heavy-flavour hadron decay  single electron\,\cite{ALICE:2019nuy} spectra. All same-sign pairs are formed and the $DCA_{\rm ee}$ is calculated.  After the $DCA_{\rm ee}$ selection, the main source of electrons contributing to the background are $\pi^{\rm 0}$-Dalitz decays. In order to reject them, a prefilter algorithm is applied on an event-by-event-basis, where tracks from the selected electron candidate
sample are combined with electron candidates from a sample with a relaxed minimum $p_{\rm T,e}$ threshold (prefilter selection tracks). All electron candidates that are part of at least one  opposite-sign combination with small invariant mass ($m_{\rm ee}$ $\leq$ 0.05\,GeV/$c^{\rm 2}$) and small opening angle ($\omega_{\rm ee} \leq 100$\,mrad), characteristic of e$^{+}$e$^{-}$ pairs from $\pi^{\rm 0}$-Dalitz decays, are rejected. The accidental rejection of two uncorrelated electrons is taken into account in the raw signal calculation. Such a prefilter strategy has already been used successfully for the analysis of the ALICE Run 1 and 2 data\,\cite{ALICE:2018fvj,ALICE:2020umb}.

The raw signal dielectron yield and the combinatorial background are shown together in Fig.~\ref{prefiltera} as a function of $m_{\rm ee}$ in central \PbPb{} collisions for three scenarios: without any prefilter algorithm (w/o PF), with prefilter electron tracks identified with the outer TOF and RICH detectors in the range $p_{\rm T,e} \ge \SI{80}{\mega\eVc}$ (PF1), and the ideal case where all true electrons with $p_{\rm T,e} \geq \SI{20}{\mega\eVc}$ are considered in the prefilter sample (PF$_{\rm ideal}$). %

The corresponding signal over background ratios $S$/$B$ and significances are shown in Fig.~\ref{prefilterb}. The prefiltering with TOF and RICH (PF1) improves the $S$/$B$ by a factor of about 2.5 and the significance by a factor of about 1.5. This is taken as input for the following performance studies. Further improvements could be achieved by tracking and identifying electrons down to lower $p_{\rm T,e}$, as shown by the ideal case (PF$_{ideal}$).

\begin{figure} [hb!]
\begin{center}
 \subfigure{\includegraphics[width=0.40\paperwidth]{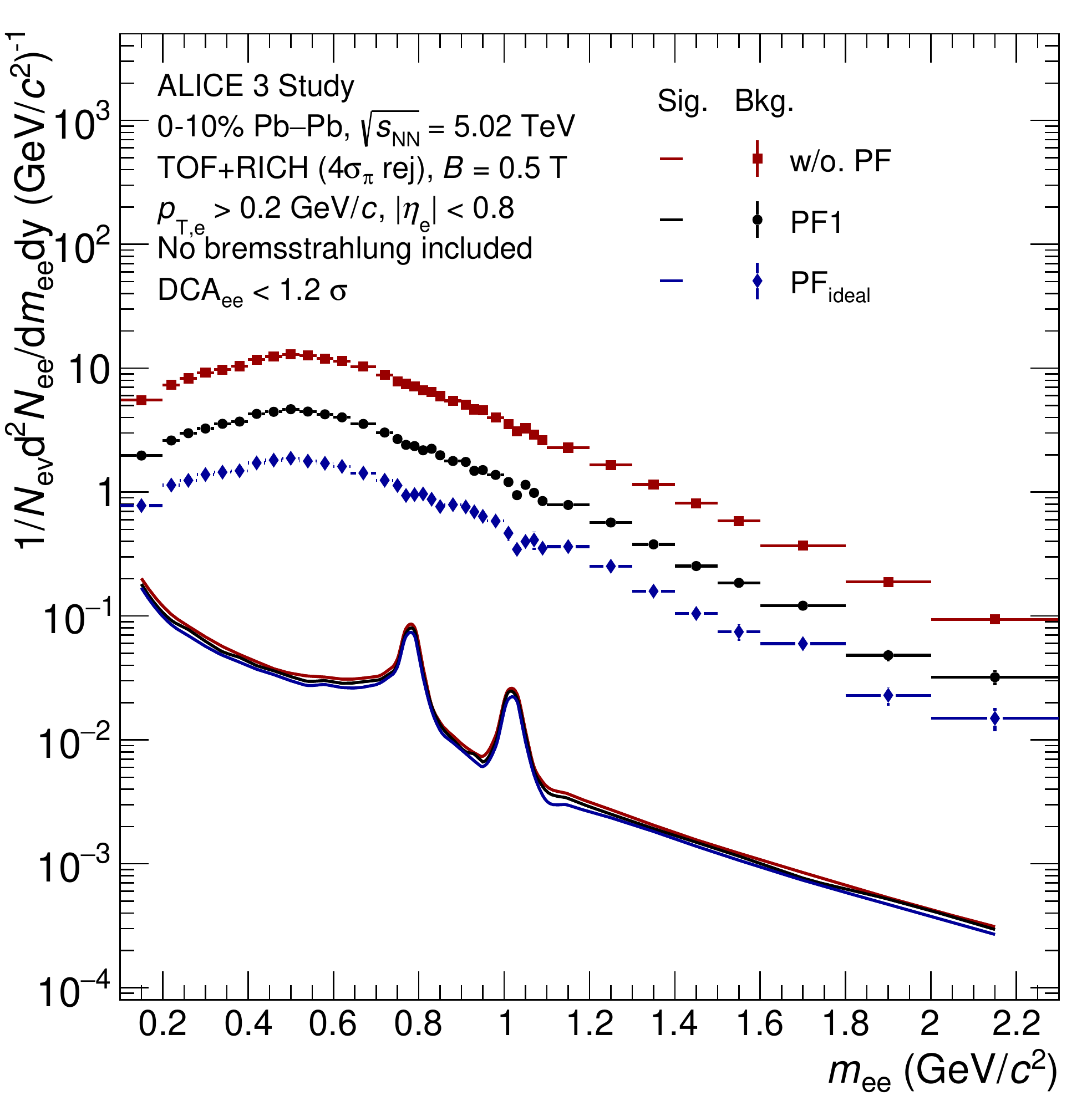}}
\caption[Expected raw signal dielectron yields and background]{Expected raw signal dielectron yield ($S$) and combinatorial background ($B$) using the outer TOF and RICH particle identification at midrapidity in central (0-10\%) Pb--Pb collisions at $\sqrt{s_{\rm NN}}$ $=$ 5.02 TeV using a TOF+RICH prefilter to reject electrons from Dalitz decays (PF1). Results without prefilter and with a prefilter based on MC information (PF$_\mathrm{ideal}$) are shown for comparison.}
\label{prefiltera}
\end{center}
\end{figure}

\begin{figure} [hb!]
\begin{center}
 \subfigure{\includegraphics[width=0.34\paperwidth]{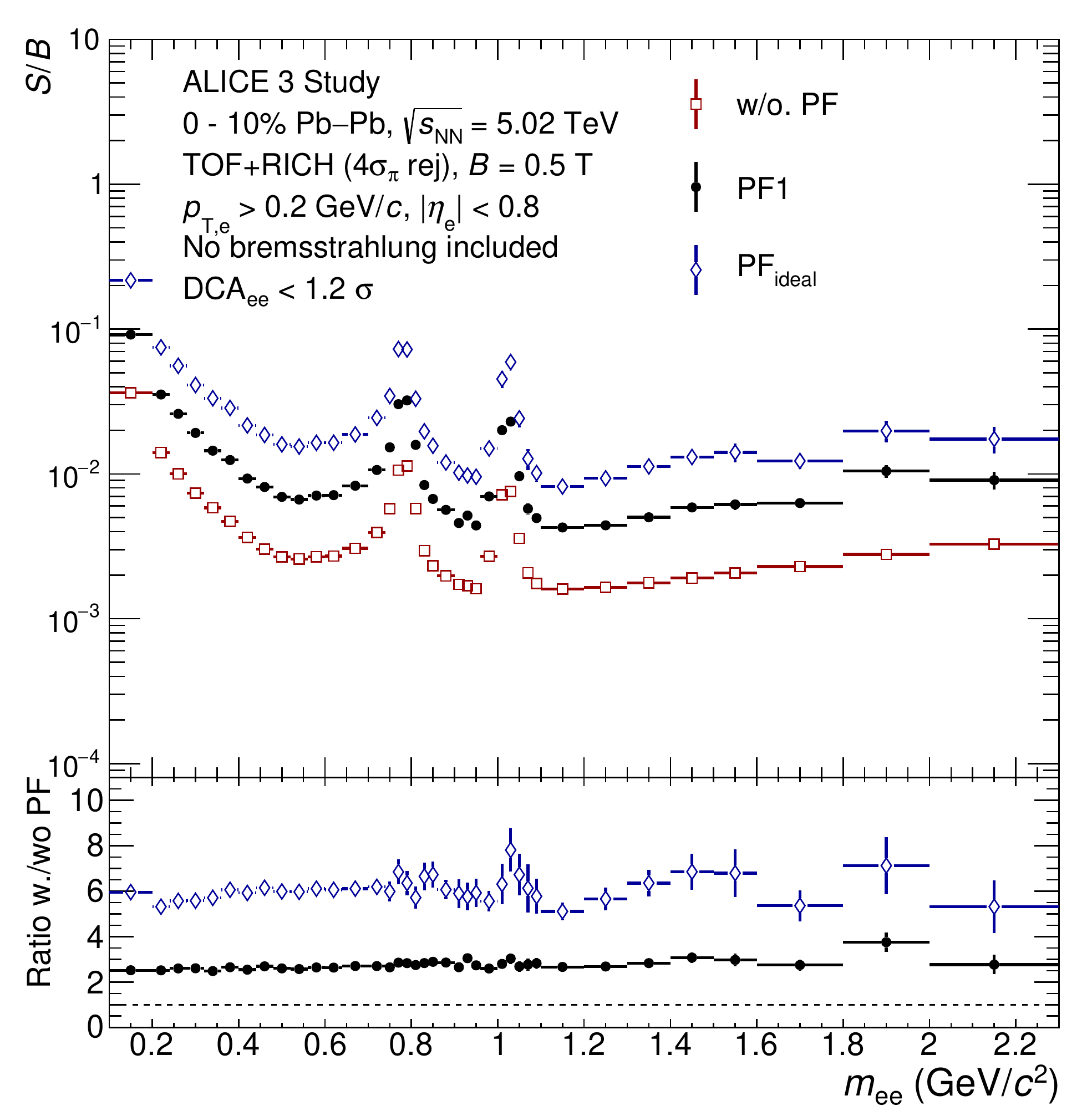}}
  \subfigure{\includegraphics[width=0.34\paperwidth]{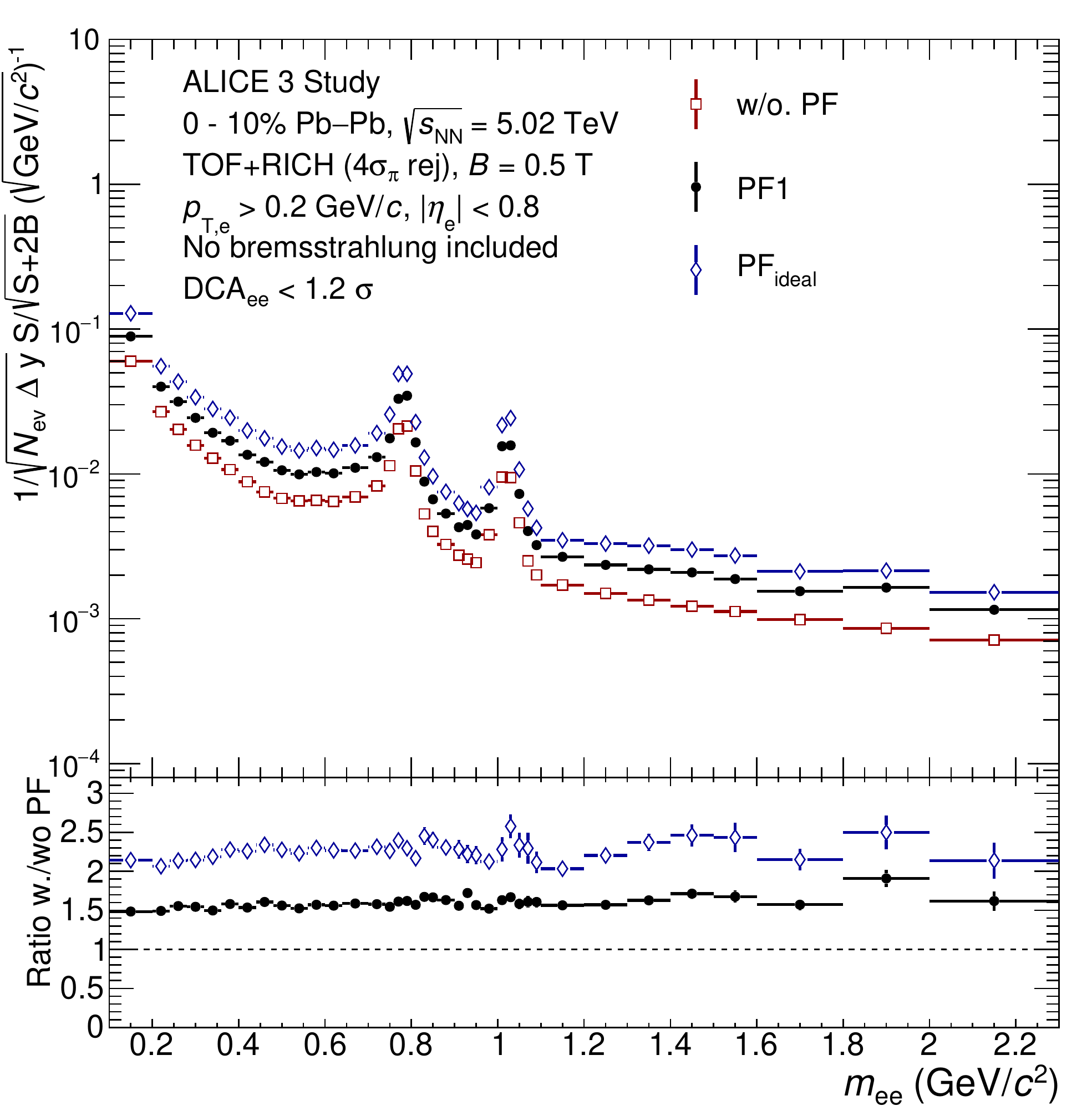}}
\caption[Expected signal to background and significance for dielectrons]{Simulated signal-to-background ratio (left) and significance ($S$/$\sqrt{S+2B}$) per event (right) as a function of $m_{\rm ee}$ using the outer TOF and RICH particle identification at midrapidity in central (0-10\%) Pb--Pb collisions at $\sqrt{s_{\rm NN}}$ $=$ 5.02 TeV using a TOF+RICH prefilter to reject electrons from Dalitz decays (PF1). Results without prefilter and with a prefilter based on MC information (PF$_\mathrm{ideal}$) are shown for comparison.}
\label{prefilterb}
\end{center}
\end{figure}

The expected raw signal dielectron spectrum in central \PbPb{} collisions at $\sqrt{s_{\rm NN}} = \SI{5.02}{\tera\eV}$, is shown in the left hand panel Fig.~\ref{dummychiralmixingplots}, with statistical uncertainties corresponding to an integrated luminosity of 5.6\,nb$^{-1}$.
Different sources of systematic uncertainties have been considered. The total uncertainty on the efficiency of tracking, track matching and electron identification is assumed to be a constant 5\% uncertainty at the pair level. We consider this value to be very conservative, based on our experience with ALICE.
A possible bias in the estimate of $B$ using the like-sign method could originate from differences in the detector acceptance for like-sign and unlike-sign pairs. To take such effects into account, a relative acceptance correction factor is calculated with dielectrons from mixed events and applied to the like-sign spectrum\,\cite{ALICE:2018fvj}. Owing to the 2$\pi$ azimuthal coverage of ALICE 3 this correction factor is very close to unity. Based on ALICE experience, an uncertainty of 0.02\% is assumed on this factor, leading to mass dependent uncertainty driven by the $S/B$ ratio.

\begin{figure} [hb!]
\begin{center}
  \subfigure{\includegraphics[width=0.34\paperwidth]{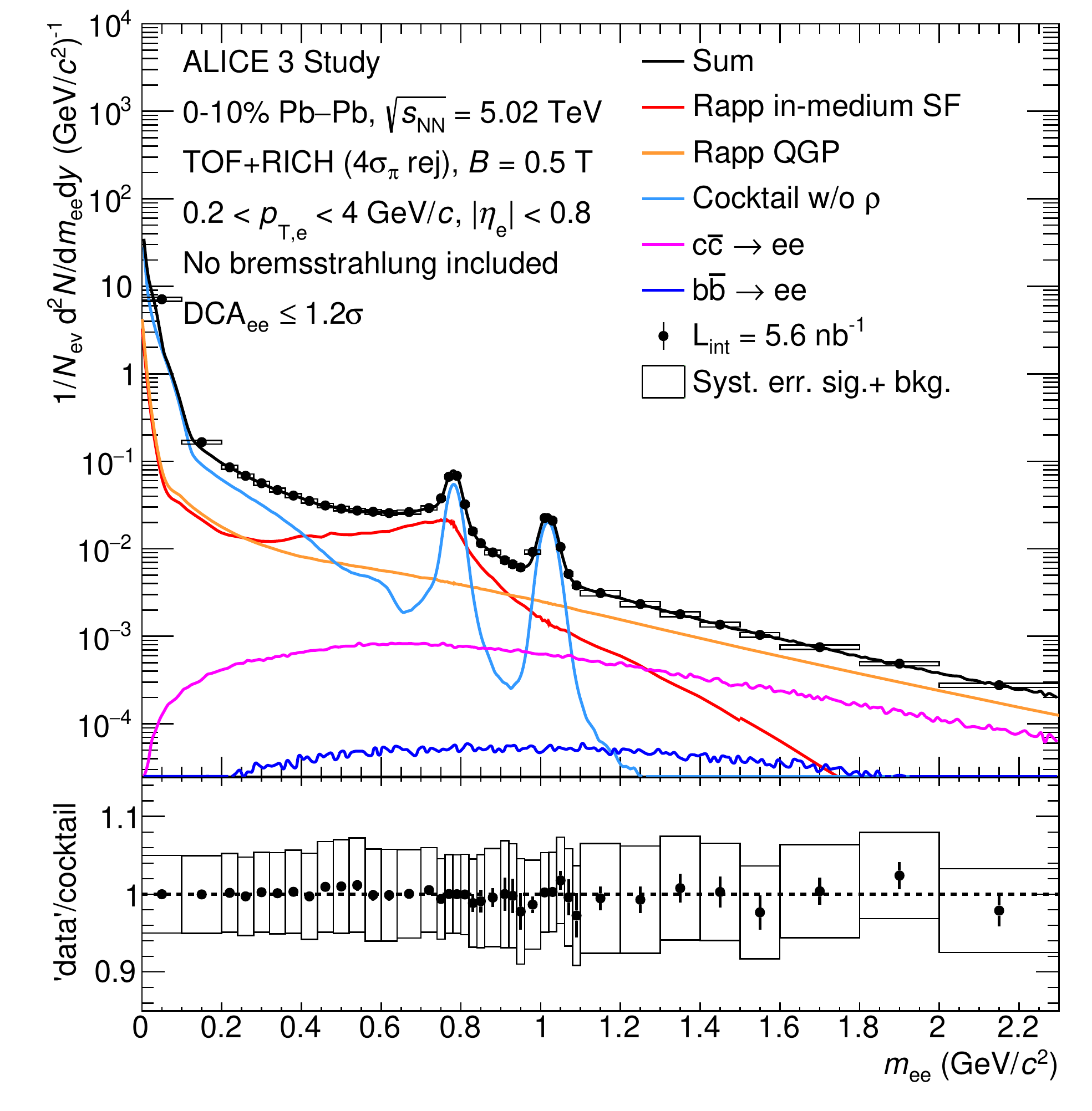}}
 \subfigure{\includegraphics[width=0.34\paperwidth]{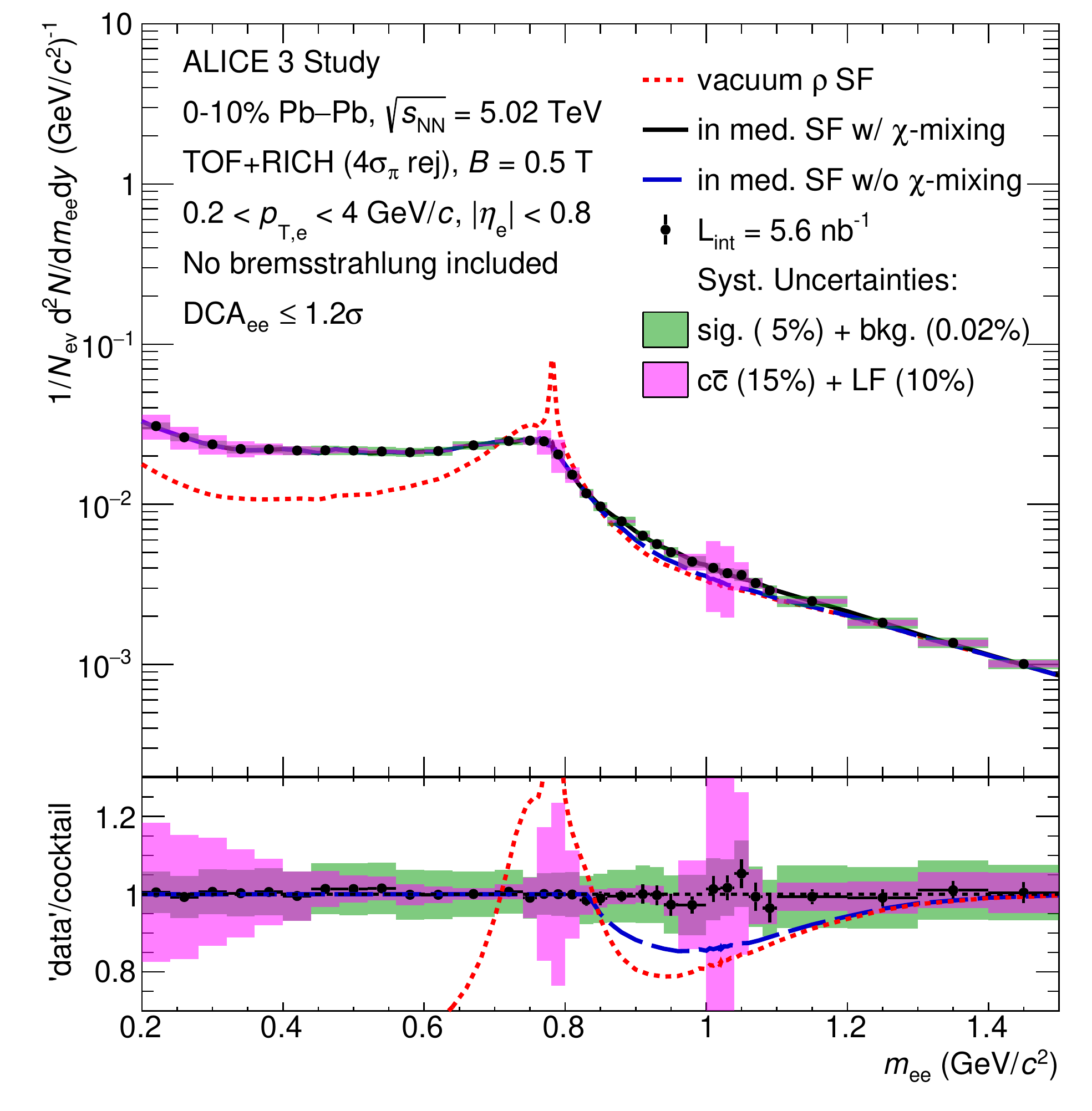}}
\caption[Simulated dielectron spectra]{Simulated raw signal spectra of inclusive dielectrons (left) and excess e$^{+}$e$^{-}$ pairs after subtraction of correlated light-hadron and heavy-flavour hadron decays (right) using the outer TOF and RICH particle identification at midrapidity in central (0-10\%) Pb--Pb collisions at $\sqrt{s_{\rm NN}} =  5.02$ TeV. The green or empty boxes show the systematic uncertainties from the combinatorial background subtraction and the tracking and electron identification. The magenta boxes (right) indicate systematic errors related to the subtraction of the light-flavour and heavy-flavour contributions. The excess spectrum is compared to predictions using different $\rho$ spectral functions (see text)\,\cite{Rapp:1999us,vanHees:2007th,Rapp:2013nxa,Hohler:2013eba}.}
\label{dummychiralmixingplots}
\end{center}
\end{figure}

In order to extract the thermal radiation yield and the $\rho$ spectral function, the hadronic cocktail and the residual background from heavy-flavour hadron decays are further subtracted from the raw spectrum to obtain an excess spectrum. 
The systematic uncertainties from the raw signal spectrum are propagated to the excess spectrum. In addition, relative systematic uncertainties of 10\% and 15\% are assumed on the light-hadron cocktail and residual heavy-flavour background, respectively. 

The expected measured thermal dielectron yield is shown in the right-hand panel of Fig.\ref{dummychiralmixingplots} around the $\rho$ mass and compared to calculations using the vacuum $\rho$ spectral function (dashed red line) and medium-modified spectral functions from a hadronic many-body approach with (black line) and without (dashed blue line) chiral mixing\,\cite{Hohler:2013eba}. 

The spectral function of low-mass dielectrons can be determined with a total uncertainty of about 6\% to 8\% for 0.4 $\leq$ $m_{\rm ee}$ $\leq$ 1.3\,GeV/$c^{\rm 2}$ in the 0-10\% most central \PbPb{} collisions, except in the $\omega$ and $\phi$ peak regions. 
Chiral mixing produces a change of 20-25\% compared to calculations performed with the vacuum spectral functions in the mass range from 0.8 to 1.2 \GeVcc{}. The high-precision measurement with ALICE~3 makes the observation of such an effect possible. Together with more differential measurements (see Section~\ref{sec:performance:physics:dileptons:azimuth} for $v_{\rm 2}$), this will provide strong constraints on the modification of the $a_{\rm 1}$ spectral function.

\begin{figure} [hb!]
\begin{center}
 \subfigure{\includegraphics[width=0.34\paperwidth]{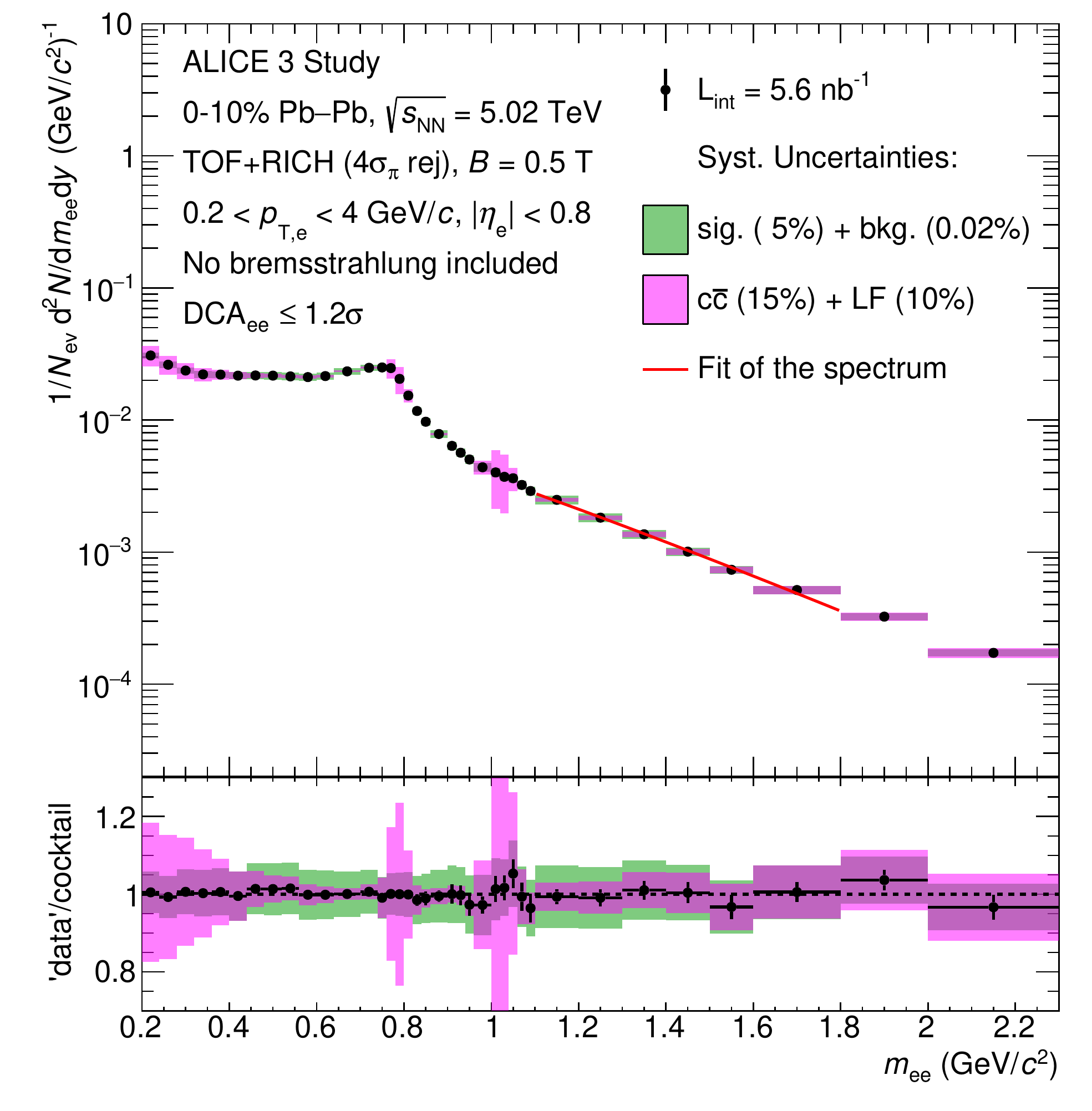}}
   \subfigure{\includegraphics[width=0.34\paperwidth]{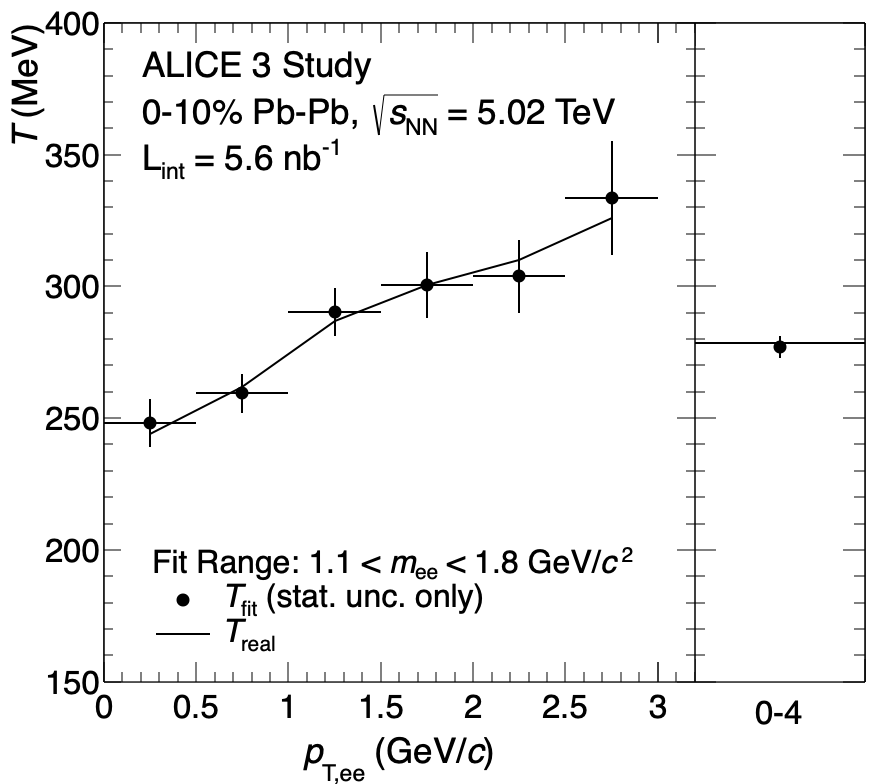}}
\caption[Simulated dielectron spectra at higher $m_{\rm ee}$]{Left: simulated raw spectra of excess e$^{+}$e$^{-}$ pairs fitted with an exponential function in the $m_{\rm ee}$ range 1.1-1.8\,\GeVcc{} to extract the early-time temperature $T_{\rm fit}$ of the medium in central (0-10\%) Pb--Pb collisions at $\sqrt{s_{\rm NN}}$ $=$ 5.02 TeV. The green boxes show the systematic uncertainties from the combinatorial background subtraction and the tracking and electron identification. The magenta boxes indicate systematic errors related to the subtraction of the light-flavour and heavy-flavour contributions. Right: extracted $T_{\rm fit}$ parameter after dielectron efficiency correction  compared to the input $T_{\rm real}$ (see text) for different selections in pair transverse momentum including the integrated case ($p_{\rm T,ee}$ $<$ 4 \GeVc). Only statistical errors are shown.}
\label{highmeeplots}
\end{center}
\end{figure}

The left-hand panel of Fig.~\ref{highmeeplots} shows the expected measurement of thermal dielectron yield over a larger mass range, up to $m_{\rm ee}$ of 2.3 \GeVcc{}. For large mass pairs, %
the systematic uncertainty arising from the remaining heavy-flavour background becomes dominant. At masses above 1\,\GeVcc{}, information on the early temperature of the medium can be extracted from the $m_{\rm ee}$ spectrum by fitting the distribution with an exponential function, ${\rm d}N_{\rm ee}/{\rm d}m_{\rm ee} \propto m_{\rm ee}^{3/2} {\rm exp}(-m_{\rm ee}/T_{\rm fit})$. To quantify the expected accuracy of the measurement, the fit parameter $T_{\rm fit}$ obtained for 1.1 $\leq$ $m_{\rm ee}$ $\leq$ 1.8\,\GeVcc{} is compared to $T_{\rm real}$ which is derived from the same fit to the thermal input spectrum. Both are shown after correcting for the dielectron efficiency in central Pb–Pb collisions in the right-hand panel of Fig.~\ref{highmeeplots} for different selections in pair transverse momentum including the integrated case ($p_{\rm T,ee} < 4$\,\GeVc). With ALICE~3, the early-time temperature of the system can be measured very precisely, with a statistical uncertainty of about 1.5\% compared to 4\% with the ALICE ITS 3~\cite{ALICE:ITS3:2019} when the thermal dielectron yield is integrated over $p_{\rm T,ee}$. Assuming fully correlated systematic uncertainties as a function $m_{\rm ee}$ for the background sources, as it was done in Ref.\,\cite{ALICE:ITS3:2019}, the total systematic error on $T^{p_{\rm T,ee} > 0}_{\rm fit}$ is expected to be of the order of 2\%. The improvement in statistical accuracy will enable a multi-differential analysis of $T_{\rm fit}$ as a function of $p_{\rm T,ee}$, as shown in the right-hand panel of Fig.\ref{highmeeplots}.

\paragraph{Azimuthal asymmetry}
\label{sec:performance:physics:dileptons:azimuth}

The elliptic flow of dielectrons in different $m_{\rm ee}$ and $p_{\rm T,ee}$ regions provides important information
to disentangle dielectron emission at early times of the collision from those produced later, once the medium already started to cool down. 

Following the strategy outlined above, the measured raw signal dielectron spectrum is simulated in semi-central (30-50\%) \PbPb{} collisions at $\sqrt{s_{\rm NN}}$ $=$ 5.02\,TeV and shown in the left panel of Fig.~\ref{mee3050figure}. For this differential study, an integrated luminosity of 35\,nb$^{\rm -1}$ was considered, corresponding to six years running. Electrons are identified with the outer TOF and RICH detectors in the rapidity range $|\eta_{e}|$ $\leq$ 1.75 for $p_{\rm T,e}$ $\geq$ 0.2 GeV/$c$. The relative contribution of thermal radiation decreases from central to peripheral collisions, and therefore only becomes dominant at slightly larger invariant mass. The elliptic flow of prompt correlated e$^{+}$e$^{-}$ pairs can be computed using the measured dielectron yields in- and out-of-plane, $N^{\rm INP}$ and $N^{\rm OOP}$, after subtraction of the residual heavy-flavour background based on the measured $DCA_{\rm ee}$ distributions, with the formula:
\begin{equation}
    v^{\rm prompt}_{\rm 2} = \frac{\pi}{4}\frac{1}{R_{\rm 2}}\frac{N^{\rm INP} - N^{\rm OOP}}{N^{\rm INP} + N^{\rm OOP}}
\end{equation}
where $R_{\rm 2}$ is the resolution of the reconstructed event plane. For small $v_{\rm 2}$ values, the absolute statistical uncertainty is quasi-independent of the value of the elliptic flow and mainly depends on the relative statistical uncertainty of the prompt dielectron yield. The expected $v^{\rm prompt}_{\rm 2}$ with its statistical uncertainty is shown with open black markers in the right-hand panel of Fig.~\ref{mee3050figure} as a function of $m_{\rm ee}$ for semi-central (30-50\%) Pb--Pb collisions at $\sqrt{s_{\rm NN}}$ $=$ 5.02\,TeV, assuming an event-plane resolution of 0.9. The absolute values of the elliptic flow are taken from the calculations in Ref.\cite{Vujanovic:2019yih}. The statistical uncertainty is smaller than 0.004 over the full $m_{\rm ee}$ range under consideration.

\begin{figure} [hb!]
\begin{center}
  \subfigure{\includegraphics[width=0.34\paperwidth]{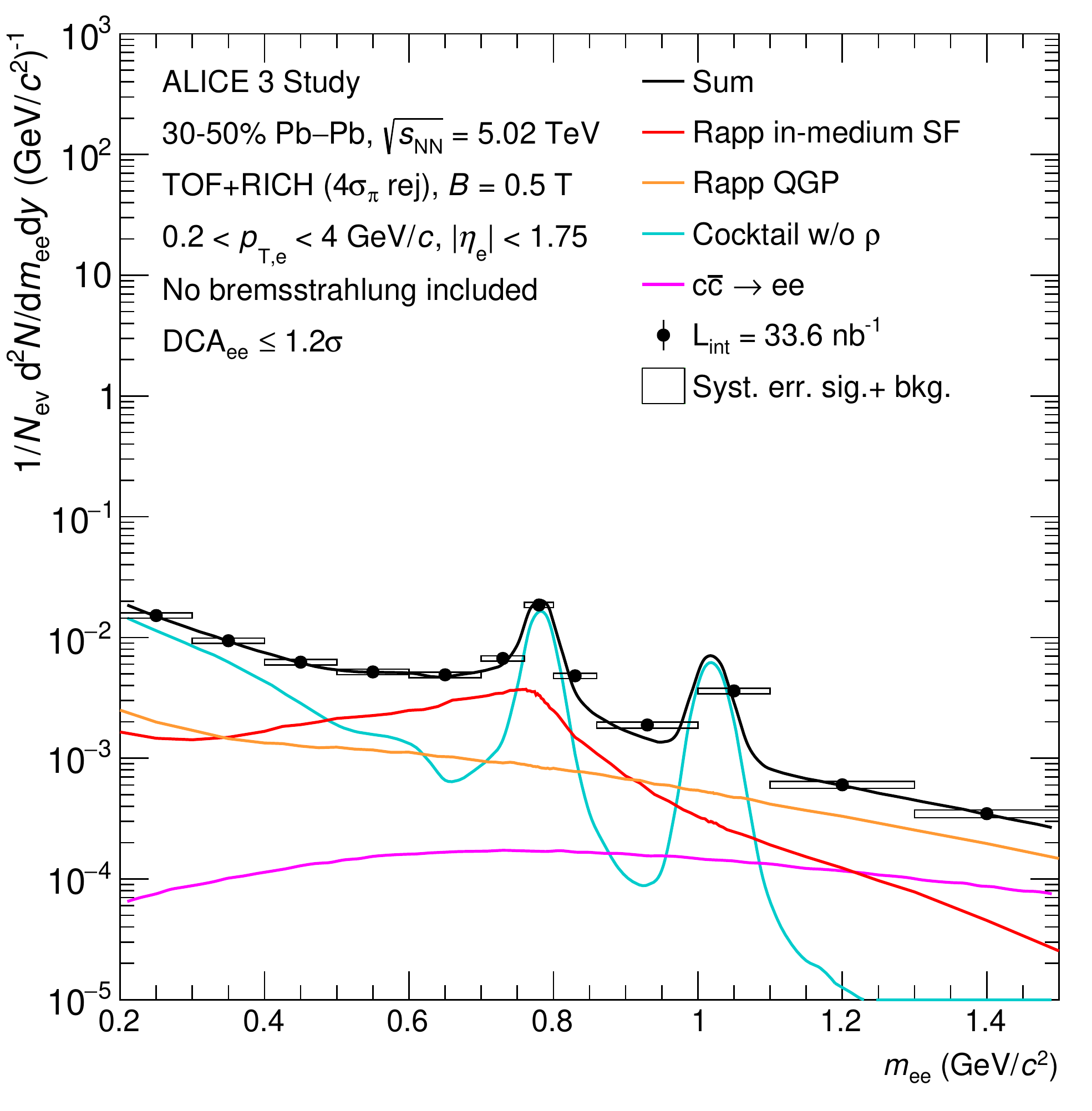}}
 \subfigure{\includegraphics[width=0.34\paperwidth]{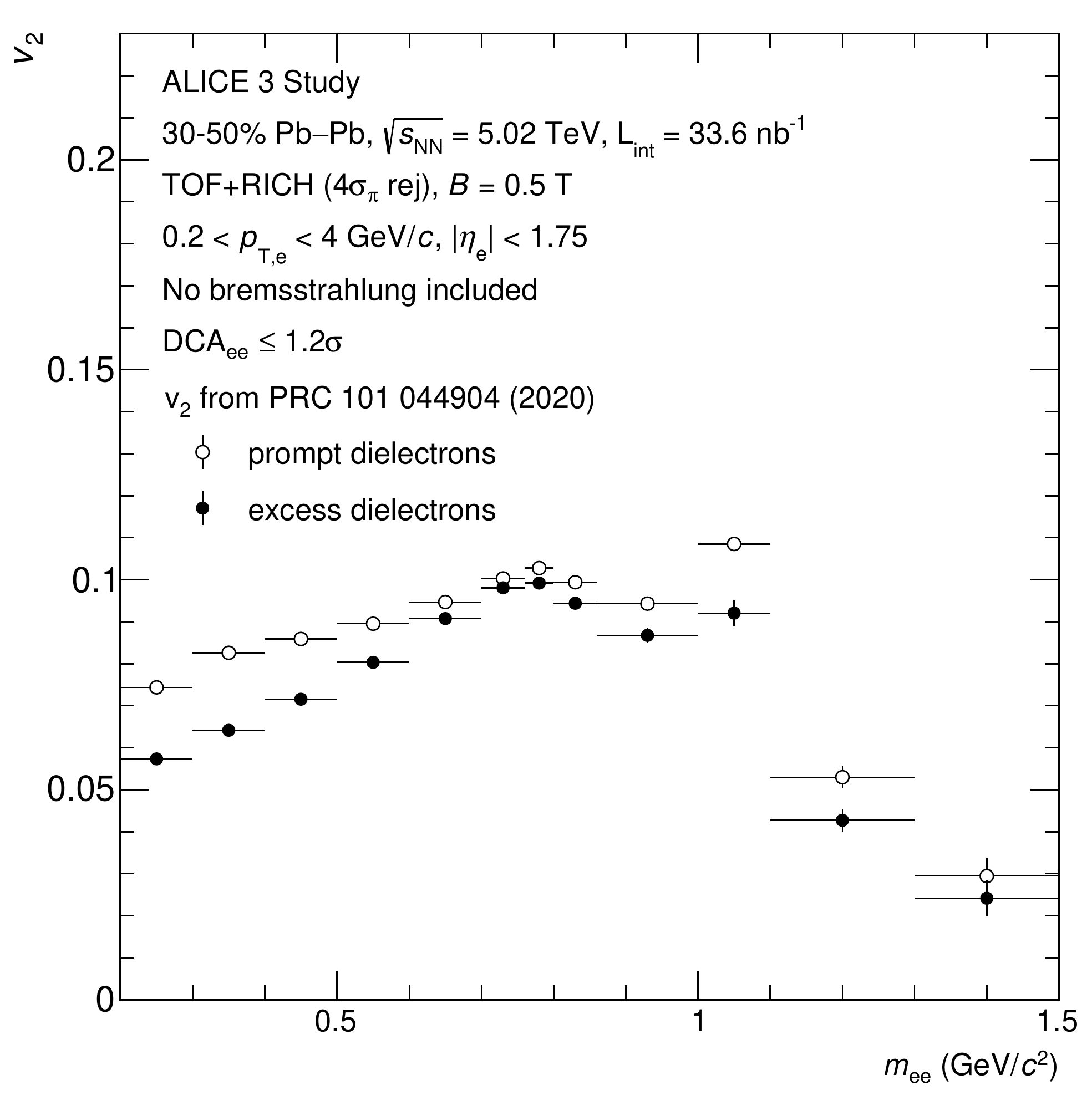}}
\caption[Expected elliptic flow of prompt and excess dielectrons]{Simulated raw signal dielectron yield (left) and expected elliptic flow of prompt and excess dielectrons (right) as a function of $m_{\rm ee}$ using the outer TOF and RICH particle identification at midrapidity for semi-central (30-50\%) Pb--Pb collisions at $\sqrt{s_{\rm NN}} = 5.02$ TeV and an integrated luminosity of $\Lint = \SI{35}{\nano\barn^{-1}}$.}
\label{mee3050figure}
\end{center}
\end{figure}

The prompt contribution from light-flavour hadron decays can be subtracted from $v^{\rm prompt}_{\rm 2}$ based on the yield and $v_{\rm 2}$ of the mother mesons from independent measurements and computing the corresponding $v^{LF}_{\rm 2}$ of decay electrons with a cocktail method. The elliptic flow of the excess spectrum is
\begin{equation}
    v^{\rm excess}_{2} = \frac{(1+N^{\rm excess}/N^{\rm LF})v^{\rm prompt}_{\rm 2} - v^{\rm LF}_{\rm 2}}{N^{\rm excess}/N^{\rm LF}},
\end{equation}
where $N^{\rm excess}$ and $N^{\rm LF}$ are the measured excess yield and calculated dielectron yield from known light-flavour hadron decays. The expected $v^{\rm excess}_{2}$\,\cite{Vujanovic:2019yih} with its statistical uncertainty is shown in solid black points as a function of $m_{\rm ee}$ in the right-hand panel of Fig.\ref{mee3050figure} for semi-central (30-50\%) Pb--Pb collisions. At low $m_{\rm ee}$ ($m_{\rm ee}$ $\leq$ 0.4 GeV/$c^{\rm 2}$), the systematic uncertainty originating from the light-flavour hadron subtraction is expected to become the dominant source of uncertainties.

\begin{figure} [hb!]
\begin{center}
  \subfigure{\includegraphics[width=0.34\paperwidth]{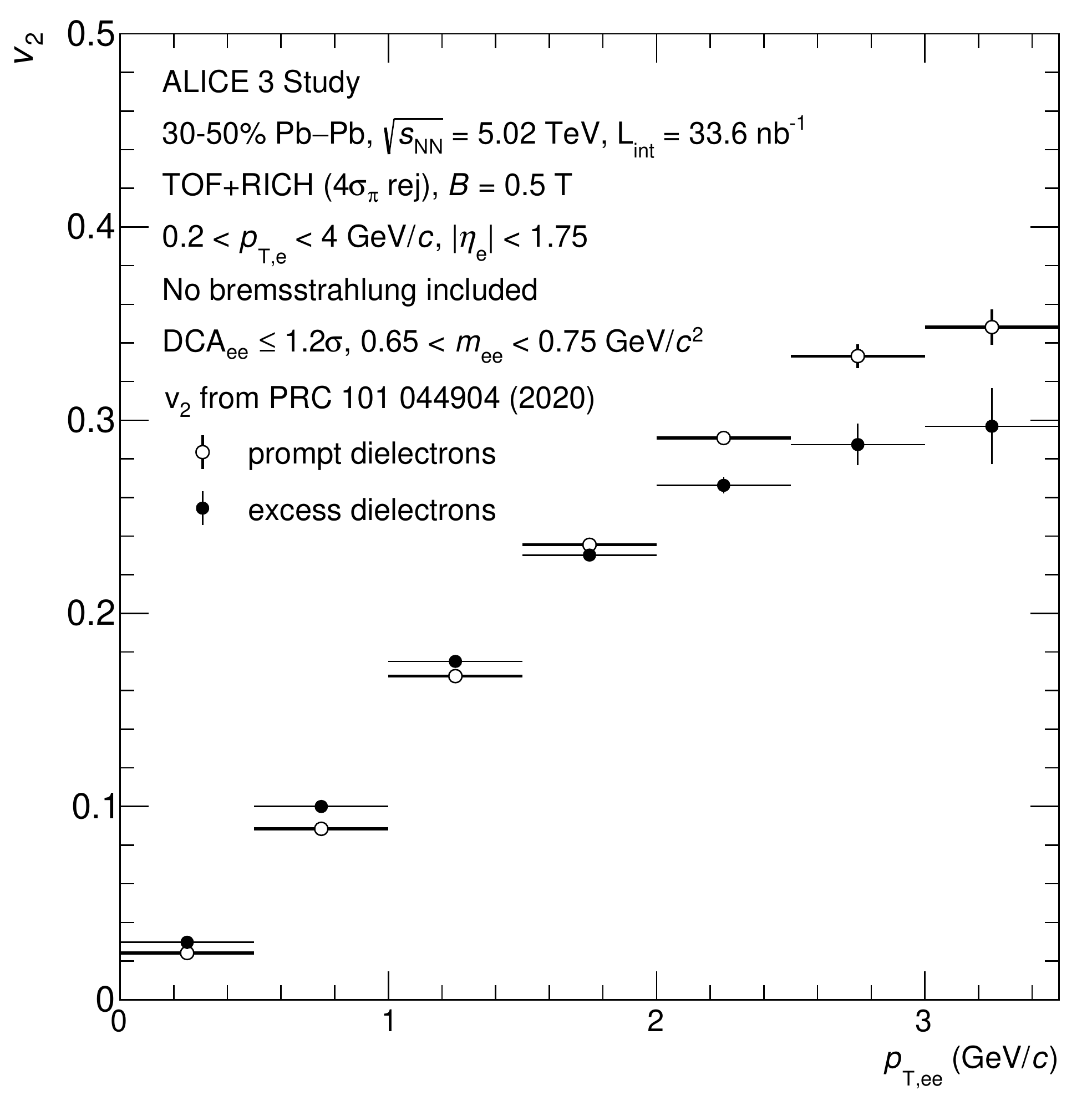}}
  \subfigure{\includegraphics[width=0.34\paperwidth]{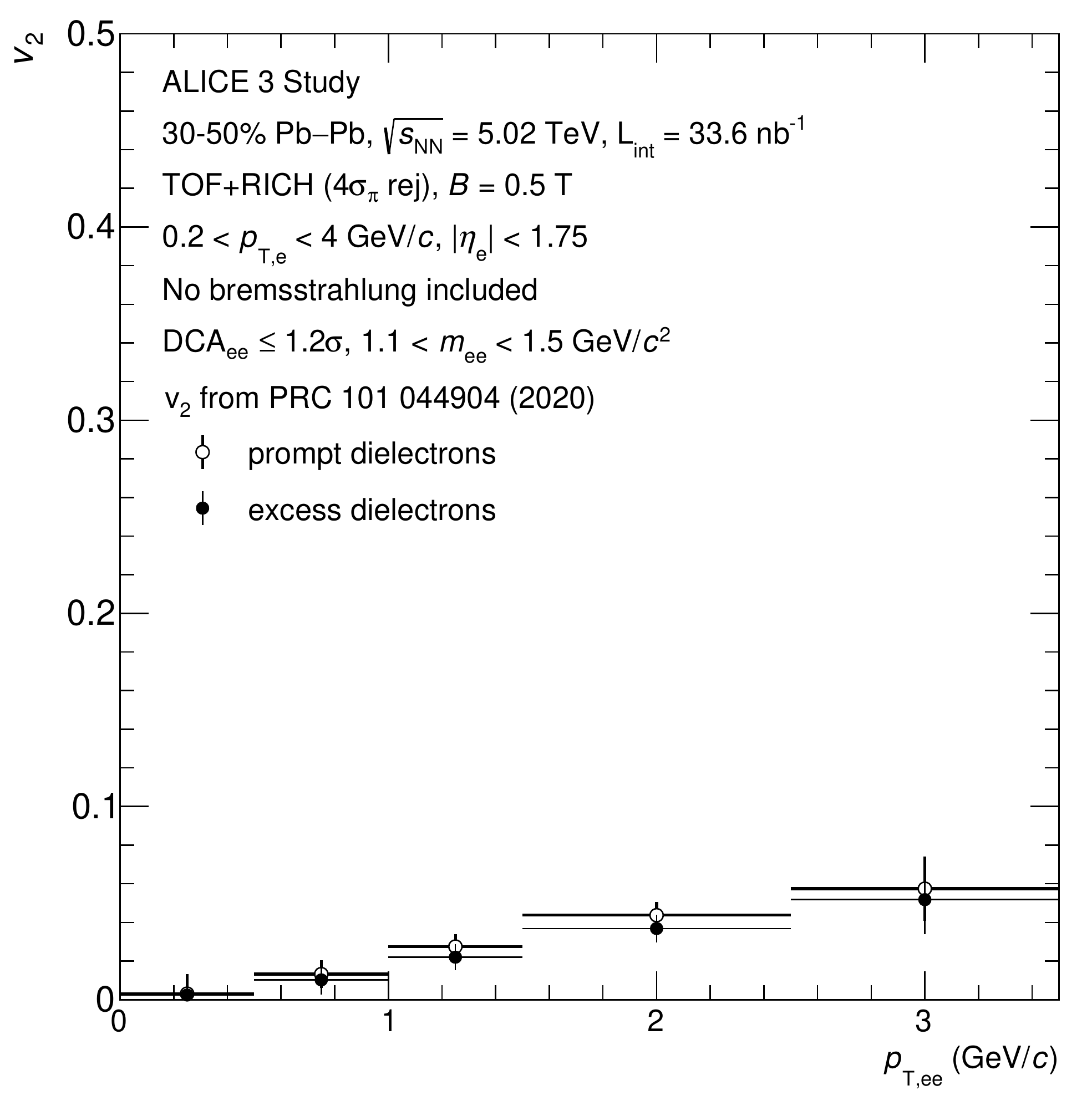}}
\caption[Expected elliptic flow of prompt and excess dielectrons]{Expected elliptic flow of prompt and excess dielectrons as a function of $p_{\rm T,ee}$ for 0.65 $\leq$ $m_{\rm ee}$ $\leq$ 0.75 GeV/$c^{2}$ (left) and 1.1 $\leq$ $m_{\rm ee}$ $\leq$ 1.5 GeV/$c^{2}$ (right) using the outer TOF and RICH particle identification at midrapidity for semi-central (30-50\%) Pb--Pb collisions at $\sqrt{s_{\rm NN}}$ $=$ 5.02 TeV and an integrated luminosity of $\Lint = \SI{35}{\nano\barn^{-1}}$.}
\label{ptee3050figure}
\end{center}
\end{figure}

For 0.65 $\leq$ $m_{\rm ee}$ $\leq$ 0.75 GeV/$c^{2}$ and 1.1 $\leq$ $m_{\rm ee}$ $\leq$ 1.5 GeV/$c^{2}$, thermal dielectrons dominate. The expected elliptic flow of prompt and excess dielectrons in these two mass ranges are shown with their statistical uncertainties as a function of $p_{\rm T,ee}$ in Fig.\ref{ptee3050figure}. The absolute elliptic flow values for 1.1 $\leq$ $m_{\rm ee}$ $\leq$ 1.5 GeV/$c^{2}$ were extrapolated from  calculations\,\cite{Vujanovic:2019yih} at $m_{\rm ee}$ = 1.5\,GeV/$c^{\rm 2}$ using the integrated $v_{\rm 2}$ as a function of $m_{\rm ee}$. Using the outer TOF and RICH detectors, the $p_{\rm T,ee}$ dependence of the excess dielectron $v_{\rm 2}$ can be measured up to about 2.5 GeV/$c$ with accuracy better than 0.005 at low $m_{\rm ee}$ and 0.01 at higher $m_{\rm ee}$.

For $m_{\rm ee} \geq 1$ GeV/$c^{\rm 2}$ and $p_\mathrm{T,ee} > 4\,\GeVc$, such measurements would profit from electron identification capabilities at higher momentum using different technologies like pre-shower or ECal detectors. In  addition, tracking and identification of very soft electrons with the inner TOF detector would improve the prefilter selection.

\paragraph{Comparison with ITS 3 and the other LHC experiments}~\\
\label{sec:performance:physics:dileptons:comparisonITS3}

The ALICE 3 detector will provide unique access to high precision measurements of dielectron production from prompt sources including thermal radiation from the medium over a large rapidity range at low invariant mass ($m_{\rm ee}$ $\leq$ $m_{\rm J/\psi}$) and low pair transverse momentum, i.e. down to $p_{\rm T,ee} = 0$ for $m_{\rm ee} \gtrsim 50$\,MeV/$c^{2}$ at midrapidity ($\eta \leq$ 1.75) or lower at forward rapidity. This phase-space region is not accessible by the CMS, ATLAS and LHCb experiments. 

The CMS and ATLAS collaborations measure dimuons and dielectrons at midrapidity ($\eta$ $\leq$ 2.5) with calorimeters and muon stations with limited reach to low \pt{} due to the amount of absorber material and the large magnetic fields (3.8 and \SI{2}{\tesla}). In ATLAS, energy loss in the calorimeters limits the range for muon detection to $p_{\rm T}$ $\geq$ 3\,GeV/$c$, which restricts the invariant masses to 6 GeV/$c^{2}$ and above for $p_{\rm T,l^{+}l^{-}} \leq 6$\,GeV/$c$\,\cite{ATLAS:muons}. The phase-space region at low mass reachable with identified electrons is similarly limited by a minimum transverse energy for electrons of 4.5\,GeV\,\cite{ATLAS:electrons}. In CMS, muons are reconstructed with two different algorithms, starting from segments in the muon chambers or from the information in the inner tracker. The latter allows a better reconstruction at low $p_{\rm T}$. Muons can be reconstructed down to lower $p_{\rm T}$ at larger rapidity compared to ATLAS\,\cite{CMS:muons}. This nevertheless still prevents the study of dilepton production at low $p_{\rm T,l^{+}l^{-}}$ for invariant mass up to $m_{\rm J/\psi}$\,\cite{CMS:jpsi}.

The LHCb collaboration has measured prompt (non-prompt) dimuons at forward rapidity 2 $\leq$ $\eta$ $\leq$ 4.5 for $p_{\rm T} \geq$ 1 (0.5)\,GeV/$c$ and $p \geq$ 20 (10)\, GeV/$c$ in pp collisions\,\cite{LHCb:dimuons}. Upgrades of the detector will allow such measurements also in colliding systems with higher charged-particle multiplicity, such as \PbPb collisions\,\cite{LHCb:upgrade}. However in the mass range $M_{\rm \rho}$ $\leq$ $m_{\rm \mu\mu}$ $\leq$ $m_{\rm J/\psi}$, backgrounds from misidentified prompt hadrons are sizable. Using same-sign dimuon candidates, the difference in the production rates of $\pi^{+}\pi^{-}$ and $\pi^{\pm}\pi^{\pm}$ leads to correction factors as large as a factor two near $m_{\rho}$\,\cite{LHCb:dimuons}. Additionally, electrons can be identified by a calorimeter system in LHCb\,\cite{LHCb:electrons}. Compared to muons, the tracking efficiency for electrons at low $p_{\rm T}$ is expected to be reduced since electrons can undergo radiative energy loss before the LHCb dipole magnet and be deflected outside of the downstream tracker acceptance.  

Finally, the ALICE upgrades in Run~4\,\cite{ALICE:ITS3:2019} (and Run~3) will significantly improve the capabilities of the ALICE detector to measure prompt dielectrons at small $m_{\rm ee}$ and $p_{\rm T,ee}$ at midrapidity ($|\eta| \leq$ 0.8) with a sizable increase in statistics with respect to the currently available data samples from Run 1 and 2. Moreover the correlated background from heavy-flavour hadron decays, which is an important background for the dielectron thermal signal in heavy-ion collisions, can be suppressed thanks to better impact parameter resolution. 

The left-hand panel of Fig.~\ref{fig:competitivitydielectron}  shows the raw signal dielectron yield in central \PbPb{} collisions with its different contributions, simulated as explained above is shown on the left-hand panel of for ALICE 3. To compare the rejection of e$^{+}$e$^{-}$ pairs from open heavy-flavour hadron decays, the contribution of such pairs for ALICE with the ITS 3 is also shown.
Correlated dielectrons from open-charm hadron decays are expected to be a dominant contribution to the measured raw dielectron spectrum for $m_{\rm ee}$ larger than about 1.5\,GeV/$c^{2}$ with ITS 3. 
The ALICE 3 tracker provides the possibility to suppress this contribution more effectively than ITS 3, so that the thermal radiation signal from the QGP and hadron gas\,\cite{Rapp:1999us,vanHees:2007th,Rapp:2013nxa} is the main source of e$^{+}$e$^{-}$ pairs up to high $m_{\rm ee}$. 
The right-panel of Fig.~\ref{fig:competitivitydielectron} shows the expected systematic uncertainty on the measured thermal spectrum arising from the subtraction of the correlated heavy-flavour background for both the ALICE 3 tracker and ITS 3 (see Section~\ref{sec:performance:physics:dileptons:chiral} for more details). 
In case of ITS 3, the precision of the measurement is still limited by the heavy-flavour background particularly at intermediate and high $m_{\rm ee}$, where thermal radiation from the QGP and the contribution from the pre-hydrodynamic phase can be studied. 
For this comparison the  same uncertainty of 15\% on the heavy-flavour contribution was assumed for ALICE 3 and ITS3. It is expected that the ALICE 3 capabilities to measure open heavy-flavour hadrons will further reduce this uncertainty, leading to a larger overall improvement of the systematic uncertainties with ALICE 3. In addition, dielectron measurements with ALICE in Run 3 and 4 are constrained to a relatively small rapidity window with a low $p_{\rm T,ee}$ coverage down to 0 for $m_{\rm ee} \geq$ 150\,MeV/$c^{2}$. The tracking and electron identification capabilities of ALICE 3 at very low transverse momenta over a large rapidity range provide the unique opportunity to study the production of very soft dielectrons and access to medium properties, such as the electrical conductivity (see Section~\ref{sec:physics:qgp_properties:el_conductivity} for more details). At the same time, it increases the effectiveness to suppress the combinatorial background with prefilter techniques as explained in Section~\ref{sec:performance:physics:dileptons:chiral}.     

\begin{figure}
\centering
\includegraphics[width=.48\linewidth]{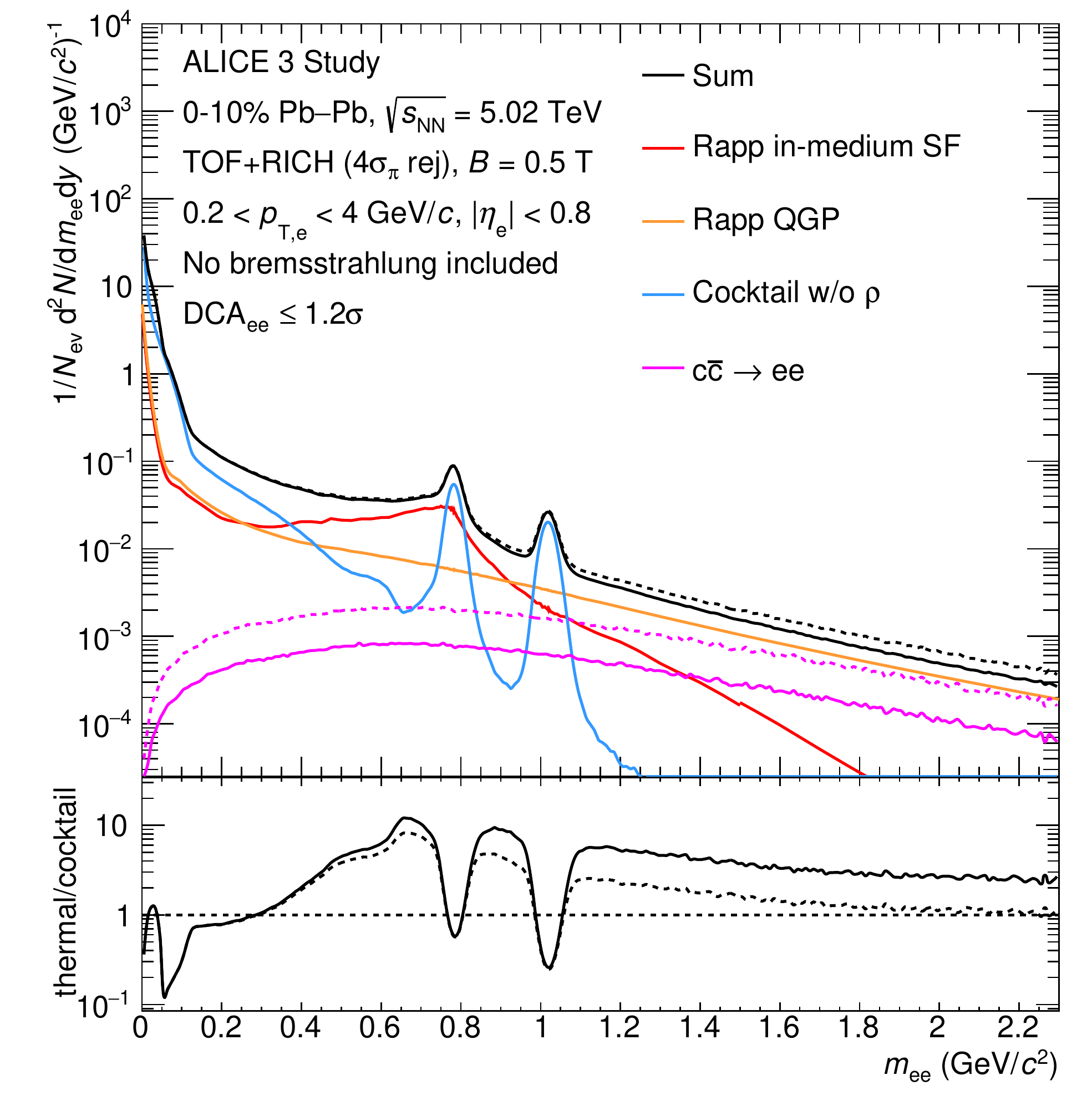}
\includegraphics[width=.48\linewidth]{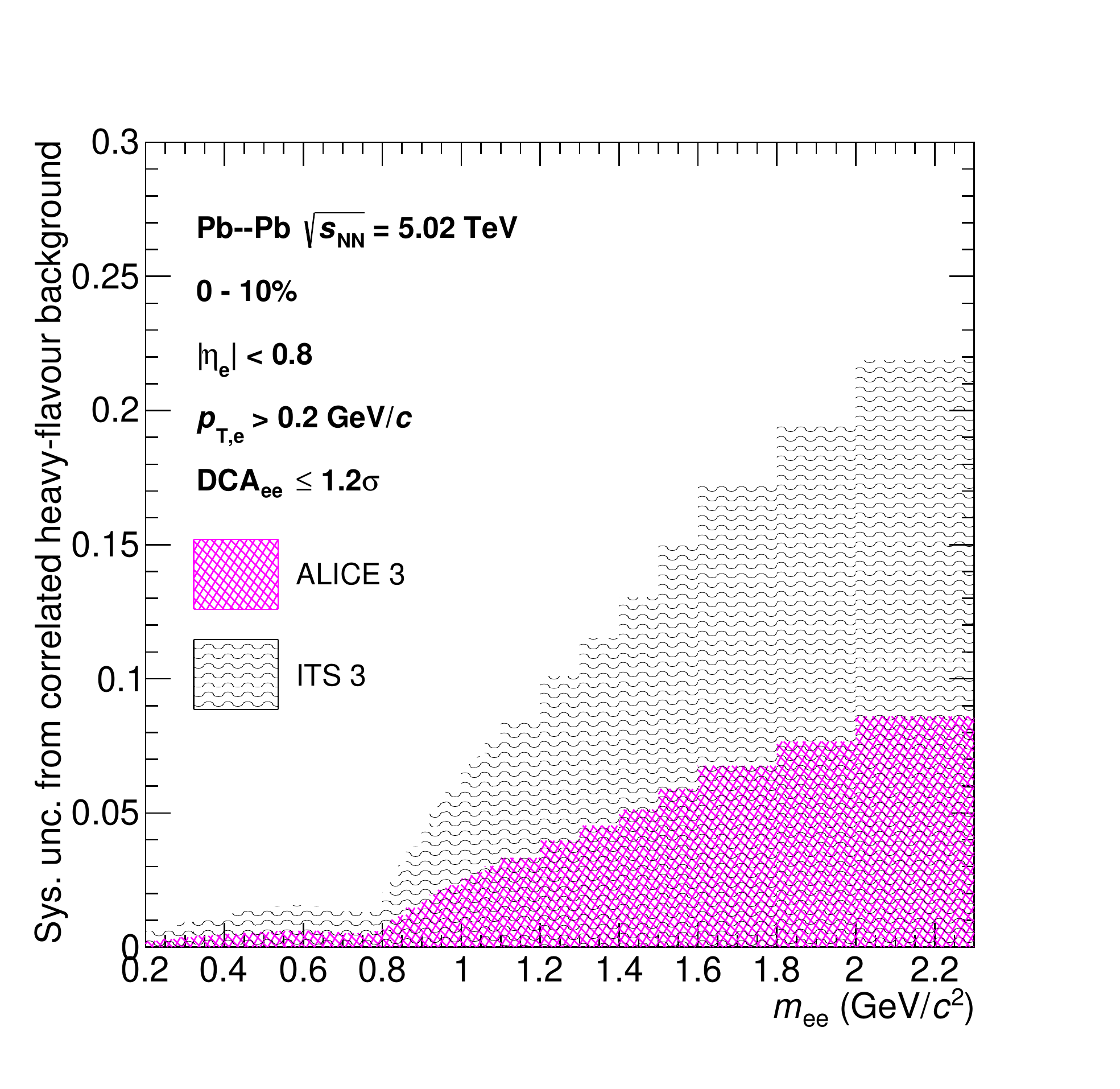}
\caption[ALICE 3 and ALICE ITS 3 performances]{Left: Expected raw dielectron yield at midrapidity in 0-10\% most central Pb--Pb collisions at $\sqrt{s_{\rm NN}}$ $=$ 5.02 TeV as a function of invariant mass $m_{\rm ee}$ using the impact parameter resolution of ALICE 3 or ALICE ITS 3 to reject the correlated open-charm background. The ratio of expected thermal dielectrons over those from hadronic decays is shown in the lower part of the figure. The dashed lines correspond to the expectations for ITS3. The performance of the ITS 3 is shown in dashed lines. Right: Corresponding systematic uncertainty on the measured thermal spectrum from hadron gas and QGP arising from the subtraction of the heavy-flavour background.}
\label{fig:competitivitydielectron}
\end{figure}

\subsubsection{Neutral pions and $\eta$ mesons at low $p_T$}

With ALICE 3 we aim at measuring both real and virtual thermal photons created in the Quark-Gluon Plasma. The reconstruction of real photons through conversions into $e^+e^-$ pairs in the detector material provides a good energy resolution at low $p_T$ ($\lesssim 3~\mathrm{GeV}/c$) where a signal of thermal photons is expected. Experimentally one reconstructs a signal of direct photons by subtracting a calculated decay photon spectrum from the $p_T$ spectrum of all photons. The largest contribution to the decay photon spectrum comes from the two-photon decays of neutral pions and $\eta$ mesons. Significantly reduced uncertainties of the measured $\pi^0$ and $\eta$ spectra at low $p_T$ are therefore crucial for a high-precision direct-photon measurement. 

ALICE 3 allows us to measure neutral mesons via photon conversions over a wide rapidity range. Of particular interest is the forward region $1.75 < |y| < 4.0$. Low-$p_T$ particles in this rapidity range have a sizeable momentum. The photon conversion measurement is therefore not affected by the drop of
the photon conversion cross section at low photon energies. Moreover, the electrons and positrons created in the photon conversions traverse the forward tracking detectors with high probability. This results in a large detection efficiency for neutral pions and $\eta$ mesons a forward rapidities.

The expected detection efficiencies for neutral pions and $\eta$ mesons at midrapidity and at forward rapidities measured via photon conversions are shown in Fig.~\ref{fig:Pi0Eta}. The drop in the efficiencies at low $p_T$ is much less pronounced in the forward region than at midrapidity for both $\pi^0$'s and $\eta$'s. Overall the efficiency is more than a factor of ten higher in the forward region. The more uniform efficiency over the complete $p_T$ range in the forward region is expected to result in smaller systematic uncertainties in the low $p_T$ region. The benefit of measuring  neutral meson in the forward region in terms of better performance and lower $p_T$ reach was also observed in~\cite{Agakichiev:1998ign}. We also expect an improved statistical uncertainty in the forward region. First studies show neglibible statistical errors in the low $p_T$ region of the order of 10$^{-4}$ and 10$^{-3}$ for neutral pions and $\eta$ meson, respectively, for an integrated Pb--Pb luminosity of 35 nb$^{-1}$. Overall, a smaller statistical uncertainties by a factor of about 2 compared to the midrapidity $|y| < 1.3$ is expected for the two meson species. In summary, this study indicates that the large rapidity coverage of ALICE 3 allows us to significantly increase the precision of low-$p_T$ $\pi^0$ and $\eta$ measurement which in turn results in a more significant measurement of direct photon in the $p_T$ range in which thermal photons are expected.
\begin{figure}[hbt]
\centerline{\includegraphics[width=0.6\textwidth]{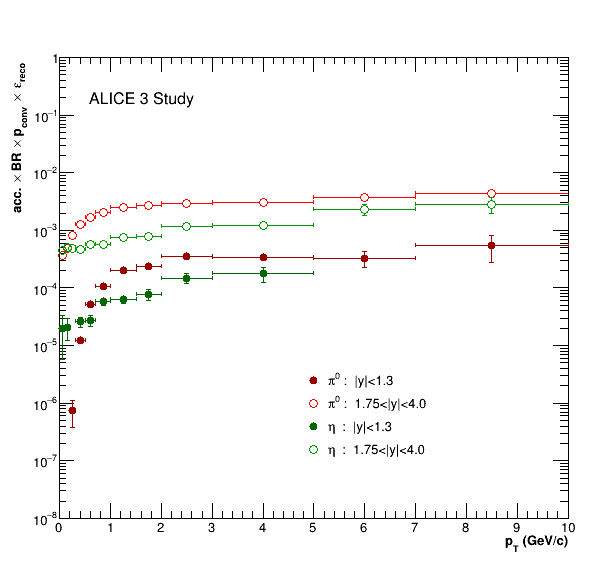}}
\caption[Total detection efficiency for neutral pions and $\eta$ mesons]{Total detection efficiency for neutral pions and $\eta$ mesons as a function of transverse momentum at mid- ($|y|<$1.3) and forward rapidity (1.75 $< |y| <$ 4.0). The total detection efficiency is defined as the product of the geometrical acceptance, the branching ratio of the two-photon decay, the probability that both photons convert in the detector material, and the reconstruction efficiency for the resulting electron and positron tracks.}
\label{fig:Pi0Eta}
\end{figure}

\subsubsection{Fluctuations of conserved charges}
\label{sec:perf:fluctuations}

In event-by-event net-charge fluctuation measurements, kinematic acceptance plays a significant role. To compare predictions from lQCD to experimental observations, the requirements of the Grand Canonical Ensemble (GCE) used in lQCD have to be mapped to experiments. Experimentally, fluctuations are investigated by analyzing the experimental data in a finite acceptance by imposing cuts on the rapidity and transverse momentum of detected particles. 
However, if the selected acceptance window is too small, Poissonian fluctuations will be come dominant and, consequently, net-baryon number will be distributed according to the Skellam distribution. On the other hand, by enlarging the acceptance, one can probe not only the critical fluctuations resulting from the phase transition but also dynamical fluctuations stemming from the different origins, such as due to baryon number conservation~\cite{Braun-Munzinger:2019yxj,Bzdak:2012an}, volume fluctuations~\cite{Braun-Munzinger:2016yjz}, thermal blurring~\cite{Ohnishi:2016bdf}, resonance decays~\cite{Arslandok:2020mda}, and so on. None of these are contained in lQCD calculations. Therefore, differential measurements of net-charge fluctuations in a wide kinematic acceptance are crucial to distinguish critical fluctuations from these contributions. Figure~\ref{fig:etadist} shows the increase in acceptance of ALICE 3 compared to that of ALICE 1-2 in view of the analysis of net-proton cumulants; from $0.6<p<1.5$~GeV/\textit{c} to $0.3<p<10$~GeV/\textit{c} and from $|\eta|<0.8$ to $|\eta|<4$, i.e.\ about a factor 10 increase in the number of measured (anti-)protons. 

\begin{figure}[htbp]
	\centering
	\includegraphics[width=9cm]{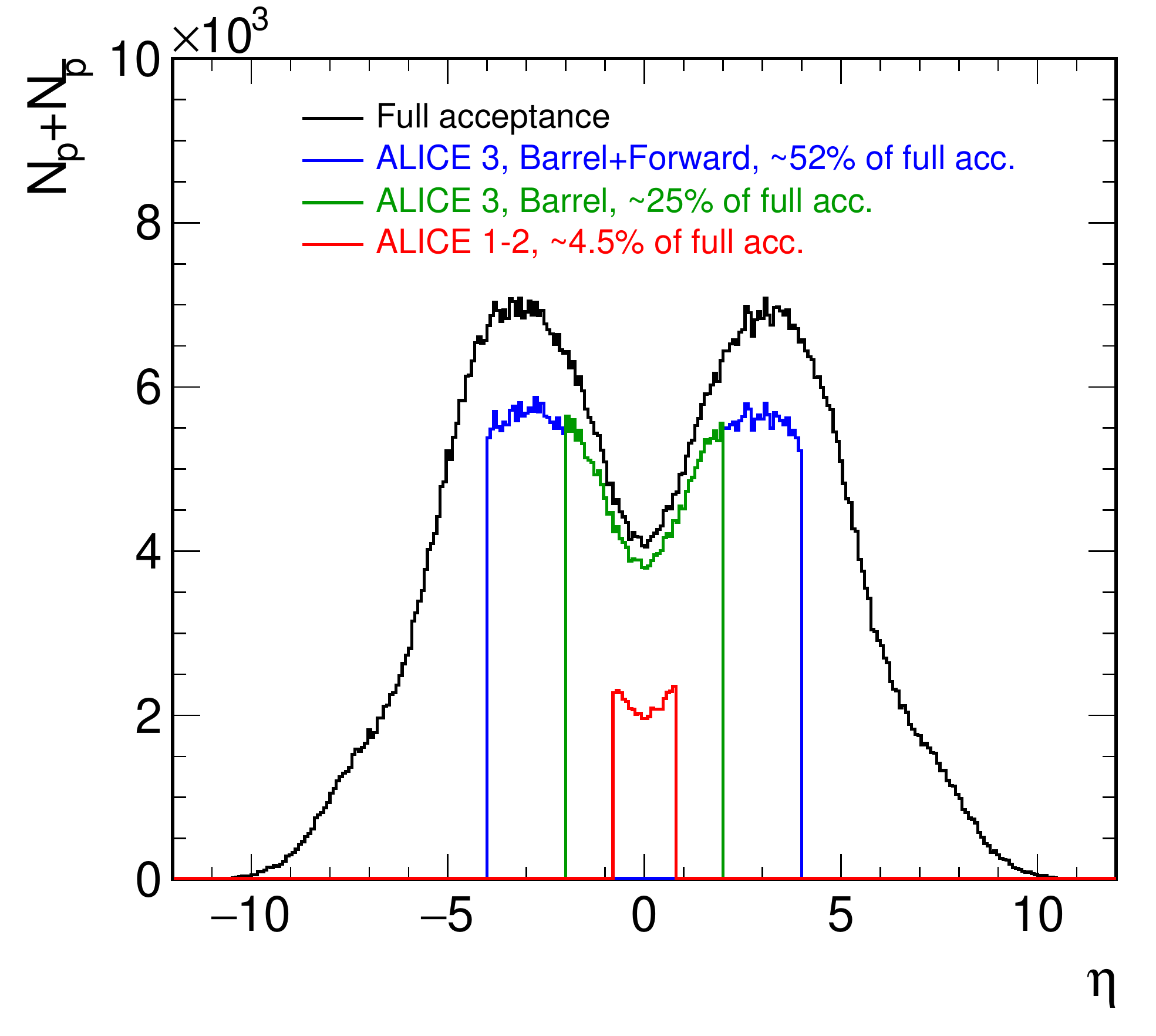}
	\caption[Pseudorapidity acceptance for protons]{(Colour online) Pseudorapidity distribution of (anti-)protons as simulated by the HIJING model~\cite{Gyulassy:1994ew} in central Pb–Pb collisions at $\sqrtsNN = 5.02$~TeV. Different lines correspond to kinematic acceptances of ALICE 1-2 and ALICE 3 periods.}
	\label{fig:etadist}
\end{figure}

One should note that, to remove the volume and temperature terms, which enter into \autoref{eq:mixed_cumulant} (in Section~\ref{sec:phys:fluctuations}), the charge susceptibilities are studied experimentally in terms of the ratios of cumulants~\cite{Gavai:2010zn}. Note that the volume term only cancels out if one first removes its fluctuation, as detailed in ~\cite{Braun-Munzinger:2016yjz}.  
The cumulant  ratio $\kappa_{n+1}/\kappa_n = 0$  while $\kappa_{n+2}/\kappa_{n} = 1$  for even $n$ as long as the fluctuations are Poissonian fluctuations, resulting an a Skellam distribution for the net baryon number. This leads to the so-called "Skellam" baseline\footnote{The Skellam distribution is defined as the probability distribution of the difference of two random variables, each generated from statistically independent Poisson distributions.}. To illustrate the advantages offered by the larger acceptance of ALICE 3, Fig.~\ref{fig:secondMomProj} shows the expected dependence of the second order cumulants of net-protons on the pseudorapidity acceptance, using the transverse momentum acceptance of ALICE 2 and ALICE 3. Due to small kinematic acceptance and large systematical uncertainties in the ALICE 1-2, it is difficult to draw a quantitative conclusion on the correlation length. The significantly larger acceptance of ALICE 3 will make it possible to establish deviations from the Skellam baseline in more detail, which will allow to quantitatively constrain the correlation length for net baryon, strangeness and - completely new - charm number.

\begin{figure}[htbp]
	\centering
	\includegraphics[width=9cm]{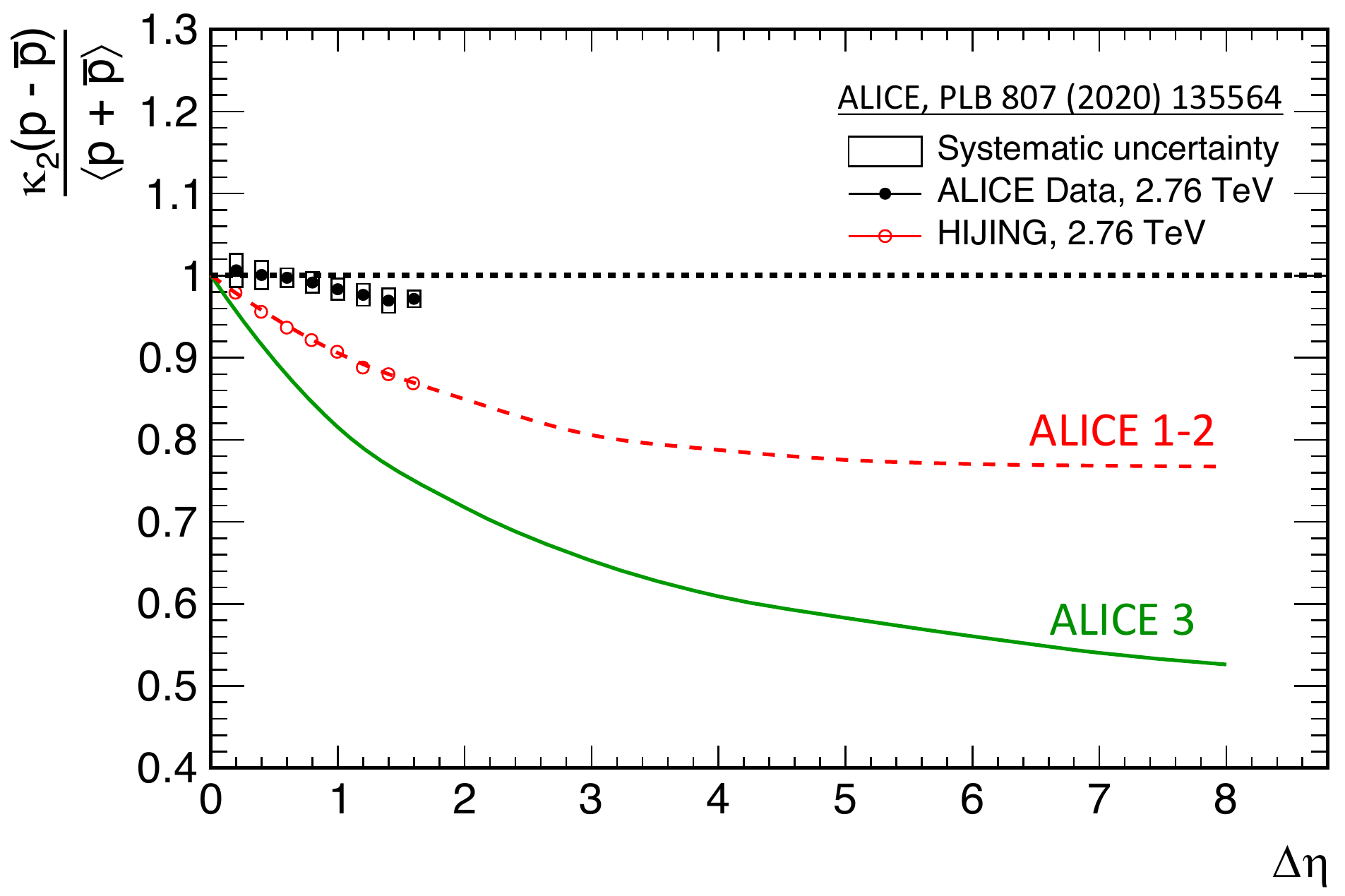}
	\caption[Pseudorapidity dependence of the normalized second cumulants of net-protons]{(Colour online) Pseudorapidity dependence of the normalized second cumulants of net-protons. The ALICE data and HIJING model calculations at $\sqrtsNN = 2.76$~TeV for 0-5\% central collisions  are shown by black and red markers, respectively, where the kinematic acceptance is $0.6<p<1.5$~GeV/\textit{c}~\cite{ALICE:2019nbs}. The solid green line indicates the HIJING model calculations with the ALICE 3 kinematic acceptance of $0.3<p<10$~GeV/\textit{c} and the dashed lines is the extrapolation of the HIJING model calculations to pseudorapidity acceptance of the ALICE 3 for the momentum range of $0.6<p<1.5$~GeV/\textit{c}.}
	\label{fig:secondMomProj}
\end{figure}

More than a factor 100 larger signal-to-background ratio and a factor 2 higher detection efficiencies for $D$ mesons, as well as the large acceptance of ALICE~3, allow the measurement of net-charm fluctuations with $D$ mesons.  These measurements are out of reach with the current ALICE detector. In the case of net-proton, the number of detected $p\bar{p}$ pairs per event is larger than one in almost all events at LHC energies. However, for the $D$ mesons, it is on the level of 0.01. Therefore, the formulas in~\cite{Luo:2011tp} are not applicable for the statistical error estimation. To calculate the statistical uncertainties, a dedicated Monte-Carlo simulation study had to be performed. Since lQCD predictions agree with the statistical hadronisation model (taking into account missing states (see \autoref{fig:BS-BC})), Poissonian statistics was assumed for the production of $D$ mesons. The average yield information was taken from~\cite{Andronic:2021erx} and events without $D\bar{D}$ pairs were not included in the cumulant calculations. The resulting relative error as a function of the event statistics is shown in Fig.~\ref{fig:K4K2_Proj_DDbar_v2}. Note that these results are based on the yields before efficiency correction, which increases the statistical errors by a factor of 5. This was studied by comparing the cumulants before and after accounting for the efficiency loss. Based on these results, the expected precision for measurements of $D\bar{D}$ fluctuations for the 4th order cumulants is of the order of a few \%. This will allow high precision measurements of the charm conservation and comparison to lQCD predictions.

\begin{figure}[htbp]
	\centering
	\includegraphics[width=9cm]{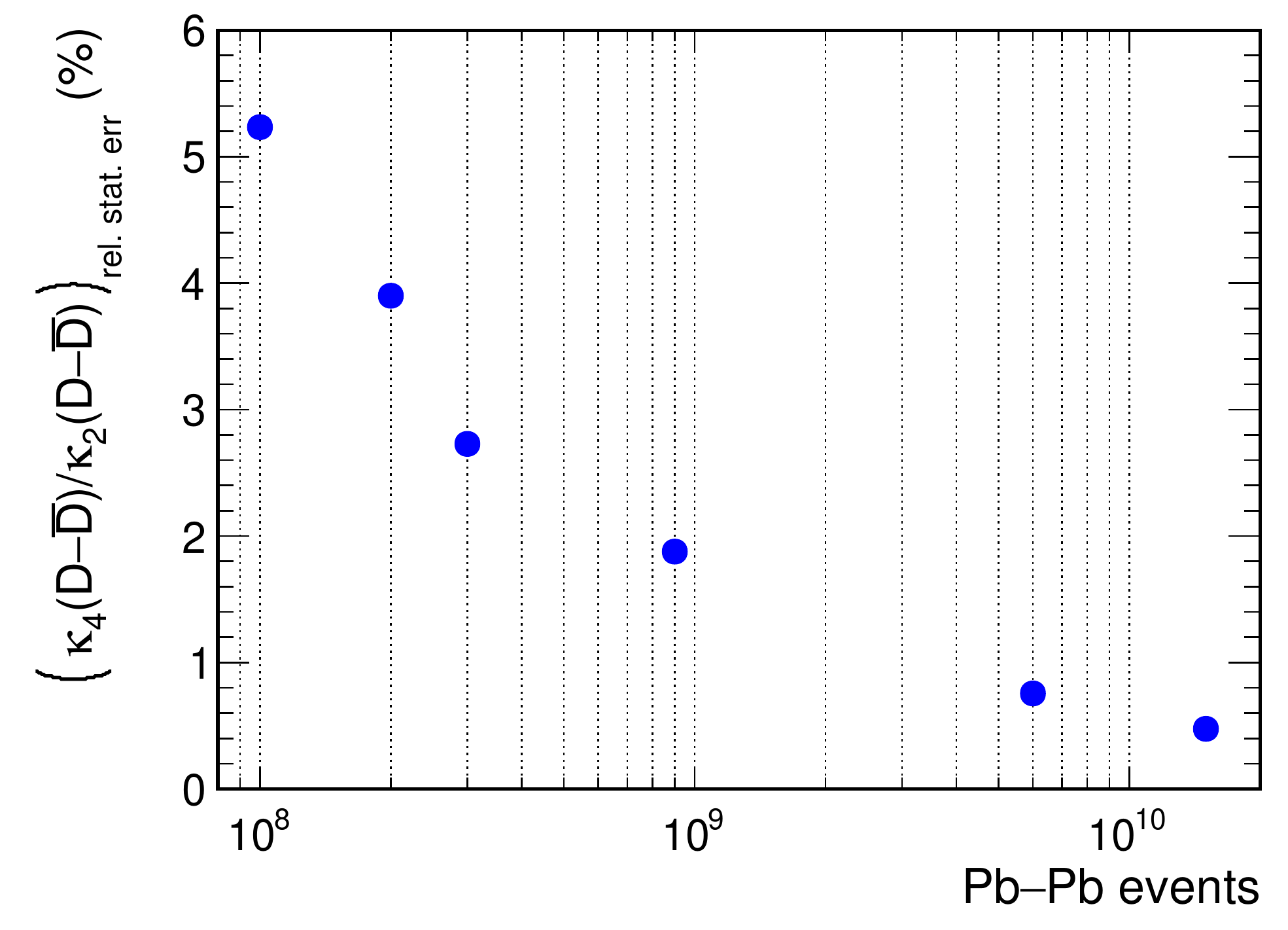}
	\caption[Uncertainty on the ratio of fourth to second order cumulants of net-D mesons]{Simulated values of $\kappa_{4}/\kappa_{2}$ as a function of the generated number of events for $D\bar{D}$ fluctuations in central Pb--Pb collisions at $\sqrt{s_{\rm NN}}$~=~5~TeV.}
	\label{fig:K4K2_Proj_DDbar_v2}
\end{figure}

\begin{figure}[htbp]
	\centering
	\includegraphics[width=9cm]{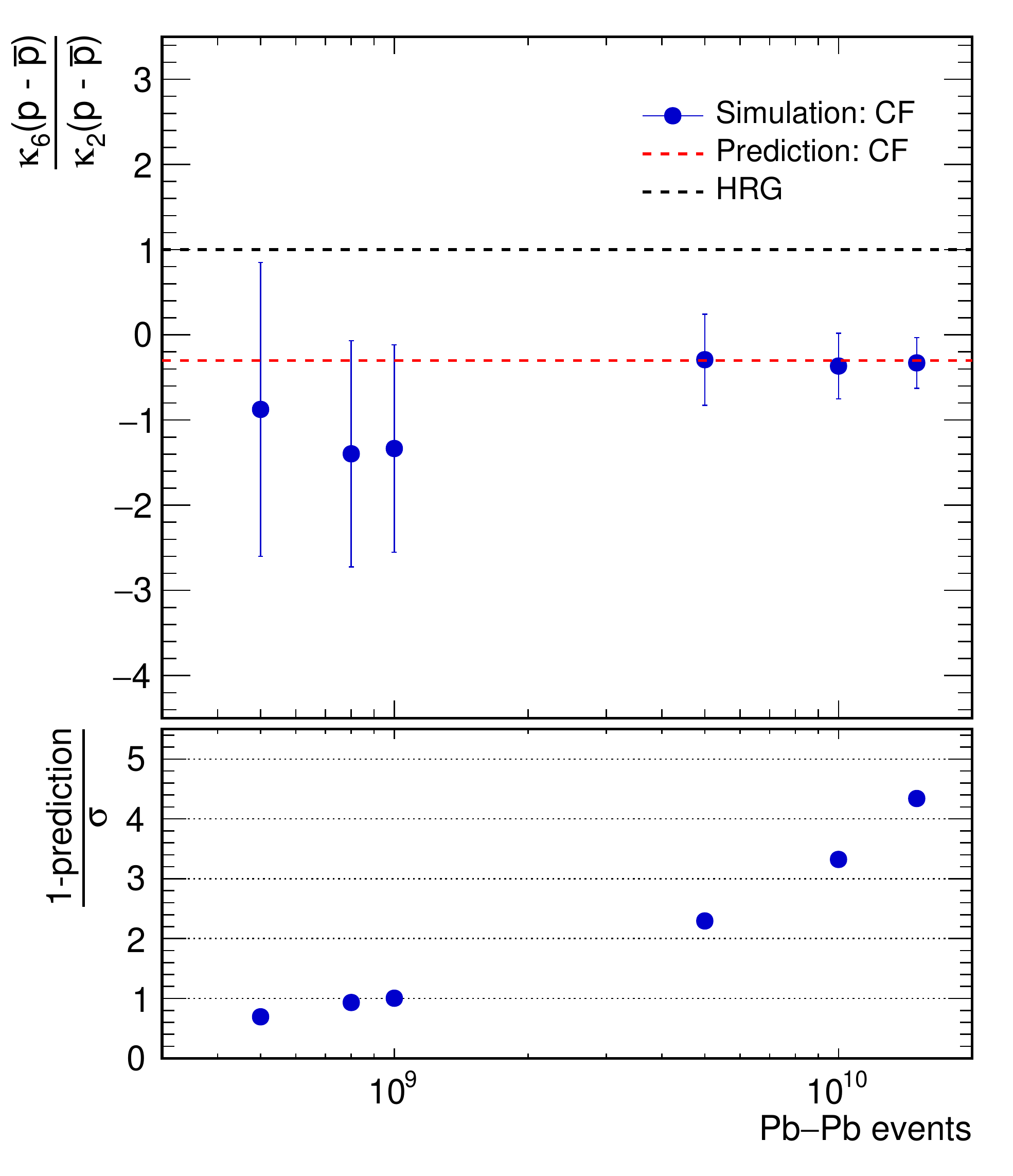}
	\caption[Uncertainty on the ratio of sixth to second order cumulants of net-protons]{(Colour online) Simulated values of $\kappa_{6}/\kappa_{2}$ as a function of the generated number of events for the ALICE~2 kinematic acceptance, $0.6<p<1.5$~GeV/$c$ and $|\eta|<0.8$. The full symbols represent results obtained with the double Gaussian approach adjusted to reproduce critical fluctuations (CF) predicted in the PQM model~\cite{Almasi:2017bhq} and the statistical hadron resonance gas model (HRG). The last two points with the largest number of events correspond to ALICE~3 statistics.}
	\label{fig:new6thmomnts}
\end{figure}

To estimate the statistics needed to address the 6th and higher-order cumulants of the net-baryon distribution with the ALICE 3, a Monte Carlo simulation is used to generate net-proton distributions based on the Polyakov-Quark-Meson (PQM) model~\cite{Almasi:2017bhq,Citron:2018lsq}. The $\kappa_{6}$ values were adjusted to account for fluctuations from participant nucleons in the 0-5\% most central Pb–Pb collisions and global baryon number conservation~\cite{Braun-Munzinger:2016yjz,Braun-Munzinger:2018yru}. The obtained results for $\kappa_{6}/\kappa_{2}$ and their corresponding statistical uncertainties are shown in Fig.~\ref{fig:new6thmomnts} as a function of the simulated event statistics. The Pb–Pb integrated luminosity of 35 nb$^{-1}$ foreseen in ALICE 3 is expected to result in more than $4\sigma$ signal of critical phenomena contained in $\kappa_{6}/\kappa_{2}$. The effect of finite efficiency for track reconstruction and particle identification are expected to be negligible in the fiducial range. Note that, a kinematic acceptance of $0.6<p<1.5$~GeV/$c$ and $|\eta|<0.8$ is used to demonstrate the gain in the statistical precision with respect to ALICE~2. Any order of net-charge fluctuation studies will largely benefit from the significant increase in the acceptance (see Fig.~\ref{fig:etadist}) in ALICE~3, particularly the differential measurements as a function of momentum, pseudorapidity and centrality.  

The main two technical difficulties in event-by-event fluctuation analyses are the understanding and control of particle detection efficiencies in the experimental acceptance and the effects of particle  mis-identification. Small efficiencies reduce the dynamical fluctuations of interest and increase experimental uncertainties arising from the efficiency correction procedure~\cite{Luo:2018ofd}. A novel experimental technique, the Identity Method (IM)~\cite{Gazdzicki:2011xz,Rustamov:2012bx,Arslandok:2018pcu}, was used to deal efficiently with overlapping energy loss distributions of different particles in the current ALICE detector, and to keep detection efficiencies as high as possible. However, the amount of overlap is still the primary source of the systematic uncertainty.
The improvement with high-purity particle identification in ALICE~3 for net-proton cumulants up to third order is illustrated in Fig.~\ref{fig:loi_error}. All in all, expected high rates and detection efficiency, large kinematic acceptance, excellent vertexing and significantly improved PID performance make ALICE~3 a unique detector for net-charge fluctuation studies. 

\begin{figure}[htbp]
	\centering
	\includegraphics[width=1\linewidth]{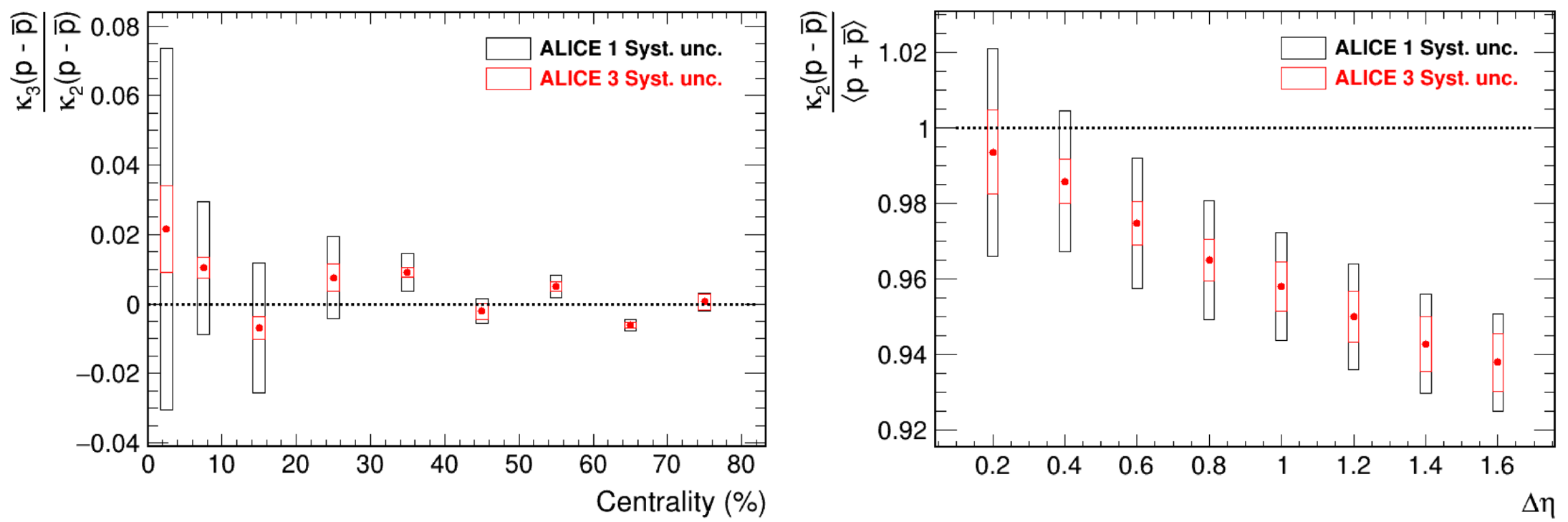}
	\caption[Systematic uncertainties on second and third order cumulants]{(Colour online) Comparison of the systematic uncertainties for ALICE 1 and ALICE 3 for the ratio of the third to second cumulants (left) and normalized second cumulants (right) of net-protons.}
	\label{fig:loi_error}
\end{figure}

\subsubsection{Soft photons}
\label{sec:performance:physics:soft_photons}

A critical aspect of a soft photon measurement is to keep the background of brems\-strahlung photons produced in the detector material, as low as possible. This background is mostly emitted by secondary electrons which are produced through photon conversions in the material in front of the FCT. In addition, background brems\-strahlung photons can also be created by electrons and positrons from $\pi^0$ Dalitz decays. The brems\-strahlung background produced in the detector material has the same $1/p_T$ form as that from internal brems\-strahlung.

We start with an analytical estimate of the contributions of electrons from conversions and $\pi^0$ Dalitz decays to the background photons produced by brems\-strahlung. Bremsstrahlung photons created by an electron passing through material are emitted at small angles with respect to the electron. For sufficiently large electron energies $E_e$, the energy distribution of the emitted photons can be approximated by 
\begin{equation}
\frac{dN_\gamma^\mathrm{bck. \, per \, electron}}{dk} \approx 
\frac{4}{3} \frac{d}{X_0}  \frac{1}{k} \quad \text{for} \;  k \ll E_e  
\end{equation}
where $k$ is the energy of the photon, $d$ the material thickness, and $X_0$ the radiation length. We use this approximation for all electrons and positrons (independent of their energy) which results in a conservative estimate. For this estimate we only consider electrons, and decay photons from neutral pion decays. We assume a rapidity density $dN_{\pi^0}/dy = 3$ in proton-proton collisions. We find that for a material thickness larger than $d/X_0 \approx 1.3\%$, the contribution of conversion electrons to brems\-strahlung dominates. At lower $\pt$ the contributions from conversions $\pi^0$ Dalitz decays are of similar magnitude.

Evaluating Low's formula for the internal bremsstrahlung signal~\cite{DELPHI:2005yew} for inelastic proton-proton collisions at $\sqrt{s} = 13\,\mathrm{TeV}$ simulated with PYTHIA, we obtain a transverse momentum spectrum given by $1/N_\mathrm{evt}\;dN_\gamma/dp_T = 0.034/p_T$ for photons in the pseudorapidity range $3 < \eta < 5$. We find that for a material thickness of 5\% $X_0$, the background of external bremsstrahlung created in the material in front of the FCT is equal to this internal bremsstrahlung signal. The experimental procedure to extract this signal involves the subtraction of the background. Assuming that the background photon spectrum is known with a precision of 5\%, we calculate the significance of the internal bremsstrahlung signal as a function of the material thickness $d/X_0$. This is shown in Fig.~\ref{fig:soft_photon_significance}. We conclude that in order to measure internal brems\-strahlung with a significance of better than $3$, the material budget of the FCT should be kept below $d/X_0 \lesssim 14\%$.

\begin{figure}
\centering
\includegraphics[width=0.6\linewidth]{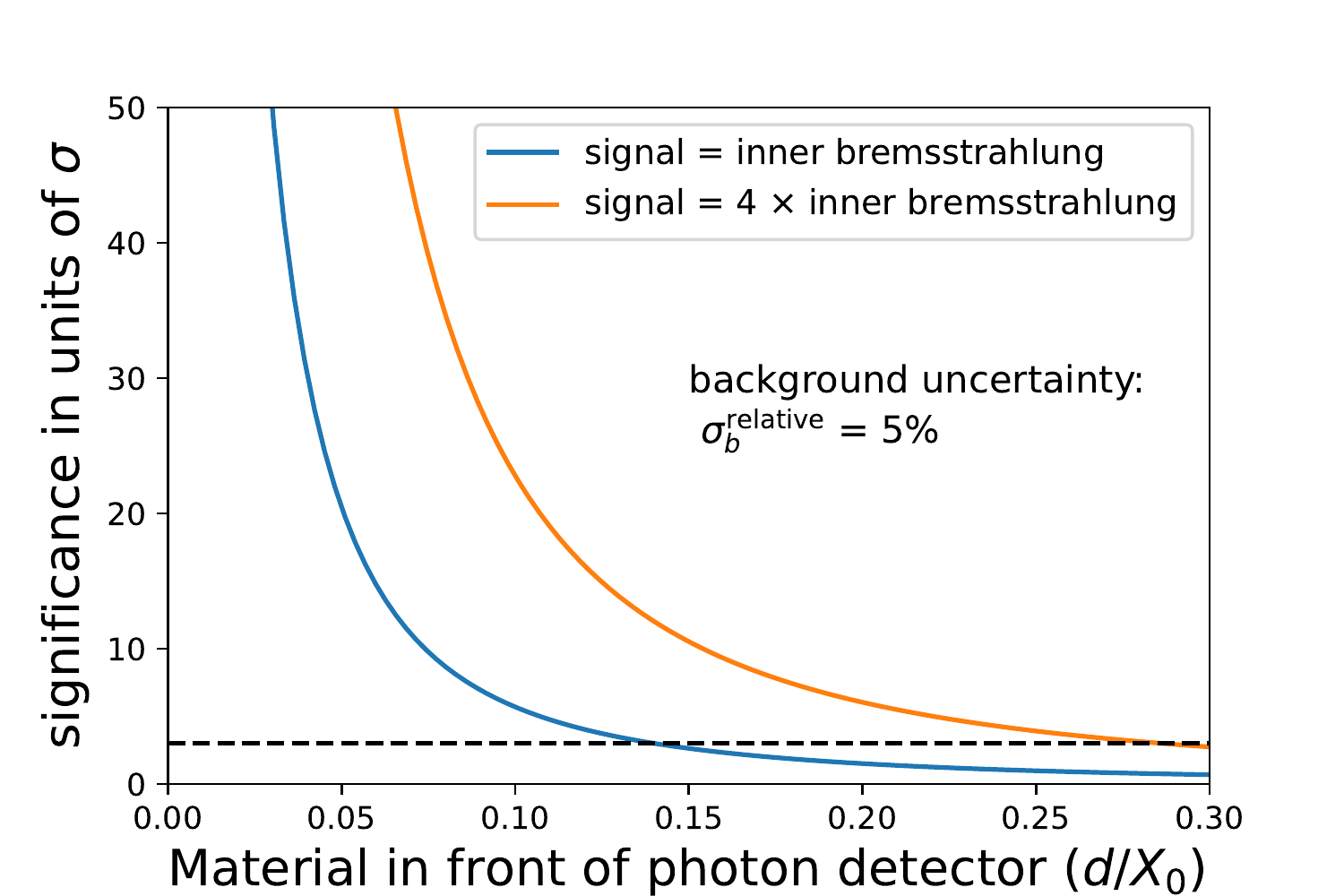}
\caption[Significance of internal bremsstrahlung]{Significance of the inner bremsstrahlung signal as a function of the material in front of the FCT assuming that the relative uncertainty of the background spectrum from external bremsstrahlung is $\sigma_b^\mathrm{relative} = 5\%$. For a material budget of less than about $d/X_0 \approx 14\%$ the inner bremsstrahlung signal can be measured with a significance of better than $3\sigma$. In previous soft photon experiments an excess of about a factor 4 above the inner bremsstrahlung was observed~\cite{DELPHI:2005yew}. This larger signal could be measured with a significance of better than $3\sigma$ for $d/X_0 \approx 28\%$.}
\label{fig:soft_photon_significance}
\end{figure}
 
 Since particles at forward rapidities cross material, such as the beam pipe, under shallow angles, they are exposed to an increased thickness of material. For example a beryllium beam pipe of \SI{500}{\um} corresponds to a radiation length of 0.14\% and 10\% at $\eta = 0$ and $\eta = 5$, respectively.

To illustrate the expected background level with an idealised setup, the left-hand panel of Fig.~\ref{fig:soft_photon_signal_and_background} shows that the brems\-strahlung photon background is small under the assumption of
0.14\% of a radiation length (\SI{500}{\um} of beryllium) in front for the FCT.  
Studies to optimise the beam pipe layout with a realistic setup are currently ongoing. 

We also used GEANT 4 to study the external brems\-strahlung created in a simplified version of the full setup consisting of a standard beam pipe, barrel silicon tracking layers, and the forward tracking disks. In this simulation photons were measured in $3 < \eta < 5$. The experimental strategy in previous soft photon measurements involved the rejection of events with electrons in the acceptance of the photon detector~\cite{DELPHI:2005yew}. We study the potential of such an analysis strategy by considering only proton-proton collisions in which there is no electron in the pseudorapidity range of the FCT. For $3 < \eta < 5$ this is the case for about 7\% of the inelastic proton-proton collisions simulated with PYTHIA 8. The background of external brems\-strahlung for these events is shown in the right-hand panel of Fig.~\ref{fig:soft_photon_signal_and_background}. We conclude that with this condition (no electrons/positron in $3 < \eta < 5$) the background level is small compared to the signal. %
\begin{figure}
\centering
\includegraphics[width=0.49\linewidth]{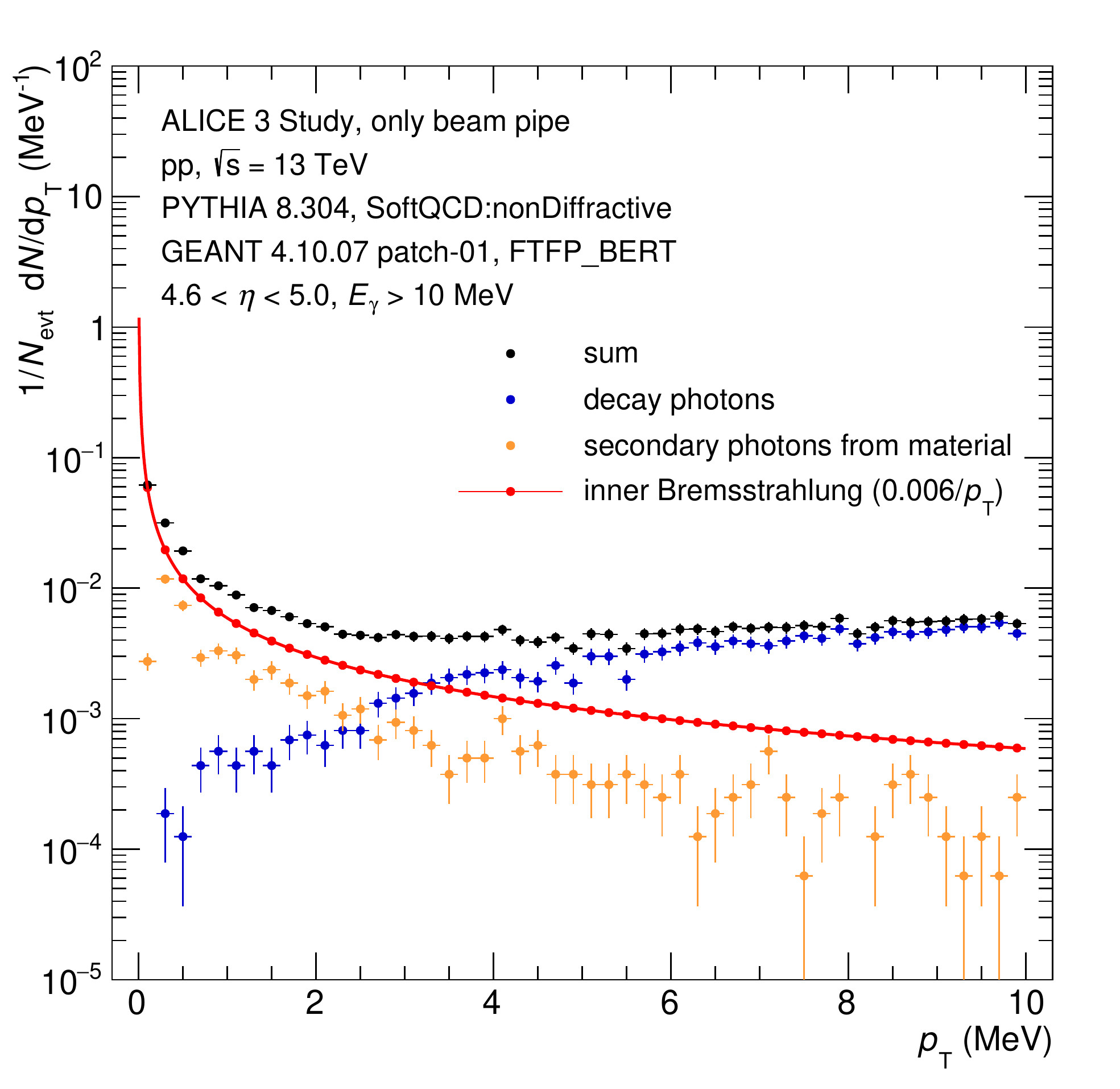}
\includegraphics[width=0.48\linewidth]{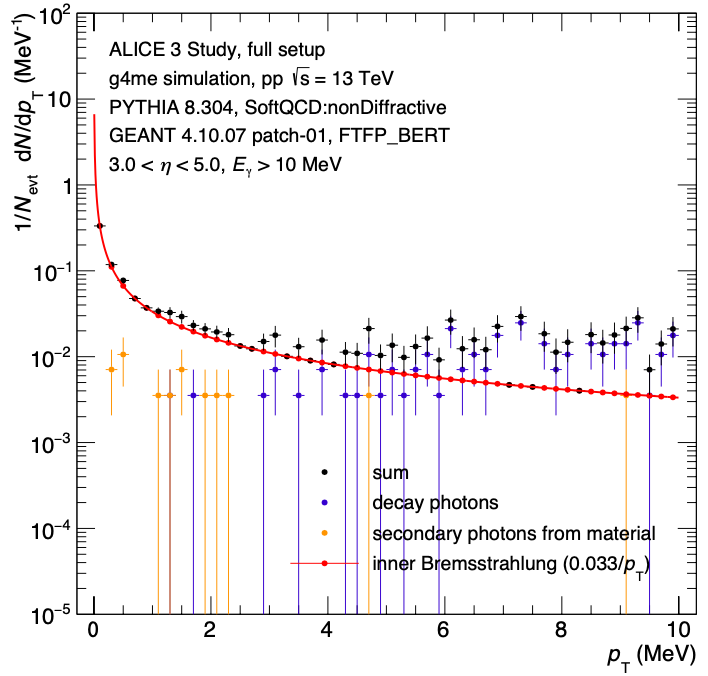}
\caption[Spectra of signal and background photons]{Transverse momentum spectra of signal and background photons at forward rapidities in pp collisions at $\sqrt{s} = 13\,\mathrm{TeV}$. The background consists of photons from the decay of $\pi^0$, $\eta$ and other hadrons and of bremsstrahlung photons produced by electrons and positrons in the material in front of the FCT. The left-hand figure shows only background photons created in the beam pipe. In the simulation the $500\,\mathrm{\mu m}$ beryllium beam pipe was shaped in a way as to avoid shallow crossing angles. Detailed studies of an optimal beam pipe design for a forward soft photon measurement are currently ongoing. The right-hand figure shows the bremsstrahlung background for a full detector setup (standard beam pipe, barrel tracking layers, forward tracking disks) for events without an electron or a positron in the pseudorapidity range of the FCT.}
\label{fig:soft_photon_signal_and_background}
\end{figure}

\begin{revised}
Previous experiments that found a significant soft photon excess above the expected signal from inner bremsstrahlung reported enhancements ranging from a factor 2 to a factor 8, quoting uncertainties in the 15\%-35\% range~\cite{BELOGIANNI:2002129,DELPHI:2005yew}. With the current level of simulations described here we are confident that the soft photon spectrum in all planned measurements can be determined  with an accuracy of 20\% or better, sufficient to confirm the predicted photon radiation from QCD and QED. This would close a long chapter of more than 30 years of experiments at different accelerator facilities. 
\end{revised}

\subsubsection{Nuclear states}
\label{sec:performance:physics:nuclear_states}

The unique particle identification capabilities of \ALICETHR{} are an ideal tool to study production of nuclei, anti-nuclei, hyper-nuclei and super-nuclei. In the following, we illustrate these capabilities with examples for nuclei and super-nuclei.

\paragraph{Nuclei}
\label{sec:performance:physics:nuclear_states:nuclei}

The excellent performance of \ALICETHR for the identification of light nuclei is demonstrated in Figs~\ref{fig:performance:detector:nuclei:btof2:BetaVsP} and~\ref{fig:performance:detector:nuclei:btof2:helium3}.
The left-hand panel of Fig.~\ref{fig:performance:detector:nuclei:btof2:helium3} shows that protons, deuterons, and tritons are clearly separated from all other species and can be identified on a track-by-track basis with negligible contamination.

For \hethree, $m/z \approx 3/2$, the TOF response is peaked between the proton and deuteron bands, resulting in a significant background that amounts to about 60\% in a $\pm 2\sigma$ window.
In order to remove this contamination, a charge-sensitive measurement is required.
Here, the feasibility of such a separation based on the cluster size in the tracking layers was studied in Section~\ref{sec:performance:detector:hadron_id}. The right panel of Fig.~\ref{fig:performance:detector:nuclei:btof2:helium3} shows the TOF distribution with a selection on the truncated mean cluster charge that retains 85\% of \hethree{}. 
In this case the contamination from protons and deuterons is decreased to about 2\%.
This study illustrates the feasibility of measurements such as the ones discussed in Section~\ref{sec:physics:exotica}. 
 
In order to distinguish \hesix and \lisix very good charge separation is required.
The \hesix peak is overshadowed by the five orders of magnitude more abundant tritons, while for \lisix the roughly $10^5$ times more abundant \hefour and the roughly 3$\times 10^7$ more abundant deuterons are the main backgrounds.
As a matter of fact, the separation between \hesix and triton is the same as that between protons and \hethree, while for \lisix and \hefour the expected separation is smaller. Nevertheless, the existing ALICE measurements of anti-alphas~\cite{ALICE:2017jmf} show that such purities in terms of charge separation are routinely achievable with precise d$E$/d$x$-sensitive measurements. If in future simulation studies the previously described cluster size method will turn out to be insufficient, three additional techniques will be considered. Firstly, the cluster size method can be further enhanced by splitting the digital readout of the chips used in the tracking layers based on an additional threshold. Secondly, time-over-threshold measurements in the timing layers would provide a measure of the charge deposit and can be used to separate the two cases.
Thirdly, equipping one of the outer silicon tracking layer with a readout of the signal amplitude would guarantee the separation of particles with different $z$ over a wide momentum range though at the cost of additional complexity of the apparatus.

In order to assess the feasibility of discovering $A=6$ anti-nuclei with \ALICETHR, we assume yields from the Statistical Hadronisation Model and a blast-wave shape for the \pt{} spectrum~\cite{Schnedermann:1993ws}.
We further use an average reconstruction efficiency of 90\% in the rigidity range above $p/z \approx 1$~\GeVc, as shown in Fig.~\ref{fig:performance:dectector:introduction:full_fast_efficiency_vs_pt_comparison}, and clean TOF identification up to $p/z \approx 5$~\GeVc. 
Based on these assumptions, a \pt-integrated overall acceptance times efficiency of about 70\% for $A=6$ nuclei produced in $|y|<1.44$ is obtained.
This will result in about 0.9 \lisix reconstructed candidates per month for Pb--Pb data taking and in about 2.6 reconstructed candidates per month for Kr--Kr collisions.
Considering that a number of 10-25 candidates is needed for a discovery, this would require 4 to 10 months of heavy-ion data taking in LHC Run~5 and 6 with the luminosities presented in Table~\ref{tab:ion_lumi}.
For \hesix the corresponding yields are about a factor of 3 lower due to the spin degeneracy and thus only upper limits could be established.
Since the uncertainty on these measurements are purely statistics dominated any further increase in the LHC luminosity will directly lead to a reduction of the required ion running time. 

\begin{figure}[t] %
  \centering
  \includegraphics[width=.59\textwidth]{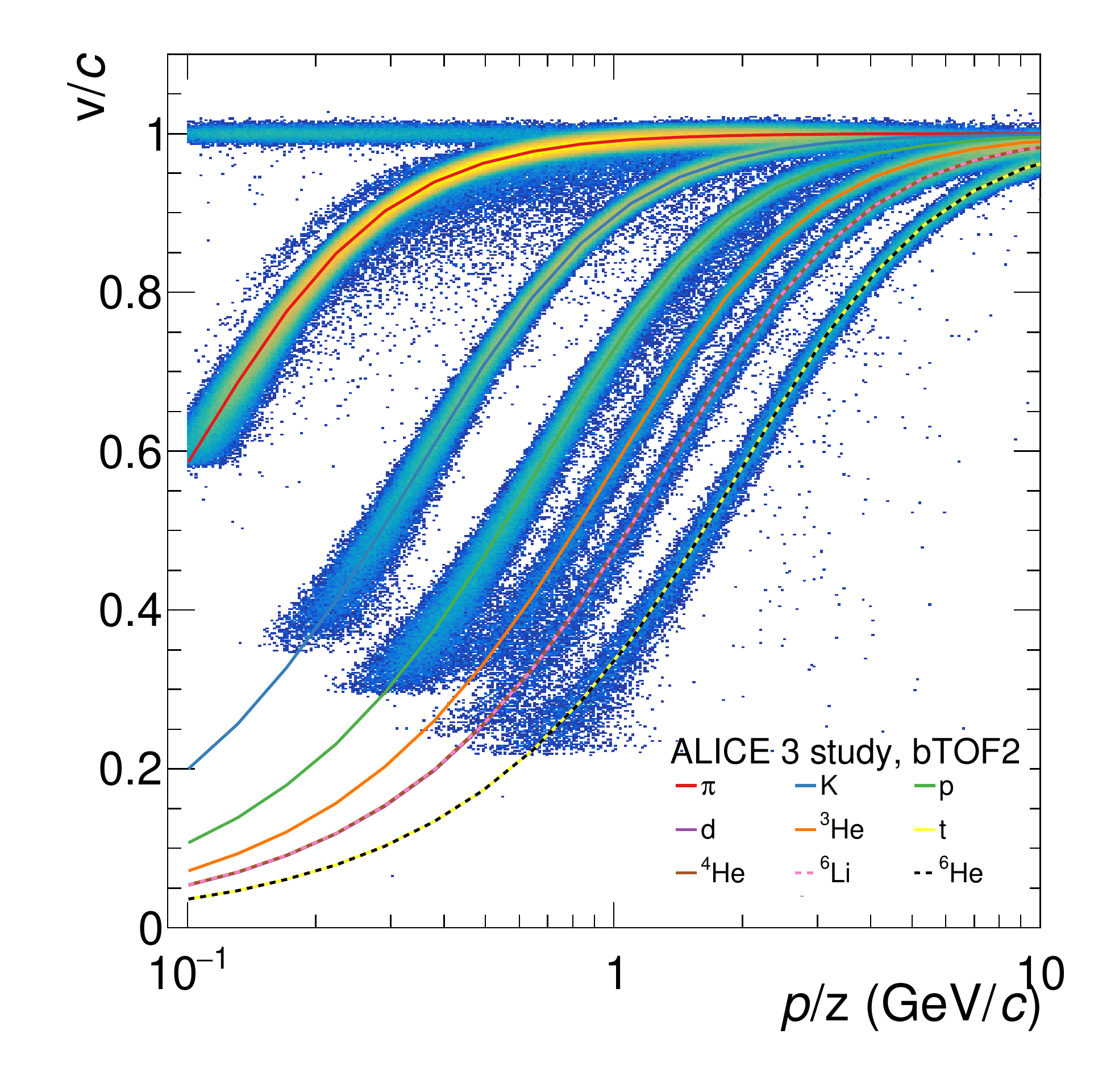}
  \caption[Particle velocity from outer TOF (nuclei)]{
  Particle velocity measured in the outer TOF. 
  The values expected for the different particle species (including light nuclei) are reported in colored bands.
  It should be noted that due to the ambiguity in the particle charge, the bands of particles with same $m/z$ overlap.
  }
  \label{fig:performance:detector:nuclei:btof2:BetaVsP}
\end{figure}

\begin{figure}[t] %
  \centering
  \includegraphics[width=.32\textwidth]{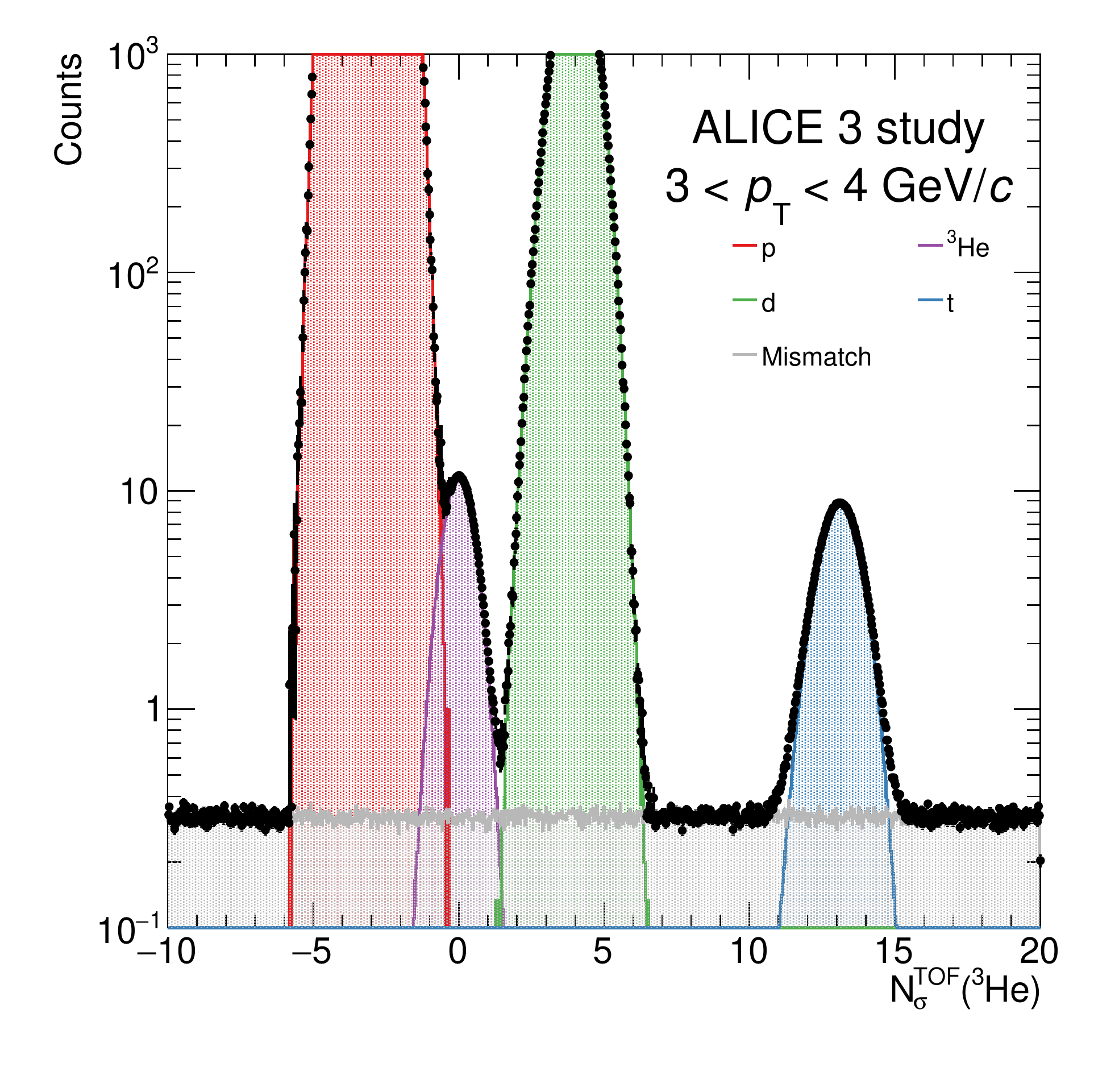}
  \includegraphics[width=.32\textwidth]{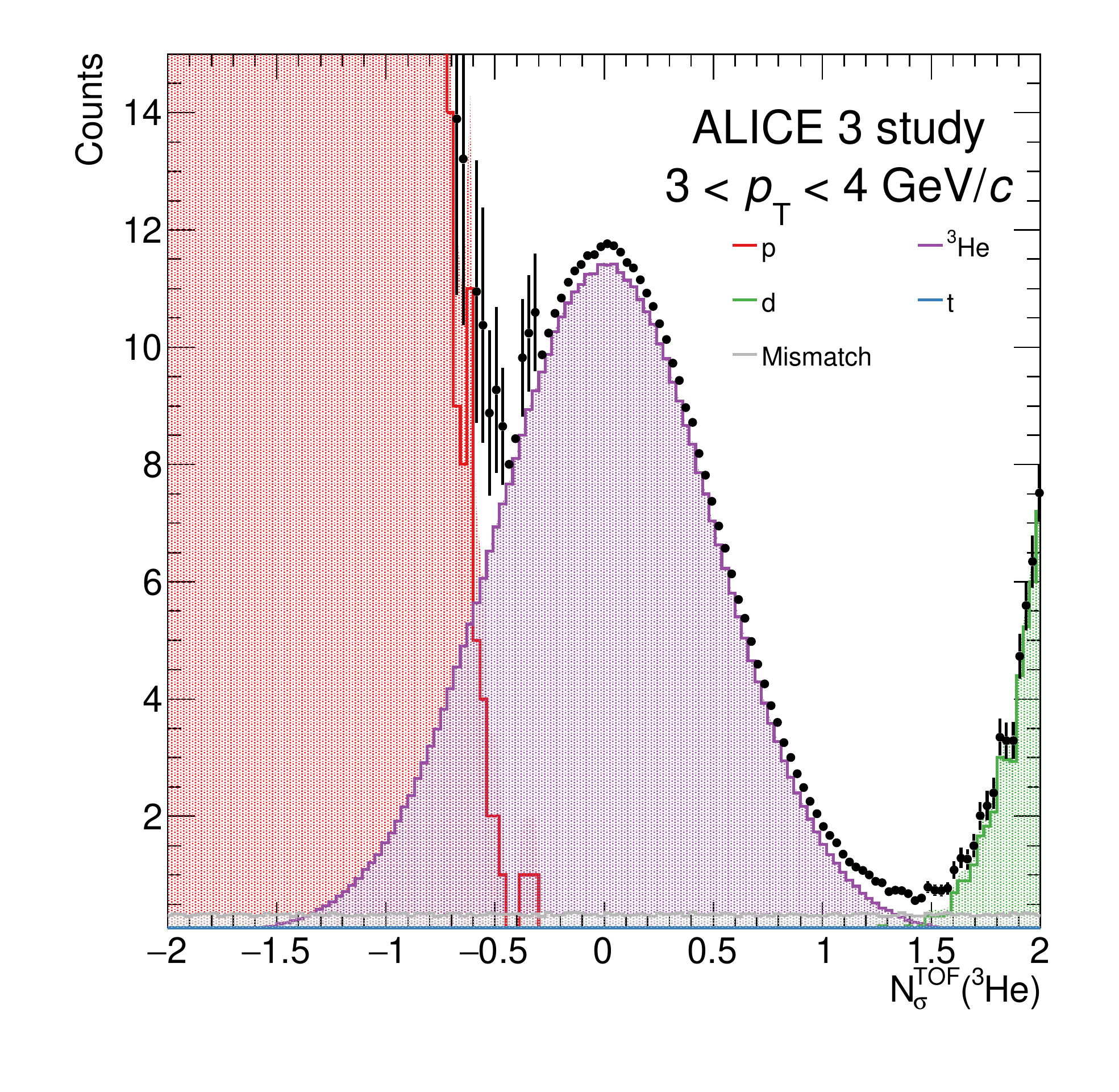}
  \includegraphics[width=.32\textwidth]{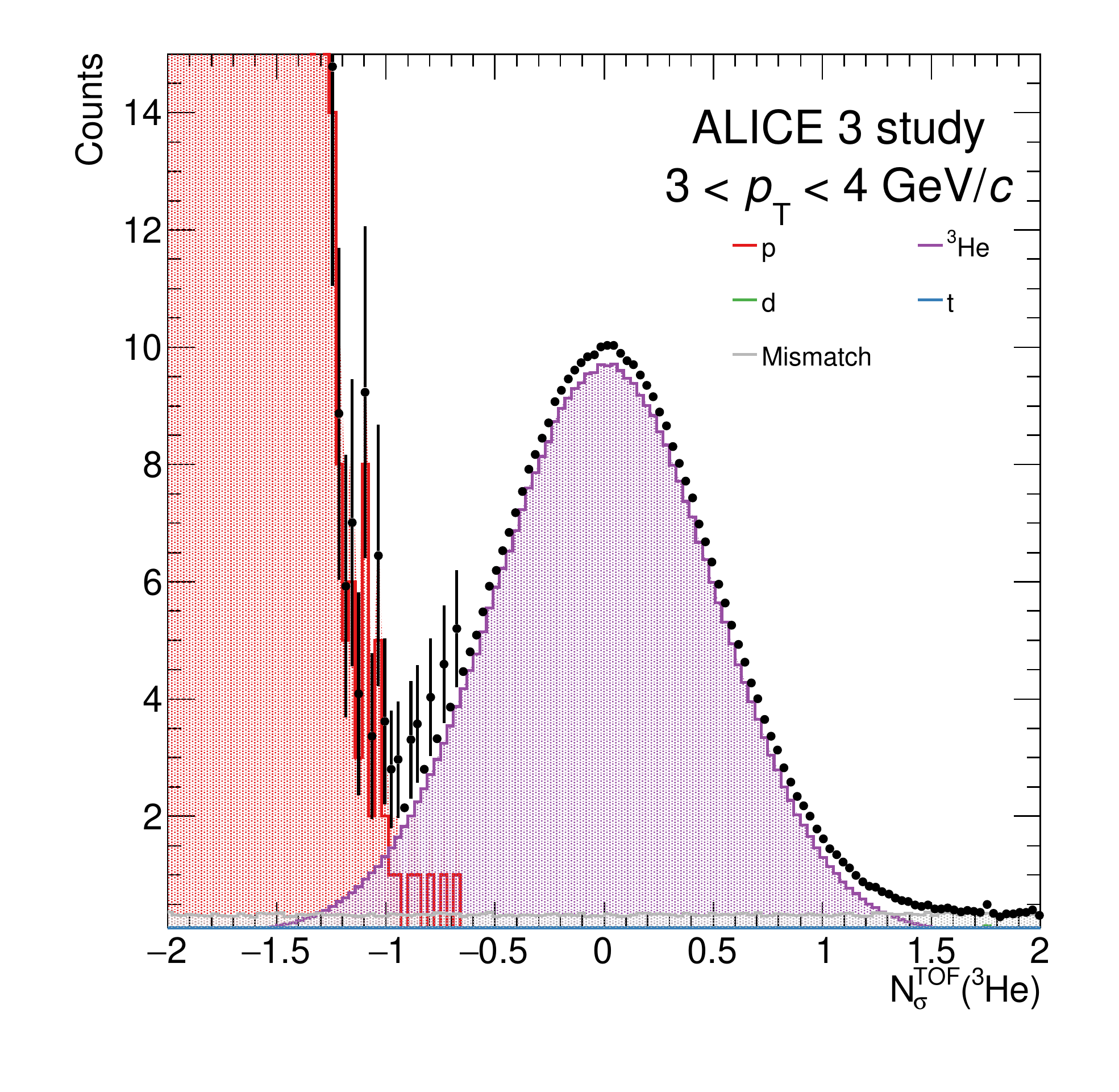}
  \caption[TOF separation for $^{3}$He]{
   (Left panel) TOF n$\sigma$-distribution for the $^{3}$He hypothesis without selection on the cluster size in the tracking layers. (Middle panel) A zoom of the same plot in the region of the $^{3}$He-peak. (Right panel) A zoom of the same region after applying an additional selection for tracks with large average cluster size.
  }
  \label{fig:performance:detector:nuclei:btof2:helium3}
\end{figure}

\paragraph{Super-nuclei}
\label{sec:performance:physics:nuclear_states:charm}

In this section we study the potential discovery of the lightest bound states of the $\Lambda_c$ baryon with one or two neutrons, the c-deuteron $c_d$ and c-triton $c_t$, using the decay of the bound $\Lambda_c$ in the $\Lambda_c \rightarrow pK\pi$ channel. The resulting decay proton subsequently binds to the remaining neutron(s) only with a probability that is estimated it to be 3-10\% using a simple coalescence model. In the model, protons with a relative momentum in the rest frame of the nucleus below a threshold of 200~MeV/$c$ are bound.
The overall branching ratio for the $c_{d} \rightarrow d+K^{-}+\pi^{+}$ and $c_{t} \rightarrow t+K^{-}+\pi^{+}$ decay is then given by the product of the branching ratio of $\Lambda_c \rightarrow pK\pi$ ($6.28\pm0.32$\%) and the aforementioned probability of the proton to bind to the remaining neutron(s).  

In the following, we discuss the \ALICETHR performance for reconstructing c-deuteron and c-triton in these decay channels. As an exemplary case, we have chosen the more abundantly produced c-deuteron. While, according to current theoretical understanding, its existence as a weakly decaying bound state is less likely than the c-triton, its investigation via a direct discovery or setting upper limits on its production is of prime interest. As a matter of fact, the strong similarities in the kinematics and reconstruction of the $c_d$ and $c_t$ decay allow a scaling of the estimated significances. Since the production of $c_t$ is suppressed by a penalty factor of about 1/350, the resulting significance of $c_t$ is expected to be about 18 times lower.

The performance for reconstructing these decays has been assessed using the fast simulation. 
The main challenge in this analysis is to suppress the combinatorial background that is formed by combining primary deuterons with pions and kaons. The excellent DCA resolution of \ALICETHR allows a very effective suppression of this background, as shown in Fig.~\ref{fig:c-deuteron}. In contrast to this, the correlated background in which a deuteron from a true c-deuteron decay picks up a wrong pion or kaon is negligible. Both background contributions have been determined and scaled to the expected luminosity for one month of \PbPb running. The resulting invariant mass distribution is shown in Fig.~\ref{fig:c-deuteron}.

Assuming the production rates of the statistical-thermal model, for the c-deuteron a centrality and $p_T$-integrated significance of about 50 per month of \PbPb running is achieved. 
This corresponds to a significance of about 2.5 per month of \PbPb running for the c-triton. Since the expected yields of alternative models are even larger~\cite{ExHIC:2017smd}, it can be concluded that \ALICETHR is well suited to either discover or rule out the existence of these states.

\begin{figure}[t]
  \centering
  \includegraphics[width=0.32\textwidth]{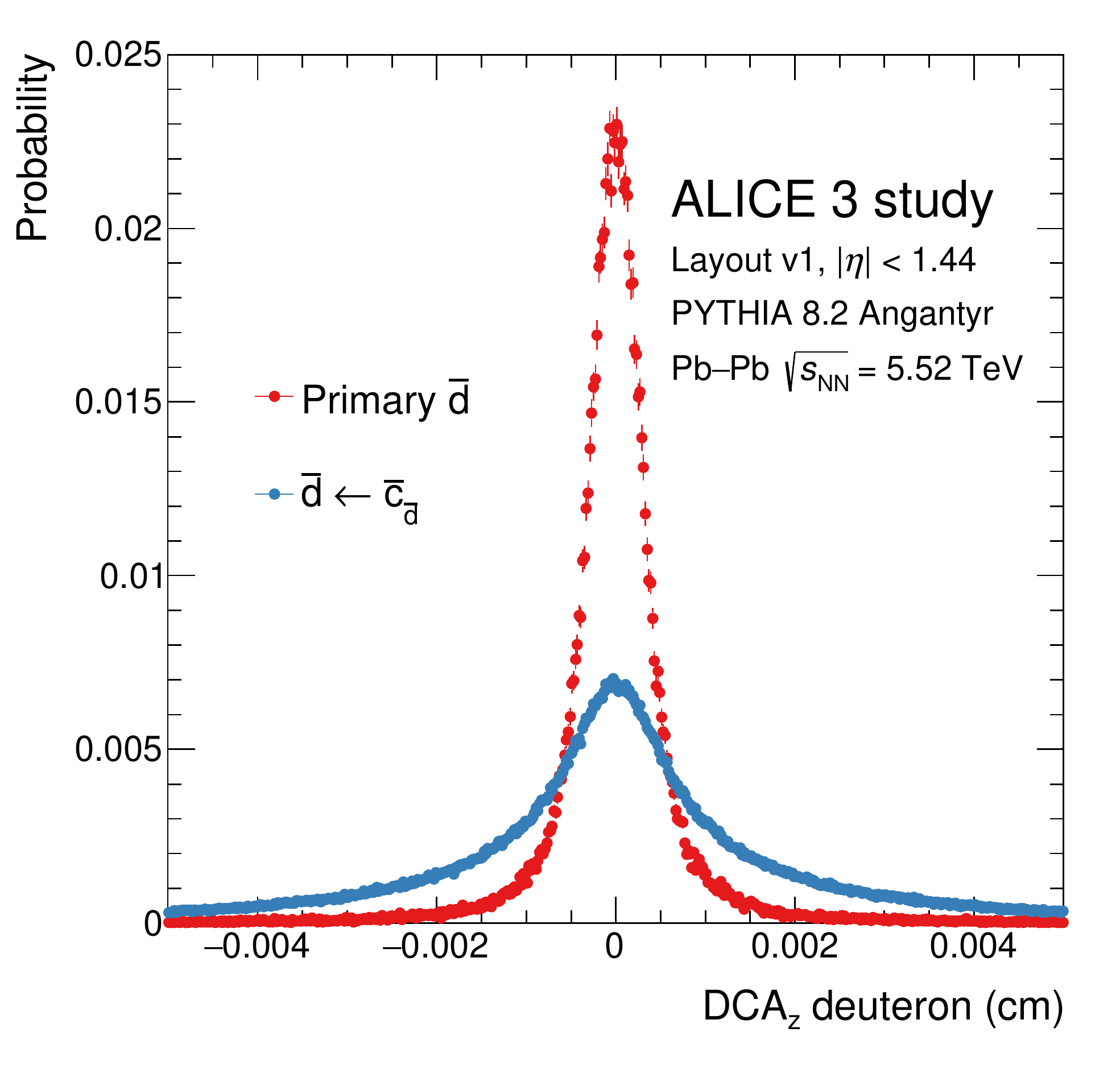}
  \includegraphics[width=0.32\textwidth]{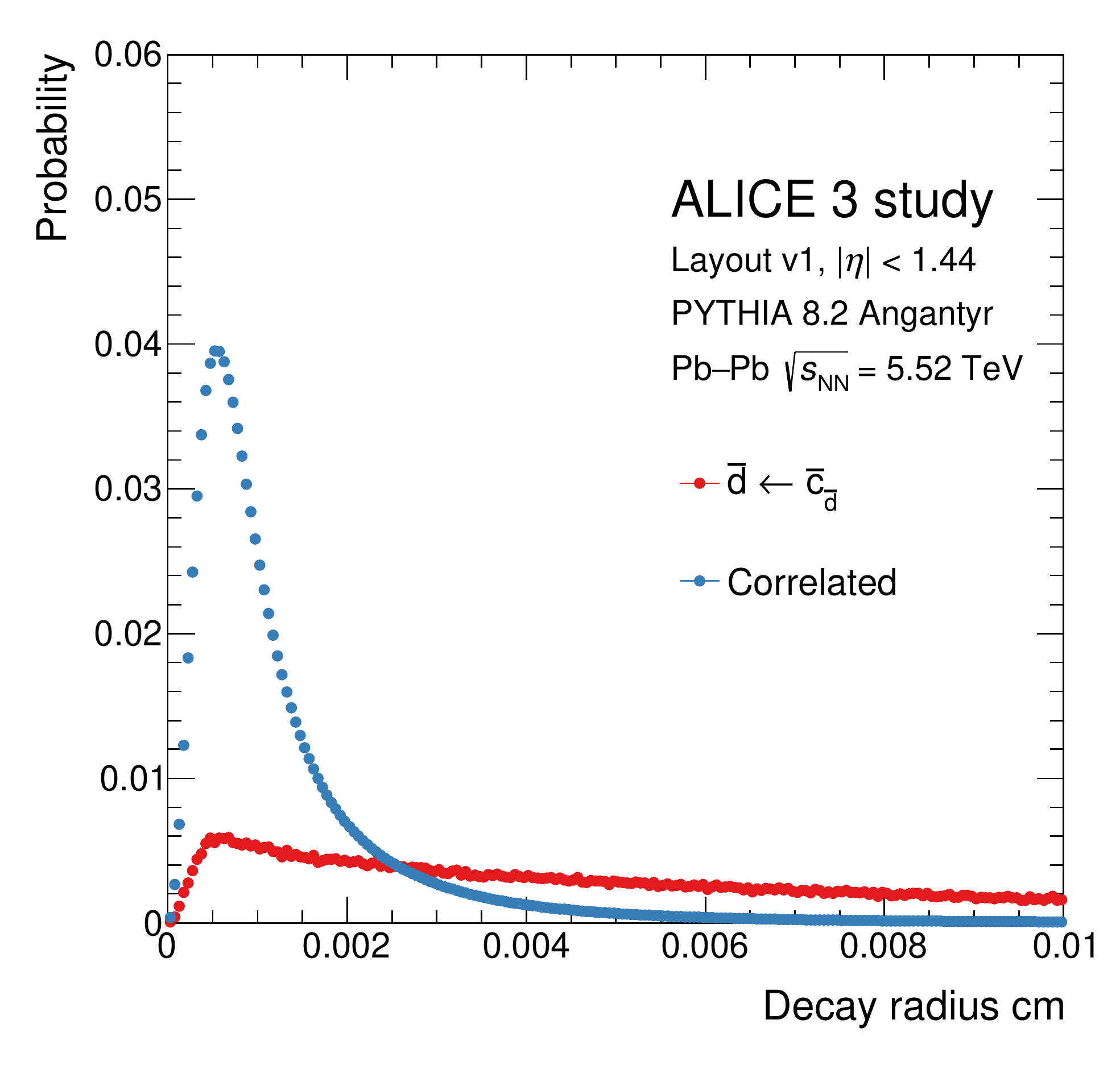}
  \includegraphics[width=0.32\textwidth]{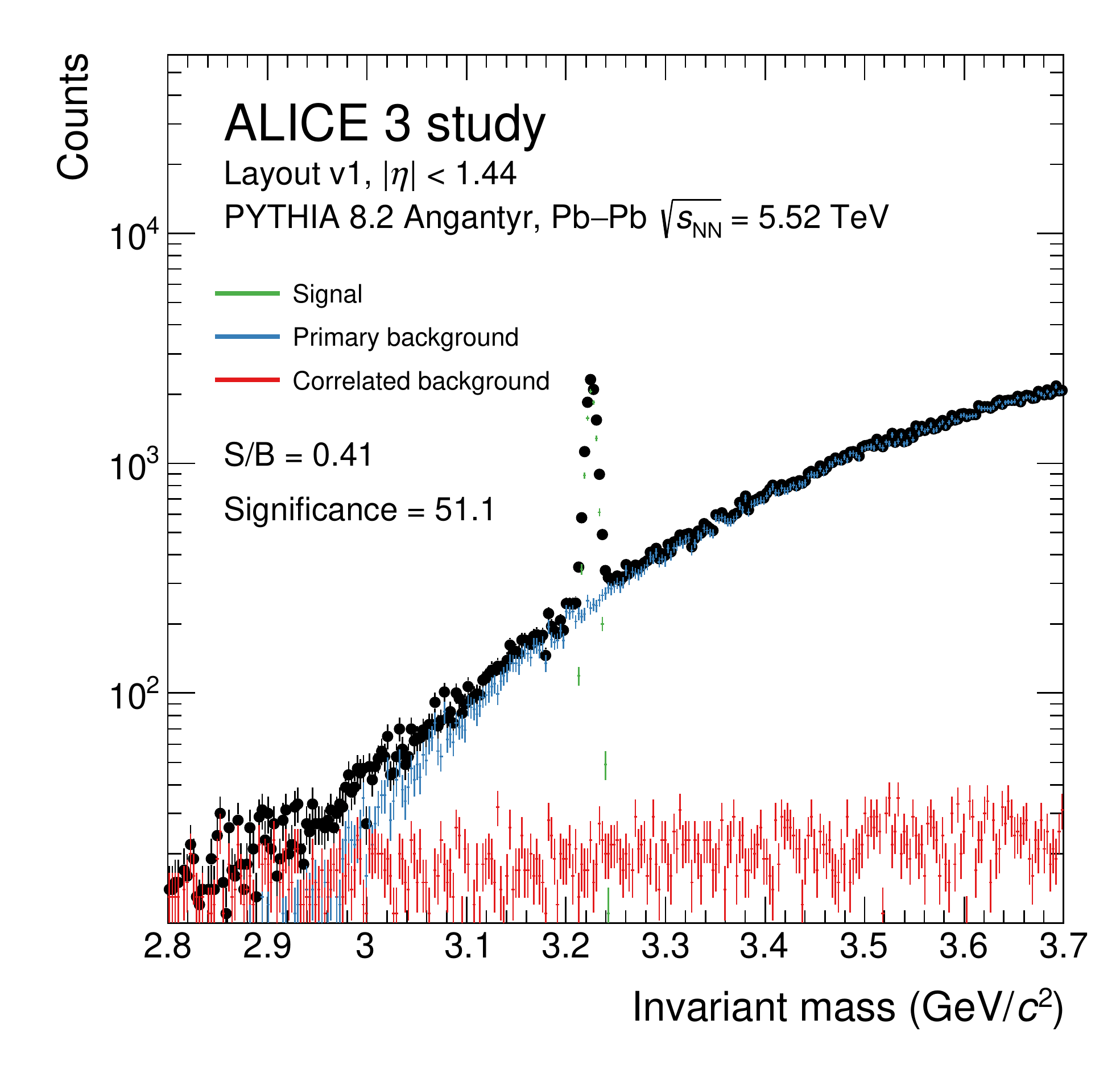}
  \caption[Distributions of DCA, decay radius, and inv. mass for deuterons] {
  Left: distribution of the distance of closest approach for primary deuterons and deuterons from c-deuteron decays.
  Middle: decay radius for signal and correlated background.
  Right: distribution of the invariant mass for signal, correlated background and background from primary deuterons for a total integrated luminosity of \SI{5.6}{\nano\barn^{-1}}.
  }
  \label{fig:c-deuteron}
\end{figure}

\paragraph{Anti-nuclei from b quarks}
\label{sec:performance:physics:nuclear_states:beauty}

The detection of cosmic-ray anti-nuclei is a promising signature of the existence of dark matter~\cite{Donato:1999gy} as discussed in Section~\Ref{sec:physics:Lb_to_Antinuclei}.
In addition, these studies give insight into the mechanisms of hadronisation.
In order to assess the performance expected for these measurements with the ALICE~3 detector, the fast simulation tool described in Sec.~\ref{sec:performance:introduction} was used.
To illustrate the experimental sensitivity to reconstruct anti-nuclei from the decay of beauty baryons, the \antiLambdab was taken as an example.
\antiLambdab baryons were generated by sampling the cross section obtained from FONLL calculations~\cite{Cacciari:1998it,Cacciari:2001td,Cacciari:2012ny} multiplied by the beauty-quark to \Lambdab baryon fragmentation fractions measured by the LHCb Collaboration in \pp collisions at $\sqrts = 13~\tev$~\cite{Aaij:2016avz}.
The decays were simulated using the PYTHIA~8 decayer, followed by the formation of \antiHe by coalescence of nucleons emerging as decay products.
As in~\cite{Winkler:2020ltd}, nucleon coalescence is modeled by merging all particles within a sphere of radius $2^{1/6}p_\mathrm{c}/2$ in momentum space, where $p_\mathrm{c}=200~\mev/c$.
The dominant channel for the formation of \antiHe is the \antiLambdabToDominant decay which has a branching ratio of 1.2\% in PYTHIA~8.

\begin{figure}[t]
  \centering
  \includegraphics[width=0.49\textwidth]{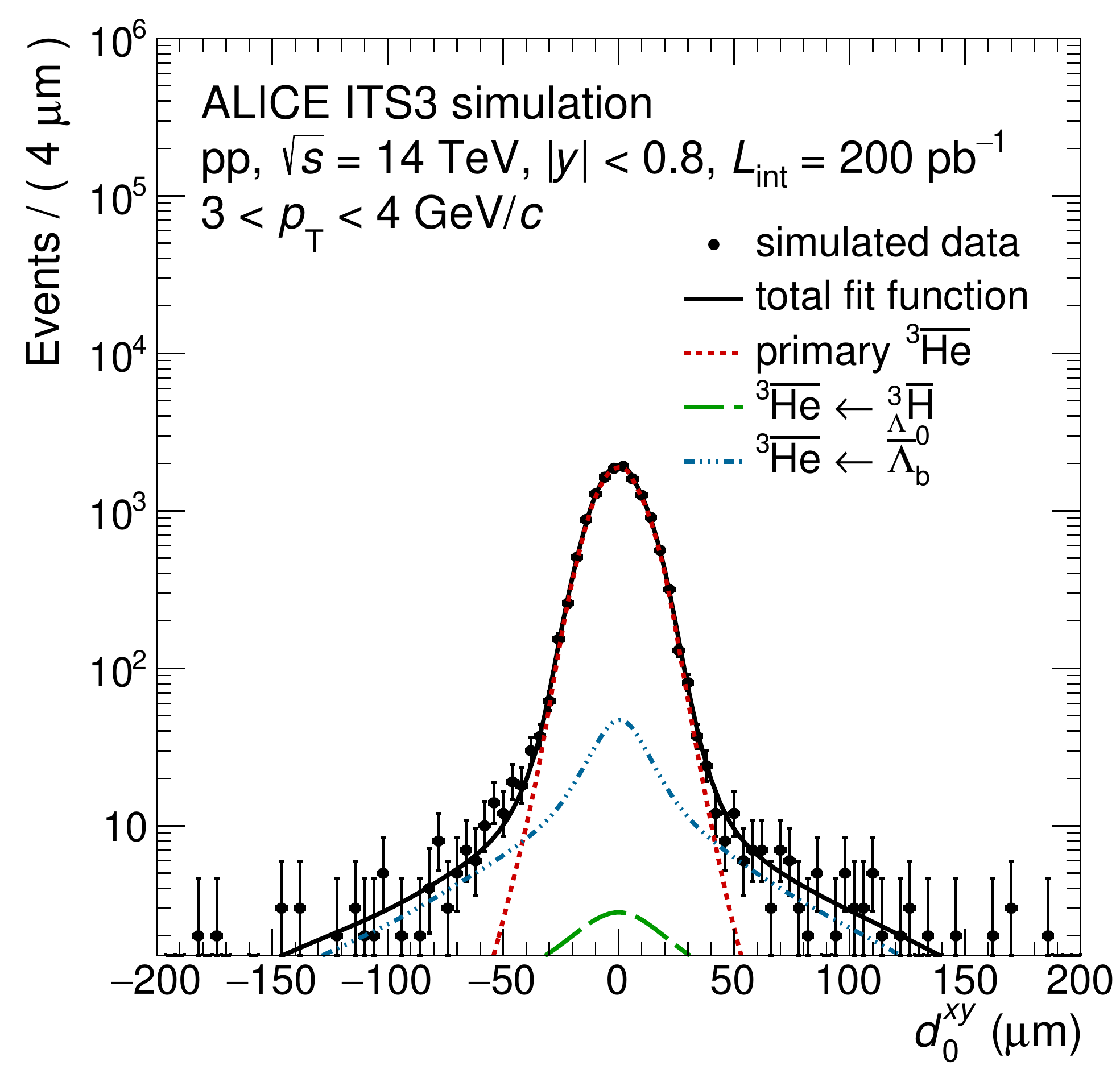}
  \includegraphics[width=0.49\textwidth]{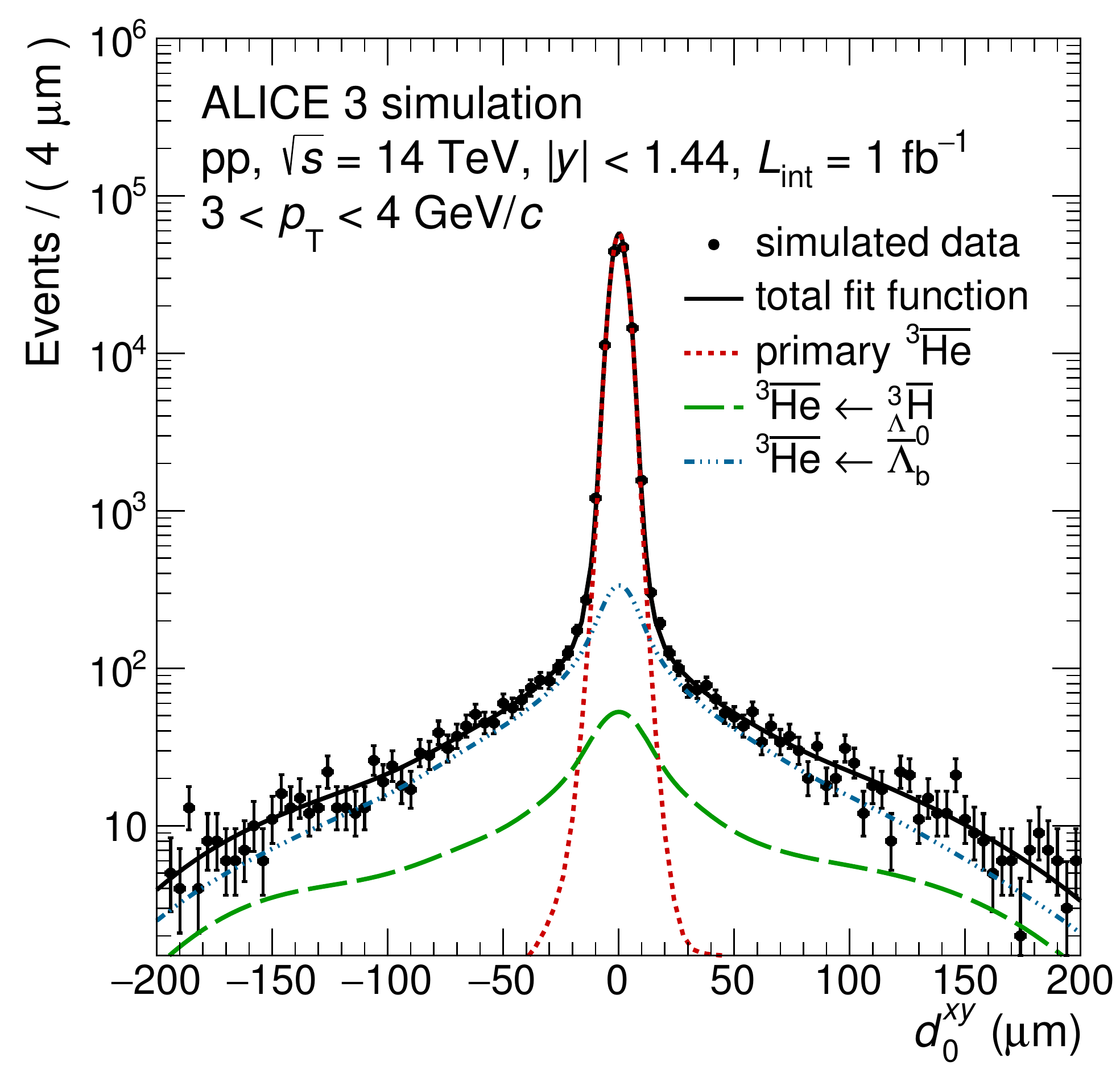}
  \caption[DCA distributions of \antiHe with ITS3 and ALICE~3] {Examples of fits to the impact-parameter distributions of \antiHe in $3<\pT<4~\gev/c$ expected with the ITS3 and ALICE3 detectors for $\Lint = 200~\invpb$ and $\Lint = 1~\invfb$, respectively.}
  \label{fig:physics:lb:imp}
\end{figure}

In \Fig{fig:physics:lb:imp} we show the simulated impact parameter distributions of reconstructed \antiHe candidates with $3<\pT<4~\gev/c$ that were identified via a combination of time-of-flight and cluster size selections as described in Sec.~\ref{sec:performance:detector:hadron_id}.
In the case of the current ALICE detector, the TPC \dedx{} is used to identify the \antiHe.
The impact parameter distribution of primary \antiHe is only driven by the detector resolution and corresponds to about \SI{4}{\um} in ALICE~3 and about \SI{12}{\um} with the ITS3.
The distribution for \antiHe stemming from \antiLambdab decays is significantly wider, reflecting the lifetime of the \Lambdab
baryons of \SI{441}{\um}~\cite{Zyla:2020zbs}.
A combined template fit to the resulting distributions is used to determine the relative abundances of primary \antiHe and of those from \antiLambdab decays.
The sensitivity of the fits depends on the one hand on the integrated luminosity times the branching ratio of \antiLambdabToHe{} and on the other hand on the background from \antihyperT decays.
For the ITS2, ITS3, and ALICE~3, this background was estimated to be about 1.5\%, 1\%, and 0.5\% of the primary \antiHe yields, based on the $\hyperT/\He$ production ratio measured in \pPb{} collisions~\cite{ALICE:2021sdc}, the BR of the \hyperTtoHe decay, and the rejection factor obtained by requiring two hits in the innermost two layers of each detector for the \antiHe.
The shape of the impact-parameter distribution was conservatively modelled to be similar to that of low-momentum \antiHe from \antiLambdab decays.

Current state-of-the-art Monte Carlo generators predict branching ratios for the \antiLambdabToHe decay that vary significantly from about $10^{-9}$ to $10^{-6}$ as reported in \Fig{fig:Lb2He3CL}.
In particular, PYTHIA~8  with the so-called \Lambdab-tune predicts values at around $2.6\times 10^{-6}$, large enough to allow the explanation of the measured flux of \antiHe observed by AMS-02 as coming from dark matter annihilation.
In~\cite{Kachelriess:2021vrh} it is however suggested that this branching ratio might be overestimated by a factor ranging from 5.6 to 17 due to the use of an increased probability for di-quark formation. Interestingly, an equivalent study performed with the HERWIG~7 event generator~\cite{Winkler:2020ltd} leads to a branching ratio that is 3~orders of magnitude lower than the one predicted by all PYTHIA~8 scenarios.
This result is partly due to the 5~times lower BR(\antiLambdabToDominant) compared to the PYTHIA~8 predictions and partly due to the different hadronisation mechanism in HERWIG~7.

The sensitivity of the branching ratio measurement was evaluated as a function of the integrated luminosity by computing the expected upper limit at 90\% confidence level with the asymptotic formula for the profile likelihood~\cite{Cowan:2010js} with different scenarios and is presented in~\Fig{fig:Lb2He3CL}.
The expected upper limit on the branching ratio is $10^{-8}$ for ALICE~3 and $10^{-6}$ with ALICE~2. 
The outstanding performance of ALICE~3 would allow to discriminate between the different prediction for the branching ratio, and thus not only provide astrophysically relevant constraints but also additional insights into the dynamics of b-quark decays.

\begin{figure}[t]
  \centering
  \includegraphics[width=0.49\textwidth]{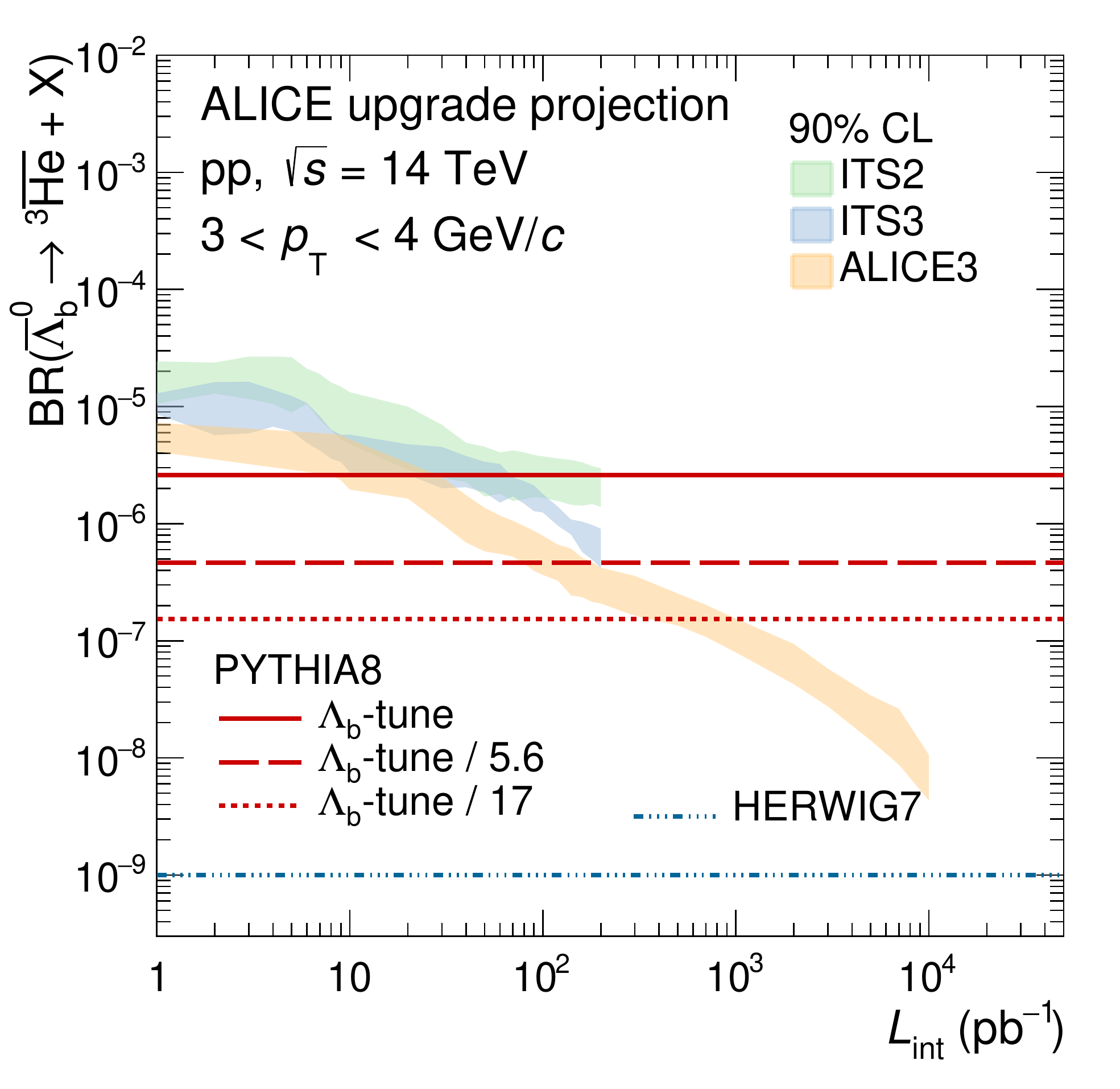}
  \caption[Limits for branching ratio of \LambdabToHe] {Expected upper limit for the branching ratio of \LambdabToHe decay at 90\% with the ITS2, ITS3, and ALICE 3 detectors.
  The predictions from PYTHIA 8 with $\Lambdab$-tune and HERWIG 7 Monte Carlo generators are also shown.}
  \label{fig:Lb2He3CL}
\end{figure}

\subsubsection{Ultra-peripheral collisions and BSM searches}
\label{sec:performance:physics:bsm}
Ultra-peripheral collisions (UPCs) are characterized by impact parameters larger than the sum of the radii of the incoming nuclei. Ions accelerated to LHC energies are intense sources of virtual photons, with a flux proportional to the square of the electric charge of the ion, therefore UPCs are dominated by photon-photon and photon-nucleus interactions providing a clean environment for vector meson photoproduction studies, light-by-light measurements and searches for axion-like particles. 

\paragraph{Photoproduction studies}
\label{sec:performance:physics:bsm:upc:photoproduction}

As discussed in Section~\ref{sec:physics:vectormesons}, ultra-peripheral photoproduction processes provide a clean environment for the investigation of the nature of hadronic states. A key requirement for studying photoproduction is large angular acceptance.  For an $n$ particle final state, the efficiency goes roughly as $\epsilon^n$, where $\epsilon$ is the single particle efficiency.  The example of the $\rho'$ decay into four pions is discussed in this section. Figure~\ref{fig:rhoprime} compares the expected geometric acceptance for $\rho'\rightarrow\pi^+\pi^-\pi^+\pi^-$, simulated in STARlight~\cite{Klein:2016yzr} as a function of rapidity, in ALICE and ALICE 3.  The $\rho'$ is simulated as a single resonance with mass $M=1.54$ GeV and width $\Gamma=0.570$ GeV, as observed by STAR~\cite{STAR:2009giy}. The ALICE and ALICE 3 acceptances are simulated with cuts on the charged particle pseudorapidity and $p_T$.  The ALICE acceptance was taken to be $\eta|<0.9$, while ALICE 3 was assumed to cover $|\eta|<4$. Both were assumed to have acceptance for $p_T > 100$ MeV/c, but the exact $p_T$ cut is relatively unimportant here.

For $\sigma(\rho'\rightarrow\pi^+\pi^-\pi^+\pi^-)$=730 mb~\cite{Citron:2018lsq}, 25 billion $\rho'\rightarrow\pi^+\pi^-\pi^+\pi^-$ are produced in 35 nb$^{-1}$ of integrated luminosity.  The signal covers a broad rapidity range.  The current ALICE detector can only reconstruct 0.4\% of the signal, in a narrow rapidity range.  In contrast, ALICE 3 has a geometric acceptance of 8.4\%, 19 times larger, corresponding to 2.1 billion events in  35 nb$^{-1}$.  ALICE 3 can use the signal rapidity dependence to determine the photon energy dependence of the cross section.   The efficiency gains should be similar for other 4-prong final states, including double-meson production. With this same efficiency and with the cross sections in Ref.~\cite{Klein:1999qj}, ALICE 3 would observe 21,000 $\rho^0\rho^0$ pairs and about 500 $\rho J/\psi$ pairs, after accounting for the branching ratios.

\begin{figure}
\centering
\includegraphics[width=0.7\textwidth]{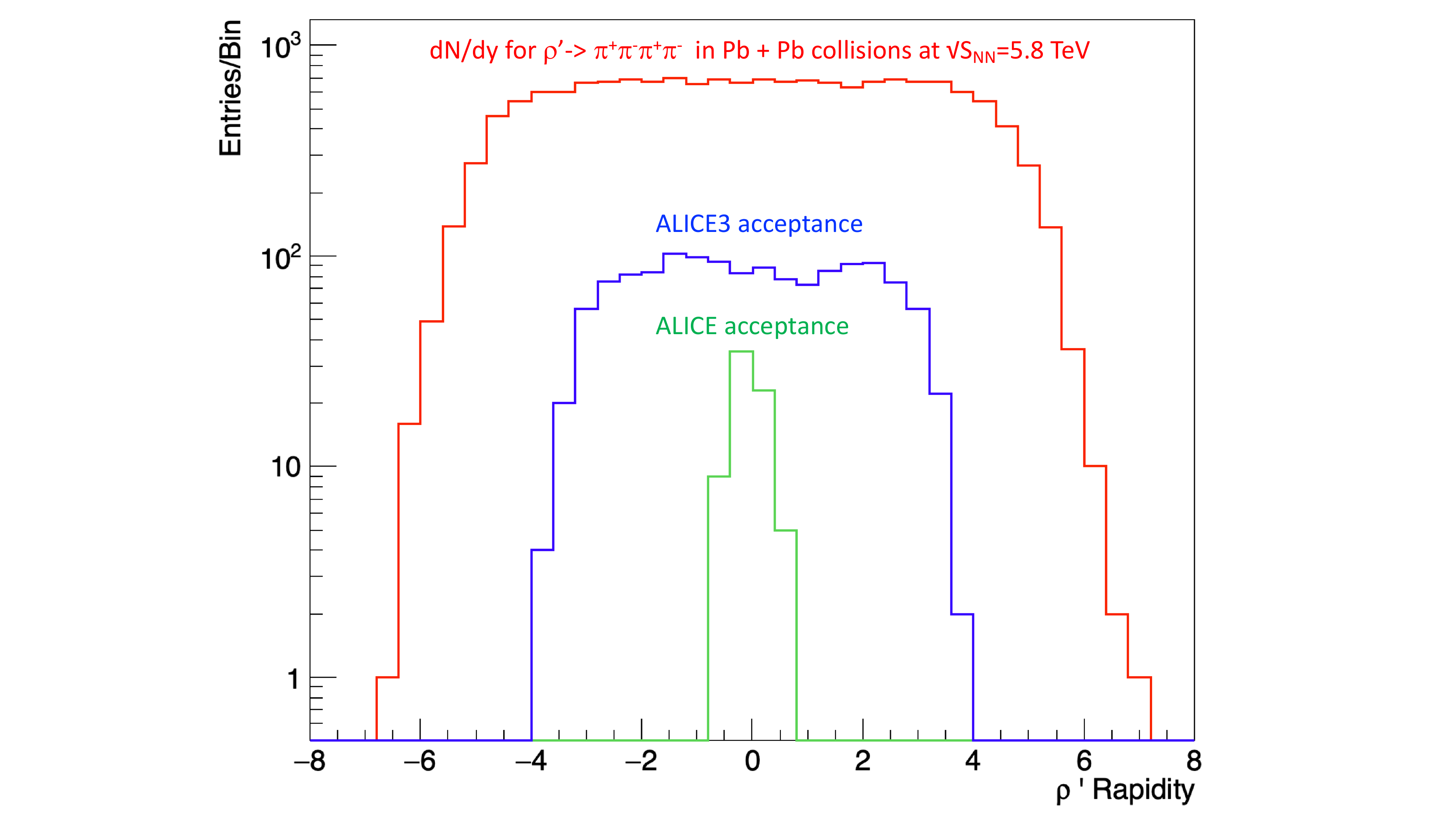}
\caption[Pseudorapidity acceptance for $\rho'\rightarrow\pi^+\pi^-\pi^+\pi^-$]{\label{fig:rhoprime}$dN/dy$ for $\rho'\rightarrow\pi^+\pi^-\pi^+\pi^-$, as generated in STARlight (red curve), within the ALICE 3 acceptance (blue), and within the current ALICE acceptance (green).}
\end{figure}

\paragraph{Light-by-light scattering measurements}
\label{sec:performance:physics:bsm:upc:lbl}
UPCs provide a clean environment for light-by-light scattering measurements~\cite{dEnterria:2013zqi}. The final state of interest is a diphoton event in an otherwise empty detector, with the two photons emitted back-to-back. Final state photons can be reconstructed either via photon conversions or via ECAL measurements. Two cases are considered in the following. In the first, we assume a photon reconstruction efficiency of 5\% in the pseudorapidity range $|\eta|<4$ which could be achieved with a dedicated photon converter in the central barrel detectors and in the forward and backward direction. In the second, we consider an ideal scenario with 100\% photon reconstruction efficiency that could be approached with ECAL measurements.

Light-by-light events in \PbPb UPCs were generated with the SuperChic event generator~\cite{Harland-Lang:2018iur}. 
The dominant source of background at low diphoton invariant masses below 2 \GeVcc is expected to come from random pairs of photons from $\pi^0\pi^0$ photoproduction~\cite{Klusek-Gawenda:2019ijn}. ${\rm PbPb} \to {\rm PbPb}\pi^0\pi^0$ background events were obtained using the $\gamma\gamma \to \pi^0\pi^0$ cross section from~\cite{Klusek-Gawenda:2013rtu} folded with the effective photon-photon luminosity in \PbPb UPCs. Decays of photoproduced $\pi^0$ pairs may result in a final state with four photons of which only two are detected, while the other two escape. This background can be reduced by imposing cuts on the transverse momentum asymmetry of the two photons. 

The expected yields of light-by-light and $\pi^0\pi^0$ background processes are shown in Fig.~\ref{fig:bsm:lbl} as a function of diphoton invariant mass after asymmetry cuts on the transverse momentum of two photons. The $\pi^0\pi^0$ remains the dominant background source in the diphoton mass region from 0.5 to \SI{3}{\giga\eVcsq}, exceeding the light-by-light spectrum by a factor of 10. However, at low invariant masses below 0.5 \GeVcc, the light-by-light signal is expected to be dominant. The strategy based on the photon conversion method can be used at low masses only, while ECAL measurements may provide high statistics in a wide range of invariant masses. 

\begin{figure}
\centering
\includegraphics[width=0.49\linewidth]{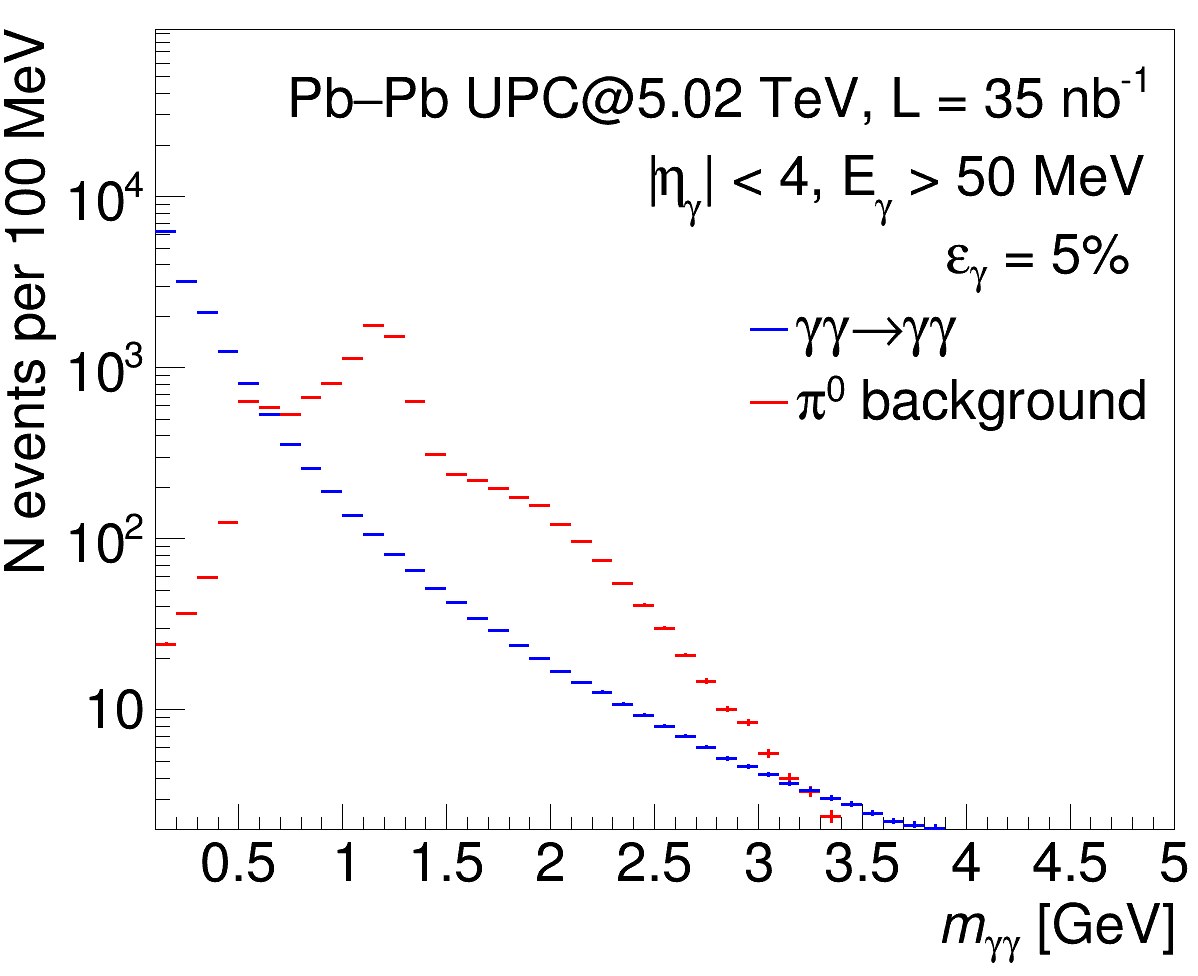}
\includegraphics[width=0.49\linewidth]{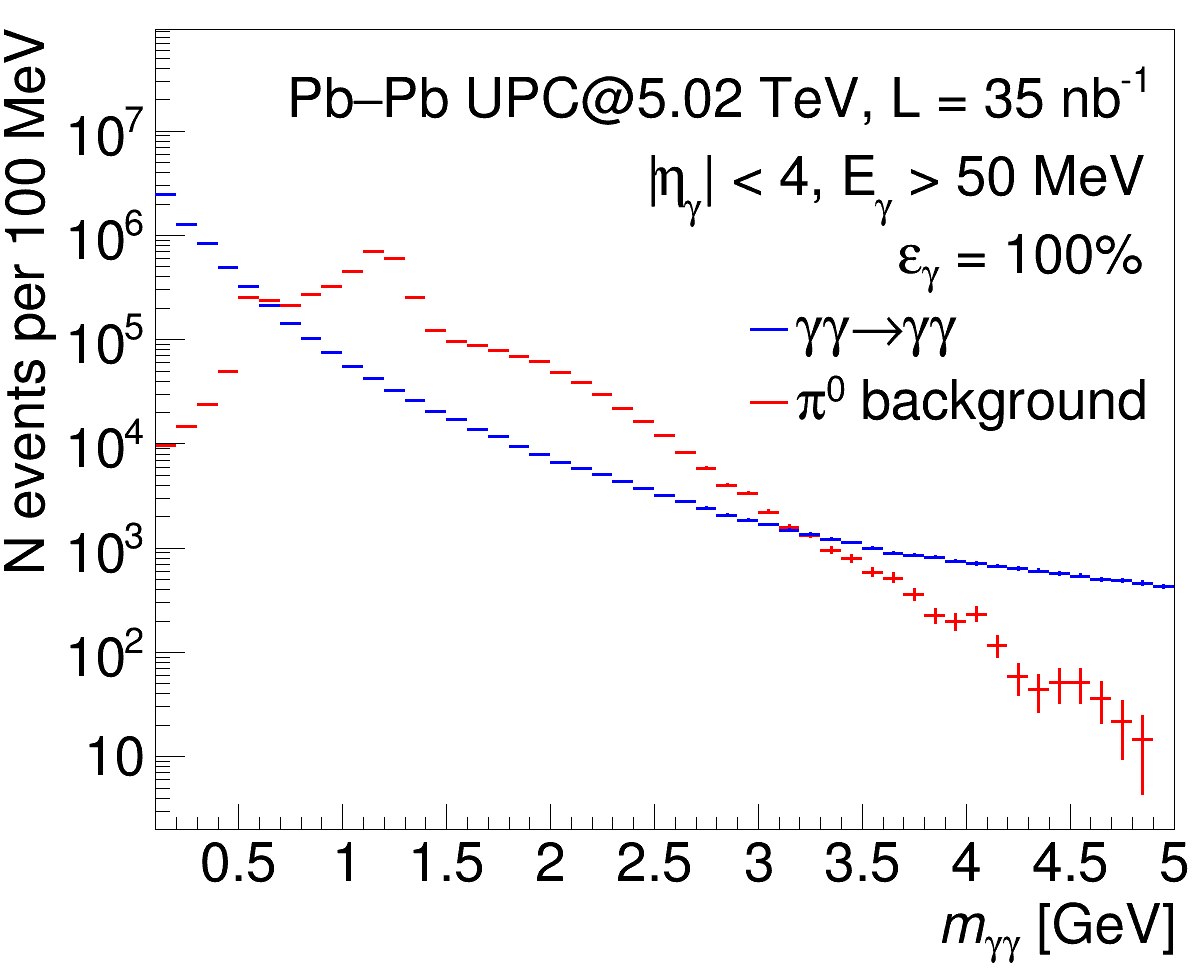}
\caption[Expected yields of light-by-light scattering]{Expected yields of light-by-light and $\pi^0\pi^0$ background processes as a function of diphoton invariant mass in \PbPb UPCs in two scenarios: assuming a photon reconstruction efficiency of 5\% (left) and the ideal case with 100\% photon reconstruction efficiency (right).}
\label{fig:bsm:lbl}
\end{figure}

\paragraph{Search for axion-like particles}
\label{sec:performance:physics:bsm:upc:alp}

Light-by-light scattering measurements can be also used for axion-like particle searches in the $\gamma\gamma \to a \to \gamma\gamma$ process~\cite{Knapen:2016moh}.  The ALP signal would be visible as a peak in the diphoton invariant mass distribution on top of light-by-light continuum and other background processes. This signature was used by the CMS~\cite{CMS:2018erd} and ATLAS~\cite{ ATLAS:2020hii} collaborations to set limits on ALPs over the mass region from $5$ to $100$\, GeV.

ALICE~3 would provide a unique opportunity to improve the limits on ALP-photon coupling in UPCs in the mass region from \SI{50}{\mega\eVcsq} to 5 \GeVcc. In this study, $a \to \gamma\gamma$ peaks were generated with Starlight~\cite{Klein:2016yzr} and injected into the continuum $m_{\gamma\gamma}$ distribution, discussed in Section~\ref{sec:performance:physics:bsm:upc:lbl}, assuming that the ALP peak width is dominated by the experimental diphoton mass resolution. The expected performance for a limit on the ALP production cross section was estimated as a function of mass and recast into limits on the ($m_a$, $1/\Lambda_a$) plane as described in~\cite{Knapen:2016moh}. 

The expected performance of ALICE~3 in 5\% and 100\% photon reconstruction efficiency scenarios is shown in Fig.~\ref{fig:bsm:alps}  together with existing limits (from~\cite{ATLAS:2020hii}) and projections for other LHC experiments~\cite{Goncalves:2021pdc,Knapen:2016moh}. As can be seen in the figure, the ALICE~3 experiment is expected to fill the gap between beam-dump and ATLAS/CMS constraints and push the limits on ALP-$\gamma$ coupling well below $1{\,\rm TeV}^{-1}$ in the intermediate mass range from \SI{50}{\mega\eVcsq} to \SI{5}{\giga\eVcsq}.

\begin{figure}
\centering
\includegraphics[width=0.65\linewidth]{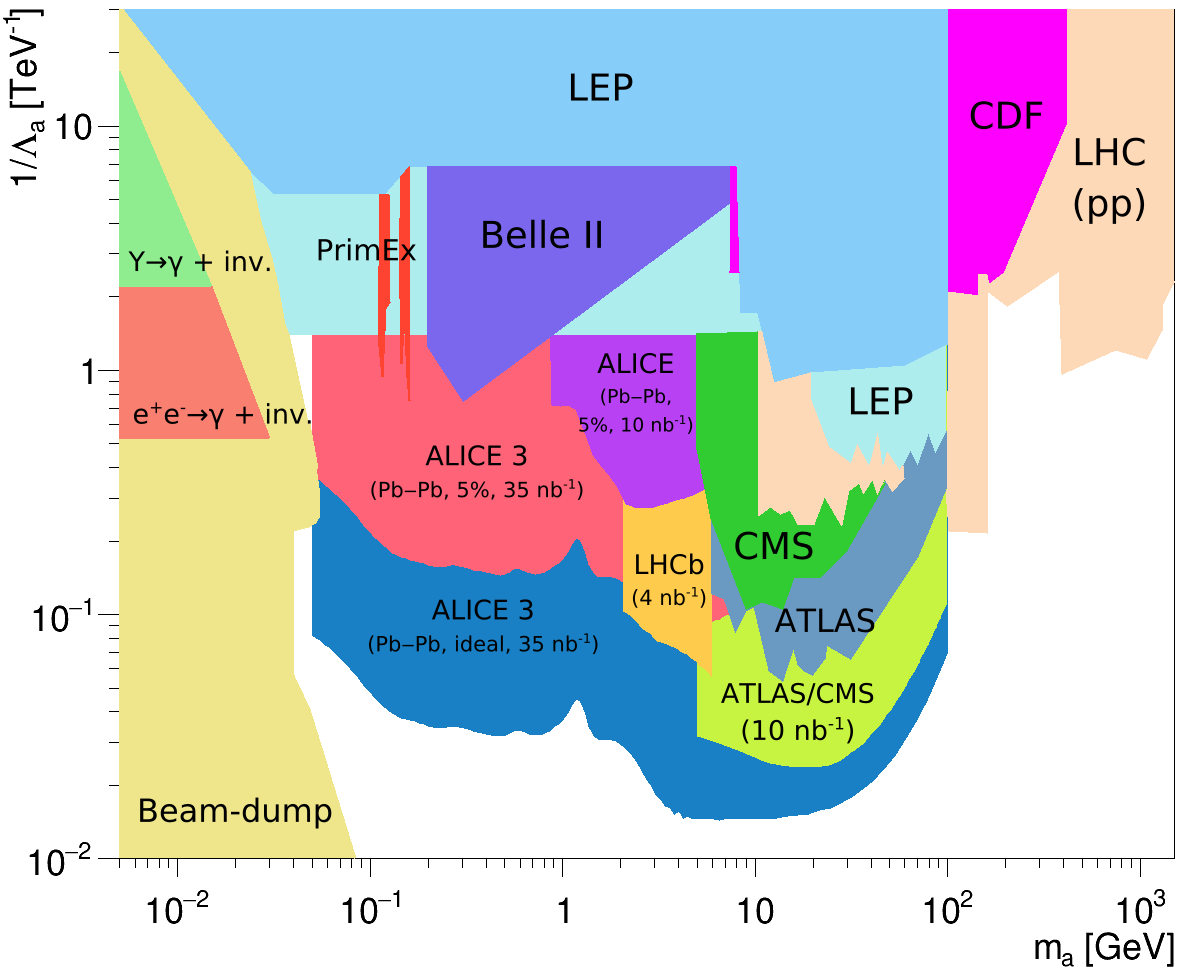}
\caption[Limits from present and future ALP searches]{Bounds in the ($m_a$, $1/\Lambda_a$) plane from existing and future ALP searches.}
\label{fig:bsm:alps}
\end{figure}

\cleardoublepage
\section{Detectors and systems}
\label{sec:systems}

In this chapter, we present studies on the implementation of the detector concept shown in Fig.~\ref{fig:alice3_cross}. 
The detector design is driven by the requirements for the measurements discussed before (cf. Tab.~\ref{tab:intro:observables}):
tracking and particle identification over a large acceptance, excellent vertexing, continuous readout. 
For each system, we discuss technologies suitable to meet the requirements.
For viable technologies, we present the state-of-the-art and point out areas which require dedicated R\&D.

\begin{figure}
    \centering
    \includegraphics[width=.95\textwidth]{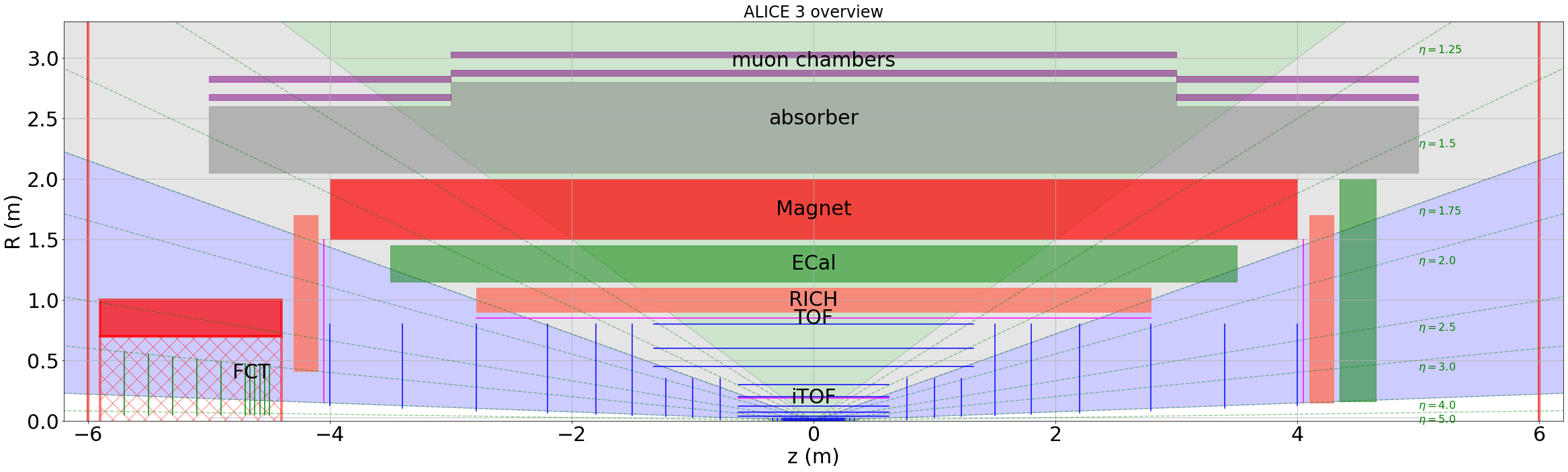}
    \caption[Longitudinal cross section of ALICE~3 detector]{Longitudinal cross section of the ALICE 3 detector: The MAPS-based tracker is complemented by PID detectors (inner and outer TOF, RICH), all of which are housed in the field from a superconducting magnet system. In addition, the electromagnetic calorimeter (ECal), the muon identifier, and the Forward Conversion Tracker (FCT) are shown.}
    \label{fig:alice3_cross}
\end{figure}

\begin{revised}
Like for the LS2 upgrades, the development, construction, and installation of this large upgraded project shall be carried out in parallel with the operation of the experiment and upgrades planned for LS3, which are well-contained and considerably smaller in scope: while the ITS3 requires dedicated R\&D activities, the actual installation and operation relies on infrastructure (e.g. readout) already used with ITS2. In addition, there is sufficient time for the installation and commissioning of the new components during LS3.

\end{revised}

\subsection{Magnet system and infrastructure}
\label{sec:systems:magnet}
The ALICE detector is installed in the cavern that hosted the L3 experiment during the operation of LEP. The L3 magnet, which is a structural part of the cavern, is being used by the ALICE experiment and it provides a solenoidal magnetic field of \SI{0.5}{\tesla} at a power consumption of approximately \SI{4}{\mega\watt}. In addition to this solenoid, a dipole magnet was installed by the ALICE collaboration in 2006 and it provides a field integral of \SI{4}{\tesla\metre} at a power consumption of \SI{4}{\mega\watt}. ALICE will continue to use this configuration for Run~3 and 4.

\begin{figure}
  \centering
  \includegraphics[width=0.45\textwidth]{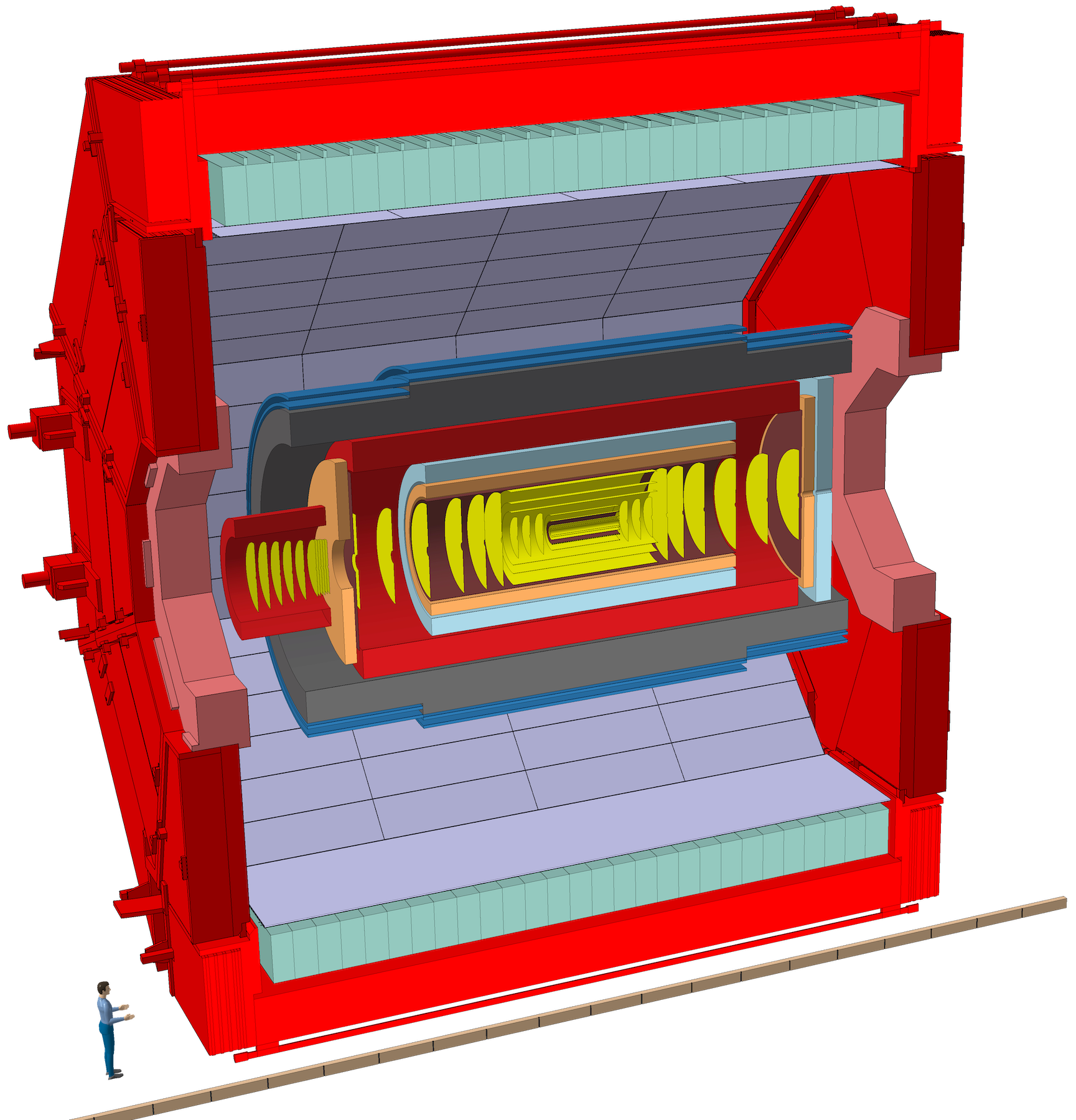}
  \hfill
  \includegraphics[width=0.45\textwidth]{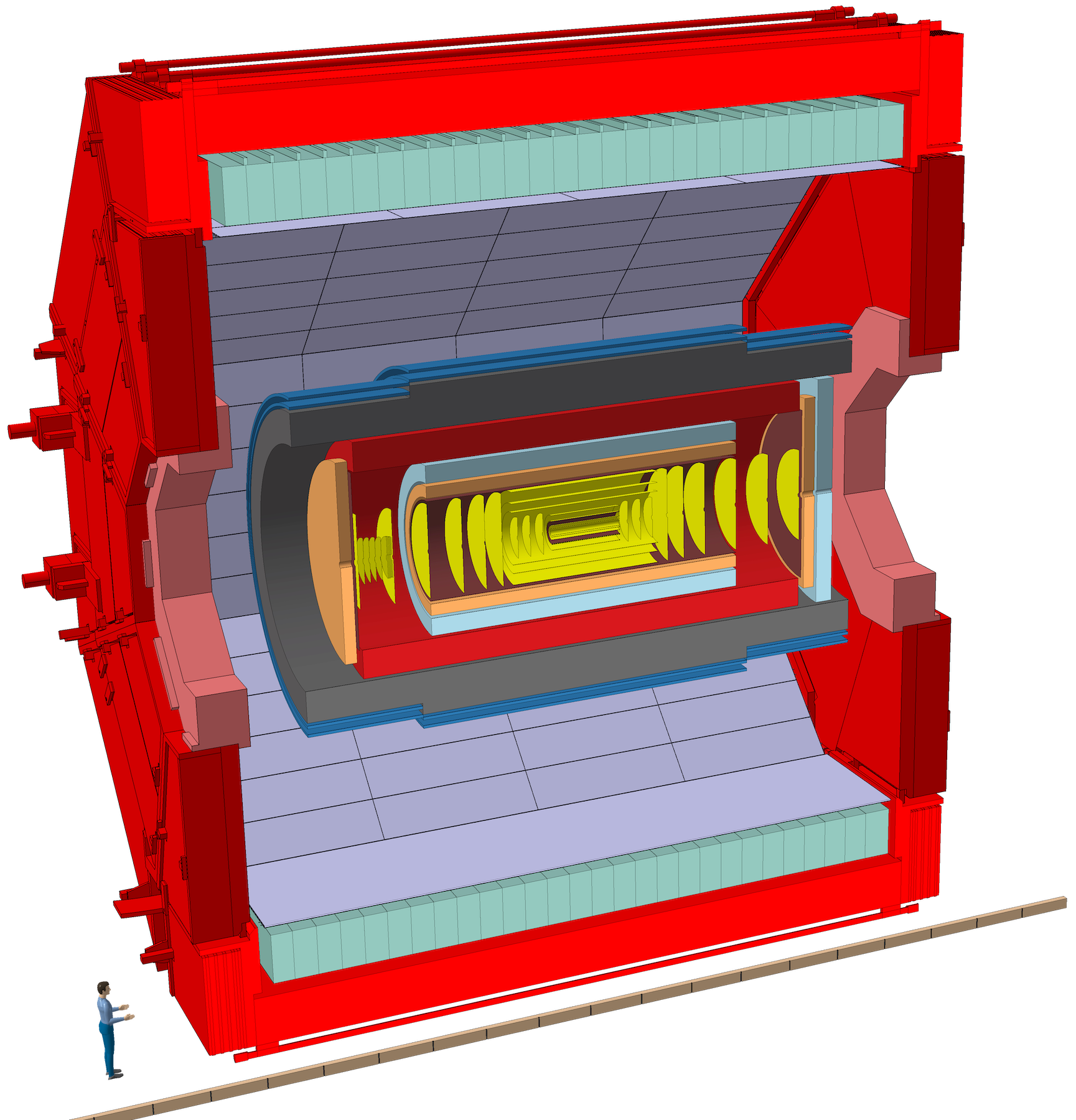}
    \caption[ALICE 3 installation]{The ALICE3 detector installed inside the L3 magnet yoke. The left figure shows the detector layout with a solenoid and a dedicated dipole magnet for the FCT. The right figure shows the detector layout with a solenoid and two dipoles integrated in the main magnet system.}
    \label{fig:det:alice_cavern}
\end{figure}

For the installation of the ALICE~3 detector, the present ALICE detector, together with the dipole magnet, will be removed and only the L3 magnet yoke will remain in the cavern. 
A superconducting magnet will be installed inside the volume of the L3 magnet and the L3 yoke will just act as a shield for the magnetic field towards the cavern, as shown in Fig.~\ref{fig:det:alice_cavern}.
The possibility of opening both magnet doors allows a symmetric detector installation and assembly with a central barrel and two end caps. 
The baseline material for the absorber of the muon system is non-magnetic steel, but magnetic steel is also a possibility.

Two options of the superconducting magnet system are shown in Fig.~\ref{fig:det:sc_magnet} and some key parameters are given in Tab.~\ref{tab:det:magnet_parameters}. 
The baseline configuration consists of a solenoid coil over the full length of \SI{7.5}{\metre} with additional windings at the coil ends that represent 50\% higher current density. The second configuration has a central solenoid of 2\,m length with a dipole magnet on either side.
The main motivation for the dipole system is improved momentum spectroscopy in the rapidity range $2<\eta<4$.
Having the solenoid and the dipoles at the same radius inside the same cylindrical volume allows easy installation and maintenance of the detectors without the need of displacing parts of the magnet system. 
It also allows us to treat the forces between dipoles and solenoid inside the cold mass, which avoids difficulties with thermal contacts.

\begin{figure}
\includegraphics[width=\textwidth]{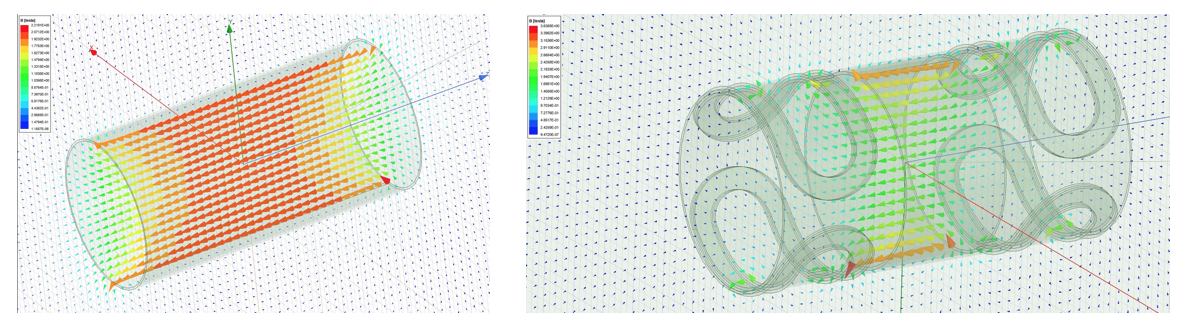}
    \caption[Superconducting magnet system]{Superconducting magnet system: Solenoid (left) and solenoid + dipoles (right).}
    \label{fig:det:sc_magnet}
\end{figure}

\begin{table}
\centering
\renewcommand{\arraystretch}{1.3}
   \begin{tabular}{l l *{2}{S[table-format=3.2]}}
   \toprule
    & {Unit} & {Solenoid} & {Solenoid + Dipole} \\
     \midrule
Central magnetic field& T &	2     &    2 \\
Cold mass length& m	& 7.5 &    7\\
Length of coil & m & 7.5 & {2 + 2 + 2} \\
Free bore radius  & m & 1.5 & 1.5 \\
Stored magnetic energy & MJ &	144 & 86  \\
Operating current& kA	& 20 & 20 \\
Inductance& H&0.6 & 0.43 \\
Cold mass weight estimate& t	& 10 & 20 \\
Vacuum vessel, radial thickness & m  &	0.5 & 0.5\\
Peak field on  conductor (excl. self-field) & T  &	2.5 &  3.9 \\
 \bottomrule
   \end{tabular}
    \caption{Parameters of the two superconducting magnet configurations. The dipoles provide a peak field of \SI{0.5}{\tesla}.}
    \label{tab:det:magnet_parameters}
\end{table}

Both magnets provide a solenoid field of up to \SI{2}{\tesla} and therefore a field integral of up to \SI{2}{\tesla\metre} at low values of $\eta$. 
Along the beam axis, i.e. at high values of $\eta$, the dipoles provide a field integral of \SI{1}{\tesla\metre}, with a peak field of \SI{\sim 0.5}{\tesla}. 
Figure~\ref{fig:det:bfield} shows the field map of the solenoid+dipole magnet system in a vertical plane through the beam axis, together with the expected performance of both magnet systems. 
For muons of $\pt = \SI{1}{\giga\eVc}$, i.e. at the multiple scattering limit, the solenoid provides a momentum resolution between 0.6 and 1\% up to $\eta=2$ and the resolution then deteriorates to about 5\% at $\eta=4$. 
The solenoid+dipole magnet system provides a  momentum resolution of 1\% at $\eta=4$, with a slight deterioration in the transition region between the solenoid and dipole field in the region of $1.2<\eta<2.2$.

These magnet configurations represent two 'classic' spectrometer geometries. It should be noted that variants of the solenoid geometry with tilted windings could represent a cost effective way to realise a performance that lies between the two geometries. 

\begin{figure}
    \centering
    \includegraphics[width=.45\textwidth]{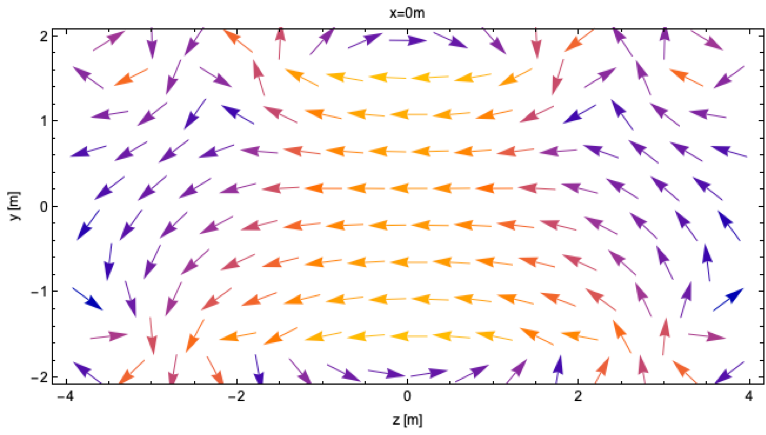}
    \hfill
    \includegraphics[width=.45\textwidth]{systems/Figures_tracker/momentum_resolution-2022-03-01}
    \caption[Fieldmap and resolution of magnet system]{Fieldmap of the solenoid+dipole magnet system (left) and momentum resolution for a muon of $\pt = \SI{1}{\giga\eVc}$ for different field configurations.}
    \label{fig:det:bfield}
\end{figure}

\ifcost
A preliminary cost estimate for the two magnet systems outlined in Tab.~\ref{tab:det:magnet_parameters} is  18\,MCHF for the solenoid option and 36\,MCHF for the solenoid+dipole option.

The cost of the power for the present ALICE magnet system, which consumes \SI{8}{\mega\watt} (for the solenoid and dipole), is 1.6\,MCHF for a full operational year. 
This amounts to 9.6\,MCHF for 6 years of operation and 16\,MCHF for 10 years of operation. The expected power consumption of $< \SI{400}{\kilo\watt}$ for the superconducting magnet system will result in 95\% power saving and therefore a significant energy and cost saving. 
\fi

\subsection{Vertex detector and outer tracker} 
\label{sec:systems:tracking}

To reach the performance goals discussed in previous chapters, 
the conceived detector is based on \num{11}~barrel layers and \num{2x12}~forward discs. 
It is divided into a \emph{Vertex Detector} made of the first \num{3}~layers and \num{2x3}~discs, retractably mounted inside a secondary vacuum, and the \emph{Outer Tracker} (cf.~Fig.~\ref{fig:systems:tracking:layout}). 
It will cover the pseudorapidity interval of $\abs{\eta}<4$, with longitudinal and radial extensions of \SI{\pm 400}{\cm} and \SIrange{0.5}{80}{\cm}, respectively. 
Table~\ref{tab:table-name} details the position, material thickness, and intrinsic resolution of each layer. 
The resulting active surface sums to around \SI{60}{\metre \squared}. 
Accounting for an overlap of sensors of about \SI{10}{\percent} to cover inactive periphery and to achieve hermeticity, a total amount of \SI{66}{\metre \squared} of silicon has to be installed.

\begin{figure}
\centering
\includegraphics[width=\textwidth]{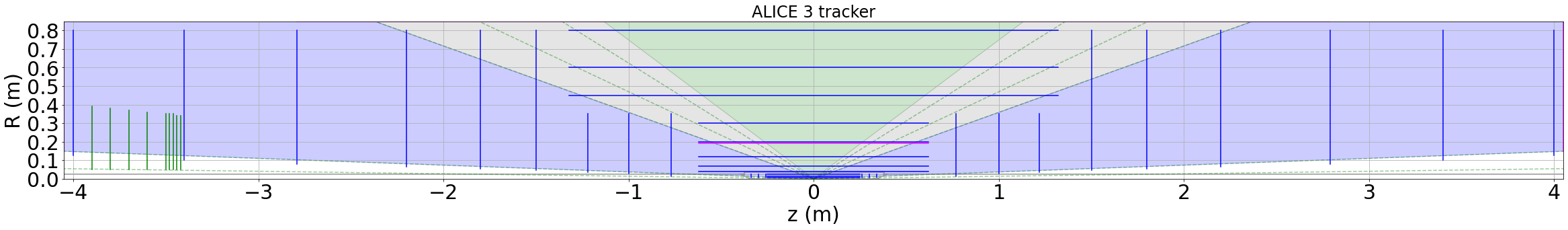}\\
\includegraphics[width=\textwidth]{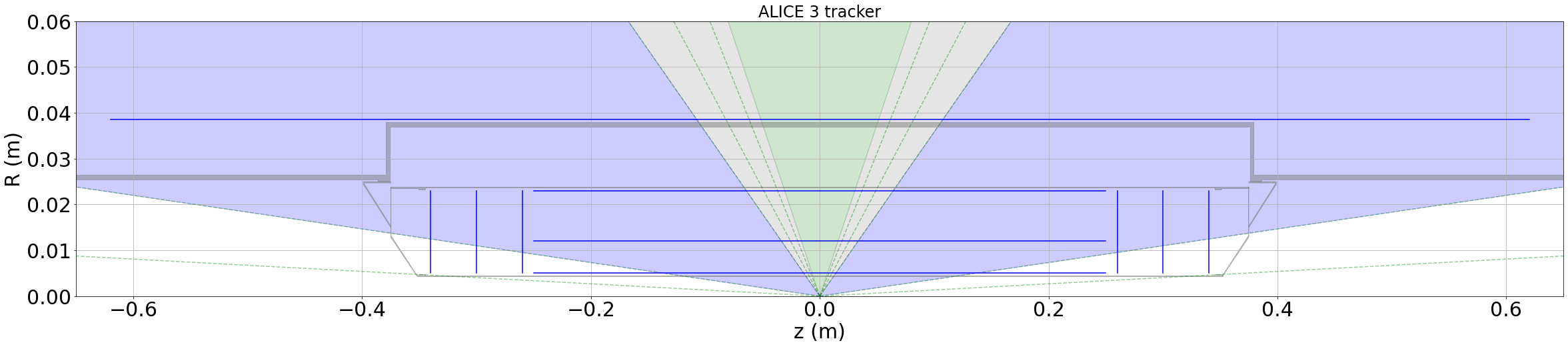}
\caption[Geometry of Vertex Detector and Outer Tracker]{Schematic $R-z$ view of the full tracker (top) and of the vertex detector separately (bottom). The blue lines represent the tracking layers. The FCT disks are marked in green. In addition, the beampipe and vacuum vessel of the vertex detector are shown in grey.}
\label{fig:systems:tracking:layout}
\end{figure}

\begin{table}
    \centering
      \begin{tabular}{%
      S[table-format=2]
      S[table-format=1.1]
      S[table-format=2.1]
      S[table-format=3]
      S[table-format=2.2]
      S[table-format=3]
      S[table-format=1.3]
      S[table-format=2]
      }
      \toprule
      {Layer} & {Material} & {Intrinsic} & \multicolumn{2}{c}{Barrel layers}
      & \multicolumn{2}{c}{Forward discs}\\
      \cmidrule(lr){4-5}
      \cmidrule(lr){6-8}
      &
      {thickness} &
      {resolution} &
      {Length ($\pm z$)} &
      {Radius ($r$)} &
      {Position ($\abs{z}$)} &
      {$R_\mathrm{in}$} &
      {$R_\mathrm{out}$} 
      \\
      &
      {($\si{\percent}X_0$)} &
      {(\si{\um})} &
      {(\si{\cm})} &
      {(\si{\cm})} &
      {(\si{\cm})} &
      {(\si{\cm})} &
      {(\si{\cm})} \\
      \midrule
       0 & 0.1 &  2.5 & 50  & 0.50 &  26 & 0.50 &  3\\
       1 & 0.1 &  2.5 & 50  & 1.20 &  30 & 0.50 &  3\\
       2 & 0.1 &  2.5 & 50  & 2.50 &  34 & 0.50 &  3\\
       \midrule
       3 & 1   & 10 & 124 & 3.75 &  77 & 5 & 35  \\
       4 & 1   & 10 & 124 & 7    & 100 & 5 & 35  \\
       5 & 1   & 10 & 124 & 12   & 122 & 5 & 35  \\
       6 & 1   & 10 & 124 & 20   & 150 & 5 & 80  \\
       7 & 1   & 10 & 124 & 30   & 180 & 5 & 80  \\
       8 & 1   & 10 & 264 & 45   & 220 & 5 & 80  \\
       9 & 1   & 10 & 264 & 60   & 279 & 5 & 80  \\
      10 & 1   & 10 & 264 & 80   & 340 & 5 & 80  \\
      11 & 1   &    &     &      & 400 & 5 & 80  \\  
     \bottomrule 
    \end{tabular}
    \caption{\label{tab:table-name} Geometry and key specifications of the tracker.}
\end{table}

\subsubsection{Specifications}
\label{sec:systems:tracking:specs}

The detector design has to fulfil the following criteria:

\begin{enumerate}
    \item\textbf{Radial distance of first detection layers.} 
    The radial distance of the first measured hit position must be as close as possible to the interaction point, which is fundamentally limited by the required aperture for the LHC beam ($\sim \SI{5}{\mm}$).
    
    \item\textbf{Material budget.}
    The vertex detector and the outer tracker target low material thicknesses of \SI{0.1}{\percent} and \SI{1}{\percent} of a radiation length, respectively.
    
    \item\textbf{Intrinsic spatial resolution.} 
    The layers of the vertex detector and of the outer tracker shall provide intrinsic position resolutions of \SI{2.5}{\um} and \SI{10}{\um}, respectively.
    Depending on the amount of charge sharing between neighbouring pixels, this translates to pixel pitches of about \SI{10}{\um} and \SI{50}{\um}, respectively.
   
    \item\textbf{Hit time resolution.}
    To achieve a time binning of \SI{500}{\ns} in the vertex detector and the outer tracker, the sensors must provide a timing resolution (r.m.s) of \SI{\sim 100}{\ns}.
     
    \item\textbf{Rate capability.}
    The sensors in the most exposed region of the vertex detector must be able to read out average hit rates of~\SI{35}{\MHz\per\cm\squared} in order to record all events in continuous readout.
    In the outer tracker, the expected rates are significantly lower, e.g. \SIrange{1}{5}{\kHz\per\cm\squared} in the outermost layers.

    \item\textbf{Data throughput.}
    Assuming an encoding with \SI{2}{bytes/hit} and a fake hit rate of \num{\sim e-8} per pixel and event, a total data rate of \SI{\sim 1}{\tera\bit\per\second} is expected.

    \item\textbf{Power consumption and powering scheme.}
    In order to keep the material thicknesses within budget, the power consumption of the sensors must stay below~\SI{70}{\mW\per\cm\squared} for the Vertex Detector and around~\SI{20}{\mW\per\cm\squared} for the Outer Tracker. 
    
    \item\textbf{Radiation hardness.}
    The maximum radiation load per operational year will be about \SI{1.5e15}{\nequiv} on the first tracking layer at a radial distance of~\SI{5}{\mm} from the interaction point.
\end{enumerate}

\begin{revised}
  The progress of ALICE~3 relies on the combined progress in effective statistics (luminosity $\times$ acceptance) and pointing resolution.
  The latter scales linearly with the radial distance of the first hit from the interaction point and with the square root of the material thickness of the first layer (multiple scattering):
  \begin{equation}
      \sigma_\mathrm{DCA} \propto r_0 \cdot \sqrt{X/X_0 \cdot \cosh{\eta}} \cdot \frac{1}{p}.
  \end{equation}
  From this scaling, it is evident that the envisaged performance can only be achieved by an ultra-thin layer as close as possible to the interaction point. The latter is fundamentally limited by the aperture required for the colliding beams. Since a much wider aperture is required at injection energy, the only way to get to \SI{5}{\mm} from the beam is a retractable detector design. 
  A static design would be limited to having the first layer at $\SI{\sim 15}{\mm}$ from the interaction point. 
  It is further important to have the first hit always in the first radial layer, whose material thickness, thus, determines the pointing resolution (also for very forward tracks).
  The pixel pitch of \SI{10}{\um} is chosen such that the resulting position resolution of \SI{\sim 2.5}{\um} is negligible with respect to the multiple scattering for tracks up to a transverse momentum of \SI{1}{\giga\eVc}.

The outer radius of the tracker defines the lever arm in the bending plane, which determines the momentum resolution.
The number of layers and their positions have been chosen as to limit the track finding inefficiency due to assignments of fake hits.
The cost of the detector is driven by the outer layers and has a weak dependence on the number of layers.
The current layout was prepared for concrete performance studies for the Letter of Intent.
In preparation of the TDRs, we will carry out a detailed optimisation, which takes into account all relevant aspects, i.e. fake hit probabilities, tracking of secondaries (strangeness tracking), and the impact of dead zones in a layer. 
\end{revised}

\comment{reinclude hit density figures here}

\comment{Reinclude data rate and power tables}

\subsubsection{Technology options}
\label{sec:systems:tracking:options}

CMOS Monolithic Active Pixel Sensors (MAPS) are considered the baseline technology for the vertex detector and the outer tracker. 
This is based on their proven performance and the existing know-how within the collaboration developed during the ITS2 project.
Figure~\ref{fig:maps} shows the principle of a MAPS, indicating the sensitive volume, which is a high-resistivity epitaxially grown layer. 
The electric field within this layer can be engineered to optimise for fast charge collection (for radiation hardness or timing) or for a certain amount of charge sharing (for better space point resolution).
The readout can include complex CMOS circuitry not only at the periphery of the chip, but also within the active pixel matrix, which allows highly optimised readout schemes. 
ALPIDE, for example, integrates a few hundred transistors within a single pixel of $\SI{\sim 30x30}{\um}$ in the Tower Semiconductor \SI{180}{\nm} CMOS imaging technology. 

\begin{figure}[!h]
    \centering
    \includegraphics[width=0.6\textwidth]{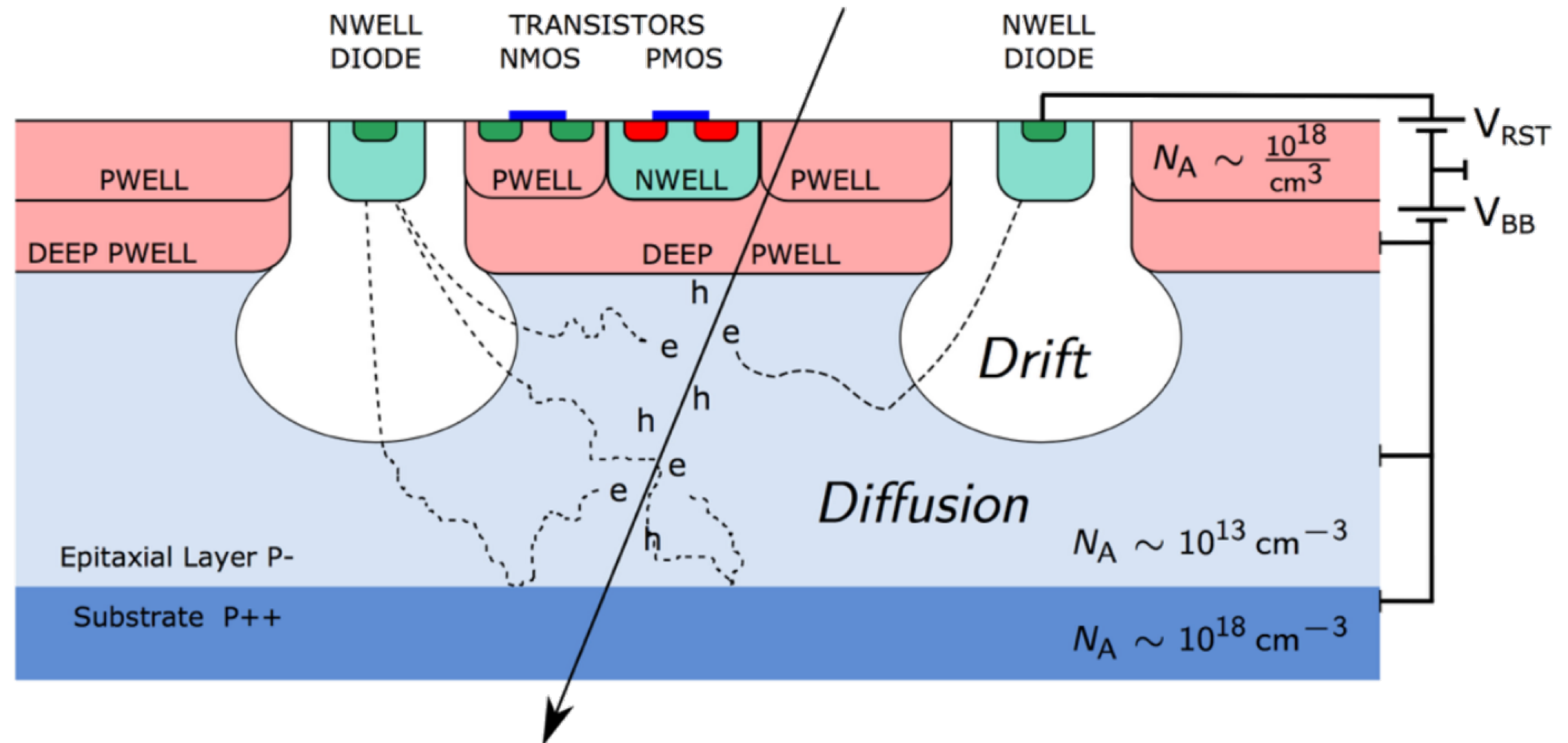}
    \caption[Cross section of ALPIDE]{Schematic cross section of MAPS as used for the ALICE ITS2 detector.}
    \label{fig:maps}
\end{figure}

In the ongoing development of ITS3, the CMOS technology node was changed from~\SI{180}{\nm} to~\SI{65}{\nm}, which gives further advantages in integration density and power consumption.
For ALICE~3, the Tower Partners Semiconductor Co. Ltd. (TPSCo) \SI{65}{\nm} CMOS imaging process is taken as reference but other processes will be considered as well.

\subsubsection{Development status}
\label{sec:systems:tracking:dev}

During LHC LS2, the ALICE collaboration successfully constructed and installed a \SI{10}{\metre\squared} MAPS-based vertex and tracking detector.
The core of the detector is the ALPIDE chip, which was developed by the collaboration within the last decade.
A number of prototypes were built within the R\&D activities of different groups focusing on improving different aspects of the technology:
\begin{description}
    \item\textbf{Radiation hardness.}
    Through optimisations of the CMOS process, MAPS can withstand \SI{\sim 1e15}{\nequiv} with mild performance degradation when cooled to \SI{\sim -30}{\celsius}~\cite{Snoeys:2017,Cardella:2019,Munker:2019,Dyndal:2020}.
    
    \item\textbf{Timing performance.}
      In the ALPIDE, the time resolution of about \SI{2}{\us} is determined by the frontend, which was optimised for power consumption at the expense of a slower response. 
      It has been demonstrated that the actual charge collection is not limiting and it has been further optimised much below \SI{1}{\ns} in other sensors~\cite{Snoeys:CMOS:2017,Ballabriga:2021,Kugathasan:2017,Buschmann:2021}. 
     
    \item\textbf{Module integration.}
    Industrial integration of chips on modules is available with standard techniques, such as wire bonding or flip-chip bonding. 
    This requires the detector sensors and electronics to be designed for manufacturability in close collaboration with industrial partners.

    \item\textbf{CMOS Technology feature size.}
    The ALICE ITS3 project, together with CERN EP R\&D, is presently testing the \SI{65}{\nm} node with the first tests in the laboratory and in-beam being encouraging.   
\end{description}

\subsubsection{R\&D challenges and plans}
\label{sec:systems:tracking:rnd}

\begin{revised}
Most sensor requirements can be met with technologies available today, but significant R\&D efforts are required in several areas. 
Two variants of MAPS are needed for the vertex detector and the outer tracker. For the former, wafer-sized sensors with excellent position resolution are needed. For the latter, reticle-size sensors of a few \si{\cm^2} with a more relaxed position resolution can be used. 
These developments can be seen as a continuation of the ITS2 and ITS3 activities, which constitute the first large-scale application of MAPS (\SI{10}{\m\squared}) at the LHC, from which the ALICE groups have vast experience.
Now, the developments for ITS3 are pioneering the usage of the \SI{65}{\nm} technology and the usage of stitching techniques for particle detectors. 
These developments of pixel sensors in CMOS technology are of broad interest for particle detectors and synergies are expected in particular with the accelerator-based experiments at the LHC, FCC, EIC, and FAIR. These activities also fit in the plans for generic R\&D at CERN and within other groups, e.g. the silicon consortium of the EIC. 
  
Another important area of R\&D is the modularisation, i.e. a standardized mounting of sensors on modules, for which we plan to explore the possibilities for industrialisation. 
To this end, different techniques for bonding should be explored, also taking into account the industrial availability of such techniques. 
Also these studies are of relevance for any large-scale detector and, thus, synergies are expected with other projects.

In addition, dedicated R\&D activities are required for the development for the mechanics of the vertex detector, which is at the limit of several existing technologies (vacuum, 3D printing, cooling, manufacturing of ultrathin walls, \dots). Here, synergies are expected with LHCb, EP R\&D, and ECFA.
\end{revised}

\paragraph{Power consumption.}
The total power consumption of a MAPS-based detector can be differentiated into three main contributions: the front-end, the on-chip data handling, and the off-detector data transmission. The importance of these contributions depends on timing requirements (front-end), hit rate (on-chip data handling and off-detector transmission), and on the chip area (on-chip data handling vs.\ off-chip transmission). 

\subparagraph*{Front end.}

\begin{revised}
The ALPIDE front-end with a rise time on the order of \SI{1}{\us} operates with \SI{20}{\nA} per pixel or \SI{20}{\milli\watt\per\centi\metre^2} assuming a pixel pitch of \SI{10}{\um}.
In order to achieve the time resolution (r.m.s.) of \SI{100}{\ns}, a faster frontend is needed, leading to an increase in power density.
In combination with the reduced supply voltage of \SIrange{1.0}{1.2}{V} in the \SI{65}{\nm} process (already implemented in the ITS3 MLR1 test structures), power densities on the order of \SI{50}{\milli\watt\per\centi\metre^2} are considered within reach through R\&D in the coming years.
This will require the optimisation of the front-end circuit for a different operating point.
\end{revised}

\subparagraph*{On-chip data handling.}
Scaling the power consumption observed for the aggregation of data from the pixel matrix in the ALPIDE with the expected data rates, a power density of about \SI{5}{\milli\watt\per\centi\metre^2} can be expected.

In a stitched sensor as planned for the vertex detector, additional high-speed data links to the edge of the sensor are required. 
Their power consumption strongly depends on the properties of the data lines (metal layers). 
With dedicated efforts (including the implementation of low-voltage signalling on-chip~\cite{Mensink:2010}), \SI{20}{\milli\watt\per\centi\metre^2} are considered realistic for the required data transfer rates.

\subparagraph*{Off-chip data transmission.}

The \SI{10}{\giga\bit\per\second}~transmitter, developed for TIMEPIX4 in a \SI{65}{\nm} TSMC process, consumes \SI{30}{\mW}~\cite{Ballabriga:Campbell:2020}, which approaches the state-of-the-art values achieved in industry.
It is planned to port the transmitter to the TPSCo~\SI{65}{\nm} process and to further improve it.

At a hit rate of~\SI{100}{\mega hit\per\cm\squared\per\second} and \SI{16}{\bit\per hit}, one~\SI{10}{\giga\bit\per\second} transmitter would be sufficient for an active area %
of~\SI{6}{\cm\squared} of the innermost layer of the vertex detector,
corresponding to a power density of~\SI{5}{\mW\per\cm\squared}.

The aforementioned bandwidth estimates are based on the assumption that the cluster size for tracks of normal incidence will be close to a single hit per cluster. This is expected due to the shallow epitaxial layer ($\sim \SI{8}{\micro\meter}$) compared to ALPIDE ($ > \SI{20}{\micro\meter}$) and the need for full depletion to achieve good radiation tolerance leading to reduced charge sharing.

For the thermal simulations, a total power density of \SI{\sim 70}{\milli\watt\per\cm\squared}, which is consistent with the sum of the contributions from the front-end (\SI{50}{\milli\watt\per\cm\squared}), in-matrix aggregation (\SI{5}{\milli\watt\per\cm\squared}), and on-chip data handling (\SI{20}{\milli\watt\per\cm\squared}) was assumed for the vertex detector.

For the outer tracker, the pixel pitch is increased from \SI{10}{\um} to \SI{50}{\um}, leading to a power density of \SI{2}{\milli\watt\per\cm\squared} for the front-end. For the outer tracker, there is no need for high-speed long-distance on-chip data transport. The data will be aggregated per module and sent of by a high-speed link, whose transfer speed can be optimised to match the required bandwidth. Assuming \SI{30}{\milli\watt} for such a link, this leads to an additional power density of \SI{0.3}{\milli\watt\per\cm\squared}.
A further step of data aggregation will then happen in the ‘Readout Units’ located close to the detector where the electrical to optical transition takes place.

\paragraph{Radiation tolerance.}
\begin{revised}
As discussed above, sensors implemented in the \SI{180}{\nm} process can achieve the required radiation tolerance, after optimising the process.
Similar modifications are already under study for the \SI{65}{\nm} process based on the experience gained previously.
It is foreseen to implement sub-chip configurability of the biasing settings to adapt to inhomogeneities in the response.
\end{revised}

\paragraph{Mechanics of the vertex detector.}
To be as close as possible to the interaction point, the tracker is placed into a secondary vacuum ``inside the beampipe'' and, in addition, it must be mounted such that it can be retracted during LHC injection (minimum required aperture $R_\mathrm{min} = \SI{16}{\mm}$) and placed close to the interaction point for data taking ($R_\mathrm{min} = \SI{5}{\mm}$). 
A similar concept is followed by the LHCb VELO~\cite{Bediaga:2013yyz}, but the application to a tracker covering a large acceptance including the midrapidity region, requires a design that minimises the amount of material in all directions.
This is new terrain and will require dedicated R\&D activities. 

Since apertures, impedance, and vacuum stability for the vacuum chambers at the interaction points inside the LHC experiments are of utmost importance to the stable operation of the LHC, severe engineering challenges are imposed.

\subsubsection{Technical implementation}
\label{sec:systems:tracking:implementation}

\begin{revised}
Figure~\ref{fig:trk_overview} shows an overview of the vertex detector and outer tracker assembly.
In the following, we give an overview of the technical implementation.
\end{revised}

\begin{figure}
  \centering
  \includegraphics[width=.9\textwidth]{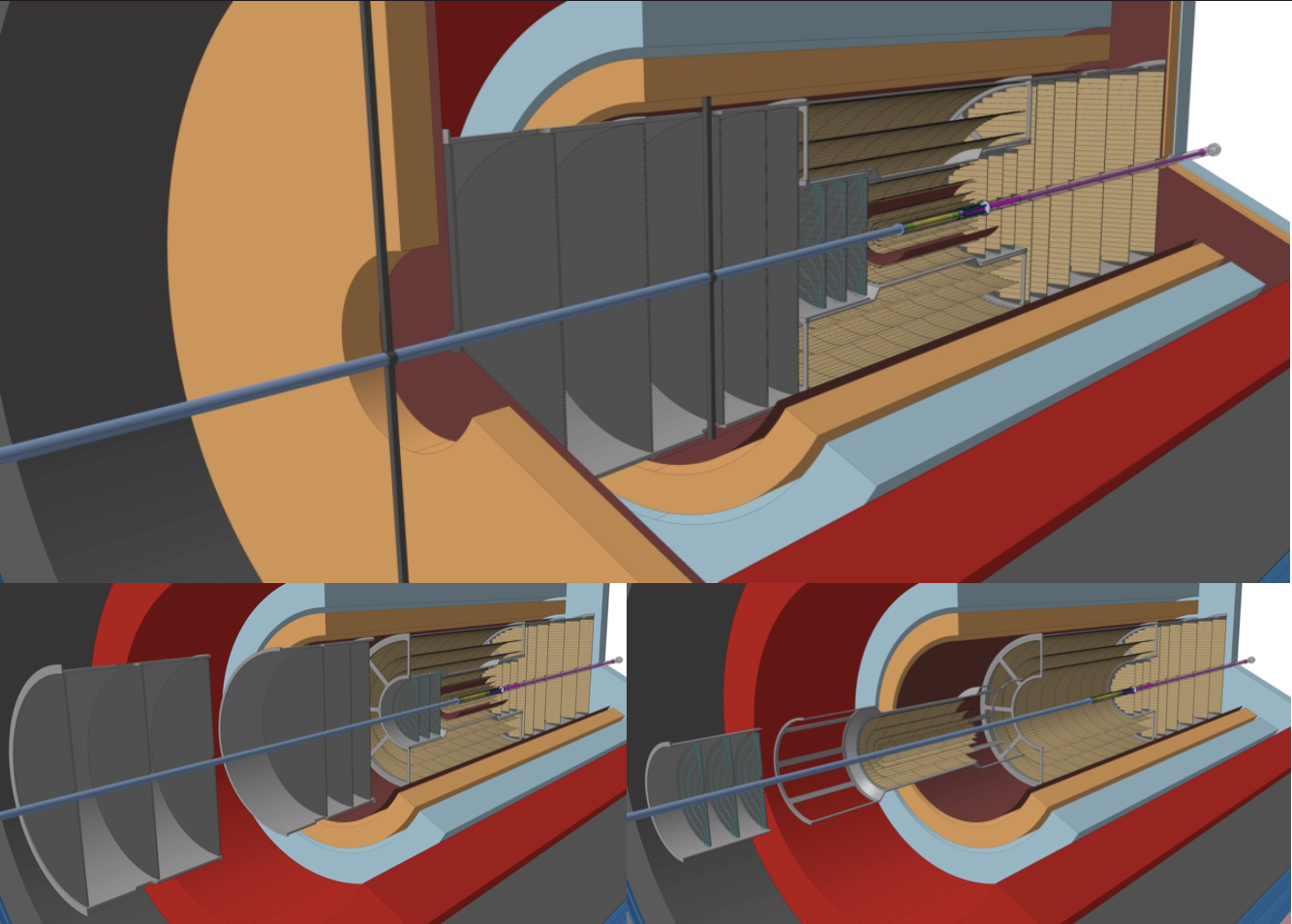}
  \caption[Overview of tracker assembly]{Overview of the vertex detector and outer tracker assembly}
  \label{fig:trk_overview}
\end{figure}

\paragraph{Vertex detector mechanics.} 

In the following, a conceptual study of a retractable vertex detector within the beam pipe is presented. 
It is built on the idea of an assembly of 4~petals, which can simultaneously rotate and, like in an iris optics diaphragm, close to leave a minimum passage of about~\SI{10}{\mm} in diameter for the beam, see Fig.~\ref{fig:IRIS}. 
The petal walls, which separate the detector from the primary LHC vacuum, dominate the material and their thickness must be minimised (see Tab.~\ref{BudgetIRIS}).

\begin{figure}
  \centering
  \includegraphics[width=10.cm]{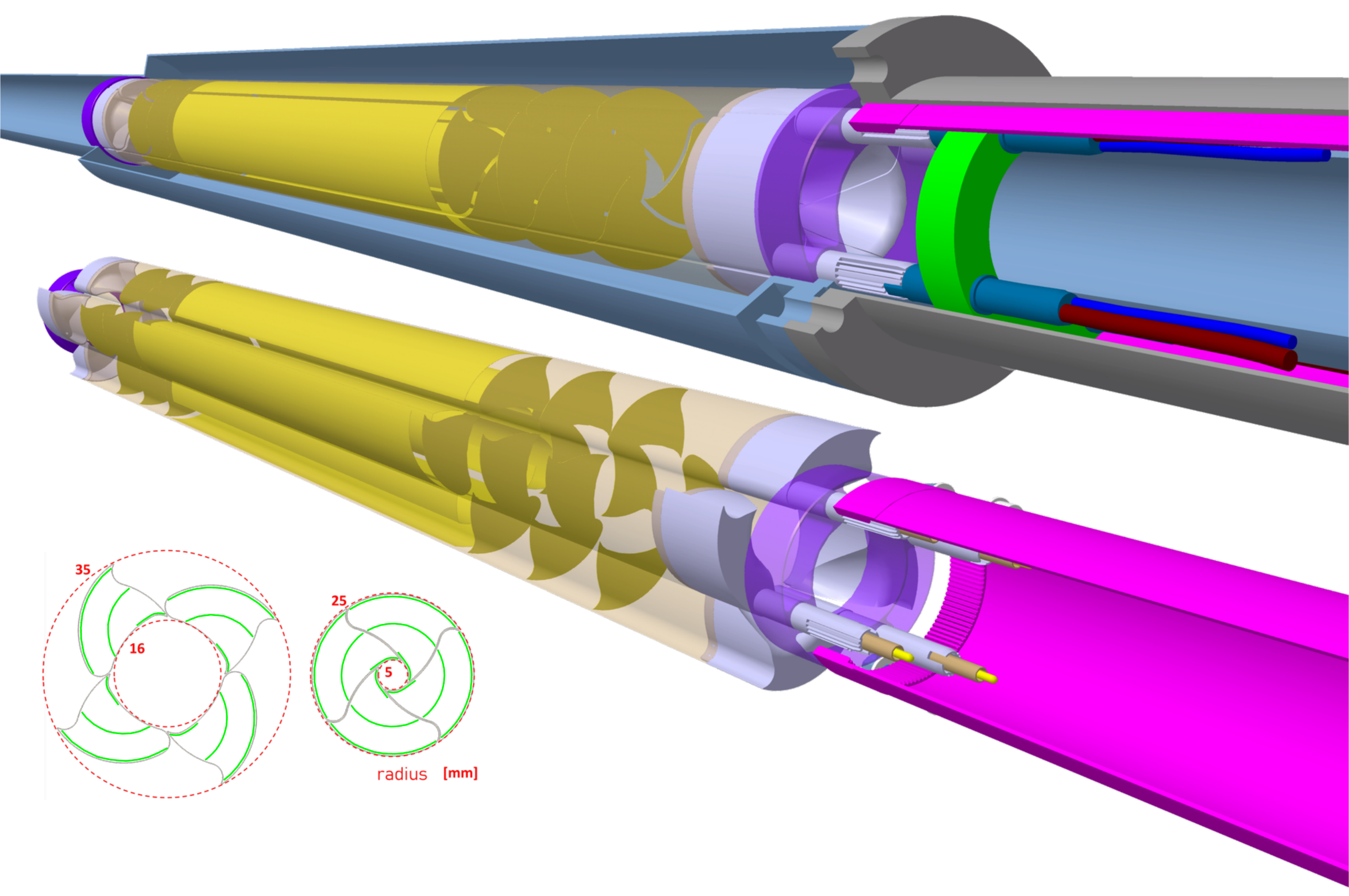}\\
  \includegraphics[width=.85\textwidth]{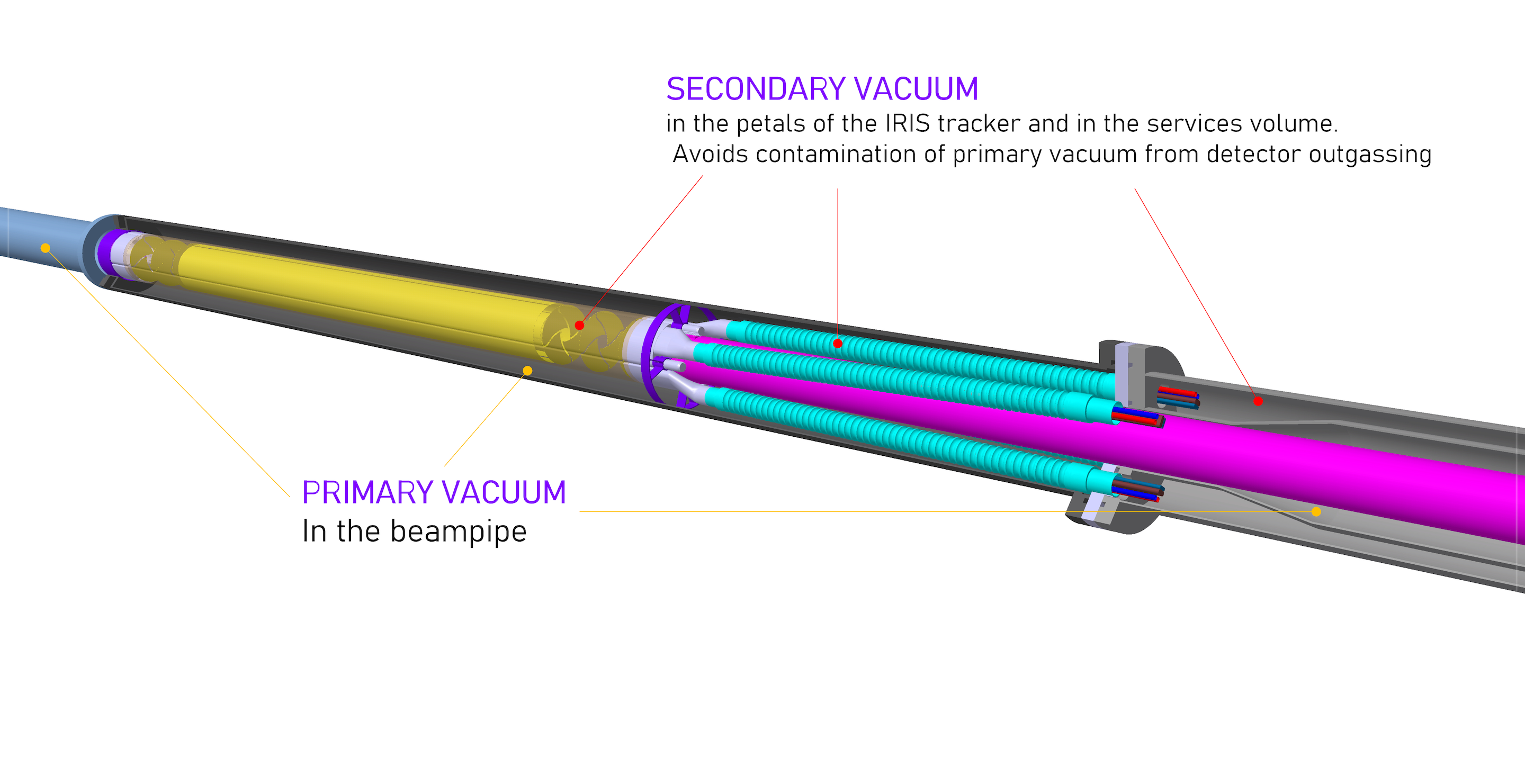}\\
  \includegraphics[width=.85\textwidth]{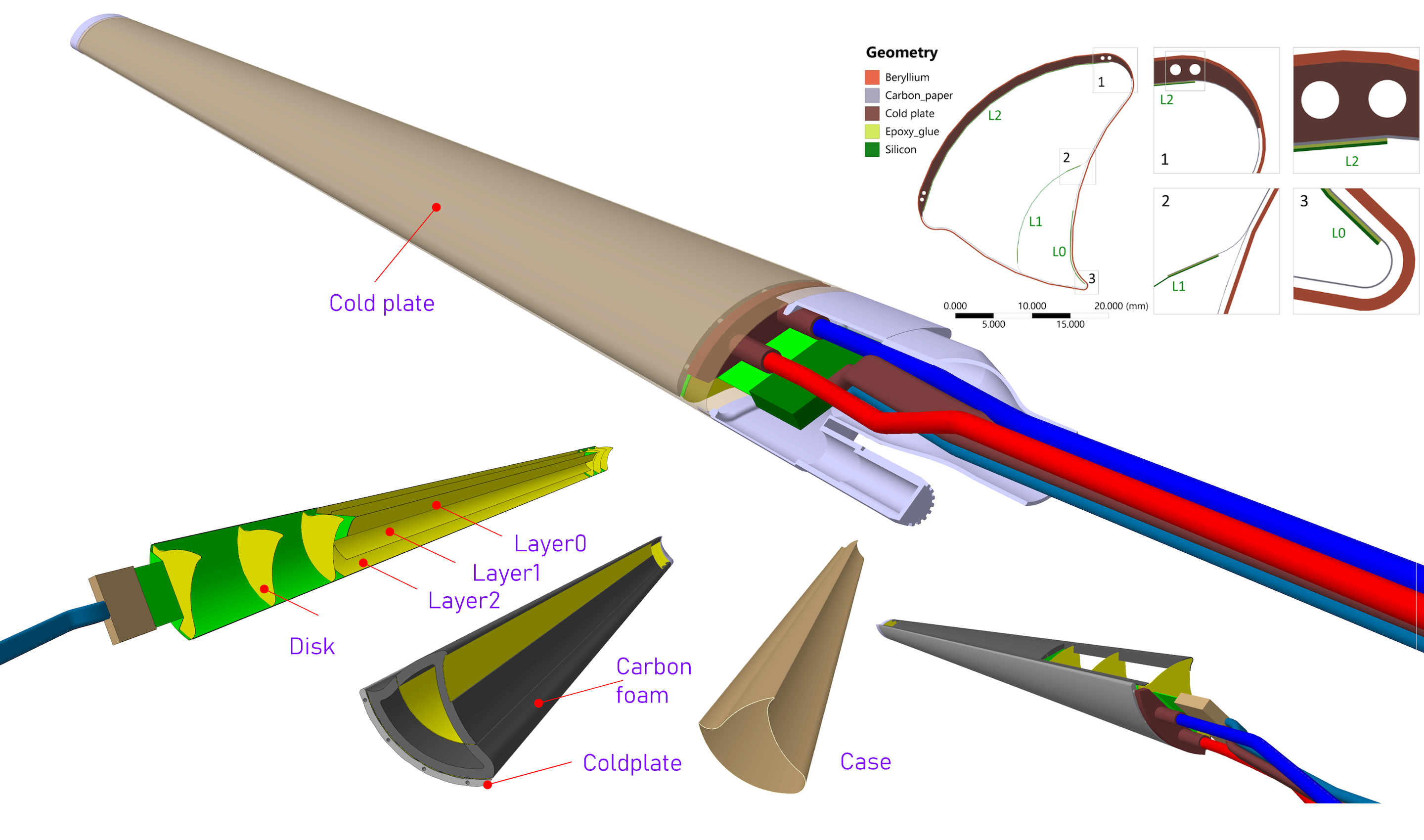}
  \caption[Study of retractable tracker]{Sketch of the vertex detector. The 3D drawings show the in-vacuum layers housed inside the beam pipe and secondary vacuum.}
  \label{fig:IRIS}
\end{figure}

\begin{table}
\centering
\begin{tabular}{%
  l
  l
  S[table-format=3]
  S[table-format=2.2]
  S[table-format=1.3]
} 
\toprule
  Component & Material & {Thickness}  & \multicolumn{2}{c}{Radiation length}\\
  \cmidrule{4-5}
            &          & {(\si{\um})} & {(\si{\cm})} & {($\si{\percent}X_0$)} \\
\midrule
    Sensor  & Si       &  30 &  9.37 & 0.032 \\ 
    Support & Be       & 250 & 35.28 & 0.071 \\
    Glue    &          &  50 & 35    & 0.014 \\
\midrule
    Total & & & & 0.117 \\
\bottomrule
\end{tabular}
\caption{Material for the first layer of the vertex detector.}
\label{BudgetIRIS}
\end{table}

The inner wall of the petals also acts as an RF foil, which is crucial to control the electromagnetic fields induced by the beam.
Since this is equally relevant for the open and closed positions, the petal geometries are designed to achieve an almost closed round bore with a diameter of about \SI{32}{\mm} and \SI{10}{\mm} when opened or closed, respectively.
A further critical mechanical challenge is to design the mechanics and vacuum equipment such as to preserve the possibility of access for maintenance.

\begin{revised}
  The iris tracker represents a very elegant concept to place a detector inside the beam pipe and its development could serve different applications. Nevertheless, if the R\&D phase shows that some of the proposed solutions are not feasible, the backup will be to go to a more standard approach, closer to LHCb concept, with the entire Vertex and Inner Tracker layer inside a two halves retractable vacuum vessel contained inside a primary vacuum vessel, with the evident drawback of large amount of sensors and cabling inside the secondary vacuum, to be qualified to an acceptable outgassing rate. This design will not compromise the placement of the first layers close to the beamline, but would have some impact on the tracker outer layers that will have to provide the needed room for the vacuum vessel.
\end{revised}

An on-detector active cooling is required to cool the sensors (\SI{70}{\mW\per\cm\squared}) and the heat generated in the RF foil by the LHC beam (\SI{90}{\mW\per\cm\squared}).
The attachment of a mini/micro channel cooled coldplate to the last layer inside the vacuum is under study.
Preliminary thermal simulations indicate that such a design allows the operation of the vertex detector at \SI{-25}{\celsius} with a temperature spread across the silicon surface below \SI{5}{\celsius}.

Apertures, impedance, and vacuum stability for the vacuum chambers at the interaction points inside the LHC experiments are key parameters for the stable operation of the LHC and the physics performance.
In a standard tracker configuration, the minimal beampipe radius sets the limit for the distance of the first layer. 
In the present LHC experiments, the beampipe wall goes down to a minimum radius of \SI{18}{mm} with a thickness of \SI{0.8}{mm}. For LS3, ALICE tests the manufacturing limits to achieve an inner radius of \SI{16}{mm} with a wall thickness of \SI{0.5}{mm}. 
A \SI{16}{mm} beampipe maintains an aperture ($> 7\sigma$) at injection, even if typically, this value is considered for the regions outside of the experiment volume. 
In the detailed analysis, possible failure scenarios at injection as well as better estimates on manufacturing and installation tolerances will be considered.

To get closer to the interaction point, the next option is to place the innermost tracker layers inside the beampipe vacuum. 
This option is being investigated in ALICE, starting from an LHCb VELO-like design and going to the completely new concept of an iris tracker, applicable to a hermetic experiment. 
In LHCb, the aperture has been reduced from 5 to \SI{3.5}{mm} in the last LS2 upgrade, while  staying well within the aperture limits defined by the LHC [LHCb-PUB-2012-018.pdf (cern.ch)].
There are several critical challenges for the iris tracker that require dedicated R\&D programmes.
The first challenge is that the iris tracker must retract from its normal (physics) position during injection and at the end of every fill to be compatible with the larger aperture required at injection energy.
In addition, to minimize the contamination of the primary (LHC) vacuum by the outgasing  from some of the detector components, the iris tracker must be contained in a secondary vacuum, which is separated from the primary vacuum by a thin wall case. 
This casing is the single largest contributor to the material as its wall is before the ﬁrst point of measurement.
Therefore, its thickness and material must be designed to achieve the minimum possible thickness. 
For reference, the LHCb VELO the RF foil, fulfilling the the same function, has a wall thickness of 0.5/\SI{0.25}{mm}, locally etched down to \SI{0.15}{mm} along the beamline (Aluminum 5083). Such a thin wall cannot withstand a pressure difference of \SI{1}{bar}. 
Consequently, the design must include a protection scheme against a possible increase of the pressure difference across the thin wall, in case of a failure.

The iris tracker is conceived as a 4-petals assembly inside the primary vacuum, in which the petals can rotate simultaneously and, like in an optical iris diaphragm, close to leave a minimum passage of about \SI{10}{mm} in diameter for the beam (see Fig.~\ref{fig:trk:iris_implementation}). 
Each petal houses a quarter of detector with the first detection layer at a minimum distance of about 5mm from the beam axis in the closed position.

\begin{figure}
    \centering
    \includegraphics[width=.75\textwidth]{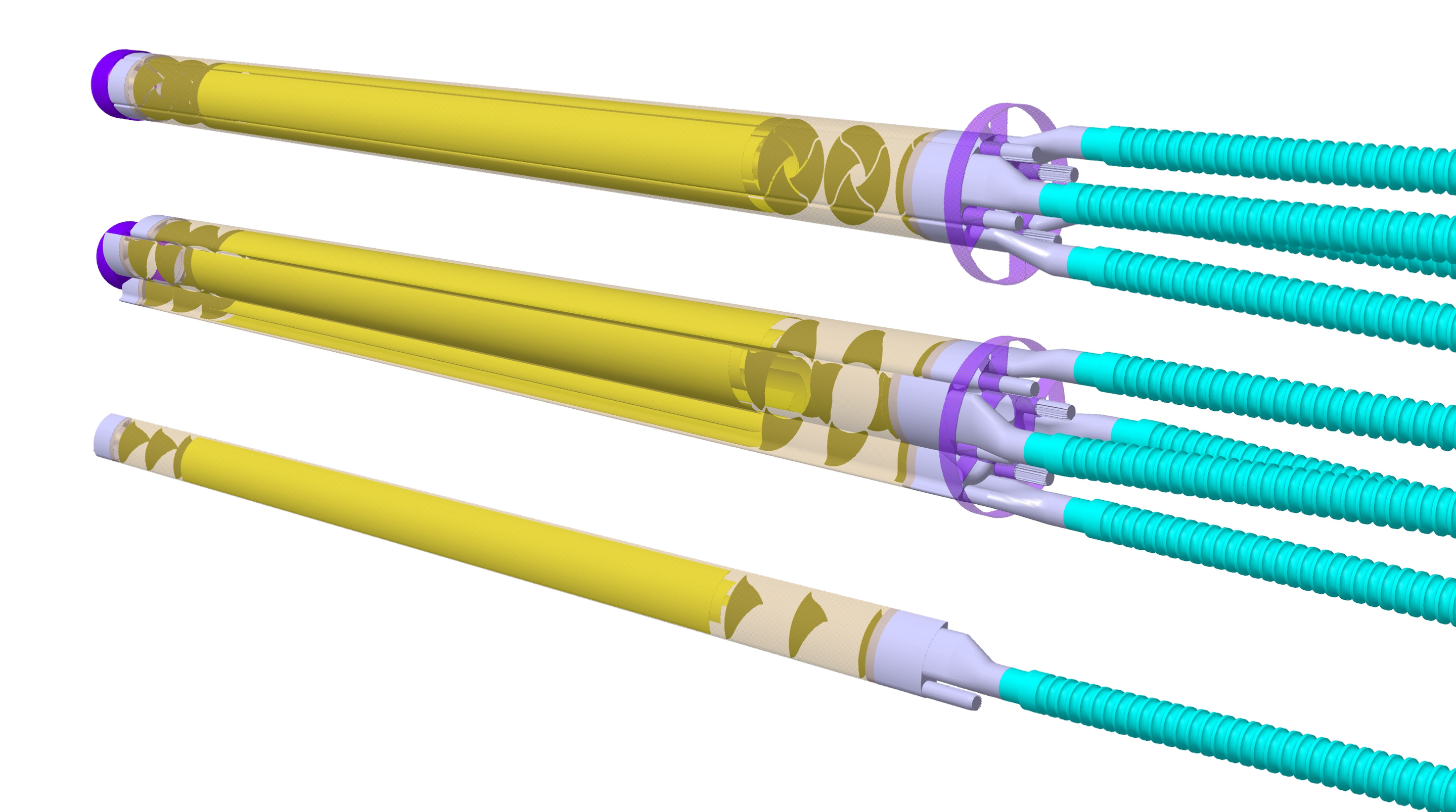}
\caption{Vertex detector implementation}
    \label{fig:trk:iris_implementation}
\end{figure}

The thin wall case of the iris petal not only provides a secondary vacuum, but also acts as an RF foil. 
When a beam of particles traverses a vacuum chamber which is not smooth or is not a perfect conductor, it will produce electromagnetic RF fields that will perturb the beam particles, described as wakefields (in the time domain) or impedance (in the frequency domain). 
These electromagnetic fields act back on the beam and influence its motion. 
Such an interaction of the beam with its surroundings results in beam energy losses, alters the shape of the bunches, and shifts the betatron and synchrotron frequencies. 
At high beam current the fields can even lead to instabilities, thus, limiting the performance of the accelerator in terms of beam quality and current intensity.  
The wall surrounding the beam should offer a continuous homogeneous surface such that there is no unbalance in the current flow induced by them.
The thin wall of the iris petal casing indeed minimizes the electrical coupling between iris components and the LHC beams and provides a surface free of abrupt changes in geometry which would generate heat and perturb both the IRIS electronics and the beam parameters. 
This holds for both the open and closed configurations, which are equally relevant.
The iris petals are designed to achieve an almost closed round bore with a diameter of about \SI{32}{mm} when open.

The reduction of the central opening diameter at the interaction point, set by the closed iris tracker petals also affects the LHC beam vacuum system in terms of achievable pressure and vacuum stability due to dynamic effects induced by beam particles hitting the surface. 
To mitigate this problem, the strategy adopted at CERN foresees the use of NEG coating deposited inside the beam pipes, which absorbs residual molecules after activation through heating. 
The NEG coating consists of a thin layer of titanium-zirconium-vanadium alloy and acts as a distributed pumping system, which is effective for the removal of all gases except methane and  noble gases, which are removed by the 780 ion pumps in LHC. 
The NEG coating of the iris petals and the activation at high temperature (\SI{>180}{\celsius}), with the silicon detectors already installed inside the petals, pose another challenge.
The alternative of using ion pumps close to the interaction point is also being evaluated.

The next critical mechanical challenge is to design the iris system and its vacuum equipment so as to preserve relatively easy access and giving the possibility to replace it. 
The radiation environment is such that the replacement of a detector could be required after some years of operation. 
The insertion and extraction of the IRIS tracker through the beampipe, by an airlock section limited by two gate-valves, is an ambitious concept that further pushes the challenge of the interface between the secondary vacuum (IRIS tracker) and the primary vacuum (beampipe) as well as the connectivity of the services including cooling. 

The iris tracker represents a very elegant concept for placing a detector inside the beampipe and its development could serve different applications.
Nevertheless, if the R\&D phase will show that some of the proposed solutions are not feasible, the backup will be to go to a more standard approach, closer to LHCb concept, with the entire vertex and inner tracker layers inside a two halves retractable vacuum vessel in turn contained in primary vacuum vessel, with the evident drawback of a large amount of sensors and cabling inside the secondary vacuum to be qualified to an acceptable outgassing rate. 
Such a design would not compromise the placement of the first layers close to the beamline, but would have some impact on the tracker outer layers that will have to provide the space required for the vacuum vessel and its connectivity.

\subparagraph{Thermal study}

For the preliminary assessment of the cooling, two-phase evaporative CO$_2$ cooling has been assumed. 
Alternatives like Novec will be considered. 
Preliminary CFD studies assuming a heat dissipation of \SI{70}{mW/cm\squared} from the sensors and \SI{90}{mW/cm\squared} from the beam impedance show that the sensors can be cooled to \SI{-25}{\celsius} with an inlet of CO$_2$ at \SI{-35}{\celsius}. 
The same analysis also shows a $\Delta T$ of 1-2 degrees within a single sensor and 5 degrees between the coldest (outermost) and hottest sensor (innermost). 
Further analyses are ongoing and the most recent results are reported in Fig.~\ref{fig:trk:thermal_sim}.
This two-dimensional thermal analysis has been performed based on the following conservative boundary conditions:
\begin{itemize}
\item The silicon sensors have a thermal conductivity of \SI{140}{W/mK} and a thickness of \SI{40}{\um}. 
\item A coldplate on the outer side of layer 2 provides a sink for the heat dissipated by the sensors and induced by the beam impedance. The coldplate is in direct thermal contact with layer 2 and connected through thermal bridges to layers 1 and 0. A high thermal conductive pyrolytic graphite foil (carbon paper) realizes the thermal contact and has a thermal conductivity of \SI{1500}{W/mK} and a thickness of \SI{20}{um}. The coldplate is also in thermal contact with the beryllium petals case through which the heat coming form the beam  is conducted.
\item Epoxy glue with a thermal conductivity of \SI{0.15}{W/mK} and a thickness of \SI{40}{\um} between the carbon papers and sensors.
\item The carbon foam is not considered in the analysis.
\item The carbon cold plate has an average thermal conductivity of \SI{100}{W/mK} and a thickness of \SI{1}{mm}.
\item The engraved channel in the carbon cold plate has a diameter of \SI{0.5}{mm}. 
\item The beryllium cover has a thermal conductivity of \SI{190}{W/mK} and a thickness of \SI{0.15}{mm}.
\item The distance between the Be wall and the carbon paper is \SI{0.15}{mm}.
\item The sensor dissipation is \SI{70}{mW/cm\squared}.
\item The beam-induced heat dissipation  has been assumed equal to \SI{100}{mW/cm\squared}. This has  been estimated by rescaling values reported in the analysis performed for nominal HL-LHC pp beam parameters (Benoit Salvant, Rainer Wanzenberg and Olga Zagorodnova,  \url{https://indico.cern.ch/event/341817/contributions/1736772/attachments/671606/923025/Impedance_for_new_ALICE_beam_pipe.pdf}).
\item The convective heat transfer coefficient between CO$_2$ and the coldplate wall has been assumed to be equal to \SI{1}{kW/m\squared K} and the CO$_2$ boiling temperature is \SI{-35}{\celsius}. It is based on the nominal HL-LHC pp beam parameters and the computation of the dissipation due to the resistive wall of the ALICE beam pipe.
\item The radiation emissivity is 0.8 for both the beryllium wall and carbon paper wall.
\end{itemize}

\begin{figure}
    \centering
    \includegraphics[width=.5\textwidth]{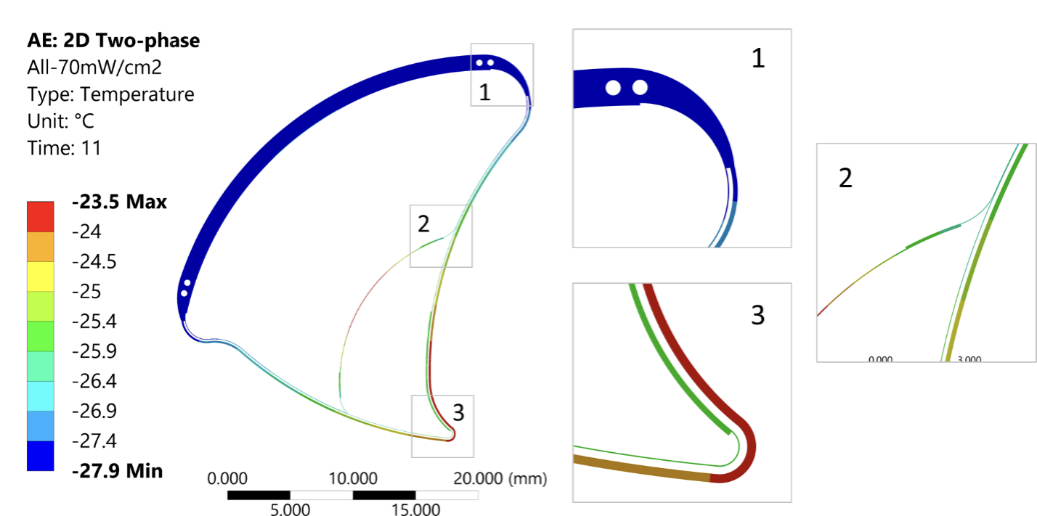}
    \caption{CFD study for vertex detector}
    \label{fig:trk:thermal_sim}
\end{figure}

\paragraph{Outer Tracker mechanics.} 
\label{par:systems:mechanics}

A sketch of the design of the tracker is shown in Fig.~\ref{fig:OuterTr}. 
It relies on modules and support elements as main building blocks.
In a module, several sensors are interconnected and bonded to a high thermal conductive substrate.
Each individual module is then interfaced to the functional support that provides mechanical support, alignment, cooling and electrical connectivity. 
For operation at ambient temperature the coolant choice is water, and the system is operated in leakless mode, below atmospheric pressure.
An R\&D programme has been launched (within EP R\&D) to characterise an ultralight coldplate design also for high pressure applications in case of two-phase cooling for operative temperatures below~\SI{0}{\degreeCelsius}.
The functional support is provided by longitudinal staves and flat disks in the central barrel section and the forward region, respectively.

\begin{figure}
\centering
\includegraphics[width=.9\textwidth]{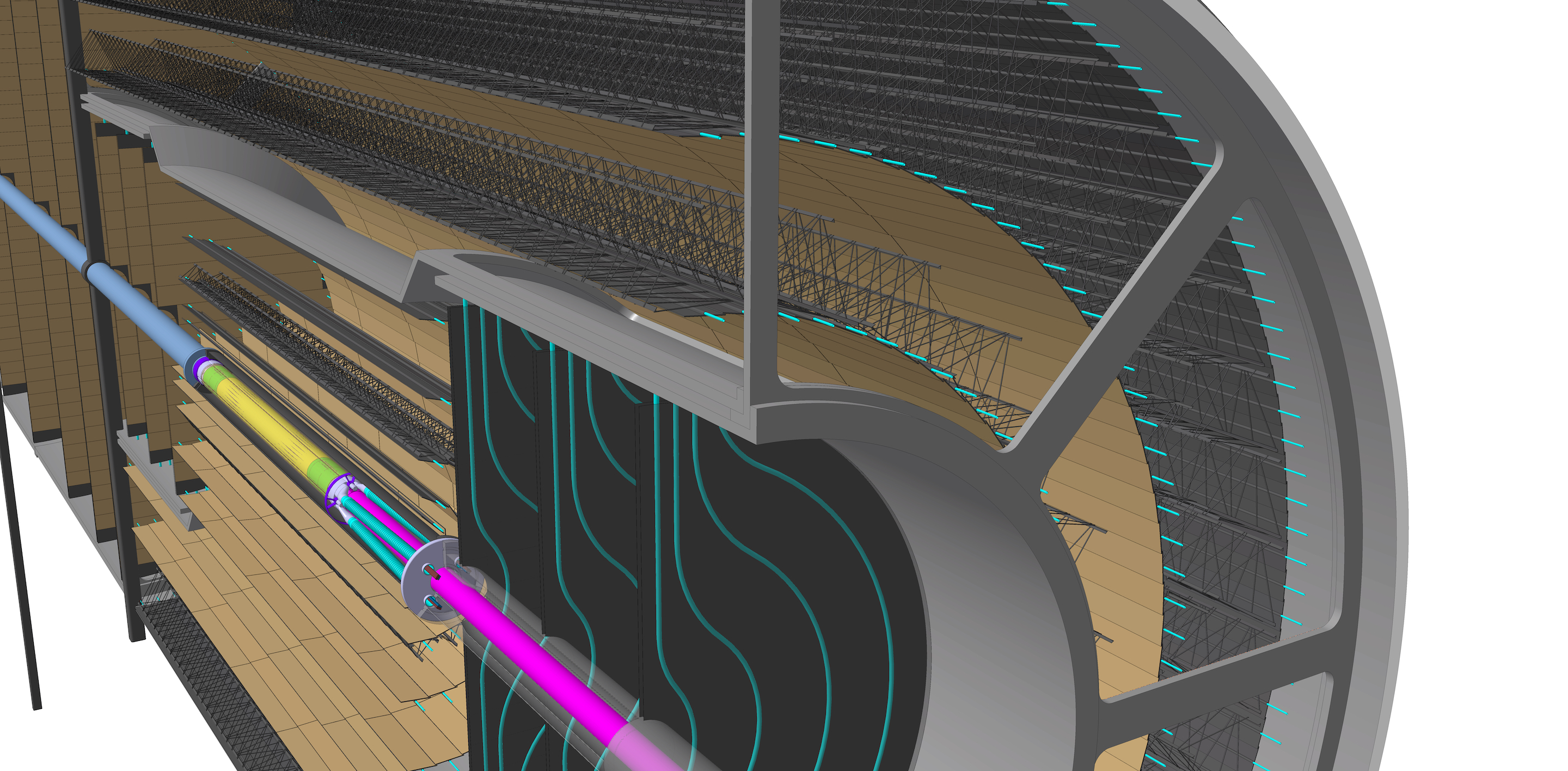}\\
\includegraphics[width=.9\textwidth]{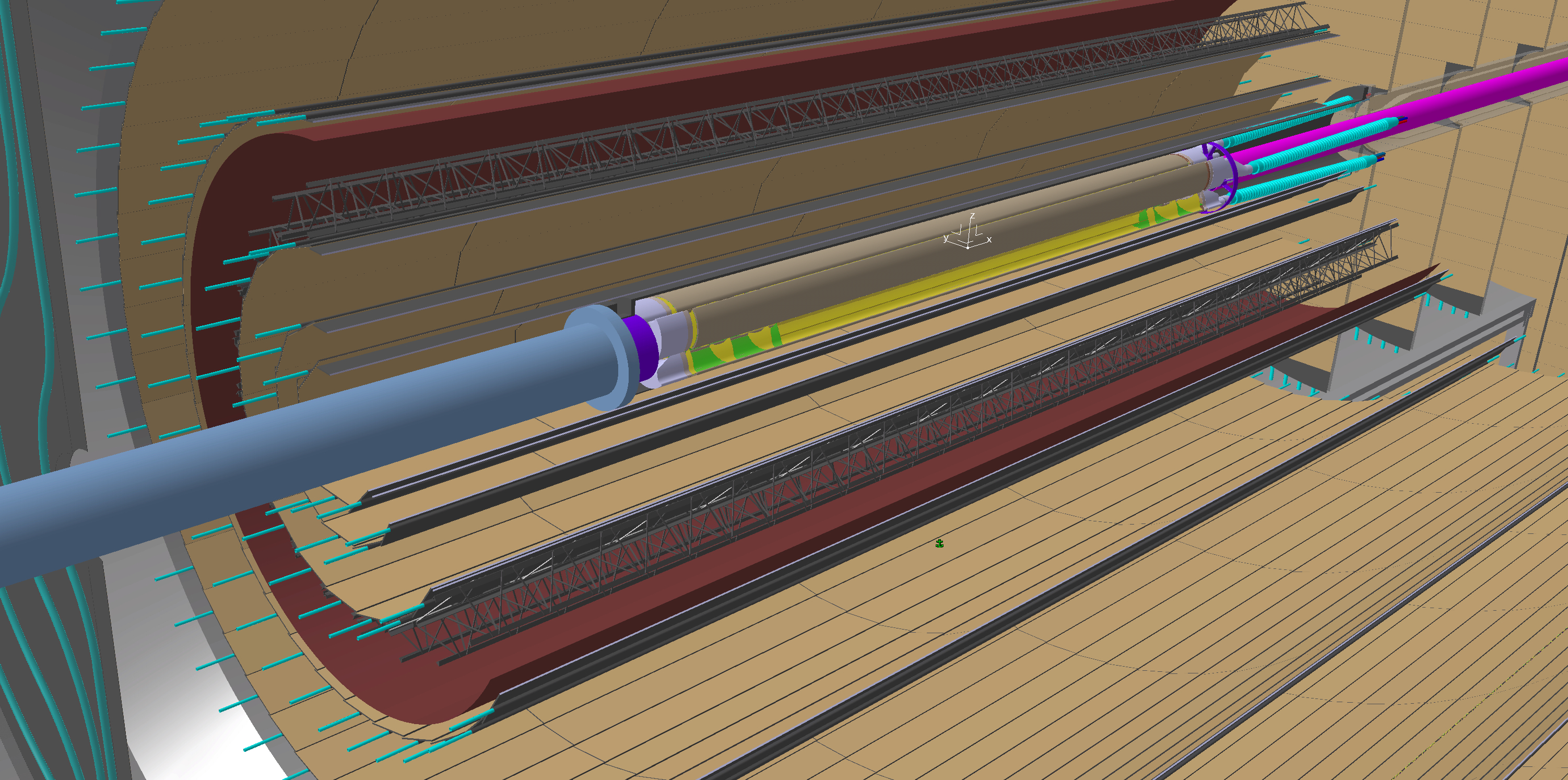}\\[.5cm]
\includegraphics[width=.9\textwidth]{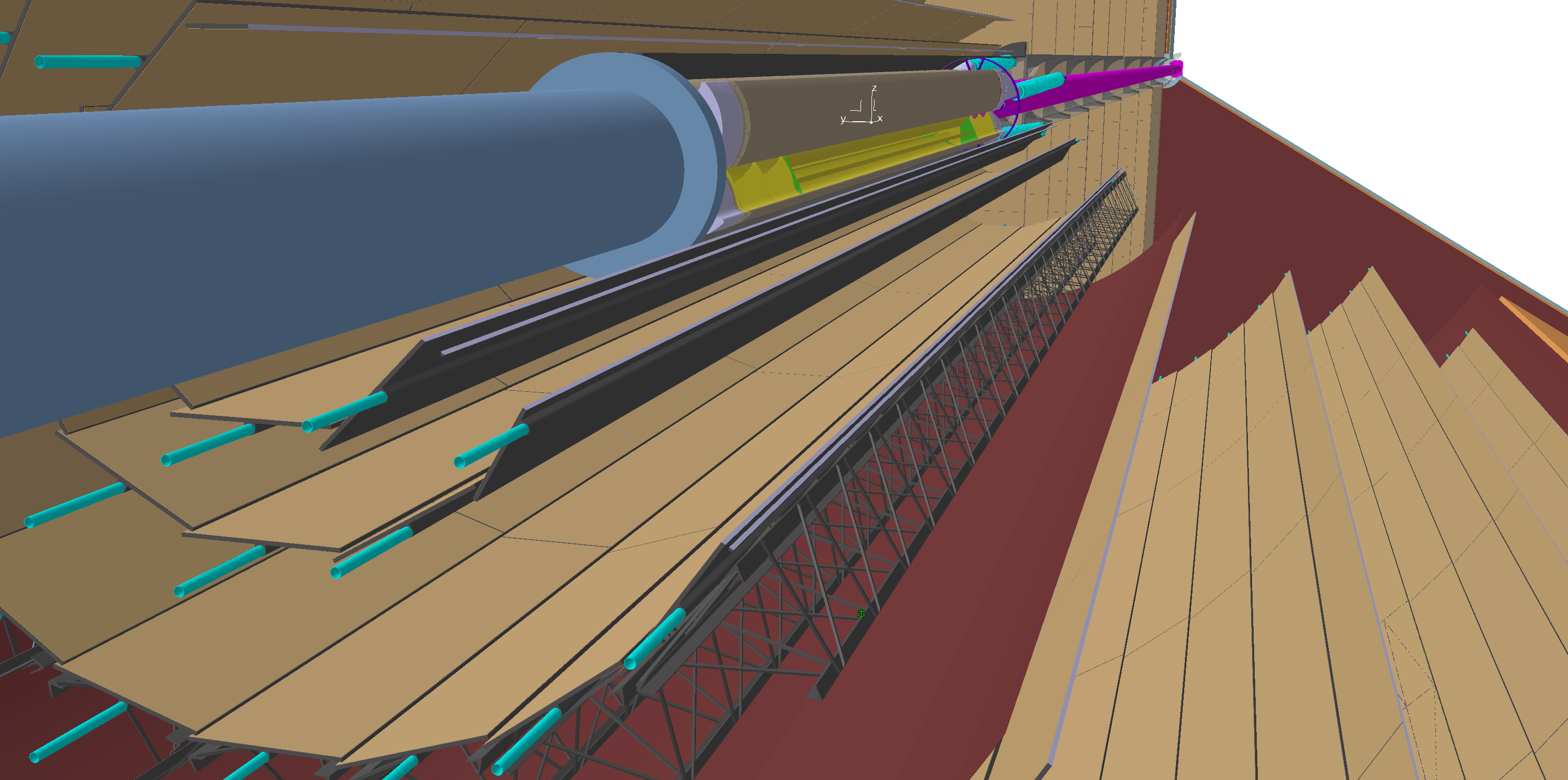}
\caption[Outer Tracker mechanics]{Sketch of the outer tracker mechanics. Modules assembled in staves structures are visible as well as services and power lines. Furthermore, the overlap of the staves can be seen.} 
\label{fig:OuterTr}
\end{figure}

Staves and discs are supported in position by large mechanical support structures with the shape of barrels that extend over the whole length of the RICH and outside, covering an overall length of \SI{8}{\metre}. 
The tracker is segmented into several barrels to allow for a sequential extraction from both sides of the experiment.%
The total material thickness per layer is expected to be $\SI{1}{\percent}X_0$, the sources contributing to it are presented in Tab.~\ref{BudgetTRK}. 
The material required for service connections has an impact on the production of secondary particules in nuclear interactions, which results in background and inefficiencies. 
This will have to be studied with full simulations and the routing of the services will have to be optimised.

\begin{table}
\centering
\begin{tabular}{%
  l
  l
  l
  S[table-format=3]
  S[table-format=2.2]
  S[table-format=1.2]
} 
\toprule
  Element & Component & Material & {Thickness}  & \multicolumn{2}{c}{Radiation length}\\
  \cmidrule{5-6}
          &           &          & {(\si{\um})} & {(\si{\cm})} & {($\si{\percent}X_0$)} \\
\midrule
  \multirow{5}*{Module}
    & Sensor        & Si           &  50 &  9.37 & 0.05 \\ 
    & Plate         & Carbon fibre & 200 & 26.08 & 0.08 \\
    & FPC layer     & Cu           &  50 &  1.34 & 0.37 \\
    & FPC isolation & Polyimide    & 200 & 28.41 & 0.07 \\
    & Glue          & Glue         & 100 & 40    & 0.03 \\
\midrule
  Space Frame
    &               & Carbon       &     &       & 0.16 \\
\midrule
  \multirow{5}*{Cold Plate}
    & \multirow{4}*{Plate}
      & Carbon fibre & {\multirow{4}*{200}} & {\multirow{4}*{mixed}} & {\multirow{4}*{0.06}} \\
    & & Carbon fleece \\
    & & Thermal Pyrolytic Graphite \\
    & & Polymide pipes \\
   & Cooling & Water & & 37.76 & 0.11\\
\midrule
Misc  & Connectors & & & & 0.10\\
\midrule
\textbf{Total}  &  &&&  & \bfseries 1.02 \\
 \bottomrule
\end{tabular}
\caption{Material budget breakdown: out-of-vacuum layers.}
\label{BudgetTRK}
\end{table}
The discs of a given endcap are mounted on the inner side of a Cylindrical Support Shell. The services and cooling manifolds for the endcap are supported by the shells, which feature machined openings to make the necessary electrical and fluidic connections. The endcap fits into a Service Cylindrical Shell that carries services in the cylindrical layers and allows for an independent extraction of the different barrel sections.

The design of the tracker relies on two main building blocks, namely modules and their support.
In a module, several sensors are interconnected and bonded to a high thermal conductive substrate, featuring machined locators for positioning purposes. 
Each individual module is then interfaced to the functional support that provides mechanical support, alignment, cooling and electrical connectivity. 
Two different types of functional supports are employed: longitudinal staves for the modules in the central barrel section and flat disks for the forward region’s modules.
The stave has a main carbon fibre spaceframe-like structure providing the positioning and the necessary stiffness. 
The spaceframe holds the coldplate in position, a sheet of high-thermal conductivity carbon layers with embedded polyimide cooling pipes that run along the longitudinal direction. 
The heat is conducted into the cooling pipes by the carbon layers and is removed by the coolant flowing in the pipes. 
A similar coldplate concept is used for the disks where the high thermal conductive carbon plate and polyimide pipe is integrated in a carbon sandwich panel that provides support and positioning.
Both cold plate designs, in smaller dimensional scale, have been developed and used for the ALICE ITS2 and MFT detectors installed in LS2. 
For operation at ambient temperature the coolant choice is water and the system is operated in leakless mode, below atmospheric pressure.
An R\&D programme has been launched in EP to characterise this ultralight coldplate design also for high pressure applications in case of two-phase cooling for operative temperatures below \SI{0}{\celsius}.
The module and the functional support are coupled at the level of the coldplate, relying on a precise locator system to control their relative positioning and on a dismountable interface for the mechanical connection.  
A thermal interface material is placed between the module substrate and the coldplate to reduce the thermal impedance. 
This solution gives the tracker concept a distinctive advantage in terms of re-workability, as a single module cell can be easily replaced or temporarily removed for repair during the construction phase.

Stave and disk are supported in position by large mechanical support structures in the shape of barrels that extend over the whole length of the RICH and outside, covering an overall length of \SI{8}{\metre}. 
The tracker is segmented in several barrels so as to allow for sequential extraction from both experiment sides: an inner and outer Barrel and an inner and outer endcap.

To minimise the material between layers, an external cylindrical support shell will provide the main support for the inner and outer barrel while the single layers will be interconnected at the barrel extremities, without any cylindrical intermediate layer structural shell.
In each layer, the stave will be fixed at both ends to the corresponding end-wheels, which are light composite end-rings, that ensure the precise positioning of the staves in a layer. 
They provide the reference plane for the fixation of the two extremities of each Stave.

The disks of a given endcap barrel are mounted on the inner side of a cylindrical support shell. 
The services and cooling manifolds for the endcap are supported by the shells, which feature machined openings to make the necessary electrical and fluidic connections. 
The endcap barrel fits into a service cylindrical shell that carries services for the staves and allows an independent extraction of the different barrel sections.

The management of the on-detector services within the entire tracker volume is key for the minimisation of the material budget in the sensitive area.
To reduce the number of cables and material within the detector volume, the sensors will adopt a serial powering scheme. 
In this solution, the modules are grouped in serial powering chains supplied from a constant current source. 
Within the module, the individual front-end chips are powered in parallel. 
The service layout will be optimised through their routes from the barrels to the main patch panels located immediately outside of the magnets at both ends of the detector.

\paragraph{Powering scheme for the outer tracker.}

To reduce the number of cables and the material within the detector volume, two suitable measures are the usage of DC-DC converters and serial powering. 
With both solutions, detector partitions of a segment are grouped in serial powering chains (see Fig.~\ref{fig:tracker:powering}). These segments are either powered via DC-DC converters and power switches located at the layer edge providing a constant current to the segments or by another cascaded serial powering chain. Within each partition, individual front-end chips are powered in parallel. 
To avoid thick conductors and large capacitors, active local power stabilisation options will be investigated. 
The power regulation can be implemented in several ways, e.g.~integrated in the pixel chip or as a separate power regulator mounted in strategic locations. For the DC-DC conversion, \SI{20}{\percent} regulation loss are included in the calculations.
\begin{revised}
In this scheme, the total power consumption of \SI{\sim 20}{\kilo\watt} can be provided over about 400 channels.
\end{revised}

FEASTMP is an example of an existing DC-DC converter for use in high-energy physics experiments~\cite{Ripamonti:2020,DCDC}. 
It would be sufficiently radiation tolerant for use in the outer tracker, but it is designed for lower voltages not sufficient for supplying a serialized power domain.
In the absence of a DC-DC converter, a longer chain of partitions or a cascaded serial powering scheme could be employed.

Another important way to reduce the mass of the services is the reduction of its granularity, which can be generated locally on the partition using dedicated switching and monitoring circuitry. 
An optimisation process of the powering concept and the power regulation circuitry has started.
The inclusion of the regulators into the pixel chips themselves would reduce the amount of material for the final detector further. In the ALICE collaboration, the first study of such a solution was performed in the context of the ITS2 using the Tower Semiconductor \SI{180}{\nm}~technology~\cite{Gajanana:2016}. This circuit was recently ported to the \SI{65}{\nm}~technology and is currently under test. In addition, recent developments in the context of the RD53 collaboration have demonstrated the feasibility of a large scale serial powering scheme using regulators in ASICs~\cite{RD53:Monteil,RD53:Padras}. This development will be used for the CMS~\cite{RD53:CMS:LaRossa, RD53:CMS:Orfanelli} and ATLAS~\cite{RD53:ATLAS:Meng,RD53:ATLAS:Matheson} upgrades. This circuit could be investigated for the use in MAPS for ALICE. 

\begin{figure}
  \centering
  \includegraphics[width=0.95\textwidth]{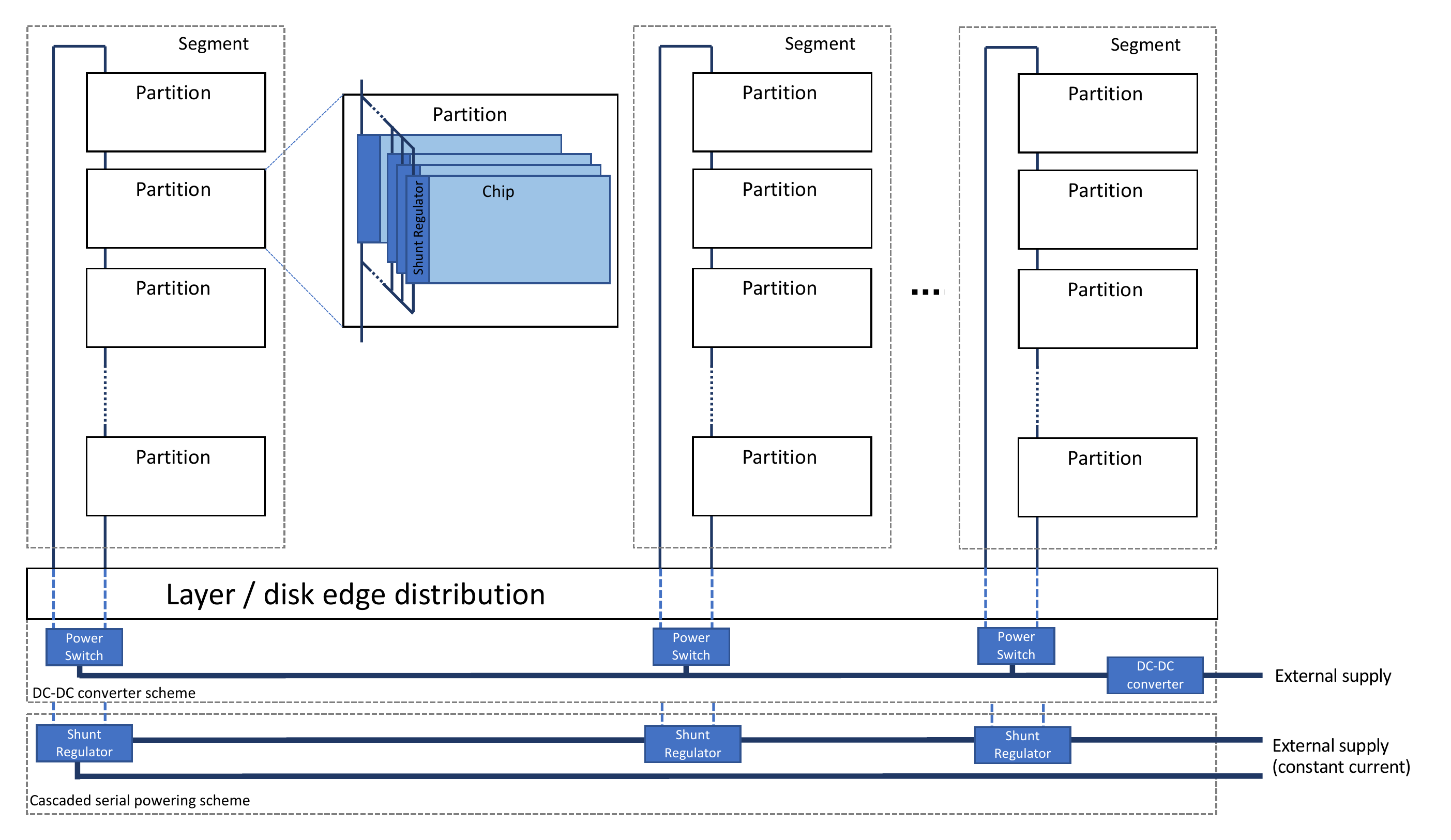}
  \caption[Powering scheme]{Powering scheme for the tracker based on serial powering of partitions and DC-DC converters or an additional chain of serial powering at the layer or disk extremity.}
  \label{fig:tracker:powering}
\end{figure}

\paragraph{Modules.}
For the outer tracker modules, the crucial aspect is the volume production. 
To factorise the production, the module is conceived as the main basic building block of the Tracker. 
It is a stand-alone unit, whose integration is thought to be fully commercialised and done with commercial partners. 
The precise dimensions as well as the number and size of the sensors they house are subject to an optimisation process, that will be carried out in close collaboration with the industry (``design for manufacturability'') from the start. 

\ifcost
\subsubsection{Cost estimates}
\label{sec:systems:tracking:cost}

The project cost estimate is given in Tab~\ref{CostTRK} for the \SI{60}{\metre\squared} silicon tracker option based on MAPS technology. The activities described in the table represent the material costs, excluding institutes’ personnel costs and basic infrastructures. 
The voice \textit{sensor} includes final production and post-production while \textit{Module assembly} account for the industrialization, fabrication and testing. 
The voice \textit{Cooling} includes the cost of the foreseen micro-channel layer in the IRIS tracker and water cooling in the tracker.

\begin{table}
\centering
 \begin{tabular}{lS[table-format=2.1]S[table-format=2.1]S[table-format=2.1]}
 \toprule
 {Component} & {Vertex} & {Tracker} & {Total} \\
 \midrule
 Sensor          &  0.1 &  3.2 &  3.3 \\ 
 Module assembly &      &  4   &  4   \\
 Mechanics       &  6   &  6.5 & 12.5 \\
 Cooling         &  1   &  2   &  3   \\
 Read-out        &      &  2   &  2   \\
 Power           &      &  2.6 &  2.6 \\
 Services        &      &  1.5 &  1.5 \\
\midrule
 \textbf{Total}  & 7.1 &  21.8 & \bfseries 28.9 \\
 \bottomrule
\end{tabular}
\caption{Cost estimate in MCHF. The item \textit{Vertex} includes the costs projection for the in-vacuum tracker layers/discs while the item \textit{Tracker} refers to the rest of the ALICE~3 silicon tracker.}
\label{CostTRK}
\end{table}

\fi

\subsection{Time of flight}
\label{sec:systems:tof}

The time-of-flight system provides particle identification over the full acceptance ($\abs{\eta} < 4$). It comprises an inner and outer TOF layers in the central barrel and forward disks on both sides of the experiment.

\subsubsection{Specifications}
\label{sec:systems:tof:specs}

The specifications and dimensions of the TOF system are given in Tab.~\ref{tab:tof_specs}. 
A time resolution of \SI{20}{\pico\second} r.m.s. together with low material budget of 1-3\%$X_0$ and a power density of 50\,mW/cm$^2$ are the key requirements. 
In recent years, impressive improvements in the timing performance of silicon sensors have been achieved by several groups~\cite{RSD,Iacobucci2019,Pancheri2020,Snoeys2017,CARTIGLIA2019350, Anderlini_2020}, which makes them an ideal candidate for our purpose. 
Large area systems as needed for ALICE~3 have not yet been built and a dedicated R\&D programme is required. 
Compared to the general-purpose detectors at the LHC, the requirements in terms of hit rate and radiation hardness are moderate, which opens several technology opportunities that can be exploited. 

\begin{table}[!ht]
    \centering
    \renewcommand{\arraystretch}{1.3}
    \begin{tabular}{l *{3}{S[table-format=2.2]}}
        \toprule
        & \textbf{Inner TOF} & \textbf{Outer TOF} & \textbf{Forward TOF}\\
         \midrule
         Radius (m) & 0.19 & 0.85 & \numrange{0.15}{1.5} \\
         $z$ range (m) 
         & \numrange{-0.62}{0.62} & \numrange{-2.79}{2.79} & 4.05 \\
         Surface (m$^2$) &   1.5 &  30 & 14 \\
         Granularity (mm$^2$) & {$1 \times 1$} & {$5 \times 5$} & {$1 \times 1$ to $5 \times 5$}\\
         Hit rate (kHz/cm$^2$) & 74 & 4 & 122 \\
         NIEL (\nequiv) / month 
         & \num{1.3e11} & \num{6.2e9} & \num{2.1e11} \\
         TID (rad) / month & \num{4e3} & \num{2e2} & \num{6.6e3} \\    
         Material budget ($\% X_0$) &  \numrange{1}{3}  & \numrange{1}{3} & \numrange{1}{3} \\
         Power density (mW/cm$^{2}$) & 50 & 50 & 50 \\
         Time resolution (ps) & 20  & 20 & 20 \\
         \bottomrule
    \end{tabular}
    \caption{TOF specifications.}
    \label{tab:tof_specs}
\end{table}

\subsubsection{System considerations and technology options}
\label{sec:systems:tof:options}

The TOF layers can be equipped with the same sensor and front-end electronics, possibly with dedicated optimization in terms of integration and readout strategy.
As an example for the outer layer, where the rate per unit area is smaller, more sensor modules can be grouped to reduce the number of data transmission lines. 
The required cell size is $\SI{1}{\mm} \times \SI{1}{\mm}$ for the inner TOF and $\SI{5}{\mm} \times \SI{5}{\mm}$ for the outer TOF in order to provide efficient matching of the tracks.
The choice of the physical cell size will be mainly determined by the sensor capacitance and related noise. 
Several of these cells are then grouped inside the front-end electronics to form the required granularity for further processing.
To limit the material budget and cost, the goal is to implement each TOF system with a single layer of sensors.

The TOF system can be partitioned into a few key building blocks: the sensor, the front-end, the Time-to-Digital Converter (TDC), the clock management system and the readout system. 

The last decade saw an impressive improvement in the performance of TDCs. 
Converters with bin size in the picosecond range, sampling rate of several tens of MHz and power consumption below 10~mW per channel have become a commodity (see, for instance,~\cite{8051262} and references therein). 
A bin size of 10~ps resulting in a r.m.s. quantization noise of 2.9~ps is negligible with respect to the specified system resolution of \SI{20}{\ps}.

High performance clock management systems have also seen continuous improvement in recent years, as very low-jitter clocks are needed in state-of-the-art serializers, offering transmission speeds well above 10~Gbit/s. 
Still, the design of a TDC and a clock management for such a large system is challenging. 
This system is largely independent of the specific choice of sensor and frontend electronics. 

The readout will be based on an asynchronous communication scheme between the pixels and the periphery to avoid the large power dissipation stemming from the distribution of a clock signal over the full matrix. 
The readout architecture, in particular the power network and the pixel-to-periphery data path, will be designed in a modular way and assembled with a digital-on-top design flow, allowing a straightforward scaling of the design.  
The data payload generated by the matrix will be collected at the chip periphery, serialised and sent off-chip through high-speed serial links. 
Time-over-Threshold will be adopted as the method to correct for time-walk. 
The low hit rate makes the data overhead associated with a double measurement per hit easily affordable. In addition, taking into account an upper limit of 100 kHz/cm$^{2}$ for the hit rate, for a full size sensor of 8 cm$^{2}$ the hit rate will be 800 kHz. Assuming a 32 bit word per event a total bandwidth of 1 Gbit/s will be more than sufficient and it can be
easily accommodated in a single differential link.

\begin{revised}
  The calibration strategy will be modelled following the scheme adopted for the present ALICE TOF detector~\cite{TOF-calibration}, for which we can calibrate, in a very effective way, more than 153k channels. 
  The calibration constants are calculated at each LHC run period, so perhaps once per month. The calibration is done with reconstructed tracks by selecting pions of a given momentum and by fitting channel-by-channel the dependence of the time difference ($t_\mathrm{TOF} - t_\mathrm{tracking}(\pi)$) as a function of the Time-over-Threshold, which is used as a proxy for the signal charge amplitude. The calibration accounts for the offset between the LHC clock and the actual collision time (common to all channels), a channel-by-channel offset which compensates for the different time delays introduced by the cable lengths and delays in the electronics (which should be constant over the time) and the time-slewing correction.
As it is done for the present TOF, we can parameterise this correction with splines and check the variation of the calibration constant at every new calibration run.
\end{revised}

Concerning the support mechanics, the plan is to adopt a strategy very similar to the one for the outer layer of the tracking systems.

Three sensor technologies have been identified as candidates for dedicated R\&D in the coming years: fully depleted CMOS sensors, Low-Gain Avalanche Diodes (LGAD) and Single Photon Avalanche Diodes (SPAD). 
Fully depleted CMOS sensors could provide both time resolution and complete coverage with a single layer, while allowing a true 2D monolithic design. 
This will reduce significantly the fabrication costs, offering at the same time a simpler and cheaper assembly. 
A vigorous R\&D is needed here as the time resolution of CMOS sensors needs to be pushed significantly beyond present values. 
Probably, a thin gain layer~\cite{Iacobucci2021} must be implemented to reach the timing resolution of 20~ps while keeping the power consumption within reasonable bounds. 

State-of-the-art LGADs offer a time resolution of 30~ps rms, which is already close to what is needed by the ALICE 3 TOF~\cite{CMSMTDTDR}. 
Very recent works on thin LGAD have reported resolutions down to 20~ps~\cite{SantaCruz1}. 
LGADs are produced on sensor-grade wafers and require dedicated readout electronics. 
This leads to higher production cost and more complex assembly procedures as compared to monolithic sensors. 
LGADs are thus considered a fall-back solution. 

SPADs fabricated in a CMOS technology could also offer a monolithic solution, although the use of a single layer seems less obvious because of dark count (sensor firing in absence of a photon) and fill-factor (fraction of area that is sensitive) issues. 
However, a SPAD-based system may allow merging the TOF and RICH readouts. 
SPADs will thus be considered only in case it is decided a viable option to pursue.
In the following, the MAPS and LGAD sensor technologies are discussed in more detail, while the SPAD option is treated in the RICH detector Section~\ref{sec:systems:rich:options}.
Independent of the technology that will be adopted, there are not stringent constraints on the temperature homogeneity along the TOF detector and also a variation of few Celsius degrees will not affect the timing performance.

\subsubsection{MAPS}
\label{sec:systems:tof:mapsdev}

The recent developments of MAPS were focussed on small pixel size, low noise and low power consumption for tracking applications, while their potential for timing has not yet been fully exploited. 
Good timing performance can only be achieved by fast charge collection, requiring the use of fully depleted sensors. 
In addition, a careful optimization of the device geometry and related field configuration is necessary. 
Sensors with small collection electrodes offer small capacitance but the electric fields are very non-uniform which limits the time resolution. 
Large electrodes show better field uniformity with the drawback of a larger capacitance. 
The introduction of avalanche gain in MAPS has therefore been identified as a promising solution to increase the signal-to-noise ratio and, thus, to reduce the effect of noise on the time resolution. 

MAPS with partial or full depletion have been realised in different process technologies~\cite{Hirono2016, Snoeys2017}. 
Several research projects targeting the development of Fully Depleted MAPS (FDMAPS) with improved timing resolution are ongoing.
A CMOS sensor achieving 105~ps time resolution is described in~\cite{Degerli2020}. 
The prototype was fabricated in a commercial High Voltage CMOS process with a minimum feature size of 150~nm. 
Wafers with a resistivity of 2~k$\Omega\cdot$cm were employed for the substrate. 
This time resolution was achieved with a sensor having diodes of $\SI{1}{\mm} \times \SI{0.5}{\mm}$ and a thickness of \SI{200}{\um}, at~\SI{-300}{\volt}. 

The FASTPIX project has explored the use of small pixels in combination with a specific process modification. 
The aim is to shape the field lines at the pixel border in order to speed up charge collection~\cite{Kugathasan2020}. 
The SEED and ARCADIA projects~\cite{Pancheri2020} have developed fully depleted CMOS sensors in a 110~nm CIS image sensor process provided by LFoundry. 
The technology uses a high-resistivity n-type substrate with a surface n-type epitaxial layer and a buried p-well to allow the integration of in-pixel electronics and to enable fast charge collection (see Fig.\ref{figure:maps}). 

\begin{figure}
\centering
\includegraphics[width=.8\linewidth]{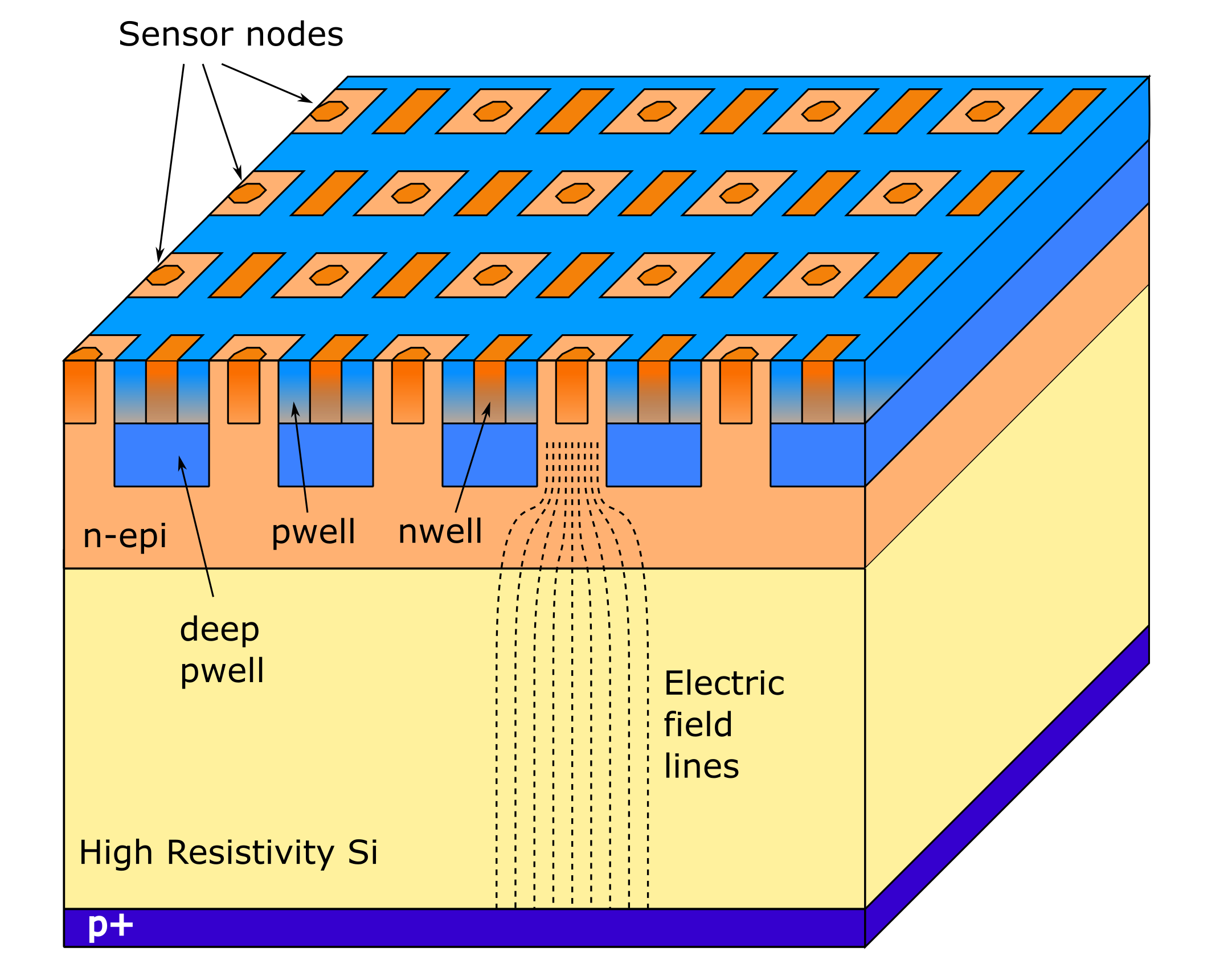}
\caption[Schematic view of ARCADIA MAPS]{ARCADIA MAPS schematic view.}
\label{figure:maps}
\end{figure}

Alternatives to a CMOS-only implementation have also been considered. 
Recently, a monolithic pixel detector prototype was produced in a SiGe BiCMOS technology, and a timing resolution between \SIrange{50}{83}{\ps} (depending on the pixel dimension) was demonstrated for $\beta$ particles emitted by a $^{90}$Sr source~\cite{Iacobucci2019}. 
With a bandwidth of more than \SI{500}{\giga\hertz}, SiGe bipolar transistors offer an attractive option for the design of low-power, wide-bandwidth, low-noise analogue circuits, which are critical for high resolution timing systems. 

At present, CMOS sensors do not reach the required time resolution of 20~ps and it is unclear whether this goal can be achieved with sensors that do not use internal gain in order to reduce the effects of electronics noise. 
CMOS integrated monolithic sensors with internal gain optimized for visible near infrared light detection have been demonstrated in the last decade~\cite{Gaberl2014,Mori2016}. 
The introduction of avalanche gain in the sensor substrate for timing applications has been proposed~\cite{Iacobucci2021}. 

\paragraph{System design}

A key challenge in the implementation of a monolithic timing sensor is that the collection electrode occupies a large fraction of the pixel. 
The pixel size will thus result as a compromise between the goal of achieving a small sensor capacitance and the need of having a sufficient area to accommodate the electronics itself. 
We envisage a hierarchical design, where each sensing diode is equipped with a dedicated front-end amplifier and discriminator. 
Larger macro-pixels are obtained by combining the signals generated by the small pixels in a single data line (see Fig.~\ref{figure:pixels}).
In this way, a single time-tagging and data processing circuit can process the full stream of data from a macro-pixel, thus keeping the power consumption and area occupation at a manageable level.
Thus, it is expected that a complete readout chain can be accommodated on the chip with a moderate integration density. 
This will favour a good yield also on very large sensors. 
The chip can hence be as large as the reticle size, or even larger if stitching techniques are applied. 
The systems will then be assembled with an approach similar to the one used in the outer tracking layers.
The R\&D will target the design of sensors achieving a time resolution of 20~ps per hit or better. 
In the outer layer, a double-layer arrangement could be employed to improve the overall resolution by $\sqrt{2}$. 
Although not favored, this option could still be competitive with LGADs in term of cost.

\begin{revised}
  The sensor is expected to be of reticle size, i.e. in the order of 2.6 cm x 3.2 cm for a typical CIS process. 
  Even though in a sensor optimized for timing the collection electrode must occupy most of the pixel area, enough space will be left to accommodate the preamplifier and discriminator directly in the active area without compromising efficiency and timing performance. 
  The TDC and the readout logic will be located in the periphery. 
  Given the moderate event rate in the TOF, a peripheral strip 1 mm wide is expected to be sufficient to accommodate the required functionality. 
  The maximum particle hit rate is expected to be smaller than \SI{100}{\kilo\hertz\per\cm\squared}. 
  On the basis of the experience acquired with the ALPIDE sensors one can assume a noise hit rate in the order of \SI{2}{\mega\hertz\per\cm\squared} for sensors with a geometry optimized for the TOF conditions. 
  The noise hit rate for a full size sensor will be \SI{16}{MHz} in this case. Assuming a 32~bit word per event this requires a total bandwidth of \SI{500}{Mbit/s} that can be easily accommodated in a single differential link.
\end{revised}

\begin{figure}
\centering
\includegraphics[width=.8\linewidth]{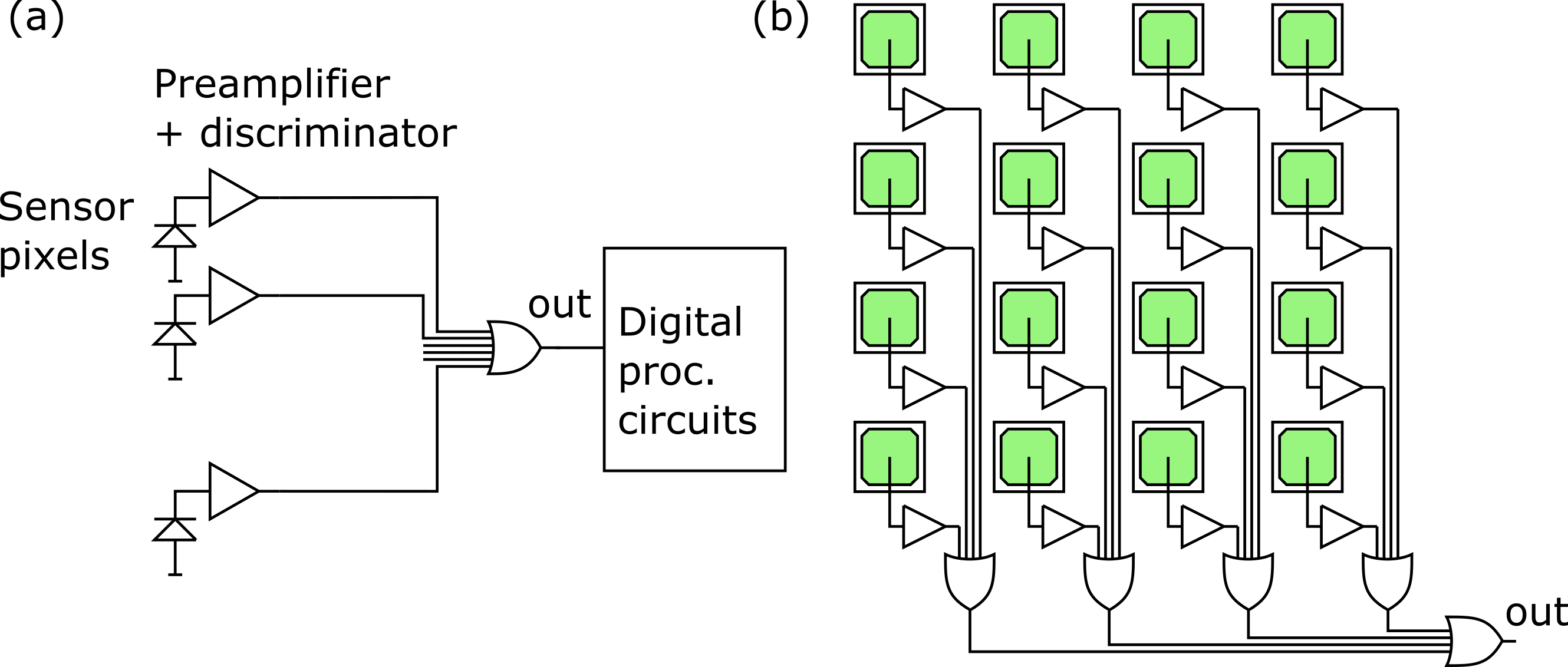}
\caption[Schematic diagram of macro pixels]{Macro-pixel schematic diagram (a) and sample layout (b).}
\label{figure:pixels}
\end{figure}

\paragraph{R\&D plans}
\label{sec:systems:tof:mapsrnd}

An improved timing resolution requires both a very uniform and fast charge collection, and fast readout electronics with a very high Signal-to-Noise Ratio (SNR)~\cite{Kramberger2019}. 
While the first requirement can be fulfilled only by minimizing the collection time of the charge carriers generated at the pixel edges, the improved SNR also requires the optimization of the sensor capacitance. 
The timing performance of a fully depleted MAPS will result from the optimization of the sensor thickness and collection electrode area. 
Thicker sensors produce larger signals, thus allowing for better SNR, but the time resolution is worse due to longer charge collection times. 
Larger sensors have a more uniform electric field, but their higher capacitance increases the contribution of the electronics noise. 

A detailed simulation work is already ongoing to optimize the sensor design. 
The detectors proposed for this R\&D will leverage the previous experience gained, in particular, with the SEED and ARCADIA projects. 
The design of the sensors will be modified to meet the very strict timing requirements of the TOF layers. 
The geometry of the in-pixel sensors will be optimized with the help of TCAD simulations, with the goal of obtaining a uniform and fast charge collection and minimizing the sensor capacitance.  
Monte Carlo simulations performed on $\SI{50}{\um} \times \SI{50}{\um}$ pixels in ARCADIA technology indicate that, for an active thickness of \SI{25}{\um}, an intrinsic sensor timing resolution as low as \SI{20}{\ps}, limited by Landau fluctuations and non-saturated carrier velocity, are within reach (see Fig.~\ref{figure:tres50um}). 
The very low electronics noise required for these thin sensors will result in significant power consumption of the frontend. 
The introduction of a moderate gain, of the order of \numrange{5}{20}, would greatly reduce this power consumption.  
Therefore, strategies to introduce a gain layer by adding dedicated implants or using engineered substrates will be evaluated through TCAD/MC simulations and their feasibility in production will be discussed with the silicon foundries.
\begin{revised}
  The addition of a gain layer will require an extra implant and mask, which is expected to increase the cost of the mask set by less than 5\%.
  Since the improvement of monolithic timing sensors is of great relevance for particle detectors, synergies are expected on various fronts.
\end{revised}

\begin{figure}
\centering
\includegraphics[width=.8\linewidth]{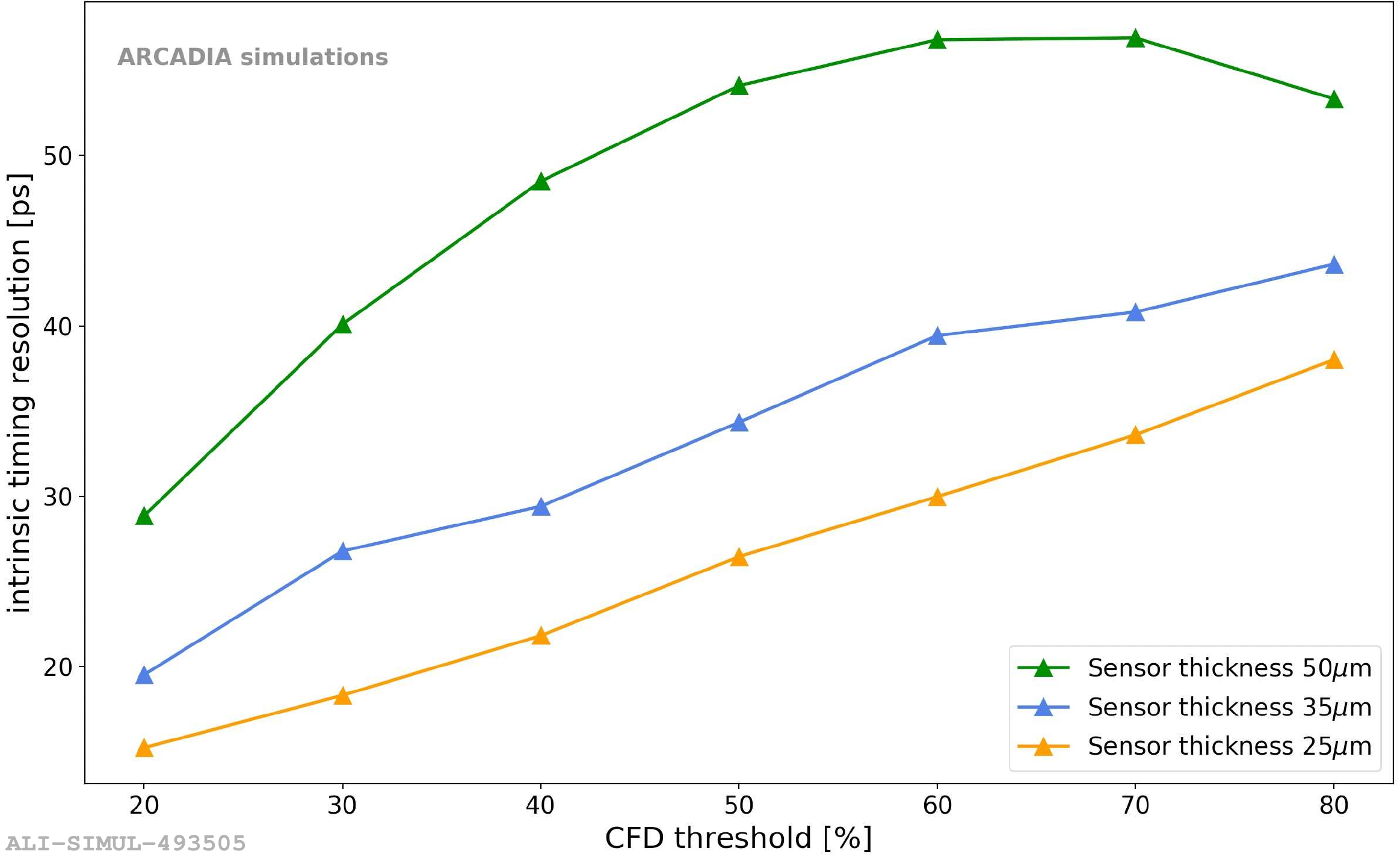}
\caption[Time reoslution simulated for ARCADIA technology]{Simulated intrinsic timing resolution for $\SI{50}{\um} \times \SI{50}{\um}$ pixels in ARCADIA technology. The pad electrode size used in the simulations is $\SI{40}{\um} \times \SI{42}{\um}$.}
\label{figure:tres50um}
\end{figure}

About a long term plan, in the first three years (2022-2024) the R\&D will be focused on two activities, which are fully complementary and can, therefore, be carried out in parallel. The first activity is the study of CMOS sensors with an intrinsic gain layer. Three submissions with a cycle of one submission per year are expected to be sufficient to provide a conclusive answer on the feasibility of the sensor itself. In this phase, the performance of the sensors can be studied using already available electronics in a similar way to what has been done for LGADs. The development of a readout architecture can proceed in parallel. It must be pointed out that the readout architecture is weakly coupled to the particular technology chosen for the sensor. Two additional years (2025-2026) are expected to develop full scale chips ready for mass production.

\subsubsection{LGAD sensors}
\label{sec:systems:tof:lgaddev}

Low Gain Avalanche Diodes (LGAD)~\cite{lgad1} are silicon avalanche pad detectors based on Avalanche Photo Diodes (APD) that demonstrated excellent timing capabilities and radiation hardness. 
LGADs have been extensively studied in recent years and represent the most mature silicon detector technology for timing applications in the HEP environment. 
They have been chosen by ATLAS and CMS for timing layers, which are part of the Phase-II upgrades~\cite{atlas_tdr, CMSMTDTDR}.

LGADs exploit internal gain by means of an additional highly-doped p-layer, called multiplication layer, close to the n-p junction to create a high field. Unlike SPADs operating in Geiger mode, LGADs operate below the breakdown voltage. 
The advantages are reduced cross-talk, easier segmentation, smaller dead time and smaller dark count rate. 
Simulation and experimental results show that the optimum gain for a \SI{50}{\um} thick LGAD is around 20--30~\cite{CARTIGLIA2015}. 
Thinner sensors show better time resolution when reducing the sensor thickness. The length of the signal is reduced proportionally and the effect of Landau and gain fluctuations on the time resolution is therefore also reduced proportionally. 
However, since the detector capacitance increases when reducing the sensor thickness the electronics noise is also increased. 
Therefore, an optimization between sensor capacitance, internal gain and front-end power consumption is essential. 

In recent years LGAD sensors have been produced by different manufactures like Centro Nacional de Microelectr\'onica (CNM, Spain), Hamamatsu Photonics K.K. (HPK, Japan) and Fondazione Bruno Kessler (FBK, Italy) with different doping levels, active thicknesses, pad sizes and inter-pad gaps. 
The design of LGAD sensors is evolving rapidly, mostly driven by the requirements of the ATLAS and CMS collaborations. 
ATLAS needs to produce $\sim$ 9k sensors of 2 $\times$ 4 cm$^2$, each sensor with 450 1.3 $\times$ 1.3 mm$^2$ pads while CMS aims at producing $\sim$ 3k sensors of 4.8 $\times$ 9.6 cm$^2$, each sensor with 1536 1 $\times$ 3 mm$^2$ pads. 
Sensors with a \SI{50}{\um} active region, within a normal \SI{300}{\um} thick silicon wafer demonstrated to provide a time resolution of 30-35~ps up to fluences of $5 \times 10^{15}$ 1-MeV n$_\text{eq}$/cm$^2$~\cite{Cartiglia, Carnesecchi, Zhao_2019}.

The time resolution of LGADs with \SI{50}{\um} thickness as a function of the threshold of a Constant Fraction Discriminator (CFD) is shown in Fig.~\ref{figure:lgad}~\cite{Carnesecchi} (left) for different values of the applied voltage. 
The different voltages, ranging from 250 V to 290 V, correspond to a gain of 19 and 31, respectively.

The time resolution improves as the CFD threshold is increased due to the larger signal slope at higher thresholds and therefore a smaller effect of the electronics noise. At larger CFD thresholds the time resolution is dominated by the Landau fluctuations. 
\begin{figure}
\centering
\includegraphics[width=.48\linewidth]{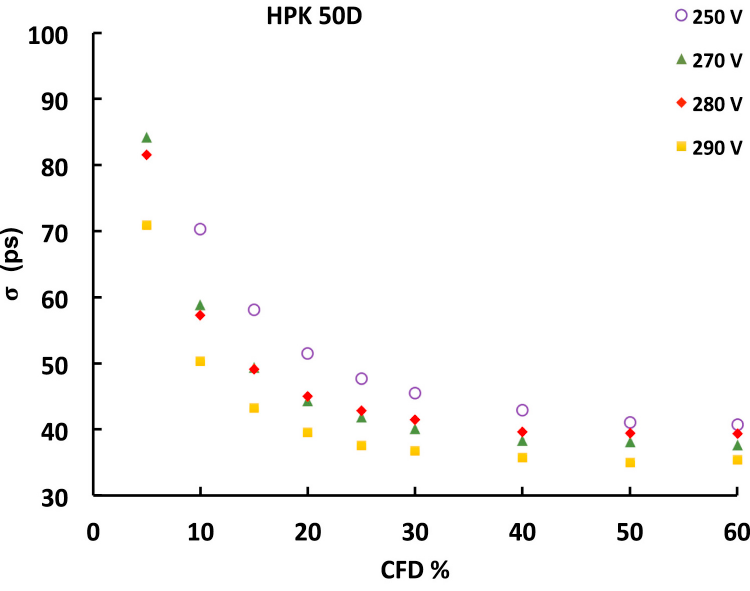}
\raisebox{0.25cm}{\includegraphics[width=.49\linewidth]{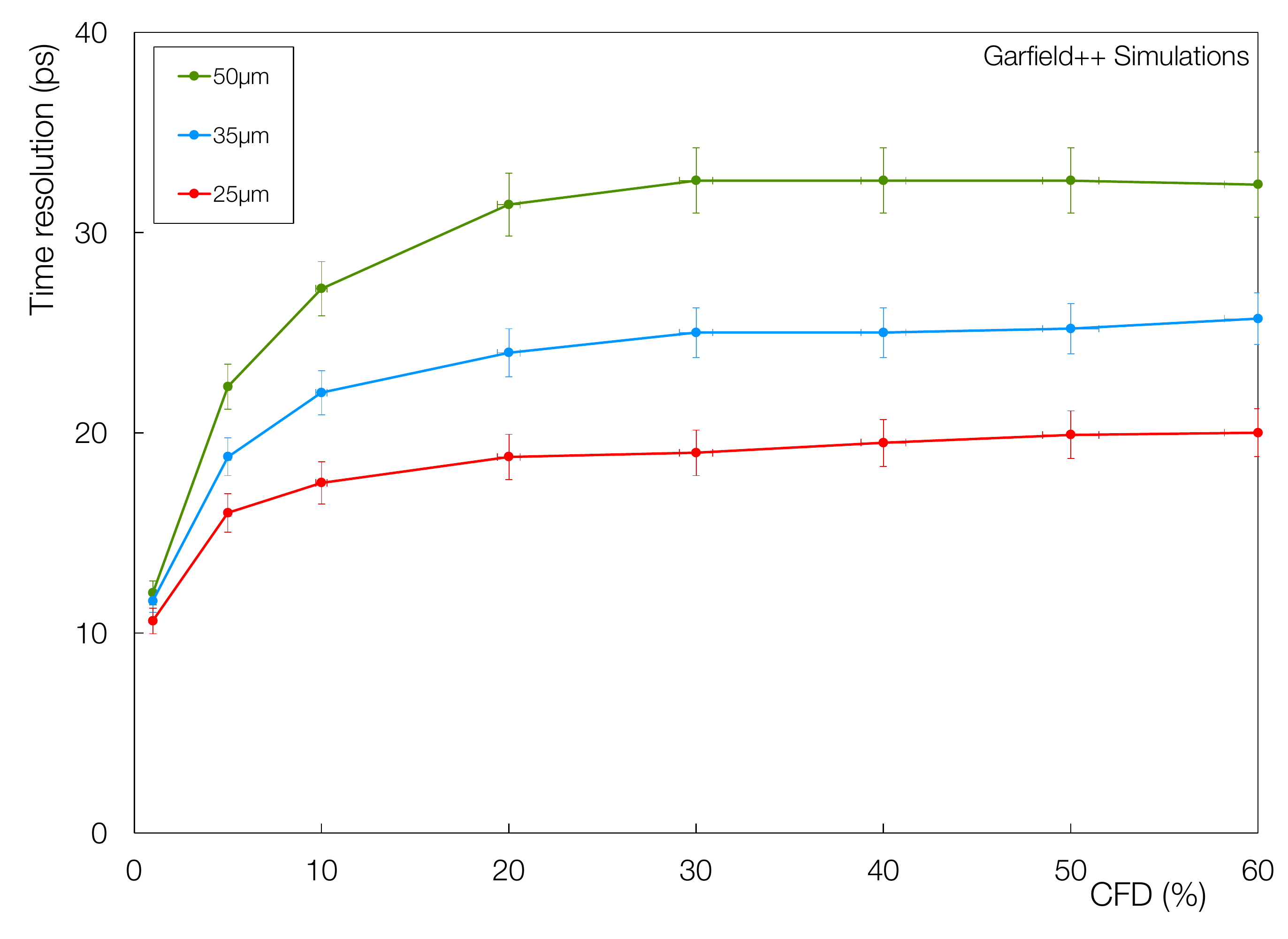}}
\caption[Time resolution for LGADs]{Left: Test beam measurements of the time resolution of \SI{50}{\um} thick LGADs from HPK (Hamamatsu Photonics K.K., Japan) as a function of Constant Fraction Discriminator (CFD) for different voltages~\cite{Carnesecchi}. Right: Garfield++ simulations of the LGAD time resolution as a function of CFD for different sensor thicknesses.}
\label{figure:lgad}
\end{figure}

Simulations carried out with Garfield++\cite{garfield++} have confirmed that very thin LGADs of \SI{25}{\um} or \SI{30}{\um} active thickness have a better intrinsic performance. 
This is shown in Fig.~\ref{figure:lgad} (right) where the simulated time resolution is reported as a function of CFD for different sensor thicknesses. The simulation is performed assuming a gain of 20 and electronics noise is neglected.

Another critical parameter is the sensor fill factor, corresponding to the fraction of the detector which is efficient for particle detection.
The high electric field generated by the gain layer needs to be properly contained to avoid cross-talk between pixels and possibly even breakdown. 
Lateral components of the electric field are controlled by deep n-doped implants
that surround the gain layer and extend onwards for a thickness of a few microns. 
However, this solution introduces a low-gain region at the pad edges thus reducing the geometric efficiency. 
An interesting isolation strategy, already used in the design of CMOS image sensors and SiPMs, has been proposed recently~\cite{trenches-lgad}, which exploits shallow trenches in the silicon substrate with the termination implants within the trench. 
Prototypes of trench-isolated LGADs have been produced, demonstrating an effective inter-pad dead area of only $\sim \SI{10}{\um}$, allowing a fill factor close to 100\%.
Another solution is represented by an evolution of LGADs called Resistive AC-Coupled Silicon Detectors (RSD or AC-LGADs). 
There, the voltage that creates the drift field is applied to a thin resistive layer that is insulated from the readout pads. 
At high frequency this layer is 'transparent' and the signal is directly induced on these pads. 
Latest results obtained with a \SI{50}{\um} thick RSD, designed at the INFN of Torino, have demonstrated a 100\% fill-factor. 
By using a TCT pulsed laser, a time resolution of less than 14 ps was measured~\cite{RSD}, which, however, does not include the Landau fluctuations that are present when detecting charged particles. 

Finally, radiation damage in silicon sensors causes a reduction of the charge collection efficiency, increase in leakage current, and changes in the sensor doping profile, thus decreasing the gain of the amplification layer.  
These effects can be usually mitigated by operating at low temperature ($\sim \SI{-30}{\celsius}$), which however has a big impact on cost and material budget which, especially, for the inner TOF layer, which has to be as small as possible. 
However, given the moderate radiation level expected in ALICE~3 these effects are expected to be negligible and LGADs will thus be operated at room temperature.

\paragraph{System design}
\label{sec:systems:tof:lgadrnd}

For the LGAD solution, the sensors and the front-end electronics are fabricated on different wafers. 
For the interconnection between the sensor and the front-end electronics two options can be considered. 
The first one, chosen by ATLAS and CMS, foresees the same approach of traditional hybrid pixel detectors. 
The channel layout in the front-end ASIC matches the pixel pattern and both the ASIC and the sensors are connected through bump-bonding. Given the pixel pitch  required for the TOF, well established and low-cost industrial flip-chip technologies can be used. 
However, the readout chip needs to have the same size as the sensor, thus increasing the fabrication cost and not making use of the high integration density offered by modern microelectronics technology. 
Fully-fledged timing front-end channels with an area of $\SI{440}{\um} \times \SI{440}{\um}$ and $\SI{55}{\um} \times \SI{55}{\um}$ have been reported  in 110~nm and 28~nm CMOS technologies.  
While a direct bump-bonding between the sensor and the front-end channel reduces the parasitic capacitance and inductance of the interconnection, the best timing performance obtained so far with LGADs has been obtained with set-ups using standard ultrasonic bonding wires between the sensor and the front-end ASICs. 
A careful designed fan-out may thus allow us to connect sensors with large pixels to much smaller front-end chips without deteriorating the system time resolution. 
This reduces significantly the prototyping and production cost of the front-end electronics, making advanced CMOS technologies, such as \SI{28}{\nm}, affordable. 
In \SI{28}{\nm}, a 256~channels chip could be implemented in an area of $\SI{3}{\mm} \times \SI{3}{\mm}$. 
Such a small chip can be prototyped through Multiproject Wafer services.

The fan-out can be implemented as a metal redistribution layer on the sensor itself or as part of the flex cable that is needed to interface the module to the outside world. 
In either case no significant cost increase is expected. 
In the inner layer a sensor of $\SI{1}{\cm} \times \SI{2}{\cm}$, (similar to the ones adopted by ATLAS and CMS) segmented in 256~pixels will provide the needed granularity. 
The moderate cost of a mask set needed to fabricate LGADs allows a straightforward optimization of the sensor design.
In the outer layer, the use of larger sensors (e.g. $\SI{10}{\cm} \times \SI{5}{\cm}$) can thus be considered in order to facilitate the assembly procedure. 
Larger pixels could also be used to reduce the number of channels. 
However, in this case the LGAD gain should also be increased in order to keep the contribution of the electronics noise to the overall jitter under control.
The option of realising a monolithic sensor using a CMOS process was already discussed above under the term 'MAPS with internal gain'. 

\paragraph{R\&D plans}
\label{sec:systems:tof:lgadimplementation}
LGADs can be considered matured technology, with the availability of broad expertise in the community for the design and production of these sensors. 
The R\&D will focus on the design of sensors optimized for the TOF requirements, exploring in particular the performance of modules based on the fan-out concept described above. 
In this early phase, a 32~channel ASIC in 110~nm CMOS already available at INFN, will be adapted to study the timing performance of the modules. 
A dedicated front-end development will be initiated only in case the LGAD solution becomes the baseline scenario. 
For the fan-out option we envisage the use of a 28~nm CMOS process, which is now a well established technology in the industry. 
Furthermore, it is expected to be available for a very long time in the future. 
The use of the 28~nm node does not pose critical design challenges. 
Analogue blocks can be implemented with performance comparable to those offered by 110~nm and 65~nm technologies, while the performance of digital circuits is obviously improved.

In case the direct bump-bonding between the sensors and the front-end ASIC is preferred, a much larger ASIC will be needed, but the component density on chip will be small. 
This will favour the use of a less aggressive technology (65 nm or even 130 nm), in which larger ASICs can be prototyped at a much lower cost.

\ifcost
\subsubsection{Cost estimates}

Table~\ref{tab:tof:cost} shows the cost estimates for the TOF components.

\begin{table}
  \centering
  \renewcommand{\arraystretch}{1.3}
  \begin{tabular}{lp{4cm}S[table-format=2.1]}
    \toprule
    Component & Comment & {Cost (MCHF)}\\
    \midrule
    Sensors & monolithic & 7.5 \\
    Mechanics & & 1.2 \\
    Read-out & & 1.6 \\
    Power & & 2.1 \\
    Cooling & & 1 \\
    Services & & 1.5 \\
    \midrule
    Total & & 14.9 \\
    \bottomrule
  \end{tabular}
  \caption{Cost estimates TOF}
  \label{tab:tof:cost}
\end{table}
\fi

\subsection{RICH detector}
\label{sec:systems:rich}

Cherenkov radiation can be used to extend the particle identification capabilities to transverse momenta out of reach for the outer TOF detector. 
The refractive index of the radiator determines the \pt window, in which the detector provides effective separation.
Table~\ref{tab:cher_radiators} shows the main properties of common materials (from gaseous to solid).
As discussed in Sec.~\ref{sec:performance} and shown in Fig.~\ref{fig:performance:detector:hadron_id:summary:5kgauss}, a refractive index $n = 1.03$ ensures continuity in the PID capabilities beyond the TOF in the central barrel.
This can be conveniently achieved by using aerogel as the radiator material.
Because of the Lorentz boost, a smaller refractive index is needed to achieve adequate \pt coverage in the forward region.

\begin{table}
    \centering
    \renewcommand{\arraystretch}{1.3}
    \setlength{\tabcolsep}{9pt}
    \begin{tabular}{l S[table-format=1.6] S[table-format=2.5] S[table-format=2.3] S[table-format=2.3] S[table-format=2.3] S[table-format=3.0]}
    \toprule
    Material
    & {Refractive}
    & \multicolumn{4}{c}{p$_\mathrm{threshold}$ ($\GeVc$)}
    & {$\lambda_\mathrm{cutoff}$}
    \\ 
    \cmidrule{3-6}
    &{index}
    &{e}
    &{$\pi$}
    &{K}
    &{p}
    &{(nm)}
    \\
    \midrule
    Ar & 1.000283 & 0.021 & 5.87 & 20.75 & 39.44 & 124 \\
    CO$_2$ & 1.000449 & 0.017 & 4.66 & 16.47 & 21.48 & 175 \\
    C$_4$F$_{10}$ & 1.0015 & 0.009 & 2.5 & 9.0 & 17 & 136 \\
    & 1.01 & 0.004 & 0.98 & 3.48 & 6.62 \\[-5pt]
    Aerogel & 1.03 & 0.002 & 0.57 & 2.0 & 3.8 & 300 \\[-5pt]
    & 1.26 & 0.001 & 0.18 & 0.64 & 1.22 \\
    C$_6$F$_{14}$ & 1.3 & 0.0006 & 0.168 & 0.594 & 1.13 & 165 \\
    NaF & 1.41 & 0.00051 & 0.140 & 0.497 & 0.944 & 125 \\
    Fused silica & 1.47 & 0.00047 & 0.129 & 0.458 & 0.87 & 158 \\
    \bottomrule
    \end{tabular}
    \caption{Main characteristics of common Cherenkov radiators}
    \label{tab:cher_radiators}
\end{table}

\subsubsection{Specifications}
\label{sec:systems:rich:specs}

The specifications of a Ring-Imaging Cherenkov (RICH) system installed following the TOF detector are derived from the requirement to extend the e/$\pi$ separation from \SI{500}{\mega\eVc} (TOF limit) to $\sim \SI{2}{\giga\eVc}$ and the charged hadron identification ($3 \sigma$ $\pi$/K separation) up to $\sim \SI{10}{\giga\eVc}$.
To this end, a RICH detector with a \SI{2}{\cm} thick aerogel tile and a photo-detection layer at \SI{20}{\cm} from the radiator, purely relying on proximity focusing, has been implemented in Geant4, see Fig.~\ref{RICH-layouts}.
The performance of SiPMs (Hamamatsu 13360 3050CS) has been assumed for the photon detection, specifically a peak photon detection efficiency (PDE) of 40\% and a dark count rate (DCR) of \SI{50}{\kilo\hertz\mm^{-2}}.
Figure~\ref{simu_single_results} shows the average number of detected photons and the Cherenkov angle resolution obtained in events with a single particle.
In addition, Pb--Pb collisions at $\sqrt{s}$= 5.76 TeV have been studied, based on Pythia~8 simulations. 
The angle reconstruction algorithm is based on the Hough Transfom method~\cite{VOLPE2014259} applied in combination with a cut on the time of \SI{1}{\nano\second}. 
The application such a time gate reduces the DCR and the related data volume by a factor 25. A more precise (i.e. $\sim 50$ ps or better) time stamping of detected Cherenkov photons allows a further oﬄine reduction of dark counts and correlated noise (in particular afterpulses). In addition, in high multiplicity events from heavy-ion collisions, most of the background for each reconstructed Cherenkov ring is related to track and photon hits produced by other charged particles. Depending on their momentum, the arrival time can differ from tens up to several hundreds of picoseconds. Therefore, knowing the time of each type of hit in the SiPM with high precision, allows better matching of particle hits with extrapolated tracks, and overall a reduction of background contribution to the Cherenkov angle error.
For a further DCR online reduction, cooling of the SiPMs can be envisaged.
An example of a single event, before and after the described processing, is shown in Fig.~\ref{simu_PbPb}, 
while Fig.~\ref{ThetsVSp-PbPb} shows the reconstructed Cherenkov angles as a function of the particle momenta.  
The specifications are summarized in Tab.~\ref{tab:RICH_specs}.

\begin{table}
\centering
\small
\begin{tabular}[t]{l r}
    \renewcommand{\arraystretch}{1.3}
    \begin{tabular}[t]{l S[table-format=4.2]}
         \toprule
         Inner Radius (m) &  0.9 \\
         Proximity gap (m) & 0.2 \\
         Length (m) &  5.6 \\
         Aerogel surface (m$^2$) &  32 \\
         NIEL (\nequiv) / month & \num{4.5e9} \\
         TID (rad) / month &  \num{1.4e2} \\ %
         Material budget (\% X$_0$) &  3 \\
         Power density (mW/cm$^2$) & 50 \\
         Cherenkov angle resolution (mrad) & 1.5 \\
         \bottomrule
    \end{tabular}
&
    \renewcommand{\arraystretch}{1.3}
    \begin{tabular}[t]{l l}
         \toprule
         PD area (m$^2$) & 39  \\
         PDE in visible range &  $ \geq 40\%$ \\
         Pixel size (mm$^2$) &  3x3  \\
         Integration fill factor &  $ \geq $ 90\% \\
         Time jitter (ps) & $\leq$ 100 \\
         Occupancy (dark count) & \num{5e-4} \\
         Occupancy (photons) & \num{5e-3}\\
         Hit rate (kHz/cm$^2$) & $1.5 \times 10^2$ \\
         \bottomrule
    \end{tabular}
\end{tabular}
    \caption[RICH specifications]{(Left) RICH and (Right) photon detector specifications}
    \label{tab:RICH_specs}
\end{table}

\begin{figure}
\begin{center}
\includegraphics[width=0.48\textwidth]{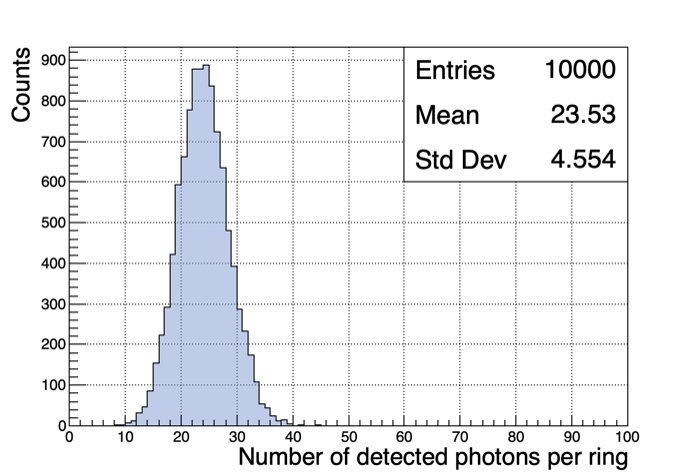}
\includegraphics[width=0.48\textwidth]{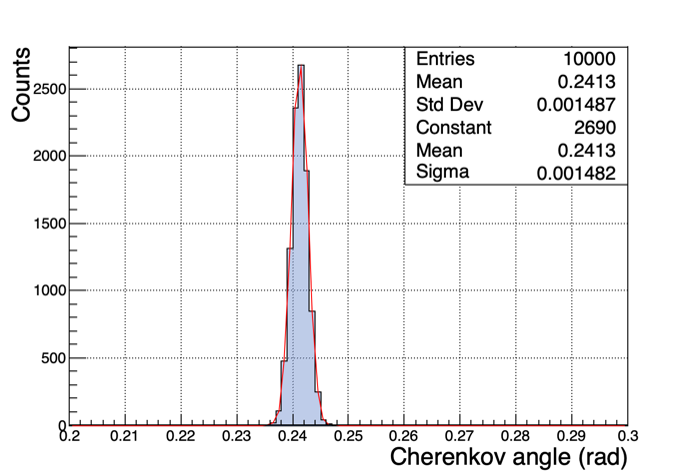}
\caption[Photon detection and angular distribution in RICH]{(Left) Distribution of the amount of detected photons/event and (Right) distribution of the reconstructed Cherenkov angle for particles at saturation in Monte-Carlo single particle events.} 
\label{simu_single_results}
\end{center}
\end{figure}

\begin{figure}
\begin{center}
\includegraphics[width=0.48\textwidth]{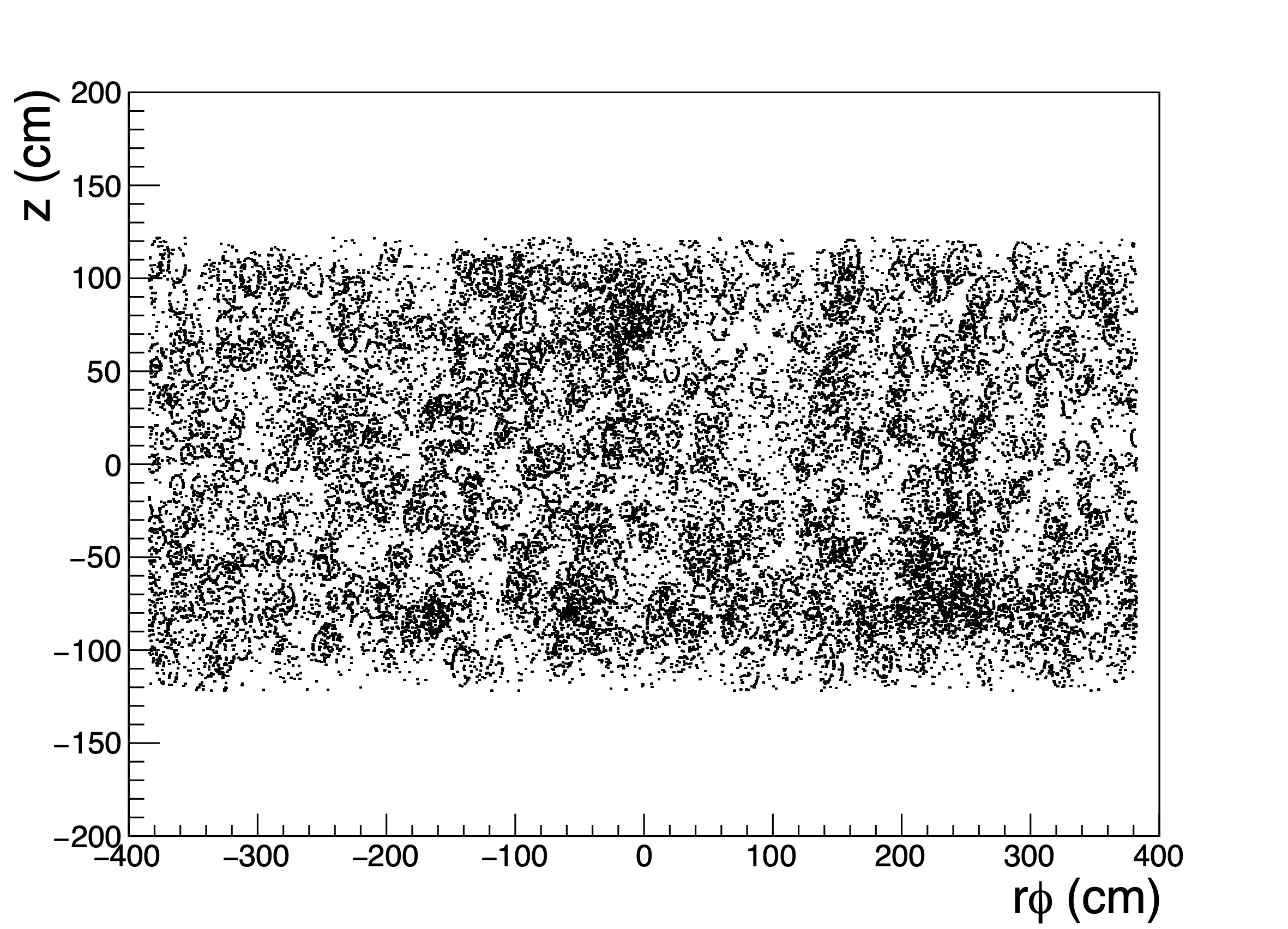}
\includegraphics[width=0.49\textwidth]{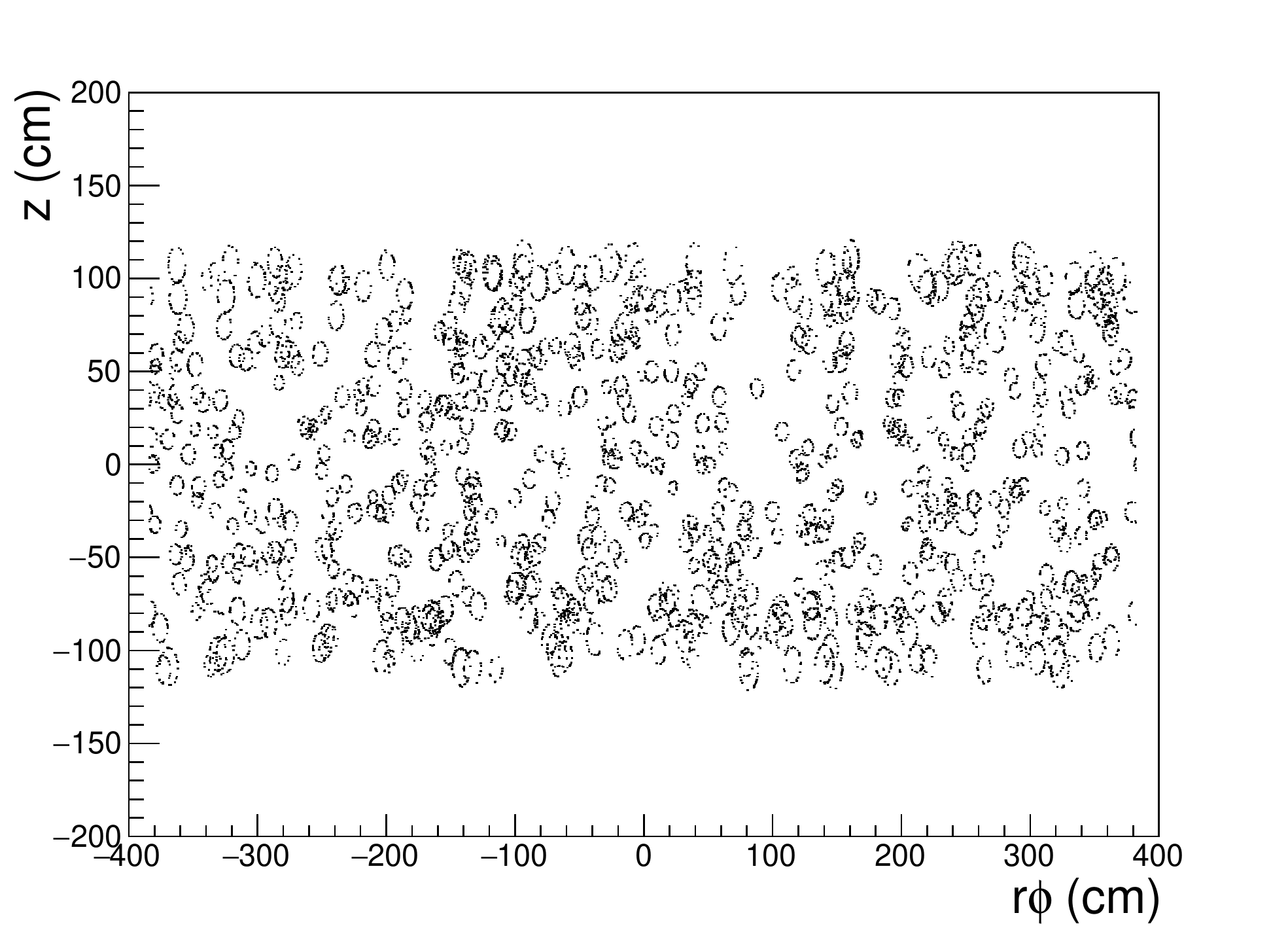}
\caption[Pb--Pb event in the RICH detector]{(Left) Example of one Monte Carlo Pb--Pb raw event in the barrel RICH and (Right) the same event after applying 1 ns time gate and Hough Transform for Cherenkov rings pattern recognition.} 
\label{simu_PbPb}
\end{center}
\end{figure}

\begin{figure}
\begin{center}
\includegraphics[width=0.7\textwidth]{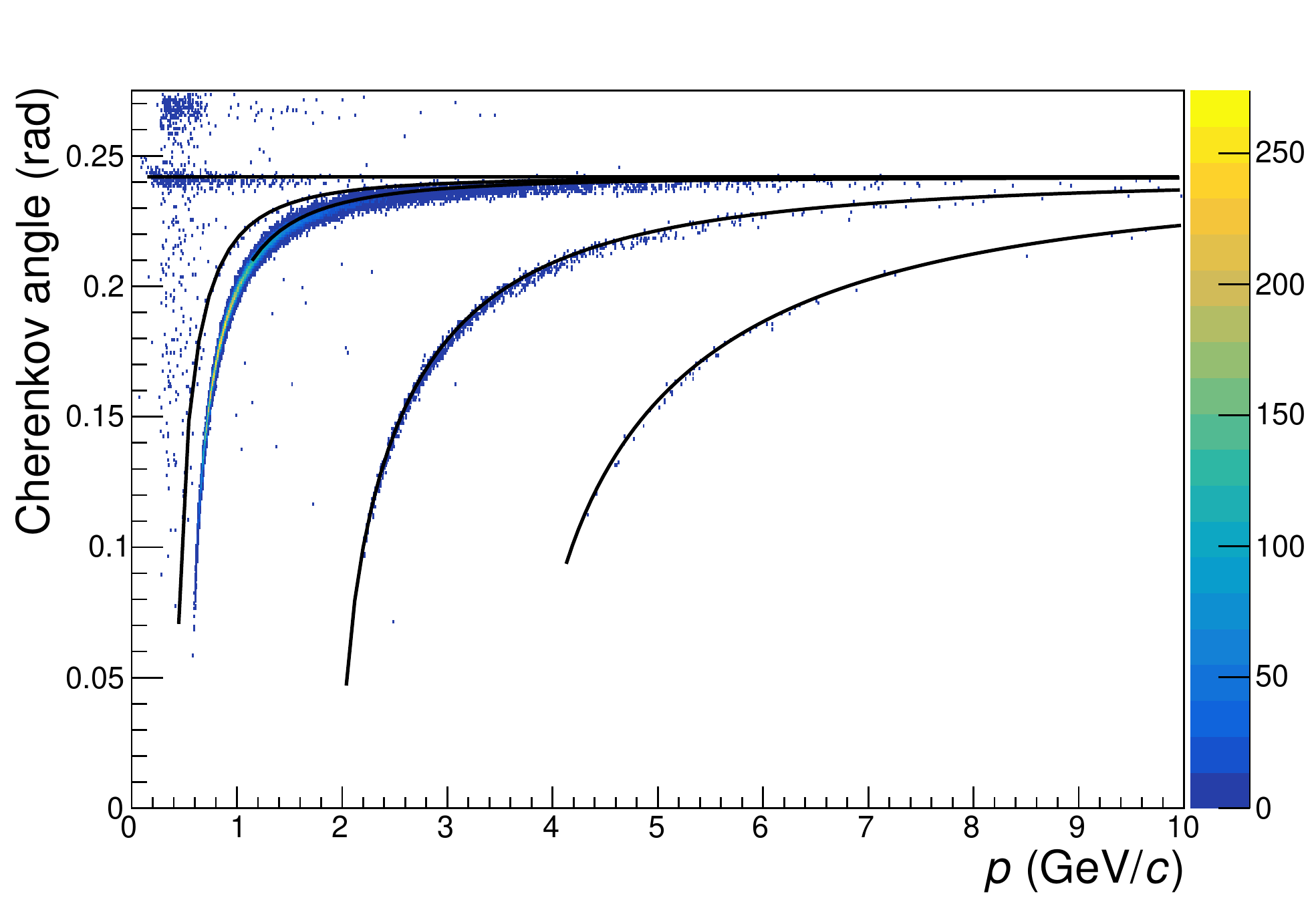}
\caption[Cherenkov angle reconstruction in \PbPb events]{Reconstructed Cherenkov angle as a function of momentum in Monte Carlo Pb--Pb events.} 
\label{ThetsVSp-PbPb}
\end{center}
\end{figure}

\subsubsection{Technology options}
\label{sec:systems:rich:options}

The design choices are mostly driven by the Cherenkov angle needed to achieve the required particle separation, operation in a (strong) magnetic field, and the need to instrument a large surface. 
The separation power is determined by the single-photon angular resolution $\sigma_{\theta_c}^\gamma$ and the number of detected photons N$_{\gamma}$ (as well as the tracking resolution):
\begin{equation}
    \sigma_{\theta_c}^\mathrm{tot} = \frac{\sigma_{\theta_c}^\gamma}{\sqrt{N_{\gamma}}} \oplus \sigma_{\theta_c}^\mathrm{track} .
\end{equation}
In turn, $\sigma_{\theta_c}^\gamma$ comprises the contributions from chromatic aberration,
geometric aberration,
spatial resolution 
and uncorrelated hits (noise):
\begin{equation}
    \sigma_{\theta_c}^\gamma = \sigma_{\theta_c}^\mathrm{chr} \oplus 
    \sigma_{\theta_c}^\mathrm{geo} \oplus 
    \sigma_{\theta_c}^\mathrm{pos} \oplus 
    \sigma_{\theta_c}^\mathrm{noise} .
\end{equation}
In the following, the contributions from tracking and noise are considered to be negligible. 
In particular, considering the performance of the proposed tracker performance, one could expect a very small contribution to the Cherenkov angle reconstruction from the tracking error. Indeed, the only tracking information necessary for the Cherenkov angle reconstruction is the particle track angle relative to the radiator plane, which can be estimated with a precision better than all dominant contributions to the Cherenkov angle error (see Table~\ref{tab:cher-errors}). The charged particle hit is relevant to match the extrapolated track and could be added to improve the tracking up to the photon detector layer. Simulation studies on the impact of alignment have shown a very small degradation ($\sim$2 \%) of the Cherenkov angle resolution for a misalignment up to 0.5 mm.

Using the procedures described in~\cite{Ypsilantis:1993cp,DiMauro:2000in}, each contribution can be estimated analytically and verified by detector simulation. Table~\ref{tab:cher-errors} shows the values of chromatic, geometric and spatial errors, as well as the total Cherenkov angle resolution for different detector configurations.
Increasing the gap $d$ results in an improvement of the geometric and pixel resolutions. 
A setup with a mirror also reduces the uncertainties from focusing and geometric aberration.

\begin{figure}
    \centering
    \includegraphics[width=\textwidth]{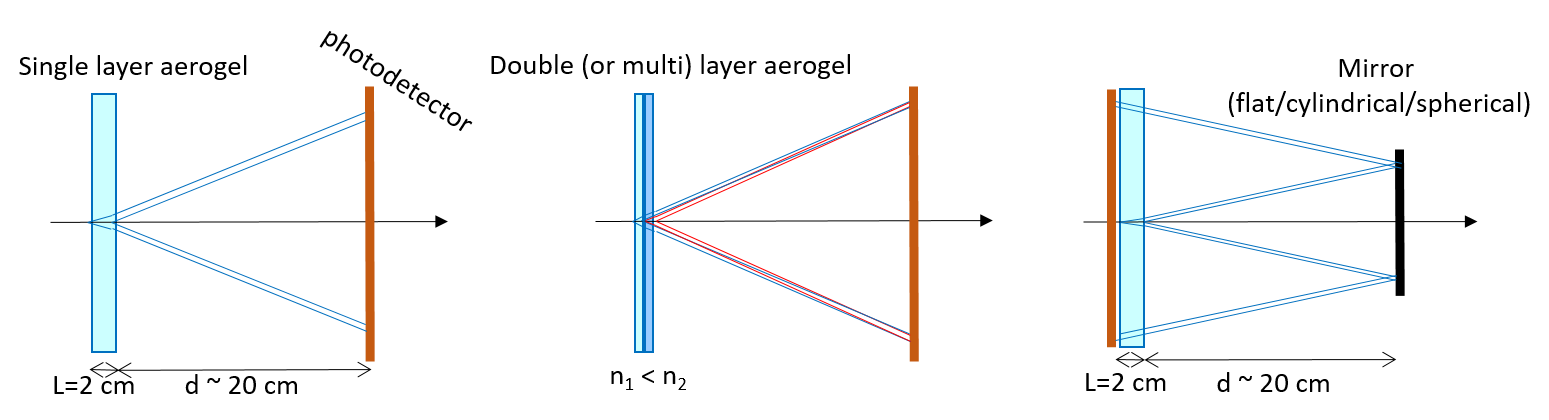}
    \caption[RICH concepts]{Schematic view of the aerogel RICH baseline layout (left), multi-layer (focsuing) aerogel layout (centre) and mirror layout (right).} 
    \label{RICH-layouts}
\end{figure}

\begin{table}[!ht]
    \centering
\scriptsize
    \renewcommand{\arraystretch}{1.5}
\begin{tabular}{l *{9}{S[table-format=2.1]}}
        \toprule

\textbf{layout} & \textbf{radiator} & \textbf{gap} & \textbf{pixel} & \textbf{chromatic} & \textbf{geometric} & \textbf{spatial} & \textbf{$\sigma_{\theta_c}$(phot.)}  & \textbf{no. of} & \textbf{$\sigma_{\theta_c}$(total)} \\[-5pt]
                & \textbf{L} & \textbf{d} & x    & \textbf{error} & \textbf{error} & \textbf{error} &  & \textbf{detected} &  \\[-5pt]
                & \textbf{(cm)} &\textbf{(cm)} &\textbf{(cm)} &\textbf{(mrad)} &\textbf{(mrad)} &\textbf{(mrad)} &\textbf{(mrad)} & \textbf{photons} & \textbf{(mrad)} \\
        \midrule
        baseline & 2 & 20 & 0.3 & 1.4 & 6.1 & 3.9 & 7.4 & 24 & 1.5 \\ 
        larger gap & 2 & 30 & 0.3 & 1.4 & 4.1 & 2.6 & 5.1 & 24 & 1.0 \\
        cyl. mirror & 2 & 40 & 0.3 & 1.4 & 4.6 & 1.9 & 5.2 & 20 & 1.1 \\
        2 aerogel & {1+1} & 20 & 0.3 & 1.4 & 3.1 & 3.9 & 5.3 & 24 & 1.1 \\
        3 aerogel & {1+1+1} & 20 & 0.3 & 1.4 & 3.1 & 3.9 & 5.3 & 32 & 0.9 \\
        \bottomrule

    \end{tabular}
    \caption[Cherenkov angle resolutions]{Contributions to the single photon Cherenkov angle resolution and overall resolution for various detector layouts.}
    \label{tab:cher-errors}
\end{table}

As baseline, we consider a radiator composed of hydrophobic aerogel tiles (\SI{15}{\cm} $\times$ \SI{15}{\cm} $\times$ \SI{2}{\cm}) from Aerogel Factory Co. Ltd (Chiba, Japan). 
Alternatively, a multi-layer radiator consisting of 2 to 4~tiles (Fig.~\ref{multi-aerogel}) with increasing refractive index and thicknesses of about \SI{1}{\cm} would improve the focusing and, thus, the angular resolution. 
Hence, a multiple refractive index aerogel radiator allows an increase in Cherenkov photon yield due to the larger overall radiator thickness, while avoiding the simultaneous degradation in single photon angular resolution associated with the uncertainty of the emission point (geometric aberration)~\cite{IIJIMA2005383}. 

\begin{figure}
\centering
\includegraphics[width=6cm]{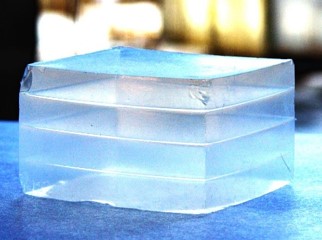}
\caption{Example of a 4-layer monolithic aerogel} 
\label{multi-aerogel}
\end{figure}

In an alternative layout, a mirror is placed outside the aerogel layer, which reflects the Cherenkov radiation onto a photo detector located upstream of the radiator, see Fig.~\ref{RICH-layouts}. 
This results both in an improvement of the angular resolution, by decreasing the geometric aberration, and in a reduction of the photosensitive area to be instrumented. 
If spherical, the mirror further focuses the photons onto a fraction of the surface, which further decreases the photosensitive area. 
The integration of a mirror in the design renders the mechanics and, in particular, the alignment more complex.
Considering the advantages in terms of performance and cost, this option is under detailed study.

While transparent in the visible range, aerogel becomes opaque for wavelengths in the UV~range, where most of the incident radiation is diffused by Rayleigh scattering.
Therefore, the photon sensors used to detect the Cherenkov radiation must be sensitive to visible light, which rules out gaseous photon detectors. 
Instead silicon-based photon sensors are an attractive and commercially available option.
They are based on Single Photon Avalanche Diodes (SPAD), reverse-biased diodes operated in Geiger mode.
Typical sizes of SPAD cells range from $\SI{10}{\um} \times \SI{10}{\um}$ to $\SI{100}{\um} \times \SI{100}{\um}$. 
The SPAD concept forms the basis of three classes of sensors, namely the analogue silicon photomultiplier (SiPM), the 2D digital silicon photomultiplier (2DdSiPM) and the 3D digital silicon photomultiplier. 
These devices have been developed mainly for photon detection, where they have demonstrated to be extremely effective in applications demanding very good timing performance~\cite{Bruschini:2019arv}.
SiPMs provide peak photon detection efficiency (PDE) up to 60\% in the visible range (Fig. ~\ref{SiPM-PDE}, ~\cite{Gundacker_2020}), insensitivity to magnetic fields, and integration in very compact structures with active area even larger than \SI{90}{\percent} ~\cite{RENSCHLER2018257}. 

\begin{figure}
\centering
\includegraphics[width=12cm]{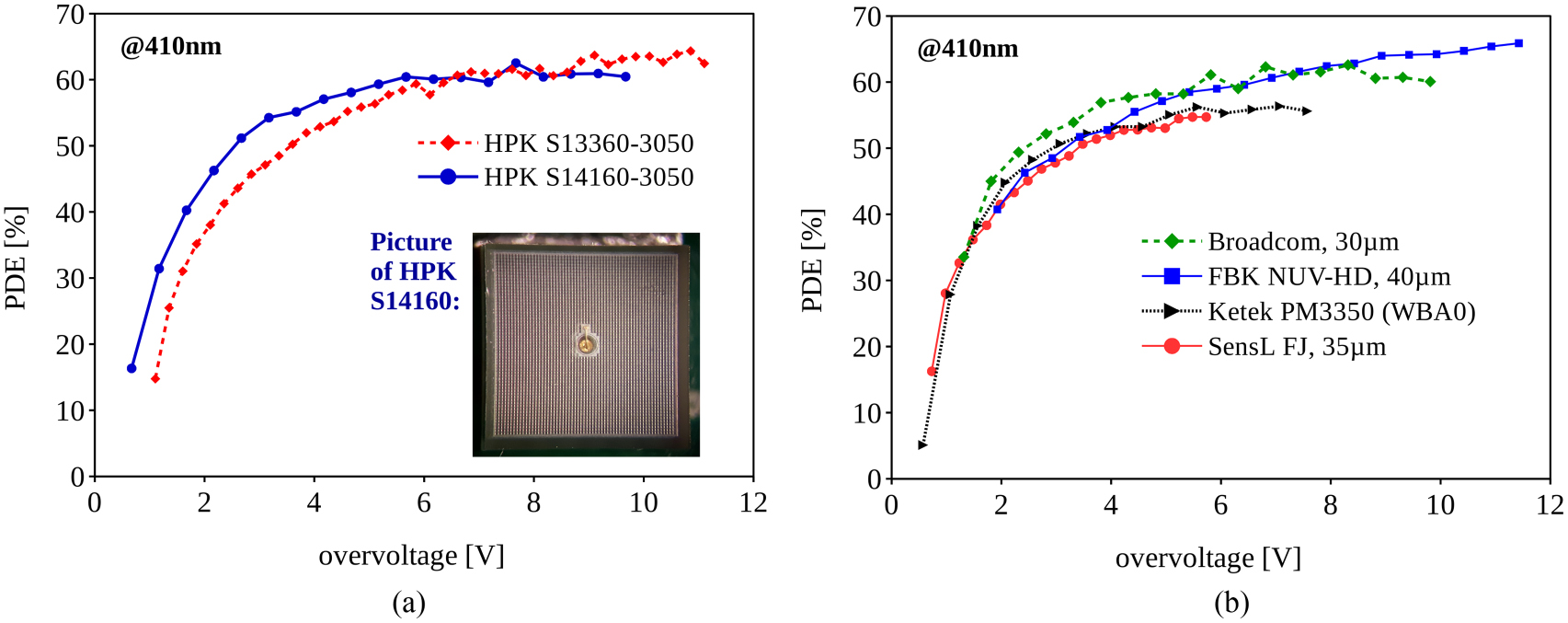}
\caption[Photon detection efficiency of SiPMs]{PDE as a function of SiPM overvoltage measured at 410 nm, for sensors from Hamamatsu (a) and other vendors (b).} 
\label{SiPM-PDE}
\end{figure}

Digital SiPMs with integrated circuitry realised in a standard CMOS process could provide the possibility to mask hot cells ~\cite{Mandai-Charbon:2012} and offer larger form factors.

Vacuum-based devices like MCP-PMTs offer a good response to visible light ($\sim 30\% $ peak quantum efficiency) and excellent time resolution ($\sim \SI{30}{\pico\second}$ FWHM). An interesting evolution of the MCP is represented by Large Area Picosecond Photon Detectors (LAPPD,~\cite{Lyashenko:2019tdj, Minot:2020hlb}), which has a similar working principle and can be produced in modules of 20x20~cm$^2$. 
This is clearly beneficial for large area applications, and is expected to reduce the cost by a factor 4 with respect to commercial MCPs. 
 Their performance is excellent in terms of dark count rate (DCR) and radiation tolerance, and interesting improvements have been obtained with small radius micro-pores ($\sim 10\;\mu$m) in terms of tolerance to magnetic field~\cite{Xie:2020yat}. 
However, for both detectors, the efficiency and gain loss due to magnetic field is still too large for the configuration of the barrel RICH where the photon detector plane is parallel to B field lines (worst situation). In addition, considerations on cost and integration of such rather bulky devices make the vacuum based PD not a primary option.

\subsubsection{Development status}
\label{sec:systems:rich:dev}

The use of silica aerogel for RICH detectors is well established. 
It is available with refractive indices in the range \numrange{1.006}{1.25} from the Budker Institute of Nuclear Physics in Novosibirsk and the Aerogel Factory Co. Ltd (Chiba, Japan) in Japan.
The latter uses a method that allows the production of hydrophobic aerogel.
As the overall angular resolution depends on the number of detectable photons, an important figure of merit is the transmittance.
The current manufacturing methods succeeded in improving the attenuation length $\Lambda$($\lambda$= 400 nm) from \SI{20}{\mm} (aerogel slabs used in HERMES) to \SI{50}{\mm} (aerogel slabs for CLAS12 and BELLE II).
The majority of Cherenkov photons are produced at short wavelengths ($\mathrm{d}N/\mathrm{d}\lambda \propto \lambda^{-2}$), which are absorbed in an aerogel radiator, and, thus, efficient detection of the photons in the visible range is crucial.

Commercially available SiPMs are a natural choice for the photon detection. 
As of today, they are available with pixel sizes up to $\SI{10}{mm}\times \SI{10}{mm}$, and arrays with up to 64 pixels are commercially offered by several vendors. 

SiPMs have not yet been employed in a RICH system, although they have been successfully tested with an aerogel radiator in studies for the RICH detector for Belle~II~\cite{KORPAR2014107} and for the ALICE VHMPID upgrade studies~\cite{BARNYAKOV2013352, Barnyakov:2014tra}. However, their effective and reliable use has been pioneered by the MAGIC and FACT experiments as photon sensors of their IACT cameras~\cite{FINK}. Combined with light-collecting Winston cones, more than 60,000 SiPM modules are currently installed in the near detector of the Tokai-to-Kamioka long-baseline neutrino experiment~\cite{NAKAYA2020}. Moreover, SiPMs are of great importance for the future Cherenkov Telescope Array (CTA) facility~\cite{AMBROSI201937}, and the DARKSIDE experiment at LNGS aiming to deploy 20~m$^2$ of SiPMs in liquid argon at 86~K for the direct search of WIMPs~\cite{DarkSide-20k:2017zyg}.
At present, SiPMs offer the most mature and established example of a SPAD-based sensor. A drawback is the large capacitance that they present to the front-end electronics. A SiPM with a size of $\SI{3}{\mm} \times \SI{3}{\mm}$ has a capacitance of about \SI{300}{\pico\farad}, which limits the achievable time resolution. SiPMs with an area of $\SI{1}{\mm} \times \SI{1}{\mm}$ and an integrated passive quenching have shown a timing jitter on photon as good as \SI{20}{\ps} FWHM~\cite{Acerbi2015}.
The cost effectiveness of modern CMOS technologies has driven the research towards digital SiPMs. In this approach, each SPAD cell is read out individually. Given the large signal and the small capacitance of a single SPAD, a simple CMOS inverter is sufficient to capture and digitize the signal. This allows a front-end stage with zero static power. Any further processing occurs in the digital domain. 
As already mentioned, one of the main advantages of this solution is the possibility to individually mask noisy SPADs~\cite{Mandai-Charbon:2012} and, thus, to increase the yield while maintaining acceptable dark count rates. In the 2D approach, the SPAD and the processing electronics are fabricated on the same CMOS wafer, which results in a fully monolithic solution analoguous to MAPS. 2DdSiPM have two major drawbacks. 
Firstly, the standard microelectronics-grade wafers are not optimal for SPAD design. 
Best performing 2D SPADs are produced in modified CMOS processes with optimized junction profiles and/or additional dedicated fabrication steps. 
This is particularly critical in order to keep the dark count rate to an acceptable level. 
The best SPADs reported so far in a standard CMOS technology achieve a time jitter of \SI{12.1}{\ps} FWHM and a dark count rate of \SI{2}{\mega\hertz\per\cm^2} at room temperature, which reduces to \SI{40}{\kilo\hertz\per\cm^2} at \SI{-65}{\celsius}~\cite{Gramuglia2021}. 
The second drawback is the reduction of the sensitive area required for the implementation of the front-end electronics, resulting in reduced fill factors. 
3DdSiPMs address these issues by fabricating the SPADs and the digital readout logic on two different wafers, which are then superimposed and interconnected using micro-bumps and/or silicon vias (TSV) ~\cite{NOLET2020162891} . 
Comprehensive reviews of SPAD-based sensors can be found in~\cite{KLANNER201936} (analogue SiPM),~\cite{Pratte20213DPC} (3DdSiPM) and in~\cite{Gundacker_2020} (all topologies).
SPADs are prone to radiation damage resulting in a significant increase of the dark count rate. It must however be considered that the radiation level expected in the RICH layer is moderate. Progress has also been made in handling in the front-end electronics the dark current which results after irradiation. 
The dependence of DCR on accumulated dose and the impact of cooling on DCR are well known and largely documented in literature~\cite{GARUTTI201969,MUSACCI2019695}. 
Cooling allows the reduction of the DCR (speciﬁcally, thermally generated noise), e.g. from \SI{e5}{Hz/mm\squared} at \SI{300}{K} to \SI{e1}{Hz/mm\squared} at \SI{200}{K}), and, therefore, the mitigation of radiation damage (see Figs.~\ref{DCR-T} and~\ref{DCR-irrad}). With the present state-of-the art devices, the implementation of cooling seems necessary to ensure the required performance throughout the planned experiment duration. In addition, a modular design will be implemented to allow accessing detector components for interventions. Considering that the performance of SiPMs damaged by irradiation can be recovered to a large extent by annealing, a speciﬁc assembly and dismantling procedure could be studied to perform thermal treatment of system components showing signs of radiation damage. The same procedure would clearly allow also replacing damaged sensors or modules.

\begin{figure}
\centering
\includegraphics[width=9cm]{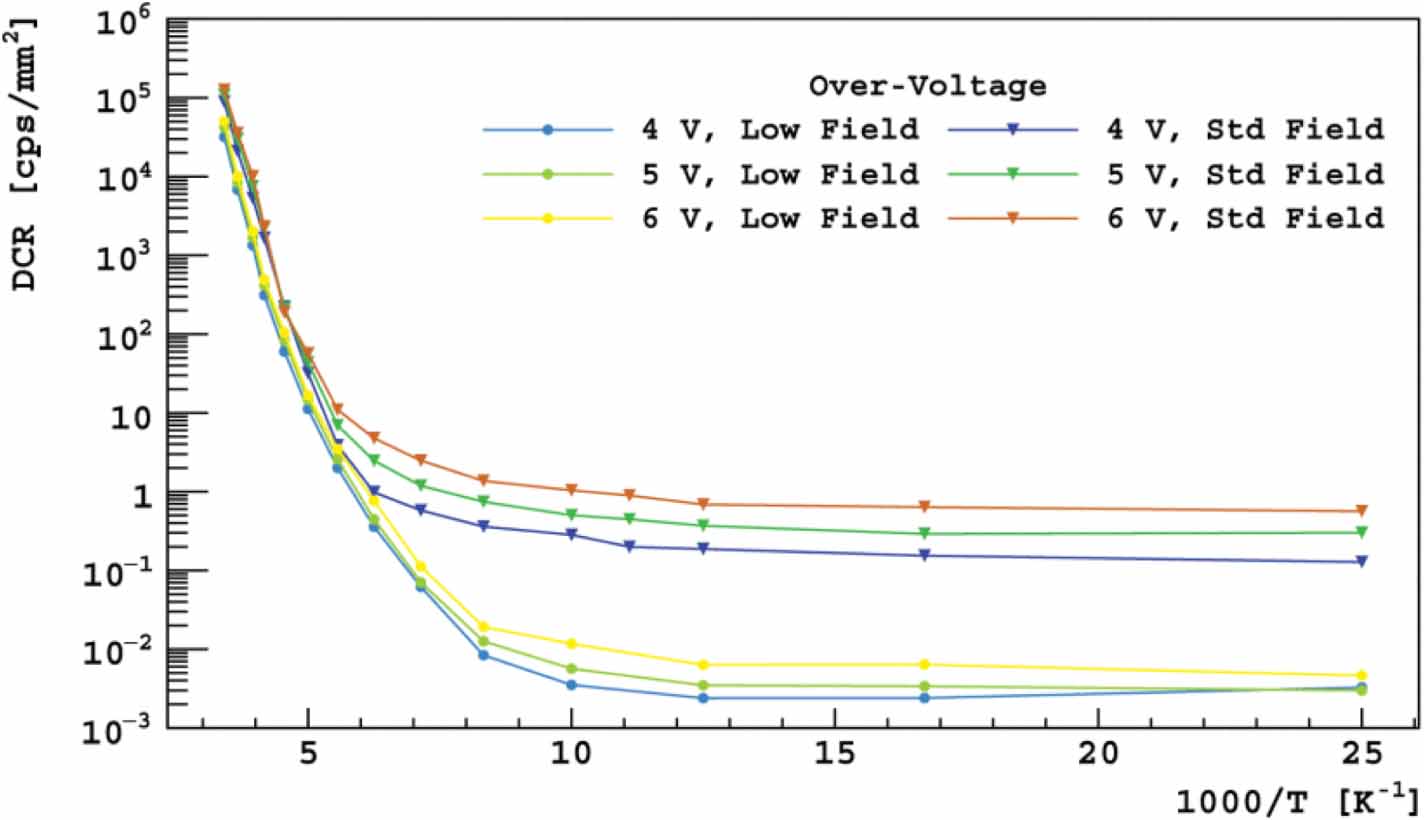}
\caption[Dark count rates of different photon sensors]{DCR as a function of 1/T for different sensors types and over-voltage~\cite{Gundacker_2020} }
\label{DCR-T}
\end{figure}

\begin{figure}
\centering
\includegraphics[width=9cm]{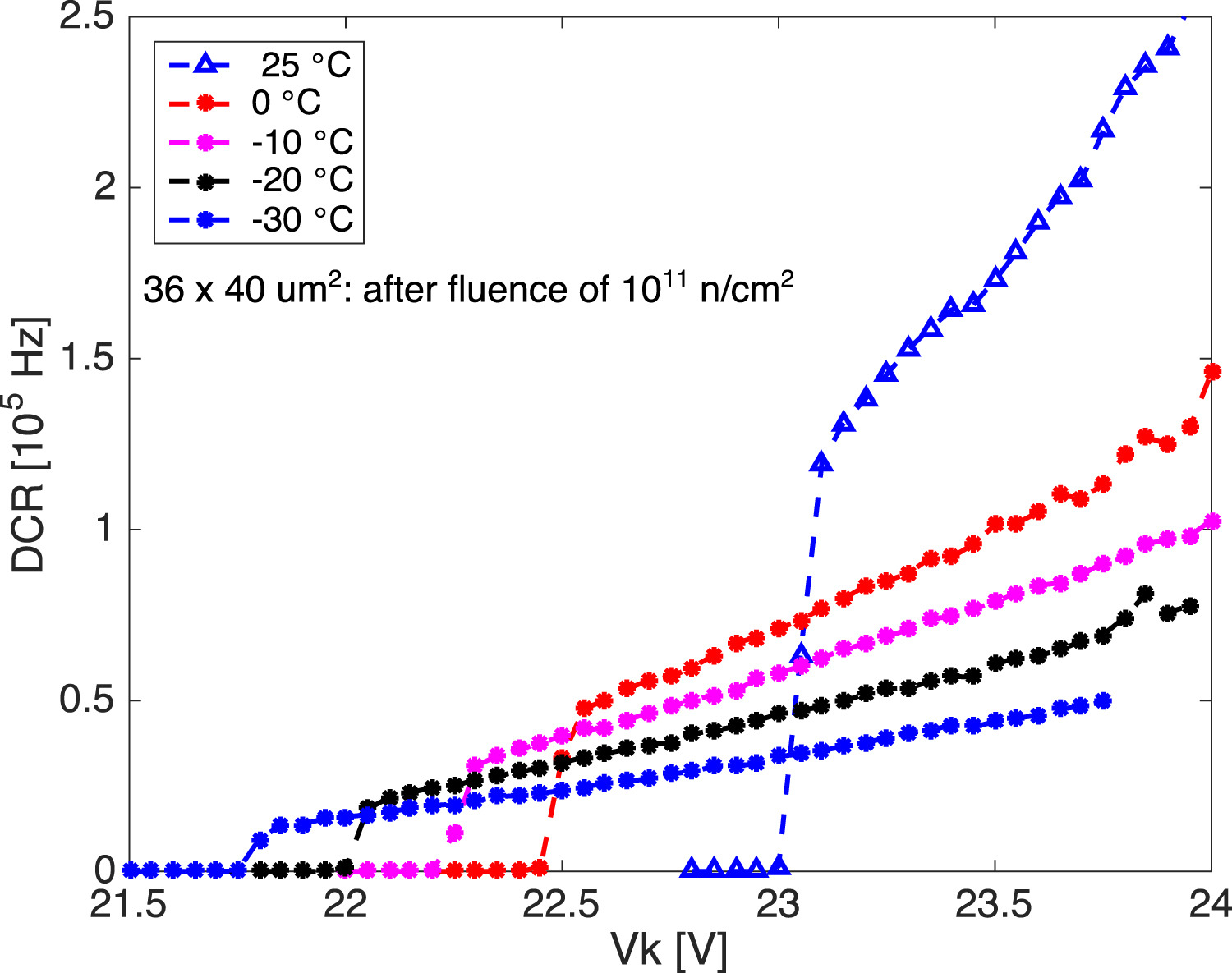}
\caption[Dependence of dark count rate on $V_\mathrm{bias}$ and $T$]{DCR as a function of biasing voltage at different cooling temperatures, of a sensor irradiated with a fluence of $10^{11} \nequiv$~\cite{MUSACCI2019695}} 
\label{DCR-irrad}
\end{figure}

\subsubsection{R\&D challenges and plans}
\label{sec:systems:rich:rnd}

While aerogel with the required properties is already available, some studies could be required on the reproducibility of the refractive index and the transparency.
Previous productions for RICH detectors like BELLE~II or HERMES are not comparable in terms of volume with the scale of the ALICE 3 RICH. It will be important to guarantee the required high quality of the aerogel optical parameters (refractive index, transparency) all along the production duration (estimated to last 20-24 months). Specific R\&D could also be devoted to the manufacturing of larger tiles (150 mm $\times$ 150 mm) and monolithic multi-index aerogel for focusing.
A collaboration on this has already started with the Aerogel Factory Co. Ltd (Chiba, Japan) and with LHCb.

The R\&D on SiPM is mainly related to the development of digital SiPMs (or SPADs)
which is motivated by the following arguments: 
\begin{itemize}
    \item The rapid evolution of SPADs (the fundamental element
of SiPM) based on CMOS imaging sensor (CIS) technology in relation to the increasing demand for
consumer and automotive applications; the sensor cost in standard CIS process is at least a factor 10
smaller than commercial analogue SiPMs.
\item The possibility to integrate circuitry for time gating logic and DCR 
reduction by masking ”hot” SPADs.
\item The “monolithic” approach and the implementation of
electronics inside the sensor simplifies the integration and minimizes the material budget, with clear
advantages also on construction costs.
\item Possibility to develop a fully customized sensor and optimize the
key performance parameters (PDE, DCR and correlated noise, radiation tolerance). The availability of
backside illumination in the so-called 3D stacking increases the chances of maximizing the ﬁll factor of
the 1st Tier and integrate the required electronics in the 2nd Tier.
\end{itemize}

To improve the radiation tolerance of SiPMs by a factor 10 dedicated efforts are needed as this aspect will likely not be of relevance for commercial applications.
Since this is a general issue for the application of SiPM in HEP, we believe synergies with other experiments, for instance, LHCb and EIC will naturally develop.

Using devices incorporating several SiPMs would be instrumental in easing the assembly of the photo-detector panels required and improve the fill factor, reducing the dead area between devices and simplifying connectivity.

An appealing possibility would be a sensor to detect both MIPs and single photons with high efficiency and good timing capability, such that it can be used both for Cherenkov detection and Time of Flight (TOF) applications.

In spite of the tremendous potential offered by the development of digital sensor in CMOS, likely the design of one or more dedicated ASICs may be needed in order to control the sensors, to manage the large amount of data and to interface the detector with the experiment DAQ and DCS systems.          
The ASIC will possibly be located off-detector, in a less harsh radiation environment, where space and power dissipation are less a concern.

\subsubsection{Technical implementation}
\label{sec:systems:rich:implementation}

Given the size of the barrel RICH system, a segmentation in modules is considered necessary for installation and maintenance.
Even though the selected aerogel is hydrophobic, operation under an inert gas flow is mandatory to preserve its properties and to control the composition of the atmosphere. 
If a SiPM-based photon detector is cooled down to temperatures well below \SI{0}{\celsius}, the relative humidity has to be maintained at few ppm to prevent condensation. 
Therefore, the system needs to be enclosed in a gas-tight vessel.

In order to minimize the acceptance losses, the simplest approach is to consider a cylindrical tank made of very light carbon fibre walls, containing the radii from \SI{\sim 0.9}{m} to \SI{\sim 1.10}{m}.
Inside the vessel, the mechanical support structure will be segmented in 20~sectors of about \SI{30}{cm} in width, corresponding to two adjacent aerogel tiles. 
A system based on rails and sliding trays could allow the independent insertion/extraction of the aerogel layer and the SiPM layer at the inner and outer shell of the vessel, respectively.

If the layout with a mirror is adopted, the layer of SiPMs and aerogel will rest on the inner shell, while the mirror layer will be supported by the outer shell. 
In addition, the large cylindrical tank could be replaced by smaller gas tight vessels enclosing the aerogel and SiPMs of each sector separately.

Depending on the development of the SiPMs to be used for the photon detector, the sensors will be assembled on support structures implementing the necessary interfaces with front-end electronics, for analogue SiPMs, or directly with the readout system, for digital SiPMs. 
The performance in terms of dark count rate and radiation hardness will define the requirements on the cooling system, which could be embedded in the SiPM support acting as cold plate.

The location at the outermost layer of the barrel makes rather straightforward the routing of services with minimal conflict with other detectors and related support structures.

For the instrumentation of the forward regions, an analogue design could be implemented in forward disks.

\ifcost
\subsubsection{Cost estimates}
\label{sec:systems:rich:cost}

A preliminary cost estimate has been made for the three layouts illustrated in Fig.\ref{RICH-layouts}; in addition, two solutions have been considered for the photon detector: a) commercially available analogue SiPM; b) development of digital SiPM (cost estimate based on present production of 8" wafers in 180 nm CIS technology). Table~\ref{tab:RICH_cost} shows a summary of the cost estimate for each subsystem of the various solutions. For the option with mirror, the reduction of the photon detector subsystem cost is related to the smaller area to be covered, since the SiPM layer will rest on the inner shell. Also, as explained in Sect.~\ref{sec:systems:rich:implementation}, a simpler and smaller gas tight vessel can be designed to enclose the aerogel and SiPM layer, thus allowing some savings on the mechanics.  

\begin{table}
\centering
    \begin{tabular}{@{}l *{3}{S[table-format=2.1]}@{}}
        \toprule
        \textbf{Component} & \multicolumn{3}{c}{\textbf{Cost (MCHF)}} \\
        \cmidrule{2-4}
        & \textbf{basic} & \textbf{focusing} & \textbf{mirror} \\
        & \textbf{layout} & \textbf{aerogel} & \textbf{layout} \\
        \midrule
        mechanics & 0.8 & 0.8 & 0.6\\
        aerogel & 1.2 & 1.8 & 1.8 \\
        analogue SiPM+FEE & 12.0 & 12.0 & 9.5 \\
        digital SiPM & 7.0 & 7.0 & 5.7 \\
        power supply & 1.0 & 1.0 & 0.8  \\
        CO2 cooling system &  1.0 &  1.0 &  0.8\\
        gas system & 0.4 & 0.4 & 0.4\\
        mirror system & & & 1.2 \\
        \midrule
        \textbf{Total - analogue SiPM} &     &  & \\
        \textbf{Total - digital SiPM} &     &  &  \\
        \bottomrule
    \end{tabular}
    \caption{Summary of main subsystem costs of the barrel RICH detector.}
    \label{tab:RICH_cost}
\end{table}

\fi

\subsection{Electromagnetic calorimeter}
\label{sec:systems:ecal}

\begin{revised}
The Electromagnetic Calorimeter (ECal) is planned to cover the full central barrel region and one forward region, i.e. an rapidity range of $-1.6 <\eta < 4$. 
Most of the rapidity range will be instrumented with a sampling calorimeter.
A fraction of the central barrel will be covered by the existing PbWO$_4$ crystals for the measurement of $\chi_\mathrm{c}$ and soft direct photons. 
These probes do not require back-to-back acceptance and the detector layout (narrow ring or wider half-ring) can be optimised based on practical aspects of the integration. 
\end{revised}

\subsubsection{Specifications}
\label{sec:systems:ecal:specs}

\begin{figure}[ht]
    \centering
    \includegraphics[width=0.49\hsize]{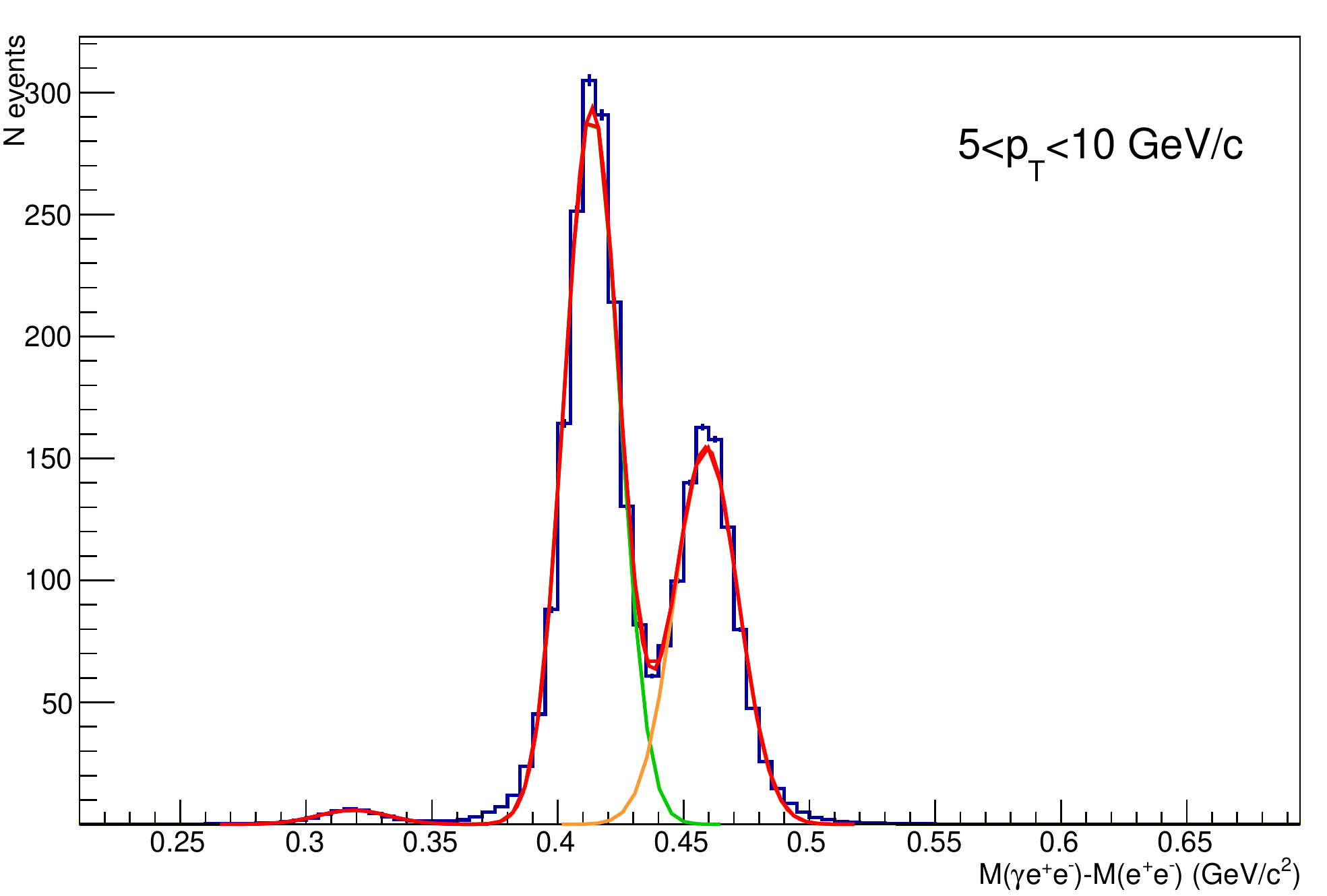}
    \hfil
    \includegraphics[width=0.49\hsize]{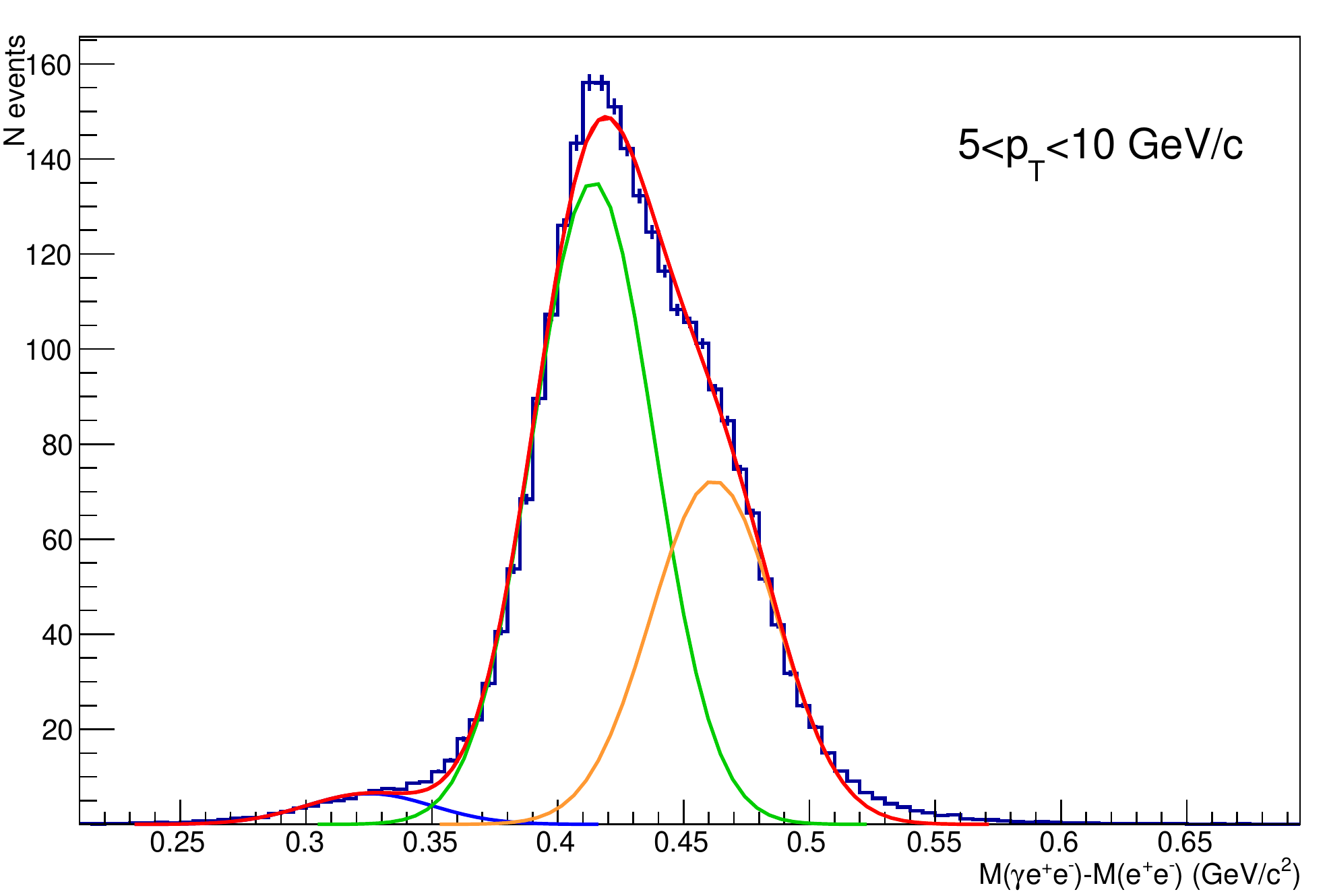}
    \caption[Invariant mass spectrum of $\chi_{cJ}$ states with ECal]{Invariant mass difference spectra of decay $\chi_{cJ}\to J/\psi \gamma$ with a photon detected in the ECal at midrapidity assuming different stochastic term of the photon energy resolution: $b=0.02~\mbox{GeV}^{1/2}$ (left) and $b=0.05~\mbox{GeV}^{1/2}$ (right).}
    \label{fig:ECAL_chic_masspeaks}
\end{figure}

The dependence of the calorimeter energy resolution on the photon energy is parametrised as
\begin{equation}
    \frac{\sigma_E}{E} = \frac{a}{E} \oplus \frac{b}{\sqrt{E}} \oplus c.
    \label{eq:v2:eresparam}
\end{equation}

At low energy it is dominated by the noise term $a/E$ related to the readout electronics and the stochastic term $b/\sqrt{E}$, which is related to the shower fluctuations in the calorimeter cells. 
A set of observables, such as direct photon production at low \pt, measurements of radiative decays of strange hyperons (e.g. $\Sigma^0\to\Lambda + \gamma$) and quarkonium states (e.g. $\chi_{cJ} \to J/\psi + \gamma$), require the detection of relatively soft photons in the energy range from $\sim10$~MeV to $\sim1$~GeV, see Fig.~\ref{fig:ECAL_chic_masspeaks}.
This calls for a stochastic term $b \sim 0.02$~GeV$^{1/2}$, with the other terms being subdominant, and a cell size of $\sim \SI{2}{\cm} \times \SI{2}{\cm}$.
For other measurements, the large acceptance is key and a moderate energy resolution $b\sim 0.07 \div 0.11$~GeV$^{1/2}$ is sufficient.
Cell sizes of about $\sim \SI{3}{\cm} \times \SI{3}{\cm}$ are required to limit shower overlaps.

The central barrel ECal is located between the RICH detector and the magnet cryostat with an inner and outer radius of \SI{1.15}{\m} and \SI{1.45}{\m}. 
The length of \SI{7}{\m} covers the pseudorapidity range $|\eta| < 1.6$. 
The endcap is realised by a disk of inner radius of \SI{0.16}{\m} and outer radius \SI{1.8}{\m}. 
It is installed downstream of the forward RICH at $z_\mathrm{ECal}=4.35$~m and covers the pseudorapidity range $1.6 < \eta < 4$.

The ECAL sector providing the precision energy measurement will cover a region of $\Delta\varphi=2\pi$ in azimuthal angle and $|\eta|<0.33$ in pseudorapidity, which corresponds to a length of $|z|< \SI{0.64}{\m}$. 

\subsubsection{Technology options}

The precision energy measurement will be provided by the existing PbWO$_4$ crystals with SiPM readout. The remaining part of the acceptance will be covered with a Pb-scintillator sampling calorimeter.

\subsubsection{Development status}
\label{sec:systems:ecal:dev}

Electromagnetic calorimeters of sampling type are widely used in high-energy experiments. 
Detection elements or cells of such calorimeters are constructed as a stack of alternating layers of a passive absorber, made of elements with high nucleus charge $Z$, and active layers of plastic scintillators. 
Scintillation light is collected from the active layers and propagated towards the photodetector via wave-length shifting fibers. 
The construction of sampling calorimeters is flexible, well developed and can be reproduced in many facilities or workshops~\cite{Li:2020uoq}. 
Optimising the sampling ratio, an energy resolution of $\sigma_E/E = 0.03/\sqrt{E}$ can be reached~\cite{Kharlov:2008tw}. 
However, the construction always requires a compromise between energy resolution and Moli\`ere radius $R_M$. 
The best energy resolution can be achieved in fine-sampling calorimeters, in which the Moli\`ere radius is large.
Compact electromagnetic showers can be achieved by a smaller sampling ratio, resulting in an inferior energy resolution. 
The ALICE EMCal is an example of an calorimeter with compact transverse shower profiles ($R_M=\SI{3}{\cm}$). 
It shows a stochastic term of $\sigma_E/E = 0.11/\sqrt{E}$~\cite{Kharlov:2018ljg}, which fulfills the needs of ALICE~3 for physics tasks involving hard photons.

Experience on the usage of PbWO$_4$ crystals for high-resolution calorimetry exists from the ALICE PHOS and the CMS ECal. 
A challenge arises from the relatively low light yield.
The light yield ($Y$) of PbWO$_4$ features a strong temperature dependence.
For this reason, the ALICE PHOS is operated at a temperature of $t = \SI{-25}{\celsius}$ to increase the light yield by a factor of 3 compared to operation at room temperature. 
Apart from the light yield, the stochastic term of the energy resolution also depends on the light collection efficiency. 
In the ALICE PHOS, light is detected by avalanche photodiodes with an area of $5 \times 5~\mathrm{mm}^2$, which cover only 5\% of the crystal cross section ($22 \times 22~\mathrm{mm}^2$). 
This leaves ample room for improving the calorimeter performance by increasing the light collection, e.g. by using larger photodetectors. 
Recent developments of semiconducting photodetectors, in particular silicon photomultipliers (SiPM), resulted in the availability of photodetectors at affordable prices.
A candidate device is the SiPM S14160-6015PS by Hamamatsu~\cite{Hamamatsu-S14160} featuring a total area of $6 \times 6~\mbox{mm}^2$ with 160,000 pixels of $15 \times 15\si{~\um^2}$. The choice of this photodetector is driven by the pixel size which should be consistent with the scintillation photon flux in a wide dynamic range of the calorimeter, and the sensitive area to ensure high photon collection efficiency.
Extrapolating from the observed PHOS energy resolution $\sigma_E/E = 0.033/\sqrt{E}$~\cite{Kharlov:2018ljg}, the specification of $\sigma_E/E = 0.02/\sqrt{E}$ can be achieved by installing 2~such SiPMs and cold operation or 6 SiPMs and operation at room temperature. The latter is favoured for the reduced complexity and the need to install the calorimeter in a limited space between the PID detectors and the cryostat. While the stochastic term $b$ of the energy resolution can be reduced by a larger light collection efficiency, improving the noise term $a$ can be pursued by deploying advanced low-noise photodetectors and front-end electronics. In the ALICE PHOS the noise term is $a=13$~MeV~\cite{ALICEPHOScalorimeter:2005lby} which is driven by the noise excess factor of the avalanche photodiode used in PHOS as a photodetector, by the induced noise of the FEE circuit and by the digization of the ADC. Recent progress in semiconductor photodetectors, such as silicon photomultipliers, also known as MPPC, multipixel photon counter, allows for a further noise suppression with respect to the ALICE PHOS performance. In order to satisfy physics requirements for soft photon detection in ALICE~3, the noise term $a \sim \SIrange{2}{5}{\mega\eV}$ is a feasible compromise between technological challenges and needed performance.

The readout electronics of the ECal should provide high energy resolution in a wide energy range and good timing resolution comparable with the intrinsic timing response of detection elements based on inorganic and organic scintillators. The most advanced energy resolution $\sigma_E/E$ is required for the precision central barrel where a target noise term of the energy resolution can be achieved with the ADC digitization precision of 1~MeV, and an energy range from 10~MeV to 100~GeV would assume the front-end electronics with two gains with a ratio of 32. Photon identification determines requirements for timing resolution $\sigma_t$ of the order of few hundred picoseconds in the sub-GeV energy domain. Recent R\&D shows that $\sigma_t \sim 200$~ps can be achieved for photons of energy $E=0.5$~GeV. Front-end electronics will be unified for all 3 sections of the ECal.

\subsubsection{Technical implementation}
\label{sec:systems:ecal:implementation}

The geometry and other parameters of the ECal are summarized in the Table~\ref{tab:v2:ecalspec}.

\begin{table}[ht]
    \centering
    \renewcommand{\arraystretch}{1.3}
    \begin{tabular}{lp{3.3cm}p{3.3cm}p{3.3cm}}
    \toprule
    ECal module & Barrel sampling & Endcap sampling &  Barrel high-precision\\
    \midrule
    \multirow{2}*{acceptance} 
    & $\Delta\varphi=2\pi$, \newline $|\eta|<1.5$ 
    & $\Delta\varphi=2\pi$, \newline $1.5<\eta<4$ 
    & $\Delta\varphi=2\pi$, \newline $|\eta|<0.33$
    \\
    \multirow{2}*{geometry}
    & $R_{\rm in} = 1.15$~m, \newline $|z|<2.7$~m 
    & $0.16<R<1.8$~m, \newline $z=4.35$~m 
    & $R_{\rm in} = 1.15$~m, \newline $|z|<0.64$~m 
    \\ 
    technology 
    & sampling Pb + scint. 
    & sampling Pb + scint. 
    & PbWO$_4$ crystals
    \\
    cell size 
    & $30 \times 30~\mbox{mm}^2$ 
    & $40 \times 40~\mbox{mm}^2$ 
    & $22 \times 22~\mbox{mm}^2$ 
    \\
    no. of channels 
    & 30\,000
    & 6\,000
    & 20\,000 
    \\
    energy range 
    & $0.1 < E < 100$~GeV 
    & $0.1 < E < 250$~GeV
    & $0.01 < E < 100$~GeV
    \\ 
    \bottomrule
    \end{tabular}
    \caption{ECAL parameters.}
    \label{tab:v2:ecalspec}
\end{table}

As a baseline, the sampling barrel calorimeter is built as a stack of alternating layers of lead and plastic scintillator of thickness 1.44~mm and 1.76~mm, respectively. 
A stack of 76 layers provides 20 radiation lengths ($l = 246$~mm) and a stochastic term of $b = \SI{0.11}{~\giga\eV^{1/2}}$. Further sampling optimization is possible at the stage of implementation, if improved energy resolution is feasible within the same geometry envelope.
The \num{30000} required cells will be constructed with a cross section of $30 \times 30~\mbox{mm}^2$ with a tapered shape to achieve projective geometry. 

The cells for the endcap are constructed from 95~layers of \SI{1.44}{\mm} lead and \SI{1.76}{\mm} scintillator sheets, which provides $25X_0$ at a thickness of \SI{304}{\mm} and a stochastic term $b = \SI{0.11}{~\giga\eV^{1/2}}$.
The transverse cell size is assumed to be $40 \times 40~\mbox{\si{\mm}}^2$, resulting in a total of \num{6000} cells. 

The high-resolution segment relies on lead tungstate crystals $\mbox{PbWO}_4$ with a cross section of $22 \times 22~\mbox{mm}^2$ and of length $l=180$~mm, corresponding to 20~radiation lengths. 
The number of cells of this precision segment will be \num{\sim 20000} forming a cylindrical matrix of 320~cells in azimuth and 62~cells in longitudinal direction. 

\ifcost
\subsubsection{Cost estimates}
\label{sec:systems:ecal:cost}
Cost estimation of various components of ECal, including new production or refurbishing of old materials from previous LHC experiments are summarized in Table~\ref{tab:systems:ecal:cost}.
\begin{table}
\centering
\renewcommand{\arraystretch}{1.3}
\begin{tabular}{llS[table-format=2.1]}
    \toprule
    Component & Comment & {Cost (MCHF)}  \\
    \midrule
    PbWO$_4$ crystals & reprocessing & 0.2 \\
    Pb-Sci sampling modules (barrel) & new production & 1.0 \\
    Pb-Sci sampling modules (forward) & new production & 0.4 \\
    Photodetectors & Hamamatsu MPPC & 1.2 \\
    FEE and readout & new production & 3.9 \\
    low-voltage power supplies & Wiener LVPS & 0.2 \\
    high-voltage power supplies & Iseg HVPS & 0.1 \\
    mechanics & ECal modules and support frame & 0.4 \\
    cooling & thermostabilisation system & 0.2 \\
    services & monitoring system, env.sensors, cables & 0.1 \\
    \midrule
    Total & & 7.7 \\
    \bottomrule 
\end{tabular}
\caption{ECal cost estimates}
\label{tab:systems:ecal:cost}
\end{table}
\fi

\subsection{Muon identifier}
\label{sec:systems:muon}

The goal of the muon identifier is the reconstruction of quarkonia, in particular charmonia, down to $\pt = 0$ in the muon channel, which complements the electron identification capabilities.

The Muon Identifier will provide muon tagging for particles reconstructed in the tracker. 
Installed outside of the magnet system, it is composed of chambers to track  charged particles that pass through the hadron absorber of about \SI{1}{kt}.
\begin{revised}
Whether the latter will be constructed from iron or non-magnetic steel will be decided based on practical considerations, e.g. mechanical stress on support structures. 
The physics performance, i.e. the matching of muons, can be achieved either way.\end{revised}

Ensuring good efficiency for charmonia down to $\pt = \SI{0}{\giga\eVc}$ over a large rapidity interval results in the requirement for reconstructing muons down to momenta of $p \sim \SI{1.5}{\giga\eVc}$ at $\eta \approx 0$.  
Simulations of the hadron background in the 0-10\% most central \PbPb collisions at $\sqrt{\sNN} = 5.5$~TeV were performed. 
An absorber thickness of 70\,cm at $\eta=0$ together with muon chambers with a granularity of $\Delta \varphi \Delta \eta = 0.02 \times 0.02$ was found to be sufficient for efficient muon identification.

\label{sec:systems:muon:options}

The moderate charged particle rate of $3$ Hz/cm$^2$ \comment{Insert value} in the muon system together with the required granularity of $50-60$\,mm pad size allows the use of Resistive Plate Chambers (RPCs) for the muon system, which represent a cost effective solution for the $\sim 40\,000$ channels.
However, the use of scintillator bars equipped with wave-length shifting fibres, which are read out with SiPMs, appears even more attractive and is considered the baseline option. We foresee two layers of crossed scintillator bars (\SI{5}{\cm} wide) with a \SI{20}{\cm} gap.

\label{sec:systems:muon:dev}

\label{sec:systems:muon:implementation}

\label{sec:systems:muon:cost}

\subsection{Forward conversion tracker}
\label{sec:systems:fct}

The aim of the Forward Conversion Tracker (FCT) is to measure photons down to transverse momenta of $\sim \SI{2}{\mega\eVc}$, which are only accessible by exploiting the Lorentz boost in the forward direction resulting in photon energies down to $E_\gamma \sim 50~\mathrm{MeV}$ at $\eta = 4$.
The photons are then measured through their conversion to \epm pairs, which requires the reconstruction of electrons with momenta down to a few \si{\mega\eVc}.
A magnetic field perpendicular to the (forward) direction of flight is required for sufficient \pt resolution in the \epm measurement.
The performance of the FCT has been studied at a position $z \approx \SI{3.4}{\metre}$ with a dipole field component $B_y \approx \SI{0.3}{\tesla}$. The exact position will be optimised taking into account the final design of the magnet system.

\subsubsection{Specifications}
\label{sec:systems:fct:specs}

In a magnetic field of \SI{0.3}{\tesla}, electron momenta of a few MeV/$c$ result in bending radii of around \SI{10}{\cm}.
They can be measured with an array of tracking layers (silicon disks) with a spacing between layers of about 2~cm. 
To achieve good resolution also for electrons with larger momenta, further disks are foreseen with a larger spacing.
This leads to the concept of 9~silicon tracker disks installed around the beampipe in the forward direction to cover the pseudorapidity range $3 < \eta < 5$.
The proposed layout is shown in Fig.~\ref{fig:systems:fct:layout}. 

As the measurement of low-\pt photons is not limited by statistics, the silicon disk of the ALICE~3 tracker directly in front of the FCT can serve as an active photon converter, which also provides tracking information at the conversion point.
Thus, a dedicated converter is not strictly necessary and, in fact, would degrade the energy and pointing resolution through multiple scattering and energy loss.
Such a converter could be considered to increase the photon conversion probability if required for the study of rare channels.

For radiation levels and particle rates, we refer to Sec.~\ref{sec:systems:tracking}. 

\begin{figure}
    \centering
    \includegraphics[width=\linewidth]{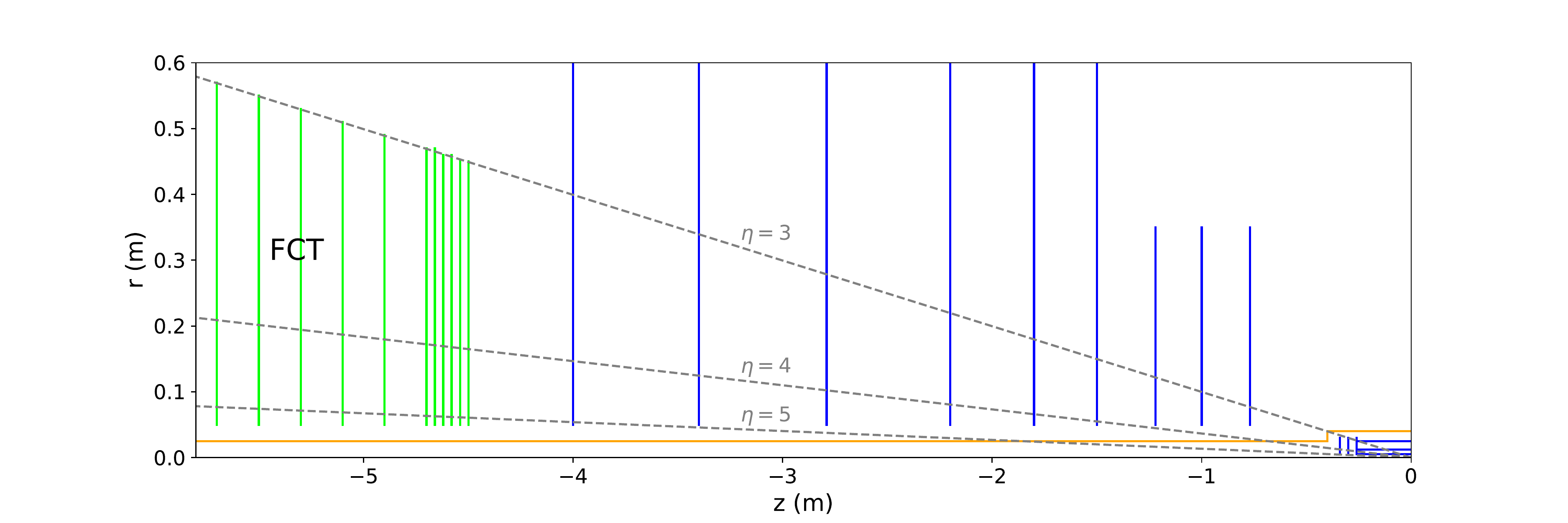}
    \caption[Layout of the FCT]{Layout of the Forward Conversion tracker (FCT). The first silicon disk of the FCT (in green) is located at $z \approx -4.5\,\mathrm{m}$. The FCT consists of silicon disks similar to those used for the ALICE 3 tracker (shown in blue). In the default design no special converter in front of the FCT is foreseen in order to obtain an optimal momentum resolution. Instead, the tracker disk in front of the FCT and/or the first FCT layer act as an active converter.}
    \label{fig:systems:fct:layout}
\end{figure}

\begin{table}
  \centering
  \begin{tabular}{l *{3}{S[table-format=+1.2]}} 
    \toprule
    Layer
    & {$z$ (m)}
    & {$r_\mathrm{min}$ (m)}
    & {$r_\mathrm{max}$ (m)} \\
    \midrule
    0 & -4.50 & 0.05 & 0.45 \\
    1 & -4.54 & 0.05 & 0.45 \\
    2 & -4.58 & 0.05 & 0.46 \\
    3 & -4.62 & 0.05 & 0.46 \\
    4 & -4.66 & 0.05 & 0.47 \\
    5 & -4.70 & 0.05 & 0.47 \\
    6 & -4.90 & 0.05 & 0.49 \\
    7 & -5.10 & 0.05 & 0.51 \\
    8 & -5.30 & 0.05 & 0.53 \\
    9 & -5.50 & 0.05 & 0.55 \\
    10 & -5.70 & 0.05 & 0.57 \\
    \bottomrule
  \end{tabular} 
  \caption[Specifications FCT]{\label{tab:systems:fct}Specifications of the silicon disks of the Forward Conversion Tracker (FCT).}
\end{table}

\subsubsection{Technology options}
\label{sec:systems:fct:options}

The tracking disks can be based on the same silicon pixel sensors used for the ALICE~3 tracker (see Sec.~\ref{sec:systems:tracking:specs}). 
Assuming the geometry listed in Tab.~\ref{tab:systems:fct}, the FCT requires instrumenting an area of $8.3~\mathrm{m^2}$.
To minimise the material and the impact of multiple scattering, the layers should be as thin as possible, ideally a few \si{\permille} per layer.
The requirements on hit measurements are met by the pixel sensors used for the ALICE~3 tracker.
Thus, they are considered the baseline option for tracking stations of the FCT.
The development status of pixel sensors is reviewed in detail in Sec.~\ref{sec:systems:tracking}.

\subsubsection{R\&D challenges and plans}
\label{sec:systems:fct:rnd}

\begin{revised}
The essential requirement is to limit the amount of material in front of the FCT to less than \SI{10}{\percent} of a radiation length.
If more material is used, the background from bremsstrahlung processes becomes prohibitively large, see Sec.~\ref{sec:performance:detector:photon_id}.
Therefore, optimising the design of the beam pipe and the vertex detector is of the utmost importance, in order to avoid shallow crossing angles. 
The obstruction of the FCT by the vertex detector could be reduced by selecting collisions with a large distance from the nominal interaction point, i.e. only analysing collisions from the tail of the Gaussian $z$ vertex distribution. 
For a collision at the nominal interaction vertex ($z = 0$), only particles with a pseudorapidity larger than $\eta=5$ do not cross the containment vessel of the vertex detector. 
Here, we assume that the $z$-vertex distribution is a Gaussian with a standard deviation of \SI{5}{\cm}. 
At the expense of reducing the available statistics by about a factor 50, the FCT analyses could be limited to events with a $z$-vertex shifted by more than \SI{10}{\cm} in the direction towards the FCT. 
For these events, particles from the primary vertex do not cross the containment vessel already for pseudorapidities larger than $\eta = \text{arcsinh} \frac{z}{r} = 4.6$. Here, $r = \SI{0.5}{\cm}$ is the radial distance of the containment vessel to the beam axis and $z = \SI{25}{\cm}$ is the distance of the edge of the containment vessel to the shifted vertex. The evaluation of the importance of this improvement for the planned soft photon measurements requires more detailed simulations.
The background from external bremsstrahlung can also be rejected at the analysis level by requiring the absence of an electron track in the vicinity of the reconstructed photon, see Sec.~\ref{sec:performance:physics:soft_photons}.
\end{revised}

The mechanical realisation of large tracking stations with material thicknesses on the order of a few \si{\permille} of a radiation length per layer poses a major challenge, and requires dedicated R\&D.
In contrast to the vertex detector (and the ITS3), for which stability arises intrinsically from the bent sensors, dedicated studies will be needed on how to achieve stable flat geometries. 
The possibility to deviate from flat geometries to increase mechanical stability, could be considered. 
It should further be noted that the geometry currently being considered requires covering areas which exceed those of individual wafers and, thus, requires interconnections of several sensors.

\ifcost
\subsubsection{Cost estimates}
\label{sec:systems:fct:cost}

A rough cost estimate for the conversion tracker is presented in Tab.~\ref{tab:systems:fct:cost}.

\begin{table}
    \centering
    \renewcommand{\arraystretch}{1.3}
    \begin{tabular}{llS[table-format=2.1]}
        \toprule
        Component & Comment & {Cost (MCHF)}  \\
        \midrule
        Sensors & wafer-scale sensors? & 0.5\\
        Mechanics & ??? & 0.5\\
        Readout & ??? & 0.5\\
        Power \\
        Cooling \\
        Services \\
        \midrule
        Total & & 1.5 \\
        \bottomrule 
    \end{tabular}
    \caption{FCT cost estimates}
    \label{tab:systems:fct:cost}
\end{table}
\fi

\subsection{Data acquisition, processing, and computing}
\label{sec:systems:data}

The aggregated data rate for ALICE 3 at the input of the data acquisition farm is estimated to about 20 Tbps.
\begin{revised}
For the detector links, we will follow the developments and evaluate available options.
\end{revised}
Given the expected evolution in detector readout links, PCIe and network cards bandwidth, a standard ALICE 3 readout node based on commodity hardware will be able to handle comfortably a data rate of 100 Gbps. 
Therefore, a farm of equivalent size to the current O2/FLP farm located in CR1 will be sufficient to perform the data acquisition of ALICE 3, with potential for further optimizations.

Analyis Object Data (AOD) usable for physics analyses must be reconstructed from the raw data through clusterization, tracking, and matching of the PID information. 
To estimate the computing needs, a two-stage processing scheme similar to Run~3 is assumed~\cite{o2tdr}.
An online processing phase during data taking provides data compression and calibration, while the final reconstruction is obtained in a second phase, which will also be run on the online computing farm.
While all data from ion collisions will be stored, an analysis and event selection process will reduce the data volumes from \pp collisions.
Separate computing resources are assumed for simulation and analysis jobs.

\subsubsection{Online processing and computing architecture}
\label{sec:systems:data:online}

The baseline for the online processing is a heterogeneous computing farm like the ALICE EPN farm built for Run~3, in which Graphics Processing Units (GPU) are used as accelerators and provide the majority of the compute power.
The performance observed in the Run~3 O$^2$~\cite{o2tdr} computing scheme demonstrates that this is a feasible and cost-effective solution.
A trend towards integrated computing, which would avoid separate memories between CPU and GPU and the resulting data transfers, would be beneficial, but already a scheme with standard PCI Express interconnect has sufficient bandwidth as shown in the following.
By the time of Run 5, a speedup factor of four compared to the fastest GPU available today is assumed \comment{check what other exp. assume}, which is a conservative estimate considering the GPU developments of the last decade.
Since the core of the computing happens in the GPUs, an increase in the number of cores per CPU will be less relevant.

The extrapolation of the required compute power is based on the GPU-based ITS tracking for Run~3~\cite{run3itsgpu}, scaled linearly with the number of hits.
With the cellular automaton seeding combining only hits from adjacent layers, the complexity does not increase with additional layers and the processing time rises linearly.
Central \PbPb collisions are used as a reference, which leaves margin for smaller collision systems.
The current implementation of the ITS tracking does not yet include all features foreseen for the future, in particular it lacks the reconstruction of looping tracks with low transverse momentum.
Therefore, 20\% of the Run~3 TPC tracking tracking time, which includes this feature, is added to the ITS tracking baseline.
The fraction of 20\% is chosen to reflect the smaller number of hits in the ITS compared to the TPC. 
For the time being, all numbers for storage and compute requirements include a flat 20\% increase over the estimates for the tracking alone.
The computing, bandwidth, and memory requirements assume peak data rates since the online processing farm must be able to handle the peak loads, while the storage requirement assumes the average data rate.

The TPC GPU tracking was recently benchmarked on new fast GPU models~\cite{chep2021GPU}.
As the corresponding numbers on these fast GPU models are not yet available for the ITS tracking, the ITS estimate is scaled by the relative performance difference of the TPC tracking on the GPU model used for the ITS measurements and on the fastest GPU model available.

\begin{table}
    \centering
    \renewcommand{\arraystretch}{1.3}
    \begin{tabular}{@{}p{3cm} *{9}{S[table-format=3.1]}@{}}
        \toprule
        & {pp} & {O--O} & {Ar--Ar} & {Ca--Ca} & {Kr--Kr} & {In--In} & {Xe--Xe} & {Pb--Pb} & {Pb--Pb}\\
        & & & & & & & & & {Run 3}\\
        \midrule
        No. of GPUs
        & 940 & 850 & 690 & 630 & 530 & 500 & 450 & 360 & 2000\\
        Raw data rate \newline (GB/s) 
        & 360 & 330 & 270 & 240 & 200 & 200 & 170 & 140 & 3500 \\
        Compressed raw \newline data rate (GB/s) 
        & 65 & 60 & 50 & 40 & 35 & 35 & 30 & 25 & 95 \\
        AOD size \newline (PB/month)
        & 180 & 95 & 80 & 75 & 55 & 45 & 40 & 20 & 10 \\
        \bottomrule
    \end{tabular}
    \caption{Extrapolated data processing requirements (computing and storage) for different collision systems.}
    \label{tab:computing}
\end{table}

Table~\ref{tab:computing} gives an overview of the estimated number of required GPUs in the computing farm and on the estimated data rates.
These projections show that a compute farm of comparable size to the EPN~farm installed for Run~3 datataking, equipped with hardware available in the 2030s, will be adequate to cover the processing needs of the experiment.

\subsubsection{Data rates and throughput}
\label{sec:systems:data:rates}

The handling of the estimated data rates and the throughput will be easily possible.
In particular, the raw data rates are of the same order of magnitude as those during Run~3.
\begin{revised}
Table~\ref{tab:det_ro} shows the estimated data rates and the number of high-speed links required for the data transfer from the detectors.
\end{revised}
Thus, a network similar to that for Run~3 will be able to handle the flow of raw data and of compressed raw data.
Data of multiple events will be processed in buckets, in a similar way to the timeframes in Run~3, such that the increased interaction rate will not complicate the event building. 
The AOD data rate will increase compared to Run~3 but will constitute a small fraction of the total rate and will not pose a challenge for the network.
The required average PCI Express transfer speed to and from the GPUs is assumed to be in the order of 1$~{\rm GB/s}$, which would increase with faster GPUS.
The current PCI Express standard already provides sufficient margin.

\begin{table}
  \begin{revised}
  \renewcommand{\arraystretch}{1.5}
  \centering
  \begin{tabular}{@{}lrrr@{}}
  \toprule
       Part & Data rate (Tbit/s) & No. of links & No. of PCIe card \\
       \midrule
       Vertex Detetor & 0.7 & 72 & 9 \\
       Outer Tracker & 4.8 & 532 & 61 \\
       TOF & 3.8 & 384 & 48 \\
       RICH & 5.3 & 630 & 66 \\
       ECal & & 1750 & 62 \\
       MID & & 72 & 4 \\
       FCT & 0.8 & 110 & 10 \\
       \midrule
       Total & 15.4 & 3550 & 260 \\ 
       \bottomrule
  \end{tabular}
  \caption{Data rates and no. of links for detectors.}
  \label{tab:det_ro}
\end{revised}
\end{table}

\subsubsection{Storage and memory}
\label{sec:systems:data:storage}

The memory requirement will actually decrease compared to Run~3, during which long time frames of about 500~collisions have to be processed in parallel because of the drift time in the TPC. 
Thus, the Run~3 processing farm would already have both sufficient GPU and host memory.

\begin{revised}
As a consequence of the larger data samples, the data volume of the AODs will grow significantly compared to Run~3.
For comparison, in Run 3 we foresee \SI{\sim 10}{PB} for a month of Pb--Pb running, and \SI{4.7}{PB} for a month of pp running at \SI{500}{kHz}. 
By online triggering the pp data volume shall be suppressed by two orders of magnitude.
Based on the numbers of Table~\ref{tab:computing}, it is deemed possible to store all data for the one month of ion data taking per year, while an event selection or skimming strategy will be required for \pp data.
While the exact skimming strategy still has to be worked out, a reduction of the \pp{} data volume by at least 2~orders of magnitude can be certainly expected, based on studies for Run~3 demonstrating factors of 4~orders of magnitude when storing heavy-flavour candidates.
\end{revised}

The storage sizes for the disk buffer and for AODs are calculated using the expected average data rates, assuming the AOD Run~3 format, which has been optimized.
The compressed raw data size assumes entropy encoding (ANS) with a similar efficiency to that for Run~3~\cite{Lettrich:2021uet}.
In addition, it is assumed that the online reconstruction can remove noise by rejecting hits not attached to tracks.
The raw data sizes are estimated using the final Run~3 ITS raw data format as it arrives in the computing farm, including all protocol overheads, which are not yet fully optimized and leave much room for improvements.

\ifcost
\subsubsection{Cost estimates}
\label{sec:systems:data:cost}

\comment{needs to be written consistently with current funding schemes \dots}

\fi

\clearpage
\section{Planning and cost}
\label{sec:planning}

The start of data taking with ALICE~3 is planned for LHC Run 5, currently scheduled to start in 2035, see Fig.~\ref{fig:alice3_timeline}.
With the goal to produce Technical Design Reports by 2026/27, R\&D programmes will be set up for the coming years, with the main activities and spending expected for 2026/27.
As discussed in the detector sections, synergies with other experiments are being explored and expected to materialise in the coming years, e.g. with LHCb and EIC.
The detector construction and pre-commissioning is foreseen for 2028--2031, which leaves one year of contingency before Long Shutdown~4.
The dismantling of the present ALICE detector followed by the installation of the ALICE~3 detector will require a shutdown of two years, in line with the current LHC schedule.

\begin{figure}
    \centering
    \includegraphics[width=.9\textwidth]{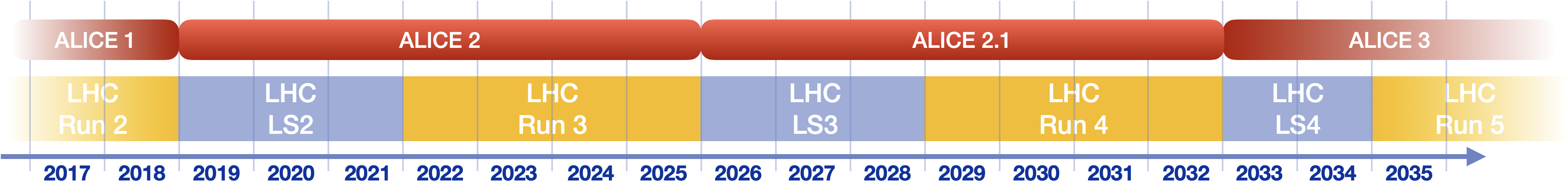}
    \caption[ALICE timeline]{Timeline towards ALICE 3}
    \label{fig:alice3_timeline}
\end{figure}

Table~\ref{tab:plan:cost_overview} shows an overview of the estimated core cost,
\begin{revised}
    which does not include R\&D activities or labour.
\end{revised}
The cost of the detector systems comprises all components including sensors, mechanics, integration, readout electronics, power distribution, cooling, and services.
The individual contributions are calculated based on the area to be instrumented with sensors, the number of data links to provide the readout bandwidth, and the required number of power connections. 
\begin{revised}
In addition to the production of the actual detector, we expect an additional cost of about 10\% for spares, which leads to about \SI{10}{MCHF} for all detector systems.
\end{revised}
For all detectors, we list the technology choice considered for the cost estimates (see detector sections for more details on the options).
For the time-of-flight detector, the baseline option are monolithic sensors with an integrated gain layer, for which an extensive R\&D programme will be required over the coming years to meet the timing requirements, see Sec.~\ref{sec:systems:tof:mapsrnd}.
Falling back to a hybrid concept as already established today, i.e. the usage of LGADs and readout chips, leads to a cost increase by about \SI{11.6}{MCHF}.
In case of the RICH detector, we aim at using monolithic SiPMs with integrated readout, which will also require a dedicated development effort. 
While analogue SiPMs are easily available commercially, their usage requires separate readout electronics and increase the system cost by about \SI{13.1}{MCHF}.

The cost of the magnet system depends on the magnetic field configuration.
The performance studies are based on the momentum resolution with a combination of a solenoid and two dipoles, see Sec.~\ref{sec:performance:detector}. 
A longer solenoid alone ($L \approx \SI{8}{\metre}$) leads to a deterioration of the momentum resolution only for $\abs{\eta} > 3$.
This is considered to have a moderate impact on the physics performance, e.g. the measurement of \DDbar{} correlations was verified to be only mildly affected.
Therefore, the combination of a longer solenoid and a smaller magnet for the Forward Conversion Tracker is presented as baseline option with an estimated cost of about \SI{25}{MCHF}.
Further studies are ongoing to evaluate the physics gain from maintaining the momentum resolution up to $\abs{\eta} = 4$.
The combination of a solenoid and two dipoles required to achieve this would increase the cost to about \SI{40}{MCHF}.

As discussed in Sec.~\ref{sec:systems:data}, the computing and processing requirements can be met by computing infrastructure very similar to the one installed for Run~3.
Given the continued evolution, a good part of the cost will be covered from the M\&O budget.

\begin{table}
    \centering
    \renewcommand{\arraystretch}{1.3}
    \begin{tabular}{@{}lp{6.5cm}S[table-format=3.1]@{}}
        \toprule
        {System} & {Technology} & {Cost (MCHF)} \\
        \midrule
        Tracker & MAPS & 30.5 \\
        TOF & Monolithic LGADs & 14.8 \\
            & \textcolor{gray}{Hybrid LGADs\footnotemark} & \textcolor{gray}{26.4}\\
        RICH & Aerogel and monolithic SiPMs & 20.9 \\
             & \textcolor{gray}{Aerogel, analogue SiPMs + readout\footnotemark[\value{footnote}]} & \textcolor{gray}{34.0}\\
        ECal & Pb-scintillator + PbWO$_4$ & 17.0 \\
        MID & Steel absorber, scintillator bars, SiPMs & 7.0 \\
        FCT & MAPS (solenoid + separate magnet) & 5.3 \\
        & \textcolor{gray}{MAPS (solenoid + dipoles)} & \textcolor{gray}{2.3} \\
        Magnet system & Superconducting solenoid + FCT magnet & 25.0\\
        & \textcolor{gray}{Superconducting solenoid and dipoles} & \textcolor{gray}{40.0}\\
        Computing & Data acquisition and processing & 6.0\\
        Common items & Beampipe, infrastructure, engineering & 15.0 \\
        \midrule
        Total & & 141.5 \\
        \bottomrule
    \end{tabular}
    \caption[ALICE 3 core cost]{Estimates of ALICE~3 core cost.}
    \label{tab:plan:cost_overview}
\end{table}
\footnotetext{Very conservative estimate based on current sensor pricing.}

\fi
\ifdraft
\else
\clearpage
\bibliography{bibliography}
\bibliographystyle{utphys}
\ifdraft
\else
\clearpage
\appendix
\end{document}